\documentclass[a4paper,twoside]{book}

\usepackage{graphics}
\usepackage{textcomp}
\usepackage{times}
\usepackage[german,english]{babel}

\usepackage[text={16.9cm,24.8cm}]{geometry}

\title{A Modular and Fault-Tolerant Data Transport Framework \\ Version $ $Id: dr-arbeit.tex,v 1.249 2003/12/11 21:19:15 timm Exp $ $}
\author{Timm M. Steinbeck}

\begin{document}
\pagestyle{empty}
\begin{titlepage}
{ \fontsize{11}{13}
\begin{center}
{\large Dissertation\\ 
submitted to the\\
Joint Faculties for  Natural Sciences and  Mathematics\\
of the Ruperto Carola University of\\
Heidelberg, Germany,\\
for the degree of \\
Doctor of Natural Sciences\\}
\end{center}
\vspace{1cm}\vspace*{1.5cm}
\begin{center}
{\huge A Modular \\
\vspace{0.5em}
 and Fault-Tolerant  \\
\vspace{0.5em}
 Data Transport \\
\vspace{0.5em}
 Framework \\} 

\end{center}
\vspace{5cm}
\begin{large}
\begin{center}
 presented by\\
\end{center}
  \begin{tabbing}
 \hspace{4cm} Diplom--Physiker: \=Timm Morten Steinbeck\\
 \hspace{4cm} born in: \>Aachen\\
  \end{tabbing}
\begin{center}
%Heidelberg,   August 8, 2003
Heidelberg,   \underline{ \hspace*{3.8cm}}
\end{center}
\vspace{1em}

  \begin{tabbing}
    \hspace{4cm}Referees: \=Prof. Dr. Volker Lindenstruth\\
                           \>Prof. Dr. Peter Bastian\\
  \end{tabbing}
\end{large}
}
\end{titlepage}
%%% Local Variables: 
%%% mode: latex
%%% TeX-master: "dr"
%%% End: 

\clearpage{\pagestyle{empty}\cleardoublepage}
%\begin{titlepage}
%\parindent{0em}

\vspace{5cm}
\begin{large}
\mbox{}\\
{\selectlanguage{german}
\begin{center}
{\bf  Ein modulares und fehlertolerantes Daten-Transport Software-Ger\"ust}
\end{center}
%Das High Level Trigger (HLT) System des zuk\"unftigen Schwerionen-Experiments ALICE am Large Hadron Collider
%des CERN mu\ss{} Daten mit einer Rate von bis 25~GB/s empfangen und handhaben k\"onnen. Diese
%Datenraten sollen auf h\"ochstens 1.25~GB/s reduziert werden, bevor die Daten gespeichert werden.
%Das System das f\"ur den HLT geplant ist um diese Datenraten zu verarbeiten ist ein gro\ss{}er PC Cluster.
%Er soll aus bis zu mehreren tausend Knoten bestehen, die \"uber ein schnelles Netzwerk untereinander
%verbunden sind. 
Das High Level Trigger (HLT) System des zuk\"unftigen Schwerionen-Experiments ALICE muss seine Eingangsdatenrate
von bis zu 25~GB/s zur Ausgabe auf h\"ochstens 1.25~GB/s reduzieren bevor die Daten gespeichert werden.
Zur Handhabung dieser Datenraten ist ein gro\ss{}er PC Cluster geplant, der 
bis zu mehreren tausend Knoten skalieren soll, die \"uber ein schnelles Netzwerk verbunden sind. 
F\"ur die Software, die auf diesem System eingesetzt werden soll, wurde ein flexibles Software-Ger\"ust
zum Transport der Daten entwickelt, das in dieser Arbeit beschrieben wird. Es besteht aus einer Reihe 
separater Komponenten, die \"uber eine gemeinsame Schnittstelle verbunden werden k\"onnen. Auf diese Weise 
k\"onnen verschiedene Konfigurationen f\"ur das System einfach erstellt werden, die sogar zur Laufzeit 
ge\"andert werden k\"onnen. 
Um ein fehlertolerantes Arbeiten des HLT Systems zu gew\"ahrleisten, enth\"alt die Software
einen einfachen Reparatur-Mechanismus, der es erlaubt ganze Knoten nach einem Fehler zu ersetzen.
Dieser Mechanismus wird in Zukunft unter Ausnutzung der dynamischen Rekonfigurierbarkeit des
Systems weiter ausgebaut werden. Zur Verbindung der einzelnen Knoten
wird eine Kommunikationsklassenbibliothek benutzt, die von den spezifischen Netzwerkeigenschaften, wie
Hardware und Protokoll, abstrahiert. Sie erlaubt es, dass eine Entscheidung f\"ur eine bestimmte Technologie
erst zu einem sp\"ateren Zeitpunkt getroffen werden muss. Die Bibliothek enth\"alt bereits funktionierende Prototypen
f\"ur das TCP-Protokoll sowie SCI Netzwerkkarten. Erweiterungen 
k\"onnen hinzugef\"ugt werden, ohne dass andere Teile des Systems ge\"andert werden m\"ussen. 
Mit dem Software-Ger\"ust wurden ausf\"uhrliche Tests und Messungen durchgef\"uhrt. Ihre Ergebnisse sowie
aus ihnen gezogene Schlussfolgerungen werden ebenfalls in dieser Arbeit vorgestellt. Messungen zeigen 
f\"ur das System sehr vielversprechende Ergebnisse, die deutlich machen, dass es beim Transport
von Daten eine ausreichende Leistung erreicht, um die durch ALICE gestellten
Anforderungen zu erf\"ullen. 
}

\mbox{}\vspace{1cm}\\

\begin{center}
{\bf A Modular and Fault-Tolerant Data Transport Framework}
\end{center}
%In the future ALICE heavy ion experiment at CERN's Large Hadron Collider
%input data rates of up to 25~GB/s have to be handled by the
%High Level Trigger (HLT) system. It has to scale the rate of the data down to at most 1.25~GB/s
%before it is written to permanent storage. 
%The HLT system that is being designed to cope with these data rates 
%consists of a large PC cluster, up to the order of a 1000 nodes,
%connected by a fast network. 
The High Level Trigger (HLT) of the future ALICE heavy-ion experiment has to reduce its input data rate 
of up to 25~GB/s to at most 1.25~GB/s for output before the data is written to permanent storage. 
To cope with these data rates a large PC cluster system is being designed to scale
to several 1000 nodes, connected by a fast network.
For the software that will run on these nodes a flexible data transport
and distribution software framework, described in this thesis, has been developed. 
The framework 
consists of a set of separate components, that can be connected via a common
interface. This allows to construct different configurations
for the HLT, that are even changeable at runtime. To ensure 
a fault-tolerant operation of the HLT, the framework 
includes a basic fail-over mechanism that allows to replace whole nodes after a failure.
The mechanism will be further expanded 
in the future, utilizing the runtime reconnection feature of the 
framework's component interface. To connect 
cluster nodes a communication class library is used that abstracts from
the actual network technology and protocol used to retain flexibility in the hardware choice. 
It contains already two working prototype versions for the TCP protocol as well as SCI network adapters. 
Extensions can be added to the library without modifications to
other parts of the framework. 
Extensive tests and measurements have been performed with the framework. Their results as well
as conclusions drawn from them are also presented in this thesis. 
Performance tests show very 
promising results for the system, indicating that it can fulfill
ALICE's requirements concerning the data transport.

\end{large}
%\end{titlepage}

\clearpage{\pagestyle{empty}\cleardoublepage}
%\maketitle

%\begin{abstract}
%
%\end{abstract}

\pagestyle{headings}

\tableofcontents
\vfill\eject
\listoffigures
\vfill\eject
\listoftables
\vfill\eject

%%%%%%%%%%%%%%%%%%%%%%%%%%%%%%%%%%%%%%%%%%%%%%%%%%%%%%%%%%%%%%%%%%%%%%%%%%%%%%%%%%%%%%%%%%%%%%%%%%%%%%%%%%%%%%%%%%%%%%%%%%%%%
%%%%%%%%%%%%%%%%%%%%%%%%%%%%%%%%%%%%%%%%%%%%%%%%%%%%%%%%%%%%%%%%%%%%%%%%%%%%%%%%%%%%%%%%%%%%%%%%%%%%%%%%%%%%%%%%%%%%%%%%%%%%%

\chapter{Introduction}

In high-energy and heavy-ion physics, as in many other scientific and academic
applications, compute clusters made up of standard PCs using the Linux operating system 
have emerged as one of the predominant type of computer systems for data analysis
and other tasks requiring large amounts of processing capabilities. 
The primary reason for this is their very good
price vs. performance ratio, owing to the usage of widely available and cheap mass market 
components. In newest developments, such as the experiments for the future Large
Hadron Collider (LHC) at CERN, large clusters will not only be used for offline data 
analysis but also for online data processing and acquisition. In these types of systems
large amounts of data of up to tens of gigabytes per second will be transported through clusters in a data flow fashion, passing 
through several stages in the processing chain. Due to the flexibility afforded by the building-block like
construction of such systems from basically identical components and taking into
account the insecurity in the predictions for the future development of that market,
a similar flexibility in the software architecture and configuration of these
systems is highly desirable. A further prime requirement for these systems is that the transport of the
data in a system, both in each node as well as from one node to another, has to be as efficient as possible.
The necessity for this requirement arises from the fact that the purpose of these systems is 
the processing of data from the experiments, for which massive amounts of CPU power are needed.
Any CPU cycles used just for transport of the data, without producing any analysis results,
increase the total number of CPUs and consequently also PCs in the system needed for the analysis, causing a higher
cost. Some overhead for the transport of data is unavoidable but it should obviously be kept
to a minimum. Next to the flexibility and efficiency, the reliablity of such a system is a natural
third primary requirement. Since the single PCs as elements of a cluster do not possess the reliability
necessary for such a system, mainly due to their low cost, a system as a whole must be tolerant
with regard to the fault of at least a number of its elements. Also, measures must be taken 
to ensure that the system either has no parts whose failure disables the whole system, called 
single points of failure, or that these points consist of especially reliable and thus more expensive
components. 

This thesis describes a framework that has been developed to be used in the type of online
data processing systems described above. It has been designed to consist of a number of independent
software components that communicate via a specified interface. They can thus be plugged together 
as needed to form a data processing chain conforming to the requirements and boundary conditions
presented through other characterics of the system, either from detector properties or from the
hardware configuration. During the framework's design and development the focus has been on 
an architecture and implementation  to minimize the processing overhead from 
the communication of the components and the transfer of the data for a minimum impact on the
processing capability of a system as a whole, as described above. Utilizing the dynamic reconfiguration
ability inherent in the pluggable component concept together with a number of specialized components, 
the framework can support setups able to tolerate faults in its software components or
hardware parts of nodes as well as even the failures of complete cluster nodes. 

The following Chapter~\ref{Chap:Requirements} provides an overview of computing technology background, 
%that should be 
helpful in understanding 
%some of the 
design decisions made for the framework. Also contained in this chapter
are a number of sample applications for which the framework can or will be used.
Chapter~\ref{Chap:Overview} details some of the higher level design decisions and choices made
for the framework and gives an architectural overview of it. In the following chapters
classes contained in modules of the framework are presented in more detail. Chapter~\ref{Chap:UtilityClasses} 
presents utility classes providing  basic functionality for the framework. The next
chapter 
%\ref{Chap:ComClasses} 
describes classes for communication between the nodes in a cluster. These 
communication classes are based on an abstract interface with implementations available for
two different networking technologies and they are used to connect framework components on
different nodes. Main parts of the framework, consisting of the interface between the 
components and a number of components and templates, are detailed in Chapter~\ref{Chap:PubSubInterface}
and~\ref{Chap:Components} respectively.
The following chapter~\ref{Chap:BenchmarksTests} presents benchmarks and system tests of the framework
and some of its constituent parts, while the final chapter~\ref{Chap:Conclusion} contains the conclusions from the
development and tests as well as an outlook for future development possibilities.
%Appendix~\ref{Chap:Installation} contains installation instructions for the modules making up
%the framework and appendix~\ref{Chap:ComponentUsage}  usage instructions for the framework components. 
Additional information for the benchmarks from chapter~\ref{Chap:BenchmarksTests} is contained 
in appendix~\ref{Chap:BenchmarkStuff}. Tables with results presented in chapter~\ref{Chap:BenchmarksTests}
are located in appendix~\ref{Chap:BenchmarkTables}. 
Descriptions of a number of developed components which became obsolete later can be found in appendix~\ref{Chap:ObsoleteComponents}.
A glossar of frequently used abbreviations can be found in appendix~\ref{Chap:Glossar}.

%I would like to thank my supervisor Volker Lindenstruth for giving me the opportunity to 
%{\bf do} this thesis and many valuable suggestions and informations. Markus Schulz I thank for
%many informational and fruitful discussions and Arne Wiebalck for much help and discussions. I 
%also would like to thank the rest of the group in Heidelberg for making up a group that
%was much fun to work in. {\bf ... }

I would like to say many thanks to several people without whom this thesis would not
exist as it does:
\begin{itemize}
\item My supervisor Volker Lindenstruth for giving me the opportunity to 
work on this thesis and the HLT project as well as many valuable suggestions, 
information, and advices. 
\item My second referee Peter Bastian for his willingness to 
read and assess my dissertation and take part in my exam.
\item Markus Schulz for
many informational and fruitful discussions, the good cooperation, as well as his assistance and advice.
\item Arne Wiebalck and Heinz Tilsner for much help, discussions, the good cooperation during our common time
at the institute, and in particular for proof-reading parts of the thesis.
%\item ??? Christoph Scholl for reading the thesis even during stressfull lecture times.
\item The whole group at the Chair for Computer Science in Heidelberg for the comfortable 
atmosphere and cooperation in the group. It was and is a fun time.
\item My parents in general for supporting me during my time of studies and for allowing me to pursue this career,
and in particular my father for proof-reading this complete work.
\item But most particular and deeply I want to thank my wife Heike for supporting and helping me during the
whole time of work on the thesis, especially during the last few months of writing. Without her love, endurance, and
support I would not have been able to do this.
\end{itemize}

\clearpage

%%%%%%%%%%%%%%%%%%%%%%%%%%%%%%%%%%%%%%%%%%%%%%%%%%%%%%%%%%%%%%%%%%%%%%%%%%%%%%%%%%%%%%%%%%%%%%%%%%%%%%%%%%%%%%%%%%%%%%%%%%%%%
%%%%%%%%%%%%%%%%%%%%%%%%%%%%%%%%%%%%%%%%%%%%%%%%%%%%%%%%%%%%%%%%%%%%%%%%%%%%%%%%%%%%%%%%%%%%%%%%%%%%%%%%%%%%%%%%%%%%%%%%%%%%%

\chapter{\label{Chap:Requirements}Background}
\section{\label{Sec:ComputingBackground}Computing Background}

\subsubsection{Overview}
Due to the continuously increasing use of clusters made up of commodity PC hardware in high-energy and nuclear physics,
for offline as well as for online purposes, the characteristics of this architecture play an increasingly
important role in computer architecture for that field. 
In the following section an overview 
of the computer technology and architecture in the PC cluster
area is given to detail the characteristics and pecularities that influence the design
of cluster systems as well as the development of software to be run on them.
Emphasis is given to clusters used for scientific tasks, particularly in the use as online data
analysis farms for the readout and triggering of high-energy physics experiments, the focus of the software framework
described in this thesis. Components of such a system come
predominantly from the PC mass market due to the good price-performance ratio present there. On the other hand this
necessity for low prices often induces compromises in technology compared to custom solutions, which have to be
taken into account when designing a cluster system's hardware and software. 

\subsubsection{Introduction to Cluster Technology}

Data analysis and other scientific applications have used and relied on computers for a long time and the amount of
processing power needed has been rising steadily. 
Stimulated
by recent increases in available computer
speed the applications have become more sophisticated, raising in turn the demands 
for processing power required by those applications. Many of these scientific problems are too large to be 
handled efficiently by one single processor and thus parallel computers are needed to run these
problems efficiently. Prices for most commercially available parallel computers, most of which 
fall into the high performance computing (HPC) category, are typically rather high for academic budgets. Many 
institutions have therefore turned to assembling comparatively low cost networks or clusters of workstations 
\cite{IEEENOW}, \cite{COW1} (NOWs or COWs)  or 
clusters of PCs, frequently called Beowulfs \cite{BeowulfPaper95}, running Linux \cite{Linux}, \cite{kernel.org}, \cite{kernel.org.mirrors} 
or another of the free Unix flavors. 
These clusters mostly consist of a number of computers made up of commodity-off-the-shelf
(COTS) components, and are typically connected either via Fast or GigaBit Ethernet 
\cite{EthernetPaper1},\cite{EthernetPatent},\cite{EthernetBluebook},\cite{Ethernet1982},\cite{IEEE802},\cite{IEEE802.3} or via a dedicated System-Area-Network (SAN), 
%??\cite{}, 
like the Scalable Coherent Interface (SCI) \cite{SCIIEEE}, 
Myrinet from Myricom \cite{MyricomWeb}, or the future ATOLL \cite{ATOLLWeb}, \cite{ATOLLPaper1}, \cite{ATOLLPaper2}. 
These SAN technologies typically have one or more of the
following characteristics compared to lower cost technology such as Ethernet: lower communication latencies,
higher bandwidth, and smaller processing overhead. The last point can be of particular importance, as 
more CPU time is available for doing actual processing instead of being used to transfer data. 

\subsubsection{CPU and Memory Development}

The above mentioned COTS components have a very competitive and advantageous
price to performance ratio due to their mass market nature. In addition they also allow to take direct advantage of the 
quickly developing increases in absolute performance in this market. The increases seen here closely follow Moore's law \cite{MooresPaper},
which in its original form states that the density of circuits on chips will increase by a factor of 2 every
year. Derived forms state that the same behavior, with different factors, applies not only to the density but also
to the performance of the chips. The most visible aspects for mass market PCs are the increase in CPU clock frequency, which by now roughly
doubles every 18 months, as well as the increase in memory size. 

The usage of mass market components however has some disadvantages as well. 
While processor
performance and memory size closely follow Moore's law and thus increase by 60~\% every year,
the memory access time only increases by 2~\% per year.
Special purpose high-performance computing hardware can implement
more elaborate measures to work around this problem than commodity hardware, as the latter is typically optimized for a low cost. 
%As a result the gap between raw processor performance and the speed of accessing data in memory is growing every year \cite{PattersonMemGap}. 
This causes the gap between raw processor performance and the speed of accessing data in memory to widen every year \cite{PattersonMemGap}. 
As a result it is increasingly costly when a processor cannot access data
in its cache but has to load it from memory. The processor has to perform several wait cycles, during which it cannot
perform any processing. The cost of memory access is most
obvious in applications that access large memory blocks in irregular patterns, which mitigates the utility
of caches. 
%While processor and memory
%performance as well as memory size closely follow Moore's law and thus increase by a certain factor every year,
%the factor by which processor performance increases is larger than the one for memory performance.
%Special purpose high-performance computing hardware can implement
%more elaborate measures to work around this problem than commodity hardware, as the latter is typically optimized for a low cost. 
%As a result the gap between raw processor performance and the speed of accessing data in
%memory is growing in size every year \cite{PattersonMemGap}. As a result it is increasingly costly when a processor cannot access data
%in its cache but has to load it from memory. The processor has to perform several wait cycles, during which it cannot
%perform any processing while it waits for the data to be loaded from memory. These costs of memory accesses are most
%obvious in applications that access large memory blocks in irregular patterns, which negates the use
%of caches. 
%{\bf Explanation???} 
%If most memory accesses are spaced further apart than the size of a cache line, 
%then a new cache line will have to be loaded for practically every access resulting in an execution speed limited by the 
%actual memory bandwidth and latency. 

\subsubsection{Busses and Networks}

A similar situation arises concerning extension busses and network interfaces. The bandwidth and 
latency of 
these parts also have been unable to keep up with the advances in CPU speed. In the bus area the Peripheral Component 
Interconnect (PCI) bus \cite{PCISIG}, \cite{PCISpec} with 64~bit and 66~MHz has only become established in the high end 
PC server/workstation market.  PCI-X with up to 133~MHz is just starting to appear there. 
The bandwidth of the 64~bit/66~MHz version of this bus reaches a theoretical peak performance of 528~MB/s, with the slower
64~bit/33~MHz and 32~bit/33~MHz versions reaching 264~MB/s and 132~MB/s respectively. Even though the 64~bit/66~MHz bus has a factor of 
4 advantage over the 32~bit/33~MHz version still dominant in the home PC segment, its peak performance is still at least a factor of 4 
lower compared to contemporary memory interfaces. 
The current predominant network technology in the PC market is 100~Mb/s Fast Ethernet with Gigabit Ethernet establishing itself
especially for servers. In cluster systems SANs are used for interconnects as well, sometimes coupled with proprietary network protocols. 
But as these technologies can be several orders of magnitude
more expensive compared to Ethernet, especially the 100~Mb/s variety, many clusters are constructed using the more cost-effective interface
choice. The communication protocol used on these Ethernet adapters is practically always the standard Internet TCP/IP protocol suite \cite{RFC793}.
This network protocol/interface combination has the advantage of being widely available, cheap, and reliable. However, 
neither of its parts was designed for the task of a cluster interconnect or SAN. Next to the obvious disadvantages of relatively
low bandwidth and high latency, this combination is not the optimal solution for this task because of another drawback. The TCP/IP protocol
consists of a protocol stack with several layers of protocols inside the operating system kernel. Data sent from a user 
application first has to be copied from the user level memory into the privileged kernel space (or system) memory. It then passes
through the protocol layers where the data is often copied from layer to layer. These copy stages have to be done by the computer's CPU,
preventing it from executing actual processing tasks. Additionally the CPU needs to access memory twice (read and write) to copy the data.
These memory accesses first slow the CPU down as it now has to wait for the memory while copying and second they take up much
of the already precious memory bandwidth in the system. For systems sending large amounts of data this can have a quite detrimental
effect on other applications running at the same time. These influences are due to several factors: the memory bandwidth
being used by copying processes, the filling up of cache space with the copied data, and the pure CPU usage itself. 
But even on better TCP/IP implementations where the data is not copied between the layers, the first copy stage
is practically always present, and the protocol stages with their required processing have to be passed as well. So in addition 
to being a comparatively slow network, both as far as latency and bandwidth are concerned, coupled with the most widely used protocol
Ethernet  also uses up more precious system resources than other technologies for connecting clusters.
A rule of thumb is that for every Megabyte of data transferred per second with TCP/IP over Ethernet depending on block size at least 
%one percent 
1~\%
CPU usage is incurred.

\subsubsection{Commodity-Off-The-Shelf and High-Performance-Computing Hardware}

As the COTS market relies on interoperability of its components, especially in the area of
memory and extension busses with their respective add-on cards, the technological advances in this area and the market acceptance are comparably 
slow. HPC system vendors on the other hand, have no such compatibility constraints and are free to use tailored and 
tuned interfaces and components in their systems. These special purpose components give them a performance advantage compared
to the cheaper clusters and earn them the classification of high performance computers.
Despite these technological and economical differences a lot of recent HPC and cluster-type systems
share a principal similarity. Both are composed of relatively cheap and standardized building blocks connected by 
a network. But whereas for clusters the nodes are single PCs, sometimes even dual CPU PCs, 
HPC systems are often composed of Symmetric Multi Processor (SMP) systems, sometimes with more than 100 CPUs. 
Similiarly, where most clusters are connected via Fast or Gigabit Ethernet and some with specialized SANs, many HPC systems
feature specially developed interconnects between the nodes with bandwidths comparable to the internal busses in PCs.

Comparing the price/performance ratio of typical clusters and HPC systems for a single CPU, clusters
rank much better than their more expensive counter-part. For easily parallelized problems which
feature a high ratio of computation on each node to communication between the nodes, a cluster offers much
better overall performance for the same price or a comparable performance for a much lower price. Most problems in
high-energy and nuclear physics are of that type and are thus well suited for clusters. 

\subsubsection{\label{Sec:ClusterSoftware}Cluster Software}

On the software side the widespread use of clusters has been primarily made possible by the rise in popularity
and support of the Linux (or GNU/Linux) \cite{Linux} operating system. This clone of the Unix operating system, freely available in source code, has begun its life
on PC systems and is now available for a wide range of hardware. Due to its popularity a wide spectrum of PC hardware extensions,
e.g. network adapters or graphics cards, is supported with drivers. Also due to its widespread use coupled with
the source code availability many people have been able to search for errors in it. As a result
bugs are usually found quickly and Linux thus has a reputation as a very stable and reliable system. Since the scientific 
and especially the academic area has for a long time been involved with Unix and has been using it widely, Linux has enjoyed an
especially quick acceptance in this area. Coupled with other 
software freely available in source code, and thus adaptable, it has established itself as a well suited
operating system companion for cost-effective clusters made up of PC components. Recently other free Unix like
systems, e.g. FreeBSD \cite{FreeBSD} or OpenBSD \cite{OpenBSD}, some of them actually older than Linux, have also gained popularity, but Linux was the starting point
for cheap Unix like cluster systems and is still the most widespread operating system there. 

Motivated by the increase in cluster usage a number of software packages have been developed to ease
the administration of a cluster. Most of these packages follow the principle of allowing the administration of 
the cluster as a single system and not as a collection of systems. The most extreme of these systems like 
Mosix \cite{MosixWeb}, \cite{MosixPaper1}, \cite{MosixPaper2} and its derivative openMosix \cite{OpenMosixWeb} treat the whole cluster as a single system by allowing process
migration over a network between the nodes for a cluster-wide load-balancing. Other systems support the 
creation of batch-queue systems for separate jobs, to be dispatched to available cluster nodes for processing, or
monitor the cluster nodes from a central location. Examples of such packages are the open source Compaq \cite{CompaqWeb} (now 
Hewlett-Packard (HP) \cite{HPWeb}) Single System Image (SSI)
Clusters (SSIC) package \cite{SSICWeb}, the Load Sharing Facility (LSF) \cite{LSFWeb} from Platform Computing \cite{PlatformWeb}, 
and the Condor package \cite{CondorWeb}, \cite{CondorPaper1}, \cite{CondorPaper2}. 

A number of packages also exist for communication inside a cluster. The most well-known of these are the two Message Passing Interface (MPI) 
\cite{MPIWeb}, \cite{MPIForum}, \cite{MPIStandard}, \cite{MPI2Standard}
implementations MPICH \cite{MPICHWeb} and LAM/MPI \cite{LAMMPIWeb} and the Parallel Virtual Machine (PVM) \cite{PVMWeb}. These three packages are designed for parallel applications
that run distributed on multiple PCs and frequently exchange data with each other. Data exchanges are primarily done between iterative calculations,
processed data is sent to other processes and received data is used as the basis for new calculations. These packages therefore are
typically not optimized for an efficient communication but rather one with a low latency. Fault tolerance also is not one of the prime foci of these
packages, as calculations can easily be restarted with the same input data. 
%Due to these differing requirements these packages have not been considered to be used for communication 

\subsubsection{Interfaces to Readout Hardware}

Another important segment for computers in sciences is the readout of data from experiment setups. For a long
time computers in this area have been equipped with the busses used for the connection of instruments, e.g. VMEbus 
\cite{IEEEVME}, \cite{VITAWeb}, \cite{VMEIntro} or 
CAMAC \cite{IEEECAMAC}, \cite{CAMACIntro}. These computers are typically based on CPUs used in the desktop market, e.g. Intel x86 or PowerPC, and often
run real-time operating system like VxWorks \cite{VxWorksWeb} or LynxOS \cite{LynxOSWeb}. Both hardware and software are very specialized and as a result have 
a small market, making them relatively expensive compared to common desktop hardware and operating systems. 
For this reason, a trend similar to the cluster tendency for parallel computing has set in to replace these special 
systems with standard PCs as well. Where instrument connections are needed interface cards for PC busses (mostly PCI)
provide the necessary connections to other equipment. In other cases special hardware is being developed
to interface experiment equipment with read-out computers with PCI or another of the PC system standard busses on the computer side. 
Despite its age and comparably low performance the old Industry Standard Architecure (ISA) bus introduced
with the first IBM PCs still enjoys some popularity here, especially in industry applications. But for new developments in the scientific
area PCI is now very often used, enabling the use of COTS systems for readout as well as for calculation. 

As these special hardware readout devices have to be accessed and are in general not supported by operating systems due to their
custom nature, specific software for them has to be provided as well. Development of device drivers for them is often too complicated and
due to the frequently required rapid development not feasible as well. Therefore mostly normal programs are used that require some special
features or privileges to gain direct access to the readout hardware. To faciliate the development of these programs, packages or
drivers exist that provide generic and easy access to any hardware in a system.
With this principle, development of a driver is required only once. It can be utilized in user-space programs afterwards.

%logp

\section{Applications for a Data Transport Framework}

%Possible fields where the developed framework can be used are 

\subsection{High-Energy and Heavy-Ion Physics Experiment Trigger Systems}

A major area of application for a data flow framework are readout and especially trigger systems for experiments in high-energy and heavy-ion physics,
as these are inherently of a data driven nature. Data arrives in chunks, the detector's events, which have to be processed. Often one event
arrives in multiple parts, which have to be assembled before, after, or as a part of processing. Depending on the exact nature of the experiment 
the analysis might have to be executed in a number of steps. Each step requires the data from previous ones, mostly the directly 
preceeding. Data is thus passed or flows from station to station, possibly being merged with data from other stations, until the desired
result is obtained or it is written to permanent storage. Due to the high rates required most often in the lower
levels of such trigger systems, a relatively generic software framework is not very well suited there. Instead more specialized
software or even hardware is required for these levels. 

Concerning the upper level trigger systems very high data rate requirements are currently found in the new generation of (relativistic) heavy-ion 
physics experiments. With their high occupancies, the resulting large event sizes coupled with still considerable event rates of at least
several hundred Hertz, and consequently very high data rates, they present one of the biggest challenges in online data processing for PC 
clusters. Among these one of the most advanced is the ALICE experiment planned for the 
heavy-ion running mode of the future Large Hadron Collider (LHC) at CERN in Geneva. Two other projects, not progressed as far as ALICE,
are the future Compressed-Baryonic-Matter (CBM) and Proton-ANtiproton-at-DArmstadt (PANDA) projects at the Gesellschaft f\"ur 
Schwerionenforschung (GSI) in Darmstadt.

\subsection{The ALICE Detector and the ALICE High Level Trigger}

\subsubsection{The ALICE Detector and the Quark-Gluon-Plasma}

\begin{figure}[hbt]
\begin{center}
\resizebox*{0.90\columnwidth}{!}{
\includegraphics{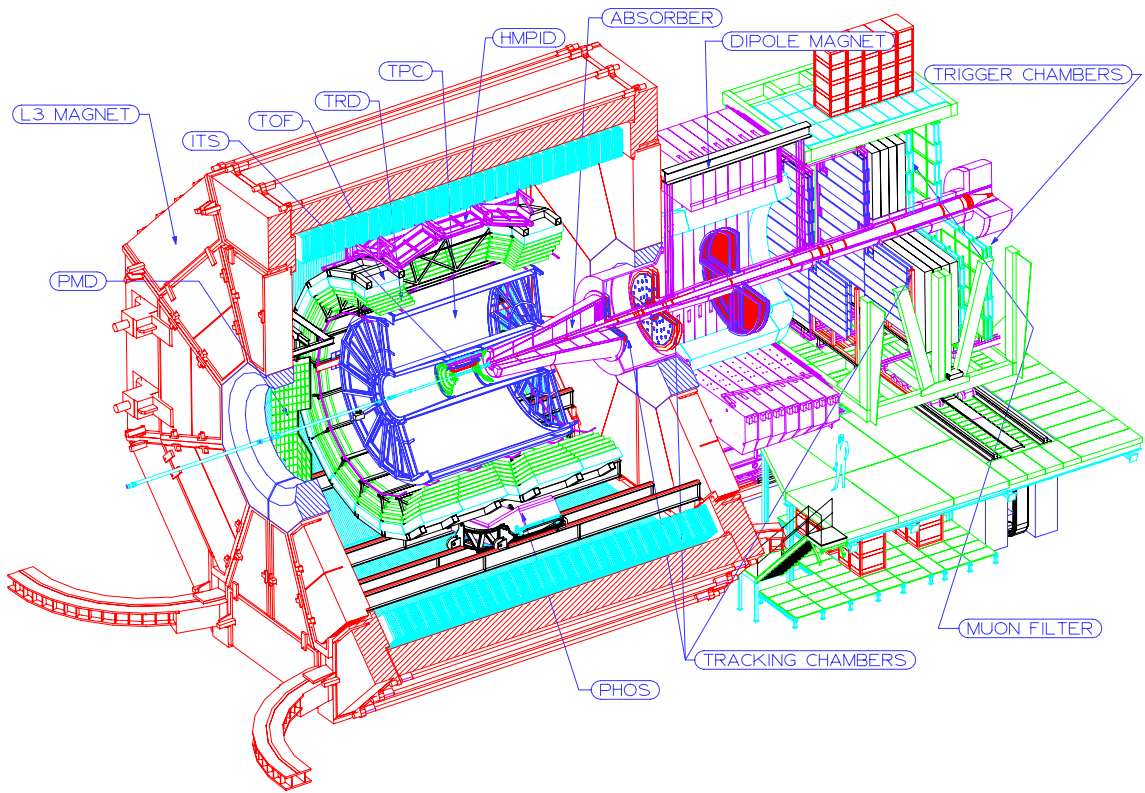}
}
\parbox{0.90\columnwidth}{
\caption[The ALICE experiment.]{\label{Fig:ALICEDetector1}The ALICE experiment. }%, the TRD detector is not shown in this picture.
}
\end{center}
\end{figure}

The primary application for which the framework has been developed is the ALICE experiment's \cite{ALICEWeb}, \cite{ALICEWebPhys}, 
\cite{ALICEGraybook}, \cite{ALICETP}
last trigger stage, the High Level Trigger (HLT). ALICE, shown in Fig.~\ref{Fig:ALICEDetector1},  is a detector for relativistic heavy-ion collisions currently 
being developed and built for the future 
Large Hadron Collider (LHC) \cite{LHCWeb} at CERN \cite{CERNWeb}. The LHC will be operated in two modes: proton-proton (pp) mode and heavy-ion (HI) mode.
In the primary proton-proton mode LHC will collide bunches of protons every 25~ns corresponding to a
collision rate of 40~MHz. In heavy-ion mode bunches of lead or calcium nuclei will collide every 100~ns or at
10~MHz respectively. ALICE will operate both in pp and HI mode although the detector is primarily designed
for heavy-ions, where its main purposes are the search for and investigation of quark-gluon-plasma (QGP). QGP is a new state of matter in which 
quarks and gluons can move freely in a volume and are not subject to the usual confinement
% \cite{} 
for strongly interacting
particles. Overviews of QGP can be found in \cite{QGPOverview1} to \cite{QGPOverview6}. The temperature or energy density in the corresponding 
volume has to be high enough to enable this state 
to be established. The minimum required energy density is predicted to be around $1~\mathrm{GeV}/\mathrm{fm}^3$, roughly 7 times the
density of normal nuclear matter. Comparable energy densities have existed during the 
first few microseconds after the Big Bang. It is expected that by colliding highly energetic lead nucleons it will be possible
to reproduce conditions which feature high enough energy densities to allow the formation of quark-gluon-plasma.
When the highly energetic fireball produced by the collision expands, it simultaneously cools down and the energy density decreases
again. The quarks and gluons in the plasma then have to recombine again to hadrons undergoing the normal color confinement.
From these hadrons and additionally produced leptons observed in the detector, one has to extract information about the 
system that existed during the collision. The number of particles produced in these collisions is very large. For ALICE
of the order of $10^4$ particles are expected for a collision in the covered pseudo-rapidity range of $\mid \eta \mid \le 0.9$.
The minimum bias collision rate at the LHC heavy-ion design luminosity of $10^{27}~\mathrm{cm}^{-2}\mathrm s^{-1}$ is 
%expected to be of the order of 
several 10~kHz. Combining event rate and particle multiplicity leads to a very high amount of data produced.
Together with the preceeding trigger stages the HLT's task is to reduce that data volume to a rate more manageable for
mass storage and also to make the most efficient use of the available output bandwidth by storing only the most interesting events
and compressing each event's data to reduce its size.

\subsubsection{ALICE's Subdetectors}
The detector consists of a number of sub-detectors, most of them arranged in a layered shell structure around the beam pipe 
covering a solid angle of almost $4 \pi$ and a pseudorapidity range of  $\mid \eta \mid \le 0.9$. 
Closest to the beam pipe is the Inner Tracking System (ITS) \cite{ALICETP-ITS}, whose primary purpose is the detection
and reconstruction of the primary and secondary vertices and track-finding for charged particles with a low transversal momentum that 
do not 
enter
the Time Projection Chamber (TPC) (see below). In addition it will also be used to improve the momentum resolution of particles
with high momentum as well as for the reconstruction of low energy particles and their $dE/dx$ identification.
% and other tasks.
Six cylindrical layers of detectors make up the ITS with a high enough space point resolution to cope with the
expected high particle densities. The layers will be placed at radii from 3.9~cm to 45~cm and will extend
from the interaction point (IP) where the collisions occur in both directions along the beam pipe. They extend from 12.25~cm for the innermost layer to 50.4~cm 
for the outermost one. For the two innermost layers, where the particle tracks are most dense, silicon pixel detectors (SPDs) have been chosen
as they provide the best possible granularity and resolution at these small radii. Furthermore, they can be operated at high rates
and will be used to determine an event's vertex together with the muon spectrometer. The two middle layers have a lower
track density due to their larger distance from the IP. Silicon drift detectors (SDDs) are used here as they are cheaper than SPDs and
have the ability to provide additional $dE/dx$ information for particle identification. In the two outermost layers silicon strip detectors (SSDs)
are sufficient to satisfy the less severe resolution requirements and the comparatively large area here makes it desirable to
use this proven, reliable, and especially cheaper technology. 

Outside of the ITS is the Time Projection Chamber (TPC) \cite{ALICETP-TPC}, \cite{ALICE-TPC-TDR}, \cite{PrivCommAVestbo}, \cite{PrivCommUFfrankenfeld}, both ALICE's physically largest sub-detector as well as the 
one producing the largest amount of data. It is the primary detector used for track finding and momentum measurements as well 
as for identification of particles by their specific energy loss $dE/dx$. The TPC is a cylinder around the beam pipe measuring 5~m in length with a 
central high voltage plane. Its inner and outer radii are 88~cm and 2.5~m respectively. At the endcaps are readout chambers consisting 
of Multi Wire Proportional Chambers (MWPCs) to amplify and read out the signals of the particle tracks. Most of the TPC's characteristics are results
of the expected high particle multiplicity and the resulting problems of distinguishing separate particle tracks.
Its  inner radius has been chosen so that the expected particle density on the inner surface is around
$0.1 - 0.2~\mathrm{particles}/\mathrm{cm}^2$. For the outer radius the criterion was to obtain a length of tracks inside the
TPC that will allow a $dE/dx$ measurement with a precision of 6 - 7~\%. The length finally is determined by ALICE's design
coverage of  $\mid \eta \mid \approx 0.9$ with a drifttime chosen to be $88~\mu \mathrm s$. Due to the high 
particle multiplicity a very fine granularity of $3 \times 10^8$ pixels has been chosen to achieve a good two track separation 
ability. This fine granularity is the reason why the data volume produced by the TPC is the largest part in ALICE. 
Expected event sizes for the TPC, already zero-suppressed and runlength encoded, are about 60 - 70 MB for central HI events. 
The readout chambers are arranged in 159 pad-rows in each slice, with pad sizes of $4~\mathrm{mm}\times 7.5~\mathrm{mm}$,  
$6~\mathrm{mm}\times 10~\mathrm{mm}$, and $6~\mathrm{mm}\times 15~\mathrm{mm}$. 

Directly adjacent around the TPC and primarily designed to complement its electron identification capability is
the Transition Radiation Detector (TRD) \cite{ALICE-TRD-Prop}, \cite{ALICE-TRD-TDR}. %, depicted in Fig.~\ref{Fig:ALICETRD}. 
Its primary purpose is to provide electron identification capabilities
for the central barrel region for momenta beyond $1~\mathrm{GeV}/c$. 
%In this range the $dE/dx$ capabilities of the TPC are no longer sufficient to provide a good enough rejection of pion background particles. 
Additionally, the TRD should
enable a thorough research of the dilepton continuum found in the central barrel region. As the TRD is a fast
detector and especially a fast tracker it also contributes an effective triggering capability for particles,
particularly electrons, with a high transverse momentum ($p_t$). Another trigger type of the TRD is the
selection of hadronic jets with a high transversal energy. With the tracking information being available a few microseconds
after each event, it becomes possible to select events with high $p_t$ particles and activate the TPC's gating grid
for readout only for those events. To optimally cooperate with the TPC the TRD has been designed to have the same acceptance
of $\mid \eta \mid \le 0.9$.
% although only half of the detector's financing has yet to be approved. 
It consists of 
six layers of chambers between radii of 2.9~m and 3.7~m. Each layer is divided into five segments along the beam axis and
18 segments around the detectors circumference. The total number of chambers is thus 540 with each chamber consisting of 
a combination of foil stacks to produce transition photons, Xenon filled MWPCs 
to detect them, and front end electronics for readout. A chamber uses between 12 and 16 pad rows in the direction of
the beam axis and each pad row consists of about 144 pads read out. The total number of channels in 
the TRD is about $1.16 \times 10^6$, 
making the TRD the second largest producer of data among the ALICE detectors. 
%another large contribution to the amount of data that has to be read out from ALICE. 

%\begin{figure}[hbt]
%\begin{center}
%\resizebox*{0.60\columnwidth}{!}{
%\includegraphics{trd_pic-1.ps}
%}
%\parbox{0.90\columnwidth}{
%\caption{\label{Fig:ALICETRD}ALICE cross sections including the TRD Detector.}
%}
%\end{center}
%\end{figure}

Outside of the TRD is a large Time-Of-Flight (TOF) array dedicated to provide particle identification
information for particles of average momentum \cite{ALICETP-PID}. It is designed to have a large acceptance
and covers a barrel area of about $100~\mathrm m^2$. Momentum coverage for hadrons is between 
about $0.5~\mathrm{GeV}/c$ and $2~\mathrm{GeV}/c$. $0.5~\mathrm{GeV}/c$ is the upper limit for which the TPC is still able to separate
Kaons and Pions based on its $dE/dx$ information and $2~\mathrm{GeV}/c$ is the limit for sufficient particle statistics
in single event analysis. The overall timing resolution of the system is designed to  be around 100~ps, which would allow $3\sigma$
separation of Kaons and Pions up to $2.1~\mathrm{GeV}/c$ momentum. For electrons the momentum range to be covered
is between $140~\mathrm{MeV}/c$ and $200~\mathrm{MeV}/c$, where $dE/dx$ information is not sufficient to distinguish electrons
and pions. In this application the timing resolution is of a lesser significance. 
As the overall inefficiency of the TOF is to be below 20~\%, the occupancy of the detector is required to be
less than 10~\% at the highest particle multiplicities expected, resulting in more than $10^5$ channels being used in the TOF.
%This restriction on the occupancy
%in turn results in more than $10^5$ channels being used in the TOF.
%, another contribution to the large event size. 
The baseline technology choice for the TOF are Pestov spark counters, which have a number of advantageous
features. Foremost among these is their very good time resolution reaching up to 25~ps. In addition they feature a 
lifetime corresponding to a running time of more than 20~years as well as an intrinsic efficiency of more than 96~\% and do not require 
preamplifiers due to their high signal output. Major drawbacks, however, are a lack of experience with systems
on a similarly large scale and a time comparable to the projected operation of ALICE. Therefore a fallback solution of Parallel
Plate Counters (PPCs) is intended, an established technology that would have a lower resolution
but would still fulfill the requirements from the physics goals.

Complementing the identification capabilities for particles outside of the momentum range covered
by the TOF is the High Momentum Particle Identification (HMPID) detector \cite{ALICETP-PID}, a Ring Image \v{C}erenkov (or Cherenkov) Counter 
(RICH). Its area is small compared to the TOF, only $10~\mathrm m^2$ and it is placed at the top of the 
detector at a radius of 4.7~m where the particle density is low. As the particle density of $50~\mathrm{particles}/\mathrm m^2$
and event rate of around 10~kHz expected to be encountered are nonetheless still high for a detector of this type,
a fast-RICH layout 
%\cite{} 
implementation is used. 
%The conversion gap for the \v{C}erenkov photons therefore has to be kept as small as possible, for which the 
%{\bf fast-RICH ??} layout can be used. 
Another advantage of this technology is the ability to operate at much higher
rates than the ones intended for heavy-ion operations at the LHC so that it can be used in pp-mode as well.
The drawback of this techology is a cathode segmented into many small pads which requires a pixel like readout 
with highly integrated electronics and a large number of readout channels.

Located opposite of the HMPID on the bottom side of the detector is the Photon Spectrometer (PHOS) \cite{ALICETP-PHOS} whose
primary purpose is the search for direct photons produced during heavy-ion collisions, which reflect
to a high degree the initial conditions found in these collisions. As a secondary task the PHOS also has to
measure the production of the neutral mesons $\pi^0$ and $\eta$, for which the momentum resolution at $25~\mathrm{GeV}$
is approximately an order of magnitude better than what the tracking detectors can achieve for charged particles.
Unfortunately, there is a large background of photons being produced in decays of hadrons with a ratio of direct 
to decay photons of approximately 5~\%. Due to the required detector sensitivity of about 5~\%
it becomes necessary to measure the rates and transversal momentum spectra of photons as well as $\pi^0$ and $\eta$ mesons
in the same detector. The resulting very high multiplicity in turn necessitates a fine segmentation of the
calorimeter, a large distance from the vertex, and a material with a small Moli\`{e}re radius to reduce the transversal 
extension of the produced showers. 
To achieve the intended
particle occupancy of less than 3~\% in heavy-ion collisions at a radius of 4.6~m $\mathrm{PbWO}_4$ has been chosen as
the material for the PHOS because of its Moli\`ere radius of about 2~cm. 
In order to prevent charged hadrons from producing unwanted signals in the PHOS, a veto-detector will have to be placed
in front of the PHOS to reject them. Similarly a system of applying time-of-flight cuts is considered to suppress
the signals from neutral hadrons other than $\pi^0$ and $\eta$. 

%{\bf TP 5.2.5, charged-hadron-veto; TP 11.7 neutral hadron rejection}

One of the most promising signatures for the production of the quark-gluon-plasma 
is the suppression of the heavy quarkonium resonances, whose decays can be 
well detected by muon-pair production. To distinguish the signature for QGP from other processes that 
could also cause this suppresion, it becomes necessary to measure the relative suppression of the
different states as well as the  ratios to unsuppressed reference processes such as the inclusive
heavy quark production. In addition the suppression has to be measured as a function of the transversal
momentum spectrum down to its low regions. The search for the produced muon pairs will be done
only in the forward direction of ALICE, along the beam pipe, and outside of the normal barrel
construction for a number of reasons. Because of the shielding
required for the background photons and hadrons only muons with a momentum of at least $4~\mathrm{GeV}/c$ can be
detected and the muons in forward direction will have a higher momentum due to Lorentz-boosting. 
The pion and kaon background is also reduced in that direction due to the higher momenta required to
penetrate the absorbers and the generally lower particle multiplicity per rapidity unit. 

The first part of the dimuon arm \cite{ALICETP-DiMu}, \cite{ALICE-DiMu-Add}, \cite{ALICE-DiMu-TDR}, \cite{ALICE-DiMu-TDR-Add} 
after the detector barrel %  designed for muon pair detection
are the absorbers already described. They are needed to reduce the particle background consisting especially
of photons and hadrons coming from the vertex to acceptable levels, resulting in the momentum cut for muons
at $4~\mathrm{GeV}/c$. But even with the absorbers in place the main problem for the dimuon arm is still the
particle multiplicity in each event rather than the event rate. To cope with this problem the tracking system
after the absorbers uses a high granularity so that the maximum 
expected occupancy is around 5~\%. The tracking system is made up of five stations consisting of two chamber 
planes per station. Each chamber plate in turn has two cathode planes read out to provide two dimensional track information.
Intermixed with the tracking stations is a large dipole magnet, with two stations each in front and behind it
and one inside the magnet. After the fifth tracking station is another passive muon filter wall in front of 
four planes of Resistive Plate Chambers (RPCs) as muon identification and trigger detectors. The RPCs are arranged
in two stations, each one providing $x$ and $y$ coordinates. Coordinate differences of the two chambers
are used to determine the muons' transversal momentum for the trigger decision. The whole muon arm covers an angle range
from 2\textdegree{}  to 9\textdegree{}, a compromise between detector acceptance and cost. It is shielded from the beam pipe by an
inner shield for protection from particles produced at large rapidities.

One of the pieces of information needed to determine the collision types that occured is the impact
parameter of a collision, a measure for the distance between the centers of colliding nuclei.
The observable allowing the best conclusion to the actual impact parameter is the energy carried
away by spectator nucleons from the beam that did not take part in the collision. Spectator neutrons and
protons have a different charge to mass ratio compared to the ions in the beam. Therefore they will be separated
from the beam by the same LHC dipole magnet that also separates the two colliding beams after the interaction point. 
Detection of the spectator nucleons will be done in two Zero Degree Calorimeters (ZDCs) \cite{ALICETP-Forward} on each side of
the interaction point at a distance of 92~m. Each of the two neutron calorimeters (ZN) will be placed between two beam pipes 
and will have to fit in the free space between them. The ZN transversal dimensions are therefore restricted 
 by this free space and have been set at $8~\mathrm{cm} \times 8~\mathrm{cm}$. Their depth of $100~\mathrm{cm}$
corresponds to 10 interaction lengths $\lambda_{int}$ of the chosen shower
material, which must have a high density to place enough absorption capability in the restricted space. 
By contrast the two proton calorimeters (ZP) are placed to the side of one of the beam pipes and do not suffer from
such tight space restrictions. Less denser and cheaper materials can be used in them and the dimensions selected
here are $16~\mathrm{cm} \times 16~\mathrm{cm} \times 150~\mathrm{cm}$, the chosen depth also corresponding to
$10~\lambda_{int}$. Quartz fibres have been chosen as the active material both for ZN and ZP in which
shower particles produce \v{C}erenkov light read out by photo-multipliers. The primary reason why quartz will
be used instead of conventional scintillators is its radiation hardness. In addition, it also
provides a good energy resolution even with small calorimeters, like the ZN, and is insensitive to
radioactive background whose particles do not produce \v{C}erenkov light. Nonetheless to avoid unnecessary radiation
exposure that might damage them, the ZDCs will be removed from the beam pipe during proton-proton mode, when their operation
is not required

Another detector covering the forward direction outside of the barrel is the Forward Multiplicity Detector (FMD) \cite{ALICETP-Forward},
measuring the $d \mathrm N /d\eta$ distribution of particles per unit of pseudo-rapidity outside the central acceptance region
of the barrel detectors. Additionally, it will provide information for the Level 0 (L0) and Level 1 (L1) triggers (see below) after an event as early as possible. 
Currently two possible choices exist for the FMD, either Micro Channel Plate (MCP) or Silicon multipad  detectors.
To determine the multiplicity in a pp event using the MCP one would simply read out and count the number of digital hits, while
in heavy-ion mode it would be necessary to sum up the analogue information of the charge collected on each of the anode pads.
MCPs have the advantage of very good timing properties with a resolution of about 50~ps. With this resolution MCPs could provide multiple types 
of information for the TOF and trigger systems, like event time $T_0$, a first z-coordinate measurement of the event vertex, identification of
beam-gas reactions, as well as a measure of pile-up protection for slower detectors, like the TPC. 
By contrast, the Si multi-pad detectors feature a time resolution of about 20~ns to 40~ns compared to less than 100~ps required.
Advantageous for the Si detectors is the fact that they are well suited to multiplicity measurements in heavy-ion collisions
and are often used for that purpose already. One possible approach therefore is the use of a combination of MCP and Si detectors. 
The FMD consists of seven disks arranged around the beam pipe at distances from the vertex from 42~cm to 225~cm. Inner radii of the 
disks range from 42~cm to 80~cm and outer ones from 105~cm to 175~cm. Together the disks cover the pseudo-rapidity ranges of 1.6 to 3.6 
on the side where the muon arm is located and 1.4 to 4.7 on the opposite side. Each of the disks will be divided into several pad segments
with the total number of pads on all seven disk being about 780. Due to the analogue summation for the MCPs and the general Si characteristics
it will not be necessary to have a high granularity to cope with the high multiplicity.

Contrary to most other detectors ALICE does not feature any large calorimeter detector in the central barrel region. Due to the large
radii that the tracking detectors need to handle the high particle multiplicities, any such calorimeter would have to cover
a very large surface and be very expensive as a result. Instead ALICE relies on multiplicity measurements of charged particles which
on average show a good correlation with the transversal energies in events. To provide some additional information about the 
transversal energies the Photon Multiplicity Detector (PMD) \cite{ALICETP-Forward} has been added to ALICE. The PMD is a preshower detector that distinguishes 
photons from charged particles, especially hadrons, and measures the energy depositions of transversing particles. 
While the energies deposited by individiual photons show large fluctuations, these are considerably reduced when the depositions
of a large number of particles are added for each event. In this way the PMD takes advantage of the high multiplicities that present
a problem for most of the other detectors. The forward region was chosen for the PMD as the photon energies are higher
in that direction, again due to the rapidity boost, and the area that needs to be covered is comparably small and manageable. 
It will be mounted on the door of the L3 magnet (see below) at a distance of 5.8~m from the vertex and covers the pseudo-rapiditiy range 
$1.8 \le \eta \le 2.6$. 

For the magnetic field ALICE will reuse the magnet \cite{ALICETP-Magnet} of the L3 detector \cite{L3Web}, \cite{L3Graybook}, \cite{L3-TP}. 
This detector will have been dismantled by the time 
the LHC will start to operate and ALICE will actually be assembled and operated in the same underground cavern as L3. 
The magnet's coil has an inner radius of 5390~mm while the yoke's outer radius reaches 7900~mm and its length
14100~mm and the magnet's total weight is about 7800~t. 168 separate octagonal turns make up the solenoidal coil, each with a
conducting section of $540~\mathrm{cm}^2$. 
At the intended magnetic field strength of 0.6~T the magnet will draw several Megawatt of electrical power.
After its operation in the L3 detector concerns exist about its continued operation in ALICE, primarily regarding the magnet's cooling system.
Investigations of the current state are in progress and a number of possible modifications for the cooling systems are already
under discussion to assure the magnet's continued functioning during ALICE's operation. 

%0.5 T, coil inner radius 5390mm, yoke outer radius 7900mm
%length of yoke: 14100mm
%electrical power (0.5T)
%total weight 7.8t
%coil: solenoid form w. 168 octagonal turns, each conducting section of 540cm^2.
%Modification of colling system

%{\bf \large Magnet }

\subsubsection{The ALICE Trigger System}

One major problem of high enery and nuclear physics experiments is that the different types of events as well as the respective
underlying physical processes occur statistically distributed. Processes with a higher probability therefore produce events more
often than rarer processes and are better researched and understood already. As a result rare events are of much more interest 
to current experiments like ALICE. To increase the number of events available for analysis, trigger systems are used to select
events indicated by certain signatures to belong to interesting rare processes. 
ALICE's trigger system \cite{ALICETP-Trigger} before the HLT is made up of three stages: Trigger Level 0 (L0), Level 1 (L1) and Level 2 (L2).
The system will be active both in pp as well as in HI mode. 
%The first two stages of the system, L0 and L1, have a fixed latency of 1.2~$\mu \mathrm s$ and 6~$\mu \mathrm s$ respectively while 
The primary objective of ALICE is the study of the hot dense matter
created during central collisions of heavy-ions. Correspondingly the main emphasis of the trigger system's design 
has been on these collisions of ions with relatively small impact parameters. The impact parameter
can be determined to a given extent from the particle multiplicity in the FMD and the
energy deposited in the ZDC. As central collisions occur relatively frequently the trigger does not need to be
very selective and the most common event types can even be scaled down so that only a subset of them is processed.

Dimuon events with particles in the mass range to be observed on the other hand occur very rarely and are therefore processed with a high priority.
The first trigger for these events
% occurs about 600~ns after an interaction and 
signals that two particle tracks above a specific 
momentum threshold have reached the dimuon RPC trigger chambers. For these events it is desirable to read out the full data
to perform correlations with other observables. 
%Due to possible pile-up of overlapping events in other sub-detectors and especially
%the TPC it might be impossible to read out the full data correctly. For this reason events with a dimuon trigger and fully usable TPC 
%data are treated with a very high priority during  readout. 
In case of pile-up for a dimuon
event it will be tried to read a reduced subset of detectors with usable event data. Readout time for such a
subset will be about $200~\mu \mathrm s$ compared to about 2~ms for a full set of detectors.

The purpose of the first trigger stage L0 is to signal an event as soon as possible after occurance using
exclusively data from the FMD. The latency of this signal is fixed to $1.2~\mu \mathrm s$. To verify that an event occured three items are checked:
\begin{itemize}
\item Whether the interaction point reported is close to the nominal collision point.
\item Whether the forward/backward distribution of particles in both directions of the beam pipe is consistent with a 
beam collision.
\item Whether the multiplicity reported from the FMD is above a given threshold.
\end{itemize}
To prevent non-central events with a muon pair being discarded at this stage the L0's requirements on centrality are not very strict. 
%and thus the output rate of the trigger is only reduced to between less than $10^4~\mathrm{Hz}$ and several $10^4~\mathrm{Hz}$. 
A positive L0
signal is used by some of the detectors, amongst them the RICH and the PHOS, to strobe their front end electronics for readout. 

The L1 trigger's decision latency time is fixed to $6.5~\mu \mathrm s$ \cite{ALICE-TRD-TDR} after an event has taken place. 
Its decision is based on FMD and ZDC information about the centrality of the event, TRD data about high momentum electrons
and information from the dimuon system. 
Since more complex correlations have to be examined for the dimuon system its data is only available at this stage. 
%Due to the more complicated correlations necessary the dimuon data is only available at this stage. 
Any event for which the dimuon trigger reports two particles with a high transversal momentum is then
classified as a priority event. 
%{\bf TRD??}
An L1 accept signal is issued in that case to all detectors, in order to
activate the readout of all detectors' front end electronics. 
% by the Data Acquisition System (DAQ) to activate the readout from all detectors' front end electronics. 
For the TPC the gating grid is activated, its maximum gating rate is 200~Hz for HI and 1~kHz for pp mode.
% on this signal which imposes an upper bound on its frequency of at most 1~kHz. 

After the gating has been activated, the drifttime of $88~\mu \mathrm s$ has to pass
before the TPC can be read out. During this time further processing of data already available from fast detectors
can be performed in the L2 trigger stage. 
Among the possibilties for processing are a mass cut on the dimuon system or fast analysis of data from the FMD.
%but this has not been finalized yet. 
Unlike the first two stages L2 does not have a fixed latency but an upper bound as defined
by the drifttime given above. Due to the variable latency the decisions cannot be synchronous anymore.
%Therefore a trigger processor pool will be available from which one processor will be used
%for each L2 trigger algorithm to be executed. 
%The algorithms will not have the full event data available as it will be read out by the DAQ, but each 
%detector participating in the L2 decision will have to provide a special stream of trigger
%data to be used. 
After a positive L2 decision the data from all detectors' Front End Electronics (FEE) will be read out.

Another part of ALICE's trigger is a past-future protection system keeping track of 
pile-ups of overlapping events in each sub-detector. Readout of data is then restricted
by the system to a subset of detectors with non-piled-up data. Using the output
of the past-future protection unit, an identifier describing the class of event that occured is generated. 
This is distributed to each of the  sub-detectors 
which then decide what to do with their data for this event. 
%which can then decide based upon it what it should to with its data for this event. 

\subsubsection{\label{Sec:ALICEDAQ}The Data Acquisition System in ALICE}

As the trigger system the Data Acquisition system (DAQ) of ALICE, DATE, \cite{ALICETP-DAQ}, \cite{DATEWeb}, \cite{DATECHEP95} 
will have to run both in proton-proton as well as in heavy-ion mode. 
As the data rate in pp-mode will be only one fifth of what is expected for HI the main requirements on the DAQ system
are given by heavy-ion operation, although this will only be active for a few weeks every year.
In general the DAQ system will have to cope with two types of events in this mode. The first 
 will contribute most of the total data stream and consists of central events at a relatively low rate
but with a large event size. In contrast, the second type of events contains a muon pair that has been reported by the trigger
and is read out with a reduced detector subset, including the dimuon arm. 
Events of this type can occur at a rate of up to 1~kHz. 
These two requirements of a large data stream and a rather high event rate will both have to be 
handled by the DAQ system. In summary the system will have to cope with an aggregate data stream to the permanent data
storage (PDS) of 1.25~GB/s.

% coming from several data sources and being distributed to several data destinations.

%On average these are 40~MB large events occuring with
%a frequency of about 50~Hz, but  size and/or rate have to be decreased by processing in the HLT
%before the data is passed to DAQ. 
%The second type of events contains a muon pair that has been reported by the trigger and for which
%only the dimuon arm, the two inner pixel planes of the ITS for vertex calculation, and the trigger detectors are read out.
%This type's typical event size is only up to approximately 250~kB but these events can occur at a rate of up to 1~kHz. 

The Architecture of the system is based on PCs connected by TCP over Gigabit Ethernet. 
Local Data Concentrator nodes (LDCs) are connected to the detector Front End Electronics (FEE) via optical links,
the Detector Data Links (DDL). Each DDL link ends in a PCI Read-Out Receiver Card (RORC) located
inside an LDC. One or more RORCs can be placed into each LDC and data from each
sub-detector may be read out over several DDLs so that data from one event can be scattered over multiple
RORCs and LDCs. Subevent data read out from the detectors in the LDCs is sent to Global Data Concentrator nodes (GDCs)
 for global event building. A GDC destination for a particular event is determined by the Event Destination Manager
(EDM) which communicates this decision to the LDCs. Fully assembled events are shipped to permanent
storage for archiving and later offline analysis.

\subsubsection{\label{Sec:ALICEHLT}The ALICE High Level Trigger}

Together with the preceeding trigger stages the task of the High Level Trigger (HLT) \cite{HLTCDR}, \cite{HLTWeb}
is to maximize the physics output that can be attained
by ALICE with the specified bandwidth to tape. To achieve this goal two approaches of online filtering and analysis are possible, which may also be
used in combination. The first approach is the customary selection of the most interesting events as
described already for the other trigger stages. In the HLT this selection will be performed
by an online analysis of events to determine the amount and type of particles that passed through 
the detector. Among the detectors whose data will be analysed at this time are the ITS, TPC, TRD, and
dimuon arm. 
The second possibility is to compress the events so that a greater number of them can
be written to tape. Best compression results are achieved by this approach if the compression
method used is adapted to the underlying data. This is similar to the approach taken for MP3 \cite{FIISWeb}, \cite{FIIS-MP3-Web} 
or Ogg Vorbis \cite{OggVorbisWeb} audio files,
where the results achieved when sound is compressed adapted to human hearing
characteristics are much better than the results from general purpose compression algorithms.
For the TPC data for example the underlying data model consists of tracks of charged particles passing
through the detector and being curved by the magnetic field in the detector. A very good compression
ratio should thus be possible to achieve if online tracking is performed on an event and only
the parameters of the found tracks are stored. For a better offline analysis capability one would 
also store the space point coordinates of clusters of deposited charges in the TPC's gas volume 
in addition to the tracks. To minimize the amount of additional data, the space point coordinates would be 
stored as distances from their associated tracks and the charges deposited would also be stored as differences
from calculated averages for the corresponding particle.

For both of the presented data reduction approaches it
is necessary to perform online tracking and charge clustering of the data read out. The two different
methods could then be combined by storing the compressed/analysed data of the most interesting events. 
To properly analyse all the events it becomes necessary that the High Level Trigger has access
to the complete data from each event or at least to the complete data from the sub-detectors whose
data is needed for the HLT decision. It is the first subsystem where
all this data can be available fully, allowing a global view of an event if desired.

% Really irrelevant???
%For the operation of the HLT three modes are planned. In the first mode the HLT will not be active
%at all. After an L2 Accept decision all data will be read out by the DAQ's LDCs and will be 
%processed by the DAQ for permanent storage. In the second mode the HLT will be active and will perform
%analysis of the events to decide which ones to read out and store. Data to be stored in this case
%will be the raw data that has been read out from the detectors directly and will not be manipulated
%by the HLT. The third mode will use the full potential of the HLT where also all events will be
%analysed to decide which ones to read out and additionally it is up to the HLTs software to decide which
%data to read out for a given event. This could be the raw detector data as in mode 2 or 
%some processed data as was detailed in the preceeding paragraphs. It might also be possible that only a certain
%part of the detector contains any interesting data, in which case only the corresponding part of the
%event would be read out at all. 

To perform the necessary processing for online analysis of all events' data, 
the HLT is planned to consist of a large PC cluster farm with a number of nodes
of the order of several hundred up to about a thousand. The connection between the nodes has to be 
made by a high performance network, possibly with a network topology adapted to the necessary flow of data
through the system. Candidates for the networking technology to be used are not yet fixed,
but for the required bandwidth at least Gigabit Ethernet (GbE) or a System Area Network (SAN) dedicated to 
communication between systems in a cluster is necessary. Possible choices for this may be the
ATOLL \cite{ATOLLWeb}, \cite{ATOLLPaper1}, \cite{ATOLLPaper2} networking technology currently being developed at the University of Mannheim, the 
shared memory (ShM) interconnect Scalable Coherent Interface (SCI) \cite{SCIIEEE} from Dolphin Systems \cite{DolphinWeb} or Myrinet \cite{MyricomWeb}.
All of these technologies are available as high performance PCI cards. 
If GbE is used then it is very likely that a protocol other than the default
TCP/IP is used to avoid the problems related to its use for high performance applications
described in section~\ref{Sec:ComputingBackground}.

Similar to the DAQ's LDCs a number of HLT nodes will be equipped with RORCs in which the DDLs coming
from the detector's front end electronics end. DDLs are connected to the RORCs via mezzanine daughter cards that 
contain the interface to the optical link. These RORC equipped nodes, designated Front End Processors (FEPs), 
are the first place where ALICE data arrives in the HLT. As for the LDCs one or more RORCs may be placed in each
FEP and apart from the addition of the RORCs, the FEPs are in no
respect different from the other nodes in the cluster. 
The RORCs in the HLT FEP node themselves, however, may well differ from the ones used in the DAQ LDCs. 
%This is not the case for the RORCs in the HLT FEPs that may be different from the ones used in the DAQ LDCs. 
In addition to the DAQ baseline
funtionality the HLT RORCs will be equipped with additional co-processor functionality
to already perform (pre-) processing of data. The intention of this preprocessing performed by the RORCs
is to take load off the HLT nodes' CPUs by performing analysis steps well suited to
such hardware co-processing. For the processing the HLT-RORCs will be equipped with FPGAs on which different
analysis tasks, e.g. a Hough transformation \cite{HoughPat}, \cite{HoughPic} for cluster finding, can be loaded as necessary.

To detail an example of the amount of RORCs needed for the TPC, it will be divided into 36 sectors called slices. Each
of these slices is again subdivided into six patches. One DDL will thus be used to read out the data
from one patch. Due to the size of the data transferred from the TPC each RORC attached to a patch DDL 
will end in its own FEP. So for the TPC alone there will be 216 FEPs, each receiving the data
from one of its 216 patches.

%Since there is obviously some overlap between the DAQ's LDCs' and the HLT's FEPs' access to the RORCs connected
%to the front end electronics, a clear separation has to be made depending on the mode of operation as
%described above. In mode one the RORCs will be read out by the LDC software and only its baseline
%functionality will be used. In modes 2 and 3 the RORCs will be read out by the HLT FEP software making
%use of the full additional processing functionality as needed. To pass the raw data from the detectors as
%well as any data produced by the HLT in mode 3 to the DAQ software 
%an interface will have to be defined so that the readout data can be passed on to the DAQ for 
%permanent storage. This interface has not been decided upon at the time of writing.

Data that has been passed via DDL and RORC from the detector into an FEP's main memory may undergo
some further local processing on that node in addition to any processing done on the RORC itself. 
After this processing the data is shipped to a node of the next group, that consists of as many nodes as
are necessary to be able to perform the analysis
of the data in real time. The output data produced by each group of nodes is again shipped to the corresponding next group
of nodes for the next processing step up to the final stage. Each of the processing stages in the
system may also receive the output data from multiple groups of the next lower level, performing
some additional merging or only merging the groups' input data without any additional processing.
After the data has passed through the system and has been successively processed and merged in this 
way, a synopsys of the whole analysed event is received by the final stage. Using this fully analysed event it is 
able to make the trigger decision about the event, whether to read it out and depending on the mode of operation
also which parts to read out. This decision and the corresponding data is then
passed to the DAQ for readout and storage. The interface between the DAQ and the HLT will consist of
10 DDL links between a set of HLT event merger nodes and a number of DAQ LDCs. To 
the DAQ the HLT will therefore appear as another detector, simplifying the interface between
the two systems. In the HLT event merger nodes PCI DDL output cards have to be used.
These cards are functionally different from the RORCs but will use the same board type
with just another FPGA configuration and a different DDL daugher card.

The exact processing sequence with the distribution of data and workload has to be kept flexible.
%, which steps of the analysis are performed where, 
%has not been layed down. It has not yet been determined which part or parts of the analysis
%can already be performed on the HLT-RORC. The type of data arriving
%in the FEPs and correspondingly the exact analysis that has to be performed there depends on this decision. 
%These facts in turn influence the
%data and processing steps to be executed in the later stages on the HLT nodes. 
The design of the data flow and processing also determines much of the architecture and network topology used
for the system. It drives the requirements on the communication between each pair of nodes,
%which node must communicate with which other nodes 
which has a very strong effect on the networking topology that can be used. In the case of switching
networks the most easy and flexible approach would be to use a topology where every node can 
communicate simultaneously at full bandwidth with every other node. However, as this would require an enormous amount of unnecessary bandwidth in the 
switch(es) it would also make this topology probably the most expensive one. If each group 
of nodes only sends their data to a specific other group of nodes, then switches could be used that connect
only those respective groups of nodes. The switches required in this case would have a much smaller number of ports and 
require much less internal bandwidth and should be cheaper as a result. As the necessary hardware components, i.e. the PC nodes
and networking hardware, are basically of the commodity type, it is no problem to postpone the decision about the
workload distribution and network topology. Because of the continuously decreasing prices it is desirable to buy these
components as late as possible in any case. With the flexibility built into the framework presented in this thesis,
for this exact purpose, the software configuration and architecture can be specified at a late point in time, before the 
start of ALICE's operation.

\begin{figure}[hbt]
\begin{center}
\resizebox*{0.95\columnwidth}{!}{
\includegraphics{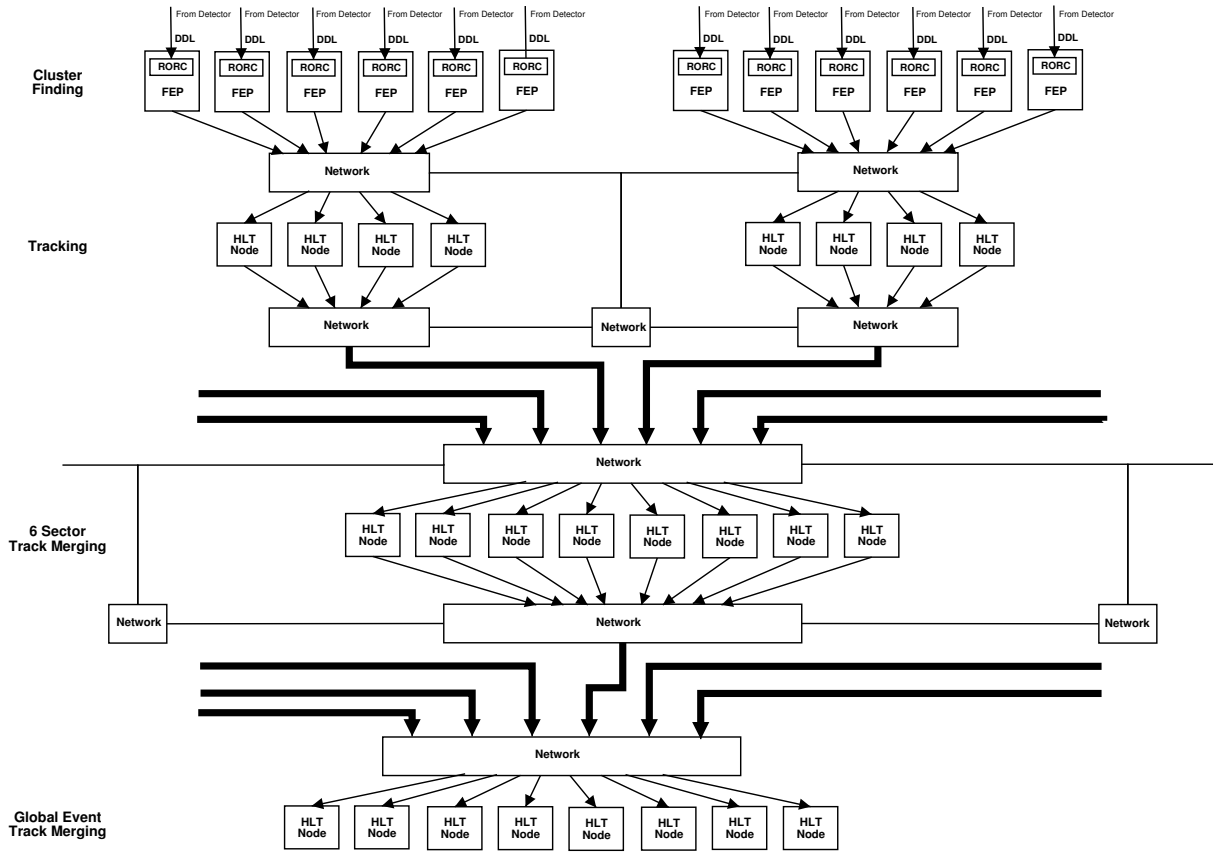}
}
\parbox{0.90\columnwidth}{
\caption[Sample architecture and topology of the HLT.]{\label{Fig:HLTArchitecture1}Sample architecture and topology of the HLT. The connections shown
signify data flow and do not specify network topology or architecture. The different network boxes
 shown need not be separate and a node might be equipped with one or more network interface cards. As shown, there will always be connectivity
between all nodes, although not necessarily with full cross-section bandwidth. The arrows depict the direction of the event data flow.}
}
\end{center}
\end{figure}

%{\bf \LARGE CDR figure 48 page 62}

A sample architecture based on the assumption that all the analysis will be executed in software is shown in
Fig.~\ref{Fig:HLTArchitecture1} for data from the TPC alone. The assumptions made for the amount 
of CPUs required for each step of processing are based on one hand on interpolations \cite{PrivComMSSTARAna} from data of the 
Level 3 trigger at the 
Solenoidal Tracker at RHIC (STAR) \cite{STARWeb} detector
at the Relativistic Heavy-Ion Collider (RHIC) \cite{RHICWeb} accelerator at Brookhaven National Lab (BNL) \cite{BNLWeb}. 
On the other hand they are based on detailed simulations of the expected detector response in ALICE \cite{ALICE-DAQ-HLT-TDR}. 
As can be seen in the figure, the cluster's topology is built in a tree-like structure where successively 
larger parts of an event are processed and merged as one approaches the tree's root. At the root of the tree
the trigger decision is made based on the derived physics quantities of the given event. This tree 
structure is a natural choice given the segmentation of the TPC and the hierarchical nature of the analysis,
which can easily be divided in multiple separate steps. 

At the top one can see the FEPs for two slices with the DDLs and RORCs for the 12 patches needed. 
Two processing steps are performed on the FEPs. The run length encoded raw data is unpacked
and then cluster-finding is done on the unpacked data to determine space-points of charge cluster 
depositions along particle tracks. This spacepoint data is then transported over a network to the next processing stage. 
On these next nodes an analysis is made to find track segments in each patch's spacepoints that
describe the particle tracks going through the TPC. 
%As there are also six tracking nodes corresponding 
%to the six patches, in principle one node could work on the data from one patch. 
For fault tolerance
reasons with regard to the failure of nodes the spacepoints from one patch are distributed among the four
nodes in the tracking group. Segments of tracks produced by six neighbouring tracking
groups are then distributed to the next group of eight nodes. In this group the track segments from
the six groups are merged to longer track segments over the respective six slices in the TPC. 
The data produced from the six groups of track merging nodes is sent to one last group
of eight nodes where the data from all track merger groups is again merged to form the data of the complete
event in the TPC. Based on this data these global mergers can make the HLT trigger decision.
In this setup 216 FEPs would be present with an additional 144
nodes for tracking. Six additional groups of eight nodes are needed for track merging of
a slice sextett and a final group of eight nodes for global event merging. In total this setup would 
thus require 416 nodes for the HLT.

All data from ALICE sub-detectors is read out upon an L2 trigger accept decision and
is subsequently present in both the DAQ and the HLT. There is thus no need for a fixed latency or an
upper bound on it for the HLT decision. The main memory of the FEPs will be 
used as derandomizing buffers for the events and event fragments. With memory sizes of several 
gigabytes expected for PCs when ALICE and the HLT will be activated, one PC will be able to store several
thousands of event fragments read out from a TCP patch via one RORC/DDL. Nonetheless 
an average latency over all events will be enforced, determined by the event buffers, the average processing time,
and the input data rates. 

The HLT will consist of a farm with a large number of commodity PCs.
Each of these individual PCs must be regarded as a relatively unreliable component and 
can fail at any moment. Experience with a small cluster in Heidelberg and elsewhere \cite{Rel1} \cite{Rel2}
suggests a failure of 
one node at least once a week in a system of that size. Therefore the HLT needs a fault tolerant
architecture that can cope with the loss of any node and still continue working. 
For the processing nodes a good approach is to distribute each task among a group of several
nodes. An example for this are the track finding nodes in the sample setup in 
Fig.~\ref{Fig:HLTArchitecture1}. Each FEP distributes its data among multiple nodes in the track finding 
group by sending incoming events on a round-robin basis to them. If one of the nodes fails 
this is noticed by a supervising instance informing the FEPs, which then can distribute data
sent to that node among the remaining target nodes. Any new incoming data would also be distributed 
among the remaining nodes only, until the FEPs are notified that the failed node is available again.
This node failure would thus cause no total system failure of the system but just a higher load on the remaining tracker nodes
and maybe a slightly reduced event rate corresponding to the loss in processing power of the failed
node. The capabilities of the HLT system as a whole would not be significantly influenced.

For the FEPs a different approach is necessary as each DDL ends in exactly 
one FEP. One simple solution to this problem is a kind of standby node equipped with RORCs, into which
DDLs from failing nodes can be plugged. This
of course would have to be done by manual intervention by a technician.
But this approach would not prevent the loss of the raw data on the FEP at the time of the failure.
An extension of the previous approach
is to copy the raw data from an FEP immediately after it has been received to one
of the other nodes in its patch group. In principle it would even be possible to use a device-to-device copy
in which the RORC directly communicates with the networking adapter connected to the second node.
The viability of this and other approaches will have to be analysed before a decision is made regarding this.
But one major feature of any architecture chosen for the HLT must be the tolerance with respect
to the failure of single nodes in the system and the lack of single points of failure in it. 

The transport of the data through the HLT will be orchestrated by the software framework presented in this thesis. 
Due to the flexibility necessary  
with regard to different setups and changing analysis requirements, the framework must be
very flexible and should allow easy changes in its configuration of the data flow. Similarly it should
take into account the unreliable nature of the single nodes and be prepared to recover from the loss
of any of them as detailed above. Furthermore, as the task of the system is the analysis of large 
amounts of data that will require large amounts of CPU power, the framework should be as efficient
as possible and not use up too much CPU time for just the transport of data through the system.

\subsection{CBM Project}

The Compressed-Baryonic-Matter experiment \cite{CBMWeb} is a detector intended for the future High-Energy-Storage-Ring  \cite{HESRWeb} (HESR)
accelerator at the Gesellschaft f\"ur Schwerionenforschung (GSI) in Darmstadt \cite{GSIWeb}. Its primary research goal is the investigation of highly 
compressed nuclear matter, that can be found for example in neutron stars and supernova explosion cores. 
The HESR is designed to provide a dedicated
heavy-ion accelerator with a number of parameters exceeding those of existing dedicated HI accelerators, like beam intensity, quality,
and energy. The aim is to investigate new regions in the baryon-phase-diagram such as the quark-gluon-plasma and the areas
of higher baryon densities. For this purpose the energy range between $10~\mathrm{GeV}$ to $40~\mathrm{GeV}$ per nucleon is investigated for a number
of criteria, like exotic states of matter or the critical point indicating a phase transition from the quark-gluon-plasma to 
higher densities. 

For the CBM detector the general HESR research goals are addressed by the simultaneous measurement of several observables
sensitive to high baryon density effects and phase transitions. Amongst others particular attention is given to the following areas: 
\begin{itemize}
\item The parameters of penetrating probes, like light, short lived vector mesons decaying into electron-positron pairs, able to 
carry undistorted information from the dense hadronic fireball
\item Strange particles
\item The collective flow of all event observables
\item Event-by-event fluctuations of particle multiplicities, particle phase-space distributions, the collision centrality, and
the reaction plane
\end{itemize}

\begin{figure}[hbt]
\begin{center}
\resizebox*{0.50\columnwidth}{!}{
\includegraphics{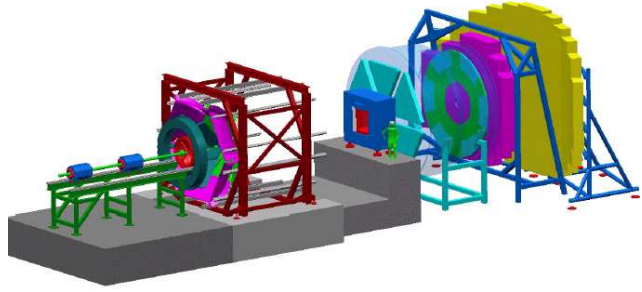}
}
\parbox{0.90\columnwidth}{
\caption[The CBM Detector.]{\label{Fig:CBM1}The CBM Detector. The beam enters from the left hand side, in front of CBM is the HADES detector. 
The setup consist of a dipole magnet (blue) with the silicon tracker mounted inside (red), a RICH Detector (turquoise), 
a TRD detector (pink) and a RPC TOF wall (yellow), the target is at the entrance of the magnet.}
}
\end{center}
\end{figure}

The CBM detector will basically consist of a magnet, silicon pixel and strip detectors, a RICH detector,
TRD detectors, and an RPC Time-Of-Flight (TOF) wall detector, placed in line behind a fixed target in the beam as shown in 
Fig.~\ref{Fig:CBM1}.  Unlike ALICE CBM is a fixed target experiment and thus does not need to provide a $4 \pi$ coverage 
around the collision point. The setup is designed to measure hadrons as well as electrons for beam energies up to 
$40~\mathrm{GeV}$ per nucleon with a large acceptance. Particle tracking and momentum determination will be performed
by the seven layers of silicon strip and pixel detectors inside the magnetic field located close to the target. In the remaining three detectors
(RICH, TRD, RPC TOF) located downstream of the magnet, particle identification will take place, with the RICH and the TRD 
being used for general electron and high-energy electron identification respectively. 

For the CBM readout a
high level trigger or event filter farm using similar principles as for ALICE (PCI readout, a large Linux PC-cluster) may
be used. In such a setup the use of the framework presented in this thesis is a possibility due to the framework's
generic design and the flexibility of its ``pluggable'' component approach. 

%\texttt{HESR: http://www-new.gsi.de/zukunftsprojekt/beschleunigeranlage\_e.html}

%\texttt{http://www-new.gsi.de/zukunftsprojekt/experimente/CBM/index\_e.html}

\subsection{PANDA Project}

The Proton-ANtiproton-at-DArmstadt (PANDA) experiment \cite{PANDAWeb} at GSI is designed to study collisions of protons ($\mathrm p$) and 
anti-protons ($\overline \mathrm p$) with three primary physics goals:
\begin{enumerate}
\item To study the behaviour of the strong force binding gluons and quarks together in hadrons, the $\mathrm p \overline \mathrm p$ collisions
will be monitored for charmonium and other short-lived particles. A detailed spectroscopic analysis will then be performed on these
particles with the aim of obtaining new results for the characteristics of the strong force at medium and longer distances as well as the 
origin of the quark and gluon confinement in hadrons.  
\item By studying high-energetic $\mathrm p \overline \mathrm p$ collisions it is expected to generate new data to determine the origin of the hadron
masses, of which only a small part, e.g. 2~\% in the nucleon, can be attributed to the valence quarks in each hadron.
\item The third main goal is the search for exotic new forms of matter predicted by strong force theories, such as glueballs that consist 
only of gluons or hybrids that contain two valence quarks and one gluon. 
\end{enumerate}

For the PANDA readout the same statement for a potential
high level trigger or event filter use and architecture applies as for CBM. This includes a corresponding use of the presented framework in such a system. One specific 
application could be searching for and selecting events in an online analysis that contain a charmonium particle. 

%\texttt{http://www-new.gsi.de/zukunftsprojekt/experimente/hesr-panda/index\_e.html}

\subsection{Relation To Other Experiments}

Other high-energy and heavy-ion experiments also use a system providing high level trigger functionality. 
The ones most related to ALICE are the ATLAS \cite{ATLASWeb}, \cite{ATLASGreybook}, \cite{ATLASTP}, 
CMS \cite{CMSWeb}, \cite{CMSGreybook}, \cite{CMSTP}, and LHCb \cite{LHCbWeb}, \cite{LHCbGreybook}, \cite{LHCbTP} detectors, 
currently also being built for operation at the LHC \cite{LHCWeb}, and
the STAR \cite{STARWeb} heavy-ion detector at RHIC \cite{RHICWeb}. 

STAR is in operation at RHIC since 2000 and therefore belongs to a different
generation of detectors compared to ALICE. It is, however, the newest heavy-ion detector currently in operation.
The architecture of its Level 3 Trigger \cite{STARL3RT99} \cite{PrivCommStarL3} is characterized by a separation into Intel i960 processors on receiver boards
and a PC farm with Alpha CPUs, all connected by Myrinet. Cluster finding is performed already on the i960 processors, and
tracking is performed on Alpha PCs using the clusters received from the i960s. The L3 trigger's 
task is to reduce the raw events of approximately 15~MB occuring with a rate of about 100~Hz to a rate of  roughly 1~Hz. 

ATLAS and CMS are two general purpose detectors whose main task is the search for the Higgs particle and signatures of physics beyond the 
Standard Model of particle physics. The ATLAS High Level Trigger \cite{ATLASHLTTDR} is separated into a Level 2 trigger and an Event Filter farm,
both consisting of standard PCs.
Together these two systems have to reduce the HLT input rate of 100~kHz events to the order of 100~Hz. A full event is between 1~MB and
5~MB in size, resulting in an output data stream of a few hundred MB/s. Data rate into the Event Filter is between about 600~Hz and 
3.3~kHz, depending on the details of operation. An event switching network is located between the Level 2 trigger and the detector's front end
electronics so that a Level 2 node can request any fragment of an event needed. Between the front end electronics and the Event Filter farm an event building network
is present so that the Event Filter farm operates on completely assembled events and does not have to perform any
event merging itself. The disadvantage of this approach is the requirement of a high bisection bandwidth between the front end electronics and the Level 2
and Event Filter farms. As the network technology Ethernet has already been chosen for most parts of the system. 

Expected event sizes for CMS are also about 1~MB. Input and output event rates for its HLT \cite{CMSHLTTDR} are 100~kHz and 100~Hz,
similar to ATLAS. An event builder network is used here as well to connect approximately 700 modules attached to the detector's front end electronics
with the HLT nodes. The HLT therefore will operate also on completely assembled events and will not perform any event merging or building itself.
For the network technology the focus at the moment lies on Ethernet or Myrinet, both of which are investigated more closely. 

The last of the three other LHC experiments, LHCb, operates with very small event sizes of around 100~kB. Its
High Level Trigger \cite{LHCbHLT} has to reduce the event rate from 40~kHz to 200~Hz. Since the output of the HLT are raw
event data as well as event summary data the resulting output data rate is 40~MB/s, relatively low compared to the other three LHC experiments. 

As can be seen from the above descriptions, these four experiments' HLTs differ in at least one crucial
parameter (required input or output data rates, event rate, or architecture) from what is required
for the ALICE HLT. 

%There will therefore be very little overlap in the systems' software so that a 
%cooperation with these experiments did not seem to offer much benefits when work on the framework presented 
%in this thesis was started. The new documents available for the three LHC experiments indicate that
%the conceptual gaps between ALICE's and the other experiments' HLTs have not closed very far,
%confirming the decision. 

\subsection{Online Video Processing \& Image Generation}

Next to the use in high-energy and nuclear physics trigger systems other applications for the presented framework are
also possible. Tasks that can be split up into sequences of parts which can be processed in parallel and/or in a pipelined 
manner are areas where the software is well suited to be used. Examples of such applications are online video processing and 
image generation, with four major uses:
%Further possible areas of application for the presented framework are online video processing and image generation, with four major
%uses:
\begin{enumerate}
\item Decompression of a received, highly compressed video stream for display
\item Compression of a video stream before transmission or storage
\item Application of one or more filters to a video data stream before it is displayed
\item On-the-fly rendering of 3D graphics into video streams for display or transmission
\end{enumerate}

The general principle for the use of the framework follows a similar pattern for all of these four applications.
A node 
receives data from an external origin and acts as a data source for the required number of processing units by distributing the data among them. Each of these
processing units sends the output data resulting from its operation to one data destination. This destination node collects the data, assembles it
into the correct order, and performs the desired action with it. It would even be a possibility to use two processing nodes for each
sub-task, one of which performs a lower quality form of the operation that can be completed in a significantly shorter time. 
If the normal, higher quality data is not received on time at the destination the low-quality backup version can be used instead. 

\subsubsection{Video Decompression}

In the first of these applications, online video decompression, the data input is the compressed video stream, 
received in the data source either by a specialized or commodity network or a specific readout device. The produced output data
takes the form of a sequence of images, to be displayed on an output media. 

\subsubsection{Video Compression}

For video compression the input is a raw video stream received most likely by a special purpose video readout device or receiver adapter. 
Output is a compressed video stream that can be written to disc, broadcast, or sent over a network to a number of receivers. 

\subsubsection{Video Filter Application}

In the application of filters to a video stream input and output can both be a stream of either compressed or uncompressed
video data. This stream is received and (re-)transmitted in the corresponding form as described for the previous two entries.

\subsubsection{3D Image Generation}

For image generation the source data are 3D scene descriptions either stored and read from a file or received from a 
generator system, either based upon a preset programm or following an operator's input. The output is a video stream for storage or
immediate display.

\clearpage

%%%%%%%%%%%%%%%%%%%%%%%%%%%%%%%%%%%%%%%%%%%%%%%%%%%%%%%%%%%%%%%%%%%%%%%%%%%%%%%%%%%%%%%%%%%%%%%%%%%%%%%%%%%%%%%%%%%%%%%%%%%%%
%%%%%%%%%%%%%%%%%%%%%%%%%%%%%%%%%%%%%%%%%%%%%%%%%%%%%%%%%%%%%%%%%%%%%%%%%%%%%%%%%%%%%%%%%%%%%%%%%%%%%%%%%%%%%%%%%%%%%%%%%%%%%

\chapter{\label{Chap:Overview}Overview of the Framework}

\section{Introduction to the Framework}

\begin{figure}[hbt]
\begin{center}
\resizebox*{0.40\columnwidth}{!}{
\includegraphics{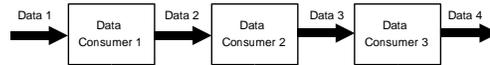}
}
\parbox{0.90\columnwidth}{
\caption{\label{Fig:DataDrivenArchitectureElements}Principle of a data driven architecture.}
}
\end{center}
\end{figure}

This thesis describes a framework that has been written for distributed online data processing in clusters as described in the
previous chapter. One of its main characteristics is the focus on data driven architectures and applications in which elements
can receive input data from other preceeding elements and produce output data for consumption by succeeding ones, as depicted in 
Fig.~\ref{Fig:DataDrivenArchitectureElements}. Further emphasis during the framework's development has been placed on
efficiency, flexibility, and fault tolerance. For efficiency in this context the focus is primarily 
set on the reduction of CPU cycles used in the framework for the transport of data, to keep as much CPU power
available for processing of data according to an application's requirements. The requirement for flexibility is implemented such that the framework
consists of a number of independent software components that can be connected together, without recompilation and even during runtime of the system,
supporting a high adaptability in its configurations. Finally, fault tolerance of the software means that
the framework has to be able to handle and recover from errors as autonomously as possible and that it should not contain any 
single points of failure. Instead the framework should be able to reconfigure itself during
runtime to work around the faulty spot, aided by the dynamic connection ability described previously. 
In the following three sections of this chapter the most central design decisions and emphasises of the framework are detailed, followed by 
overviews of the components making up the framework and its software architecture.

\section{\label{Sec:Design}Framework Design Considerations}

A major point of attention and optimization during the design and implementation of the framework
has been to avoid unnecessary overhead, mainly usage of 
too much CPU time and memory bandwidth. The suppression of these effects as far as possible is
important since the framework is basically a means of getting data to the right place at 
the right time for its first intended use in the ALICE High Level Trigger.
Considering the HLT's purpose, i.e. the analysis of that data, and the size of the data involved,
any amount of processing saved can substantially reduce the size of the whole system needed to process 
the required quantity of data, 
and a reduction of a system's size also implies a reduction of its cost. 

Furthermore the software has been optimized for a high throughput rate of events in a
system. It has not been optimized for latency, i.e. the time elapsed between the event's entering of the system
and the HLT's trigger decision for that event. The reason for this decision
is that the HLT will be operated in a stream mode where new events will come in continuously.
In such a system latency can be balanced by sufficiently large buffers to hold events that have been processed
by one stage while the next stage processes preceeding events.
This argument is made with the background that the recent and projected future development of memory size
shows a steady increase following Moore's Law. 
The increase in size is accompanied by a development of memory cost so that the price of the doubled amount of memory
is at most only slightly more than the price for the original amount. 
%The increase in size comes with at most a moderate increase
%in price for a doubled memory size. 
Memory bandwidth in contrast can be scaled by the same amount, but this
comes at a much increased cost and is not easily available for COTS PCs.

Another important point that has to be considered for such a system 
%Another problem of such a system 
is the available memory bandwidth. As described in 
section~\ref{Sec:ComputingBackground}, the available bandwidth of memory in PCs has not increased by
the same factor as the CPU power. In theory CPUs are able to operate much faster than in reality
where they are slowed down waiting for memory accesses. Caches have helped to resolve the situation partially
and, depending on a program's memory access pattern, effects may be more or less pronounced.
In order to partially compensate for and work around the problem, trade-offs have been made in the framework
that sacrifice amount of memory used over memory bandwidth. The
motivation for this is the same as in the previous paragraph, memory size is cheap while memory bandwidth is not. 

% {\bf motivation/argumentation/??} for this is that the recent and projected future development of size
%shows a steady increase following Moore's Law. This increase in size comes with at most a moderate increase
%in price for a doubled memory size. By contrast memory bandwidth can be scaled by the same amount but this
%comes at a much increased cost and is not easily available for COTS PCs. 

%(Low) Overhead
%Rate, nicht Latenz; 
%Problem: Memory bandwidth,
%large amounts of data, not only streaming access, cache misses; -thrashing??
%Trade memory size for bandwidth; (Size is cheap, bw isn't)

To allow a flexible and easy operation and configuration of a system using the framework as well as
an easy customization, the framework should be composed of relatively small components,
that can be connected together using a defined common communication interface between components.
As such a system has to run distributed over the nodes in a cluster,
a mechanism to allow components to communicate across nodes is needed as well. For efficiency
reasons the communication between components has not been generalized to use a communication mechanism
that would also work between nodes. Instead an efficient communication mechanism has been chosen
that works only locally and special bridging components have been developed to connect components across nodes.
%These components are described in section~\ref{Sec:DataFlowComponents}.

\begin{figure}[h]
\begin{center}
\resizebox*{0.25\columnwidth}{!}{
\includegraphics{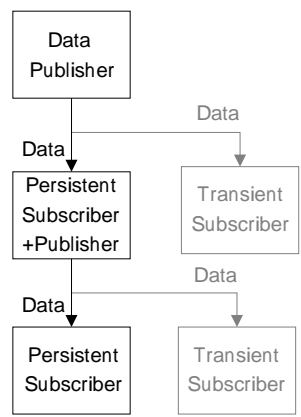}
}
\parbox{0.90\columnwidth}{
\caption[Example of persistent and transient subscribers.]{\label{Fig:PersistentTransientSubscribers1}Example of persistent and transient subscribers.
The figure shows five processes with two publisher and four subscriber objects. Two of the subscriber objects are transient subscribers in
components attached to monitor the publishers' data streams.}
}
\end{center}
\end{figure}

\label{Sec:PubSubInterfaceRequirements}
For the interface between the components a number of requirements have been specified:
\begin{itemize}
\item A data producer should be able to feed multiple consumers to allow easy 
monitoring of both the framework's and the analysis code's correct functionality.
\item For efficiency reasons, only data descriptors should be exchanged between local processes, with the 
data itself kept in shared memory. As the data is potentially very large, especially in the case of the 
ALICE HLT, this requirement serves the purpose of preventing copy steps of the data between components.
\item Two kinds of data consumers should be supported:
  \begin{itemize}
  \item  Blocking consumers, called persistent subscribers, which need to access the input data until they have  
  finished processing it. These are the actual analysis components that need to work on event data
  until the analysis is finished.
  \item Monitoring consumers, called transient subscribers, that do not need to process every event 
  and have to tolerate overriding event releases by the producer. 
  %These are components foreseen for monitoring   for which it is not vital that every event can be completely processed.
  These components can be attached as taps at any point in the data stream, statically as well as dynamically.
  An example of this is shown in Fig.~\ref{Fig:PersistentTransientSubscribers1}. 
  \end{itemize}
\item Where possible producers and consumers should be identifiable by name rather than by numbers to ease setting up and debugging. 
\item The design should be object-oriented to make use of the advantages of object-oriented software development like reuse
and encapsulation.
\item Actual communication between processes should be hidden behind an abstract interface to allow an easy exchange of the underlying
communication mechanism without having to change the upper layers.
\item It should run at least on the Linux operating system on Intel compatible CPUs as a baseline.
\end{itemize}

Since there is currently much development in the area of cluster interconnect and networking technology suitable for 
a system using the developed framework, 
such as the ALICE HLT, no decision has been made yet for a specific networking technology
to be supported. Instead it has been decided that any inter-node communication is to be hidden behind an abstract communication
interface that supports communication implementations using different networking technologies and protocols. This approach
is also desirable with respect to flexibility, as it allows to setup a cluster for the framework with any supported
network technology. It also allows easily to write implementations for new additional network interfaces that can 
be used by any system using this framework without changing the framework components. 

Concerning the usage of network communication in the framework, two separate types can be distinguished.
The first of these two types, the transport of the actual data to be processed, in general produces the bigger data volume. 
%As far as data volume transported is concerned the first is in general the bigger of the two, this is the transport of the
%actual data to worked on, that is expected to be present in larger blocks. 
For ALICE this is the event data and the minimum block size for TPC raw data would be of the order of several hundred kilobytes. %512~kB. 
The second type of communication consists of small messages being exchanged
for control, setup, and handshaking purposes, with a typical size from several tens of bytes up to a few hundreds
of bytes. Taking this distinction between the two communication types into account, the abstract communication interface has been 
designed to provide optimized functions for the transfer of small data, like messages, as well as large blocks of data.

%Network communication not yet decided; Hide behind abstract interface
%Can be separated typically into two types of communication: 
% - Small message like transfers for control, setup, and handshaking communication 
% - Large data blocks for transfer of actual data.
%Provide two kinds of communication for these types.

\section{Components Overview}

The framework consists of a number of separate components, that can be combined into a running system
to allow maximum flexibility in its configuration. This flexibility necessitates that the components communicate via a defined 
interface that supports connection during runtime. To allow a maximum communication efficiency and usability, this interface 
is based on shared memory for the data exchange, as described above, and named pipes for control messages, like the exchange of 
descriptors. Named pipes make it possible to address each process by a unique name without the need to construct and manage a separate
namespace. Instead the operating system file namespace is used. At the same time pipes provide an efficient operating system mechanism
 for a process to wait for incoming data without polling and thus without consuming CPU cycles while waiting.
Two major kinds of components currently exist in the framework, generic components, not 
specific for a particular task, and components developed for use in the ALICE High Level Trigger. 

With the generic components one can again distinguish four types: data flow components, worker component templates, worker, and fault tolerance components. 
Data flow components do not modify the data passing through a system but are responsible for routing the flow of data in it. These include
components to merge parts of events into one part, to scatter and gather data among multiple nodes, e.g. for load-balancing purposes, and to
transport data between nodes over a connecting network. Worker component templates are provided in the form of three sample programs that can easily be extended for components 
which respectively produce data (data sources), receive data (data sinks), and receive, process, and produce new data (data processing or analysis components).
The generic worker components are a number of components, partly based on the template components, that act as data source, sink, or processing components.
Finally, the set of fault tolerance components is responsible for 
making a framework system tolerant against faults of framework components, hardware parts, or even complete nodes. Some
overlap exists between the fault tolerance and data flow components as a number of the fault tolerance components are extended 
versions of data flow components, performing the same tasks with added functionality. 

ALICE specific components are
analysis components that execute the different stages of the detector data processing. Starting at the raw data read out from the detector, each
component represents one step in the analysis process and  accepts a specific type of input data. This input data is processed and another type of output data is produced.
The new ouput data is in turn  made available to the next
step in the chain for further processing. After the last step has been executed, a fully reconstructed event is available as the base
for the trigger decision. 

\begin{figure}[hbt]
\begin{center}
\resizebox*{0.60\columnwidth}{!}{
\includegraphics{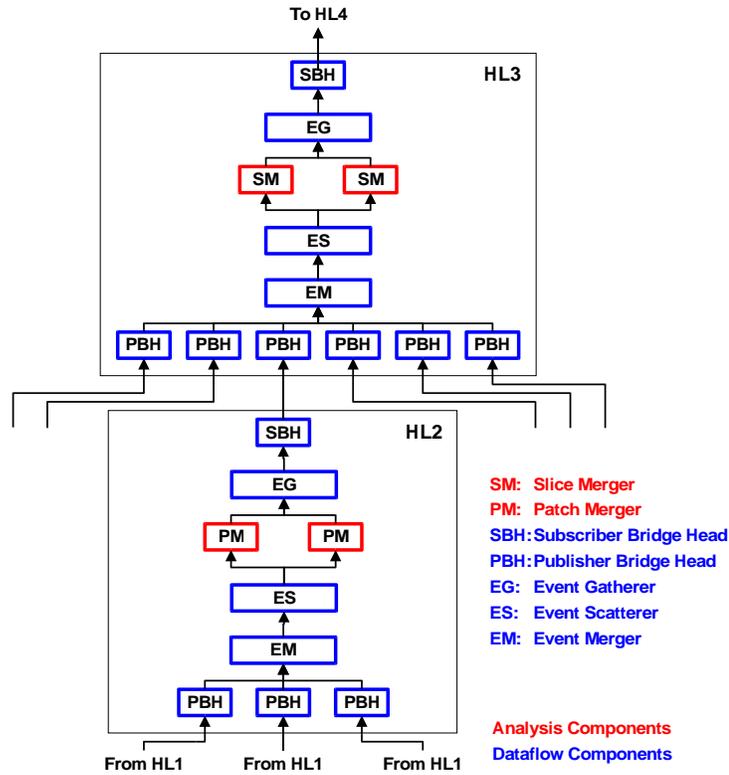}
}
\parbox{0.90\columnwidth}{
\caption[Sample HLT Component Configuration.]{\label{Fig:HLTComponentSetup1}Sample component configuration in two hierarchy levels (HL) in the HLT.
Each small box represents one process in the chain. Each large box represents a node.}
}
\end{center}
\end{figure}

%{\bf \LARGE CDR figure 73 page 96}

Fig.~\ref{Fig:HLTComponentSetup1} shows a number of the framework's components in a possible setup as it might be used in the ALICE High Level Trigger for 
the processing of TPC data. The figure shows
two nodes in the central two hierarchy levels (HL) of an HLT configuration with several components running on each of them. 
A detailed description of the components is provided later in chapter~\ref{Chap:Components}. 
Components shown in blue are generic data flow components, while red ones are ALICE HLT processing
components. One can see the data flow components that connect multiple nodes (SBH and PBH), merge parts of
the same event (EM), and that scatter and regather event streams for load balancing (ES, EG). The two types of processing components
in the figure perform different levels of merging, the first (PM) on the level of the subsector patches and the second (SM) merges
a number of the sector slices. On the following hierarchy levels more merging components are present to reach
a fully merged event at the end of the processing chain. As can be seen, the configuration makes use of the inherent structure
and hierarchy in the TPC and its data analysis to arrive at a natural decomposition and distribution of the different processing tasks.

\section{\label{Sec:FWArchitecture}Software Architecture}

\begin{figure}[hbt]
\begin{center}
\resizebox*{0.7\columnwidth}{!}{
\includegraphics{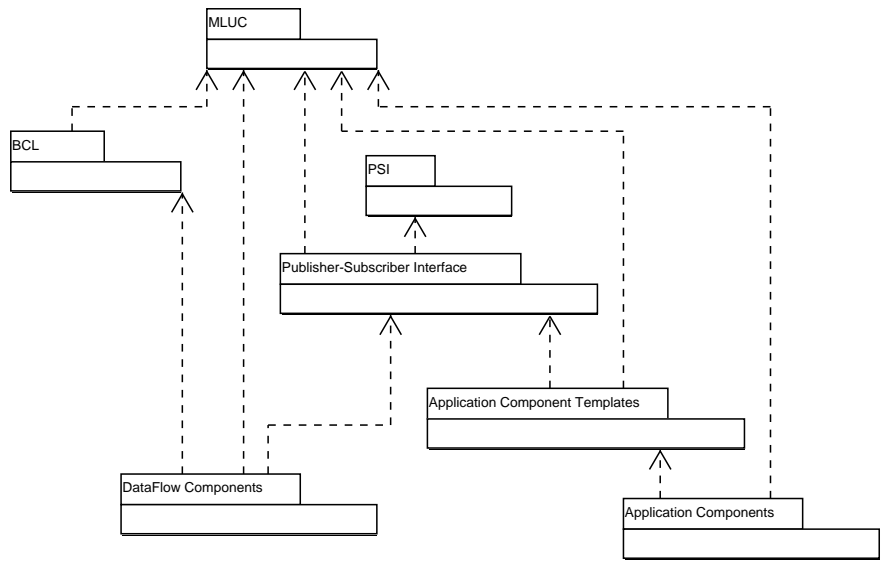}
}
\parbox{0.90\columnwidth}{
\caption[The modules making up the framework and their dependencies.]{\label{Fig:FrameworkModules}The modules making up the framework and their 
dependencies shown in UML notation. The boxes denote the modules and the arrows their mutual dependencies.}
}
\end{center}
\end{figure}

From the software architectural side, the framework is divided into a number of distinct packages or modules, some of which are dependent
on others. The relationship between the modules is shown in UML notation in Fig.~\ref{Fig:FrameworkModules}. 

At the top of the figure are two basic modules, MLUC and PSI, that
% that are described in chapter~\ref{Chap:UtilityClasses}.
provide basic functionality and do not rely on any other module. The More or Less Useful Class library 
(MLUC) package is a
C++ class library providing generic classes used in the other packages. One of the classes 
provided by MLUC is a thread class to encapsulate operating system functionality for creating multi-threaded programs. 
Also included is a string class, that actually provides less functionality compared to the string class
present in the standard C++ library available with the GNU C++ (g++) compiler. The GNU class however is not thread safe
and as all but the most trivial programs in the framework are multi-threaded it could not be used. Another
major class in the package to be included here is a new vector class for handling of dynamic arrays. In most cases
where the vector class is used, it is used in an almost queue like functionality, with items being added at the end
and removed from near the beginning. With the Standard Template Library (STL) vector class, also distributed with the g++
compiler, this access pattern causes all elements after the one removed to be copied one element forward. Already for moderately
large arrays this copying process takes up a lot of CPU time and obviously also uses up a lot of memory bandwidth.
%However in the removal is not always strictly the first element and in general it does not even have to be an element
%at the beginning of the array which makes an STL queue class unsuitable 
To change that situation the new dynamic array class was written, that trades off memory
size used for the array for a reduction in used memory bandwidth, as outlined in the previous design section~\ref{Sec:Design}. 
The last major functionality contained in the MLUC library is a logging facility for programs. This logging facility
is designed to have multiple severity levels of log messages and to have a negligible overhead when a severity level
is turned off. Furthermore it features a modular system of logging targets to which messages are dispatched, that can be
configured completely at runtime.

The second basic module, the PCI and Shared memory Interface (PSI), 
%is an interface for PCI bus hardware and shared memory, that 
provides user level access to PCI bus devices as well as a shared memory interface.
PCI devices that need to be accessed can be special readout hardware while
shared memory is used to exchange data between framework components. 
The module consists of a library
providing an Application Programmer's Interface (API) together with a driver that performs the actual hardware accesses
and operating system interactions. As the name suggests PSI's primary purpose is to provide access to
PCI adapter cards and other devices from normal user space programs. More specifically it allows to access the Configuration
Space Registers (CSRs) of any PCI device, including bridges, as well as access to the memory and I/O regions described by the Base
Address Registers (BARs) of PCI devices. Memory BARs as well as arbitrary physical memory addresses can be 
mapped into the address space of user space programs and can then be used like any normal program memory. Shared memory access is possible to
two kinds of shared memory, the first of which is ordinary shared memory that can be located anywhere in the physical memory of
the computer and does not need to be physically contiguous. The second type of shared memory is located in a memory
area obtained by using the big physical area memory patch \cite{bigphysareapatchWeb}. This patch reserves a single large physical memory region on
system start and can allocate chunks of this memory to programs and drivers. The special characteristic provided by this
patch is that any memory allocated from it is actually contiguous in the physical memory, making it ideally suited for
streaming access and for access by Direct-Memory-Access (DMA) capable PCI devices. This last point is especially important,
as many SAN adapters are actually DMA capable and could transfer data from and to this memory without using the CPU. 
A tool library also contained in the package  makes use of the basic functions in the driver to provide more complex higher
level functionality together with a number of utility programs. 

%In addition
%to the basic library/driver the package also contains another tool library providing higher level functionality building on
%the basic instructions in the library as well as some small utility programs that allow to do some simple actions without
%writing programs. 

At the left side of the figure is the Basic Communication Library (BCL) module providing the abstract interface for communication outlined
in the preceding section. This module is also a C++ class library, which contains an abstract base class defining general communication
functionality. Function names in the library have been chosen to follow the widely known and used socket API of the POSIX \cite{IEEEPOSIX} 
or Single Unix specification \cite{SingleUnixWeb}, \cite{SingleUnixOnline}, \cite{SingleUnix}
where appropriate. The base class provides functions for performing a bind to an address that remote programs
can connect to as well as functions for connecting to remote program's addresses. Two further abstract classes are derived from this class,
one for each of the two communication patterns from section~\ref{Sec:Design}, transfers of small message-like data or
large blocks of data. Each of the two classes provides its own API adapted to its specific task. The message like API has functions for
directly sending and receiving small structures to any address. With the block API a user first has to request memory in the 
remote buffer to store the data before it can actually be sent. From these two base classes in turn the 
classes are derived that contain the implementations of the APIs for actual network protocols. Currently both communication types are supported for the
TCP protocol \cite{RFC793} as the most widely available baseline and for the shared memory SCI interface \cite{SCIIEEE} as an example of a
low latency and low overhead SAN technology. More details about this module can be found in chapter~\ref{Chap:ComClasses}.

While the previous three packages provide functionality that can be used in many projects, the remaining four packages, 
covered in chapters~\ref{Chap:PubSubInterface} and~\ref{Chap:Components}, contain the code of the actual data 
flow framework itself. Utilized by the other three modules, the 
Publisher-Subscriber-Interface module contains the implementation of the interface
used by the framework components for communicating locally on one node. The interface is based on 
shared memory to hold the data to be transported between components and named pipes to send descriptors
holding the location of the data to be exchanged as well as its size and type. It makes use of
the publisher-subscriber pattern \cite{Pattern}, also known as producer-consumer principle. The Publisher-Subscriber-Interface
package provides a set of classes that encapsulate the interface and that can be easily used in programs that want
to communicate with components in the framework or in components themselves. Components using this interface
are supplied by the remaining three modules. 
Of these modules the Data Flow Components module also relies upon the BCL module. Components in this module 
are used to shape the flow of data through a system built with the framework. One component merges data streams 
containing different fragments of the same event so that one event stream with larger fragments or complete
events is produced. Two other components can be used to split up a stream of events evenly into a number
of streams of lower rates and to later reunite them into one large stream. This can be used for load balancing purposes
with each of the smaller streams being handled by a separate node or CPU. The final pair of components is used to form
a bridge between two nodes to transparently connect components on the nodes using the publisher-subscriber interface.
For this purpose the first of these components uses the common interface for component communication to obtain the data to be sent. 
This received data is then
transmitted to its counterpart component on the peer node, using communication classes from the BCL library. 
Data received by the component running on this node is again made available to other components, 
also by way of the component interface.

The second module using the Publisher-Subscriber-Interface module is the Application Components Template module that provides  templates
for application specific or user components. These components are needed so that a user building a system with the framework can incorporate
components with functionality specific to the setup and tasks of that system. Three basic types of these user components
can be distinguished: data sources, data sinks, and analysis or processing components. Data sources are components that 
accept or read out data from a source external to the framework. Sources can be simple files, network daemons,
or special hardware such as the ALICE RORCs. This data is made available to further framework components, these data source components
thus feed a system with data to be processed. Data sinks in analogy are components that only accept data from the framework
to perform a specific task with it. Possible tasks are writing data to files on disk or a database or sending it via a network to another system.
%An example are the components that receive HLT processed ALICE events and send them to the DAQ system for storage. 
Sinks and sources are in principle the opposite endpoints for data in the framework. Analysis components in contrast are located
in the middle of the framework. They receive data from other components and process it, either producing and outputting new data
 or just outputting the same data again. Its output data can again be used as the input for
other components. In the Application Components Template module templates for all of these three types of application components 
are contained, providing most of the functionality needed to include specific components in a system.
A user only has to add custom code to provide the data for sources, handle the received data for sinks,
and produce new output data from received one for analysis components. 

The Application Components package is making use of the Template module and supplies a number of working user
components based on the templates. Some of these components can be used directly in production systems while 
others are intended for development and testing. An example of the second category is a data source file publisher
component that uses a set of files specified on the command line and publishes them round-robin into a system's
data flow chain. A contained component that can be used in a production system is a data sink that accepts
events and, after a configurable number of events have been received, calculates the average rate of received events
and reports that rate using the logging facility of the MLUC package.

\clearpage

%%%%%%%%%%%%%%%%%%%%%%%%%%%%%%%%%%%%%%%%%%%%%%%%%%%%%%%%%%%%%%%%%%%%%%%%%%%%%%%%%%%%%%%%%%%%%%%%%%%%%%%%%%%%%%%%%%%%%%%%%%%%%
%%%%%%%%%%%%%%%%%%%%%%%%%%%%%%%%%%%%%%%%%%%%%%%%%%%%%%%%%%%%%%%%%%%%%%%%%%%%%%%%%%%%%%%%%%%%%%%%%%%%%%%%%%%%%%%%%%%%%%%%%%%%%

\chapter{\label{Chap:UtilityClasses}Utility Software}
\section{\label{Sec:PSI}PCI and Shared Memory Software}

There are two types of applications
requiring a special device
driver in the context of a data flow framework: Access to specific readout hardware, including prototype and development boards, and use
of large shared memory areas for interprocess exchange of data. For the last usage the operating system
supplied shared memory, such as System V Shared Memory \cite{SingleUnixWeb}, \cite{SingleUnixOnline}, \cite{SingleUnix}, 
\cite{LinuxIPCmanpage}, could in principle also be used. However, under Linux
at least there seems to be a restriction to a maximum segment size of about 32~MB. This restriction makes the approach unfeasible
for the use in the desired application, where buffer sizes of several hundreds of megabytes are needed
to store a sufficiently large number of data blocks. As far as the first
application is concerned, programming a separate device driver for every piece of hardware is the more
elegant approach, but is the most undesired too, due to the complexities and overhead involved in the programming of device drivers
compared to ordinary user space programs. 

The PCI and Shared memory Interface (PSI) software described in this section is used to provide the handling of shared memory for
the publisher-subscriber interface and the framework components described in chapters~\ref{Chap:PubSubInterface} and
\ref{Chap:Components} respectively. 
For the ALICE HLT the module will additionally be used for the development of programs 
accessing the RORC readout cards.
%, as it cannot be relied on that the people who will develop them are proficient with writing device drivers.
%In addition it is easier and quicker to write a userspace program, even one accessing hardware
%devices, than it is to write a device driver. 
An item that at the moment is not supported by the 
software is interrupt handling.
%, for which there currently exist two unsolved problems. The first problem for an interrupt handler is how 
%to inform a user space process about the occurance of an interrupt, and the second problem
%is that a triggered interrupt should be turned off as soon as possible. Turning it
%off in the user space program might leave it on too long and cause problems for the system. 
%This makes it necessary to be able to turn off interrupts of any device in a generic interrupt handler, and
%it is not yet clear if and how this can be done. 

To provide a background for the explanations of PSI software functions, parts
of this section contain a brief overview of some characteristics of the PCI bus. For more in-depth
explanations or specifications please refer to e.g. \cite{PCISpec} or \cite{PCISysArch}.

One consequence associated with the interface are the security risks it presents, since it allows any user of a
system who has privileges to access the driver unrestricted access to any memory area of the system, even 
operating system memory. This memory access could be used to gain full access privileges
to the system and access any desired data. One way to prevent or restrict this, is to make the
device node used to access the driver available only to a trusted group of users and regulate access in this
way. At a later stage, the driver might be modified in such a way, that it allows access
only to a certain set of devices and shared memory segment areas. This set of PCI devices might be specified with the PCI
vendor and device id, unique to each PCI device. Both restriction sets, memory and devices,  could be either compiled
into the driver itself or for somewhat greater flexibility could be specified as parameters when the
driver is loaded.

\begin{figure}[h]
\begin{center}
\resizebox*{0.75\columnwidth}{!}{
\includegraphics{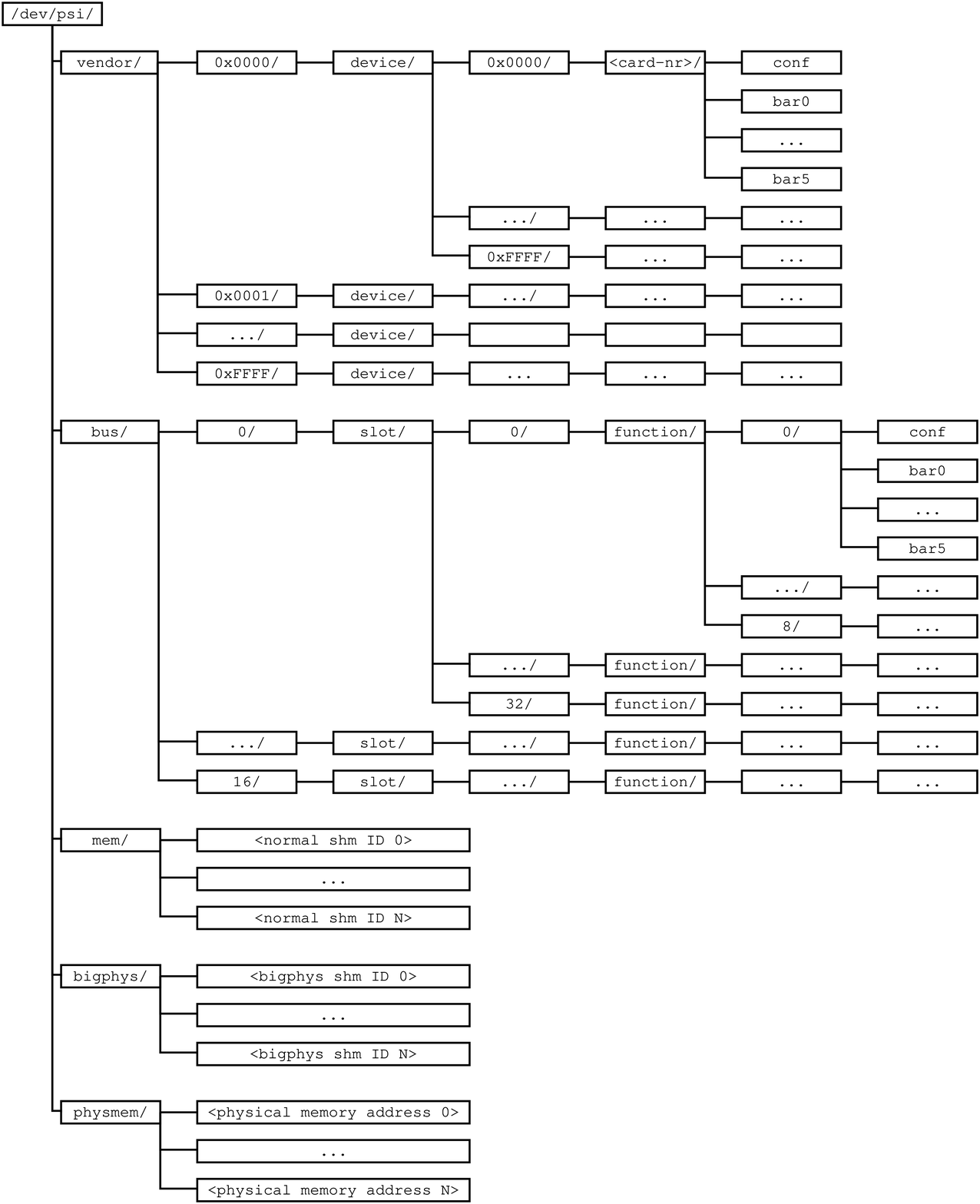}
}
\parbox{0.90\columnwidth}{
\caption{\label{Fig:PSITreeGraph}The virtual device tree structure of the PSI driver.}
}
\end{center}
\end{figure}

The interface operates on the principle of regions and a virtual device tree. A region corresponds to each type
of access one wants to perform and is analogous to a file handle. It is necessary to open a region before being
able to access any device or memory area. Each region allows access to only one device or area of memory associated with it. 
Which device or memory area has to be opened is specified using a string that describes a node in the virtual device tree.
A graphical sketch of the tree structure is shown in Fig.~\ref{Fig:PSITreeGraph}. Fig.~\ref{Fig:PSITree} shows
the strings used to specify the corresponding device regions. 

\begin{figure}[hbt]
\begin{center}
\begin{verbatim}

/dev/psi/vendor/<PCI vendor id>/device/<PCI device id>/<card nr>/conf
/dev/psi/vendor/<PCI vendor id>/device/<PCI device id>/<card nr>/bar<bar nr>

/dev/psi/bus/<bus nr>/slot/<slot nr>/function/<function nr>/conf
/dev/psi/bus/<bus nr>/slot/<slot nr>/function/<function nr>/bar<bar nr>

/dev/psi/mem/<shared memory id>

/dev/psi/bigphys/<bigphys shared memory id>

/dev/psi/physmem/<physical memory address>

\end{verbatim}
\parbox{0.90\columnwidth}{
\caption[The strings for the PSI driver region types.]{\label{Fig:PSITree}The strings for the PSI driver region types that can be specified using its virtual device tree.}
}
\end{center}
\end{figure}

PCI devices that have to be addressed can be specified in two ways, logically and physically. Logical addressing
is done using a device's vendor and device ID, unique 16~bit numbers. The vendor ID is assigned by the PCI group and the device ID
is assigned by a device's vendor so that each manufacturer and each device have its unique ID. With the combination of vendor and device ID
it is ensured that no two types of devices in a PCI bus can be mixed up. But it is of course still possible that multiple 
cards of the same type are inserted into a system. For such a case a card index is available to specify which
of the present cards should be used. This is the format shown in the first two lines of Fig.~\ref{Fig:PSITree}. 

Physical addressing, also called geographical addressing, of a card makes use of the architecture of the PCI bus. A system can actually contain
multiple PCI busses, up to a maximum of 16, numbered starting at 0. Each bus in turn can have a number of slots for 
plug-in cards, also numbered from 0 up to a maximum of 32 slots. Fixed built-in devices are still assigned a number in the 
same range, called slot/device number. Finally, each device can contain multiple functions, it can basically 
be divided into multiple sub-devices up to a maximum of 8. Addressing a device with these three parameters is done using the 
syntax shown in the third and fourth lines in the figure. 

Once a certain device or card is specified, the part of the device that has to be accessed must be specified. 
A PCI device's configuration can be written or read through its configuration space, a 64~byte region present for
each device's separate function or sub-device. In addition to this configuration space, each device's function
can offer up to six address regions for accessing its specific functions. These address regions can be either memory mapped regions,
addressed like computer's RAM, or they can be I/O regions. As the I/O address range is very limited many devices 
use memory mapped address regions. The six accessible address ranges of each device are specified in registers in the 
device's configuration space called Base Address Registers (BARs). Frequently the term BAR is also used to specify the
actual address areas that a register points to. For both of the above methods of addressing a PCI device, it is 
possible to either specify the configuration space to be accessed, shown in the first of each pair of lines, or
to access the area pointed to by one of the available BARs. 

Next to PCI hardware, the interface allows also access to shared memory areas. Two different types of shared memories are
supported: Ordinary shared memory and memory accessible using the big physical area patch \cite{bigphysareapatchWeb}. 
Ordinary shared memory does not differ much from other interprocess memory, like e.g. System V ShM.
It is located in normal RAM and is available on every system without modifications to the operating system kernel.
A shared memory region of this type is opened by using the syntax of the \texttt{/dev/\-psi/\-mem} line in Fig.~\ref{Fig:PSITree}. The
shared memory ID in this line can be any string, unique for each shared memory region to be
opened. To share a memory range programs have to open a region using the same ID.
One drawback of this shared memory type however is that it can be limited in the maximum size possible for a segment. 
System V shared memory for example has failed to be allocated beyond a size of around 32~MB in tests. Another potential disadvantage
is the fact that although this type of shared memory appears contiguous in the virtual address space
of a user program it may be scattered in the actual physical RAM. 

Bigphys shared memory in contrast does not suffer from these two problems. It is in principle similar to the standard
shared memory and its regions are also identified using a string ID. The major difference is that bigphys shared memory is allocated 
via a kernel patch, the big physical memory area patch. This patch allocates a specified number of memory pages at the start 
of the system, with the advantage that the allocated memory actually exists as one large block in physical memory. Other drivers,
in this case the PSI driver, can request a certain amount from that allocated memory. If this request can be
fulfilled the returned memory range will be contiguous as well and will neither be swapped out nor removed by the system
from its physical address for any other reason. The continuity of that type of memory has the advantage 
of being well suited for data streaming purposes and for accesses from Direct-Memory-Access (DMA) capable hardware.
DMA capable hardware typically receives a pointer to a memory area and then either reads data from
or writes data into that location. One example of DMA capable hardware is the ALICE RORC, that receives a pointer
to a large data buffer in the computers RAM and copies received event data into that buffer on its own without copying being done
by the system's CPU. 

Immediate physical memory addresses that can be anywhere inside the system's memory address space --- 32~bit for typical PCs --- are the
last type of region that the PSI driver can access. 
The driver performs no check on the presence of  actual
hardware accessible at that address. Write accesses to unused addresses will succeed, without the data being written
anywhere. Data read attempts will return invalid data, in most cases data with all bits set. 
%One type of hardware access that has
%not been implemented in the PSI driver at the time of writing are direct I/O port accesses, where an absolute port number is specified in analogy 
%to the physical memory regions. This may be implemented in future versions of the software.

After a region has been opened, its size needs to be set for the region types where this cannot be determined
automatically. This is the case for the three memory region types, normal shared, bigphys shared, and physical memory. 
PCI BAR regions and configuration space regions are sized automatically. The respective BAR size to be used is determined
conforming to the PCI specification from the device's configuration space. Configuration space regions are sized to the default 
value of 64~bytes. For a sized region it is possible to read data from and write data to it. These read and write accesses
can be done in units of 1, 2, or 4 bytes and can be any arbitrary multiple of the unit size. 
For all memory types, including memory BAR regions, it is also possible to map them into the user program's address 
space. The mapping returns a pointer to the program that can be used like a normal pointer variable in a C program.
It provides a very direct and fast access to the memory region concerned, as there is no operating system call associated
with each region access, making this the most effective and easiest way of hardware access.

Built on top of the basic functions provided by the PSI library is a further library, the PSI Tool Library,
that makes some higher level functions available. Included functions address various tasks
associated with PCI configuration spaces. The most basic ones of these are for reading out configuration
spaces directly into data structures with variables corresponding to the decoded elements of the configuration
space. These structures can be printed out in a human readable form using further available functions. 
Other functions in this higher level library deal with the sizes of regions pointed to by PCI devices' BARs, which can be
determined by specific read and write accesses to a device's configuration space. These
BAR region sizes, together with flags that can be set for a BAR, can be read out and printed by another
set of functions. Defined flag values can specify whether two 32~bit BAR addresses have to be combined to a
64~bit address and whether a certain memory BAR is prefetchable (cacheable) or not. 

In addition to determining the size of a region pointed to by a BAR it is also possible to configure
a device's BAR by assigning an address under which that region will be accessible. This assigned
address has to point to a free memory region of a size large enough to accomodate the 
needed window in order to avoid conflicts. In order to ensure this the library function first determines the sizes of the 
memory regions needed and scans all devices on the system's PCI bus(ses) to determine the
address ranges already in use by other devices. 
If the device is located behind one or more PCI bridge devices,
the library will take further steps to configure the region so that it will be inside the address
window passed by the bridges to devices behind it. 
%If the device is located behind a PCI bridge device
%it will also mask out any address ranges that outside of the windows which the bridge can be configured to pass
%through to devices behind it. 
Once the used address ranges are determined, the remaining free ones will be 
scanned for the best fitting window, the smallest one still large enough to take up a region of the
necessary size. During this process the necessary alignment of the address according to the PCI specification 
is also taken into account. Finally, when the device itself is properly configured, the function will again
scan the bus for bridges in front of the  device and configure their window sizes accordingly so that
accesses to the device will pass through them.

Another function contained in this library can be used to test open regions
by writing a number of patterns to it, reading back the data, and comparing it with the written data. 
If the data read back does not correspond to the data written, it is read a second time to get an 
indication of whether an error occured on reading or on writing. Four patterns can be written by this
function:
\begin{itemize}
\item {\em Walking Ones}, where one set bit is shifted once for each word written,
e.g. 0x00000001, 0x00000002, \dots, 0x80000000, to each 32~bit memory location.
\item {\em Walking Zeroes}, where one unset bit is shifted once for each word written,
e.g. 0xFFFFFFFE, 0xFFFFFFFD, \dots, 0x7FFFFFFF, to each 32~bit memory location.
\item {\em Full Bits}, where alternatingly data words with all bits set and all bits unset are written,
e.g. 0x00000000 and 0xFFFFFFFF, to each 32~bit memory location.
\item {\em Flipping Bits}, where alternating data words with every odd or even bit set are written,
e.g. 0xAAAAAAAA and 0x55555555, to each 32~bit memory location.
\end{itemize}

The final part of the PCI and Shared Memory software is a set of small user programs for access
to most of the region types described above from a command line shell. This avoids the necessity of writing
separate programs for small infrequent accesses, e.g. basic testing.
The type of region to be accessed, the parameters required for that region type, and the data to be written are all specified via
command line parameters. Data to be read 
is printed out normally. The first program in this set allows read and write accesses to a PCI device's configuration
space or its BARs as well as to physical memory addresses. Data read can be dumped in a format
suitable for input as write data so that it can be written to a file and later to the same
or another region. A second program can perform the test routine described in the previous paragraphs 
on the same region types as the read/write program. The final utility program reads 
an arbitrary data file and dumps it into a format that can be used as write data input for the 
read/write program. This allows any file, for example configuration data, to be written
directly into a specified region.

\section{\label{Sec:MLUC}The Utility Class Library}

During the development of the framework the necessity arose for a number of utility classes with a sufficiently generic functionality
to place them into a separate library, the More or Less Useful Class library (MLUC), so that they can be used in other projects as well. 
Some of these classes encapsulate
existing system functionality with an object oriented (OO) interface while others encapsulate and supply new functionality. A third set of classes was written
to replace or enhance classes from standard libraries either for performance reasons or because the 
standard library class did not work in a multi-threaded environment. 

\subsection{Function Encapsulation Classes}

\subsubsection{\label{Sec:LoggingSystem}The Logging System}

The first set of classes is designed to offer a fast logging system for programs to easily and flexibly 
dispatch informative messages to one or more destinations. Different levels of severity for
messages are supported by the classes, ranging from fatal errors to debugging aids. Three main criteria have been set for
the development of these classes: Normal operation of a system should be influenced by the logging system as little as possible,
therefore the overhead of a message with a deactivated severity level has to be very low. 
Secondly, the system should support multiple message destinations in a manner transparent to the code performing
the actual logging. Destinations should also be changeable at runtime, again transparently to the code using the logging system.
The last requirement for the system was that the severity levels should be selectable independently of each other. In many 
systems of this kind deactivating a severity level causes all less severe levels to be deactivated as well. 

\begin{figure}[hbt]
\begin{center}
\begin{verbatim}

LOG( MLUCLog::kDebug, "LogTest", "Msg 1" ) << "Log test message 1." << ENDLOG;
\end{verbatim}
\parbox{0.90\columnwidth}{
\caption{\label{Fig:LogUsage}A sample usage of the logging system.}
}
\end{center}
\end{figure}

%\begin{table}[hbt]
%\begin{center}
%\begin{tabular}{|l|c|l|}
%\hline 
%Level Identifier & Numerical & Notes \\
%                 & Value     & \\
%\hline \hline
%kBenchmark & 1 & For benchmark results \\
%\hline
%kDebug & 2 & For debugging purposes, should be turned off \\
% & &         in production systems \\
%\hline
%kInformational & 4 & Infrequent information about the normal \\
% & &                 operation of system \\
%\hline
%kWarning & 8 & Signals abnormal conditions that can be \\
% & &           handled by the system but {\bf restrict} \\
% & &           its operation \\ 
%\hline
%kError & 16 & Abnormal condition that cannot be handled by \\
% & &          the system and severly limits its operation \\
%\hline
%kFatal & 32 & A fatal condition that could cause the program to \\
% & &          abort itself or be aborted by the system in the \\
% & &          immediate future. \\
%\hline
%\end{tabular}
%\end{center}
%\caption{\label{Tab:LogLevels}A list of available severity levels for the logging system, sorted by severity.}
%\end{table}

A code example of a logging message is shown in Fig.~\ref{Fig:LogUsage}.
% as an overview of how the system is actually used in programs. 
The first parameter to the \texttt{LOG} call 
specifies the severity of the given message, in this case debugging severity, a list of the available levels 
follows below. 
%is shown in table~\ref{Tab:LogLevels}. 
Following the severity 
are two strings, the first specifying the origin of the message, the second holding keywords categorizing the message.
After the \texttt{LOG} call the C++ stream operator \texttt{<<} is used to pass the actual contents of the message
to the system. To signal the end of the log message the \texttt{END\-LOG} constant is streamed into the system.
%{\bf More details about the background of the log call used will be presented in the rest of this section. }

The requirement for individually selectable severity levels of the system has been implemented by 
using a single bit in a 32~bit number for each level, restricting the number of levels to 32.
%but this is not seen as a problem. 
Currently six levels are used in the system. In order of decreasing severity these are
{\em fatal}, {\em error}, {\em warning}, {\em informational}, {\em debug}, and {\em benchmark}. As these six levels should 
cover most of the occuring applications, the remaining 26 allow enough extension possibilities. Bits and thus levels can 
be set, unset, or queried using the standard bit operations in the C/C++ language. Most of these operations can be translated
directly to one or two low-level machine instructions so that especially the frequently used query operation can be executed
effectively. 
%This effectiveness is important for the first of the design criteria mentioned earlier.

As stated in the first paragraph, the first design criterion of the system was a very low overhead for logging calls with 
deactivated severity levels. The rationale for this requirement comes from the fact that the less severe messages, like debugging messages, are 
frequently used and as a result can occur very often during the runtime of the system. 
These message levels are typically activated only during the development and testing of a
system or in case of errors, and not during normal production, as logging a message is generally very slow on the typical timescales of such systems.
But to retain the ability to diagnose runtime problems it is not desirable to remove them from a production
system completely, and so one takes the compromise of deactivating the severity levels concerned. In case of a problem they
can be activated again during the running of the system. Although it will be impossible to completely eliminate any impact of 
messages with deactivated levels on system performance, it still should be kept as low as possible. This includes
even the avoidance of a function call in such a situation if possible. 
The overheads resulting from various methods of calling a logging system are presented later in 
section~\ref{Sec:LoggingBenchmark}. In a preceeding summary it can be stated that function calls have a much higher overhead than the
method chosen here. However, despite all these efforts to make the system effective it should still be easy to use in programs.

\begin{figure}[hbt]
\begin{center}
\begin{verbatim}

#define LOG( lvl, origin, keyword ) if (gLogLevel & lvl) \
                                    gLog.LogMsg( lvl, origin, keyword, \
                                    __FILE__, __LINE__, __DATE__,  __TIME__ )
\end{verbatim}
\parbox{0.90\columnwidth}{
\caption{\label{Fig:LogMacro}The definition of the main logging macro.}
}
\end{center}
\end{figure}

To achieve the aims of low overhead for unused severity levels and ease of use, a combination of C preprocessor macros, C++ class 
methods, and overloaded operators has been chosen, partly hidden from users. The first important part is the macro 
definition for the \texttt{LOG} call,
shown in Fig.~\ref{Fig:LogMacro}, that hides an \texttt{if}-statement and a method call executed when the condition
in the \texttt{if}-statement is true. This \texttt{if}-statement is mainly responsible for achieving the required effectiveness. 
In the \texttt{if} condition it is tested whether the bit corresponding to the message's severity is set in the \texttt{gLog\-Level} variable.
\texttt{gLog\-Level} is a global variable specifying the activated severity levels in a program. If the respective severity bit is not set,
the rest of the logging statement will not be executed at all. For disabled severity levels the overhead of a logging call thus 
amounts to an \texttt{if}-statement with a test for a set bit. On Intel compatible processors the GNU C Compiler (gcc) (Version 2.95.3)
translates this into the four processor instructions for an i686 (Pentium-Pro or later) processor shown in Fig.~\ref{Fig:LoglevelTest}.

\begin{figure}[hbt]
\begin{center}
\begin{verbatim}

        movl gLogLevel,%eax
        addl $16,%esp
        testb $2,(%eax)
        je .L2612
\end{verbatim}
\parbox{0.90\columnwidth}{
\caption[Processor instructions generated for the log severity level test.]{\label{Fig:LoglevelTest}The four processor instructions generated for the log severity level test.}
}
\end{center}
\end{figure}

The \texttt{gLog} object used in the \texttt{LOG} macro is not an object itself but a global reference. It points to an instance of the 
\texttt{MLUC\-Log} class, the actual interface to the logging system. This object also handles the dispatching of logged 
messages to the different message present destinations. By using a reference instead of the
\texttt{MLUC\-Log} instance directly it is possible to transparently change the log message interface. 
For reasons of brevity the global \texttt{MLUC\-Log} instance will be referred to 
as the {\em \texttt{gLog} object}. 

In the \texttt{gLog} object's \texttt{Log\-Msg} method, called in the macro when a message is logged, the message is prepared. 
Four preprocessor macros are passed to the function in addition to the three parameters passed to the \texttt{LOG} macro: 
The name of the code's originating file together with the line number in the file, and the time and
date when it was compiled. Additionally the current date and time as well as the name of the host on which the program runs are 
stored. To ensure thread safety a mutual exclusion semaphore (mutex) is locked so that only one thread at a time can access the
logging system. As the locking is only performed for messages actually logged, the impact on a running system should be minimal. 
The message content streamed into the logging system is stored in the \texttt{gLog} object until the \texttt{END\-LOG} 
identifier has been streamed. This causes the full message to be assembled and passed to the active logging destinations.
After this has been done the logging mutex is unlocked again, releasing the logging system for access by other threads. 

%The final requirement for the logging system was for the possibility of using multiple destinations for the log messages.
%Activation of these log targets and switching between them should additionally be possible at runtime in a manner that is
%transparent to the application using the system and its programmer. 

\begin{figure}[hbt]
\begin{center}
\resizebox*{0.6\columnwidth}{!}{
\includegraphics{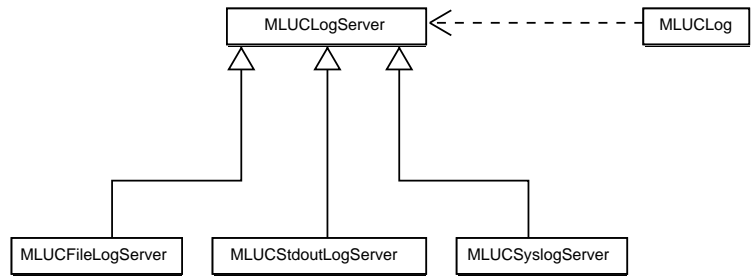}
}
\parbox{0.90\columnwidth}{
\caption{\label{Fig:LoggingClasses}The classes used in the logging system.}
}
\end{center}
\end{figure}

To achieve the final goal of multiple transparent logging destinations for the logging system, it has been divided
into multiple classes: The \texttt{MLUC\-Log} class as the primary interface, the \texttt{MLUC\-Log\-Serv\-er} class as the abstract interface for 
message destinations, and classes derived from \texttt{MLUC\-Log\-Serv\-er} with the actual destination implementations. This division is shown
in Fig.~\ref{Fig:LoggingClasses}.

\begin{figure}[hbt]
\begin{center}
\resizebox*{0.85\columnwidth}{!}{
\includegraphics{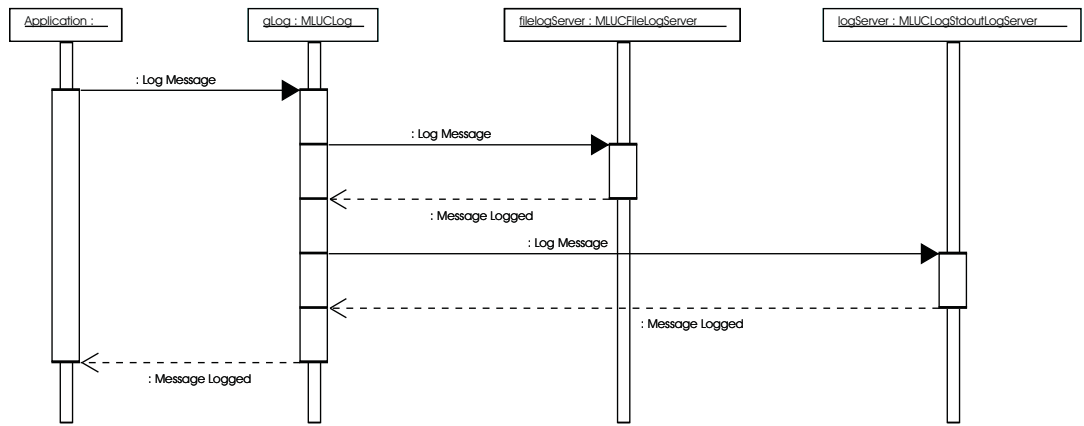}
}
\parbox{0.90\columnwidth}{
\caption{\label{Fig:LogSequence}Sample sequence of calls in the logging system.}
}
\end{center}
\end{figure}

The global \texttt{gLog} object contains a list with pointers to instances of \texttt{MLUC\-Log\-Serv\-er} derived classes. Each of these 
derived classes is responsible for delivering log messages to a certain destination, examples of which are the standard 
output channel of programs, sets of files, or the syslog facility present on Unix computers. 
Instances of the classes corresponding to desired log destinations are registered
with the \texttt{gLog} object, which enters them in its list of destinations. When the program makes a call to the logging system
the complete message is assembled in the \texttt{gLog} object. Having received the message the object iterates over its
list of registered \texttt{MLUC\-Log\-Serv\-er} destination classes, dispatching the message to each of them. After this the destination classes 
are responsible for delivering the message, e.g. by printing it to the standard output or writing it to a file.
The interface to the destinations, the \texttt{MLUC\-Log\-Serv\-er} class, is hidden inside the
logging system. Only when a class for a new destination has to be written 
%does it have 
has it
to be used directly. 

A sample sequence of a logging call is shown in Fig.~\ref{Fig:LogSequence}. In this sequence an application makes a logging call
for a message to the global \texttt{gLog} object. Two instances of \texttt{MLUC\-Log\-Serv\-er} derived classes are registered in
the \texttt{gLog} object: an \texttt{MLUC\-Stdout\-Log\-Serv\-er}, printing messages to the program's standard output, and
an \texttt{MLUC\-File\-Log\-Serv\-er}, writing the messages to a file. The \texttt{gLog} object sends the received message first to the 
\texttt{MLUC\-File\-Log\-Serv\-er} object and then to the
\texttt{MLUC\-Stdout\-Log\-Serv\-er} object. After the message has been dispatched to both registered \texttt{MLUC\-Log\-Serv\-er} objects 
the logging call returns control to the application's code.

Depending on the implementation of the logging servers used logging a message might block a program, e.g. when a disk is full or when a 
network error for a log target node occurs. The currently provided servers will block in such situations, corresponding to their respective
destination. To avoid blocking a log server implementation could make use of a background thread to for sending or writing of the data. Such a solution
however could require the discarding of log messages when this background thread is blocked for too long.

\subsubsection{Multi-Threading Classes}

Most of the programs used in the framework consist of multiple threads for two reasons. Primarily, the motivation is
to prevent programs from blocking completely when one part of a program blocks. This might occur when a system function used to 
communicate with another program blocks, for example because the communication partner has terminated. 
%Such a situation might occur when a program
%communicates with another program and a system function used for this blocks for example because the communication partner has
%been forcefully terminated. 
If on the other hand the program is multi-threaded, only the thread executing the communication can block.
% and not the complete program as in the single-threaded case. 
Care has to be taken of course that no other thread in the program will block 
because of waiting for a certain condition in the communication thread. 

To facilitate multi-threaded programming and especially communication between threads in one program, several classes have been
implemented in the MLUC library. \texttt{MLUC\-Thread}, the first of these, has the purpose of handling of the threads themselves, e.g. starting and 
stopping. It uses the POSIX threads (\texttt{pthreads}) API \cite{IEEEPOSIX}, \cite{SingleUnixWeb}, \cite{SingleUnixOnline}, 
\cite{SingleUnix} and encapsulates it into a class providing 
methods for starting and aborting threads. An abstract (pure virtual) member function \texttt{Run} declared in the class is called when a thread
is started, serving as the actual thread function, so that the thread is terminated when the \texttt{Run} function ends. 
%The \texttt{Run} function has to be overwritten in derived classes with the code that the thread is to execute. 
Creating a new thread as a consequence involves deriving a class from \texttt{MLUC\-Thread} and overwriting
the \texttt{Run} method with the code to be executed by the thread. For integration with functionality in other classes, the template class \texttt{MLUC\-Ob\-ject\-Thread} 
has been derived from \texttt{MLUC\-Thread}. It uses another class type as its template parameter, accepting an object and a member function of that class 
in its constructor. In its \texttt{Run} method the specified member function of the given object is called, making that method the actual thread function.

For purposes of signalling between threads the \texttt{MLUC\-Con\-di\-tion\-Sem} class implements a condition semaphore, also called a signal.
%, has been added to the class library. 
Waiting for signals from other threads is supported by the class either with a specified timeout or without.
When a timeout has been specified, the \texttt{Wait} function returns even when no signal has been received.
Otherwise it will wait indefinitely for a signal to arrive before returning.
To prevent race conditions between waiting for a signal and signalling, the \texttt{MLUC\-Con\-di\-tion\-Sem} uses an internal mutual exclusion semaphore (mutex). It is
acquired by default and is released atomically before a wait is entered. When an attempt is made to signal a thread waiting on this object, the 
mutex  is tried to be locked as well. If no thread is waiting for a signal on the object, this will cause the signalling thread to block until another thread calls
one of the object's \texttt{Wait} functions. To support longer processing sections between waits, without blocking signalling threads, 
the class supplies two member functions for manual locking and unlocking of its internal lock. 

In addition to these signalling features the \texttt{MLUC\-Con\-di\-tion\-Sem} class also supports a notification data structure. This queue consists of a list of 64~bit data
items. Items are added to the end of the list and queried or removed from its beginning. Using the queue makes it possible to provide a thread
with a list of items to be processed by signalling it whenever a new item has been added to the list. To ensure thread safety while
maintaining efficiency the notification data is protected by a separate lock, distinct from the internal lock associated with the signal/wait 
functionality. Internally the class uses the \texttt{MLUC\-Vec\-tor} class covered later in this section, making use of the provided efficiency features of
that class.

For inter-thread communication where exchanging single 64~bit values is not sufficient, a second First-In/First-Out (FIFO) communication class is available.
%providing FIFO functionality for .
The \texttt{MLUC\-Fifo} class offers an interface for the exchange of data of any size between multiple threads. For efficiency reasons the interface is 
optimized so that it is not necessary to have the data available for a write call to be copied into the FIFO. Instead a location of a
specified size is allocated in the FIFO and the pointer to that location is returned. The thread writing into the FIFO can then write its data directly
without having to store it in a temporary location and copying it from there into the FIFO. 
After writing the thread calls a commit function that updates the FIFO's internal tables, 
exposing the written data into the FIFO as available for reading. For thread safety the allocation and commit member functions of the class also perform
locking/unlocking respectively so that only one thread at a time is able to write into it. 

Reading works in a similar manner: When data is available for reading the responsible member function returns a pointer to the start of the available data.
After the reading thread has finished processing the data it calls another function to free the accessed data. The free call also updates the object's internal
tables,  marking the freed space as being available again for writing new data. Similar to writing, reading also involves a locking mechanism ensuring that no two
threads can access the data simultaneously. To allow concurrent reading and writing separate read and write mutexes are used. 

An \texttt{MLUC\-Fifo} object contains two buffers for data, an ordinary and an emergency one, where the emergency buffer is typically rather small. 
On writing data it is possible to specify into which of the two buffers the data should be stored. On reading this is not possible, instead the emergency buffer
is always checked first for the availability of data, and only if it is empty is the ordinary buffer checked. This mechanism ensures that it will
always be possible to send high priority messages to a thread. 

A FIFO's size is initially set to a power of 2 and can be resized by doubling its size if necessary. If the amount of the buffer used drops below a specified 
threshold, e.g. a quarter of its size, the buffer is reduced again to half its size. These measures ensure that a buffer will not suffer from a yo-yo
effect of constant expanding and shrinking when it is used around a resizing threshold. If the resize ability of a FIFO's buffer is not desired, then it is
also possible to disallow resizing completely. In such a case writing to a full buffer will fail.

\subsubsection{Timer Classes}

In multi-threaded systems it can be necessary that threads wait for a specified time while still being interruptible during the wait.
When the thread itself knows for how long it needs to wait, then the timeout \texttt{Wait} method from 
the \texttt{MLUC\-Con\-di\-tion\-Sem} class 
can be used. If however the thread itself does not know how long it needs to wait, then some other method must be used. To solve this problem MLUC provides
the \texttt{MLUC\-Timer} and \texttt{MLUC\-Timer\-Call\-back} classes. 

The \texttt{MLUC\-Timer} class allows to set timeouts associated with a specific instance of a class derived from \texttt{MLUC\-Timer\-Call\-back}. 
\texttt{MLUC\-Timer\-Call\-back} is an 
abstract base class consisting of just one abstract member function, \texttt{Timer\-Ex\-pired}. When a time set in \texttt{MLUC\-Timer} has passed the 
\texttt{Timer\-Ex\-pired} function in the specified instance of the \texttt{MLUC\-Timer\-Call\-back} derived class is called. To provide more information about the timer 
that has expired, a
64~bit value, that can also hold a pointer, can be passed to \texttt{MLUC\-Timer} when the timeout is started. This value is then subsequently passed to the 
\texttt{Timer\-Ex\-pired} function as well. Additionally, it is possible
to remove set timeouts before their expiration and to set new waiting times for active timeouts. 

Internally the \texttt{MLUC\-Timer} class uses a thread class in which the main timer loop runs. This thread also calls the \texttt{Timer\-Ex\-pired} functions of the objects 
registered for each timeout. Implementations of these functions as a consequence have to fulfill two requirements.
Firstly, since the function is called in most cases from a thread different from the one in which the timeout was set, it has to be ensured that the function is 
thread-safe and
that all data accesses are properly synchronized by mutex locks. Secondly, the function should not take too long or even block, as this could slow or even stop
the complete timer loop and its functionality. 

To address both of the above issues and also make the timer functionality better accessible, a third specialized class has been
developed. The \texttt{MLUC\-Timer\-Signal} class is derived both from \texttt{MLUC\-Timer\-Call\-back} and \texttt{MLUC\-Con\-di\-tion\-Sem} described earlier.  
Its implementation of the \texttt{Timer\-Ex\-pired} 
function, inherited from \texttt{MLUC\-Timer\-Call\-back}, adds the supplied 64~bit data value to the notification list inherited from \texttt{MLUC\-Con\-di\-tion\-Sem} and calls that
class's \texttt{Sig\-nal} function. By using the class it thus becomes easy to have a thread wait for events from the timer and other sources at the same time.

\subsubsection{\label{Sec:MonitoringClasses}Monitoring Classes}

Monitoring of a cluster node's parameters is a functionality not especially needed for a data transport framework, but is useful in many other applications. 
A class hierarchy in MLUC allows to monitor many relevant system parameters through a common interface, e.g. CPU load, network throughput, or hard disk 
throughput. At the base of this hierarchy is the \texttt{MLUC\-Value\-Mon\-i\-tor} class that declares an abstract method \texttt{Get\-Value} to read out a
64~bit large system parameter and allows to specify a description for that parameter. Derived classes overwrite the \texttt{Get\-Value} method to read out
and return specific system parameters. Implementations exist, amongst others, for reading out different CPU usage values, incoming and outgoing network traffic,
separately on each network interface or globally for all interfaces, and for measuring the amount of data read from and written to hard disks. 

In addition to the basic functionality of reading out these parameters the \texttt{MLUC\-Value\-Mon\-i\-tor} class hierarchy also contains methods to calculate averages of the
last values read for each parameter, to print the values to standard output, and to write each read value to a file together with a timestamp. Especially
the last capability has been very valuable for performance and correlation analysis of programs used in the framework.

\subsubsection{\label{Sec:FTHandler}The Tolerance Handler Class}

%Used internally in the \texttt{Ali\-HLT\-Tol\-e\-rant\-Round\-Robin\-Event\-Scat\-terer} class, the \texttt{MLUCTol\-e\-rance\-Han\-dler}
%class from the MLUC library is responsible for managing the different paths in the event scatterer.
The \texttt{MLUCTol\-e\-rance\-Han\-dler} class is used internally in some of the fault tolerance components presented in section~\ref{Sec:FTComponents}.
It is able to manage a given number of items that can be either functional or non-functional, 
for example corresponding to processing nodes in the HLT. For a task, identified
by a 64~bit index number, one of the items to which it is assigned can be determined. When all managed
items are functional the worker item is obtained by a simple modulo operation on the task's index number with
the total number of items available (functional and non-functional). 

\begin{figure}[hbt]
\begin{center}
\resizebox*{0.50\columnwidth}{!}{
\includegraphics{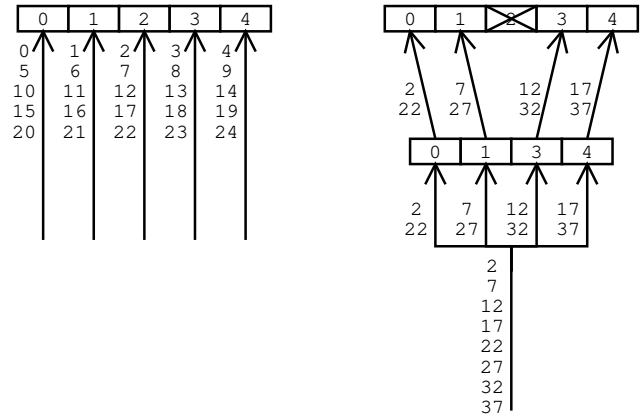}
}
\parbox{0.90\columnwidth}{
\caption[Dispatching principle in the fault tolerance handler class.]{\label{Fig:FaultToleranceHandler1}The dispatching principle in the fault tolerance handler class
for five items. On the left side the task sequence is shown for all five items functional. On the right
item 2 is non-functional, here only the tasks that would have been assigned to item 2 are shown with
their distribution among the remaining nodes.}
}
\end{center}
\end{figure}

When one or more of the items are non-functional, additional steps have to be taken. Using the above modulo 
operation's result, a check is always made whether the item found is functional or not. If it is functional
the task is assigned to it. Otherwise the next step is taken. The task's index number is divided by 
the total number of items available. A second modulo operation
is performed on this division's result, based upon the number of functional items. The number obtained
from this operation is used as the index for a map array. In this array the indices of the available 
functional items are contained. From the array the item to which the task is assigned is determined by the second index. 
Fig.~\ref{Fig:FaultToleranceHandler1} shows the principle for five items, on the left with all items 
functional and on the right with item 2 non-functional. 

\begin{figure}
\begin{center}
\resizebox*{1.0\columnwidth}{!}{
\includegraphics{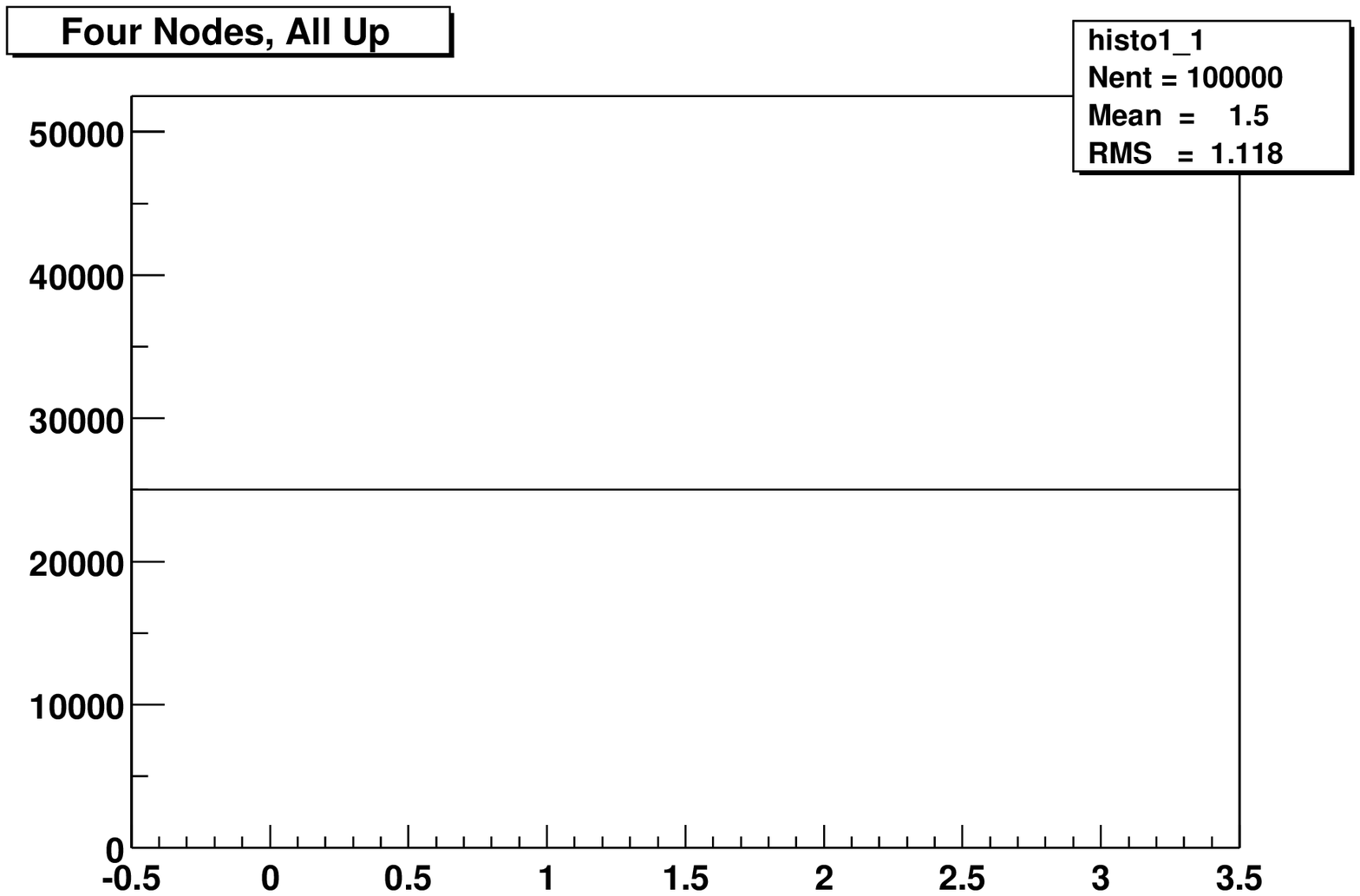}
\hfill
\includegraphics{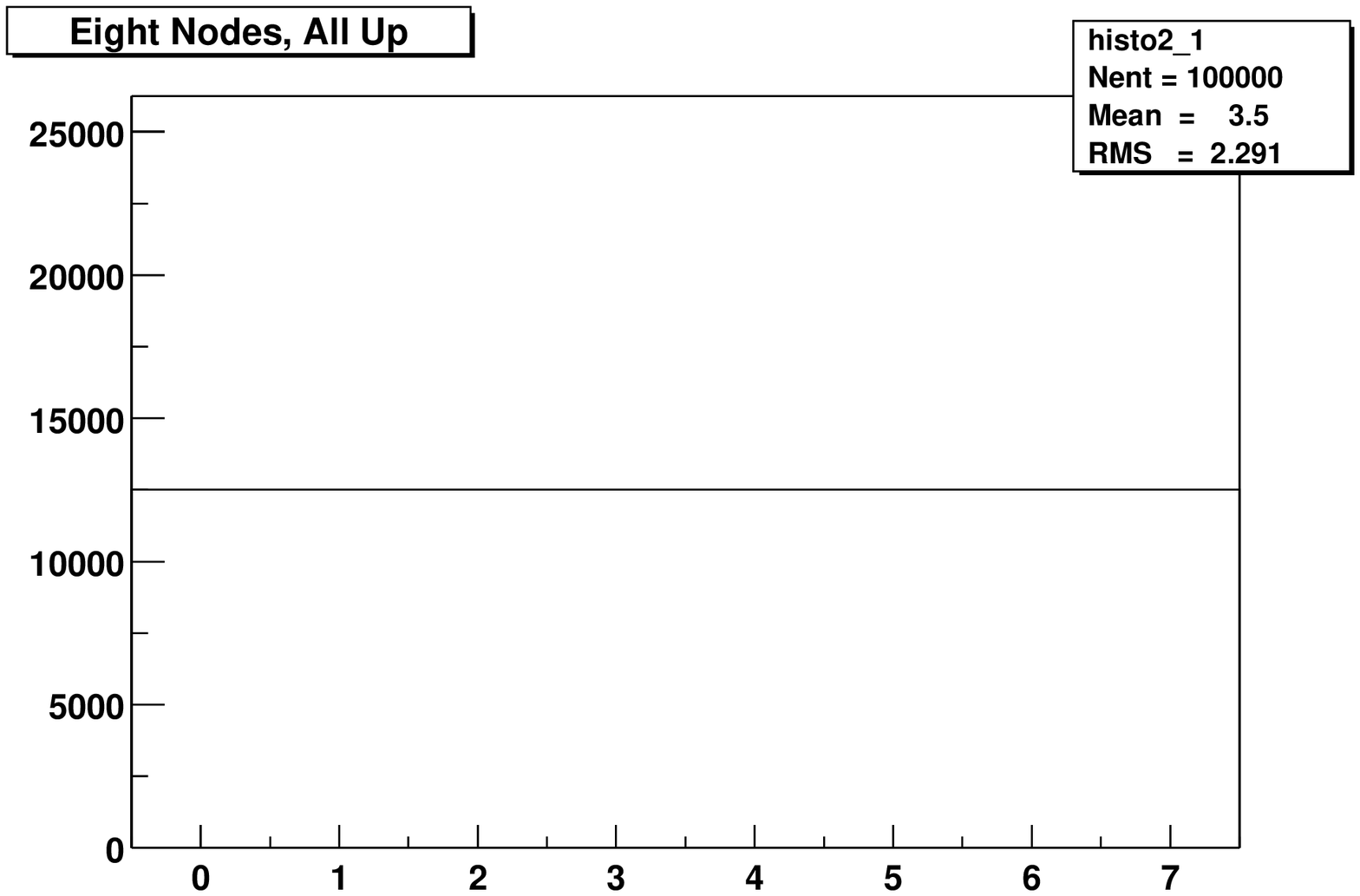}
}
\resizebox*{1.0\columnwidth}{!}{
\includegraphics{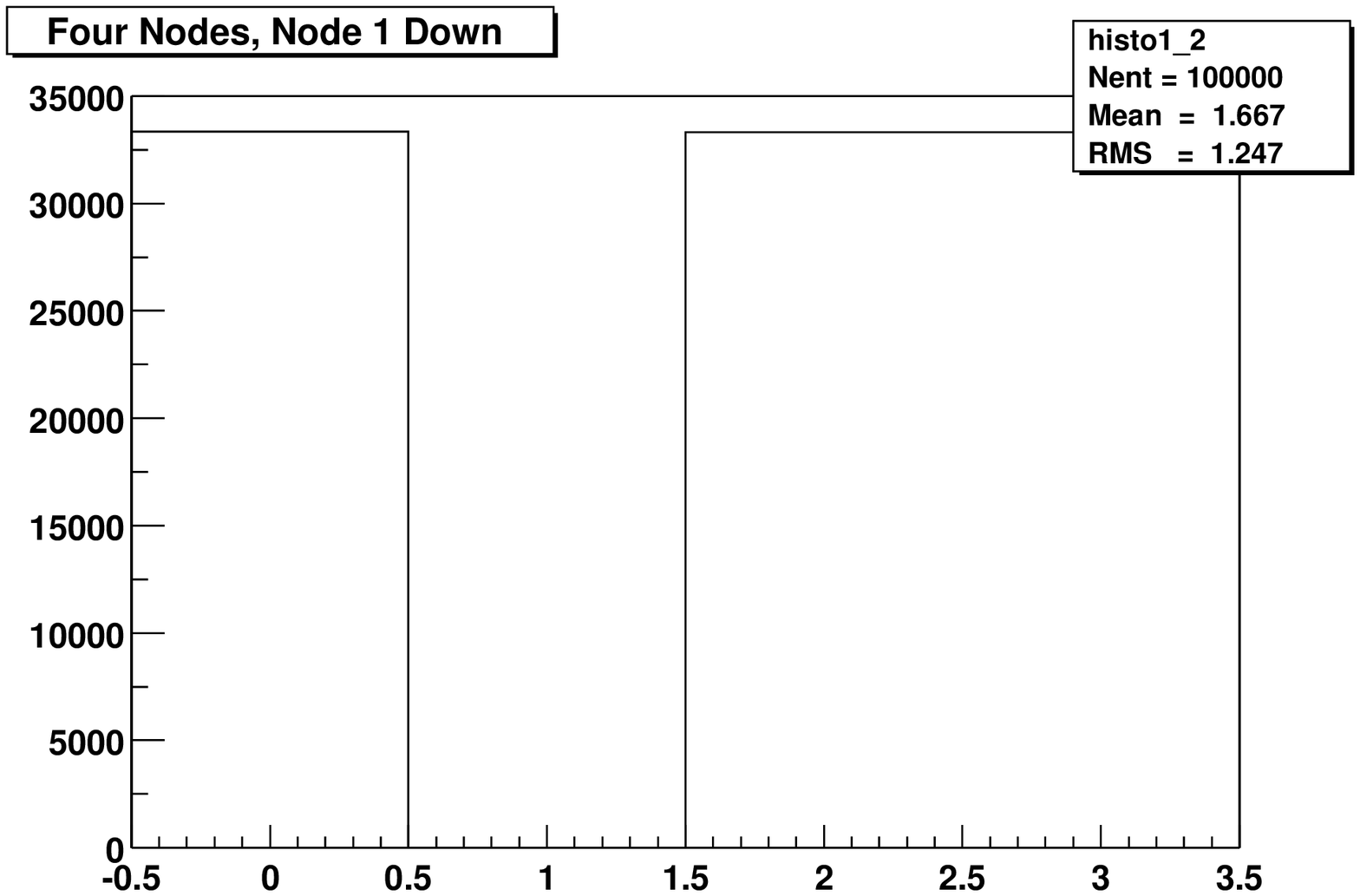}
\hfill
\includegraphics{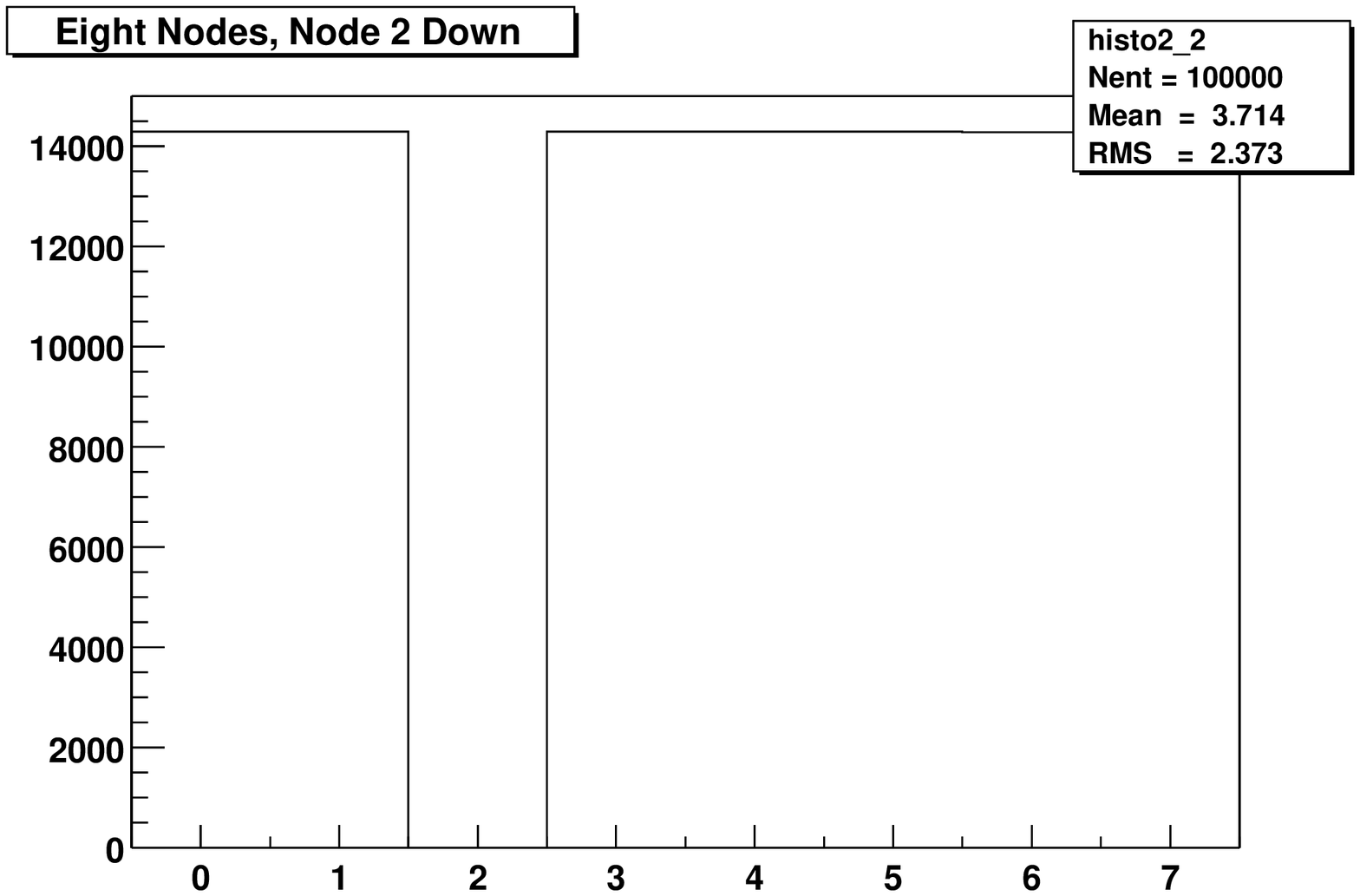}
}
\resizebox*{1.0\columnwidth}{!}{
\includegraphics{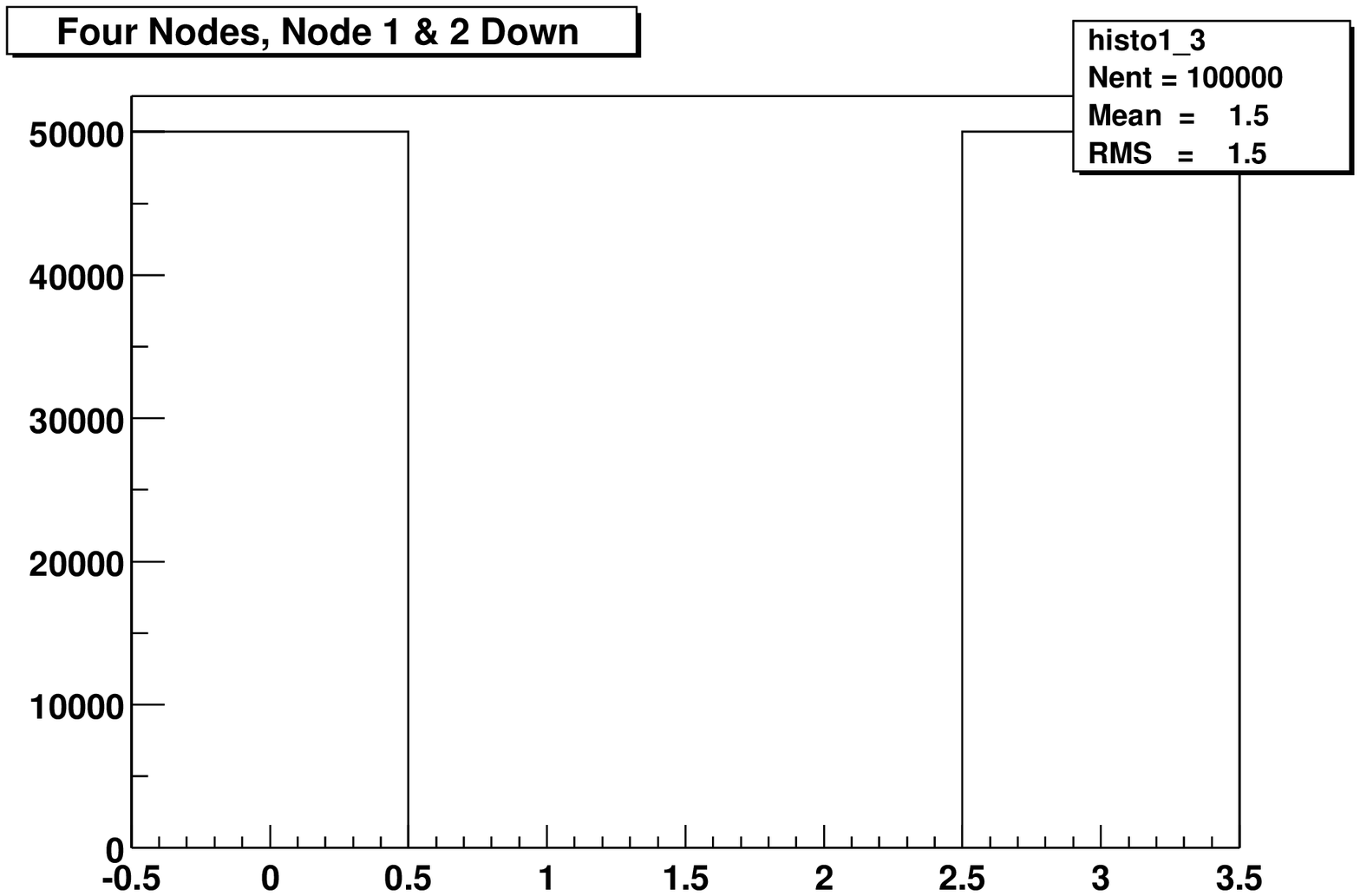}
\hfill
\includegraphics{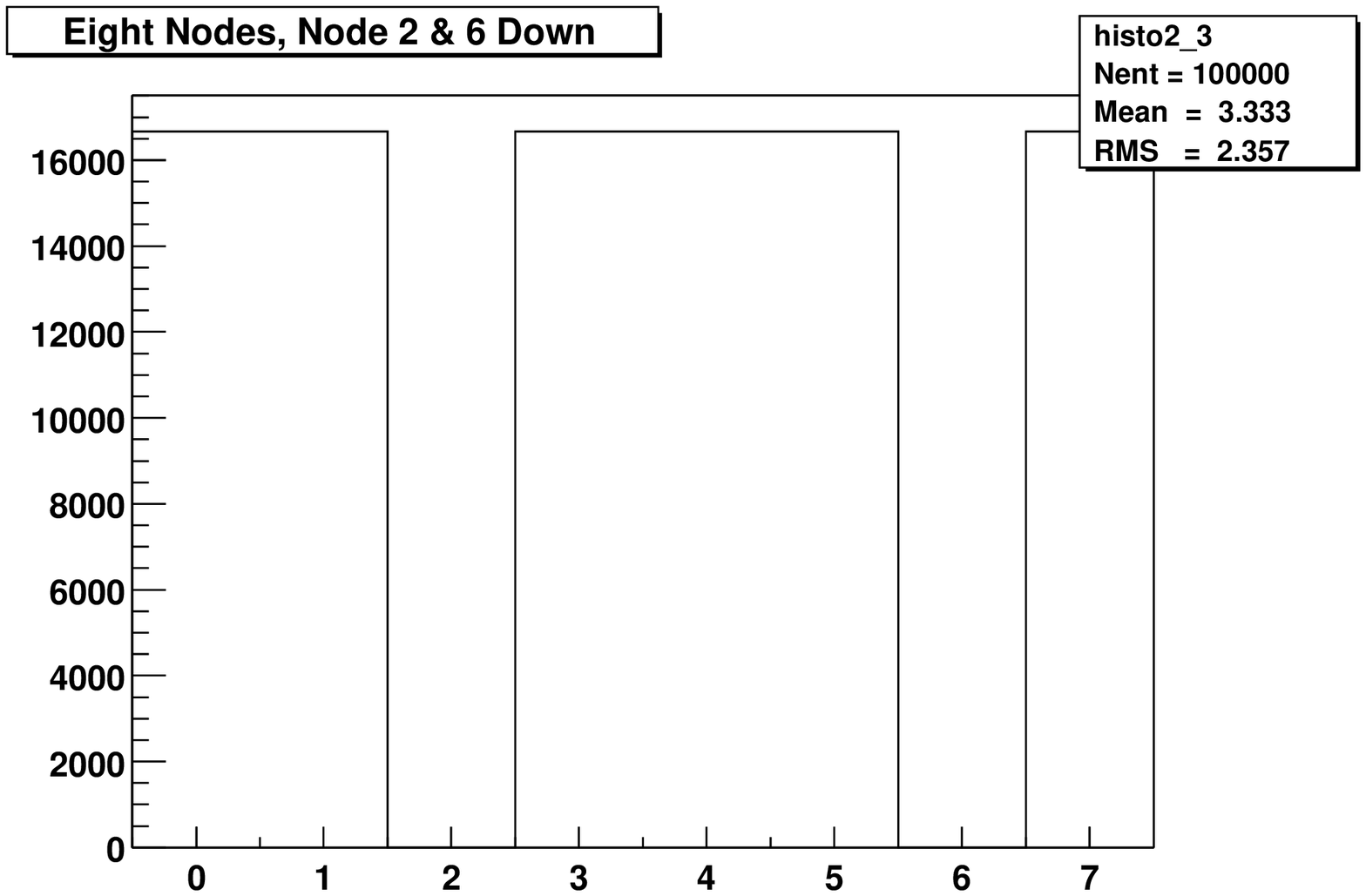}
}
\parbox{0.90\columnwidth}{
\caption[Event distributions of the fault tolerance handler class.]{\label{Fig:FTScattererDistributions}Sample event distributions of the fault tolerance handler class for four nodes on the left and eight on the right.
In the two top histograms all nodes are available, in the middle ones one node has failed, and in the bottom one two nodes have failed. The x-axes show the node index number
and the y-axes the number of assigned tasks.}
}
\end{center}
\end{figure}

With the above rules, tasks assigned to functional items are not affected, while all tasks that would have
to be processed by non-functional items are redistributed among the available ones. %{\bf \LARGE Problems??}
To show that in the case of errors the distribution is done evenly, a small test program has been written
that simulates a number of item errors and fills histograms for each item with the number of tasks
assigned to it. Sample distributions for a number of parameters are shown in Fig.~\ref{Fig:FTScattererDistributions}. 

%\begin{figure}
%\begin{center}
%\resizebox*{1.0\columnwidth}{!}{
%\includegraphics{FTScattererDistribution-4-4.eps}
%\hfill
%\includegraphics{FTScattererDistribution-4-3.eps}
%\hfill
%\includegraphics{FTScattererDistribution-4-2.eps}
%}
%\resizebox*{1.0\columnwidth}{!}{
%\includegraphics{FTScattererDistribution-8-6.eps}
%\hfill
%\includegraphics{FTScattererDistribution-8-7.eps}
%\hfill
%\includegraphics{FTScattererDistribution-8-8.eps}
%}
%\end{center}
%\caption{\label{Fig:FTScattererDistributions}}
%}
%\end{center}
%\end{figure}

\subsection{Functionality Replacement Classes }

\subsubsection{\label{Sec:MLUCVector}The MLUCVector Class}

In the framework many uses of a dynamic array class involve an almost FIFO-like behaviour where new values are added to the end
of an existing list. Access and removal of values, however, is not always strictly from the beginning but can in extreme cases be from the end as well.
Typically though, removal is done from the first few elements of a list.  Due to this non-strict removal from the head of the list, the queue class from
the C++ Standard Template Library (STL) is not suitable, nor is any other queue class. A dynamic array class, allowing random
access, is used instead. Although the STL list class is also usable in principle, tests have shown its performance to be slower than the vector class, probably because
of the large number of element allocation and deallocation operations performed.

Unfortunately the STL vector class has a major drawback in this usage pattern. Whenever an element is removed
all elements located after it in the list are shifted one slot
forward. With a potentially very large number of events in the system coupled with a high rate this leads to a large
number of copying operations that have to be executed in a node. For example, with event sizes of 8~kB, an event buffer size of 256~MB,
an event descriptor size of 32~byte, and an event rate of 200~Hz, about 200~MB/s will be copied in memory just as a result of handling
a list of event descriptors. To overcome this problem the new dynamic array class \texttt{MLUC\-Vector} has been designed as a replacement 
class for the MLUC library.

Like the STL vector class the \texttt{MLUC\-Vector} class is a template class with the template parameter defining the type of data stored.
Unlike the STL class the \texttt{MLUC\-Vector} uses a preallocated array as a ring buffer with a number of elements equal to a power of 2. 
In addition to the array for storing the contained elements themselves another array for a similar number of boolean elements is used
to specify the validity of each element slot in the primary array. If the valid flag for a corresponding element slot is false, the slot
is unused. 

The advantage of using powers of 2 as sizes for a ring buffer is an easy wrap around handling. All operations on indices, e.g. the increment
of an index for looping over all contents, are followed by applying a bitwise \texttt{AND} operation with a specific mask. This mask is the number of available
elements minus one, corresponding to set bits for all valid indices in the buffer. Indices that have become too high or low by a previous operation are thus
automatically truncated back to valid values. As the \texttt{AND} operation is typically very cheap, it often can be executed in one clock cycle by current
processor types. It often is even cheaper to apply than to use the ordinary check for index wrap around.  The combination of a comparison (or subtraction) 
operation coupled with a conditional jump that corresponds mostly to this check make these instructions typically at least as
expensive as the \texttt{AND} operation applied in this class. 

When an element is added to an \texttt{MLUC\-Vector} object, it is inserted at the position specified by the end index of the ring buffer which 
is subsequently increased by one. In addition, the valid flag for that location is set to true, indicating that the slot is now used. For read accesses the index
of a specific slot may be given, allowing random access to each element. To search for elements corresponding to specific
criteria search functions are available that iterate over all valid elements using a caller supplied callback function for element
comparison. 

Removal of elements is done by specifying ring buffer indices, whose corresponding valid flags are then set to false. If the element to
be freed is the first or last element, the appropriate index is increased or decreased respectively. When invalid elements are adjacent to 
such a freed boundary element the corresponding index is increased until a valid element is encountered. 

When no free space remains between the ring buffer's end and start indices two different actions are possible. If the ring buffer contains invalid elements,
the buffer is compacted by shifting the valid elements together. After this operation the ring 
buffer consists of two separate blocks, containing all used and all unused element slots respectively. A buffer without invalid elements can be resized
by doubling the size of its internal buffer arrays. Moving of elements can be necessary after a resize if the used element block
wraps around the end of the buffer. In this case it has to be moved to the new end of the enlarged buffer. 
As soon as the number of used slots in a previously enlarged \texttt{MLUC\-Vector} object drops to less than a quarter of the 
available slots the buffer is compacted again and its size is halfed. A buffer is never shrunk below its originally specified size. 
If resizing is not desired or necessary, then it is possible to set a flag in the constructor that inhibits resizing, both growing and shrinking, for the
object concerned.

With the described features of the \texttt{MLUC\-Vector} class it is ensured that the available resources, especially memory bandwidth, are used optimally for the dominant
access pattern specified above. Correspondingly the change from the STL vector class to the MLUCVector class has brought a significant increase in the framework's speed. 

\subsubsection{Allocation Cache Classes}

To avoid copying large amounts of data, or small amounts very often, many parts of the framework only store pointers to data instead of the 
data itself. Only these pointers are passed between functions or different threads. This approach, however, has another problem of 
frequently issued allocation and deallocation calls, that usually are costly as well if used in such large numbers.
In response to this problem two more classes, \texttt{MLUC\-Al\-loc\-Cache} and \texttt{MLUC\-Ob\-ject\-Cache}, have been introduced into the MLUC library to
prevent these frequent calls to the memory subsystem. 

Both classes allocate a specified amount of elements on creation and store them in a pool of available elements.
Instead of calling the normal memory allocation routines, e.g. \texttt{malloc} for C or \texttt{new} for C++, a program calls the allocation
routine of one of these classes. This routine checks whether there is at least one element available in its pool and returns 
a pointer to the first available element if this is the case. The element's pointer is then removed from the available pool and stored in a list of used elements.
If the pool of preallocated elements is exhausted, the allocation objects use the system allocation routines to 
obtain the requested element. A pointer to this allocated element is stored in another list for additionally allocated elements. 

When an element can be freed again the allocating object's release function is called. If the element was allocated additionally, because the 
preallocated pool was exhausted, it is removed from the list it was stored in and is freed again, using 
the appropriate system deallocation call. For elements that originated from the preallocated pool, the pointer
to the element is removed from the list of used elements. Subsequently it is reinserted into the pool to make it available for further use. 

For both of these classes the amount of elements to preallocate is specified as a power of 2 and the \texttt{MLUC\-Vector} class is used internally 
to store all element lists. The lists of available and used pool elements are both presized to the number of preallocated elements so that
from the beginning both lists have enough space available to store all pool elements. No resize will be necessary for them. By using the 
\texttt{MLUC\-Vector} class in this manner, the two classes benefit from its low overhead features and can handle operations on their internal lists
efficiently. 

The \texttt{MLUC\-Ob\-ject\-Cache} class is a template class with the template parameter defining the type of elements for which the allocation object
functions as a cache. Objects of the given type are preallocated, including executing their default constructor, and are stored in the allocation object's pool.
When a pool object is released again, it is not directly reinserted into the pool. Before the insertion one of its methods, \texttt{Re\-set\-Cached\-Ob\-ject}, is called.
This method's task is to reset the object into a clean state, corresponding to its state immediately after creation, 
ensuring that a used object can be reused by the calling program without a check for a usable state. A consequence of this mechanism is that the \texttt{Re\-set\-Cached\-Ob\-ject}
method must be present in each class to be managed by an \texttt{MLUC\-Ob\-ject\-Cache} instance. 

In contrast the \texttt{MLUC\-Al\-lo\-ca\-tion} object manages only blocks of memory of a specified size. The block size is specified in the object's constructor together
with the amount of elements to be stored. Elements contained in an \texttt{MLUC\-Al\-lo\-ca\-tion} instance are neither explicitly overwritten with zeroes on 
creation nor upon release, again to save memory bandwidth.
A program using an object of the \texttt{MLUC\-Al\-loc\-Cache} class thus can make no assumptions about the content of memory blocks received from it. 

\subsubsection{The String Class}

Many pieces of code in the framework have to store a name or another type of textual information. Compared to the traditional C
handling of the string type C++ string classes allow a significantly easier handling of these texts. In conjunction with the STL string class the 
discussed multi-threading of the framework
poses a serious problem, as this class uses internal static members, globally shared between all objects of its type. 
Presumably for performance reasons these members furthermore are not protected by mutex semaphores.
These global data items can thus be accessed and even changed simultaneously by multiple threads in a program. This behaviour has led
to several very hard to trace bugs during the development of the framework until the real cause of the problem was found.

To work around the problem a primitive replacement string class, \texttt{MLUC\-String}, has been written and included in MLUC. This class is very simple, providing
only the most basic functions to ease handling of textual data. Its function names also do not conform to the standard string class functions,
as the aim was not to provide a complete reimplementation of the STL standard \texttt{string} class. Instead the decision was made for consistency reasons
to have this string class conform to the naming conventions used in the MLUC class library as a whole.

\clearpage

%%%%%%%%%%%%%%%%%%%%%%%%%%%%%%%%%%%%%%%%%%%%%%%%%%%%%%%%%%%%%%%%%%%%%%%%%%%%%%%%%%%%%%%%%%%%%%%%%%%%%%%%%%%%%%%%%%%%%%%%%%%%%
%%%%%%%%%%%%%%%%%%%%%%%%%%%%%%%%%%%%%%%%%%%%%%%%%%%%%%%%%%%%%%%%%%%%%%%%%%%%%%%%%%%%%%%%%%%%%%%%%%%%%%%%%%%%%%%%%%%%%%%%%%%%%

\chapter{\label{Chap:ComClasses}The Communication Class Library}

\section{Overview}
One of the
% prime
concepts of the ALICE High Level Trigger is to purchase the necessary components rather late, to take
advantage of new technological developments. This concept should not only be applied to the cluster nodes themselves but to the network
used for the interconnection of the nodes as well. To be able to support this aproach and, of equal importance, to maintain the generality of the 
framework, the communication technology and protocol used for communication between processes on different nodes have not been fixed. 
Instead a C++ class library has been developed that
exports an abstract communication API with implementations for two network technologies. The API provided by this
Basic Communication Library (BCL) has been designed to be generic and independent of any specific network technology. 
Despite this generality the API has also been designed so that implementations are able to make use of
low-overhead, high performance, or efficiency capabilities present in the respective underlying network technology or protocol
used. After evaluation the communication packages described in section~\ref{Sec:ClusterSoftware} have not been considered as a 
basis for communication in the framework due to their different requirements, characteristics, and intended uses. 

The API is split up into two parts, each optimized for a different communication pattern. Its first part 
is designed for the transfer of small amounts of data corresponding to the sending and receiving of short
messages. For these small amounts of data the use of special transfer types like DMA is typically too much overhead so that they
should be sent via Programmed I/O (PIO) transfers. The other API part is designed for the transfer of Binary Large OBjects (BLOBs),
large blocks of data, in one
transfer. For these block transfers any overhead needed for special transfers, e.g. DMA, is considered to be negligible compared
to the actual transfer of the data. These special transfer mechanisms are thus acceptable and even desirable if they provide
a lighter load on the host CPU and/or memory system. 

To demonstrate the actual generality of the library's API and also to provide a usable communication subsystem, two implementations
providing the API's functionality have been written. The first is based on the Transmission Control Protol (TCP) \cite{RFC793} as the most widespread 
network protocol. Its prime advantage is its availability on practically every computing platform and that most Unix variants, 
including Linux, contain a very robust and efficient implementation. For the hardware used with this protocol, Fast or Gigabit
Ethernet are very widespread,  with very cost-effective adapters being available for standard PC nodes. TCP's disadvantage is its 
high overhead compared to dedicated System Area Network adapters, partly due to its design as a Wide Area Network (WAN) protocol over 
unreliable connections. 
The second communication implementation is provided, although only in a prototype form, for the SISCI API \cite{SISCI} on top
of Dolphin SCI SAN \cite{SCIIEEE}, \cite{DolphinWeb} interface cards. These SCI adapters are shared memory interface cards with a network bandwidth of more than 650~MB/s and
latencies of below $2~\mu \mathrm s$. Their primary disadvantages are the comparatively high price, which almost doubles the cost of a node compared
to Gigabit Ethernet, and their weaker default reliability. Unlike TCP the SISCI API does not provide a reliable data delivery and
packet loss has been observed although the physical layer specification guarantees packet delivery. 

Various computers exchanging data can organize that data in different formats. The most common problem
being the byte-order of stored multi-byte integer values. To avoid this type of problem when transporting data a number of helper classes and structures
have been created that enable automatic translation of data between different storage formats. To take advantage of these capabilities the 
data types concerned have to be declared using a special type definition language. This language is then translated into normal C++ code 
by utility programs provided in the library. 

\section{\label{Sec:CommunicationParadigms}Communication Paradigms}

Several of the design decisions and paradigms chosen for the Basic Communication Library are different from design
characteristics found in common network APIs, notably the socket API used for TCP. To avoid misunderstandings and 
confusion in later chapters these design decisions will be presented here.

\subsection{\label{Sec:BCLGeneralDesign}General Design Features}

%{\bf \Large Address URLs??}

A primary general design feature of the BCL library is that all data transport operations, both for message and
block communications, can be executed with or without a previously established connection. In the library support is provided
for establishing connections between two communication objects, but its use is not mandatory. If a data transport
to a remote partner is performed without an explicitly established connection, then a connection will be set up implicitly for that
transfer if required by the underlying protocol. After the transfer has completed the connection is terminated again. 
For a transfer to a remote partner, to which a
connection is already established, this connection will be used to transport the data. 
The motivation behind this scheme is that in large systems it might be necessary to exchange messages with a large number 
of communication partners at a low rate. For these infrequent exchanges it would be too much overhead to establish open connections to all potential
partners, if it would be possible at all and would not be restricted by the operating system. A user application should not be required  
to establish a connection manually for each of these transfers as that would complicate the application's program unnecessarily. On the other hand,
there may be frequent exchanges of data with other remote communication partners. For these the overhead of
establishing a connection for each transfer has to be avoided, and a connection should be established only once. Therefore both explicitly 
as well as implicitly established connections are supported by the library. 

One additional property supported for explicit connections is the on-demand connection. This means that a %explicit
connection is not actually established immediately when the connection attempt is made. Instead the connection address is entered into
an internal connection list but is marked as not yet established. When the first data transfer to this address is started, the connection
is checked and found not to be established yet. As for an implicit connection, it is then established as part
of the send operation. Unlike in the case of implicit connections, however, the connection remains established and is not terminated at the
end of the operation. Like other explicitly established connections it has to be closed explicitly by the calling program as well
when it is not needed anymore.

Another decision for the library is partially influenced by the above requirement for both 
%explicitly and implicitly established connections 
con\-nec\-tion-less and con\-nec\-tion-based data transfers. Prior to any data send operation, each
communication object must have been assigned its own receive address and must have performed a bind operation on it in order to make its address
available to external programs. There are two reasons for this demand. The first of these, derived from the
support for optional connections above, is that each program should be able to receive answers to messages it transmits. For the connection-less
send mode the receiver cannot use the back-channel of a connection established from a remote object to it. %The second reason is of a more general nature. 
In some instances sending of 
 data requires a unique identifier in the system, e.g. to regulate access to a resource on a remote node.
This remote identifier is trivially obtained by using a valid receive address for a specific network technology, which has to be unique 
by design. To ensure that a given address is actually valid and therefore unique a successful bind operation has to be performed on
each communication object's address before it can send data.

For some types of connections it might become necessary to perform some handshaking or negotiating before sending data even when using an already 
established connection. One example is SCI where multiple senders have to regulate access to a shared memory segment provided by a receiver process.
%The overhead involved in these negotiations can be unaffordable or undesired for specific types of high performance applications 
%where message exchange should be as quick as possible. 
Only in cases where a point-to-point connection is used between two
communication partners, it is possible to avoid that overhead for each send operation. In these cases an already established connection 
can be locked and later unlocked by a sending object via two library functions. Locking of a connection makes the receiver object exclusively available 
to the locking sender object. No other sender object can send data to this receiver object even if a connection
to it is established. This negotiation requirement is specific for ShM networks like SCI.
Therefore the functions for locking and unlocking do not have to contain any functionality. A further function is provided so that an application program
can determine whether a given communication object supports locking or not and can make use of the other functions as appropriate. 

With regard to the handling of errors the choice has been made for a combination of return values and error callbacks. Every function in the 
library's API returns an integer value of zero on success and a non-negative value describing the error that occured otherwise. These error
values are taken from the standard C \texttt{errno.h} header file with the advantage that the preexisting C standard functions to 
convert the integer values to error descriptions can be used directly without any additional effort. In addition to these return values, that
have to be evaluated explicitly, another method of detecting errors is available  based on error callback objects 
registered with communication objects. The callback classes are derived from one common base class,
% \texttt{BCL\-Error\-Call\-back}
that exports an interface of methods called for the various error types. If an error occurs inside a communication object
it calls the appropriate function for all its registered callback objects. Parameters passed to these error callback
methods include a pointer to the originating object and an integer value describing the error, identical to the error return value 
returned by the function. In addition to these two basic parameters further arguments are passed if necessary, e.g. the remote address
for a failed connection attempt. After all callback objects have been called, the communication object's function returns with the integer error
described above.
Exception handling has not been chosen for error handling to keep its use optional and not make it mandatory. It can be used by providing
a callback object that throws an appropriate exception when one of its error functions is called. 
Beyond the callback objects registered statically with each communication object most of the communication objects' function calls can accept
an optional argument representing a pointer to an error callback object to be used in addition to the registered objects.

\subsection{Message Communication Design Features}

For the design of the message communication classes and their interface only a few design choices have been made. The primary design 
choice is to base the design on the pattern 
of sending and receiving of messages rather than on a stream of bytes, as e.g. for the standard socket API.
%Furthermore, the primary characteristic of the message interface is that send and receive calls both exist in two versions, one
%including a timeout to be applied to the operation and the other one without, to wait indefinitely for an operation's completion.
Send and receive calls can include a timeout to be applied to the operation, which can be infinite.
In most such cases, however, a fixed timeout of the underlying communication technology used, will expire and cause a
system function to return with an error. 

Another feature is actually more a requirement than a design decision. As a calling program cannot know in advance the size
of a message received the allocation of the memory space for that message has to be made by a communication object's receive
method. By extension it also has to free the message again after the calling code has finished processing it, for which 
there are actually two reasons. The primary reason is that a calling program does not have to make any assumptions about the
memory allocation function, e.g. \texttt{new}, \texttt{new []}, or \texttt{malloc}. Therefore the library has the
liberty to choose which function to use and to change it without affecting a user's program. The second reason is that for some
network technologies it might be possible to store a message in an internal buffer which may even be done directly by the network hardware. The
object then justs returns a pointer into that buffer without the steps of allocating memory and copying the message. Such a message is not 
freed using any system function but instead by the buffer management for the object's internal buffer.

\subsection{\label{Sec:BlobParadigms}Blob Communication Design Features}

For the blob communication mechanism more characteristics have been specified than for the message classes. Initially, a user
may set the size of the buffer where received data will be stored so that user code can access it. For some of the blob communication implementations
%for the different network types
 it may in addition be possible to specify the receive buffer itself. 
However, this feature might not be supported by a specific implementation of the blob interface, one example 
is the existing SCI implementation. Although it might not be possible to specify the receive buffer directly, a user always has the 
possibility to obtain a pointer to the receive buffer from the communication object. This enables it
to write any received data into the receive buffer directly, from where it can be accessed by user programs. No additional copy step
is required to copy the data from the communication object. Instead a user can directly access 
data that has been received from a remote node via the receive buffer pointer.

To enable this type of direct transfers it is necessary that the sending node knows beforehand where the data should be stored in the 
receive buffer and whether it is not already full so that the data cannot be stored at all. For this the sending process is 
split up into two parts each contained in its own function. In the first step an ID for the transfer is obtained by specifying the
size of the data to be transferred. This transfer ID is queried from the remote receiving object, using
an optional timeout, and then passed back to the user program.
With this transfer ID the program can then use the second function by supplying it with the obtained ID and a pointer to the data to be sent. 
The communication object now has a receive buffer location associated with this transfer ID, and transfers the data to that location
in the remote node. For this sending process a gather call is available where the data to be 
transferred is collected from multiple blocks scattered in memory. The size of data for which the transfer ID is obtained must of 
course be the sum of the sizes of all data blocks.

A transfer ID obtained for such a transfer is not automatically transmitted  to the receiver object or program. Instead a user program has to pass it
explicitly to its communication partner, most likely by using a message communication object. In the receiving program it is now possible to
use the transmitted transfer ID to get access to the transferred data. Using the ID one can either obtain a direct pointer to the data or
an offset to the data from the start of the receive buffer. With these informations a user program can access 
the data and process it as required. Once this is finished another communication object function has to be called 
to free the buffer block in which the data was stored. This block is again identified by passing the transfer ID used. 

For the communication required between two blob objects, e.g. to negotiate a transfer ID, each blob communication object 
makes use of the facilities offered by the message communication mechanism instead of using an internal one. A message
communication object is assigned to each blob object to be available exclusively to that blob object, implying that this message
object must not be used by the user program. The advantage of this approach, besides avoiding duplicate
development, is that the message communication may use another network technology than the blob communication. One
example is if two technologies exist, one with low latency but comparably high overhead and the other with low overhead and higher latency. 
In such a case the 
high-overhead/low-latency technology could be used for the message exchange and the low-overhead/high-latency one for the blob transfers.

To achieve an even lower overhead of sending with the avoidance of the additional latency incurred by the negotiation phase before each
transfer a special approach can be taken. A sender can allocate a large block of a remote buffer in advance by requesting a large transfer block.
This block could be as large as the whole buffer which is made possible by using a function that queries a remote node's receive buffer size.
The transfer ID of this block is sent once to the remote object, which stores it for future use. From this point on the sending program
can perform a completely local buffer management in its obtained block and only sends the offsets in that buffer to its receiving node.
This completely avoids waiting for reply messages from the receiver and decreases the time overhead associated with each data transfer 
by twice the message sending latency. 

\section{Auxiliary Classes}

\subsection{\label{Sec:DataFormatTranslation}Data Format Translation}

One of the main problems encountered in network communication on potentially heterogeneous systems is the different
format of stored data, most often encountered in the form of different byte orders for integer
data. To work around this problem, a helper hierarchy with one base class and structure is included in the BCL. The base structure,
\texttt{BCL\-Net\-work\-Data\-Struct}, provides a header for derived data to be stored. Derived types
are used to actually store the data.  Code and meta-data required to execute the translations is contained in the
base class as well as its derived classes. 
In the header three elements are stored to provide information about a structure's original data format at creation, its current
data format, and the total length of the data structure. Since the native data format of any given system node is trivially known it is always 
possible to convert data, described by this format, into a node's native format. By also supplying the data's original format, it becomes 
possible to handle data not directly under the control of this mechanism as well. This cannot be done automatically anymore though. 
Including the total length of the structure furthermore enables all software stages to know how 
much data they have to handle without having to know its actual content. 

In the basic class \texttt{BCL\-Net\-work\-Data} a number of static member functions are provided to transform integer data of different sizes 
between data formats. For each of the integer sizes 16~bit, 32~bit, and 64~bit two functions are available to convert the data.
One can be used to convert the data in place. The other one works with separate source and destination, copying the data
during the translation process. In addition to these static functions, the class provides further functions to aid the
handling of different formats of data. Importing of data structures into a class is supported by different methods either at an object's 
creation using its constructor or by calling member functions later. 
At creation only two possibilities for copying the data into the
class's internal data structure exist: transforming it to the node's local format in the process or copying the data untransformed. For an existing
object it is possible either to copy the data structure, as on creation, or to {\em adopt} it by setting an internal pointer 
to the structure. Both approaches can optionally be combined with the same data transformation possible for the constructor. The advantage of the second approach 
is that it avoids the overhead of copying data, which for large amounts of data and/or high frequencies of transforming data can be quite 
significant.
After data from a network data structure has been imported into an object, functions exist, that allow to transform the data, either
to its original format, the node's current format, or a user specified data format. In addition it is possible to query both the data's
original as well as its current data format.

Data types to be managed by this mechanism have to be declared using a very simple type definition language
(vstdl), translated into normal C++ code by a program in the BCL library. The language
supports only plain 8, 16, 32, and 64~bit sized unsigned integer types. Each structure type must be derived from another vstdl type,
at least from \texttt{BCL\-Net\-work\-Data}, as its two respective C++ elements contain the translation functionality. A sample of the declaration
of such a datatype is given in Fig.~\ref{Fig:BCLVSTDL}, showing the three size options available for a structure element: A single scalar
type, a fixed size array, or a variable size array. Fig.~\ref{Fig:BCLVSTDLStruct} shows the C++ structure type generated from the previous
vstdl definition.  As for the base \texttt{BCL\-Net\-work\-Data} types the generated class and structure differ by the \texttt{Struct} modifier
appended to the structure's name. The base name for both is the name specified in the vstdl type definition and both are derived from
the corresponding  C++ type for the vstdl parent type. Transformation is performed in the inverse inheritance hierarchy. A derived
class first converts its own elements and then calls its parent class's transformation functions. 

Structure elements of the variable size array type can only be contained as the last element in a 
structure. Otherwise the C++ declaration of the structure would have to allow moving subsequent elements due to the array's changing size,
this however is not supported by the C++ language. This also implies that each structure can only contain one such element. For these elements the 
array is preceeded by an automatically generated member, holding the number of actual elements making up the array, to allow for the correct handling of the
changing array size.

\begin{figure}[hbt]
\begin{center}
\begin{verbatim}



#// Anything after a '#' at the beginning of the line is copied 
#// verbatim into the generated files.
#
name SampleData: BCLNetworkData
  uint32 fField;
  uint8  fArray[4];
  uint16 fVarArray[];
#
\end{verbatim}
\parbox{0.90\columnwidth}{
\caption[Sample data type declaration using the BCL type definition language.]{\label{Fig:BCLVSTDL}A sample data type declared using the simple BCL type definition language.}
}
\end{center}
\end{figure}

\begin{figure}[hbt]
\begin{center}
\begin{verbatim}



// Anything after a '#' at the beginning of the line is copied 
// verbatim into the generated files.

struct SampleDataStruct: public BCLNetworkDataStruct
    {
    uint32 fField;
    uint8  fArray[4];
    uint32 fVarArrayCnt;
    uint16 fVarArray[0];
    }

\end{verbatim}
\parbox{0.90\columnwidth}{
\caption[Data structure generated from the sample vstdl data type.]{\label{Fig:BCLVSTDLStruct}The data structure generated from the sample vstdl data type in Fig.~\ref{Fig:BCLVSTDL}.}
}
\end{center}
\end{figure}

\begin{figure}[hbt]
\begin{center}
\resizebox*{0.8\columnwidth}{!}{
\includegraphics{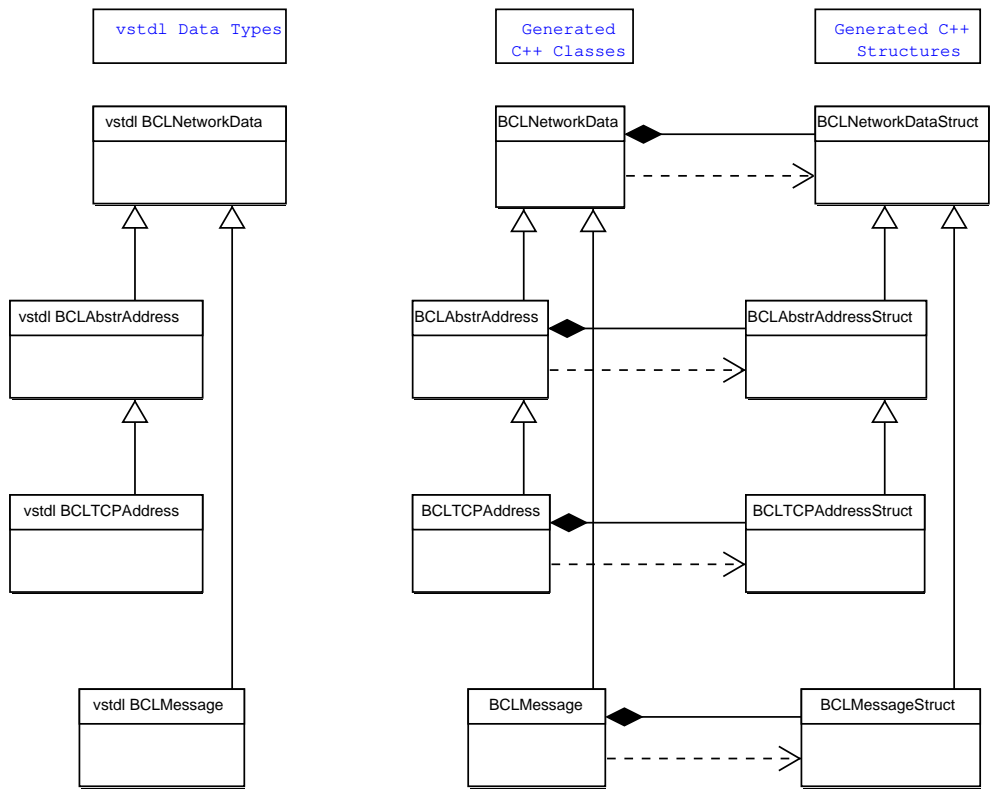}
}
\parbox{0.90\columnwidth}{
\caption{\label{Fig:DataTransformationClasses}A sample UML hierarchy of data types for data transformation.}
}
\end{center}
\end{figure}

A sample hierarchy of three generated vstdl data types from the BCL is shown in Fig.~\ref{Fig:DataTransformationClasses}, with the vstdl types 
on the left side, the generated C++ classes in the middle and the generated C++ structures on the right. The vstdl \texttt{BCL\-Net\-work\-Data} type 
displayed in the figure exists only virtually, since only C++ class and structure exist for the represented base type. 
Of the three vstdl types \texttt{BCL\-Abstr\-Ad\-dress} and \texttt{BCL\-Mes\-sage} are directly derived from \texttt{BCL\-Net\-work\-Data}
while the third type, \texttt{BCL\-TCP\-Ad\-dress}, is in turn derived from
\texttt{BCL\-Abstr\-Ad\-dress}. In the resulting C++ classes and structures the hierarchy of the respective vstdl types is reflected directly.
Also displayed in the figure is the mutual dependency of each type's class and structure with the class directly containing an embedded structure type
as well as a pointer to the structure. This pointer is used to access structures that have not been copied into a class object but that have been adopted
for efficiency reasons as discussed. 

Using the interface provided by \texttt{BCL\-Net\-work\-Data} and generated classes for other vstdl data types, it becomes possible for 
programs to handle the parts of data it needs, independent of the data format they were originally stored in. This is achieved without having to 
write the conversion code for every data type explicitly, which can instead be written as a vstdl type definition, from which the necessary C++ code
is subsequently generated. 
One drawback of the mechanism employed is that code working with C++ types generated from a vstdl data type definition
is unable to transparently handle derived data types as well. It will always handle only the data elements it was compiled for. A more 
mature and flexible data format conversion scheme might be implemented and used in later versions of the library and framework.

\subsection{\label{Sec:AddressClasses}Address Classes}

Addresses used in the BCL library are based on a vstdl type hierarchy, with the abstract address type \texttt{BCL\-Abstr\-Ad\-dress} at its root. 
\texttt{BCL\-Abstr\-Ad\-dress} is shown in Fig.~\ref{Fig:DataTransformationClasses}. It contains only an integer element that defines the type of address
in addition to the \texttt{BCL\-Net\-work\-Data} inherited header. Each 
communication implementation defines its own constant to identify its address type, e.g. \texttt{1} for SCI and \texttt{2} for TCP. 
The address types for each network technology are derived from the \texttt{BCL\-Abstr\-Ad\-dress} type, included are implementations
of addresses for SCI as well as for TCP. 

SCI addresses include three 16~bit elements, the first of which is used to identify the specific node concerned. It
is unique to each SCI adapter card, making it rather an adapter than a node ID but is 
sufficient to identify a node. Which adapter in a node is used for transmission is defined by the second number in the structure, 
this number is required if multiple adapters are present in a node and is 0 otherwise. 
The third number finally holds the ID of the shared memory segment used to receive the data in the target
node and must be a unique identifier in each node. 
For TCP the address structure contains only two elements, a 32~bit field with the target node's IP number and a 16~bit number for 
the port that the receiving communication object uses. 
Fig.~\ref{Fig:AddressTypes} shows the three different vstdl types used for address handling. At the top is the abstract address type definition with its
single element to define the network technology supported. In the middle is the SCI address type with its three described
16~bit numbers, and at the bottom the TCP address with the IP and port number required for a TCP connection. 

\begin{figure}[hbt]
\begin{center}
\begin{verbatim}

name BCLAbstrAddress: BCLNetworkData
  uint32 fComID;

name BCLSCIAddress: BCLAbstrAddress
  uint16 fNodeID;
  uint16 fAdapterNr;
  uint16 fSegmentID;

name BCLTCPAddress: BCLAbstrAddress
  uint32 fIPNr;
  uint16 fPortNr;
\end{verbatim}
\parbox{0.90\columnwidth}{
\caption[vstdl types for basic, SCI, and TCP addresses.]{\label{Fig:AddressTypes}The vstdl types for basic addresses \texttt{BCL\-Abstr\-Ad\-dress}, SCI addresses \texttt{BCL\-SCI\-Ad\-dress}, 
and TCP addresses \texttt{BCL\-TCP\-Ad\-dress}.}
}
\end{center}
\end{figure}

\subsection{Message Classes}

For the message communication mechanism the basic datatype used to define the message header, \texttt{BCL\-Mes\-sage}, is derived from \texttt{BCL\-Net\-work\-Data}.
It is thus also based on the data transformation mechanism from section~\ref{Sec:DataFormatTranslation}. The first of its three fields contains
an ID that defines the type of the message, outside the scope of the library and under the control of the application. Unlike this field the second field contains 
an ID to identify messages, reserved for use by the library itself. For the current implementations this is just a counter increased
for each message sent. The final field allows the specification of flags to affect the sending of a message. At the
moment, though, the field is not used by the library and no flags are specified, neither general message flags nor flags specific to an
implementation of the message communication interface. Fig.~\ref{Fig:BCLMessageType} shows the vstdl type definition of the
\texttt{BCL\-Mes\-sage} type with the three fields described.

\begin{figure}[hbt]
\begin{center}
\begin{verbatim}

name BCLMessage: BCLNetworkData
  uint32 fMsgType;
  BCLMessageID fMsgID;
  uint32 fFlags;
\end{verbatim}
\parbox{0.90\columnwidth}{
\caption{\label{Fig:BCLMessageType}The vstdl type defining the basic message header.}
}
\end{center}
\end{figure}

\subsection{Error Callbacks}

\begin{figure}[hbt]
\begin{center}
\resizebox*{0.8\columnwidth}{!}{
\includegraphics{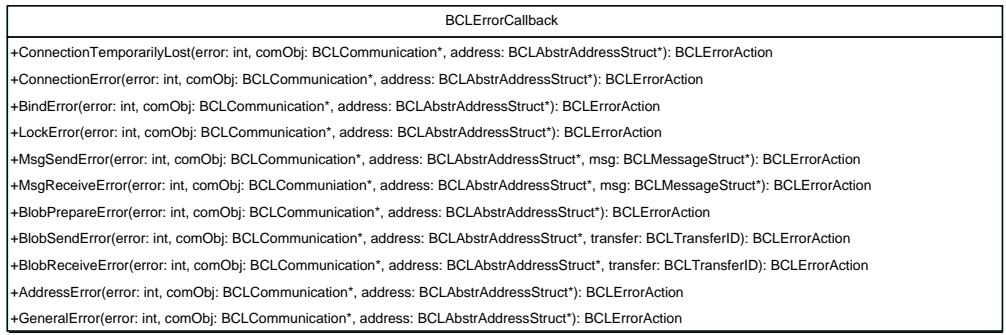}
}
\parbox{0.90\columnwidth}{
\caption{\label{Fig:BCLErrorCallback}UML diagram of the BCLErrorCallback class.}
}
\end{center}
\end{figure}

Error handling is performed partly by a set of callback object classes derived from
the abstract base class \texttt{BCL\-Error\-Call\-back} shown in Fig.~\ref{Fig:BCLErrorCallback}. Instances of this or derived classes can be registered
with communication objects and provide a number of callback functions, called by a communication object when the corresponding
error has occured.

As can be in seen in Fig.~\ref{Fig:BCLErrorCallback}, the class contains callback methods for different types of errors: 
\begin{itemize}
\item General errors, applying to any communication object
\item Send and receive errors for message communication objects 
\item Prepare, send, and receive errors for blob communication objects
\end{itemize}
All methods accept
at least a set of three common parameters: an indicator for the error that occured, a pointer to the originating communication object, and a pointer to an
address structure involved. Depending on the context this last parameter may contain either a local address, e.g. on a bind operation, or a remote address, e.g. for
a connect or send operation. In addition to these common parameters more may be accepted or required as appropriate for the error type that occured.
For message errors this
is a pointer to the message concerned by the respective error and for blob transfers it is the transfer ID. The available callback methods are not declared as abstract 
methods. Instead each is provided as a default implementation that only returns a default value described below so that derived classes have to implement only 
those methods whose functionality is needed.

The value returned by the callback functions is an action indicator containing a suggestion from the callback object how to handle the error. This action
can have one of three values indicating either to ignore the error, abort the operation, or make a retry attempt of the failed operation. 
Since the value is only treated as a suggestion the communication object can ignore the values returned by all callback objects and proceed
differently, as an action might not be possible for a specific case. In the current implementation of the communication classes, the error callbacks' return values are 
not evaluated at all, but the option to do so is already present for later implementations of the library. 

\begin{figure}[hbt]
\begin{center}
\resizebox*{0.4\columnwidth}{!}{
\includegraphics{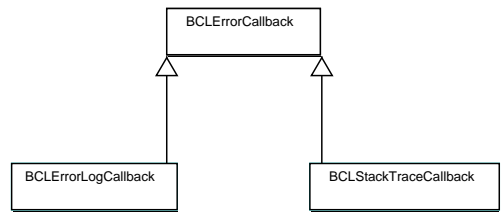}
}
\parbox{0.90\columnwidth}{
\caption{\label{Fig:ErrorCallbackClasses}The three error callback classes in the library.}
}
\end{center}
\end{figure}

Next to the base callback class two more derived classes are contained in the communication library, shown in Fig.~\ref{Fig:ErrorCallbackClasses}. 
The first of these,
\texttt{BCL\-Error\-Log\-Call\-back}, calls the logging system of the MLUC library with an appropriate error message constructed from its parameters. 
In the other class, \texttt{BCL\-Stack\-Trace\-Call\-back}, a set of 
system debugging functions is used to dynamically obtain a trace of the current call stack. This trace is then also passed 
to the MLUC logging classes. Beyond these two included classes an application can also derive its own classes from \texttt{BCL\-Error\-Call\-back} to implement
any error callback handling necessary.

\subsection{\label{Sec:BCLAddressURLClasses}Address URL Functions}

To support future additions of communication classes to the library without the need to recompile programs using the library, 
a set of functions is included in the library that allows to specify BCL communication addresses in a Uniform Resource Locator (URL) 
like format. Addresses in this format specify the network technology to be used, whether the message or blob communication mechanism is
to be used, and the specific address information required by the technology concerned. In this way a generic separation of address handling and network
technology has been introduced into the library. The supported abstract address format supported is now essentially a string type. 
Fig.~\ref{Fig:BCLURLAddresses} shows
the syntax for TCP and SCI addresses. As can be seen in the figure, the elements of the URL specifying the actual address 
correspond to the elements of the respective address structure types described in section~\ref{Sec:AddressClasses}. 

\begin{figure}[hbt]
\begin{center}
\begin{verbatim}

tcp(msg|blob)://<IP Nr>:<Port Nr>/
sci(msg|blob)://<SCI Node ID>[.<Adapter-Nr>]:<Segment ID>/

\end{verbatim}
\parbox{0.90\columnwidth}{
\caption{\label{Fig:BCLURLAddresses}TCP and SCI address URL format.}
}
\end{center}
\end{figure}

Address URLs are processed by four functions in the library, two for creating 
BCL objects and two for releasing them. Objects can be allocated for either local
or remote addresses by the \texttt{BCL\-De\-code\-Lo\-cal\-URL} or \texttt{BCL\-De\-code\-Re\-mote\-URL}
function respectively. For local addresses an address structure is returned together with an appropriate communication object
of a class derived from \texttt{BCL\-Com\-mu\-ni\-ca\-tion}. Additionally, a flag is provided indicating whether
the returned object is a message or blob communication object. To release allocated objects
two \texttt{BCL\-Free\-Add\-ress} functions are available, one to release only an address structure and the second one to also release
the communication object. A fifth helper function, \texttt{BCL\-Get\-Al\-lowed\-URLs}, is provided to aid in providing 
lists of allowed addresses to program users. It returns two list of strings containing valid message and blob
address URL formats.

\section{Communication Classes}

\begin{figure}[h]
\begin{center}
\resizebox*{0.75\columnwidth}{!}{
\includegraphics{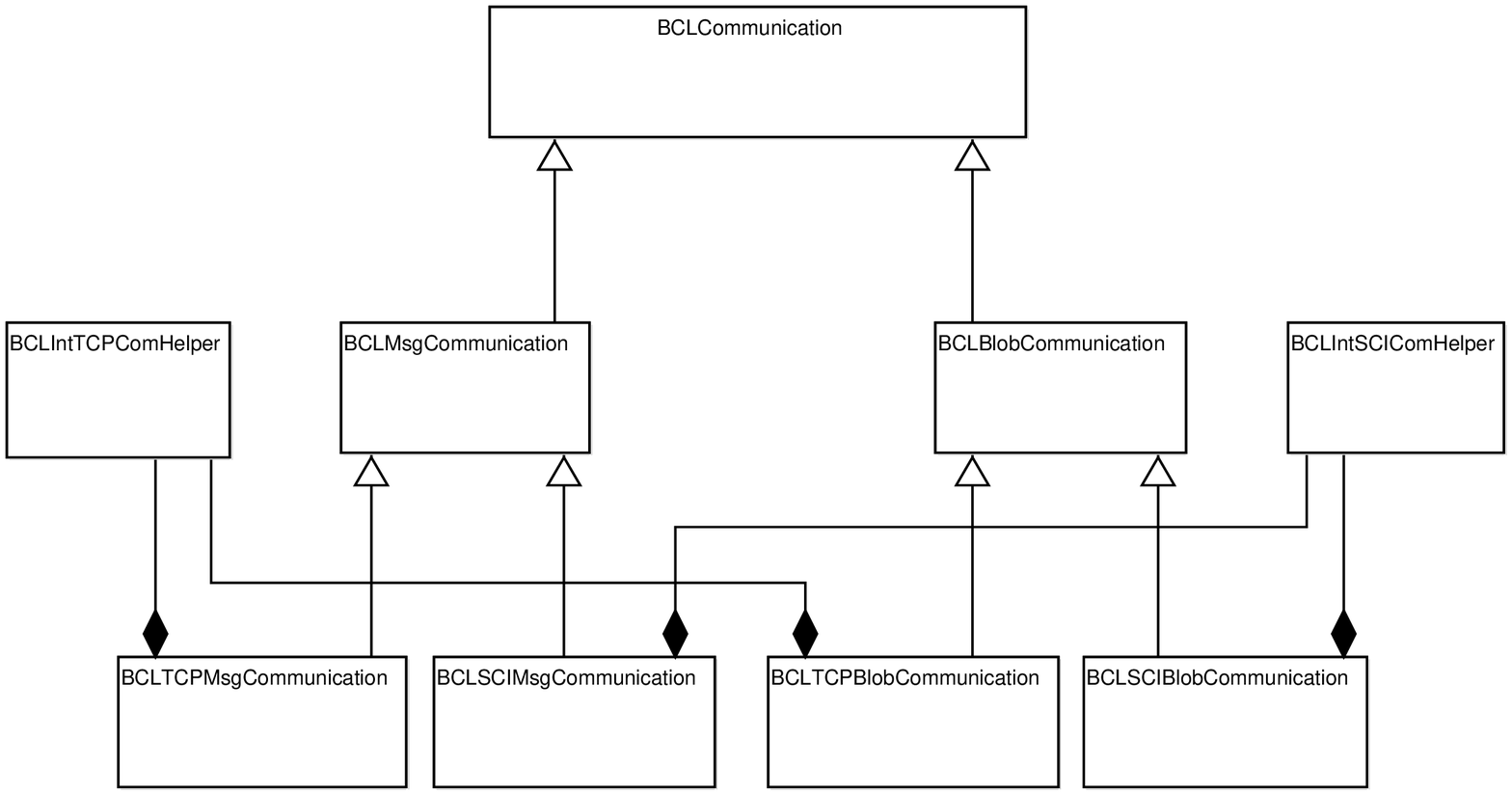}
}
\parbox{0.90\columnwidth}{
\caption{\label{Fig:MainComClasses}The UML class hierarchy of the primary communication class.}
}
\end{center}
\end{figure}

The seven primary communication classes in the library are organized in a tree hierarchy, displayed in Fig.~\ref{Fig:MainComClasses}.
At the root of this class hierarchy is the \texttt{BCL\-Com\-mu\-ni\-ca\-tion} class, providing basic services and declaring interface
functions common to both of the previously described communication types. Derived from this class are \texttt{BCL\-Msg\-Com\-mu\-ni\-ca\-tion}
and \texttt{BCL\-Blob\-Com\-mu\-ni\-ca\-tion}, which declare interfaces and implement common services for the message-like and data-block 
communication mechanisms respectively.

Two classes are derived from each of the two communication type base classes to provide implementations for the TCP protocol 
using the standard socket API as well as for the shared memory interconnection technology SCI by Dolphin using the SISCI 
API. All four implementation classes are designed to be able to make use of as many performance and efficiency optimizing features 
as possible for the specific network technology used. 
In addition to these primary communication classes two separate classes, \texttt{BCL\-Int\-TCP\-Com\-Helper} and \texttt{BCL\-Int\-SCI\-Com\-Helper}, 
are present, used by the two implementation 
classes for each network type as shown. They supply functions and variables common to both communication mechanisms
but specific to each network technology. These two classes are not intended to be used directly in a program but are for the library's internal 
use only as signified by the \texttt{Int} specifier in their names.

\subsection{\label{Sec:BCLBasicInterfaceClasses}The Basic Interface Classes}

\subsubsection{The BCLCommunication Class}

\begin{figure}[hbt]
\begin{center}
\resizebox*{0.8\columnwidth}{!}{
\includegraphics{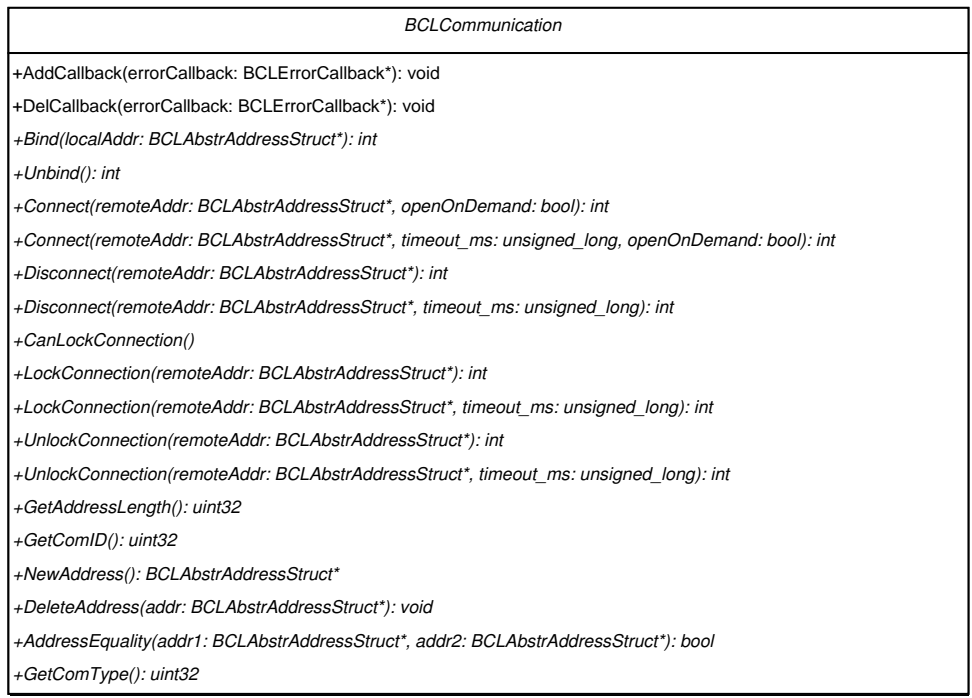}
}
\parbox{0.90\columnwidth}{
\caption{\label{Fig:BCLCommunication}UML class diagram of BCLCommunication's main features.}
}
\end{center}
\end{figure}

At the base of the communication class hierarchy is the \texttt{BCL\-Com\-mu\-ni\-ca\-tion} class containing 
functionality common to all communication types and mechanisms. Primarily, however, it defines the common interface
for the different communication types using abstract member functions. A UML diagram of the class with the main
features described in the following paragraphs is shown in Fig.~\ref{Fig:BCLCommunication}.
The main functionality contained in \texttt{BCL\-Com\-mu\-ni\-ca\-tion} is the handling of the error callback objects, that can be
registered with each communication object. Two public functions are provided, allowing to add or remove callbacks
to a communication object plus a number of protected methods for internal use by this or derived classes. Each of these functions corresponds to 
one of the different error functions exported by the callback interface. They are called when an error occurs and in turn call the appriopriate error 
function for each registered callback object as well as for the optional callback parameter object supported by most functions.

Common interface parts defined by the BCLCommunication class include functions for binding, connecting, locking connections, 
handling addresses, and one function for querying whether a communication object is a message or a blob communication object. 
For all functions any address needed must be specified in the form of a base address structure \texttt{BCL\-Abstr\-Ad\-dress\-Struct}
or a pointer to it, to keep the interface generic from a specific network technology. The actual address 
used still has to be of the type required by the corresponding communication object itself and is thus specific
to the network technology supported by that object. 

The \texttt{Bind} function defined in the interface requires one argument only, a structure pointer 
holding the address to which the communication object should be bound. Data sent to  this address must be received by the communication object,
and the address therefore must be a valid address for the network technology chosen. 
No parameter is required by the \texttt{Un\-bind} function, which releases a bind to an address, as each communication object has exactly one address it is bound 
to. Both functions return an integer value to indicate the success or error status of the operation as described previously. 
Unlike the \texttt{Bind} call the \texttt{Con\-nect} call exists in two versions: one using a millisecond granularity timeout value and the other without a timeout. 
Both calls require a remote address to connect to and a boolean stating whether the connection should be established
immediately or as an on-demand type connection, as explained in section~\ref{Sec:BCLGeneralDesign}. The supplied default value of false for the boolean
parameter specifies 
that the connection has to be established immediately. As a communication object can be connected to multiple remote communication partners the 
\texttt{Dis\-con\-nect} function requires the remote address of the connection to be terminated. 
Similar to the \texttt{Con\-nect} function, the \texttt{Dis\-con\-nect} function also exists in two variants, one with a milliseconds timeout and one without.
All four connection related functions return the standard integer error indicator.  

The following set of five functions is responsible for locking connections, with the first of these functions allowing to query whether the locking
feature is supported by this object, which depends on the network technology implemented. Two functions
are available to lock an established connection, one with timeout value and one without. Like the \texttt{Con\-nect} function 
both need the address of the remote
connection partner and return an integer error value. The two unlock functions are also similar with respect to their required 
arguments. Each needs the address and one of them allows to use a timeout value. 

Support for handling network addresses in an abstract manner is provided by the final set of functions defined by \texttt{BCL\-Com\-mu\-ni\-ca\-tion}.
In conjunction with the set of helper functions that allow to specify addresses in a string similar to Internet address URLs, any user program working with 
address classes needs to know as little as possible about the format of the underlying addresses and the 
specifics of the network technology used by a communication object. The first two of the address support functions
allow to query two values relevant
for address handling, namely the actual length of an address and the value of the communication ID.
These two values allow to identify whether a given address belongs to a specific
communication implementation. The address length is also required for storing, copying, or allocating addresses. 

Allocating memory for addresses is also supported by the second set of two functions, that allow to allocate memory for an address and to
free an allocated address. The allocation function \texttt{New\-Address} returns an address structure object with header fields and the communication ID 
initialized to values appropriate for
the communication object. \texttt{De\-lete\-Ad\-dress} frees the memory allocated for an address structure by using the free call 
corresponding to the allocation call used in \texttt{New\-Address}. By using these two functions it 
is ensured, that the allocate and free functions always match, the correct amount of memory is allocated, and the basic fields are initialized correctly.
A fifth address helper function is used to compare address structures for identity, to support comparing
of addresses whose exact type and contents might not be known at compile time. Comparing two structures bytewise relying on their length is always
possible, but  addresses with different byte contents might point to the same remote address, e.g. because of different byte orders. 
The final helper function provided by \texttt{BCL\-Com\-mu\-ni\-ca\-tion} specifies whether a communication object is a message or a blob communication object so 
that it is possible to distinguish between these two sub-hierarchies in a generic manner.  

%{\bf Functions are all defined as pure virtual and do not perform/execute anything. The above text just describes/defines what implementations of
%these functions have to perform.}

\subsubsection{\label{Sec:BCLMsgCommunication}The BCLMsgCommunication Class}

\begin{figure}[hbt]
\begin{center}
\resizebox*{0.85\columnwidth}{!}{
\includegraphics{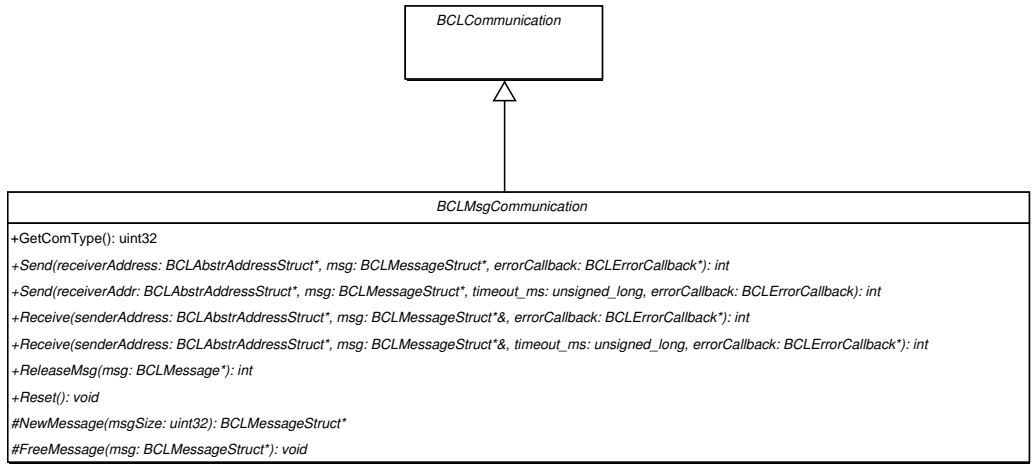}
}
\parbox{0.90\columnwidth}{
\caption{\label{Fig:BCLMsgCommunication}UML class diagram of BCLMsgCommunication's main features.}
}
\end{center}
\end{figure}

The \texttt{BCL\-Msg\-Com\-mu\-ni\-ca\-tion} class is derived from \texttt{BCL\-Com\-mu\-ni\-ca\-tion}. It defines the interface for the message communication mechanism, 
as shown in Fig.~\ref{Fig:BCLMsgCommunication}.
This class provides almost no additional functionality beyond that provided by \texttt{BCL\-Com\-mu\-ni\-ca\-tion} but defines the interface only. The one supplied functionality
is the implementation of the helper function to distinguish between message and blob classes, which returns 
the indicator for a message communication object so that derived classes for specific network technologies do not have to implement this function
themselves. 
Primarily, the \texttt{BCL\-Msg\-Com\-mu\-ni\-ca\-tion} class defines the interfaces for sending and receiving of messages on top of the basic
interface defined in \texttt{BCL\-Com\-mu\-ni\-ca\-tion}. %Both \texttt{Send} and \texttt{Re\-ceive} calls defined by the object appear in two versions,
%one using a timeout parameter for its operation and the other one operating without. 
Beyond these send and receive calls a \texttt{Re\-set} call is provided, that serves to reset a message communication 
object to a clean defined state. Resetting an object might cause some previously received data to be lost.
%and it might also cause established connection to this object or from this object to other objects to be closed. 

For all \texttt{Send} functions two parameters are needed, the remote address where to send the message to and the message itself. The address is supplied
as a pointer to a \texttt{BCL\-Abstr\-Ad\-dress\-Struct} object, and the message is specified as a pointer to a \texttt{BCL\-Mes\-sage\-Struct} structure. This message object
can be of the actual \texttt{BCL\-Mes\-sage\-Struct} type, or it can be of a structure type derived either directly or indirectly from \texttt{BCL\-Mes\-sage\-Struct}. 
An optional third respectively fourth parameter is a pointer to a \texttt{BCL\-Error\-Call\-back} object. In the case 
of an error this object's error reporting callback functions will be called prior to those of the communication
object. In order to always provide a \texttt{Re\-ceive} function with a message's originating address, a \texttt{Send} function implementation 
should send its address prior to the actual message data. 
This might not be necessary, if the remote communication partner has other methods of finding out the originating address of a received 
message, but a \texttt{Re\-ceive} function must always be able to provide a message's source address.
% to a calling application.

The \texttt{Re\-ceive} functions also accept two mandatory and one or two optional arguments. In analogy to the
\texttt{Send} functions the \texttt{Re\-ceive} function's optional parameter is a callback object whose functions will be called in addition to the ones 
of registered objects. As their first
parameter, both \texttt{Re\-ceive} functions use a pointer to a memory location where the address of the sending communication object will be stored. 
A check is executed
whether the structure has the correct length expected for addresses used by this communication object. If that size is incorrect, the \texttt{Re\-ceive} call fails.
The second mandatory parameter is a reference to a \texttt{BCL\-Mes\-sage\-Struct} pointer, used to return the received message itself to the calling
program. 
Memory to store a message is allocated by a call to \texttt{New\-Me\-ssage}, another function  
provided as the declaration for an abstract member function, to be implemented by derived message communication classes. The function is declared as 
a protected member function so that it is only available for internal use by other methods of this or derived classes and not as an interface
to programs. Allocated memory for a message is filled with the contents of the message received. The pointer to the new message is returned
via the reference parameter. It is not mandatory for the \texttt{Re\-ceive} calls to allocate memory for the message data, it could for example
also be stored in an internal buffer, or it might be written directly into a reserved buffer by the sender, like in the case of SCI. Here
it would be sufficient to return a pointer to this internal buffer memory. 

A message that has been received and its memory been allocated, has to be released again after the program has finished using the message's data. 
To retain flexibility in the way the messages are allocated in the \texttt{Re\-ceive} functions, the \texttt{BCL\-Msg\-Com\-mu\-ni\-ca\-tion} class declares the 
\texttt{Re\-lease\-Msg} method for this purpose.
It accepts a pointer to an allocated message as its argument. This pointer is then passed to the protected member function, 
\texttt{Free\-Mes\-sage}.
As is the case for \texttt{New\-Me\-ssage}, \texttt{Free\-Mes\-sage} is also declared as a pure virtual member function to be implemented by 
derived message classes.

%{\bf All Send an Receive functions call the passed error callback object if it is present as well as all registered error callback objects. The return value
%of the functions is the usual integer error indicator.}

\subsubsection{\label{Sec:BCLBlobCommunication}The BCLBlobCommunication Class}

\begin{figure}[hbt]
\begin{center}
\resizebox*{1.0\columnwidth}{!}{
\includegraphics{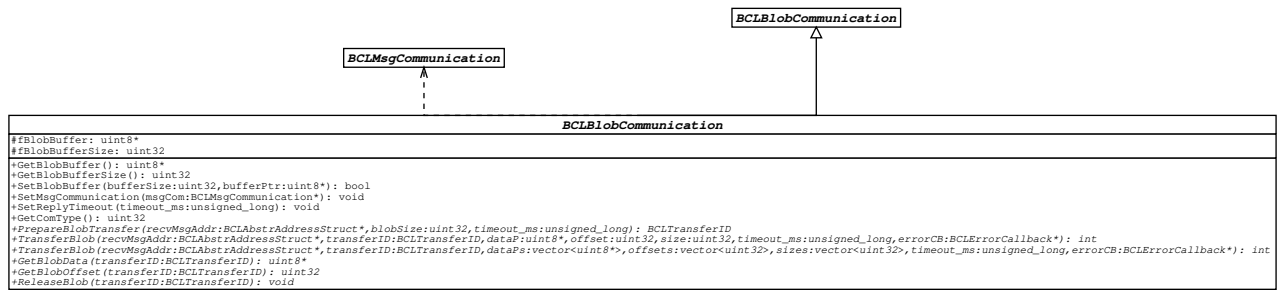}
}
\parbox{0.90\columnwidth}{
\caption{\label{Fig:BCLBlobCommunication}UML class diagram of BCLBlobCommunication's main features.}
}
\end{center}
\end{figure}

As the \texttt{BCL\-Msg\-Com\-mu\-ni\-ca\-tion} class, the \texttt{BCL\-Blob\-Com\-mu\-ni\-ca\-tion} class shown in 
Fig.~\ref{Fig:BCLBlobCommunication} is also derived from the \texttt{BCL\-Com\-mu\-ni\-ca\-tion} class. It contains functionality for the 
blob communication mechanism and also declares an abstract interface for it. 
Functionality is provided by this class mainly for the
blob receive buffer, that stores data received from remote nodes for access by the program. 
One further helper function in the class is an implementation of the function from the \texttt{BCL\-Com\-mu\-ni\-ca\-tion} class that
identifies objects of this and any derived class as blob objects.
The main function for the receive buffer is used to specify the buffer's size and optionally the receive buffer directly by using a pointer 
to the respective area of memory. Setting the buffer's size is always allowed, it is allocated immediately with the new size. If the buffer
was already allocated, the old buffer is released and a new one allocated with the given size. Specifying the buffer to be used itself in a parameter
to this function is not guaranteed to be supported. This depends on the blob implementation. For the SCI implementation, for example, this is not possible,
since received data is written directly into the buffer by the remote sender program. As the current implementation of the
SISCI API for SCI does not allow to specify arbitrary memory locations into which data can be written remotely, the allocation of the receive
buffer has to be done by the SISCI driver or library. To allow the required overriding of the original \texttt{Set\-Blob\-Buff\-er} function in \texttt{BCL\-Blob\-Com\-mu\-ni\-ca\-tion}
by blob implementation classes the function is declared as a virtual function. Two additional helper functions are available
to obtain a pointer to the buffer for accessing received data and to query the buffer's size. 

As was pointed out in section~\ref{Sec:BlobParadigms}, the blob communication mechanism has to exchange handshaking and control messages with a 
remote communication partner using exclusively reserved message communication objects. The specification of the message object to be used by a
blob object can be done either in a \texttt{BCL\-Blob\-Com\-mu\-ni\-ca\-tion} constructor or it can be done later using a separate function. Both
approaches use a pointer to the message object set in the blob object.
%, after which the message object is used by the blob object. 
A helper 
function allows to set the timeout
%of millisecond granularity, 
used to wait for  answers from the remote node via the 
associated message communication objects. When replies are not received within the specified timeout they are treated as errors. An exception is the 
special timeout value of 0 which disables timeouts, resulting in an infinite wait. 

The blob communication interface, declared by the \texttt{BCL\-Blob\-Com\-mu\-ni\-ca\-tion} class, consists of two parts, the interface for the sender to transmit
the data and the one for the receiver to access received data. Each of the declared sending and receiving interfaces consists of three pure virtual functions.
%To better understand the sending interface of the class the data transmission principle already described is {\bf quickly summarized here}: 
Before a transfer is started, a negotiation has to be performed first to determine where to store the data in the receiving buffer,
if it can be stored at all. This negotiation is performed by the \texttt{Pre\-pare\-Blob\-Trans\-fer} function, that uses the address of the remote message
communication object associated with the receiving blob object. It is not necessary for the object to know the address of the blob object 
itself. The second parameter required by  this function is the size of the data to be sent, the data itself is not passed to the blob object for the preparation of the
transfer. Optionally, a timeout can be supplied as well, in order to restrict the time that the function waits for the reply from the remote message communication
object. On a successful negotation the function returns an ID that can be used subsequently to execute the transfer. 
%A value of -1 is returned on an error. 
Using the returned transfer ID the data can be sent to the remote blob object with the help of one of the two available \texttt{Trans\-fer\-Blob} functions. 
Both functions require the remote message address as their first argument followed by the transfer ID returned from 
\texttt{Pre\-pare\-Blob\-Trans\-fer}. Two optional arguments, supported by both functions, are a timeout and a pointer to
an additional error callback object. This object is used in addition to the registered callback objects. By default the timeout is disabled through the value zero  
and the default callback pointer is empty. 

In addition to these common parameters the simpler of the two transfer functions accepts three more parameters that describe
the data to be transmitted. Of these parameters the first is a pointer to the actual data to be transmitted itself, followed by an offset
holding the location where this data block should be stored in the previously reserved receive transfer area. The offset is specified relative to the start
of the reserved transfer area and can be used to write multiple data blocks into a transfer area with multiple calls. This is needed especially when
a larger portion of the receive buffer is reserved in advance, and the sender performs its own buffer management in that area, as discussed in
section~\ref{Sec:BlobParadigms}. As the function's final parameter, the data size is specified. The sum of the size and the data offset must not exceed the size parameter
passed in the transfer preparation function call. Explicitly specifying the size parameter is done to allow splitting a transfer over multiple calls
in conjunction with the function's offset parameter. 
Unlike the first transfer call, the more complex second one allows to perform a scatter-gather-transfer with one call. Therefore the
three parameters describing the transfer, data pointers, offsets in the receive buffer, and block sizes, are not specified as scalars but instead as 
three vectors, or dynamic arrays, each containing multiple values. The arrays holding the pointers to the data and data block sizes must contain the 
same number of elements for the transfer attempt to be valid, but the array of offsets can contain less values than the other two arrays and may even be completely empty. 
If it contains less offsets than needed, the remaining offsets will be calculated so that each block will be immediately
adjacent to the previous block. In particular this means that if no offsets are specified the blocks will start at the beginning of the transfer area
and will then continue directly adjacent to form a single large block. Essentially this provides a merging or coalescing functionality for the scatter-gather-transfer.

%The receiving interface of the class consists of three functions, each of which has only one parameter, the ID of a received transfer.
Three functions make up the receiving interface of the class, each of which uses only one parameter, the ID of a received transfer.
One of the functions, \texttt{Get\-Blob\-Data}, returns a pointer to the beginning of the area of the receive buffer. With this pointer an application program can
access the data that was received for the particular transfer. \texttt{Get\-Blob\-Off\-set}, the second function, returns the offset to the transfer's block in the buffer,
relative to the buffer's start. Together with the pointer to the receive buffer itself this can also be used to get access to the received data. 
The final receive interface function, \texttt{Re\-lease\-Blob}, is used to release the buffer space occupied by a received transfer. 
No method is provided by the blob communication interface to send a transfer ID to a receiving node
or to wait for a completed transfer. This has to be done explicitly by a user program, for example using a second message communication object.

\subsection{\label{Sec:TCPComClasses}The TCP Communication Classes}

The classes presented in this section implement the message and blob facilities on top of the widely used Transmission Control Protocol (TCP) \cite{RFC793}. 
Both primary classes \texttt{BCL\-TCP\-Msg\-Com\-mu\-ni\-ca\-tion} and \texttt{BCL\-TCP\-Blob\-Com\-mu\-ni\-ca\-tion} for the two different communication mechanisms make use of the 
services provided by a third class, \texttt{BCL\-Int\-TCP\-Com\-Helper}. Communication is based upon the established POSIX socket API 
\cite{IEEEPOSIX}, \cite{SingleUnixWeb}, \cite{SingleUnixOnline}, 
\cite{Beej}, available 
on basically every Unix system as well as on Windows systems.
TCP is a connection-oriented protocol, and the use of the BCL communication functions without an explicitly established connection results 
in a TCP connection being initiated for the duration of the sending. On the other hand, the system API used to access the TCP protocol
trivially allows to receive data from multiple connections simultaneously, making the locking functionality unnecessary %, which is thus not implemented
for TCP. 

\subsubsection{\label{Sec:BCLIntTCPComHelper}The TCP Communication Helper Class}

In the \texttt{BCL\-Int\-TCP\-Com\-Helper} class functions are provided for the TCP protocol used by both communication mechanisms, and which therefore
have been placed in a separate class for reuse. 
%{\bf In order of succession} 
Most of the functions supplied by the TCP helper class follow the interface defined by the top-level communication class \texttt{BCL\-Com\-mu\-ni\-ca\-tion} and differ
primarily in the number and type of parameters. Parameters supported by this class can be made more specific for the TCP implementation compared to the generic
interface definition. For example, for the functionality to bind to a 
valid address
% so that other communication objects can initiate connections 
provided by the helper class's \texttt{Bind} function, the generic address pointer to
a \texttt{BCL\-Abstr\-Ad\-dress\-Struct} instance is replaced by a \texttt{BCL\-TCP\-Ad\-dress\-Struct} pointer. 

In the \texttt{Bind} function the socket that will be used to accept incoming connections is created using the \texttt{sock\-et} function.
The \texttt{SO\_\-RE\-USE\-ADDR} flag
is set by the \texttt{set\-sock\-opt} function to allow the socket to bind to an address that has been in use before and not been properly cleaned up. Since a bind to an address
in actual use will still fail, the option can be safely used here. After setting the flag the socket is bound to the port number from the specified address
 using the \texttt{bind} API function. If an IP number has been specified in the address the socket will be bound to that particular
address so that data can be received only from the network interface associated with that address. 
Otherwise it will not be bound to a particular IP number, and data sent to any IP address of the current node can be received. 
When the \texttt{bind} call has completed successfully,
the socket is set to a state in which the system accepts incoming connections using the \texttt{lis\-ten} system call. Finally, a background
thread is created to wait for and handle incoming connections as well as data arriving on an already  established connection. This accept thread is described in more 
detail below. 
Analogous to the \texttt{Bind} function the helper class also contains an \texttt{Un\-bind} function that ends the ability to accept incoming 
connections and data.
To unbind the socket created above the function sets an internal flag and tries to establish a short connection to the local receive socket itself to activate the created 
accept thread. Upon activation the thread checks whether the flag has been set. If this is the case, it cleans up and terminates, setting another flag in the process.
The connection established by the \texttt{Un\-bind} function is closed again immediately, and the function waits a specific time for the second flag to be set, signalling the
termination of the accept thread. If the flag is not set within the required timespan the thread is aborted forcefully. After termination or abortion of the accept 
thread, the accept socket created in \texttt{Bind} is closed to complete the unbind operation. 

The second type of functionality provided by the TCP helper class for both communication mechanisms is the ability to connect to and disconnect 
from remote communication objects. For this purpose two interfaces are available in the helper class. One of these interfaces is 
identical to the connection interface supplied by the \texttt{BCL\-Com\-mu\-ni\-ca\-tion}
class, again with the exception of TCP address structures being used instead of abstract addresses. Its parameters are the address pointer mentioned,
a flag whether a timeout is to be used together with the timeout value, and a boolean flag specifying the connection type. 
The timeout value may be ignored depending on the timeout flag's value and the connection flag indicates whether to connect
immediately or create an on-demand connection. 
This interface is used for explicitly initiated 
connects by the TCP communication classes. It makes use of the second basic connection interface in the helper class.

This second helper connection interface is used by any sending function in the TCP communication classes that needs to use a connection for the duration of 
the send operation as well as by the explicit connection interface. The function for initiating a connection, \texttt{Con\-nect\-To\-Re\-mote}, requires as its primary
arguments a pointer to the TCP address of the remote connection partner and a pointer to a structure used internally to store data for
an established connection. In this structure the remote address and the socket used for this connection are stored together with a use 
count. It is filled by the connection function and allows the code that initiated the connection to use it through the 
stored socket descriptor.
%Similar to the first connection interface, this function also accepts a flag specifying the use of a timeout and a timeout value evaluated and used 
%based upon the flag's value. 
In addition to these arguments \texttt{Con\-nect\-To\-Re\-mote} supports two more input flags and one output flag, the first of which specifies whether the connection 
should be established immediately or as an on-demand connection. The second input flag controls whether the connection
is a permanent connection initiated by an explicit user connection call or whether it is an implicitly established connection. In the first case, the connection data 
as returned by the function is stored in a list of connections if it has not been stored there already. Related to this, the output flag of the function 
indicates whether the connection was already stored in the connection list, indicating that the connection had already been established before the
\texttt{Con\-nect\-To\-Re\-mote} function call. 

If the connection has been already established, only its data is copied into the connection data structure, and the appropriate output flag is set so that
a connection can be used multiple times. 
For a connection that needs to be established immediately, but that has been stored as an on-demand connection, the call will cause the connection 
to be established.
%,  as if it had not been stored. 
The established connection's data will be stored in the slot already used. It is stored independently of whether 
the function call causing the connection to be established, specifies a permanent connection or not. 
When a connection cannot be found in the internal connection list three possibilities have to be distinguished. 
\begin{enumerate}
\item An on-demand connection that has to be stored: In this case just the remote address is stored in the list with a usage count of one and an invalid 
socket descriptor. The invalid descriptor indicates later calls to this function that the entry is an on-demand connection which still has to 
be established. 
\item An immediate connection to be made permanent as well: In this case the connection is established as 
described below and, if successful, stored in the internal list. 
\item An implicit connection from a transmission operation: This connection needs to 
be established immediately but does not have to be stored in the internal list. It is also established as described below, but contrary to the other cases
the connection data is only returned to the caller and is not placed in the list. 
\end{enumerate}

If a connection has to be established for one of the three cases listed above, certain steps are executed in \texttt{Con\-nect\-To\-Re\-mote}, starting with the creation
of the socket used as the local endpoint for the connection concerned. After creation the \texttt{TCP\_\-NO\-DE\-LAY} socket option is set as any message 
should be sent immediately without waiting for further data that might have to be sent. 
%Not setting this option can incur an additional message latency of the time
%that the TCP subsystem waits for more data before sending out a message. 
When these preparation steps are completed an attempt is made to establish the connection using the \texttt{con\-nect} system call. If a timeout has been specified, 
it is used for the connection attempt by setting the timeout with the \texttt{SO\_\-SND\-TIMEO} socket option. 
After the \texttt{con\-nect} 
function returns the old timeout value is restored. At this point the connection has been established if the \texttt{con\-nect} function signals success 
and the connection data can be stored in the connection list as required (see above). 
One combination of the two on-demand and store input flags for \texttt{Con\-nect\-To\-Re\-mote} has not been discussed in the previous paragraphs: an on-demand connection
that does not have to be stored as well. This combination of flags does not have any significance for establishing 
any kind of connection but the
function will still return whether a connection could be found in the internal list. As a consequence this input flag combination can be used to check for the existence of 
a connection with a given remote address in the list of a communication helper object. 

For terminating existing connections, two functions for the two different connection interfaces exist in analogy to the two connection functions. 
The \texttt{Dis\-con\-nect} function of the explicit 
connection interface requires three of the arguments of the corresponding \texttt{Con\-nect} function, the address of the remote connection partner and
the timeout flag and value. Using the input flag combination to the \texttt{Con\-nect\-To\-Re\-mote} function described in the previous 
paragraph, it determines whether a connection to the specified address is active. In that case it uses the disconnect function of the second interface type 
to terminate the connection. 
The second disconnect function, \texttt{Dis\-con\-nect\-From\-Re\-mote}, also accepts three arguments, but unlike for the \texttt{Dis\-con\-nect} function
the first argument is not 
the remote connection address. Instead it expects a pointer to the connection data structure returned by \texttt{Con\-nect\-To\-Re\-mote} that contains 
the data of the connection to be closed, including the remote address. Its two other parameters are again the timeout flag and value parameters
with the same meaning as for the other functions. Using the remote address in the connection data structure the function searches the connection list for a 
matching connection, and if it is found, its reference count is decreased by one. For a reference count greater than zero the function terminates without
any further action. If the reference count is zero or less, the connection's socket is closed by the \texttt{close} system call, again using the \texttt{SO\_\-SND\-TIMEO} 
option if the timeout is specified. Following this the connection data item is removed from the list of connections. Saving the socket's previous timeout value
is not necessary in this case as the socket is closed and thus unusable at the end of this block. 
If no connection data structure containing the specified remote connection address could be found, it is presumed that the connection is an implicitly 
established one that has not been stored in the list. In this case the socket in the structure is simply closed as before.

As introduced above, the \texttt{Bind} call starts a background thread with the task of managing new incoming connections as well as data arriving on established
connections. This thread consists of a loop that runs until the end flag is set by the \texttt{Un\-bind} function as described in the previous paragraph. 
In the loop the \texttt{se\-lect}
system function is used to check the socket created and bound in \texttt{Bind} as well as any socket belonging to an accepted connection for available data. 
To ensure 
regular checks of the end flag the \texttt{se\-lect} function is called with a timeout of 500~ms as a safety measure in addition to the short connection made 
from \texttt{Un\-bind} (cf. above). 

If a \texttt{se\-lect} function indicates that new data is available for the listen socket, this signifies a new connection attempt, which  is 
accepted by the \texttt{ac\-cept} system call. The new socket returned by \texttt{ac\-cept} is passed to a callback function of the helper object's 
parent, an instance of either \texttt{BCL\-TCP\-Msg\-Com\-mu\-ni\-ca\-tion} or \texttt{BCL\-TCP\-Blob\-Com\-mu\-ni\-ca\-tion}. If this function returns the boolean value false, 
the connection is rejected and the socket closed. Otherwise the connection is accepted and placed into the list of currently
established connections to be checked for new data by the background loop. 
When new data is signalled to be available on an accepted connection, a second callback function of the parent object is called. 
In this callback function the steps required to
read the received data have to be performed, and its implementations differ between the TCP message and blob communication classes. 
If this callback function returns false 
this is an indicator that an error occured while attempting to read and the socket is placed in a list of connections to be closed. This is necessary because
a receiver can only detect a connection that has been closed by a sender when \texttt{se\-lect} returns available data for a socket while a subsequent read call fails 
with no available data. 
%In this way broken or terminated connections are prevented from lingering around and eventually filling the connection list of an helper object.
Errors that occur during a \texttt{se\-lect} call have a configurable limit for the number of calls allowed to fail in a row. If this limit is
exceeded, the background thread handling the connections is terminated, as it is assumed that either a fatal error occured with the listen socket or that 
it has been closed from outside the loop. The error count for \texttt{se\-lect} calls is reset after every successful call.

During the receiving of messages many \texttt{read} calls for small amounts of data are executed, e.g. first the header of the sender's address is read to check for the correct
address length, and then the rest of the address is read. Following this, the header of the message and the message data itself are read. Consequently  four read calls 
have to be made to receive one message. To reduce this overhead of many  system \texttt{read} calls, a special read aggregation mechanism has been implemented
in the TCP helper class. In this mechanism the communication classes do not use the read system call directly but instead use a read function provided
by the helper class. When called, this function checks an internal read buffer, currently 1~kB large, for the presence of data. With data being present in that buffer,
the data requested to be read is copied from that buffer instead. If not enough or no data is present in the buffer, then 
a read system call is made. In this call it is attempted to read as much data as is available in one call, up to a maximum of the read buffer's size. 
Later read requests can then 
again be satisified from the buffer's internal memory. A read attempt that, after emptying the read buffer, requires more data than would fit in  the buffer 
will not result in a read system call to fill the buffer again, but instead the read call is made so that  the whole amount of data is 
read directly into the desired final location. This ``override'' was implemented to avoid read requests for large amounts of data being ``translated'' into
multiple small read system calls to fill the buffer followed by memory copying operations from the buffer into the final destination memory.

\subsubsection{The TCP Message Communication Class}

Making use of the functionality provided by the TCP helper class, the \texttt{BCL\-TCP\-Msg\-Com\-mu\-ni\-ca\-tion} class contains the implementation of the message communication 
functionality for the TCP protocol.
The basic functionalities for binding to a given receive address and establishing connections to remote addresses
are implemented based upon the helper class functionality described in the previous subsection. In many message class functions appropriate functions provided
by the helper class are simply called
with the appropriate type casting of the parameters from abstract to the respective TCP types. 
As written in the TCP section's introduction, TCP trivially supports receiving data simultaneously from multiple data destinations. 
Connection locking is therefore not supplied by the TCP message class. Since the pure virtual functions inherited from the communication base 
class still have to be implemented to be able to use the class, they are provided as empty functions. An error is returned by these functions to indicate that the
functionality is not supported. 

One of two primary functions of the class is the sending of messages to remote processes using implementations of the two \texttt{Send} functions declared in 
\texttt{BCL\-Msg\-Com\-mu\-ni\-ca\-tion}, both of which call the same internal member function. In analogy to functions from the helper class this function accepts a 
timeout flag and value to support both connection functions that differ in the support for a timeout. 
In addition to these two parameters it also requires the remote receiver address and the message to be sent. An optional error callback object pointer can be 
passed as well. 
Checks are performed first in this internal \texttt{Send} function on the remote address for the correct communication ID and minimum structure size of
the \texttt{BCL\-Mes\-sage\-Struct} length. A further check is 
performed on the message itself to reduce segmentation violations during sending. The first and last byte of the message data are read
once so that potential  violations occur before the sending is started at all to avoid that a sending operation is aborted while in progress.
After these checks the connection is established or data of an existing connection retrieved by calling the second connection function \texttt{Con\-nect\-To\-Re\-mote}
 in the helper class. If a timeout value has been specified, it is set here after saving the old value for restoration after the send is completed.

In the next step the address of the sending object is sent to the message recipient to inform it about the message's origin.
Like all TCP send operations in the library classes, sending itself is done in a loop where all remaining data is passed to a \texttt{write} call for sending. Depending upon
the amount of data written, as returned by \texttt{write}, the loop is either ended or a \texttt{se\-lect} call is made to wait for the connection to become available 
for writing again, using a timeout as appropriate. An additional inner loop is present around the \texttt{se\-lect} call to account for uncaught signals that cause 
\texttt{se\-lect} to exit even though the connection is not available again as required. 
When the address has been written successfully, the actual message is sent using a similar loop. 
If the connection had been established implicitly for this operation, it is terminated by calling the second disconnect function,
\texttt{Dis\-con\-nect\-From\-Re\-mote}, in the helper class. 
During message transfers it may happen that a connection to a specific receiver address is interrupted, for example due to a temporarily
severed network connection. A mechanism has been implemented in the TCP message class to hide this fact from the calling code and prevent error handling at 
that level as well as data loss.  The mechanism is present in the form of a loop in the \texttt{Send} function surrounding 
the parts of the function between and including the establishing and termination of the connection. When a lost connection is detected during a message
send operation, the connection is closed and the loop starts again, attempting to reestablish the severed connection. If this fails due to a more
severe and/or permanent network error, the transfer is aborted and the error is escalated to the calling program. 
%{\bf Such a transfer abort can either be triggered by an underlying TCP error or because of a specified timeout that runs out. }

Data sent from other communication objects is received with the help of the background accept thread started by the communication helper class. 
When new data is available from an established connection, the thread invokes a callback function of its parent message object with a connection data structure. 
Using this connection
data a check is made on an internal list whether a data receive operation for this connection is currently already in progress. If this is not the 
case, a new operation data structure is created and filled with the necessary data. 
As detailed in the TCP helper class description, the address header is read first from the connection's socket to check whether 
the received sender's address has the correct size. Like all data read operations, this is performed by attempting to first read
as much data as necessary in one call using the helper class's buffered read function. In case of insufficient amounts of data being returned in order to satisfy
the read request,
a \texttt{se\-lect} call with a zero timeout to return immediately is made to determine whether there is more data available on the connection.
Similarly to its use in the \texttt{Send} function, the \texttt{se\-lect} call is surrounded by a loop that handles interruptions by  
calling the function again. For a \texttt{se\-lect} call that signals available data a new read is attempted for the whole missing amount of data. This process is repeated
until \texttt{se\-lect} indicates that no more data is currently available for reading. In this case the receive operation data is inserted into the 
appropriate internal list, if not already present, and the callback function terminates. 

If an error occurs during a message read an operation that has been continued by this call is removed from the operations list, while
a started operation is not entered into it. Read operations are thus fully aborted upon an error. 
An address header that has been successfully read is checked for the correct length of the whole
structure. A mismatch here causes the receive operation to be aborted  as for a read error above. For a correct address size the rest
of the address itself is read by a similar read loop and the complete address is checked for the correct communication
ID type. A failure of this check again causes the termination of the receive operation. 
In the next step the reading of the message's header data is attempted, followed by a check for its correct minimum size. After this check is passed, the memory to store the 
complete message is allocated using the size specified in the header. The header already read is copied into this memory and the pointer to it is stored
in the read operation's data structure for later calls, if the operation cannot be completed in this call. Following the allocation,
the rest of the message data is read into this new memory using a similar read operation sequence.
Upon successful completion of the read operations the final receive steps are performed. The receiver's address and the received message are added to
a list of received messages. If this function call was the continuation of a receive operation,
the operation's data structure is now removed from the list of in-progress operations. Finally, a signal is sent to wake up receive functions waiting for new data.

Receiving of messages is done through implementations of the two \texttt{Re\-ceive} interface functions 
declared in the \texttt{BCL\-Msg\-Com\-mu\-ni\-ca\-tion} 
class from section~\ref{Sec:BCLMsgCommunication}. Both functions call an internal third receive function that handles the two different cases with and without
a specified timeout. If a received message is already available when the function is called, the corresponding sender address and message data pointer
is passed to the function's caller using the corresponding parameters. The address is copied into a structure provided by the caller, and the
pointer to the allocated message memory is returned as the message's address is known only after it has already been received. 
If no message is available for return to the calling code, a wait is entered on a signal object. An appropriate timeout is used if it has been specified in the
function's parameters. When the wait call returns, either because the timeout expired or because a signal has been sent by the background receive thread,
the list of messages is checked again and the first available message is returned as above. If still no message is available, the wait is entered again or the function
terminates, depending on whether the timeout has already fully expired or not. 
Messages that have been received like this must be released by the user with a call to the \texttt{Re\-lease\-Msg} method to free the 
memory allocated in the incoming data callback function described in the previous paragraphs. 

\subsubsection{The TCP Blob Communication Class}

The purpose of the \texttt{BCL\-TCP\-Blob\-Com\-mu\-ni\-ca\-tion} class is to provide the implementation of the blob communication
 mechanism on top of the TCP protocol and socket API. 
Similar to the TCP message communication class, basic functionality like binding or connecting
makes use of the facilities supplied by an internal instance of the TCP communication helper class. 
The implementations of these functions in the blob class also
perform the necessary address type checking and casting  before calling the helper object's 
corresponding function, but unlike the message class's implementations the functions perform some additional steps. 
In the \texttt{Bind} call two further steps are executed after the helper class's function has been called, initialization 
of the list of free blocks in the buffer and starting of a background thread to handle requests issued
to this communication object.  The free block list is initialized to contain one block for the whole buffer. It will later be
used to keep track of used and free areas in the buffer.

In the started request thread a wait is entered for incoming messages, using the message communication object assigned to the blob object's
exclusive use, as described in section~\ref{Sec:BlobParadigms}. Received messages are handled according to their 
type, which in general is either one of five different requests issued to this object or a reply from a remote object to an issued request. 
Of these five requests the first is the most basic one and has to be issued before any direct communication can take place between two 
blob communication objects. This request queries via its associated message object the address to which a remote blob object has been bound.
It is necessary to use the message object as only the message address is passed to blob object functions. 
To avoid unnecessary message traffic the address is queried only once when a connection is established. It is stored in the blob object
together with its associated message address for later uses. 
Two more query requests handled by the thread are a query for the blob buffer's size and a request for
a free block in that buffer. Of these two the first is handled by generating a simple reply message containing the desired
size. In order to answer the third request type a block has to be obtained from the list of free blocks. If a free block of the desired size is available,
its starting offset is sent in the reply message, otherwise a -1 is sent as a buffer full reply.
In the search for a free
block the whole list is searched to find the smallest free block of at least the requested size to reduce fragmentation of large blocks.

The remaining two request messages handled by the thread are notifications about the connection status
between the sender and receiver object. When a blob object has established a connection to a remote partner the first of these
two messages is sent to the remote object to trigger a reverse connection to the initiator object. This connection
is made both for the blob object itself and for its message object to profit from the connection for the frequent reply messages
expected. No reply other than establishing the connection is sent in response to receiving this connection message. 
A disconnect of such an established connection between two blob objects and their associated message objects is triggered by the last
request message handled by the background thread. The disconnect message is sent by the communication partner that initiates the process 
to ensure that the
connections in both directions are terminated. Like the connection notification, this message is not answered.
Replies to one of the first three request types are handled identically by placing the received reply in a list and by
triggering a signal object on which any thread expecting a reply waits. When a thread is woken up by this signal the list is
checked for the expected reply using a reference number placed in the original request. If the reply is found, it is extracted from the
list and processed as needed. 
Less additional work than in the \texttt{Bind} function is performed in the class's \texttt{Un\-bind} function, which just terminates the 
background thread prior to calling the helper object's \texttt{Un\-bind} function.

In the TCP blob class's  \texttt{Con\-nect} function several steps are executed  before and after calling the helper function's \texttt{Con\-nect} 
function. After the obligatory address type check and the check for an associated message object, the function calls the \texttt{Con\-nect} function 
of this object. Using the established message connection the remote blob partner's address is obtained using the address query message mechanism 
described above. The received address is passed to the helper object's \texttt{Con\-nect} function to establish the blob connection.
As the final step in this function, a connection notification  message is sent to the remote object
to initiate a reverse connection as discussed above. 
In the class's \texttt{Dis\-con\-nect} function  the same checks as in the \texttt{Con\-nect} function are performed initially for the address 
type and message object presence.
Using the remote blob object address from the cache list the blob object connection is terminated by calling the helper object's 
\texttt{Dis\-con\-nect} function. The message connection which is still established is used to send a disconnect notification message to abort the established reverse 
connections too. Afterward the message connection is closed as well, and finally the remote blob address is removed from the cache list. 
In analogy with the TCP message class, the locking feature is not supported
by this class. Implementations of the locking functions return the same {\em function-not-supported} error indicator as in the message class.

As detailed in sections~\ref{Sec:BlobParadigms} and~\ref{Sec:BCLBlobCommunication}, a blob data transfer is split up into
two phases. During the first phase the two communication partners negotiate where the transfer data has to be placed
in the receiver's data buffer, and the second phase is the actual transfer of the data. 
Functionality for the first phase is contained in the TCP blob class's \texttt{Pre\-pare\-Blob\-Trans\-fer} method defined in its 
\texttt{BCL\-Blob\-Com\-mu\-ni\-ca\-tion} base class. 
Like the previous functions this function also first performs the address type check together with a check for a configured message
communication object. After the checks are passed, it requests a block in the remote blob object's buffer by sending an appropriate request message with the size
specified in the function's parameter. The block's offset is then received in the request's reply 
message and is used as the transfer ID returned to the calling code. An error indicator that has been received for the
block request corresponds to an invalid transfer ID and thus can be returned directly as well. 
After the preparation phase the actual data transfer is performed by one of two \texttt{Trans\-fer\-Blob}
functions. These two functions differ in the argument types and actions they have to perform, as described for the \texttt{BCL\-Blob\-Com\-mu\-ni\-ca\-tion} class
in section~\ref{Sec:BCLBlobCommunication}. 
In the single block version of the function the block parameters are simply passed to the multiple block version by placing them into
lists containing just one element each. 
In the second, multi-block, transfer function the same address checks as in the transfer preparations function are made, followed by
additional checks of the transfer parameters, starting with the validity of the transfer ID. The two lists of block sizes and pointers
are checked for identical numbers of elements, while the offset list is not checked since any missing offsets cause the blocks 
to be placed consecutively. 

No validity checks
are made for the offsets and sizes of the data blocks in this function, instead the values
for each block are sent to the receiver prior to the data itself to be validated there. The obvious drawback of transferring 
data that will be discarded if the parameters are wrong is reasonable as this should happen rarely. However, the advantage is the avoidance of a validation
message exchange, that would otherwise increase the transfer latency. Actually the only occasion where offsets could be wrong, apart from a bug in
the library, arises when an incorrect transfer ID is passed to the transfer function. 
After the preliminary checks have been passed successfully, the remote blob communication object's address is queried, either from
the cache list for established connections or through a query message. Using this address the helper class's \texttt{Con\-nect\-To\-Re\-mote} function
is called to establish a connection to the remote blob object. A timeout is set for this connection, if one has been specified in the function's
parameters. 
Following this is the main loop of the transfer function that iterates over each block to be transferred. 
In the loop each block's data pointer, size, and offset are extracted from their respective list or calculated in the case of a missing offset.
These parameters are then placed into a structure derived from 
\texttt{BCL\-Net\-work\-Data\-Struct} described in section~\ref{Sec:DataFormatTranslation}. This structure is then sent to the receiver blob object prior to the 
data block itself. Sending of the 
header as well as the data is done in loops similar to the one used for sending message data, detailed in the
preceeding message class section. 

Errors that occur during the transfer can be separated in two classes. Broken connections, the first category of error, 
are handled by signalling
a broken connection to the helper object and then reestablishing the connection. Such an interrupted  transfer is resumed by retransmitting the block where the 
interruption occured, as it cannot be determined exactly which amount of data was successfully sent. Other 
types of errors that occur during the writing of data or during attempts to reestablish a connection are handled by aborting the function.
%Timeouts are handled as a timeout error. 
Once all blocks of a transfer have been sent the function waits indefinitely or for the specified timeout to read a 32~bit data word indicating
that the receiver has read all transmitted data successfully. After receiving this indicator value, the connection is released by calling the 
helper object's \texttt{Dis\-con\-nect\-From\-Re\-mote} function, to either decrease its usage count or terminate it.

Receiving of blob data is achieved similar to receiving of messages with the help of the background thread in the TCP helper class. From the thread a data reading
callback is invoked that functions similarly to the corresponding message class function. With the help of the connection data structure transfers are either started or
resumed as appropriate. The actual process of reading the data is also split into two parts, for the transfer header  
containing the block's destination information and the data itself. 
Using the offset and size values from the transfer header a check is made whether the transfer is valid and can be placed into a reserved block in 
the object's receive buffer. If this is the case, the read operations for the data are performed such that it is placed directly into the appropriate
receive buffer area. All reading steps are executed similarly to the reading of messages in small inner loops with \texttt{read} and \texttt{se\-lect}
calls. Uncompleted transfers for which no more data is available for reading are placed into a list to be resumed in later calls of the function
when new data is available. Errors that occur during the receive operation are signalled to the calling accept thread via the functions's return value
and result in the closing of the concerned connection. 
When all expected data has been read from the socket, the 32~bit completion indicator expected by the sending transfer function is written
 using a \texttt{se\-lect} call with a short non-zero timeout to verify that the connection is available for writing, followed
by a single 32~bit write operation. At this stage errors are ignored and not reported back to the accept thread. After writing the completion
indicator the function terminates, signalling success to the calling accept thread. 

Access to data received from a remote object is possible via the \texttt{Get\-Blob\-Data} function or a combination of the \texttt{Get\-Blob\-Off\-set} and 
\texttt{Get\-Blob\-Buffer} functions, as described in section~\ref{Sec:BCLBlobCommunication}. 
The starting offset required to access the data, both for \texttt{Get\-Blob\-Data} and \texttt{Get\-Blob\-Off\-set}, is equal to the transfer ID 
passed as the functions' only parameter. 
Both functions check the list of used blocks for a block whose starting offset is equal to
that transfer ID. If such a block is found the transfer ID is presumed to be valid. Otherwise an error is reported and no valid pointer or
offset is returned. 
When a block of received data can be released, the object's \texttt{Re\-lease\-Blob} function is called to free the used block in the receive buffer.
The block is located in the list by comparing its starting offset with the transfer ID passed as the function's parameter. If the appropriate block
could be found, it is removed from the used block list and inserted into the free block list to be used again.

\subsection{The SCI Communication Classes}

In this section the three classes are described which implement the two different communication facilities on top of the SISCI API \cite{SISCI} for 
SCI adapter cards by Dolphin \cite{DolphinWeb}. Functionality in these classes is very similar to what is provided by the three TCP classes: one 
class each for the message and blob communication facilities, \texttt{BCL\-SCI\-Msg\-Com\-mu\-ni\-ca\-tion} and \texttt{BCL\-SCI\-Blob\-Com\-mu\-ni\-ca\-tion}, and 
the helper class \texttt{BCL\-Int\-SCI\-Com\-Helper} encapsulating functions for both communication mechanisms. 

Dolphin SCI cards are a high performance system area network (SAN) technology designed to provide tightly coupled interconnects in clusters of 
PCs or workstations. SCI is an IEEE standard \cite{SCIIEEE} for a shared memory interface. Nodes connected via SCI are able to write to or 
read from a remote node's memory directly. In addition, the Dolphin SCI adapters contain DMA engines capable of copying data autonomously between nodes
without intervention by a host CPU. Low level details of accessing the adapters are hidden by the SISCI C-API, based upon supplied 
device drivers to access the adapter cards for controlling connections to remote nodes and DMA transfers. Programmed I/O (PIO) transfers, performed
by a host CPU, are executed over an established connection without any API or driver intervention. 
For the SCI communication classes any data transfers, for messages as well as for blobs, are performed via a memory segment in the receiver object
to which the remote sender writes the data to be transferred.  This memory segment, exported for remote node access, has to be allocated
by the SISCI API. It is not possible to specify a normal user allocated memory block to the SISCI system for exporting.

\subsubsection{\label{Sec:BCLIntSCIComHelper}The SCI Communication Helper Class}

Analogous to the \texttt{BCL\-Int\-TCP\-Com\-Helper} class the \texttt{BCL\-Int\-SCI\-Com\-Helper} class provides common services to the two classes implementing the 
message and blob communication mechanisms on top of the SISCI API and SCI network. Due to the more complex SISCI API being used to access the SCI cards,
this helper class needs to contain more support functions than its TCP counterpart. 
The first interaction with the SISCI API is made by this class already in its constructor, as each API function call requires a handle
to an SCI descriptor in its call. This descriptor is 
obtained using an \texttt{SCI\-Open} function call in the constructor. An error occuring here is only logged without any further action, although
subsequent SISCI calls  will fail due to the invalid descriptor. Therefore, all functions in the class check the descriptor's validity first
before performing any other action, particular prior to any API function call. 

As for the TCP helper class the first functionality supported by this class is binding to a valid address to enable an object to 
accept remote connections and data. 
For this purpose  a local memory segment is allocated and exported under a given ID for remote access using the SISCI API. 
The \texttt{Bind} function that executes this task accepts two arguments, the size of the memory segment to be allocated
and the address under which the segment is to be made available in the form of a pointer to an \texttt{BCL\-SCI\-Ad\-dress\-Struct}
structure.
In the \texttt{Bind} function the local ID of the SCI adapter specified in the address is queried. Using that ID's lower 16~bit together with the 16~bit
segment ID also specified in the address, a 32~bit ID is generated. Since each segment ID has to be unique on each node and each node ID
has to be unique in a cluster, the generated 32~bit ID is unique in a whole cluster as well. A prerequisite is that only the lower 16~bits of a node
ID are used in a cluster. Next to the node ID restriction to be 16~bit, a further requirement placed on the generated ID is that its final value 
must not be 0xFFFF, which is reserved as an invalid ID. 

A memory segment of the given size is created with the specified 16~bit segment ID. It is mapped and exported so that the creating program as well as
remote programs can access it for reading and writing. Creation of the segment is executed using the \texttt{SCI\-Cre\-ate\-Seg\-ment} call, specifying
a size one page larger than the user specified amount. This additional page, typically 4~kB, is needed to provide room for a header structure located at the segment's
start. Mapping and exporting are done via \texttt{SCI\-Map\-Lo\-cal\-Seg\-ment} and the combination of \texttt{SCI\-Pre\-pare\-Seg\-ment} followed by 
\texttt{SCI\-Set\-Seg\-ment\-A\-vai\-la\-ble}. Following the segment creation an SCI interrupt is created by calling the \texttt{SCI\-Cre\-ate\-In\-ter\-rupt} function with no
ID specified so that an available one can be selected and returned by the SCI system. This interrupt will be used to signal the availability of new data 
that has been written into the memory segment to receiving objects to avoid the need of polling for data. 
Before the segment is made fully available for remote access using the \texttt{SCI\-Set\-Seg\-ment\-A\-vai\-la\-ble} call, the header at its start is filled
with the returned interrupt ID, the segment's size, and further management data required to handle the memory as a FIFO ring buffer area. To ensure
accessibility from remote nodes the header structure is derived from \texttt{BCL\-Net\-work\-Data\-Struct} and thus allows to use the data transformation
mechanism described in section~\ref{Sec:DataFormatTranslation}. 
The mapping function
returns a standard C pointer, allowing the use of the segment's data in the local program like any other memory area. 
This pointer is used to access the messages that have been written into the segment from remote processes. 

For errors occuring during any of these steps the error number returned by the SCI subsystem is logged and transformed into a standard C
system error code passed to the calling function. Ultimately this number is reported to the program using the library from
 an SCI message or blob communication object, either as a return value or an error callback parameter. After an error has occured, 
the steps that have been taken in the binding process are reversed so that the object is brought into the same state as before the call. 
The above reversal is achieved by calling the \texttt{Un\-bind} function in the helper class, which iterates through each of the steps of
the bind process in the opposite order. For each of these steps it is checked whether it has been performed by analysing the allocated resources.
First the segment is made unvailable for external connections, and then the created interrupt is removed using the 
interrupt number stored in the segment's control structure. After this the segment is unmapped and finally
destroyed, freeing the allocated memory. API functions called in this process are \texttt{SCI\-Set\-Seg\-ment\-Un\-a\-vai\-lab\-le}, 
\texttt{SCI\-Re\-move\-In\-ter\-rupt}, \texttt{SCI\-Un\-map\-Seg\-ment}, and \texttt{SCI\-Re\-move\-Seg\-ment} in this order.

Like the TCP communication helper class \texttt{BCL\-Int\-TCP\-Com\-Helper} from section~\ref{Sec:BCLIntTCPComHelper}, the  \texttt{BCL\-Int\-SCI\-Com\-Helper} class
also offers two sets of function calls for establishing and terminating connections to 
remote objects. Both interfaces are identical to the TCP functions in their respective tasks and arguments, apart from the SCI and TCP differences,
e.g. in the addresses used. The first of the interfaces is used for explicitly established connections while the other one is used to establish implicit 
connections, as well as by the first helper class functions for explicit connections. 
In adition to the remote address and an optional additional error callback object, the explicit connection API supports two flags, 
one specifying the use of a timeout value and the other to specify whether a 
connection should be established immediately or as an on-demand connection. 
The explicit \texttt{Dis\-con\-nect} function also accepts four of these parameters: the remote address, the
error callback object, and the timeout flag and value parameters. 

Five of the eight arguments being used by the implicit connection function \texttt{Con\-nect\-To\-Re\-mote} are in principle similar to the \texttt{Con\-nect}
function's arguments. A small difference can be found in the flag that differentiates between an immediate and an on-demand connection. Since a set flag
specifies that a connection is to be established immediately, the flag's meaning is reversed with respect to the flag for the 
\texttt{Con\-nect} function. 
Concerning the parameters specific to this function, the first  is used to return information about the established connection to the calling function.
This information includes a SISCI API descriptor for the connected remote memory segment, a pointer to access the segment mapped into the local program's 
memory, and a handle used to trigger the remote segment's data notification interrupt. 
The first of the remaining two parameters is a flag that specifies whether
the connection is to be stored and thus established as a permanent connection or whether it is an implicit connection only established for a single
send operation. A return flag for  the function's caller is contained in the last parameter, indicating whether the
connection was already existing or whether it  had to be initiated by this function call. In the first case the data for the connection is obtained from an internal
list and is returned to the calling function without further communication taking place, while in the second case the connection has been established, 
if this was specified in the function call. 

In the first step of the connection attempt the unique global ID of the remote segment is constructed from the node and segment IDs in the given address as
described for the bind process. If the resulting 32~bit ID is invalid, the connection attempt is aborted with an
error. Otherwise a new SCI descriptor handler is obtained by calling the \texttt{SCI\-Open} function. This step is necessary because each SCI descriptor
is only able to handle one remote segment, one local segment, and one interrupt. Using this new descriptor a connection attempt is made to
the memory segment on the remote node with \texttt{SCI\-Con\-nect\-Seg\-ment}, specifying the remote node's ID and the ID of the segment to connect to. 
One retry is made for
two types of SCI errors reported back from this function. This has been found experimentally to be necessary.
% under certain circumstances where a first connection attempt is refused. 
After a successful connection attempt the \texttt{SCI\-Map\-Re\-mote\-Seg\-ment} function is used to map the header part of the remote segment into the local address
space. From the header the segment's total size is obtained, which allows to map the whole segment into the local address space. The segment descriptor and the
pointer returned by the two API functions are stored in the data structure for the established connection. 

After completion of the mapping, an SCI sequence is created for the remote segment. A sequence is an SCI mechanism that
allows to check for errors while accessing a remote segment and to exercise some control over local read and write buffers for a
remote segment. It is created by calling the \texttt{SCI\-Cre\-ate\-Map\-Se\-quence} function in two attempts, first for a fast sequence type, of which
only a limited number are available, and for the normal type if the fast one fails. Then the created sequence is cleared from old errors by 
calling the \texttt{SCI\-Start\-Se\-quence} API call in a loop until it indicates success or a specified timeout expires. 
In the final step of the connection process the remote segment's interrupt number is read, and a connection attempt to the interrupt is made using the
\texttt{SCI\-Con\-nect\-In\-ter\-rupt} function with the returned descriptor being stored in the connection data structure.
As for the \texttt{SCI\-Con\-nect\-Seg\-ment} call, this function may fail the first time under specific circumstances and is therefore attempted
 a second time upon failure. 
For an on-demand connection it is possible that an unestablished connection is locked by the user. Instead of establishing the connection when the lock call
is made, a flag is set in the connection data structure in the helper class, and the flag is checked in \texttt{Con\-nect\-To\-Re\-mote} when a connection has 
been fully established. If the flag is set, the class's internal \texttt{Lock\-Con\-nec\-tion} function, described below, is called with the connection's data 
to establish the lock.

Closing of an established connection is performed by the \texttt{Dis\-con\-nect\-From\-Re\-mote} function, which requires the connection data structure of the 
connection concerned. Additionally, an optional error callback as well as timeout flag and value parameters are accepted by the function. If a matching established
connection is found, its usage count is decremented. If the usage count subsequently  is zero the connection is terminated. 
To terminate a connection first a local segment created by a DMA enabled blob class (see below) is destroyed, and the connection's
lock flag is checked. If the flag is set, the remote segment is unlocked. The SCI 
calls to release the connection's resources are made next, releasing the remote interrupt, unmapping the segment and disconnecting from it.
\texttt{SCI\-Dis\-con\-nect\-In\-ter\-rupt}, \texttt{SCI\-Un\-map\-Seg\-ment}, and \texttt{SCI\-Dis\-con\-nect\-Seg\-ment} respectively are used for these
steps. Afterwards the 
SCI descriptor used for the connection is closed by calling \texttt{SCI\-Close}, and the stored connection data is removed from the object.
If no matching connection could be found in the helper object, the connection is presumed to be an implicitly established one and the
steps are performed identically, except for the removal of the connection's data structure.

All PIO read and write operations executed on remote memory segments by a node's CPU, in the helper class as well as in the message and blob classes, are contained
in an error checking loop. This loop uses the SCI sequence data created in the \texttt{Con\-nect\-To\-Re\-mote} function with the help of two internal functions of the
class. These two functions, \texttt{Start\-Se\-quence}  and \texttt{Check\-Se\-quence}, are called prior and after the access  respectively.
\texttt{Start\-Se\-quence} is used to initialize the sequence while \texttt{Check\-Se\-quence} clears any pending errors and
checks for errors that occured during the operation. The return value from \texttt{Check\-Se\-quence} can be one of three types: success, a fatal
error, or a temporary error. For the last type the operation has to be repeated until one of the two other cases occurs. 
In the two functions the appropriate SCI API calls \texttt{SCI\-Start\-Se\-quence} and \texttt{SCI\-Check\-Se\-quence} are encapsulated with the appropriate
temporary variables and error conversion required. The \texttt{SCI\-Start\-Se\-quence} call is also repeated until either success or a fatal error is reported, after
which the operation can be started or has to be aborted respectively.

Unlike the TCP communication classes, the SCI message communication class supports the locking feature 
defined in the \texttt{BCL\-Com\-mu\-ni\-ca\-tion} class, as already detailed in the preceeding paragraphs.
Like the connection API two versions of the locking functions exist. One used for explicit user initiated locking and a second one 
used internally by the first version and to handle the write arbitration necessary for normal send operations. 
The \texttt{Lock\-Con\-nec\-tion} function for explicit locking requires the connection's remote address, the usual timeout flag and value, and an optional
erorr callback parameter. For unestablished on-demand connections only the lock flag in the connection's data structure is set as discussed above.
If an established connection to the specified address exists, the internal locking function is called with the connection's data structure to obtain
the lock. When the lock arbitration in that function has completed successfully, lock flags are set in the connection data structure and in the remote segment's
initial header structure. In the last locking step a local cached copy of the remote segment's control data is created to avoid remote read or write calls. This
caching is possible, as only the local process is allowed to access that data after locking. 
In the corresponding \texttt{Un\-lock\-Con\-nec\-tion} function the same parameters as for the \texttt{Lock\-Con\-nec\-tion} function are available for use. For locked
but unestablished on-demand connections the lock flag is simply cleared without any further action. 
If the specified connection is established and locked, the locking flag in the remote control structure is cleared and the internal \texttt{Un\-lock\-Con\-nec\-tion} 
function is called to relase the granted write access to the remote segment. In a last step, the lock flag in the local connection data structure is cleared
as well. 

The internal version of the \texttt{Lock\-Con\-nec\-tion} function accepts the connection data structure in addition to the parameters required for the
explicit \texttt{Lock\-Con\-nec\-tion} function. For write arbitration two 32~bit large fields in the segment control data are used, one to request
write access and the other to indicate the current owner of the write access. By writing the sender's own unique 32~bit ID into the first field and triggering
the remote segment's interrupt, the remote receiving process is notified that a process requests permission to write. After this activation
the remote process reads the requesting ID and if it is valid and no other process currently owns the segment, the ID is written
into the field for the current owner. When the requesting process reads the owner ID and finds its own ID, it knows that it has been granted exclusive write
access to the remote segment. If it reads another ID, the process of writing its ID into the request field is repeated, with small busy waits, until 
the request is granted or a specified timeout expires. The busy wait uses a short loop to run for $2~\mu \mathrm s$ to keep the intervals short
and not delay too long. Normal system wait functions 
which actually suspend a process without 
using processing time cannot be used for this purpose as they work with a granularity of 10~ms, much too long for this purpose. 
A granted write access is released in the internal \texttt{Un\-lock\-Con\-nec\-tion} function by writing the invalid 32~bit ID with all set bits into the
current owner field of the remote segment's header structure. After writing the function triggers the remote segment's interrupt to allow the receiver to update its 
arbitration state and ends normally. 
For locked connections two helper functions exist to increase a programs efficiency by handling a local cache copy of a remote segment's control data. 
Since only the message communication class has to deal with reading and writing values from the control data structure, when it writes messages to the remote
segment,  these two functions are
only used by that class and not by the blob class. The first of these two functions, \texttt{Up\-date\-Cached\-CD}, updates a local cache copy from the remote
segment's data structure, while the second one, \texttt{Write\-Back\-Cached\-CD}, writes back a modified cache copy.

Two additional helper functions contained in \texttt{BCL\-Int\-SCI\-Com\-Helper} are only used by the SCI blob communication classes to compensate 
restrictions in the SISCI API implementation. It is not possible to execute a DMA transfer, where the SCI card copies
the data autonomously to the remote node, from normal user space program memory. Instead these transfers are only possible if a local SCI memory segment 
is used as the data source. This is an implementation issue with the supplied SISCI API which does not allow to register ordinary memory as a segment to
make it usable as a data source location. The SCI adapters themselves are in principle zero-copy DMA capable.
To retain  the desired DMA capability these helper functions allow to create a 
local memory segment as a buffer for DMA transfers. Blob Data to be transferred  is copied 
from the ordinary program memory into this local segment in chunks of up to the segment's size. Each chunk is then transferred by the DMA engine 
from the local segment to the 
remote target memory segment. This approach still requires a memory copy using the CPU, but as it is now only local it should still
be faster and more efficient than a PIO transfer of the data to the remote memory. 
Creation of this buffer segment is done in the \texttt{Cre\-ate\-Local\-Seg\-ment} function with the size specified as a parameter. The segment's
data is stored in a connection data structure to associate it with a specific connection, either an implicit or explicit one. An ID for the segment is obtained 
by using IDs from the part of the 32~bit large SCI ID space that cannot be used by user segments for which only 16~bit are allowed. 
A  call to \texttt{SCI\-Cre\-ate\-Seg\-ment} is made for each possible ID to test whether it is already used or available. If the call succeeds, the 
segment created by it is used as the local segment. Otherwise the process is repeated until a free ID is found or all available segment IDs have been tested.
In the latter case the function cannot create a segment and fails with an error. 

A successfully created segment has to be prepared so that it can be accessed by the SCI adapter card for DMA and from the program through a pointer, using 
the \texttt{SCI\-Pre\-pare\-Seg\-ment} and \texttt{SCI\-Map\-Lo\-cal\-Seg\-ment} calls respectively. Handles obtained from the three SCI API Calls
are stored in the data structure of the connection associated with the segment. 
In the \texttt{De\-stroy\-Local\-Seg\-ment} function a local segment's usage counter is decreased. If it is zero or less the segment is released. To release
a segment its user space pointer is invalidated and unmapped by calling \texttt{SCI\-Un\-map\-Seg\-ment}, and the segment itself is 
freed with \texttt{SCI\-Re\-move\-Seg\-ment}.
In addition to these helper functions a number of smaller utility functions are provided. These functions allow access to several of the object's
fields so that the parents of helper class objects can 
use SISCI API routines that require these parameters to expand the class's functionality as necessary.

%{\bf 16~bit segment IDs because of global unique 32~bit ID for node ID/segment ID combination.}

\subsubsection{\label{Sec:BCLSCIMsgCommunication}The SCI Message Communication Class}

Next to the SCI helper class the second important class for the SCI communication implementation is the \texttt{BCL\-SCI\-Msg\-Com\-mu\-ni\-ca\-tion} class
that provides the message communication mechanism using the SCI network technology. Like the TCP message class it relies on services from
its corresponding helper class but performs more actions beyond type checking and calling helper class functions in its binding and
connection functions. 
In the \texttt{Bind} function, this additional action is performed after  the address type has been checked and cast and the  \texttt{Bind} function from the helper class
has been called. A background thread responsible for the arbitration of write access requests to the local segment, as 
discussed in the previous section, is created as well. 
The \texttt{Un\-bind} function in the message class first stops this started arbitration thread and calls the helper object's \texttt{Un\-bind} function afterwards. 
Support for explicit connections, the next basic functionality provided by the message communication class, is implemented in the
\texttt{Con\-nect} and \texttt{Dis\-con\-nect} functions, which have been declared in the \texttt{BCL\-Com\-mu\-ni\-ca\-tion} base class. All of these
functions basically just call the corresponding helper \texttt{Con\-nect} or \texttt{Dis\-con\-nect} function with the appropriately converted
parameters. 
As detailed in the preceeding section, the SCI message communication class, unlike the TCP classes, supports the locking feature for established 
connections to avoid the overhead of write arbitration for each message in cases where this is possible. Both the \texttt{Lock\-Con\-nec\-tion} and the \texttt{Un\-lock\-Con\-nec\-tion} 
functions are implemented in the two versions defined in the base communication class. All four functions also call their respective counterpart in the helper
class with the required parameters. 

%{\large \bf Ringbuffer organisation?}

The first functionality specific to the message class is contained in the implementations of the \texttt{Send} functions inherited from \texttt{BCL\-Msg\-Com\-mu\-ni\-ca\-tion},
which call a third internal version that can send with and without a specified timeout. A passed message is checked in this internal \texttt{Send} function 
by reading its first and last byte to prevent segmentation violations during the process of copying the message into
the remote memory segment, similar to the TCP message class. 
In the next step the helper object's \texttt{Con\-nect\-To\-Re\-mote} function is called to retrieve an existing connection's data or to establish a new one to the
destination address. Depending on whether the connection is locked or not, the header information for the remote segment is either read from the cached
copy maintained by the helper object or from the remote memory itself. 

If the connection is not locked already, the helper class's internal \texttt{Lock\-Con\-nec\-tion} function is called to obtain the required write privileges for
the segment. After the write access has been obtained, the free amount of memory in the segment is calculated. If insufficient space is available, according to the current 
cached header copy of a locked connection, it is updated from the original remote header. If still insufficient memory is available to write
the specified message into the remote segment, the function aborts after executing its cleanup section. 
During the write process the sending object's address is written first, followed by the message data itself. Both steps are surrounded by 
the SCI sequence error checking loops described in the helper class's section. After these two write steps have been completed, the new write index
is determined and written back, both to a locked connection's cache copy as well as to the remote segment's header structure. This last step is
necessary even for locked connections so that the remote receiver can determine the amount of new data. With the write index written back 
correctly, the write privilege for an unlocked connection is released and the remote segment interrupt is triggered to inform the receiver
about the availability of a new message. Finally, the helper class \texttt{Dis\-con\-nect\-From\-Re\-mote} function is called to relase the connection,
terminating it if it is no longer used.

The counterparts to the \texttt{Send} functions are the implementations of the two \texttt{Re\-ceive} functions defined in the message class. 
Both functions also call a third internal function to perform the required actions, either with or without a timeout. In this third function the read and write indices
in the receive segment's header structure are checked to determine if data is already available. If no data is available, a wait is entered on a 
signal object triggered by the background thread when the SCI interrupt is triggered itself. 
%If a timeout has been specified to the \texttt{Re\-ceive} function, it is also used to limit the time waiting for new data. 
%proper address structure
When the function progresses past the signal, either because data was already available or after the wait, a small loop is run for a fixed number
of iterations or until the read and write indices indicate that data is available. Each loop iteration performs a $10~\mu \mathrm s$ busy wait. This waiting loop
is necessary as SCI does not guarantee in-order delivery,
and the SCI packet that triggers the interrupt can arrive slightly before the packet holding the updated write index. The function exits with a timeout error
when no data is available after the wait . 
If data is available in the receive memory segment, a number of checks are performed on the data to ensure its validity. 
For the expected address structure the correct size and type are checked first, and for the message data itself the minimum length
to store a message header of type \texttt{BCL\-Mes\-sage\-Struct} is required. When these conditions are met the address is copied into the 
function's output address parameter, and memory for the message is allocated. The message data from the segment buffer is copied into
that memory, and the pointer to it is placed in the pointer reference parameter for output. As the function's last step the read index
in the segment's header structure is updated to reflect the amount of data extracted from the receive buffer. The above copy steps are required
to be able to maintain a simple ring buffer whose management can be simply split between local reader and remote writer. If the data were to be
accessed directly in the shared memory a more elaborate buffer management would be required since an application could not be relied upon to
release the messages in the correct order. The required more complex buffer management would most likely require more (costly) remote read and write operations 
from the writing process.

The final important method in the \texttt{BCL\-SCI\-Msg\-Com\-mu\-ni\-ca\-tion} class is the interrupt handler function executed by the arbitration background
thread. It consists of a loop that runs while the object is bound to its receive address. In the first part of the loop the request field
in the receive segment's header structure is checked for a valid sender ID indicating an active write request by a remote object. If such a request 
is found and no sender is currently active, the request ID is placed into the current sender ID field, granting write access to that object. Following this
arbitration part the function enters a wait for the segment's associated interrupt to be triggered using the \texttt{SCI\-Wait\-For\-In\-ter\-rupt} API
function. When this functions returns because an interrupt has been received, the segment header's read and write indices are checked. If they are
unequal, the signal object on which the \texttt{Re\-ceive} function waits is triggered to indicate new data to the function. As the last part
of the loop, a timer that has been started when write access was granted to a sending object is checked. If the connection is not locked explicitly and 
no change has been made on the segment's write index for a specific amount of time, the current sender field in the header is reset, removing the
sender's write access. This timeout sender reset has been introduced to cope with unexpectedly terminated connections from remote objects during a
sending process.

\subsubsection{\label{Sec:BCLSCIBlobCommunication}The SCI Blob Communication Class}

An implementation of the blob communication mechanism defined by the \texttt{BCL\-Blob\-Com\-mu\-ni\-ca\-tion} class from section~\ref{Sec:BCLBlobCommunication} on top
of the SCI network technology is provided by the \texttt{BCL\-SCI\-Blob\-Com\-mu\-ni\-ca\-tion} class. Unlike 
the other three classes implementing one of the communication mechanisms it contains some network specific code already in its constructor. 
If DMA functionality is enabled, it uses the \texttt{SCI\-Query} function to determine three parameters relevant for
DMA transfers: the starting offset alignment, the block size alignment, and the maximum blocksize. 
%The two alignment values specify the base values that 
%all block starting offsets, both for data source and destination, and block sizes respectively have to be multiples of. 
As already introduced in the \texttt{BCL\-Int\-SCI\-Com\-Helper} class's section, the SISCI API does not allow to specify an arbitrary user space buffer for receiving
but only SCI segments that have been allocated by \texttt{SCI\-Cre\-ate\-Seg\-ment}. To modify the default behaviour of the \texttt{Set\-Blob\-Buff\-er} function inherited
from \texttt{BCL\-Blob\-Com\-mu\-ni\-ca\-tion} the function is overwritten to return an error when a user area is specified as a receive buffer. 

In the class's \texttt{Bind} function the helper objects \texttt{Bind} function is called initially to create a segment and make it available for 
remote connections. The pointer to the segment's data part is then passed to the \texttt{BCL\-Blob\-Com\-mu\-ni\-ca\-tion} parent class's \texttt{Set\-Blob\-Buff\-er}
function to set the receive buffer pointer stored in that class. In the next step the object's free block list is initialized to contain one block describing
the whole buffer area, and a background thread is started to handle request messages and replies for the object. 
To reverse these steps the \texttt{Un\-bind} function terminates the background request thread and calls the helper object's
\texttt{Un\-bind} function to release the allocated resources. 

%{\bf \Large CreateLocalSegment}

Similar to the \texttt{BCL\-SCI\-Msg\-Com\-mu\-ni\-ca\-tion} class, the \texttt{Con\-nect} and \texttt{Dis\-con\-nect} functions exist in two public and one protected internal version
each, with the public versions calling the respective internal one.
In the internal \texttt{Con\-nect} function the object's associated message object is first connected to its remote counterpart by calling its own
\texttt{Con\-nect} function. After this connection is established the message object is used to query the remote blob object's own address if necessary.
This address can also be specified in a parameter to the function when it is called in response to a connection request message from another object. In 
that case the address query is skipped. 
Following the successful address query the remote blob address is placed into a list of addresses to avoid further query messages. %cache
If the function is called from one of the public \texttt{Con\-nect} functions, a connection request message is sent to the remote blob object containing
the local blob object's own address. As written above, in response to such a message a reverse connection to the message's originating object
is established. %, particularly for the benefit of the query reply message that have to be sent. 

Like the internal \texttt{Con\-nect} function the \texttt{Dis\-con\-nect} function can also be called with or without the remote blob object's address.
The second case is used
if it is called from one of the public \texttt{Dis\-con\-nect} versions. Similar to a connection request, the remote blob address is specified when 
the function is called in response to a disconnect request message received from a remote object. When the address is not specified, it is obtained from the
address cache or it has to be queried using an address query message. Once the remote blob address is available, the \texttt{Dis\-con\-nect} function of the 
communication helper object is called to terminate the blob object connection.
During a locally initiated disconnect, the result of a call to a public \texttt{Dis\-con\-nect} function, a disconnect request message is sent to the remote
blob object to terminate the previously established reverse connection. After sending that message the message object's \texttt{Dis\-con\-nect} function
is called to terminate that connection as well, and the remote blob address is cleared from the object's address cache. 
Similar to the TCP blob communication class, the SCI blob class does not require the connection locking feature as the space for each data block is negotiated 
before the transfer. Therefore locking functions are implemented as empty functions only.  
An implemented helper function is \texttt{Get\-Re\-mote\-Blob\-Size}, which uses the associated message object to send a message that queries the 
remote blob object's buffer size.  If a reply is received, the size contained in the reply is returned to the function's caller.

The first function that implements functionality for the blob data transfer is \texttt{Pre\-pare\-Blob\-Trans\-fer}. It sends a query message to the remote 
destination object to request a free block of the specified size. In the received reply message the destination
object specifies the offset  of the block in the remote receive buffer segment or it indicates that no free block of sufficient size is available. 
Similarly to the TCP blob object, the offset in the buffer is equal to the transfer ID for that particular transfer, as it is used for the 
\texttt{Trans\-fer\-Blob} function. 
Using the returned transfer ID one of the two implemented \texttt{Trans\-fer\-Blob} functions can be called to transmit the block or blocks to the receiver
object. Again like the TCP blob class, the transfer function that allows only one block to be specified converts the block's parameters into parameter
lists and calls the multi-block version of the function. This function performs the same preliminary checks for the validity of the transfer ID and 
equal number of elements in the data pointer and size lists as the TCP function.
After obtaining the remote blob address, either from the cache list or by querying it with an appropriate message, the helper object's 
\texttt{Con\-nect\-To\-Re\-mote} function is called to get access to a connection to the remote object, either by establishing a new or using an existing one. 
If the class's DMA functionality
is enabled, a DMA queue is created using the \texttt{SCI\-Cre\-ate\-DMA\-Queue} API function to process DMA transfers. 
With the required  preparations completed, a loop is started that transfers each of the specified blocks to its respective remote destination.

Without enabled DMA each block is simply copied to its destination in the remote memory segment mapped into the current
processes' address space. This is done by a normal \texttt{mem\-cpy} surrounded by the SCI error checking described previously. 
For DMA enabled transfers several steps have to be executed for each block. It starts with the calculation of {\em slack} areas for the beginning
and end of the block, which result from improper alignment with the values required by the SISCI system. Such slack blocks have to be copied to the remote destination by a 
normal \texttt{mem\-cpy} call. In the framework these steps should not be necessary as all memory and buffer management functions use
a sufficiently large, matching, alignment value. Support for unaligned blocks has been included nonetheless to retain the generic usability of the class. 
The remaining block data is split into smaller blocks,
up to the size of the local DMA segment used for the DMA transfer. Each of these sub-blocks is then copied by \texttt{mem\-cpy} into 
the local helper SCI segment. After copying the block's data, a DMA transfer is initiated by calling \texttt{SCI\-Enqueue\-DMA\-Trans\-fer} to place the transfer
into the DMA queue and \texttt{SCI\-Post\-DMA\-Queue} to start the transfer. \texttt{SCI\-Wait\-For\-DMA\-Queue}  is used to wait for the transfer's completion,
after which the queue is reset for the next transfer by calling the \texttt{SCI\-Re\-set\-DMA\-Queue} function. If one of the calls used in transmitting the block
returns an error, the transfer is presumed to have failed, and the block is copied explicitly by another \texttt{mem\-cpy} call. 
When all data of a block is transferred, the last byte of the block in the remote destination is compared with the last byte in the block's source. 
For a more reliable check the destination byte is filled with the bitwise inverted value from the data block's last byte before the transfer is started. 
The check is done in a loop that runs until the bytes are equal or a timeout expires, to ensure that the data block has been transmitted
correctly when the function exits. Problems with data arriving out of order, like for the message classes have not been
detected here, presumably due to the time needed to send the explicit announce message before the data can be accessed. 
Finally, after all data blocks have been transferred to the remote buffer, the DMA queue is freed if it was created. The established connection
is released, and terminated if it is no longer used, by the communication helper's \texttt{Dis\-con\-nect\-From\-Re\-mote}.

For the receiver of a blob data transfer the \texttt{Get\-Blob\-Data} and \texttt{Get\-Blob\-Off\-set} functions allow access to the received data by returning
a direct pointer to the data or the data's offset from the blob buffer's beginning respectively. Both functions check the list of used blocks in the buffer
to verify  the specified transfer ID by comparing it with the block's offset in the buffer. 
To free a block used by transferred data, the \texttt{Re\-lease\-Blob} function is used that searches for a used block with the starting offset specified 
by the transfer's ID. A found block is removed from the used block list and merged into the free blocks list for further use.

Much of the \texttt{BCL\-SCI\-Blob\-Com\-mu\-ni\-ca\-tion} class's functionality is contained in the \texttt{Re\-quest\-Han\-dler} function that runs in the background 
thread started by the \texttt{Bind} function to handle request messages and replies. The function consists of a loop in which messages are received by
the associated message communication object and then processed according to their type. As for the TCP class, the request handler handles three query messages
with their respective replies and four other request messages. 
Of the three queries, the first one is for a blob's address, and this is answered with a reply containing the three parameters of the blob object's 
own address. In reply to the second query for the blob's receive buffer size the appropriate size is sent to the query message's sender. 
The third query request type is a query for a block in the object's receive buffer. A matching block has to be searched first by the object's buffer manager in the
free block list. If a matching block is found, it is moved into the used block list. Its parameters are placed into the reply message sent
back to the requesting object. Without a matching block a negative reply is sent back to inform the sender that no block is currently available.
Reply messages to queries sent by the object are all handled in an identical manner. A reply is placed into a list of received replies by the request handler,
and the signal object on which all functions expecting a reply wait is triggered. Upon waking up, each of these functions checks the list of received
replies for its expected reply. The list is checked by comparing a unique identifier tagging the queries upon sending and copied into a reply message by the receiver. 
A matching reply is removed from the list for processing by its corresponding function. 

All four request messages handled by the request loop are for explicit connections between blob objects and their associated message objects, 
to establish, terminate, lock, and unlock connections. For each of these four actions, a message is sent to the remote blob object concerned after the
respective action has already been performed by the sending object to initiate the same action for the reverse connection to the sender. If such a request message is
received in the request handler loop, the corresponding functions for the associated message object and the blob object's own communication helper object 
are called. Establishing and terminating of connections is handled by calling the object's \texttt{Con\-nect} or \texttt{Dis\-con\-nect} function respectively. 
This  function
in turn calls both the helper's and the message object's corresponding functions. 
To distinguish this from the case of explicitly called functions to initiate or terminate a connection, the remote blob object's address is passed
directly to the object's function. 
For connection locking or unlocking the class's internal \texttt{Lock\-Con\-nec\-tion} or \texttt{Un\-lock\-Con\-nec\-tion} function is called, which
just calls the message object's locking function, if that object supports the locking feature. As the SCI blob class itself does not support 
connection locking, no further action has to be performed in this case.

\clearpage

%%%%%%%%%%%%%%%%%%%%%%%%%%%%%%%%%%%%%%%%%%%%%%%%%%%%%%%%%%%%%%%%%%%%%%%%%%%%%%%%%%%%%%%%%%%%%%%%%%%%%%%%%%%%%%%%%%%%%%%%%%%%%
%%%%%%%%%%%%%%%%%%%%%%%%%%%%%%%%%%%%%%%%%%%%%%%%%%%%%%%%%%%%%%%%%%%%%%%%%%%%%%%%%%%%%%%%%%%%%%%%%%%%%%%%%%%%%%%%%%%%%%%%%%%%%

\chapter{\label{Chap:PubSubInterface}The Publisher-Subscriber Framework Interface}

%{\large \bf event definition?}

\section{\label{Sec:PubSubInterfacePrinciple}The Publisher-Subscriber Interface Principle}

For the exchange of data between the different components in the framework an interface has been
developed conforming to the requirements specified in section~\ref{Sec:PubSubInterfaceRequirements}. 
This interface is made up of several classes
%,shown in Fig.~\ref{Fig:PubSubClasses}, 
which together provide the communication between components
following the publisher-subscriber paradigm. This section details the actual software architecture and implementation
according to the requirements outlined in section~\ref{Sec:PubSubInterfaceRequirements}.

\begin{figure}[h]
\begin{center}
\resizebox*{0.95\columnwidth}{!}{
\includegraphics{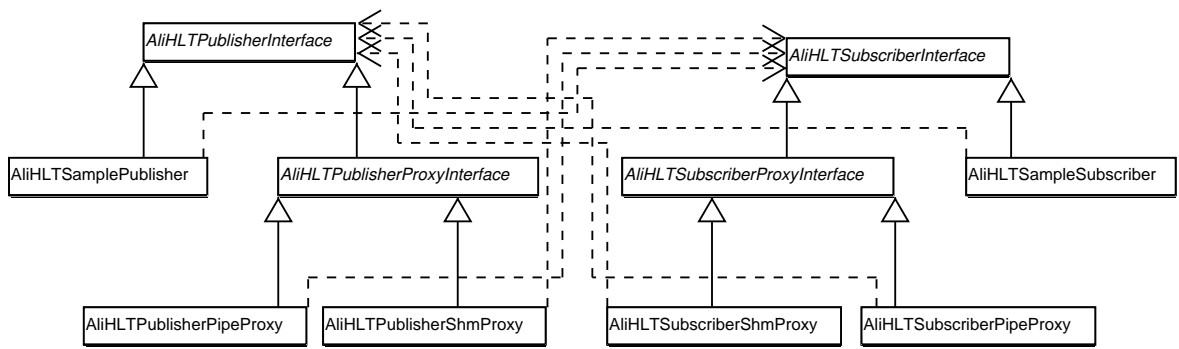}
}
\parbox{0.90\columnwidth}{
\caption{\label{Fig:PubSubClasses}Hierarchy of the classes making up the publisher-subscriber interface.}
}
\end{center}
\end{figure}

Two separate class trees, displayed in Fig.~\ref{Fig:PubSubClasses}, make up the class hierarchy for the publisher and subscriber part of the framework respectively.
At the root of each tree is an abstract base class that defines the calling interface of the corresponding part of the
component interface. 
Each class in one of the two trees addresses its counterparts in the other tree through this defined interface.
%knows about this calling interface defined for the {\bf counterpart/partner} class tree.

The model behind the component interface is a data producing component containing a publisher object through which it makes
its data available. A data consumer component  contains a subscriber object which receives published data from a producer
and performs the necessary processing on it. 
Communication between these two object types should be encapsulated so that they can communicate when they are situated in different 
processes and address spaces as well as when they are present in the same process, directly calling each others functions. For this reason the 
implementations of the two classes use the interface of their respective opposite tree and do not contain any built in communication primitives. 
%To achieve a better encapsulation of the communication between these two object types, situated in different 
%processes and address spaces, and to allow their use in the same process, directly calling each others functions, the 
%implementations of the two classes use the interface of their respective opposite tree and do not contain any built in communication primitives. 
%Publisher and subscriber components can thus be either connected directly or via 
Communication between objects in different processes is handled by proxy objects that implement the corresponding opposite interface
and do not process the calls but only forward them to their own counterpart in the remote process who calls the target object's
corresponding function. These communication classes are called publisher and subscriber proxy classes respectively. The advantage of this
approach is that publisher and subscriber object can be either situated in separate processes, calling functions of proxy
objects for communication with their counterpart, or in the same process, directly calling each other's functions. 

\begin{figure}[h]
\begin{center}
\resizebox*{0.65\columnwidth}{!}{
\includegraphics{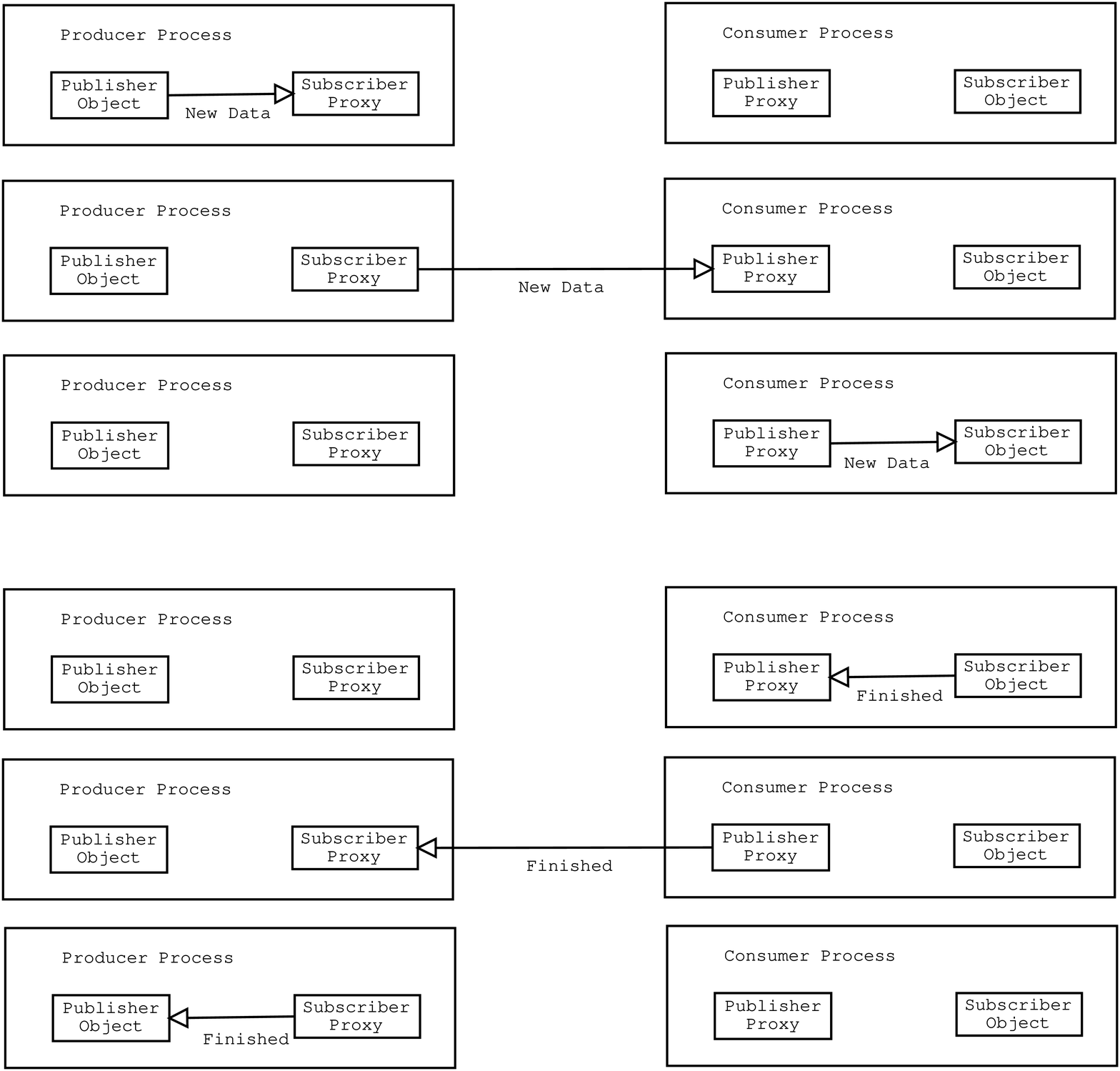}
}
\parbox{0.90\columnwidth}{
\caption{\label{Fig:PubSubComPrinciple}Publisher-subscriber principle of communication.}
}
\end{center}
\end{figure}

\begin{figure}[h]
\begin{center}
\resizebox*{0.75\columnwidth}{!}{
\includegraphics{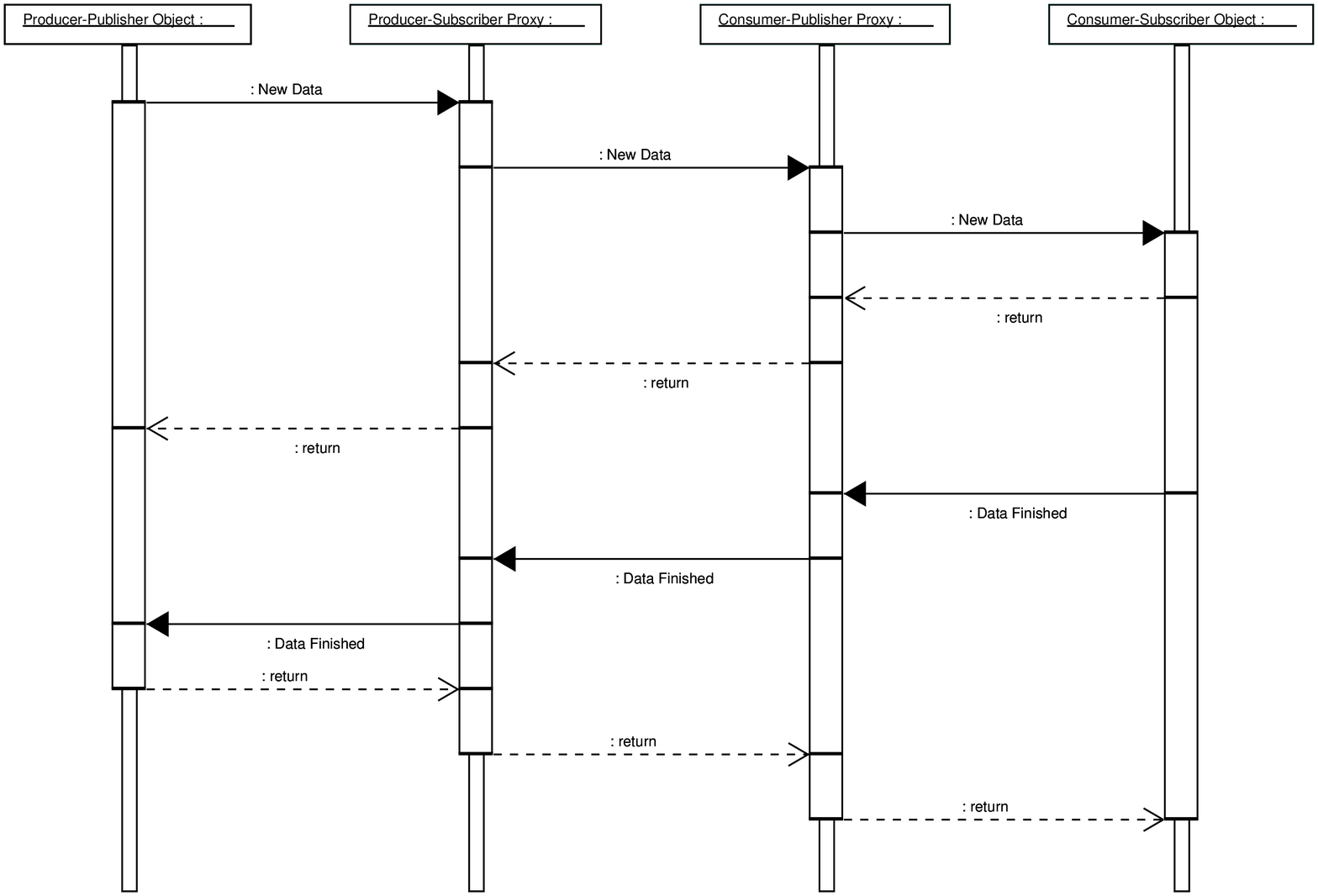}
}
\parbox{0.90\columnwidth}{
\caption{\label{Fig:PubSubComPrinciple-UML-Sequence}Publisher-subscriber principle of communication UML sequence diagram.}
}
\end{center}
\end{figure}

An example of this principle is sketched in Fig.~\ref{Fig:PubSubComPrinciple} which shows a producer process containing a publisher object and a subscriber proxy
as well as a consumer process with a publisher proxy and a subscriber object. When new data is available the publisher calls the subscriber proxy's 
new data function which collects the specified information and sends it to the publisher proxy in the consumer process.
This proxy retrieves the received information and calls the subscriber object's new data function with this data. As soon as the
subscriber object in the consumer process has finished processing the data, it calls the publisher proxy's
{\em data finished} function. The publisher proxy again uses the specified data and sends it to its subscriber proxy counterpart in the producer
process. Using this information the subscriber proxy calls the {\em data finished} function of the 
publisher object that can release the produced data and continue. Fig.~\ref{Fig:PubSubComPrinciple-UML-Sequence} shows the same process as a UML sequence diagram.

The primary purpose of the interface is to allow a data producer to announce newly available data, for example new ALICE events, to multiple subscribers. 
Placing the data in shared memory in the producer process implies that the management of the data's buffers has to be handled by the
producer process which needs to know where to place new data. In order for the producer to know when some specific data can be released again,
it has to be informed by each of its subscribers when they have finished working on each received data block. 

For efficiency reasons it should also be possible for consumers to collect finished events and inform their producer about multiple finished events in one call.
During such an aggregation process by one or multiple consumers their producer process might run out of buffer space for new event data before 
the consumers have received and finished enough events 
%so that they will 
to
inform the producer about their released events. 
In such a case a producer has to be able to send a high watermark to its consumers when it threatens to run out of buffer space.
For a non-blocking or transient type of consumer, a producer must also be able to forcibly cancel a consumer's access to
an event's data buffers. This is necessary to free a buffer when the consumer uses and thus locks the buffer too long. 

In some cases it might be desirable for a consumer to send some data about the processing of an event back to the producer along with the finished event information. 
An example for this in the ALICE HLT are the HLT trigger decisions for an event that have to be communicated back along the path that the event data
has taken. To support this ability, the calls informing about finished events allow to attach arbitrary information to each event. This information is treated
as opaque to the interface itself and is just transported from subscriber to publisher. For some consumers it might also be of interest to receive this
kind of event finished
information produced by other consumers attached to the same producer. Support for this is provided by allowing the subscriber to set a
flag in their publisher to indicate interest in this information. Whenever new event finished information becomes available afterwards it will be 
forwarded to this consumer.

For consumers that want to reduce the amount of data or events they receive from their producer two approaches are possible that can also be used in combination. 
The simpler possibility is to set a modulo restriction based on the event sequence number so that only events with a sequence number evenly 
divisible by that specifier 
will be published to the consumer. The more complex but also more flexible approach allows to use tags called trigger words associated 
with each event. A consumer has the possibility to specify a set of trigger words, and only events that have a correct trigger word set will be announced
to that consumer. 

One final necessity for the interface is that each side has to be able to query whether its partner process is still alive. This is accomplished by using {\em ping}
calls which must be answered within a predefined time by an {\em acknowledge} call. If this answer is missing, the partner can be presumed 
to be locked up or have terminated. 

In the rest of this chapter first a number of basic data types used in the interface's implementation are presented. In the following the definitions of the
central interface classes as well as the proxy classes, used for communication between publisher and subscriber processes, are described.
The final two parts respectively describe implementations of publisher and subscriber classes providing actual functionality. 

%\clearpage

\section{Auxiliary Data Types}

In the following section a number of datatypes will be described that have been defined for use in the publisher-subscriber interface. These types 
are used in the definitions of both publisher and subscriber class interfaces. The first subsection~\ref{Sec:PubSubFlatTypes} lists simple
datatypes with no or very little inner structure while the following subsection~\ref{Sec:PubSubStructureTypes} contains the descriptions for more
complex structured data types. 

\subsection{\label{Sec:PubSubFlatTypes}Flat Types}
\subsubsection{Integer Types}

To ensure type compatibility in a system that can use multiple node architectures a number of unsigned integer types have been defined
that always have the same size and thus value range on different systems. The definition is made from the basic datatypes defined for the 
C/C++ language using \texttt{\#ifdef} preprocessor directives. Four different tpes are defined with 1, 2, 4, and 8 bytes (or 8, 16, 32, and 64 bits), 
named \texttt{Ali\-UInt\-8\_t}, \texttt{Ali\-UInt\-16\_t}, \texttt{Ali\-UInt\-32\_t}, and \texttt{Ali\-UInt\-64\_t} respectively. 
For the 64~bit type an exception has to be made
as a type of that size is in general not supported by compilers on 32~bit platforms and so a GNU Compiler extension, the \texttt{unsigned long long} 
datatype, had to be used.

\subsubsection{\label{Sec:AliEventID_t}The Event ID}

For identification purposes each event has to be tagged with a unique ID. The \texttt{Ali\-Event\-ID\_t} datatype used for this task is simply defined to be 
identical to the previously declared 64~bit unsigned integer type \texttt{Ali\-UInt\-64\_t}. Depending on the application equal sized structures
can be overlaid over this flat datatype. The most simple possible use would be to just encode a unique sequence number into this type. A more
complex use could store a timestamp, as for example returned by \texttt{get\-time\-of\-day}, using the higher 32~bits for the second portion and
the lower 32~bits for the microseconds. 

\subsubsection{\label{Sec:AliHLTNodeID_t}The Node ID}

Unique identification of the nodes in a cluster running a data flow chain is provided by the \texttt{Ali\-HLT\-Node\-ID\_t} datatype. As it is assumed that
every node in such a cluster will be equipped with at least one TCP/IP network interface, the 32~bit IP address of that interface is
used as the node's ID. To store this address ID the \texttt{Ali\-HLT\-Node\-ID\_t} type has been defined to be the 32~bit unsigned integer \texttt{Ali\-UInt\-32\_t}
type. In the case of multiple interfaces and IP addresses for one node the first address returned by the 
\texttt{get\-host\-by\-name} system call will be used.

\subsection{\label{Sec:PubSubStructureTypes}Structure Types}

\subsubsection{\label{Sec:AliHLTEventDataType}The Data-Type Specifier}

To allow the specification of the type of data stored in a memory block, the \texttt{Ali\-HLT\-Event\-Data\-Type} type was introduced. 
It is basically again the 64~bit large unsigned integer
\texttt{Ali\-UInt\-64\_t} overlayed with an additional structure as an 8-character-array. This array overlay allows to print the datatype, for example for 
debugging purposes. The prerequisite to this is that the data type is specified in a format such that each of the 8 characters is actually printable. 
For the application in the ALICE High Level Trigger ``\texttt{ADC\-COUNT}'', ``\texttt{CLUS\-TERS}'', or ``\texttt{TRAC\-SEGS}'' are among the possible values. 

\subsubsection{\label{Sec:AliHLTEventDataOrigin}The Data-Origin Specifier}

In analogy to \texttt{Ali\-HLT\-Event\-Data\-Type}, the \texttt{Ali\-HLT\-Event\-Data\-Origin} type allows to specify the origin of the data stored in a memory
block. It uses the same principle of overlaying an array of 4 characters over a 32~bit \texttt{Ali\-UInt\-32\_t} ID. For the ALICE HLT this can contain
the detector where the data originated from. Possible values are ``\texttt{TPC~}`` or ``\texttt{DIMU}'' for the TPC and DiMuon arm respectively. 

\subsubsection{\label{Sec:AliHLTShmID}The Shared Memory Identifier Structure}

This structure, named \texttt{Ali\-HLT\-Shm\-ID}, holds the information required to access the shared memory areas where event data
published by a data producer process is stored. 
It contains two fields, the first of which is an \texttt{Ali\-UInt\-32\_t} member that defines the type of the shared memory segment.
Possible values currently define an invalid segment, a big physical area shared memory segment (bigphys), or a shared physical memory segment
(physmem). In the second field the actual ID of the shared memory segment is contained in an overlayed \texttt{Ali\-UInt\-64\_t}/8-character-array organization
as described above.

\subsubsection{\label{Sec:AliHLTBlockHeader}The Basic Structure Header}

Every complex data structure that will be passed between processes will contain at its beginning an instance of this \texttt{Ali\-HLT\-Block\-Header} header structure. 
To allow for an opaque
% a transparent
 transport of such structures, the header contains as its first element a 32~bit unsigned integer holding the length of 
the whole structure in bytes. The next two fields allow to specify the type of the structure as a type and subtype combination. Both
 use the unsigned integer/character array overlay principle of the two preceeding types, with a 32~bit/4 byte size for the 
type field and 24~bit/3 byte size for the subtype. The last field in the header structure is an 8~bit unsigned integer carrying a version number for each
structure type which allows to 
%transparently
 add elements to a structure. 

\subsubsection{\label{Sec:AliHLTSubEventDataBlockDescriptor}The Sub-Event Data Block Descriptor}

Information describing a block of data stored in shared memory is contained in the \texttt{Ali\-HLT\-Sub\-Event\-Data\-Block\-De\-scrip\-tor} structure type.
Since a block descriptor will not be exchanged between processes by itself but only as part of a \texttt{Ali\-HLT\-Sub\-Event\-Data\-De\-scrip\-tor}
structure described below, it does not contain an instance of the header structure discussed above. The first element of the structure is the ID of the shared memory 
segment holding the described data in the form of an \texttt{Ali\-HLT\-Shm\-ID} field. Following this there are two 32~bit unsigned integer fields that contain the starting offset
of the block relativ to the beginning of the shared memory segment and its size in bytes. 

Behind these fields required to access the data, five more fields are defined which describe the data in the shared 
memory itself. The first of these is the ID of the node that produced the data, followed by two fields with the type of the data and its origin. Each of the three
fields is of the corresponding type described above. Finally, two 32~bit unsigned integers are available, the first of 
which can contain a specification of the data under the control of each application while the second contains the byte order that the data has been stored in. 
Fig.~\ref{Fig:AliHLTSubEventDataBlockDescriptor} shows the definition of this type. 

\begin{figure}[hbt]
\begin{center}
\begin{verbatim}


struct AliHLTSubEventDataBlockDescriptor
    {
        AliHLTShmID             fShmID;
        AliUInt32_t             fBlockSize;
        AliUInt32_t             fBlockOffset;
        AliHLTNodeID_t          fProducerNode;
        AliHLTEventDataType     fDataType;
        AliHLTEventDataOrigin   fDataOrigin;
        AliUInt32_t             fDataSpecification;
        AliUInt32_t             fByteOrder;
    };
\end{verbatim}
\parbox{0.90\columnwidth}{
\caption{\label{Fig:AliHLTSubEventDataBlockDescriptor}Definition of the \texttt{Ali\-HLT\-Sub\-Event\-Data\-Block\-De\-scrip\-tor} datatype.}
}
\end{center}
\end{figure}

If the datatype of a block has the value ``\texttt{COM\-POS\-IT}'', then the data block described contains another \texttt{Ali\-HLT\-Sub\-Event\-Data\-De\-scrip\-tor} structure described
below, allowing hierarchical event descriptions.

\subsubsection{\label{Sec:AliHLTSubEventDataDescriptor}The Sub-Event Data Descriptor}

To describe the information for a whole subevent the \texttt{Ali\-HLT\-Sub\-Event\-Data\-De\-scrip\-tor} can be used. This structure's first element is a header of the 
\texttt{Ali\-HLT\-Block\-Header} type followed by an \texttt{Ali\-Event\-ID\_t} field containing the ID of the event concerned. The next two elements are 32~bit
unsigned integers holding event time information, the seconds and microseconds of the timestamp of the event's creation. After these elements another 32~bit
unsigned integer is placed that contains a timestamp specifying the maximum allowed event age in the system. This timestap is specified 
in seconds as returned by the Unix \texttt{time} function. Using it information can be broadcast
through a system to purge events from the system that have not been freed properly, preventing the slowing down or filling up of the system.

The sixth field of the structure contains the datatype of the whole event. If all the datablocks of the event are of the same type, then this field can contain
this datatype specifier. Otherwise this field contains the specifier ``\texttt{COM\-POS\-IT}'' to indicate a composite event. 
Behind this field there is another \texttt{Ali\-UInt\-32\_t}
element holding the number of data blocks contained in this descriptor followed by the corresponding number of \texttt{Ali\-HLT\-Sub\-Event\-Data\-Block\-De\-scrip\-tor} 
structures containing
the information for each of the data blocks. 
Fig.~\ref{Fig:AliHLTSubEventDataDescriptor} shows this type's definition.

\begin{figure}[hbt]
\begin{center}
\begin{verbatim}


struct AliHLTSubEventDataDescriptor
    {
        AliHLTBlockHeader                   fHeader;
        AliEventID_t                        fEventID;
        AliUInt32_t                         fEventBirth_s;
        AliUInt32_t                         fEventBirth_us;
        AliUInt32_t                         fOldestEventBirth_s;
        AliHLTEventDataType                 fDataType;
        AliUInt32_t                         fDataBlockCount;
        AliHLTSubEventDataBlockDescriptor   fDataBlocks[0];
    };
\end{verbatim}
\parbox{0.90\columnwidth}{
\caption{\label{Fig:AliHLTSubEventDataDescriptor}Definition of the \texttt{Ali\-HLT\-Sub\-Event\-Data\-De\-scrip\-tor} datatype.}
}
\end{center}
\end{figure}

Hierarchical event descriptions are supported by allowing data blocks to contain locations of further \texttt{Ali\-HLT\-Sub\-Event\-Data\-De\-scrip\-tor} structures,
placed in shared memory as described in the previous section. 

\subsubsection{\label{Sec:AliHLTEventTriggerStruct}The Event Trigger Type}

In the \texttt{Ali\-HLT\-Event\-Trig\-ger\-Struct} data is contained that characterizes a particular event and allows a subscriber to select only a particular subset of
events for processing. For applications in high-energy or heavy-ion physics this could be trigger information received from preceeding trigger
levels specifying a type of event. Since the organization of this type of data cannot be known in advance by the framework,
this type has no complex inner structure. It contains only the header field as well as an \texttt{Ali\-UInt\-32\_t} holding the number of 32~bit data words 
in the structure and another \texttt{Ali\-UInt\-32\_t} marking the beginning of the data word array. 
For determining matches between structures a comparison
function type is defined. A comparison function which performs a bytewise comparison of two structures is supplied in the framework.

\subsubsection{\label{Sec:AliHLTEventDoneData}The Event Done Data Type}

The structures of the \texttt{Ali\-HLT\-Event\-Done\-Data} type contain information transferred back from a subscriber to a publisher 
about events whose processing has been finished.
This additional information in these structures is opaque to the framework, which only transports
the data for interpretation by the higher layers of the system's components. Therefore, application specific data is implemented similarly to the event trigger
type described previously, as one 32~bit unsigned integer holding the number of data words in the structure followed by the array of 32~bit unsigned data words itself.
Preceeding these fields are the usual header structure and a field with the ID of the event concerned. In the following text the expression ``non-trivial event done data''
will be used to describe an event done data structure that contains at least one 32~bit data word. 

%\clearpage

\section{The Interface Definition Classes}
\subsection{\label{Sec:PublisherInterface}The Publisher Interface}

\begin{figure}[h]
\begin{center}
\resizebox*{0.75\columnwidth}{!}{
\includegraphics{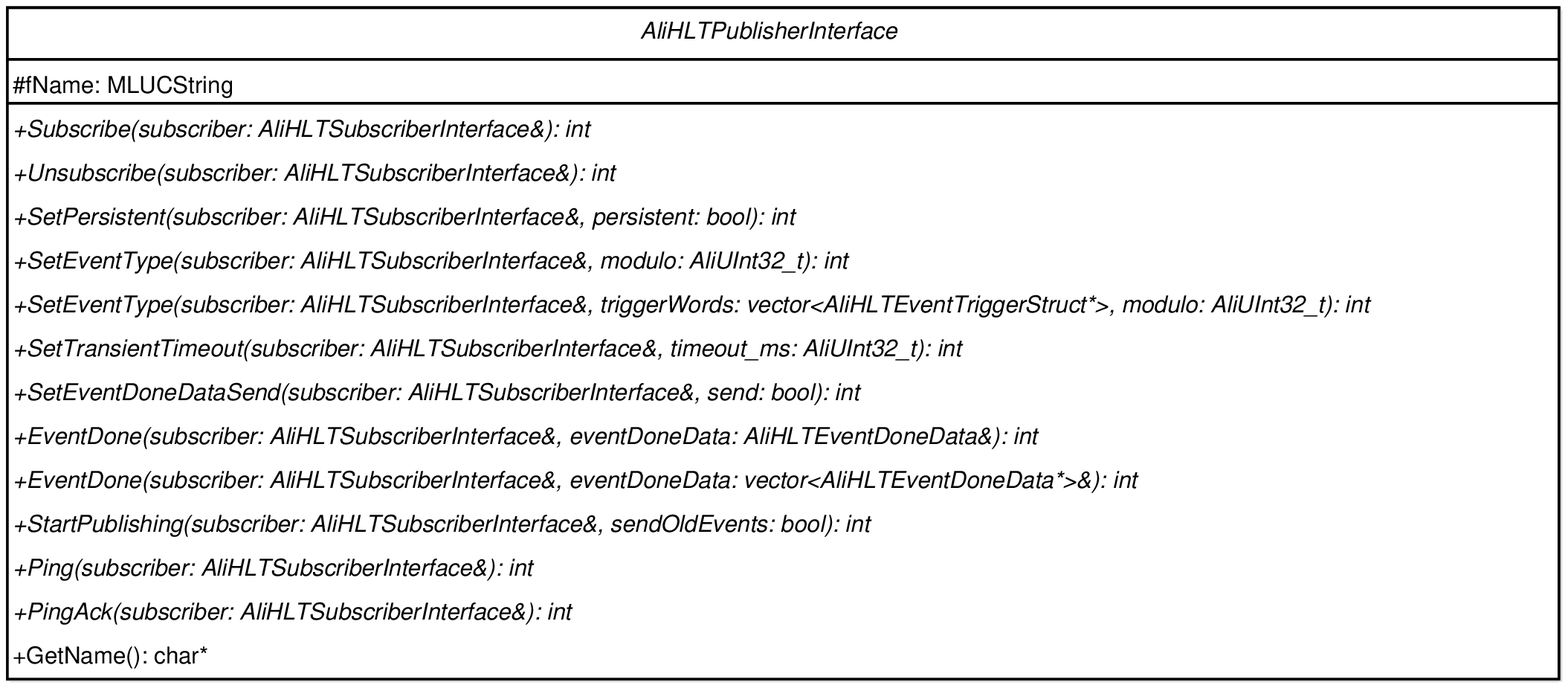}
}
\parbox{0.90\columnwidth}{
\caption{\label{Fig:AliHLTPublisherInterface}UML class diagram of \texttt{Ali\-HLT\-Pub\-lish\-er\-Inter\-face}.}
}
\end{center}
\end{figure}

Fig.~\ref{Fig:AliHLTPublisherInterface} shows the interface for publisher classes, \texttt{Ali\-HLT\-Pub\-lish\-er\-Inter\-face}, as a UML class diagram. 
This interface defines abstract methods for each of the tasks described in section~\ref{Sec:PubSubInterfacePrinciple}. 
In the \texttt{Ali\-HLT\-Pub\-lish\-er\-Inter\-face} class one data member and one non-abstract member function are contained. The data member holds
the name under 
which the publisher referred to by an object is known. It is returned by the \texttt{Get\-Name} member function. 

%{\bf \large Changed StartPublishing Call}

The first two of the methods defining the publisher interface, \texttt{Sub\-scribe} and \texttt{Un\-sub\-scribe},  handle subscribing and unsubscribing 
to a publisher. Both methods accept as only parameter a reference to the subscriber to be subscribed to.
% to the corresponding publisher object. 
Once a subscriber object is subscribed to a publisher, its type can be set with the \texttt{Set\-Per\-sis\-tent} method. It requires 
a reference to the subscriber concerned as 
its first parameter together with a boolean flag specifying whether the subscriber should be treated as persistent or transient. 
Initially after subscription,
all subscribers are marked as persistent requiring the method to be called for transient subscribers only. 

Using the two following \texttt{Set\-Event\-Type} functions, it is possible to scale down the number of events received by a subscriber from
its publisher. In the first of the functions the event sequence modulo specifier is set that scales down the event rate to the given ratio. Only events whose
sequence number is evenly divisible by the specified modulo number will be announced to this subscriber. The second variant of the functions accepts a vector of trigger word
structures, described in section~\ref{Sec:AliHLTEventTriggerStruct}, together with a modulo specifier identical to the one passed in the simpler function's version. 
Specified trigger word structures are stored associated with the subscriber, and each new event is checked for a match with one of them. 

A method of interest to transient subscribers only is \texttt{Set\-Tran\-si\-ent\-Time\-out}. 
It allows to set the minimum timeout before an event used
by a transient subscriber can be cancelled. A publisher will cancel a transient subscriber's event only when all
persistent subscriber have released it and at least the amount of time specified in this call has passed. 
To prevent transient subscribers from setting arbitrarily large intervals it is suggested that publisher implementations have a separate
allowed maximum for this timeout. Transient subscribers should thus be prepared for shorter timeouts than specified.
The last subscriber configuration method is the \texttt{Set\-Event\-Done\-Data\-Send} method, that also uses a boolean flag as its parameter. This
flag specifies whether the subscriber concerned is interested in receiving the data sent along with finished events from other subscribers. Since this information
might be needed by the subscriber to properly process the event, this information is 
forwarded immediately after another subscriber has released a specific event in the publisher. 

To inform a publisher that processing of an event has finished, a subscriber calls one of the two \texttt{Event\-Done} methods in the publisher interface that
differ only with respect to the arguments taken. In the first call one \texttt{Ali\-HLT\-Event\-Done\-Data} structure argument, described in 
\ref{Sec:AliHLTEventDoneData}, is accepted to release only one event while the second version allows to release
multiple events in one call by accepting a vector of these structures. As described previously, each of the structure arguments contains the identifier of the 
event to be released.
% and additionally a number of 32~bit data words. These data words are not interpreted or otherwise handled by the publisher but are left for processing 
%by higher levels of a component. 

By calling the \texttt{Start\-Pub\-lishing} method, a subscriber activates publishing of new events for itself. After calling this method a subscriber will receive all 
new events as they become available in the publisher. This call will not return immediately but instead will enter a loop until \texttt{Un\-sub\-scribe}
 has been called. 
The final two methods are used by a subscriber to test a publisher's availability (\texttt{Ping}) or to reply to a publisher to acknowledge the subscriber's own availability
(\texttt{Ping\-Ack}). An optional boolean flag, which can be given as the second parameter, allows to specify that
all events currently contained in the publisher, already announced to other subscribers, should be announced to the
subscriber concerned. If this flag is not set, only events which arrive in the publisher after the  \texttt{Start\-Pub\-lishing} call will be announced.

\subsection{\label{Sec:SubscriberInterface}The Subscriber Interface}

\begin{figure}[h]
\begin{center}
\resizebox*{0.75\columnwidth}{!}{
\includegraphics{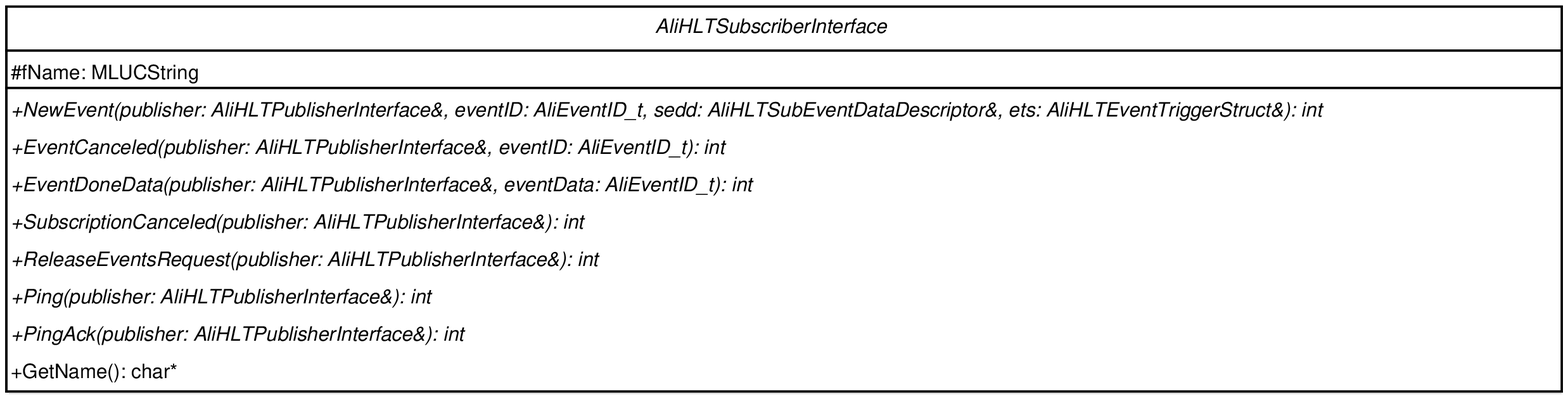}
}
\parbox{0.90\columnwidth}{
\caption{\label{Fig:AliHLTSubscriberInterface}UML class diagram of \texttt{Ali\-HLT\-Sub\-scrib\-er\-Inter\-face}.}
}
\end{center}
\end{figure}

Fig.~\ref{Fig:AliHLTSubscriberInterface} shows the interface for subscriber classes \texttt{Ali\-HLT\-Sub\-scrib\-er\-Inter\-face} as a UML class diagram. 
This interface defines abstract methods for each of the subscriber tasks described in section~\ref{Sec:PubSubInterfacePrinciple}. 
Similar to the publisher interface class \texttt{Ali\-HLT\-Sub\-scrib\-er\-Inter\-face} also contains the subscriber's name as the only data member and 
the non-abstract member function
\texttt{Get\-Name} to return that name. In analogy to the publisher interface all defined abstract subscriber interface functions require 
a reference to the publisher object calling the corresponding subscriber function as their first argument. 

\texttt{New\-Event}, the first of the abstract member functions defined for the subscriber interface, will be called when the publisher to which
the respective subscriber is attached announces new data. It is therefore the most important subscriber member function. In addition to the
calling publisher's reference it accepts three more arguments. The first of these is an \texttt{Ali\-Event\-ID\_t} containing the ID of the event
being announced. Following is an \texttt{Ali\-HLT\-Sub\-Event\-Data\-De\-scrip\-tor} holding the information about the actual data blocks contained
in the event, mostly located in shared memory. The final argument is an event trigger structure of the \texttt{Ali\-HLT\-Event\-Trig\-ger\-Struct}  type,
containing information that more closely characterizes the event concerned. Event trigger data is passed to the subscriber so that it can determine the processing of the event 
based on this trigger information. 

The second abstract subscriber interface function is \texttt{Event\-Can\-celed}. It is called for transient subscribers whenever the publisher to which they are attached
cancels an event
before the subscriber itself has finished working on it. After this function has been called a subscriber should not rely on any data located in shared memory to 
still be valid. Processing on this event should be stopped as soon as possible after this function has been called. Parameters to this function
are the calling publisher's reference and the ID of the event being cancelled. 

Next is another notification function, that can be called while any subscriber has not yet finished processing an event. 
\texttt{Event\-Done\-Data} is called when the publisher receives event done information from another subscriber and  this subscriber has 
requested to receive this kind of data using the publisher's \texttt{Set\-Event\-Done\-Data\-Send} function. The function will be called with the 
exact \texttt{Ali\-HLT\-Event\-Done\-Data} structure that the publisher received from the other subscriber, provided that the structure is non-trivial, 
containing at least one data word. 

When the publisher cancels a subscription the \texttt{Sub\-scrip\-tion\-Can\-celed} function of the subscriber concerned is called. Such a cancellation might happen
because the subscriber has called the publisher's \texttt{Un\-sub\-scribe} function or because the producer process is ending. 
Each subscriber's \texttt{Re\-lease\-Events\-Re\-quest} function is called when the publisher threatens to run out of 
buffer space to signal that the subscribers should release events as soon as possible. 
The final abstract member functions defined for the subscriber interface are the \texttt{Ping} and \texttt{Ping\-Ack} functions that have the same 
meaning as the respective functions in the publisher interface described in section~\ref{Sec:PublisherInterface}. 

%\clearpage

\section{The Proxy Classes}

\begin{figure}[h]
\begin{center}
\resizebox*{0.75\columnwidth}{!}{
\includegraphics{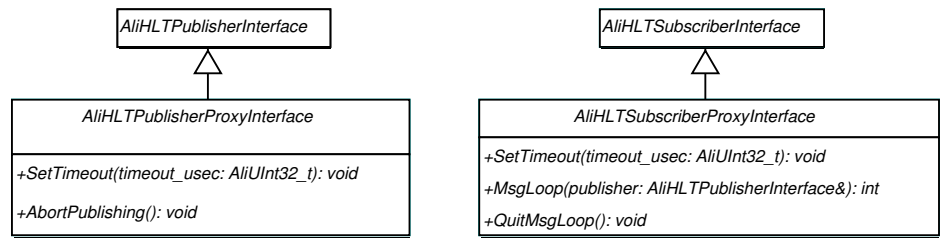}
}
\parbox{0.90\columnwidth}{
\caption{\label{Fig:ProxyInterfaces}UML class diagram of the two proxy interface classes.}
}
\end{center}
\end{figure}

Beyond the functions defined already in the publisher and subscriber interface definition classes, two derived classes exist defining additional interface
functions for the two types of proxy classes. These two interface classes are shown in Fig.~\ref{Fig:ProxyInterfaces} with their respective, additionally defined 
functions. 

Both classes define a \texttt{Set\-Time\-out} function allowing to specify a communication timeout used by 
proxy implementations in a program. Since only the proxy classes are intended for communication between processes, but not publisher or
subscriber classes in general, it is reasonable
%makes sense
 to place this function in these classes. The \texttt{Ali\-HLT\-Pub\-lish\-er\-Proxy\-Inter\-face} class 
defines one additional abstract function, \texttt{Abort\-Pub\-lish\-ing}. Its purpose is to terminate the publishing loop 
started by the proxy classes when their \texttt{Start\-Pub\-lishing} function is called as described below. Usually this function ends when a 
\texttt{Sub\-scrip\-tion\-Can\-celed} message is received from the publisher. If, however, the publisher process has died or the connection
between publisher and subscriber processes is interrupted or broken, the publisher proxy cannot determine when to leave the message loop. 

Two additional functions are defined in the \texttt{Ali\-HLT\-Sub\-scrib\-er\-Proxy\-Inter\-face} class, \texttt{Msg\-Loop} and \texttt{Quit\-Msg\-Loop}. The \texttt{Msg\-Loop} 
function is intended to contain the implementation of the proxy's loop for receiving and handling messages from the opposite publisher proxy in derived classes. 
It should be called in a separate thread from the publisher object to which the proxy is attached. When the loop has to be terminated because the subscription
has been cancelled, the publisher calls the third defined function, \texttt{Quit\-Msg\-Loop}, whose purpose is to end the message loop. 

In the following section the proxy classes present in the current framework are described. They can be divided into two types categorized by the
type of communication between publisher and subscriber proxies: pipe proxies and shared memory proxies. Communication for the two pipe proxy classes is done
via named pipe \cite{LinuxIPCmanpage} system resources while the shared memory proxies communicate via System V shared memory \cite{SingleUnixWeb}, \cite{SingleUnixOnline}, \cite{SingleUnix}, 
\cite{LinuxIPCmanpage}. In addition to these four proxy classes the final part
of this section covers the subscription loop, the mechanism for subscribers to register with publisher objects in other processes. 

\subsection{The Pipe Proxy Classes}

\subsubsection{\label{Sec:AliHLTPipeCom}The Pipe Communication Class}

\begin{figure}[h]
\begin{center}
\resizebox*{0.4\columnwidth}{!}{
\includegraphics{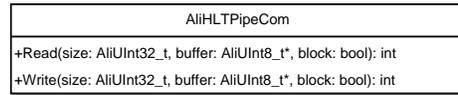}
}
\parbox{0.90\columnwidth}{
\caption{\label{Fig:AliHLTPipeCom-BlockInterface}The memory block read and write functions of the pipe communication class.}
}
\end{center}
\end{figure}

Named pipes used for communication between the pipe proxy classes are encapsulated by the \texttt{Ali\-HLT\-Pipe\-Com} class. This class supports
two pipes simultaneously, one for reading and one for writing, as the pipe communication between the proxies  is executed via
two named pipes. One of these is used for communication from publisher to subscriber and the other from subscriber to publisher. Two pipes are used 
to avoid lockup situations.
%where one side has filled the pipe completely. In such a case neither side can write anything and waits for some response from the 
%other side, which cannot be written because the pipe is blocked. 
The naming of the pipes is based on a scheme that places all pipes
in the \texttt{/tmp} directory and assigns a base name identical to the subscriber's ID to them. Each pipe has an additional suffix, either
\texttt{Publ\-To\-Subs} or \texttt{Subs\-To\-Publ} depending on the flow of communication. The full file  names for a subscriber whose ID
is {\em TestSubscriber} then are \texttt{/tmp/\-Test\-Sub\-scrib\-er\-Publ\-To\-Subs} and \texttt{/tmp/\-Test\-Sub\-scrib\-er\-Subs\-To\-Publ}. 
The pipe class offers an interface for reading and writing similar
to the system \texttt{read}/\texttt{write} function calls. This interface allows direct specification of the number of bytes to read or write together with a memory block
where the write data was stored, respectively where read data should be stored. Functions of this interface are shown in Fig.~\ref{Fig:AliHLTPipeCom-BlockInterface}. 

\begin{figure}[h]
\begin{center}
\resizebox*{0.4\columnwidth}{!}{
\includegraphics{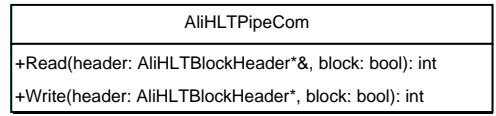}
}
\parbox{0.90\columnwidth}{
\caption{\label{Fig:AliHLTPipeCom-HeaderInterface}The header structure read and write functions of the pipe communication class.}
}
\end{center}
\end{figure}

In addition, a second interface
% type
 is supported that directly allows to read and write structures described by an element of the \texttt{Ali\-HLT\-Block\-Header}
type, detailed above in section~\ref{Sec:AliHLTBlockHeader}. This interface is shown in Fig.~\ref{Fig:AliHLTPipeCom-HeaderInterface}. For writing, this 
version of the \texttt{Write} function requires a pointer to such a block header structure as its primary argument. It will write as many 
bytes from the memory pointed to as specified in the header's length field. In the \texttt{Read} function the header
structure is read first and then the required amount of bytes is allocated as specified in the header's length field. The header already read is copied to the
beginning of this memory block and the rest of the data is read and placed directly into that memory. 
In the function's pointer reference argument the pointer to the allocated memory is then returned to the calling code that 
later has to free the allocated memory again through a call to \texttt{de\-lete []}. 

To avoid blocking, e.g in case a process attempts to write to a pipe that has been filled because the reader process has died or locked up, 
all read and write function
calls accept an additional boolean argument specifying whether the call is to be blocking or non-blocking. For non-blocking calls a member variable 
in the pipe object is used that specifies the timeout to use when a call would block. This timeout can be specified on a microsecond granularity. 

In order to reduce the  number of system \texttt{read} calls in a more efficient reading mechanism, the pipe communication class implements a caching strategy, similar
to what is implemented in the \texttt{BCL\-Int\-TCP\-Com\-Helper} class described in section~\ref{Sec:BCLIntTCPComHelper}. Each object contains a buffer of 4~kB size, the
maximum amount of data that can be read from or written to a pipe in one system call. If this buffer is empty upon a read call, the object tries to read the full
amount of 4~kB into this buffer at once. Since the pipes are opened in non-blocking mode the read attempt will read as much data as is available up to the specified amount.
Data that has been requested by the calling code will be provided from this cache until it is exhausted, saving a number of read calls.

\subsubsection{\label{Sec:AliHLTPublisherPipeProxy}The Publisher Pipe Proxy Class}

\begin{figure}[h]
\begin{center}
\resizebox*{0.2\columnwidth}{!}{
\includegraphics{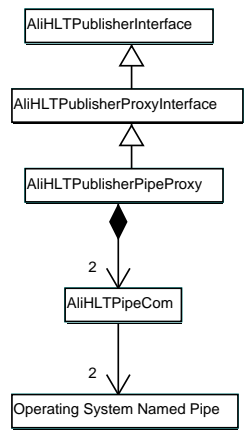}
}
\parbox{0.90\columnwidth}{
\caption[Publisher pipe proxy interface and pipe communication class UML diagram.]{\label{Fig:PublisherPipeProxyClass}UML class diagram of the publisher pipe proxy interface classes and the pipe communication class.}
}
\end{center}
\end{figure}

Handling the publisher part of the communication based on named pipes is the task of the \texttt{Ali\-HLT\-Pub\-lish\-er\-Pipe\-Proxy} class, 
implementing the abstract functions defined in the publisher interface. Two named pipes, provided by an
instance of the \texttt{Ali\-HLT\-Pipe\-Com} class described above, are used for the two communication directions. 
Additionally, the proxy uses a third named pipe, encapsulated in another
pipe object, for connection to a publisher's subscription loop, as described below in section~\ref{Sec:SubscriptionLoop}. This pipe object uses just
one of the two possible pipes as no reading will be done from it, and only the subscription request will be written. 
Fig.~\ref{Fig:PublisherPipeProxyClass} shows the UML relationship of the proxy class and the pipe communication class. 

In the \texttt{Sub\-scribe} function the publisher pipe proxy first tries to open the subscription request pipe to the producer process's subscription loop.
Then the two pipes used for the actual communication between the publisher and subscriber processes are created. It is mandatory
that these pipes are created by the publisher proxy in the subscriber process and that they are already present when the subscription request is written into
the request pipe. 
The subscription message sent to the publisher contains the subscriber's name field as its identifier, preceeded by the string ``\texttt{pipe:}'' to specify
that pipe proxies are used. After its construction this message is written into the subscription pipe and the \texttt{Sub\-scribe} function ends. 

To unsubscribe none of these steps have to be executed by the \texttt{Un\-sub\-scribe} function. Instead it starts by 
creating the unsubscription message. This message also contains the subscriber's name as the identifier. 
Again the function ends immediately after writing the message into the pipe for 
%subscriber (process with publisher proxy) to publisher (process with subscriber proxy) communication. 
subscriber to publisher communication. This communication can also be seen as communication between a publisher proxy in the subscriber process
and a subscriber proxy in the publisher process.

Both messages created in the functions described contain the block header structure as their first element. They can be
written into their respective pipes via the block header write functions. For the rest of the functions the messages follow the same principle.
Any additional parameters needed by the messages are encapsulated into data structures that contain this header structure as well and can also be 
written using the header write functions. 

For efficiency reasons, calls which accept parameters and that thus have to send multiple structures do not actually send each structure with a separate write call. 
Instead a coalescing step is taken where a block of memory is allocated with a size large enough for all data to be sent. The message, all
parameters, and any additional data are then copied into that block, and the block is passed to the pipe object for writing using a single call to the \texttt{Write} 
function. When the allocation of the buffer block fails, seperate calls to the block header \texttt{Write} function are used.

At the end of the \texttt{Start\-Pub\-lishing} function, after the message and parameters have been written, the function does not return but instead enters
the message loop function. In this loop the publisher proxy waits for messages that arrive on the publisher-to-subscriber pipe. Received messages are checked for the correct
header identification and are then handled according to their respective type. For most of the messages, reading of the necessary parameter structures and their decoding into
normal C++ parameters is done in separate functions. In these functions the parameter structures are also checked for the correct header identification.
Once the C++ parameters
 have been obtained the corresponding function in the attached subscriber object is 
called with these parameters. When the message corresponding to the \texttt{Sub\-scrip\-tion\-Can\-celed} method 
is received, the message loop is ended normally. 

To ensure the correct transport of the data through the named pipes, the publisher proxy class has the ability of performing a 32~bit 
Cyclic Redundancy Checksum (CRC) \cite{CRC}
over the data for each message including its parameters. This checksum is sent after the actual data. 
In the subscriber proxy's receiving message loop the same checksum is calculated using the received
data.  The subscriber proxy's checksum is then compared with the value calculated and sent by the publisher proxy. 
If the comparison indicates that an error occured, this is reported, and 
the received data is discarded without further action. 
The publisher proxy message loop and the subscriber proxy functions implement the identical error checking mechanism.
This capability can be activated using \texttt{\#de\-fine} statements at compile time of the classes.

\subsubsection{\label{Sec:AliHLTSubscriberPipeProxy}The Subscriber Pipe Proxy Class}

\begin{figure}[h]
\begin{center}
\resizebox*{0.2\columnwidth}{!}{
\includegraphics{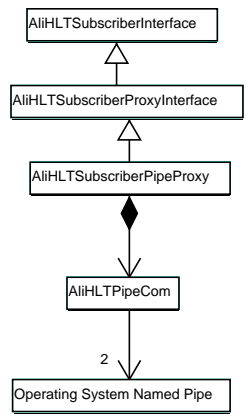}
}
\parbox{0.90\columnwidth}{
\caption[Subscriber pipe proxy interface and pipe communication class UML diagram.]{\label{Fig:SubscriberPipeProxyClass}UML class diagram of the subscriber pipe proxy interface classes and the pipe communication class.}
}
\end{center}
\end{figure}

In the producer process it is the \texttt{Ali\-HLT\-Sub\-scrib\-er\-Pipe\-Proxy}'s task to handle the subscriber part of the named pipe communication. For
this purpose it implements all abstract functions defined in the subscriber proxy interface.  This class uses only
one pipe communication object to handle the same two publisher-subscriber pipes. The meaning of the pipes with regard to reading and writing is of course
reversed with respect to the publisher proxy class. A subscriber proxy object is created in the producer process only when a subscription has taken place, and
the pipes are created at the beginning of the publisher proxy's \texttt{Sub\-scribe} function. 
Therefore the pipes will exist already upon creation of an object of this class. In the class's constructor they thus only have to be opened.
Fig.~\ref{Fig:SubscriberPipeProxyClass} shows a UML diagram of the relationship between the proxy and pipe communication classes. 

As for \texttt{Ali\-HLT\-Pub\-lish\-er\-Pipe\-Proxy}, the implemented interface functions mainly create a message object, identical to the publisher proxy's message object, 
and several parameter
objects as required to hold the necessary arguments. These data structures are then coalesced and sent utilizing the same mechanisms as used in \texttt{Ali\-HLT\-Pub\-lish\-er\-Pipe\-Proxy}. 
Only if the coalesced sending call fails, the structures are sent with separate \texttt{Write} calls for each of them. The subscriber proxy's functions and its
message loop implement the same optional CRC error checks as the publisher proxy, as has been described in the previous section.

Unlike the publisher proxy class, no call to one of the interface functions causes a subscriber proxy object to enter the message loop function defined for
subscriber proxy objects. Instead the message loop has to be called explicitly by a publisher object after a proxy object has been subscribed to it, for concurrency
preferably in a separate thread. In analogy to the publisher proxy class the message loop calls functions to handle the more 
complex messages with multiple arguments for parameter extraction. Using the extracted parameters, the proxy calls the appropriate interface functions 
in the publisher object to which it is attached. A further difference to the publisher proxy class is that the message loop does not automatically end when
a specific message is sent or received. As is the case for starting the loop, it has to be terminated explicitly by a call to the \texttt{Quit\-Msg\-Loop} function from the 
parent publisher object.

\subsection{The Shared Memory Proxy Classes}

\subsubsection{The Shared Memory Communication Classes}

In analogy to the pipe communication class for the pipe proxies, the two shared memory proxy classes also rely on a common base class, \texttt{Ali\-HLT\-Shm\-Com},
to handle the interaction with the System V shared memory functions. In addition, the class also handles the buffer management for the shared
memory blocks used for communication between the proxies. 
%It differs from the pipe communication class in that each shared memory object is only able to handle
%one shared memory and thus can be used only for one direction of communication, either reading or writing. 

Since System V shared memory segments cannot be identified by names but only by integer IDs, such an ID has to be passed to the communication
object together with the segment's size. These two arguments are passed to the object's constructor where the shared memory segment
will be created or opened.  
To support the case where a communication partner connects to a segment already created by its partner, the class's constructor
accepts a flag argument. This flag allows to specify whether the object should create the segment specified or whether it should just try to connect to an
existing segment. 

%side will only initiate and undertake communication when the other side is already present and ready

One problem encountered in communication via shared memory is that it does not support suspending a process while waiting for data, 
notifying and waking it up when data has become available. Similarly, it is not possible to wait when no space is available for writing, to be 
notified after enough space for
the attempted write operation has again become available. If one were to use continuous polling of the parameters that indicate available data,
% to read or space for write operations, 
this would result in a high CPU load on the system. As a compromise the approach chosen for this class
uses a number of read or write attempts followed by \texttt{usleep} calls. The \texttt{usleep} calls use the minimum granularity available to processes on a Linux system 
of 10~ms. To
reduce the impact of this rather high sleep time, the number of poll retries executed before sleeping can be configured for each object during runtime.

\begin{figure}[h]
\begin{center}
\resizebox*{0.4\columnwidth}{!}{
\includegraphics{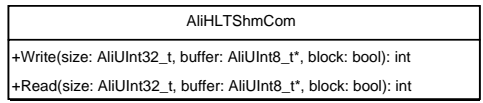}
}
\parbox{0.90\columnwidth}{
\caption{\label{Fig:AliHLTShmCom-BlockInterface}The memory block interface of the shared memory communication class.}
}
\end{center}
\end{figure}

Similar to the pipe communication class the shared memory communication class also supports multiple kinds of read and write calls, although in this case 
they do not just differ in 
ease of use but also in the degree of efficiency supported. The first function set for reading and writing 
in Fig.\ref{Fig:AliHLTShmCom-BlockInterface}  works identical to the basic functions provided
in the pipe communication class. Both functions accept a size specifier for the amount of data to read or write and a pointer to the data buffer. 
Their final parameter is a flag, specifying whether the call should block indefinitely or use a specified timeout. Using this 
interface the data is copied by the CPU from the source memory to the destination by \texttt{mem\-cpy} calls.

\begin{figure}[h]
\begin{center}
\resizebox*{0.6\columnwidth}{!}{
\includegraphics{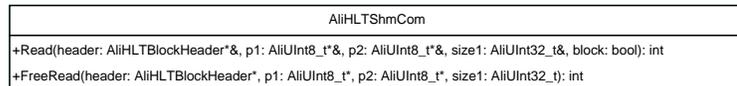}
}
\parbox{0.90\columnwidth}{
\caption{\label{Fig:AliHLTShmCom-HeaderInterface}The block header read interface of the shared memory communication class.}
}
\end{center}
\end{figure}

The second call interface, shown in Fig.~\ref{Fig:AliHLTShmCom-HeaderInterface}, only provides two functions to allow reading, while no support for data writing is present. 
These functions contain support for reading
data structures described by a block header structure at their beginning, analogous to the second interface type of the pipe communication class. 
Unlike the 
two functions from the first set, this interface does not always perform copy operations for the data. Instead the first of the two functions 
can return a direct pointer to the data
structure located in the shared memory segment, avoiding the copy steps. To prevent overwriting of data while it is still in use,
the occupied memory is not directly marked as available again. 
This is performed by the second function of this set, which can be called when the read data can be released. 
Copy steps are only necessary if the read data, described by the header structure,
is wrapped around in the buffer. In such a case the first part lies at the buffer's end and the second at its beginning. This situation is handled by 
allocating a further memory block of the required size and copying the data from the shared memory buffer into that memory by two \texttt{mem\-cpy} calls. The 
allocated memory is returned as the pointer to the block header structure. When the free function is called in this case it does not only release the memory in the buffer but also
frees the specifically allocated memory again. In the first of the two functions, \texttt{Read}, five arguments are accepted. Only the first and last of these arguments
 are relevant to the user. 
The first is a reference in which the pointer to the structure will be returned and the last is an optional flag that indicates whether or not the function should block
while waiting for data. Two pointers and a size specifier make up the remaining three arguments in which information is returned from the function
that the second function, \texttt{Free\-Read},
uses to determine whether the memory with the data has been allocated or is in the shared memory. Except for the block argument, which does not apply to the free operation,
the remaining four parameters supported by the \texttt{Read} function have to be passed in the call to the \texttt{Free\-Read} function to provide it with the required 
information to release the block or blocks.

\begin{figure}[h]
\begin{center}
\resizebox*{0.5\columnwidth}{!}{
\includegraphics{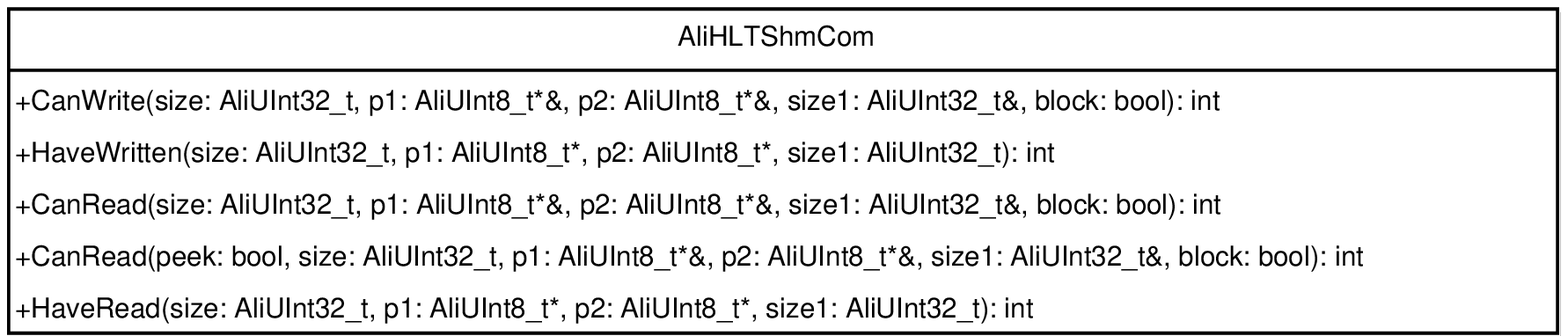}
}
\parbox{0.90\columnwidth}{
\caption{\label{Fig:AliHLTShmCom-DirectAccessInterface}The 
%efficient 
direct access interface of the shared memory communication class.}
}
\end{center}
\end{figure}

The third interface, shown in Fig.~\ref{Fig:AliHLTShmCom-DirectAccessInterface}, also allows a more efficient 
approach to communication by using a set of two functions for writing and three for reading. For writing only the 
size of the data to be written 
is passed to the first of the two functions, \texttt{Can\-Write}, together with a flag specifying blocking or non-blocking mode. Three parameters are returned by the 
function, two pointers and another size specifier. When the memory block for writing the specified amount of data is present as one block
in the shared memory buffer, then the second pointer and the returned size specifer are both zero. In the first pointer parameter
a pointer  to the target block 
in shared memory is returned. In contrast, when the block for writing wraps, then the first pointer points to the part of the block
located at the buffer's end and the second
pointer to the one at the buffer's beginning. The returned size specifier contains the size of the block's first part located at the buffer's end. Using the two
pointer arguments, the data can then be written into the shared memory buffer. To avoid copying, the data can even be directly created in the shared memory, taking
into account the two parts of the block. Once the data is present in the 
buffer, the second function \texttt{Have\-Written} can be used to commit it and make it available for reading. \texttt{Have\-Written} requires the first four of 
\texttt{Can\-Write}'s parameters, the fifth blocking parameter is not applicable. Using these parameters it determines whether the block is in one piece
or wrapped around and then accordingly sets the amount of data written in the object's internal structures. 

Of the three functions available for the read part of this interface, two \texttt{Can\-Read} functions are used to determine whether data is available for reading.
These functions only partly differ in their 
arguments, having five of them in common: a size parameter specifying how many bytes to read, two pointer arguments, a second size specifier, and a
blocking/non-blocking flag. In combination the two pointer arguments and the additional size specifier basically have the same function as in the \texttt{Can\-Write} function,
specifying the memory where the data to be read is located either as one block or as two wrapped around parts. The difference between the two functions is an
initial flag argument, available in one of the functions. The flag specifies whether the read indices should be updated and mark the data whose pointers are
returned as read, or whether the function should just peek for available data and not modify any indices. In the function without this additional flag
argument the first \texttt{Can\-Read} function is called with all its specified parameters and the peek flag set to false. This function will thus mark the data as read
so that the next \texttt{Can\-Read} call will return pointers to the next available data block. Once the data has been accessed and can be released, the third function,
\texttt{Free\-Read}, is available to release the data and make the memory blocks usable for writing new data. This function requires the two size and pointer arguments
of the \texttt{Can\-Read} functions and updates the indices of the object to mark the block concerned as free.

\subsubsection{The Publisher Shared Memory Proxy Class}

\begin{figure}[h]
\begin{center}
\resizebox*{0.5\columnwidth}{!}{
\includegraphics{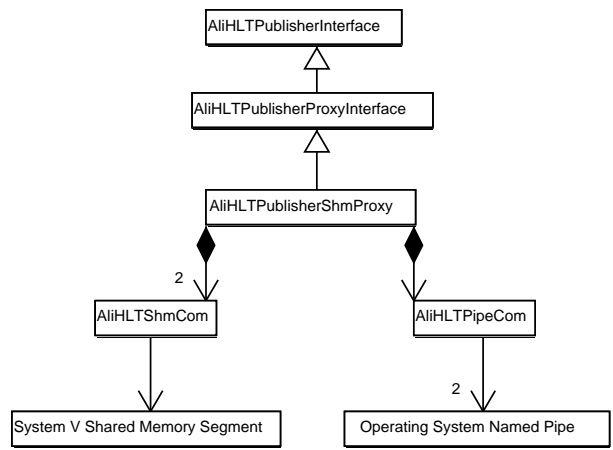}
}
\parbox{0.90\columnwidth}{
\caption[Shared memory publisher proxy and communication classes UML diagram.]{\label{Fig:PublisherShmProxyClass}The shared memory publisher proxy class and its relation to the pipe and shared memory communication class.}
}
\end{center}
\end{figure}

Similar to the \texttt{Ali\-HLT\-Pub\-lish\-er\-Pipe\-Proxy}, the task of the \texttt{Ali\-HLT\-Pub\-lish\-er\-Shm\-Proxy} class is to handle the publisher side of the shared memory communication
in the data consumer processes. It uses two shared memory communication objects for the two communication directions from publisher proxy to subscriber proxy and 
vice-versa. Each of these requires its own shared memory key, although both use the same size. The communication objects and their memory segments 
are created by the class's constructor. In addition to the two shared memory segments, one pipe 
communication object is used to execute the subscription through the publisher's subscription loop described below in section~\ref{Sec:SubscriptionLoop}. 
\texttt{Ali\-HLT\-Pub\-lish\-er\-Shm\-Proxy}'s hierarchy and its relation to the two communication classes is shown in Fig.~\ref{Fig:PublisherShmProxyClass}. 

The subscription pipe is opened in the class's \texttt{Sub\-scribe} function using the publisher's name specified in the object's constructor, in order to build the
pipe name as described for the publisher pipe proxy in section~\ref {Sec:AliHLTPublisherPipeProxy}. Since the shared memory segments have already been created the subscription 
request message can be sent directly. This message includes the subscriber's name preceeded by '\texttt{shm:}' to indicate that a shared memory proxy is used. 
Following the name the message contains the rest of the information needed by the publisher to 
establish a communication: the two shared memory keys and the segments' 
common size. 

In the implementations of the functions defined in the subscriber interface, the approach used is basically always identical. The function determines the total amount of 
data that it has to write for the message and its required parameters. This size is then passed to the call of the \texttt{Can\-Write} function of the publisher-proxy to
subscriber-proxy shared memory object to obtain the shared memory block for writing the message. If this block in the shared memory
segment consists of only one part and is not wrapped around, then the message and its parameters are created directly in the memory block
just allocated. Additional data can be copied from function parameters as necessary. If the data is not in one block but wrapped around,
an additional memory block is allocated and the message and parameter creation functions are called using this local memory block. Once all the required message data is
present in this block it is copied into the two parts of the shared memory block through two \texttt{mem\-cpy} calls. After these steps the \texttt{Have\-Written}
function is called to commit the data and update the write indices in the communication object appropriately. Once this is done all functions except for 
\texttt{Start\-Pub\-lishing} end, indicating successful completion to the caller.

The \texttt{Start\-Pub\-lishing} function does not terminate once the message has been written into the communication object, but calls the
class's message loop to process messages received from the producer processes' subscriber proxy,  
similar to the publisher pipe proxy. In the loop the messages received from the subscriber proxy 
are read using the block header read function
of the subscriber-proxy to publisher-proxy communication object described earlier. 
For the three messages \texttt{New\-Event},
\texttt{Event\-Can\-celed}, and \texttt{Event\-Done\-Data} that require multiple parameters, handler functions are called to read the necessary parameter data from the communication object.
The parameters are subsequently decoded and the appropriate function in the attached subscriber object is called with them. For the other, simpler, messages the subscriber functions can be called 
directly without the need for further parameters. After calling the appropriate subscriber's functions the data in the shared memory is released again using the
appropriate \texttt{Free\-Read} calls for the message and its parameters.

\subsubsection{The Subscriber Shared Memory Proxy Class}

\begin{figure}[h]
\begin{center}
\resizebox*{0.25\columnwidth}{!}{
\includegraphics{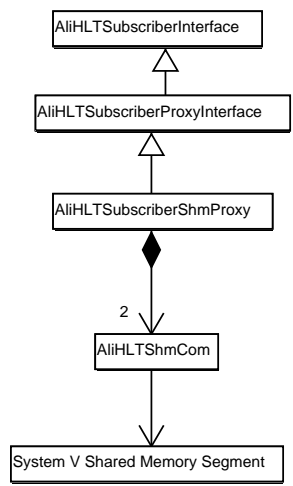}
}
\parbox{0.90\columnwidth}{
\caption[Shared memory subscriber proxy and communication classes UML diagram.]{\label{Fig:SubscriberShmProxyClass}The shared memory subscriber proxy class and its relation to the shared memory communication class.}
}
\end{center}
\end{figure}

The subscriber shared memory proxy class \texttt{Ali\-HLT\-Sub\-scrib\-er\-Shm\-Proxy} performs the subscriber proxy functions in the producer process, analogous the subscriber pipe proxy
class described in section~\ref{Sec:AliHLTSubscriberPipeProxy}. For this purpose it implements the functions defined in the subscriber interface in a similar manner  
as in the shared memory publisher proxy class's functions described in the previous section. Each function determines the size of the message to send 
to the publisher proxy together with necessary parameters and other data. Message
data is created either directly in the shared memory buffer used for the subscriber-proxy to publisher-proxy communication, or it is created in an intermediate
memory block and then copied into the shared memory segment. 

As for the pipe subscriber proxy, no interface function called will cause a message loop to be entered. Instead the publisher to which this subscriber is attached to
has to call the \texttt{Msg\-Loop} function, defined in the \texttt{Ali\-HLT\-Sub\-scrib\-er\-Proxy\-Inter\-face} class, to run in a separate thread. Once the subscription has been 
cancelled the publisher has to call \texttt{Quit\-Msg\-Loop} to exit the message loop and terminate the thread.

\subsection{\label{Sec:SubscriptionLoop}The Subscription Loop Function}

Related to the proxy classes is the subscription loop function \texttt{Pub\-lish\-er\-Pipe\-Sub\-scrip\-tion\-In\-put\-Loop} which should be called by any producer process
in a separate thread to wait for incoming subscription requests. Its only parameter is a reference to the 
publisher object whose subscription requests it should handle.
From the object it obtains the publisher's name used to create the full name of the subscription pipe so that it
is located in the \texttt{/tmp} directory and consists of the name of the publisher with the appended \texttt{Subs\-Ser\-vice} identifier. 

A pipe communication object as described in~\ref{Sec:AliHLTPipeCom} is used to create and open a pipe with the constructed name. In the loop
a blocking read is entered to wait for incoming messages with subscription requests. As each subscription request is contained in a single message
described by a block header structure, a single \texttt{Read} call is sufficient to retrieve the data needed for a subscription. When a subscription request
has been received, the function strips the type specifier, either pipe or shared memory, from the subscriber's name to determine which type of proxy to create.

For a pipe proxy the only information needed is the name so that an \texttt{Ali\-HLT\-Sub\-scrib\-er\-Pipe\-Proxy} object can be created directly. In the case of 
a shared memory proxy the function additionally has to extract the size of the two shared memory segments as well as the two keys for them from the message.
Using these three parameters and the subscriber's name an \texttt{Ali\-HLT\-Sub\-scrib\-er\-Shm\-Proxy} object is created. 

After the correct subscriber proxy object has been created the function calls its publisher object's \texttt{Sub\-scribe} function with the 
created proxy as its argument, subscribing the proxy and its associated subscriber object in the consumer process. Following this, the function releases the message
that has been allocated in the pipe communication object and reenters the read call waiting for the next request.

When the subscription loop has to exit because the producer process ends, a global flag variable is set to be evaluated in the function's loop.  Subsequently, a 
quit message is sent to the loop's named pipe. Upon reading that message and detecting the global quit flag set, the function leaves the loop and terminates.

%\clearpage

\section{The Publisher Implementation Classes}

\begin{figure}[h]
\begin{center}
\resizebox*{0.6\columnwidth}{!}{
\includegraphics{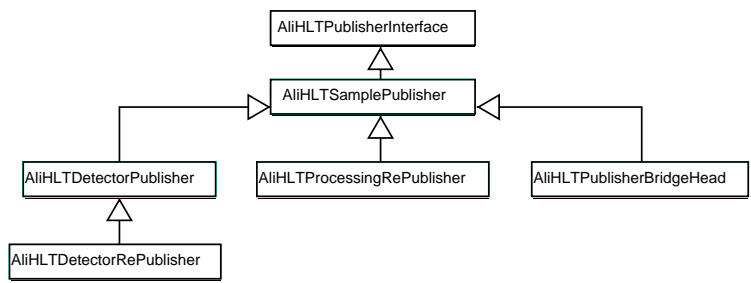}
}
\parbox{0.90\columnwidth}{
\caption{\label{Fig:PublisherClasses}UML class diagram of the publisher implementation classes.}
}
\end{center}
\end{figure}

Only one publisher implementation class directly derived from the publisher interface definition \texttt{Ali\-HLT\-Pub\-lish\-er\-Inter\-face} exists, implementing
the basic publisher functionality of managing a number of subscribers and events. Other publisher classes are in turn derived from this base class to extend
its functionality. The most important of these classes, shown in Fig.~\ref{Fig:PublisherClasses}, are described in the following section. 

\subsection{\label{Sec:AliHLTSamplePublisher}The Sample Publisher Class}

%For the different publisher class there exists
\texttt{Ali\-HLT\-Sample\-Pub\-lish\-er} is the base class for all other publisher implementation classes.
It is the only class that implements the basic functionality of managing multiple subscribers, announcing events
to them, and freeing the events again once all subscribers have released them. All other publisher classes inherit this functionality
from \texttt{Ali\-HLT\-Sample\-Pub\-lish\-er}. In addition to implementing
the required abstract methods defined in the publisher interface it provides a set of other functions that serve as 
the external API of this class. It also supports a number of callback functions that allow for further customization
of a derived publisher, e.g. by implementing an action when an event has been released. These callback functions are not defined
as abstract methods so that not every derived class has to implement all of them but instead are present
as empty virtual method bodies. 
The provided external API allows other programs or classes to use the features present in the sample publisher class, e.g. subscriber and event handling, 
management, and accounting.

\subsubsection{Internal Architecture}

Internally the \texttt{Ali\-HLT\-Sample\-Pub\-lish\-er} class makes use of a number of different threads and two main lists that store the data for each
subscriber and each event respectively. An entry in the subscriber list contains a pointer to the subscriber object or proxy,
%, most likely a subscriber proxy,
in the form of a pointer to a subscriber interface, together with pointers to two thread objects. These two threads are used for communication with the 
subscriber object. The first is used for the subscriber proxy's message loop from which the publisher interface functions are called and the second for calling the 
subscriber object's interface functions. This
second thread object also implements the subscriber interface and can thus be accessed by the publisher class similar to a subscriber. Calls to the subscriber interface
functions place the required data in a memory FIFO of the thread class. The thread itself runs a loop which waits for data from its FIFO object and
calls the interface functions in the subscriber object, decoupling the publisher's main functions from the timing behaviour of the subscriber object's functions.
This is of particular importance when the subscriber object actually is a proxy object that communicates with other processes and could block waiting
for them. 

A subscriber data structure furthermore contains a number of fields relevant to the subscriber's status corresponding to the parameters that can be set
using the respective publisher interface functions. Three flags define whether a subscriber is persistent or transient, whether it is active and receives 
events, and whether a subscriber receives event done data received from the publisher's other subscribers. Additionally, three fields related to the data that 
can be set using the publisher's \texttt{Set\-Event\-Type} functions are present. Of these three elements the first is a list holding the event trigger type structures that can be
specified. The other two are the event modulo number used to restrict the rate of events and the number of events that have been announced while the
specific subscriber has been active. These two numbers are used to determine whether a specific event is announced to a subscriber with a set event
modulo number. The final element in the subscriber structures is the number of ping calls that have been made to this subscriber.
This number is increased when a \texttt{Ping}
call is made to the subscriber object and decreased when a \texttt{Ping\-Ack} call is received from it. When the number reaches
a configurable maximum the subscriber is presumed to be unable to process any calls and is removed in the publisher. 

In each element of the event list a number of fields are stored as well. The first of these is the ID of the event whose data is stored in that particular element.
This is followed by two reference counters, one for the total number of subscribers and one for the number of transient
subscribers to which this event has been announced. Two more fields contain data regarding event timeouts, the maximum timeout used for that event and the ID 
of the currently active timeout
for that event. Two sublists hold a list of subscribers which have not released the event that has been announced to them and a list of all
event done data structures that have been received for that event. This last list is used for one of the callback functions presented below. 

The event list itself is organized in a manner similar to the principle of the \texttt{MLUC\-Vector} class described in section~\ref{Sec:MLUCVector}:
a fixed size array used as a ring buffer. 
This approach is applied since events are typically processed in an approximate first-in-first-out manner. As the releasing of events is not guaranteed to be sequential, each 
entry in the list contains a {\em used} flag that specifies whether the data contained in the element is valid. Free slots for new events are always searched from the 
end of the used space while the search for events to be freed is started at its beginning. If no size for the array is specified in the class's 
constructor, a normal dynamic array class, the \texttt{vector} class from the STL library, is used instead of the ring buffer. 

Beyond the two communication threads for each subscriber, each sample publisher object uses four more threads in addition to a program's
main thread. The first of these is used to handle expired timeouts for each event. It runs in a loop waiting for signals from the timer object for expired timeouts.
Any transient subscriber still locking an event with an expired timeout is forced to release the event. A loop waiting for timer signals is used in the second thread
as well, but these timeouts signal expirations of wait times for ping messages. After a certain number 
of ping acknowledge replies have not been received, the subscriber concerned will be removed from the publisher's list. Cleanup of subscribers  
in the process of being removed from a pulisher's list is the task of the third publisher thread. Subscriber data structures are passed to this thread using a signal
object. When the thread detects that a subscriber is not used anymore, it will free any data structures that have been allocated for this subscriber. It is necessary
to use this  
approach to prevent a thread from releasing a subscriber when another thread still accesses that subscriber's data structures. The final of these 
four threads runs the timer used for every timeout in the publisher, including event and ping timeouts.

\subsubsection{External Sample Publisher Interface}

\begin{figure}[h]
\begin{center}
\resizebox*{0.75\columnwidth}{!}{
\includegraphics{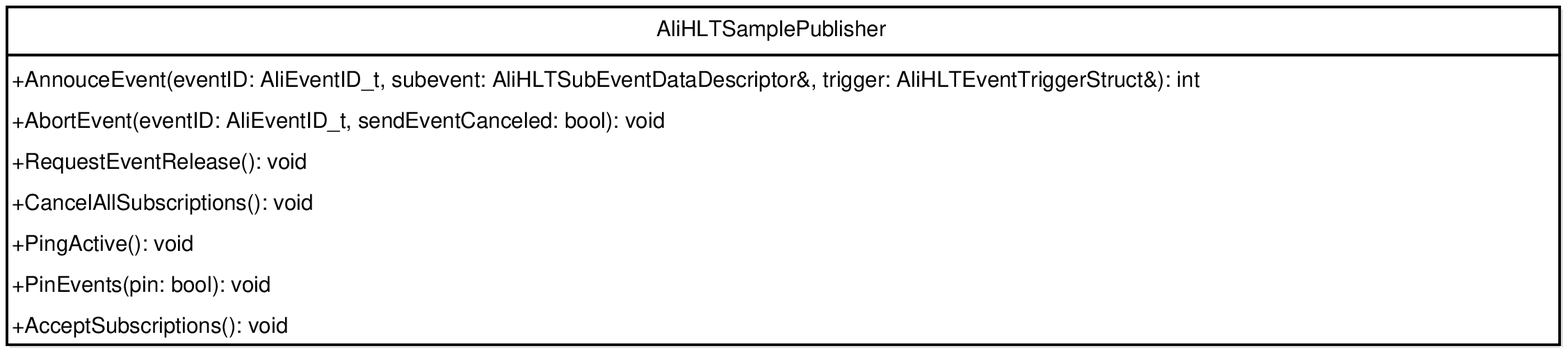}
}
\parbox{0.90\columnwidth}{
\caption[\texttt{Ali\-HLT\-Sample\-Pub\-lish\-er} external API functions.]{\label{Fig:AliHLTSamplePublisher-API}UML class diagram of the \texttt{Ali\-HLT\-Sample\-Pub\-lish\-er} external API functions.}
}
\end{center}
\end{figure}

Beyond the standard publisher interface for use by subscriber objects the sample publisher class provides a second function interface, 
shown in Fig.~\ref{Fig:AliHLTSamplePublisher-API}. This API consists of seven additional 
functions to be called from external functions, some of which correspond loosely to functions defined in either the publisher or subscriber interface. 
\texttt{An\-nounce\-Event}, the first of these seven functions, accepts the same three arguments as the subscriber's \texttt{New\-Event} function: an event ID, 
a sub-event descriptor
for the event's data, and its event trigger type structure. The event described by these three parameters will be announced to subscribers 
currently attached to this publisher object, depending on the trigger types and modulo counters set for each subscriber. 

To forcibly remove an event from a publisher's internal list
% without any publisher internal influences, e.g. all subscribers are finished or the event timeout has expired,
the publisher's \texttt{Abort\-Event} function can be called. The ID of the event to be removed is the function's first parameter. An optional second parameter is a flag 
specifying whether an \texttt{Event\-Can\-celed} call is made to all subscribers that still use the event. By default this flag is true so that the \texttt{Event\-Can\-celed} calls are made. 
When a producer program starts to run out of buffer space for events, the \texttt{Re\-quest\-Event\-Re\-lease} function can be called to make \texttt{Re\-lease\-Events\-Re\-quest}
calls to all attached subscribers informing them of the imminent buffer shortage. To terminate all subscriptions or call the ping function for all attached subscribers
respectively the \texttt{Can\-cel\-All\-Sub\-scrip\-tions} and \texttt{Ping\-Ac\-tive} functions are available, both without any parameters as for \texttt{Re\-lease\-Events\-Re\-quest}. 

In the case that releasing events in a publisher object has to be inhibited for a time, the \texttt{Pin\-Events}
function can be called. Its argument is a flag that specifies whether events are to be freed normally when all subscribers have released a particular event, or whether the 
event should be kept in the publisher nonetheless. A possible application case for the function could be a subscriber that has terminated unexpectedly and should be reattached. 
Any events still in the system should be kept available so that they can be reannounced to this subscriber once it is subscribed again. When the pinning is released,
any events not in use by at least one publisher are released immediately. 

The final of the sample publisher API functions, \texttt{Accept\-Sub\-scrip\-tions}, is called to start the subscription loop for the publisher to wait for incoming subscription
requests. It calls the function that has been specified for the publisher using the \texttt{Set\-Sub\-scrip\-tion\-Loop} function described below, in most cases
the loop function described in~\ref{Sec:SubscriptionLoop}.

\subsubsection{Configurable Functions}

\begin{figure}[h]
\begin{center}
\resizebox*{0.65\columnwidth}{!}{
\includegraphics{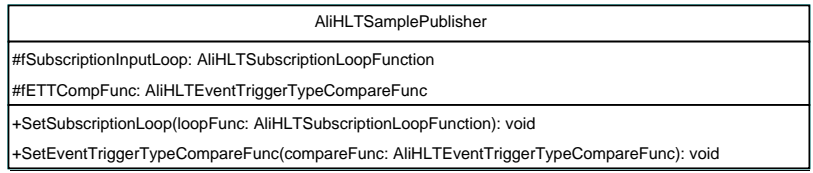}
}
\parbox{0.90\columnwidth}{
\caption[\texttt{Ali\-HLT\-Sample\-Pub\-lish\-er} configurable functions.]{\label{Fig:AliHLTSamplePublisher-ConfFuncs}UML class diagram of the \texttt{Ali\-HLT\-Sample\-Pub\-lish\-er} configurable functions and the functions used to set them.}
}
\end{center}
\end{figure}

Customization of sample publisher objects is supported by two configurable functions in the class. 
Fig.~\ref{Fig:AliHLTSamplePublisher-ConfFuncs} shows the two function pointers together with the two methods used to set them. The first of these,
\texttt{Set\-Sub\-scrip\-tion\-Loop}, allows to specify a function to be called as a subscription loop when the publisher's \texttt{Accept\-Sub\-scrip\-tions} function
is called. This lets programs use subscription loop functions different from the one described in section~\ref{Sec:SubscriptionLoop}, 
to support subscription requests through other mechanisms than named pipes.

A feature in the framework that has not been fully specified is the evaluation of the event trigger type structures defined in section~\ref{Sec:AliHLTEventTriggerStruct}. 
As these structures can be used to determine which events are announced to subscribers, the sample publisher has to be able to determine when an event's trigger type
structure is matched by a structure restricting events for a subscriber. On the other hand, to leave the relevance and interpretation to a particular
application, the meaning and content of these structures has not been specified. To work around the problem
presented by these two conflicting requirements, the sample publisher class supports a second configurable function used to compare two event trigger
type structures. The first of the two structures is used to restrict events for this subscriber and the second one is the trigger
structure that has been specified for the event concerned. When a match is found the configured comparison functions returns \texttt{true}, otherwise
\texttt{false}. The \texttt{Set\-Event\-Trig\-ger\-Type\-Com\-pare\-Func} function can be used to set this comparison function.
%, a pointer to which is passed as its argument. 

\subsubsection{Callback Functions}

\begin{figure}[h]
\begin{center}
\resizebox*{0.80\columnwidth}{!}{
\includegraphics{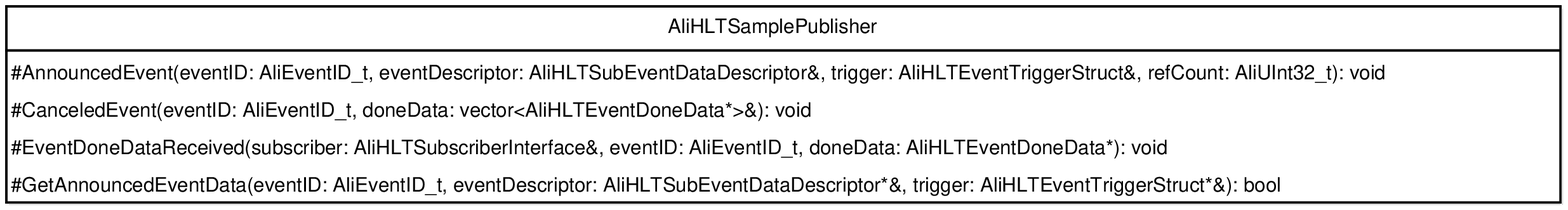}
}
\parbox{0.90\columnwidth}{
\caption[\texttt{Ali\-HLT\-Sample\-Pub\-lish\-er} callback functions.]{\label{Fig:AliHLTSamplePublisher-Callbacks}UML class diagram of the callback functions in \texttt{Ali\-HLT\-Sample\-Pub\-lish\-er}.}
}
\end{center}
\end{figure}

To allow further customization of the sample publisher class through derived classes, the class contains four callback functions invoked when specific 
events occur. These functions, shown in Fig.~\ref{Fig:AliHLTSamplePublisher-Callbacks}, are implemented as empty function bodies and can be overwritten by classes 
derived from \texttt{Ali\-HLT\-Sample\-Pub\-lish\-er} to adapt or extend the class's behaviour. 

The first two of these functions, \texttt{An\-nounced\-Event} and \texttt{Can\-celed\-Event}, are called when an event has been announced to all interested subscribers or
when it has been released from the publisher respectively. Parameters passed to the \texttt{An\-nounced\-Event} function include the three parameters used in the call to
the \texttt{An\-nounce\-Event} function that has been used to announce a particular event. Additionally, the reference count for the number of subscribers
to which this event has been announced is passed as the function's fourth parameter. \texttt{Can\-celed\-Event} is called with two parameters, the first is the
ID of the event that has been released. A vector of event done data structures is passed to the function in its second parameter. This list holds all non-trivial event done data
structures that have been received from attached subscribers for that specific event. 
When such a structure has been received from a subscriber, the third callback function,  \texttt{Event\-Done\-Data\-Re\-ceived}, is called. Arguments 
passed to this function are the name of the subscriber from which the data has been received, the ID of the respective event, and a pointer to the event done 
data structure that has been received. 

The final of the four callback functions, \texttt{Get\-An\-nounced\-Event\-Data}, is called by the publisher object when a subscriber specifies that it wants to receive
events that have already been announced to other subscribers in the \texttt{Start\-Pub\-lishing} call. To avoid the duplicate effort of
storing each event's sub-event descriptor and event trigger structure in both the sample publisher object and the calling application code, the callback is used
to obtain these two data structures from a derived class for each event. The referenced parameters to the two structures have to be filled with pointers 
to the event's actual data inside the function. 
If this is not possible, the function has to return false. Otherwise it has to return true so that the publisher knows that the data 
has been filled in and that the event can be announced.

\subsection{\label{Sec:AliHLTDetectorPublisher}The Detector Publisher Class}

\begin{figure}[h]
\begin{center}
\resizebox*{0.80\columnwidth}{!}{
\includegraphics{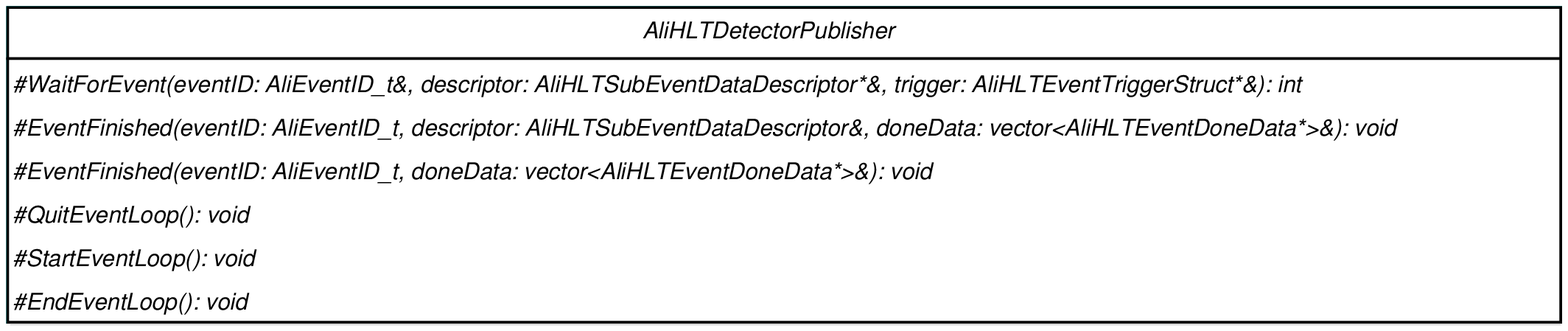}
}
\parbox{0.90\columnwidth}{
\caption{\label{Fig:AliHLTDetectorPublisher-Callbacks}\texttt{Ali\-HLT\-De\-tec\-tor\-Pub\-lish\-er} abstract callback functions.}
}
\end{center}
\end{figure}

The publisher class \texttt{Ali\-HLT\-De\-tec\-tor\-Pub\-lish\-er} is intended for producer programs that address detector hardware. It is derived from and enhances
\texttt{Ali\-HLT\-Sample\-Pub\-lish\-er} to provide a framework for programs that access a hardware device and insert its data into a
processing chain. For this purpose it implements three of the callback methods introduced in the sample publisher class and provides six additional abstract
callback methods that have to be provided by actual implementation classes. The class's main feature is an event loop that runs in a separate thread
and that calls three of the abstract callbacks at different times. Two functions, \texttt{Start\-Event\-Loop} and \texttt{End\-Event\-Loop}, are called respectively at the beginning
and end of the event loop, while a third \texttt{Wait\-For\-Event} is called repeatedly to retrieve new events for announcement. Two further callbacks are the 
\texttt{Event\-Fin\-ished} functions that differ in the arguments accepted. They are called when an event is in the process of being released under different circumstances. One is
used when the sub-event descriptor for the event could be found in a wrapper class that handles the descriptors, and the other if the descriptor could
not be found. The final callback method \texttt{Quit\-Event\-Loop} is called when the event loop has to be terminated. This call is necessary because the event loop might 
be blocked inside the \texttt{Wait\-For\-Event} method and the \texttt{Quit\-Event\-Loop} is intended to make that function return to the calling event loop. A UML diagram of the
callback functions can be seen in Fig.~\ref{Fig:AliHLTDetectorPublisher-Callbacks}. 

In addition to the event loop the class provides a number of other features intended to reduce the amount of implementation work that has to be done for each new
data producer program. It has support for a shared memory manager class that facilitates dereferencing of shared memory segments used for data exchange between
a producer and its consumers. Access is provided to a buffer manager class as well as to a descriptor handler class, as detailed in the preceeding paragraph. 
The former of these two allows to use a buffer manager class from inside the publisher with a minimum of effort while the second basically functions as a higher level
allocation cache for sub-event data descriptor structures. Producer specific code in the \texttt{Wait\-For\-Event} method can use this handler object to obtain
descriptor structures as needed in an efficient manner. 

To implement a data producer based on this class, one first has to create a derived class that implements the class's six abstract callback methods, and an
object of this class has to be created in the producer program. Properly configured instances of the shared memory and buffer manager classes as well as the 
descriptor handler class have to be specified to the publisher object. Once this is done the event loop and the publisher's normal subscription loop
have to be started, which will also cause the \texttt{Start\-Event} callback method to be called, followed by multiple calls to the \texttt{Wait\-For\-Event}
method to retrieve events as needed. The managing and accounting of events and subscribers will be handled by the sample publisher base class, as described.

\subsection{\label{Sec:AliHLTDetectorRePublisher}The Detector RePublisher Class}

The \texttt{Ali\-HLT\-De\-tec\-tor\-Re\-Pub\-lish\-er} class is derived from the \texttt{Ali\-HLT\-De\-tec\-tor\-Pub\-lish\-er} class and is intended to be used in conjunction with the
\texttt{Ali\-HLT\-De\-tec\-tor\-Sub\-scrib\-er} class presented below in section~\ref{Sec:AliHLTDetectorSubscriber}. It can be used with that class to republish events that 
have been received by a detector subscriber instance for reannouncement to other subscribers. Examples where this is used
are the \texttt{Event\-Gath\-erer}, \texttt{Event\-Scat\-terer}, and \texttt{Event\-Merger} components described in section~\ref{Sec:DataFlowComponents} below. 

To store the sub-event descriptor and event trigger structures of announced events in the class, it overwrites the sample publisher's \texttt{An\-nounced\-Event} method 
so that these structures can be reused 
when a subscriber requests already announced events. In addition it also overwrites the six abstract callbacks defined by the detector publisher class, although they are just empty 
implementations, 
except for the two \texttt{Event\-Fin\-ished} methods. The \texttt{Event\-Fin\-ished} methods first attempt to release any buffer blocks and
shared memory segments still allocated and locked for an event. Subsequently the \texttt{Event\-Done} method of the event's originating publisher is 
called to propagate the event's release through its originating producers. This call is made using the aggregated event done data structures that have been received 
from the subscribers attached to the republisher class. 

\subsection{\label{Sec:AliHLTProcessingRePublisher}The Processing Component Publisher Class}

\begin{figure}[h]
\begin{center}
\resizebox*{1.0\columnwidth}{!}{
\includegraphics{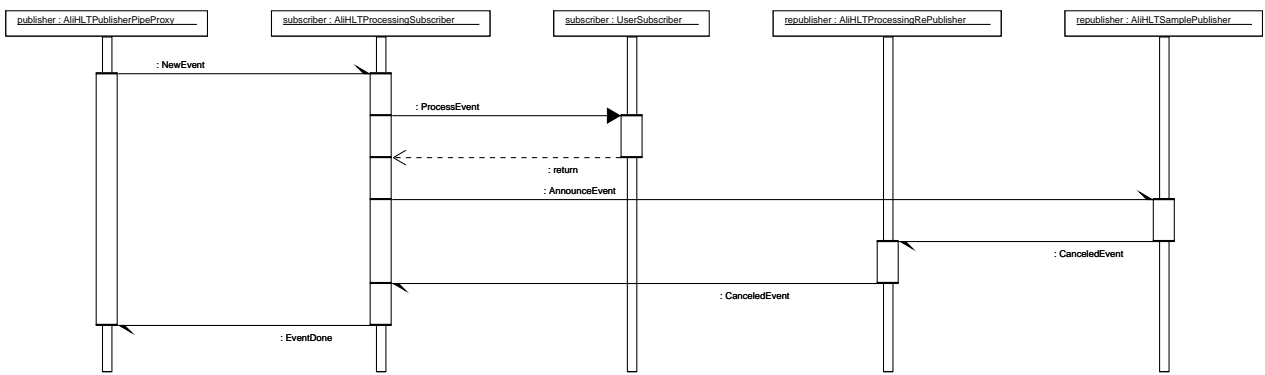}
}
\parbox{0.90\columnwidth}{
\caption[Sample calling sequence for the processing subscriber and republisher classes.]{\label{Fig:ProcessingPubSubSequence}Sample calling sequence for the processing subscriber and republisher classes. Objects with the same name but different
classes indicate functionality in the same object provided by different class definitions.}
}
\end{center}
\end{figure}

In analogy to the \texttt{Ali\-HLT\-De\-tec\-tor\-Re\-Pub\-lish\-er} and  \texttt{Ali\-HLT\-De\-tec\-tor\-Pub\-lish\-er} classes 
the \texttt{Ali\-HLT\-Pro\-cessing\-Re\-Pub\-lish\-er} class is intended to be used
together with the \texttt{Ali\-HLT\-Pro\-cessing\-Sub\-scrib\-er} class (section~\ref{Sec:AliHLTProcessingSubscriber}). 
It overrides three of the callback methods provided by the 
sample publisher class, which are \texttt{Can\-celed\-Event}, \texttt{Event\-Done\-Data\-Re\-ceived}, and \texttt{Get\-An\-nounced\-Event\-Data}.
They are forwarded to correspondingly named methods in
the subscriber class for actual processing. The two classes are intended to be used in analysis components, as described in sections~\ref{Sec:UserComponentTemplates} to 
\ref{Sec:AnalysisComponents}, that contain a subscriber for receiving data, processing it, and producing new data. This produced data is then subsequently announced by another
publisher in the process. A sample calling sequence for an event that has been announced to a program's  publisher proxy class, reannounced, and released by a processing 
republisher class is shown in Fig.~\ref{Fig:ProcessingPubSubSequence}. 

\subsection{\label{Sec:AliHLTPublisherBridgeHead}The Publisher Bridge Head Class}

Like the two preceeding classes the \texttt{Ali\-HLT\-Pub\-lish\-er\-Bridge\-Head} class is also designed to be used in cooperation with
another class, the \texttt{Ali\-HLT\-Sub\-scrib\-er\-Bridge\-Head}, introduced in section~\ref{Sec:AliHLTSubscriberBridgeHead} and described in
more detail in \ref{Sec:BridgeComponents}. 
Unlike the two other cases, however, the two bridge head classes are not 
situated in the same process, but instead each is present in its own process. In most cases these two processes will not even be running on the same node but on two 
separate nodes, as they  together provide a transparent connection between components on different nodes.
The connection mechanism as well as the publisher and subscriber bridge head classes are described more detailed
in section~\ref{Sec:BridgeComponents}. 

\section{The Subscriber Implementation Classes}

Unlike the publisher classes there is no single basic implementation class at the root of the subscriber class hierarchy. An \texttt{Ali\-HLT\-Sample\-Sub\-scrib\-er} class
also exists but its functions are mainly just empty bodies. The only function that performs any action
is the class's \texttt{Ping} method that calls the calling publisher's \texttt{Ping\-Ack} method as a response. The reason for this lack of a basic 
subscriber implementation is that unlike the publisher's event and subscriber management and accounting there is little or no general overlap of functionality
between the different subscriber classes. Therefore the \texttt{Ali\-HLT\-Sample\-Sub\-scrib\-er}
class is primarily a useful base class for subscribers that implement only some of the calls defined in the subscriber interface. Most of the subscriber classes
are derived from \texttt{Ali\-HLT\-Sub\-scrib\-er\-Inter\-face} directly, rather than from an intermediate subscriber implementation class. Fig. {\ref{Fig:SubscriberClasses}
shows the class hierarchy for the three classes described in the following sections, including the subscriber interface and the sample subscriber class. 

\begin{figure}[h]
\begin{center}
\resizebox*{0.6\columnwidth}{!}{
\includegraphics{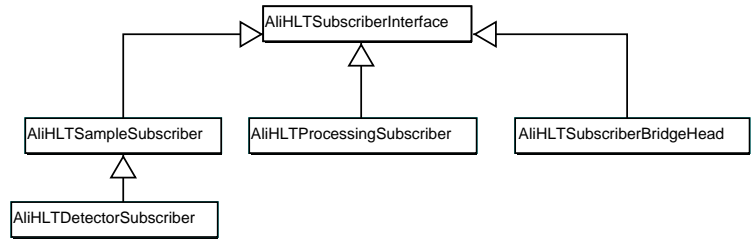}
}
\parbox{0.90\columnwidth}{
\caption{\label{Fig:SubscriberClasses}UML class diagram of the subscriber implementation classes.}
}
\end{center}
\end{figure}

\subsection{\label{Sec:AliHLTDetectorSubscriber}The Detector Subscriber Class}

One of the classes derived from \texttt{Ali\-HLT\-Sample\-Sub\-scrib\-er} is the \texttt{Ali\-HLT\-De\-tec\-tor\-Sub\-scrib\-er} class that was originally intended as the
companion to the \texttt{Ali\-HLT\-De\-tec\-tor\-Pub\-lish\-er} class in~\ref{Sec:AliHLTDetectorPublisher}. For the primary purpose as the subscriber object in data 
analysis programs of the type described in section~\ref{Sec:UserComponentTemplates}, 
it has been superseded by the \texttt{Ali\-HLT\-Pro\-cessing\-Sub\-scrib\-er} class described in the following
section~\ref{Sec:AliHLTProcessingSubscriber}. This class provides a more advanced framework for receiving, processing, and reannouncing events.
It reduces the additional tasks required for the writing of application specific programs to the implementation of one function. 
The detector subscriber class is still used as the receiving subscriber
for a number of data flow components, like the event gathererer and scatterer programs presented in~\ref{Sec:DataFlowComponents}. 

In the detector subscriber class three of the subscriber interface functions provided by the \texttt{Ali\-HLT\-Sample\-Sub\-scrib\-er} are overwritten with additional
functionality. The class also starts a thread for the class's main processing loop, that uses a signal to wait for incoming events, 
prepares them, and calls the processing function
for the event. 
This processing function is defined as an abstract method that has to be implemented by derived classes to provide actual processing functionality.
Included in the preparation is a dereferencing step to convert the shared memory ID/offset combination for each data block into an actual C pointer passed
to the processing function. When events are cancelled before they reach the processing step they are removed from the queue where they have been placed to be processed. For 
events cancelled while being processed, a flag is set that should be checked periodically in the processing function to avoid working on data that has been 
overwritten. Resources that have been allocated for these events are released after processing has finished or aborted. 
Events are added to the notification queue of the signal used in the main loop by the implementation of the \texttt{New\-Event} function.

%{\bf ??NewEvent add events to queue, queue worked of in main loop, cf. processing subscriber??}
%{\bf Processing components: process function calls appropriate processing function for purpose in component's main object}

\subsection{\label{Sec:AliHLTProcessingSubscriber}The Processing Component Subscriber Class}

The \texttt{Ali\-HL\-Pro\-cessing\-Sub\-scrib\-er} class is the successor to the 
\texttt{Ali\-HLT\-De\-tec\-tor\-Sub\-scrib\-er} class. It is designed to be used as a subscriber object in 
either analysis components, with a subscriber and publisher, or in data sink components with only a subscriber for receiving data. For this purpose it
implements all defined subscriber interface functions and starts two internal threads as well as a timer thread. Of these two internal threads one executes the class's main 
processing loop while the other one contains a cleanup loop. 

In the class's \texttt{New\-Event} function the specified sub-event descriptor and trigger structures are copied. Pointers to these copies are added to the
data queue of a signal object before it is triggered. In the main loop the processing subscriber waits for this signal to be triggered and as soon as this happens, it
retrieves these two event meta-data pointers from the signal's queue and prepares them for processing. As for the detector subscriber from the previous section
this preparation includes the conversion of the shared memory ID/block offset pairs into pointers to each data block in the event. Unlike in the detector subscriber,
a memory block for output data is also obtained from attached buffer manager and shared memory objects. The prepared and dereferenced block descriptors
as well as the output memory block are used in the call to the event processing function, which again is defined as an abstract function that has to be overwritten
by derived classes. If this function completes processing successfully and produces new output data in the output shared memory, and if 
the object is part of an analysis component and not a
data sink, a sub-event descriptor is built for this data and announced via an associated \texttt{Ali\-HLT\-Pro\-cessing\-Re\-Pub\-lish\-er} object (cf. 
section~\ref{Sec:AliHLTProcessingRePublisher}) to any interested subscriber. For subscribers in data sink components event done data produced by the
processing function is used to send the event done message to the event's originating publisher. In analysis components a flag decides when an event done message is 
sent to the originating publisher, either when the event has been processed and new output data produced, similar to the data sink case, or when the associated republisher
object informs the subscriber that the produced event data has been released by its subscribers. 
In the latter case event done messages will propagate back through a whole processing chain
from the last processing component. Any event done data produced by the processing function is stored in this case with the event's other meta-data and is 
attached to the event done data that has been received from the republisher's attached subscribers. This assembled event done data is then used in the event done 
message sent to the event's originating subscriber. Fig.~\ref{Fig:ProcessingSubscriberEventDoneSequence1} and 
Fig.~\ref{Fig:ProcessingSubscriberEventDoneSequence2} show sequence diagrams of the two cases for sending an event done message back to the originating publisher.
To prevent event losses in the system the main loop contains error detection logic at each stage of the preparation, processing, and announcing steps. This is 
coupled with retry handling that ensures that an event with an error occuring anywhere in the stages is processed until it
succeeds or until a permanent unresolvable error occurs.

\begin{figure}[h]
\begin{center}
\resizebox*{1.0\columnwidth}{!}{
\includegraphics{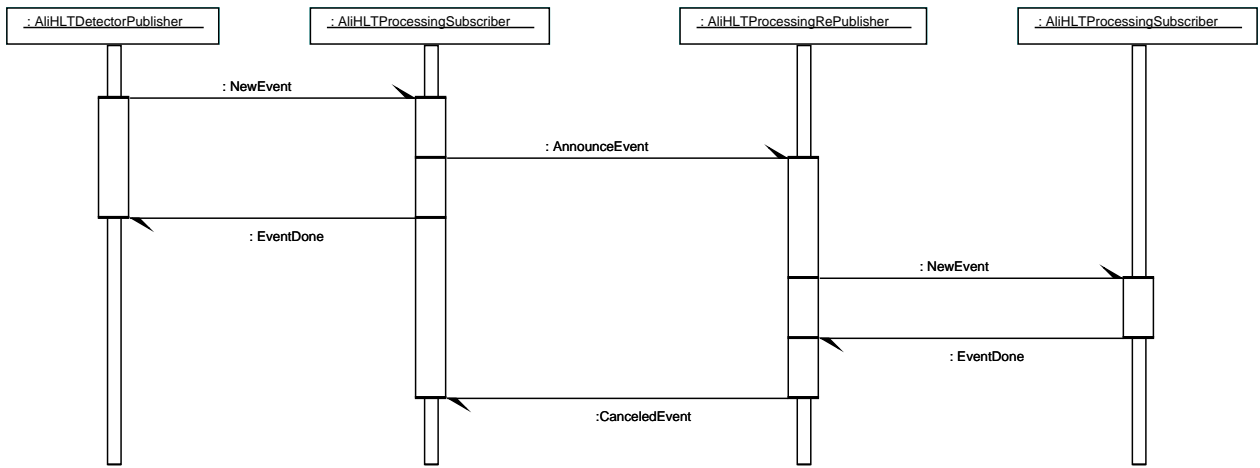}
}
\parbox{0.90\columnwidth}{
\caption[Processing subscriber object event done sequence 1.]{\label{Fig:ProcessingSubscriberEventDoneSequence1}Sequence of messages when a processing subscriber object sends an event done as soon
as it has finished processing an event. Intermediate proxy objects have been left out for clarity.}
}
\end{center}
\end{figure}

\begin{figure}[h]
\begin{center}
\resizebox*{1.0\columnwidth}{!}{
\includegraphics{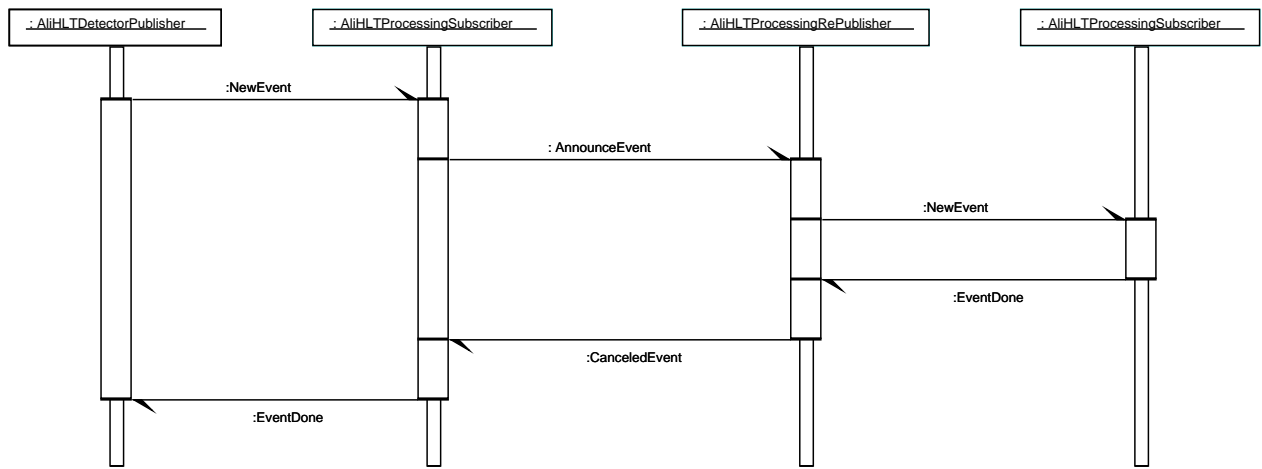}
}
\parbox{0.90\columnwidth}{
\caption[Processing subscriber object event done sequence 2.]{\label{Fig:ProcessingSubscriberEventDoneSequence2}Sequence of messages when a processing subscriber object sends an event done only when
it has been released by its associated output processing republisher object. Intermediate proxy objects have been left out for clarity.}
}
\end{center}
\end{figure}

When an event is ready to be freed, any blocks reserved for its output data are released and a pointer to its event done data is placed into the queue of a further
signal object. The subsequent triggering of this signal causes the cleanup loop running in the second thread to be activated.
%Waiting for this signal to be triggered is performed in the . 
In this loop, the \texttt{Event\-Done} call to the event's publisher is made using the assembled event done data. In addition, the event's meta-data is removed
from the internal structures of the object and further cleanup is performed as needed.

\subsection{\label{Sec:AliHLTSubscriberBridgeHead}The Subscriber Bridge Head Class}

At the sending end of a data bridge to an \texttt{Ali\-HLT\-Pub\-lish\-er\-Bridge\-Head} object from section~\ref{Sec:AliHLTPublisherBridgeHead} and 
\ref{Sec:BridgeComponents} is an instance
of the \texttt {AliHLTSubscriberBridgeHead} class implementing the subscriber interface functions. This class is described in more detail
in section~\ref{Sec:BridgeComponents} together with the other classes used in the bridging components.

\clearpage

%%%%%%%%%%%%%%%%%%%%%%%%%%%%%%%%%%%%%%%%%%%%%%%%%%%%%%%%%%%%%%%%%%%%%%%%%%%%%%%%%%%%%%%%%%%%%%%%%%%%%%%%%%%%%%%%%%%%%%%%%%%%%
%%%%%%%%%%%%%%%%%%%%%%%%%%%%%%%%%%%%%%%%%%%%%%%%%%%%%%%%%%%%%%%%%%%%%%%%%%%%%%%%%%%%%%%%%%%%%%%%%%%%%%%%%%%%%%%%%%%%%%%%%%%%%

\chapter{\label{Chap:Components}The Framework Components}

Based upon the publisher-subscriber interface classes described in the previous chapter, a number of
software components and component templates have been developed as the main part of the framework 
to allow the construction of complex data flow chains in PC cluster systems. The components
can be separated by their purpose into several categories described in the following
sections. For the configuration of the data flow in such a system a set of fully functional 
components exists, which are described in section~\ref{Sec:DataFlowComponents}. 
Section~\ref{Sec:UserComponentTemplates} details a number of template programs without actual functionality,
whose purpose it is to ease the writing of components for specific tasks. Templates are provided for data 
sink, source, and processing components. Several worker components to create, modify, or otherwise process
event data, some based upon these templates, are described in the following sections~\ref{Sec:WorkerComponents} to~\ref{Sec:AnalysisComponents}.
The second of these includes analysis components which have been written for use in the ALICE HLT or 
its prototypes. The final section~\ref{Sec:FTComponents} contains descriptions  of 
components dedicated to ensuring the fault tolerance of systems created using this framework. They
 function in conjunction with components from~\ref{Sec:DataFlowComponents}.

For the program components described below, a number of additional classes and functions beyond the 
interface classes described in chapter~\ref{Chap:PubSubInterface}, have been written
that contain some of their key functionality. 
These classes are described where appropriate.
%These classes are described in the corresponding components' section or in 
%their own section if they are used by multiple components. 

\section{\label{Sec:DataFlowComponents}Data Flow Components}

The components described below are intended to configure the flow of data in a
system constructed using the framework. Amongst others, components exist to merge parts belonging to the same event,
%into one event descriptor, 
to connect components on different computers, and to split up and rejoin a 
stream of events into multiple smaller event streams. None of these components modify the data 
specified by the event descriptors exchanged between the programs through the 
publisher-subscriber interface. Some modify the descriptors while they are forwarded
unchanged by others.

\subsection{Event Merger Component}

Since multiple data sources may exist that produce data blocks belonging to one event, the \texttt{Event\-Merger} component exists to merge
the multiple event descriptors for these parts into a single descriptor containing all blocks. For this purpose the program uses 
multiple subscribers, derived from the class \texttt{Ali\-HLT\-De\-tec\-tor\-Sub\-scrib\-er}, to receive the event parts. 
One output publisher, derived from \texttt{Ali\-HLT\-De\-tec\-tor\-Re\-Pub\-lish\-er}, is used to announce merged descriptors. 
The component's main functionality is contained in an object of the \texttt{Ali\-HLT\-Event\-Merger} class to
which the subscriber objects forward received events. Fully merged events are passed to the republisher object for announcement to attached consumer
components. Fig.~\ref{Fig:EventMergerClasses} and~\ref{Fig:EventMergerSequence} show the relation of the classes in the component and a sample calling sequence
of these classes respectively. The subscriber and publisher classes do not contain any significant functionality beyond calling the merger class's
corresponding functions. 

\begin{figure}[hbt]
\begin{center}
\resizebox*{0.4\columnwidth}{!}{
\includegraphics{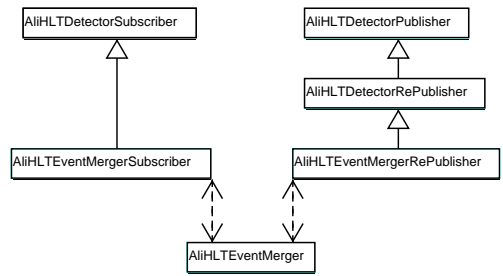}
}
\parbox{0.90\columnwidth}{
\caption{\label{Fig:EventMergerClasses}The relation of the different classes in the \texttt{Event\-Merger} component.}
}
\end{center}
\end{figure}

\begin{figure}[hbt]
\begin{center}
\resizebox*{1.0\columnwidth}{!}{
\includegraphics{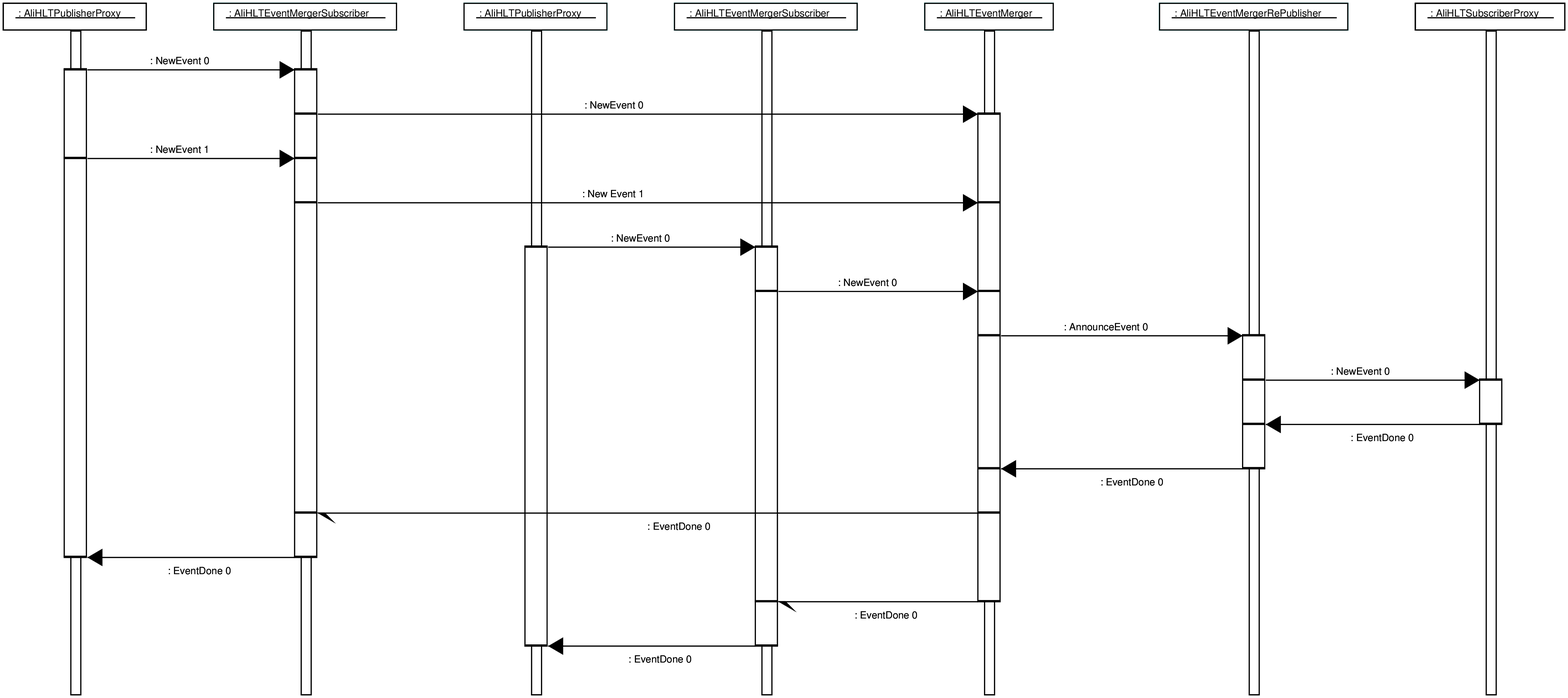}
}
\parbox{0.90\columnwidth}{
\caption[A sample calling sequence for the different classes in the \texttt{Event\-Merger} component.]{\label{Fig:EventMergerSequence}A 
sample calling sequence for the different classes in the \texttt{Event\-Merger} component. This example uses two input subscribers.}
}
\end{center}
\end{figure}

In addition to the component's main thread, one thread is started as a subscription thread for the republisher object. One more thread is started
for the message loop for each configured subscriber in addition to the two cleanup and processing threads started internally by each
\texttt{Ali\-HLT\-De\-tec\-tor\-Sub\-scrib\-er} instance. In the program's main thread a loop is entered that waits for all parts of an event to be completely received
after which the assembled event is announced again. A timeout is configurable that will cause events to be announced when one or more parts were not received within a specified amount
of time.

\subsubsection{The Event Merger Class}

The two main parts of the \texttt{Ali\-HLT\-Event\-Merger} class are its list of configured input subscribers and the list for partially received and unannounced events. 
Of these two the subscriber list is the more simple one. It just stores pointers to the configured subscriber objects of the \texttt{Ali\-HLT\-Event\-Merger\-Sub\-scrib\-er} class. 
Among the four most important elements stored in the event list structures are the number of contributing subscribers expected for this event as well as the number
of subscribers from which parts have already been received. In addition, the event trigger structure from the first received subevent is also stored for each event. Another
possibility might be to concatenate the event trigger structures from all event parts for the event's reannouncement. The fourth important
element of these structures is a list of descriptors for each data block contained in the received subevents. 

When an event part is received by one of the configured subscribers, the event list is checked whether an entry for that particular event already exists. If no existing
entry can be found, a new one is created with the event trigger data that has been received for this part, otherwise the existing entry is used. 
In both cases the number of subscribers from which
data has already been received is increased, and the block descriptions contained in the received sub-event descriptor are added to the event's data block list. As soon as the 
number of received sub-events is equal to the number of configured subscribers, the list entry for the event concerned
is placed into a signal object. The subsequent triggering of this signal object activates the merger components's main thread to retrieve the 
block list from the event data structure. A new event descriptor for the aggregated list will be constructed and announced through the 
republisher object in the program. During these steps the event data will not be removed from the event data list. It is kept in the list until the republisher
object declares that the event has been cancelled through its appropriate callback function. When the specified timeout expires, the event list is also searched for the triggered
event. If it is found, the event's data structure will be signalled to the main thread as well, irrespective of the number of sub-events that have been received
so far. 

After the republisher has released an event,  it informs the \texttt{Ali\-HLT\-Event\-Merger} object by calling its \texttt{Event\-Done} method. The merger object searches
for the event in its event list. If it is found it is removed, and all used resources are freed. Finally, an \texttt{Event\-Done} message is sent to all upstream publishers 
 the merger component is subscribed to, allowing them to release the event as well.

\subsection{\label{Sec:EventScatterer}Event Scatterer Component}

One CPU executing one analysis component will not always be sufficient to alone perform a specific processing step of a chain at the required rate. 
The processing load of the 
steps concerned will thus have to be distributed among a number of CPUs. To provide this functionality in the framework the \texttt{Event\-Scat\-terer} 
component has been created, which splits up an incoming stream of sub-events into multiple streams consisting of correspondingly lower rate of sub-events. Splitting of 
the stream is executed on an event-by-event basis, distributing whole events, and not by splitting up data from one event. In a manner analogous to the 
\texttt{Event\-Merger} component, the \texttt{Event\-Scat\-terer} component uses one input subscriber and multiple output republishers. The input subscriber is
derived from \texttt{Ali\-HLT\-De\-tec\-tor\-Sub\-scrib\-er} and is used to receive the input event stream, 
while the output publishers derived from \texttt{Ali\-HLT\-De\-tec\-tor\-Re\-Pub\-lish\-er} make the multiple output streams available to other components. 
Unlike in the case of the merger component the scatterer's main functionality is not contained in one specific class to allow the possibility of different
algorithms for the distribution of the incoming events. The scatterer base class \texttt{Ali\-HLT\-Event\-Scat\-terer} has been defined to provide parts of the required
functionality together with a number of callback functions that define an interface for scatterer classes to be used in the scatterer component. Currently 
only one derived class, \texttt{Ali\-HLT\-Round\-Robin\-Event\-Scat\-terer}, is implemented for use in this component together with one class for use in
the fault tolerance scatterer described below in section~\ref{Sec:TolerantEventScatterer}. It uses a simple round-robin 
algorithm for the distribution of the events among the configured output publishers. Neither the subscriber nor the republisher classes provide significant 
additional functionality beyond interfacing with the central scatterer object. 

In addition to the program's main thread and the two threads started by each subscriber, one subscription loop thread is started for each republisher 
object. In the main thread the subscriber's message communication loop for the publisher-subscriber interface is executed. As soon as this message loops ends the scatterer 
component will be terminated as well. Fig.~\ref{Fig:EventScattererClasses} and Fig.~\ref{Fig:EventScattererSequence} show the relationship of the different 
classes in the scatterer and a sample calling sequence respectively. 

\begin{figure}[hbt]
\begin{center}
\resizebox*{0.4\columnwidth}{!}{
\includegraphics{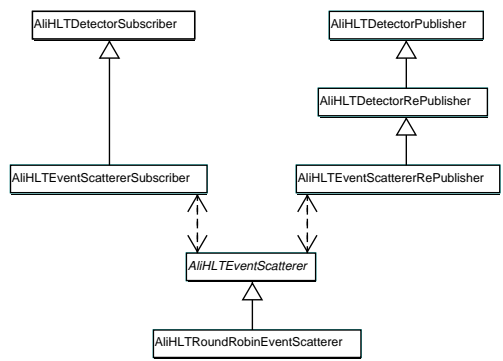}
}
\parbox{0.90\columnwidth}{
\caption{\label{Fig:EventScattererClasses}The relation of the different classes in the \texttt{Event\-Scat\-terer} component.}
}
\end{center}
\end{figure}

\begin{figure}[hbt]
\begin{center}
\resizebox*{1.0\columnwidth}{!}{
\includegraphics{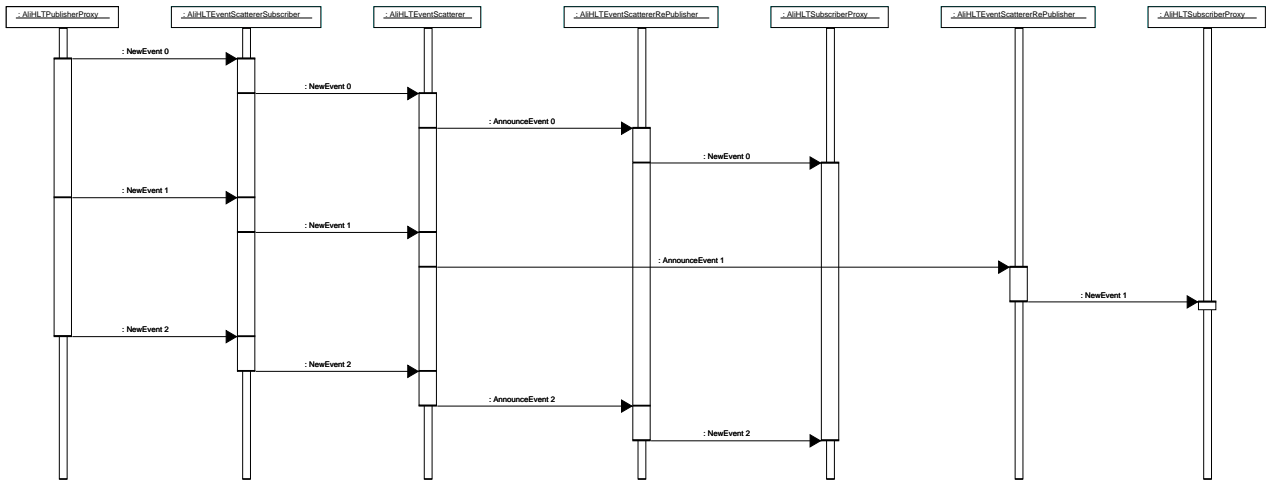}
}
\parbox{0.90\columnwidth}{
\caption[A sample calling sequence for the different classes in the \texttt{Event\-Scat\-terer} component.]{\label{Fig:EventScattererSequence}A sample calling sequence for the different classes in the \texttt{Event\-Scat\-terer} component. 
This example uses two output publishers among which events are distributed round-robin.  
%{\bf ??EventDone??}
}
}
\end{center}
\end{figure}

\subsubsection{The Event Scatterer Base Class}

\begin{figure}[hbt]
\begin{center}
\resizebox*{0.65\columnwidth}{!}{
\includegraphics{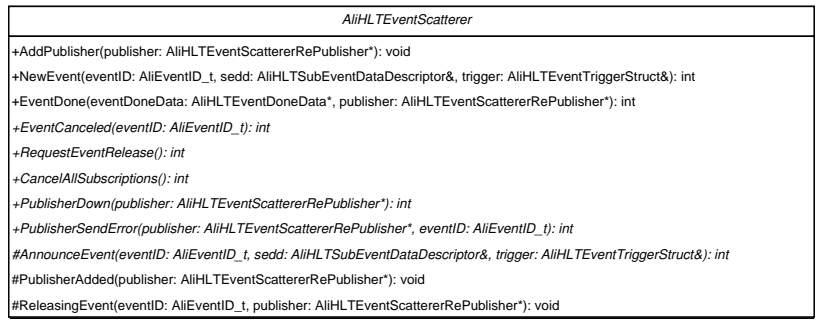}
}
\parbox{0.90\columnwidth}{
\caption[Interface functions provided by the \texttt{Ali\-HLT\-Event\-Scat\-terer} class.]{\label{Fig:EventScattererInterface}The interface functions provided by the \texttt{Ali\-HLT\-Event\-Scat\-terer} class with the three
public methods, five public abstract methods, and three internal methods, one abstract and two callbacks.}
}
\end{center}
\end{figure}

Internally the \texttt{Ali\-HLT\-Event\-Scat\-terer} class mainly consists of the list of output publishers which have been configured to be used.
The interface functions it provides and defines are shown in Fig.~\ref{Fig:EventScattererInterface}. 
Among these are three main functions for use by programs, derived base classes, and its related subscriber and republisher classes:
\texttt{Add\-Pub\-lish\-er}, \texttt{New\-Event}, and \texttt{Event\-Done}. The first of these, \texttt{Add\-Pub\-lish\-er}, has to be called to add 
an output publisher to a scatterer object to make its part of the received data available. It has to be
called during the initialization of the scatterer component in its main thread and calls the \texttt{Pub\-lish\-er\-Added} callback with
the publisher object that has been added. 
%\texttt{New\-Event} and \texttt{Event\-Done} are the other two functions. The first of these 
The next of these functions, \texttt{New\-Event}, is called by the \texttt{Ali\-HLT\-Event\-Scat\-terer\-Sub\-scrib\-er} 
object when a new event is received.
\texttt{Event\-Done} on the other hand is called by one of the \texttt{Ali\-HLT\-Event\-Scat\-terer\-Re\-Pub\-lish\-er} objects when an event is released.
%these functions call one of the abstract interface functions described below. 
Both of these functions call a further function declared or defined by this class. 
\texttt{New\-Event} calls the abstract function \texttt{An\-nounce\-Event} to dispatch an event to one of the available publishers 
to be announced, and \texttt{Event\-Done} calls the empty \texttt{Re\-leasing\-Event} notification callback. Following this notification call, \texttt{Event\-Done}
calls the \texttt{Event\-Done} method of the \texttt{Ali\-HLT\-Event\-Scat\-terer\-Sub\-scrib\-er} object to allow the event to be released in its originating
producer. 

Among the six defined abstract methods in the \texttt{Ali\-HLT\-Event\-Scat\-terer} class three are public methods which are called directly by the subscriber 
object in the component. Two more public methods are provided for the case of publisher errors in the component. The final one is called internally
by the class's \texttt{New\-Event} method. There are two public abstract methods, \texttt{Event\-Can\-celed} and \texttt{Re\-quest\-Event\-Re\-lease},
called by the subscriber object when a particular event has been cancelled or when a request to release events has been received respectively.
\texttt{Can\-cel\-All\-Sub\-scrip\-tions}, the third of these methods, is called when the subscriber's own subscription has been terminated by its publisher. 
The two publisher error methods are called \texttt{Pub\-lish\-er\-Down} and \texttt{Pub\-lish\-er\-Send\-Error}. \texttt{Pub\-lish\-er\-Down} is called from outside the class, either
by a republisher object or by an external supervising instance, in response to a non-trivial error. Its purpose is to 
mark that publisher as unavailable, preventing the scatterer from sending any data to it. In contrast \texttt{Pub\-lish\-er\-Send\-Error}  is called whenever 
an error occured announcing an event for a specific publisher object. This is not considered a severe error and does not necessitate the removal of the publisher
concerned. The final abstract method \texttt{An\-nounce\-Event} is the central method for each scatterer class. It is called by \texttt{New\-Event} whenever a new event 
is received to  determine to which output publisher an event is dispatched for publishing. This is handled according to each scatterer type's specific algorithm.

\subsubsection{The Round-Robin Event Scatterer Class}

In the basic \texttt{Event\-Scat\-terer} component the \texttt{Ali\-HLT\-Event\-Scat\-terer} interface implementation is provided by its derived class  
\texttt{Ali\-HLT\-Round\-Robin\-Event\-Scat\-terer}. It provides implementations of the six abstract methods defined in the base class.
%, whileneither modifying the default behaviour of other base class methods nor implementing any of the two provided callback methods. 
It neither overrides the default behaviour of other base class methods, nor does it implement any of the two callback methods provided by the base class. 

A simple round-robin algorithm is used by the central \texttt{An\-nounce\-Event} method to select an output publisher for each event. To ensure consistency 
for multiple parts of an event passing through different parts of a system, this algorithm is not based on the event sequence number but uses 
an event's ID instead. The same algorithm is also used by the implementation of the \texttt{Event\-Can\-celed} method to determine
the republisher to which the notification about an event's cancelation has to be forwarded. 
\texttt{Re\-quest\-Event\-Re\-lease} just forwards the release request to all publishers and 
\texttt{Can\-cel\-All\-Sub\-scrip\-tions}  cancels all publishers' subscriptions. 
Empty implementations
without any functionality are provided for the two publisher error notification functions, effectively disabling handling of errors occuring
in one of the scatterer's publishers.

\subsection{\label{Sec:EventGatherer}Event Gatherer Component}

Most event streams that have been split up with the help of the \texttt{Event\-Scat\-terer} component described in the previous section will have to be united into
a single stream again at a 
later point of a data processing chain. This task is performed by the \texttt{Event\-Gath\-erer} component, which can be seen as the inverse component to the
scatterer, with multiple input subscribers and one output publisher in place of the scatterer's multiple output publishers and single input subscriber. 
As for the merger component the subscribers' class is derived from \texttt{Ali\-HLT\-De\-tec\-tor\-Sub\-scrib\-er} and the publisher's is derived from
\texttt{Ali\-HLT\-De\-tec\-tor\-Re\-Pub\-lish\-er}. Fig.~\ref{Fig:EventGathererClasses} shows the relationship of the classes used in the \texttt{Event\-Gath\-erer} component. 
The merger and the gatherer components are very similar in their internal architecture. Their main difference
is in the gatherer not having to receive one part of an event from each of its input subscribers. Instead it just
has to forward each received event to its output publisher unchanged. Fig.~\ref{Fig:EventGathererSequence} shows a sample sequence of events for this component. 

\begin{figure}[hbt]
\begin{center}
\resizebox*{0.4\columnwidth}{!}{
\includegraphics{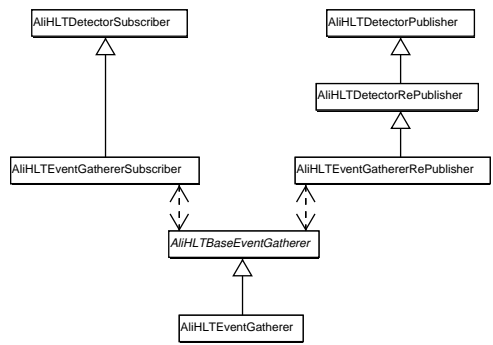}
}
\parbox{0.90\columnwidth}{
\caption{\label{Fig:EventGathererClasses}The relation of the different classes in the \texttt{Event\-Gath\-erer} component.}
}
\end{center}
\end{figure}

\begin{figure}[hbt]
\begin{center}
\resizebox*{1.0\columnwidth}{!}{
\includegraphics{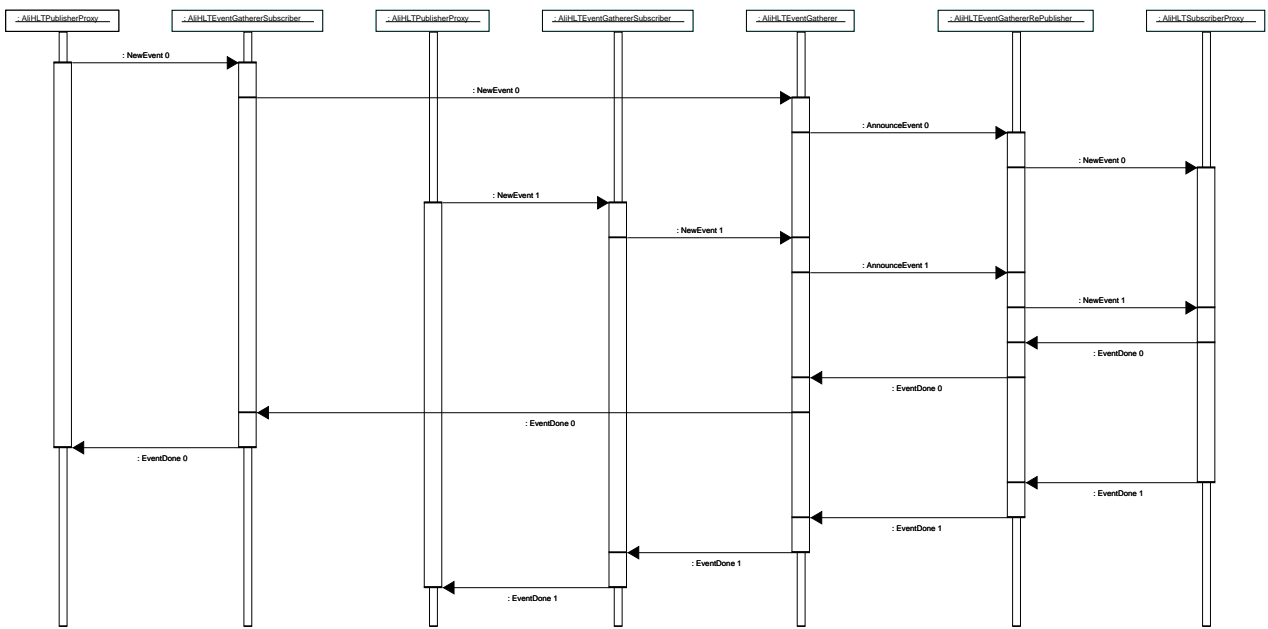}
}
\parbox{0.90\columnwidth}{
\caption[A sample calling sequence for the different classes in the \texttt{Event\-Gath\-erer} component.]{\label{Fig:EventGathererSequence}A sample calling sequence 
for the different classes in the \texttt{Event\-Gath\-erer} component. This example uses two input subscribers.}
}
\end{center}
\end{figure}

As in the \texttt{Event\-Scat\-terer} component a base class, called \texttt{Ali\-HLT\-Base\-Event\-Gath\-erer}, is used to define the central interface 
for the main gatherer class in the component with one data structure 
and five abstract methods. Actual gathering functionality is contained in a derived 
class \texttt{Ali\-HLT\-Event\-Gath\-erer} that provides implementations of these methods. As for the previous merger and scatterer components neither the subscriber
nor the republisher component class contain significant functionality beyond the forwarding of function calls to the
central gatherer object. 

Internally, the gatherer's primary data structures are its list of configured subscribers as well as a list of received and forwarded events which the ouput 
republisher could not yet release because they are still in use by at least one of its subscribers. This event list is necessary in the gatherer
as it has to keep track of the event's originating publishers, to be able to send \texttt{Event\-Done} messages for released events. 
In this respect it differs from the \texttt{Event\-Merger}, as that component receives parts of one event from each of the publishers it is attached to and thus has to send
\texttt{Event\-Done} messages to each of them as well. Allocation and work task assignment for threads in the gatherer component is identical to the merger.
One thread is started for each subscriber as the message loop for the publisher-subscriber interface communication plus each subscriber's two 
internal threads for processing and cleanup. One further thread is created as a subscription request thread for the republisher object. In the component's
main thread a loop is entered that waits for received events to announce them via the republisher to any further components. 

\subsubsection{The Event Gatherer Base Class}

\begin{figure}[hbt]
\begin{center}
\resizebox*{0.65\columnwidth}{!}{
\includegraphics{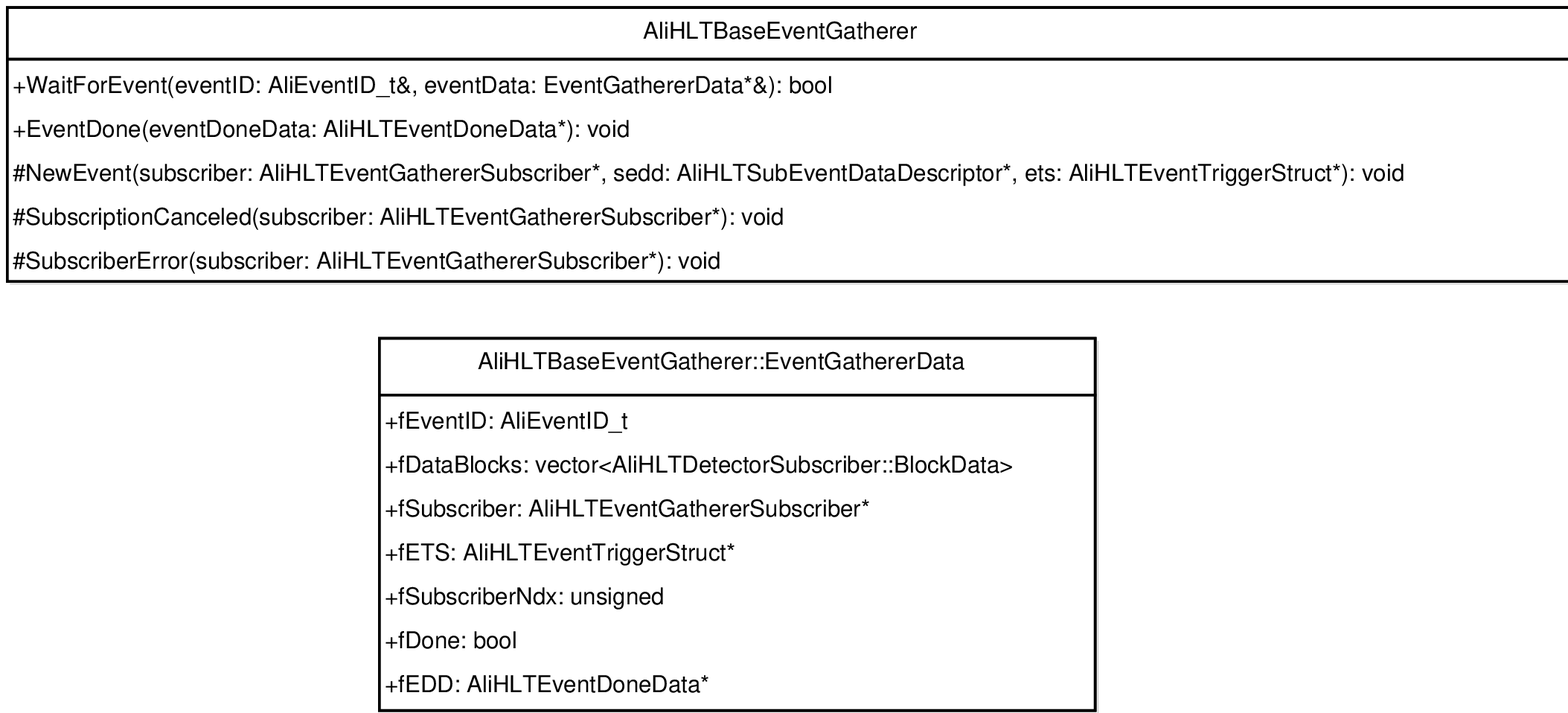}
}
\parbox{0.90\columnwidth}{
\caption[\texttt{Ali\-HLT\-Base\-Event\-Gath\-erer} functions and \texttt{Event\-Gath\-erer\-Data} structure.]{\label{Fig:AliHLTBaseEventGatherer}The functions in the \texttt{Ali\-HLT\-Base\-Event\-Gath\-erer} class and the data fields in its embedded
\texttt{Event\-Gath\-erer\-Data} structure.}
}
\end{center}
\end{figure}

In the central gatherer object base class \texttt{Ali\-HLT\-Base\-Event\-Gath\-erer} an interface is defined consisting of one structure data type 
and five abstract methods, both shown in Fig.~\ref{Fig:AliHLTBaseEventGatherer}. 
The \texttt{Event\-Gath\-erer\-Data} type is used to store the data required to associate each event
correctly with its originating subscriber. Primarily, this includes the event's ID and the index number of and pointer to the originating subscriber
object. Also available are an event's trigger data as well as any event done data structures received for the event. These last two elements, however, are not used
in the standard gatherer component. Finally, descriptors for the event's datablocks are stored as well to construct a new subevent descriptor from them.
This descriptor is used for the event's announcement by the republisher object. Constructing a new subevent descriptor is necessary, as announcing runs in a 
separate thread from the receiving thread and the original descriptor may already have been released when the event is announced. 

The abstract method \texttt{Wait\-For\-Event} is intended to be called externally to wait 
for an event to arrive. In the \texttt{Event\-Gath\-erer} component this is done in the program's main thread. One further function, \texttt{Event\-Done}, is called
by the component's output publisher when an event has been released. Two of the remaining three functions, \texttt{New\-Event} and \texttt{Sub\-scrip\-tion\-Can\-celed},
are called by the subscribers configured for the component in response to a stimulus from the publisher they are subscribed to. The stimuli are either the 
arrival of a new event or respectively the cancellation of their subscription. The last function, \texttt{Sub\-scrib\-er\-Error}, is called in response to an error
that occurs in one of the specified subscribers, e.g. when attempting to send an event done notification back.

\subsubsection{The Event Gatherer Class}

In the class \texttt{Ali\-HLT\-Event\-Gath\-erer}, derived from the class \texttt{Ali\-HLT\-Base\-Event\-Gath\-erer}, 
the two central data structures are a list of configured subscribers and
a list of data structures of the base gatherer's \texttt{Event\-Gath\-erer\-Data} structures. The second list is used to store information about each event which has been
received by one of the subscriber objects and announced through the republisher object, but is not yet released. Events are added to this list
in the class's implementation of the \texttt{New\-Event} function. In this function a pointer to the event data structure in this list is added as notification
data to a signal object before it is triggered. A wait for this signal to be triggered is entered in \texttt{Wait\-For\-Event}. Upon return from the wait the first
available event structure in the notification data is returned to the function's caller. In the \texttt{Event\-Gath\-erer} component this caller is the function's main thread, 
which uses this data to announce the event. When the provided implementation of the \texttt{Event\-Done} function is called by the republisher
to signal a released event, the list is searched for the event concerned. If the event is found, its structure is removed from the list, and an event done message is sent
to the subscriber from which the event has been received. 
A subscriber is removed from the object's list of subscribers if its subscription is cancelled
through the \texttt{Sub\-scrip\-tion\-Can\-celed} function. 
Each event that has been received through that
subscriber subsequently has to be cancelled in all subscribers attached to the republisher as well. The final of the five abstract methods defined in the 
\texttt{Ali\-HLT\-Base\-Event\-Gath\-erer} class, \texttt{Sub\-scrib\-er\-Error}, is only implemented as an empty function body with no functionality. Subscriber error handling is not
supported by this class and thus neither by the component.

%{\bf ?Forward reference to tolerant gatherer component?}

\subsection{\label{Sec:BridgeComponents}Bridge Components}

All components in the framework rely on the publisher-subscriber interface for communication between components. Due to the used mechanisms of named pipes and 
shared memory any communication in the framework is restricted to be local on one node. To lift this restriction and enable inter-node communication
and data-exchange of components a set of two specialized bridging components has been developed. In the first, the 
\texttt{Sub\-scrib\-er\-Bridge\-Head}, data is accepted from a producer component and sent via a network to its partner component, the 
\texttt{Pub\-lish\-er\-Bridge\-Head}. The \texttt{Pub\-lish\-er\-Bridge\-Head} places the received data in a shared memory segment and announces it via its publisher
object to further components subscribed to it. Fig.~\ref{Fig:BridgeClasses} shows the relation of the different publisher-subscriber and communication 
classes in these two components. A sample  of the calls that occur between the classes in the components is displayed in Fig.~\ref{Fig:BridgeCallSequence}. 
Using the standard subscriber and publisher interface objects for receiving and reannouncing data supports transparent connections of other remote framework 
components without special measures required in any of them.

\begin{figure}[hbt]
\begin{center}
\resizebox*{0.4\columnwidth}{!}{
\includegraphics{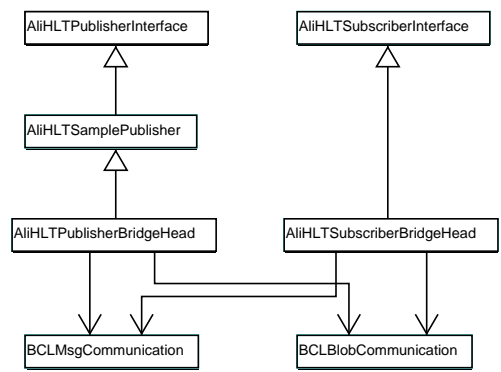}
}
\parbox{0.90\columnwidth}{
\caption{\label{Fig:BridgeClasses}The classes in the bridge components.}
}
\end{center}
\end{figure}

\begin{figure}[hbt]
\begin{center}
\resizebox*{0.9\columnwidth}{!}{
\includegraphics{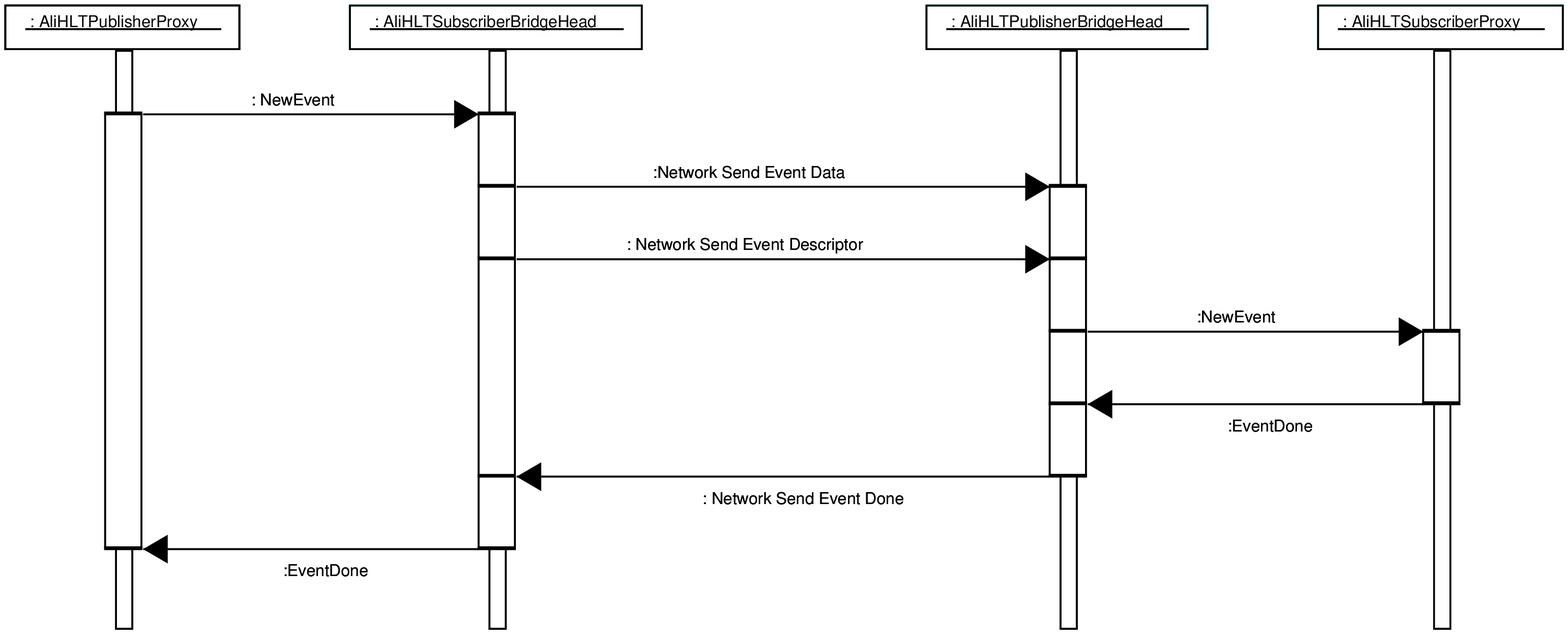}
}
\parbox{0.90\columnwidth}{
\caption[Calling sequence in the bridge components.]{\label{Fig:BridgeCallSequence}Calling sequence in the bridge components. The three network objects used on each side are not shown.}
}
\end{center}
\end{figure}

In the \texttt{Sub\-scrib\-er\-Bridge\-Head} component the major part of the functionality is provided by an instance of the \texttt{Ali\-HLT\-Sub\-scrib\-er\-Bridge\-Head} class 
together with two instances
of classes derived from the \texttt{BCL\-Msg\-Com\-mu\-ni\-ca\-tion} class and one instance of a \texttt{BCL\-Blob\-Com\-mu\-ni\-ca\-tion} derived
class, all three described in section~\ref{Sec:BCLBasicInterfaceClasses}. Of these communication 
classes one message class object is used for the application level communication between the two bridge components, and the second message object is used 
for the required communication between the blob objects in the two components. 
Internally the \texttt{Sub\-scrib\-er\-Bridge\-Head} uses two threads in addition to its main thread and any background threads that may be created internally by the 
different communication classes. In the main thread the message loop responsible for the
publisher-subscriber interface communication is run. The first additional thread is used for the network message loop that accepts and handles
network messages received from the remote \texttt{Pub\-lish\-er\-Bridge\-Head} component. In the third thread the transfer loop for events is run
that receives sub-events from the subscription loop through a signal object, accesses their data, and sends it over the network to the 
\texttt{Pub\-lish\-er\-Bridge\-Head} together with the parts of the sub-events' descriptors necessary to announce the event. 
The class uses the approach of reserving the whole receive blob
buffer and performing buffer management on it locally, as described in section~\ref{Sec:BlobParadigms}. Local buffer management in the \texttt{Sub\-scrib\-er\-Bridge\-Head} is
possible as each \texttt{Pub\-lish\-er\-Bridge\-Head} component
receives its data from only one \texttt{Sub\-scrib\-er\-Bridge\-Head}, which thus can use the receive buffer exclusively. 
This approach has been chosen to minimize the number of messages 
exchanged between the two components and thus reduce the latency time needed to transfer an event.

On the receiving side the primary constituents of the \texttt{Pub\-lish\-er\-Bridge\-Head} component are an instance of the \texttt{Ali\-HLT\-Pub\-lish\-er\-Bridge\-Head} class
together with the same three communication class instances as in the \texttt{Sub\-scrib\-er\-Bridge\-Head}. The purposes of these communication objects are identical
to the ones in the \texttt{Sub\-scrib\-er\-Bridge\-Head}: application level communication, blob message communication, and blob data transfer. Also similar to its sending
counterpart, the \texttt{Pub\-lish\-er\-Bridge\-Head} uses two additional threads beyond the main thread and the threads started internally by its communication objects. 
One of the two additional threads executes the common loop for accepting new subscriptions for the publisher object while the other is the retry thread. This
retry thread is responsible for trying to resend event done messages to the \texttt{Sub\-scrib\-er\-Bridge\-Head} where previous sending attempts
for an event have been unsuccessful. In the program's main thread the message loop to receive and handle network messages from the sending component is
executed, similar to the message loop in the subscriber bridge head.

\subsubsection{The Subscriber Bridge Head Class}

The main data element of the \texttt{Ali\-HLT\-Sub\-scrib\-er\-Bridge\-Head} class is a list of data structures for events that have been received from its publisher object. 
Pointers to the corresponding sub-event's descriptor and trigger structures are stored in each event's structure as well as a pointer to its originating publisher
proxy object. Additionally, the number of retries that have been made to send the event to the publisher bridge head component are stored together with
data about the event's destination location in the receive buffer. This last information is required to release the part of the buffer used by that event, as the receive buffer's
management is performed in the sender component as described above. An event structure's first three elements are the event descriptor,
trigger structure, and publisher interface pointer. They are set when the event has been received from the publisher in the subscriber object's \texttt{New\-Event} method
before it is added to a signal object. This signal object is then triggered subsequently, to inform the transfer loop described below that a new event is available for sending.
Buffer management data for an event is only set when the event's block in the receive buffer has been successfully allocated, which takes place during the attempt to transfer it.
The retry counter is increased every time a send attempt of the event to the remote partner fails. In addition to the event list the \texttt{Ali\-HLT\-Sub\-scrib\-er\-Bridge\-Head}
class stores pointers to the three BCL communication objects used for the network communication with the \texttt{Pub\-lish\-er\-Bridge\-Head} component. A 
pointer to the buffer manager object used for the receive buffer is also contained in the class. 

Next to the functions implemented for the subscriber interface there are two functions that perform the major tasks of the subscriber bridge head class. In the
\texttt{Msg\-Loop} function any messages received from the remote \texttt{Ali\-HLT\-Pub\-lish\-er\-Bridge\-Head} partner object are handled. These are primarily connect and 
disconnect request messages as well as event done messages. Connection messages contain the addresses of the remote program message and blob message communication
objects. If no connection is established, these addresses are extracted from the message and are used to establish a connection to the remote component.
When a connection has been established successfully, the remote blob buffer size is queried, and the whole buffer is reserved as a transfer buffer for the events. The 
buffer size is also used to initialize the buffer manager object correctly. Events already stored in the object's event list are now added again 
to the transfer loop signal object. After these additions the signal object is then triggered to activate the transfer loop. 
For disconnect requests not much action is required except for initiating the 
actual disconnection of the three communication objects. Received event done messages contain the event's ID as well as any non-trivial event done data that has been
received from the publisher bridge head object. This event done data is extracted from the message and is used to send an event done notification
to the publisher that the component is subscribed to. Further actions in response to a received event done message include the cleanup of all object internal
data related to that event, especially releasing the block occupied by the event in the buffer manager object. 

The second main function of the \texttt{Ali\-HLT\-Sub\-scrib\-er\-Bridge\-Head} class is the \texttt{Trans\-fer\-Loop} function, 
that is responsible for the transfer and announcement of 
an event and its data to the remote \texttt{Ali\-HLT\-Pub\-lish\-er\-Bridge\-Head} object. In this function a wait is entered on a signal object triggered when new events
are available for transfer, as described above. Available events are extracted from the signal object's notification queue for processing. 
For each event a first check is performed 
whether a connection to the remote publisher bridge head is established, otherwise an attempt is made to establish one. 
If that connection attempt fails as well, the event transfer attempt is aborted. 
When a transfer is aborted the event concerned is entered into the object's event retry list for a later send attempt. 
As soon as a connection is available, a block for
the event data is allocated in the buffer manager. The event data is then transferred into this block in the remote receive buffer by the blob communication object's 
multi-block transfer function described in section~\ref{Sec:BCLBlobCommunication}. After the successful transfer of the data an event descriptor message is constructed from
the event's original descriptor and the buffer manager data. This message is then sent to the remote component to announce the event. If the event has been cancelled
by its originating publisher before the send process is complete, a special abort message is sent as the validity of the transferred event data cannot be 
assurred. Otherwise the announce message is sent normally and the event is kept in the list until the event done message for it is received from the publisher bridge head. 

%{\bf Too detailed description above??, Connect/Disconnect functions}

\subsubsection{\label{Sec:AliHLTPublisherBridgeHead2}The Publisher Bridge Head Class}

In the \texttt{Ali\-HLT\-Pub\-lish\-er\-Bridge\-Head} class the two main data members are the list of events that
have been received over the network from the subscriber bridge head and a retry list of released events for which the
sending of the event done message to the remote \texttt{Ali\-HLT\-Sub\-scrib\-er\-Bridge\-Head} object has failed.
For each received event the sub-event data descriptor and the event trigger structure received from
the sender component are stored in the event list. Each event's done data obtained
from attached subscribers is stored in the retry list. This data is sent in each
attempt to the subscriber bridge head. In a retry loop failed event done data structures are attempted to be sent again when a retry timeout 
has expired. 
In addition to these two main data lists each \texttt{Ali\-HLT\-Pub\-lish\-er\-Bridge\-Head}
object also stores pointers to the three communication objects used. 

Three of the functions from the callback interface provided for derived classes by the \texttt{Ali\-HLT\-Sample\-Pub\-lish\-er} class are implemented
in the publisher bridge head class: \texttt{Can\-celed\-Event}, \texttt{An\-nounced\-Event}, and \texttt{Get\-An\-nounced\-Event\-Data}.
Of these three functions \texttt{An\-nounced\-Event} has a notification purpose only without actual functionality. \texttt{Get\-An\-nounced\-Event\-Data}'s
purpose is to obtain an event's stored data descriptor and trigger structure for the reannouncement of events. \texttt{Can\-celed\-Event} initiates the sending
of released events' done data to the \texttt{Ali\-HLT\-Sub\-scrib\-er\-Bridge\-Head}. 

Besides these three callback functions one further function, \texttt{Msg\-Loop}, contains the main functionality of 
this class. Similar to the subscriber class from the previous section, this function's purpose is to receive network messages
from its remote counterpart. The most important messages handled in this function are connection and disconnection requests as well as new event 
announcement messages. Connection request messages are
% partially
handled somewhat in the same way as in the subscriber bridge head class. The address of the remote
partner is extracted from the message, and then a connection to this component is established if it is not existing already. 
No send attempt of event done data accumulated before the connection is made, these attempts are only triggered by their
respective timeouts, unlike for the subscriber bridge head's event announcement sends.
For disconnect
requests the connection to the partner is simply aborted. \texttt{New\-Event} messages are the most complicated messages handled in the function.
An event's trigger structure and descriptor data are extracted from the message. The event trigger structure is subsequently used unchanged but 
the event descriptor is modified to use the correct shared memory segment ID, since this is not available in the sending component. When the correct
data structures are assembled, they are added to the event data list and following this the event is announced by the component's publisher to
its subscribers. 

%{\bf Too detailed description above??, Connect/Disconnect functions}

\subsection{\label{Sec:TriggerFilterComponent}Trigger Filter Component}

One further functionality that has to be executed by a component
%, at least for the framework's use in the ALICE High Level Trigger, 
is
the triggered filtering of events. This means for the \texttt{Trig\-gered\-Fil\-ter} component that it has to receive events from a publisher and store
them until a trigger decision for each is received. Based upon this trigger decision it determines which blocks of an event to forward and
announces these blocks via its own publisher object to further subscribers. The mechanism by which the trigger data is received is the one
provided by the \texttt{Set\-Event\-Done\-Data\-Send} and \texttt{Event\-Done\-Data} functions, defined in the \texttt{Ali\-HLT\-Pub\-lish\-er\-Inter\-face} and
\texttt{Ali\-HLT\-Sub\-scrib\-er\-Inter\-face} classes respectively. Trigger decisions are arrays of structures of the 
\texttt{Ali\-HLT\-Trig\-ger\-De\-cision\-Block} type described below. 

Components that make the trigger decision for a particular event encapsulate vectors of these \texttt{Ali\-HLT\-Trig\-ger\-De\-cision\-Block} 
data structures into \texttt{Ali\-HLT\-Event\-Done\-Data} structures. These structures are then transported back along the 
path that the event has been announced on. Components like the \texttt{Trig\-gered\-Fil\-ter} which have requested this
will receive event done data originating from a publisher's other subscribers. Each of the blocks in the
trigger decision is then compared to an event's data descriptor to determine which blocks are to be forwarded. Based upon this result a new 
event descriptor is constructed from the original one, and the event is announced to the filter's subscribers. Fig.~\ref{Fig:TriggerFilterSequence}
shows a schematic sequence of events in the trigger filter component, a description of the classes follows below. For events where no block
is selected through the received trigger decision two kinds of behaviour can be configured via command line options: 
Either the event concerned is not announced by the filter component at all or it can be announced as an empty event without any data blocks. 

\begin{figure}[hbt]
\begin{center}
\resizebox*{0.9\columnwidth}{!}{
\includegraphics{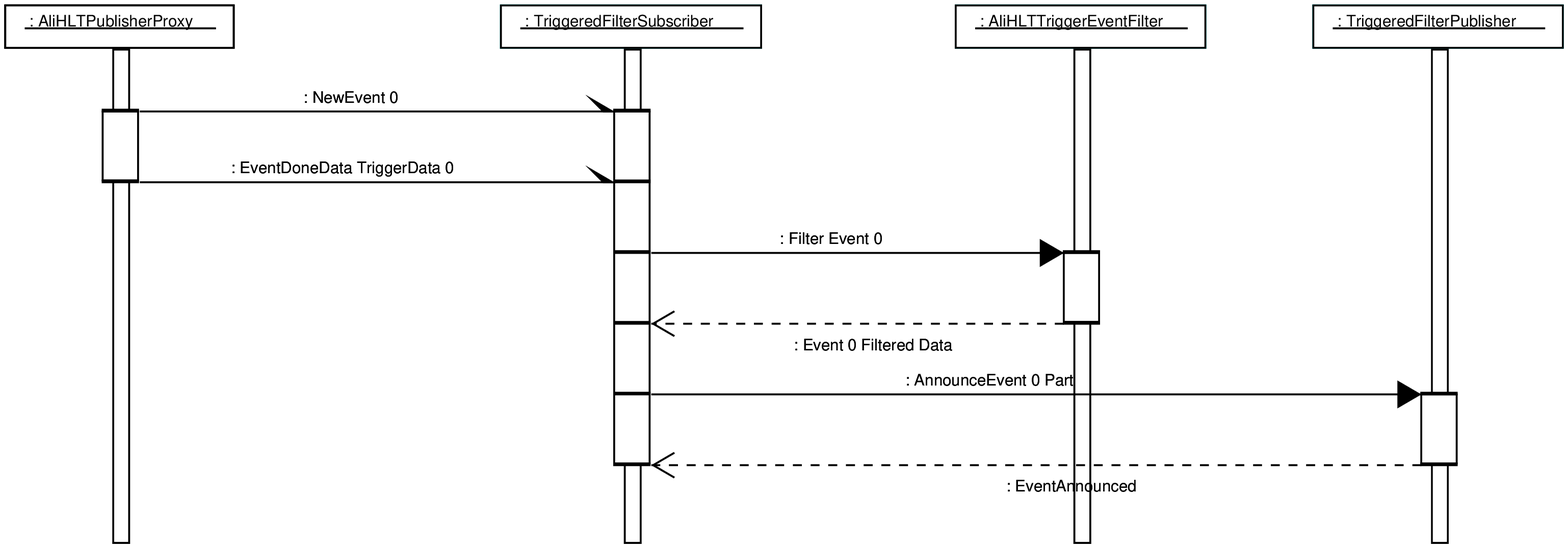}
}
\parbox{0.90\columnwidth}{
\caption{\label{Fig:TriggerFilterSequence}Sequence diagram for a \texttt{Trig\-gered\-Fil\-ter} component.}
}
\end{center}
\end{figure}

Internally the \texttt{Trig\-gered\-Fil\-ter} component consists of three main objects: a \texttt{Trig\-gered\-Fil\-ter\-Sub\-scrib\-er} object,
a \texttt{Trig\-gered\-Fil\-ter\-Pub\-lish\-er} object, and an \texttt{Ali\-HLT\-Trig\-ger\-Event\-Fil\-ter} object. Its main logic is contained
in the subscriber object which makes use of the event filter object for evaluating each event's trigger data. The publisher object
does not contain much functionality beyond the one provided by its \texttt{Ali\-HLT\-Sample\-Pub\-lish\-er} base class. It starts
a thread that contains the standard subscription loop and implements two of its base class's callback functions, \texttt{Can\-celed\-Event}
and \texttt{Get\-An\-nounced\-Event\-Data}. Calls to both functions are only forwarded to corresponding functions in the subscriber object. No 
threads apart from the mentioned subscription thread and those started internally by the \texttt{Ali\-HLT\-Sample\-Pub\-lish\-er} class
are started in this component. Fig.~\ref{Fig:TriggerFilterClasses} shows the relation of the different classes in the component. 

\begin{figure}[hbt]
\begin{center}
\resizebox*{0.40\columnwidth}{!}{
\includegraphics{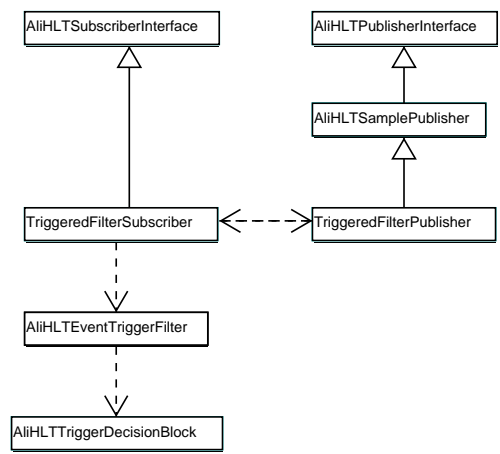}
}
\parbox{0.90\columnwidth}{
\caption{\label{Fig:TriggerFilterClasses}The classes in the trigger filter component.}
}
\end{center}
\end{figure}

\subsubsection{The Trigger Filter Subscriber Class}

The \texttt{Trig\-gered\-Fil\-ter\-Sub\-scrib\-er} class contains the component's main functionality. Its main data
parts are two lists. One of them holds events that have been announced to the component, but for which no trigger decision has been sent so far and which thus have not been
announced yet. The other contains the descriptors and trigger structures of events which have already been announced by the component's own publisher
and which have not been released yet. Both lists contain structures of the same type, storing an event's descriptor and trigger data as well as a pointer to
its originating publisher proxy. 

The class's functionality is contained primarily in the implementations of the \texttt{New\-Event} and \texttt{Event\-Done\-Data} subscriber interface
functions. Supplementary functionality is contained in the subscriber interface function \texttt{Event\-Can\-celed} as well as in the
\texttt{Can\-celed\-Event} function called by the component's publisher object. In the \texttt{Event\-Can\-celed} function the respective cancelled event
is searched in the two event lists and is removed if found. If the event has already been announced through the component's 
own publisher, it is aborted in the component itself, and the event cancelled message is forwarded to its subscribers as well. 
In the \texttt{Can\-celed\-Event} function the event is also searched
in the lists. If it is found, the event done data that has been received by the \texttt{Trig\-gered\-Fil\-ter\-Pub\-lish\-er} object is used in 
the \texttt{Event\-Done} call to the event's originating publisher. 

An event is added to the list of received events in the \texttt{New\-Event} function by placing copies of its descriptor and trigger data
 into the list. No further event processing or announcing is performed in this function as this only happens upon receipt of event done data from the 
event's publisher in the \texttt{Event\-Done\-Data} function. When event done data is received, the trigger decision for the event concerned
 is extracted from it and the event
is searched in the list of received events. The trigger decision data and the event descriptor are then passed to the
\texttt{Ali\-HLT\-Trig\-ger\-Event\-Fil\-ter} object to appropriately filter the event's descriptor. Depending on the results and the current setting the resulting descriptor 
is then used to announce the event through the component's publisher to further subscribers. If the event is not announced any further, an event done message is sent
to its originating publisher to release the event.

\subsubsection{The Trigger Decision Block}

Three data elements are contained in the \texttt{Ali\-HLT\-Trig\-ger\-De\-cision\-Block} structure that specify which data blocks of an event are to be read out: the block's
data type, its data origin and its data specification. These three fields directly correspond to the three fields of the same name and function in the 
\texttt{Ali\-HLT\-Sub\-Event\-Data\-Block\-De\-scrip\-tor} described in section~\ref{Sec:AliHLTSubEventDataBlockDescriptor} and are of the same respective type. 

\subsubsection{The Trigger Event Filter Class}

In the \texttt{Ali\-HLT\-Trig\-ger\-Event\-Fil\-ter} class the main functionality is contained in the \texttt{Fil\-ter\-Event\-De\-scrip\-tor} function. This function accepts
an event's data descriptor and a list of trigger decision blocks as its parameters. The trigger decision blocks are used to filter the data blocks from the event
descriptor to be forwarded according to the trigger decision. Upon return from this function the event descriptor only contains those blocks
that have not been filtered out so that it can be used directly to announce these events. 

Matching of an event's data blocks with the information in the trigger decision blocks is performed differently for the data type and origin and for the 
data specification field. For the type and origin a match is made if one of three conditions is met: The corresponding fields in the data block and
the decision block are identical or one of the two fields contains the wildcard pattern of all 64~bit respectively 32~bit set. 
For the event data specification field matching modes are differentiated in the class by specifying a matching function in the
filter object. 
Two predefined functions for this purpose are provided in the library. More matching modes are also possible by specifying user-defined matching functions 
instead of these predefined ones. In the first and
simpler of the existing matching modes a match is found when the specification values from the descriptor and decision blocks are identical. This is similar to the
matching for the data type and origin fields, although without the possibility for wildcards. 

The second data specification matching mode is more complex and specific to the framework's use in the ALICE High Level Trigger. It 
currently exists only as a first draft version and is still subject to modification. In this mode the
data specification field is used to indicate an event data's origin in the detector given in the data origin field. For data originating from
the ALICE TPC the data specification contains the minimum and maximum numbers of the slice and patch specifiers as defined in section~\ref{Sec:ALICEDAQ}. 
If a data block's specification overlaps with a decision block's in both slice and patch numbers, then the block is marked for readout. All four fields
(mininum and maximum slice and patch) in a trigger decision block are allowed to take the value of all 8~bit set, which corresponds to a wildcard for that
number. 
Data originating from ALICE's DiMuon arm contains the numbers of the DDLs used for readout of the data. A trigger decision
block contains the minimum and maximum number of the DDLs to be read out for an event. For both the minimum and maximum DDL number for readout
in the decision block 8 set bits again corresponds to a wildcard value for the number in that decision block.

\section{\label{Sec:UserComponentTemplates}Application Component Templates}

To ease the programming of worker components for tasks other than those currently provided, three templates have been
included in the framework. In general, application components can be divided into three types according to their position in a chain:
\begin{itemize}
\item Data source components that obtain data from a source outside of the chain and make it available via a publisher object to other 
framework components. They are located at the beginning of a chain
\item Data processing components that receive data via a subscriber object, process it to produce some new output data, and make
the new produced data available again from a publisher object. They are located in the middle of a chain.
\item Data sink components that receive data using a subscriber object and then either process the data and/or forward it to some destination outside
of a framework chain. They are located at the end of a chain.
\end{itemize}
Fig.~\ref{Fig:ComponentTemplates} shows the principle of the three application component types with their respective 
position in a chain.
For each of these three types one template is present, written and commented to be adapted easily to a particular task at the 
intended position in a data processing chain. The following descriptions of the templates also contain instructions on how to proceed in
adapting the templates to their intended tasks. 

\begin{figure}[hbt]
\begin{center}
\resizebox*{0.40\columnwidth}{!}{
\includegraphics{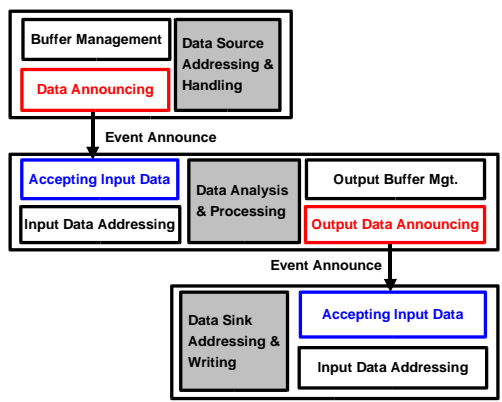}
}
\parbox{0.90\columnwidth}{
\caption[The three application component template types.]{\label{Fig:ComponentTemplates}The three application component template types
at their positions in the chain. The gray boxes indicate the parts where user or application specific functionality has to be inserted.}
}
\end{center}
\end{figure}

\subsection{Data Source Template}

Of the three component templates the data source component is the most complex one to implement due to the largest number of requirements and potential uses.
The template is mainly intended for implementations that access a specialized readout hardware, e.g. in the form of a PCI card.
Its main constituent part is an instance of a class derived from the \texttt{Ali\-HLT\-De\-tec\-tor\-Pub\-lish\-er} class described in~\ref{Sec:AliHLTDetectorPublisher}.
All six virtual functions defined in the detector publisher class are implemented by the template
publisher class, although the functionality provided by the \texttt{Wait\-For\-Event} method generates random data for publishing only. 
Functionality in the class's other methods can be used as provided for software-only components that do not have to access hardware.
An exception here are constants, like the block size for an event, which most likely differ for real tasks. 

For data sources that have to access and communicate with special hardware devices more code will have to be added to the six
detector publisher interface methods. In the \texttt{Wait\-For\-Event} and \texttt{Event\-Fin\-ished} methods the functionality of the buffer manager  
object has to be replaced, if this task is performed already by the hardware. In this case a block in the output shared memory 
will not have to be allocated using the buffer manager in \texttt{Wait\-For\-Event}. Instead the location of the data will have to be read out from the hardware.
Similarly, in \texttt{Event\-Fin\-ished} the block will not have to be released in the buffer manager but the hardware has to be informed that it
can now reuse the occupied memory. In \texttt{Start\-Event\-Loop} code has to be inserted
to initialize the hardware device, while in \texttt{End\-Event\-Loop} the device has to be deactivated. Finally, in \texttt{Quit\-Event\-Loop}
an interface between the hardware and the component could be required to abort the event loop in \texttt{Wait\-For\-Event}
while it is still waiting for the device to provide information and/or data for a valid event. 
%This is needed to terminate the event loop from the software side. 

%\subsubsection{??Command line Parameters??}

\subsection{Data Processor Template}

In the data processing template two classes are used directly, one derived from \texttt{Ali\-HLT\-Pro\-cessing\-Sub\-scrib\-er} described in
\ref{Sec:AliHLTProcessingSubscriber} and one derived from the \texttt{Ali\-HLT\-Pro\-cessing\-Compo\-nent} class described in more detail below.
Only two functions have to be implemented in the \texttt{Ali\-HLT\-Pro\-cessing\-Sub\-scrib\-er} derived class to be able to use the class in the template. 
The first of these functions is the class's constructor, which has to supply required parameters to the base class's
constructor. Additional parameters that have to be passed to the new derived class can be added to those parameters as well. 
\texttt{Pro\-cess\-Event} is the second function that has to be implemented, defined as an abstract function in the \texttt{Ali\-HLT\-Pro\-cessing\-Sub\-scrib\-er}
class. It is called by the parent class when a new event is available for processing by the object. Input parameters to this function include structures 
containing the event's data block descriptor with dereferenced pointers, the event's trigger data, and a pointer to a preallocated output shared memory block
as well as its size. Two primary output parameters of the function are a list of created output data blocks as well as a pointer to an event done data structure 
used in the \texttt{Event\-Done} message to the event's originating publisher. In the function the input data blocks can be immediately accessed and processed. Output
data can be placed directly into the provided output shared memory block. 

For the class derived from \texttt{Ali\-HLT\-Pro\-cessing\-Compo\-nent} two cases have to be distinguished, whether or not the processing subscriber class requires
a set of parameters for its constructor different from the one for \texttt{Ali\-HLT\-Pro\-cessing\-Sub\-scrib\-er}'s constructor. If the 
constructor parameters are identical for the two classes, a template class derived from \texttt{Ali\-HLT\-Pro\-cessing\-Compo\-nent} can be used with the subscriber
class's name as the template parameter. This class contains an implementation of the abstract subscriber creation method described below, 
that supplies the default parameters to the
constructor. To supply additional parameters required by the subscriber constructor a custom class has to be derived from \texttt{Ali\-HLT\-Pro\-cessing\-Compo\-nent}
that implements the abstract subscriber creation function with the necessary parameters. Both of these approaches are present in the 
sample data processor component, a \texttt{\#de\-fine} statement selects one of them. 

\subsubsection{The Processing Component Class}

\texttt{Ali\-HLT\-Pro\-cessing\-Compo\-nent} is a complex class that encapsulates almost all functionality needed to set up a processing component.
It parses the program's command line parameters to extract necessary arguments and optional specifiers. Based upon these it creates all required
objects and initializes them. Among the objects being created are cache classes for frequently needed data types, a buffer manager object, a republisher
object, and objects for accessing shared memory. Creation of the subscriber class required for processing is not directly contained in the component class.
Instead an abstract function, \texttt{Cre\-ateSub\-scrib\-er}, is defined and called with the purpose of creating and returning a new subscriber object. 
This object must be of a class derived from \texttt{Ali\-HLT\-Pro\-cessing\-Sub\-scrib\-er} to supply all functionality assumed by the component class. 
Also all necessary threads for the operation of a processing component are started so that amongst others the publisher's subscription loop, the subscriber's message handler 
loop, and a processing thread can operate without any further actions. 

To make use of the functionalities of this class, a derived class has to be defined that implements the abstract \texttt{Cre\-ateSub\-scrib\-er} function. For processing subscriber
classes whose constructors do not require any special arguments the \texttt{Ali\-HLT\-De\-fault\-Pro\-cessing\-Compo\-nent} class can be used. This template class implements a subscriber 
creation function using the template parameter as the type of class to create with the processing subscriber default parameters. With a suitable derived processing 
component class available an
object of that class has to be created with its required arguments in the component's \texttt{main} function, and %then 
the class's \texttt{Run} method has to be called to
activate the processing component and start the processing of data. 

%\subsubsection{??Command line Parameters??}

\subsection{Data Sink Template}

The data sink template is very similar to the data processing component and uses the same two primary objects of classes derived from
\texttt{Ali\-HLT\-Pro\-cessing\-Sub\-scrib\-er} and \texttt{Ali\-HLT\-Pro\-cessing\-Compo\-nent}. New events arriving are also handed to the user code in the
\texttt{Pro\-cess\-Event} function that has to be implemented in the subscriber class. The difference between the two component types is 
attained by calling the \texttt{No\-Re\-Pub\-lish\-er} function of the \texttt{Ali\-HLT\-Pro\-cessing\-Compo\-nent} derived class. This function specifies to
the component class object that no republisher object is to be created, inhibiting the publishing of any produced data. Mostly,
however, this component will not produce additional data but only perform a specific task with its received input data, e.g. writing to a file. 

%\subsubsection{??Command line Parameters??}

\section{\label{Sec:WorkerComponents}Generic Worker Components}

In the following section a number of worker components are described not dedicated to a specific task of the framework.
%purpose. 
Most of them are intended to be used in debugging new components or chain setups, although they can also be used in small chains with
limited functionality.

\subsection{Random Trigger Decision Component}

To aid in debugging the \texttt{Trig\-ger\-Fil\-ter} component's functionality, discussed in section~\ref{Sec:TriggerFilterComponent}, the
\texttt{Ali\-Ran\-dom\-Trig\-ger\-De\-cision\-Unit} component was created. It is a data sink component that does not process in any way 
the input data it receives. Instead it generates a random trigger decision consisting of multiple trigger decision blocks for each 
event. The generated trigger decision blocks are used as the event done data payload when the event is released. 
%In the current version, the random trigger 
%generator is specific to the ALICE High Level Trigger functionality and makes use of the TPC characteristics, although it can be extended
%for other trigger types as well.  

For each of the generated trigger blocks one of the available block types is chosen at random. Seven trigger block types
are available:
\begin{itemize}
\item Empty or untriggered events
\item Completely triggered events
\item A specified TPC slice region, defined by a minimum and maximum slice number
\item A specified TPC patch region, defined by a minimum and maximum patch number
\item A specific type of data
\item A specific type of data in a certain TPC slice region
\item A specific type of data in a certain TPC patch region
\end{itemize}
If a block with one of the first trigger types is selected, no other decision block is allowed for the event concerned.
The available datatypes as well as the valid slice and patch numbers are specified to the component via command line parameters.
These parameters also allow to specify the trigger types to be used as well as a statistical weight for each of them.

\subsection{Block Comparer Component}

Testing the functionality of different paths in an event chain is the purpose of the \texttt{Block\-Com\-parer} component. This component will 
compare the data of all blocks in an event it receives and will provide a detailed report of the differences found. Its most 
simple and also most important application is to attach it to an event merger component with one input subscriber attached 
directly to an event's originating publisher and the other to a publisher that publishes the same data after it has passed through a more complex 
chain setup of multiple components. If the data has been incorrectly transferred at one point of this chain, then the block comparer component
will detect and report this error. Fig.~\ref{Fig:BlockComparerPrinciple} shows a sample setup of the principle. A publisher component announces data to
a merger and a subscriber bridge head. From the subscriber bridge head the event data is sent via two publisher bridge heads and one subscriber bridge head
to the second input subscriber of the merger. The merger announces the received events to the block comparer that compares their
two blocks and thus can detect errors that have occured during the data's transmission. 

\begin{figure}[hbt]
\begin{center}
\resizebox*{0.30\columnwidth}{!}{
\includegraphics{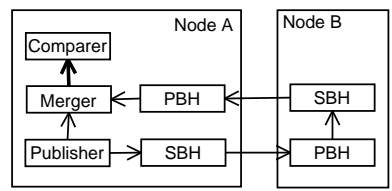}
}
\parbox{0.90\columnwidth}{
\caption[\texttt{Block\-Com\-parer} component sample setup.]{\label{Fig:BlockComparerPrinciple}Sample setup to illustrate the operation 
principle of the \texttt{Block\-Com\-parer} component (SBH: SubscriberBridgeHead, PBH: PublisherBridgeHead).}
}
\end{center}
\end{figure}

\subsection{Event Dropper Component}

By using the \texttt{Event\-Drop\-per} component it is possible to test the behaviour of components and complete chain setups when events are lost 
in the system due to an error and are thus not released. This component is a very simple program with a subscriber object that sends \texttt{Event\-Done} messages
back for most of its received events. Using  a configurable rate, e.g. every hundredth event, the \texttt{Event\-Done} message is not sent at all and the
event is just dropped. For the producing component to which the \texttt{Event\-Drop\-per} is attached this means that the event will never be
released by one of its subscribers and can only be removed when timeouts expire to force its release. 

\subsection{Event Keeper Component}

In a manner similar to the event dropper, the \texttt{Event\-Keeper} component is designed to simulate the behaviour of components that require a specific
amount of time to process events. It contains a subscriber object that starts a timer for each received event. When the timer expires
an event done message is sent back to the event's publisher. A constant amount of time to wait between the receiving of the event and its \texttt{Event\-Done} message 
can be specified via a command line parameter.

\subsection{\label{Sec:EventSupervisor}Event Supervisor Component}

During the testing of components, especially in complex setups, events can become lost due to errors. 
%Detection of these errors is more complicated due to the fact that events might arrive out of order due to the complex nature of chain setups. 
Detection of such lost events is the task of the \texttt{Event\-Super\-visor} component. 
It keeps track of every event received, and when a certain number of events has been received after a missing event a warning is issued.
The number of events before the warning is configurable through command line parameters. If for instance the alarm intervall has been set to 50
and if event 100 is not received, an error will be reported when event 150 is received. Due to the fact that the component requires event IDs 
to be consecutive numbers, it cannot be used in configurations that encode other information in the ID.

\subsection{File Publisher Component}

In order to be able to simulate chains without having any special readout hardware available, a functionality was created to publish data contained in
normal system files into a chain.
The \texttt{Mult\-File\-Pub\-lish\-er} component reads data from multiple files and publishes events using that data. Each event contains data
from one file. Files are alternated in a round-robin fashion. IDs of the events are numbered consecutively, starting with a configurable offset.
The number of events to publish as well as the time interval to wait between the publishing of events can also be specified 
on the command line as well as the three data characteristics:
type, origin, and specification.

In order to make the component more efficient, the data is read from the files directly into shared memory from which the events are published. This avoids
file I/O and/or copy steps into shared memory for each event, reducing the CPU load. The component is mostly used in the simulation and testing of chain configurations
without special readout hardware available.

\subsection{\label{Sec:DummyLoad}Dummy Load Component}

Simulation of chain setups without actual processing components is the purpose of the \texttt{Dummy\-Load} component 
based on the data processing template with \texttt{Ali\-HLT\-Pro\-cessing\-Compo\-nent} and \texttt{Ali\-HLT\-Pro\-cessing\-Sub\-scrib\-er} based classes. It simulates 
a processing component that receives input data and publishes new output data. To simulate different analysis components, a number of parameters in the 
\texttt{Dummy\-Load} can be configured via command line arguments. The most important of these parameters are the size of the output data and the 
simulated processing time for the data. Specification of the output data's size is made as the percentage of the input data's size,  
the value of this can be greater 
than 100~\%, inflating the original data. Processing time can be specified in two ways, either as a constant value or proportional to the size
of the input data. Similar to the file publisher component, it is also possible to specify the output data's three characteristics type, origin, and 
specification. 

Main parts of the component are two classes derived from \texttt{Ali\-HLT\-Pro\-cessing\-Compo\-nent} and \texttt{Ali\-HLT\-Pro\-cessing\-Sub\-scrib\-er}. The processing
subscriber class implements the \texttt{Pro\-cess\-Event} method to copy the necessary amount of input data into the output shared memory and
simulate processing for the specified amount of time. In the processing component class the main task is the evaluation of the additional command line
arguments to extract the parameters that specify the simulation parameters.

\subsection{Data Writer Component}

Data that has been produced by components in a chain may be required to be stored permanently for later access. A very simple method for this is
provided by the \texttt{Data\-Writer} component that creates a file using a configurable name prefix for each block in each event. Files are
enumerated by the event's ID and the block number in the event. For a large number of events this results in a correspondingly large number of files so
that a periodic means of reducing the amount of files, e.g. by creating archives of events, becomes necessary. But for short and/or slow running 
setups this approach is sufficient. 

\subsection{Event Rate Component}

The final generic worker component, the \texttt{Event\-Rate\-Sub\-scrib\-er}, is a very simple data sink component. It receives events and immediately sends event done messages
back to its publisher. After a configurable amount of events has been received, the component calculates the rate of events averaged over this number as well as
the global average rate over all received events and prints these results to the logging system. This component is intended as a simple way to monitor the performance
of a system in the absence of a more complete control and supervision system.

\section{\label{Sec:AnalysisComponents}TPC Analysis Components for the ALICE High Level Trigger}

There exist a number of analysis components ready to be used for the application of the framework in the ALICE High Level Trigger. 
Together these components allow to process run-length encoded ADC values, as read 
out from ALICE's TPC, via space-points and tracklets to complete tracks of the whole TPC. 
After running a properly defined chain with these components the result is a completely reconstructed event of 
the whole TPC with all available particle tracks. The processing components in appropriate order
 are the ADC Unpacker, the ClusterFinder, the VertexFinder, the Tracker, optionally the patch internal track
merger, the patch track merger, and the slice patch merger, all of which are described below. The analysis parts of these components have been written 
by collaborating partners from the University of Bergen, Norway \cite{AndersRT2001Paper}, \cite{PARA02Paper}, \cite{AndersRT2003Paper}, and
have been integrated into the framework in Heidelberg.
%as part of a separate Ph.D. thesis by Anders Strand Vestbo from the University of Bergen, Norway \cite{AndersRT2001Paper}, \cite{PARA02Paper}.

\subsection{The ADC Unpacker}

The initial component in a processing chain for the ALICE TPC data is the \texttt{ADC\-Un\-packer} component. It accepts input data in the form of zero-suppressed
and run-length encoded ADC values as they are read out from one TPC patch.
% {\bf\cite{}}. 
This data is uncompressed by filling in the suppressed zero values
to create the component's output data. During the uncompression process the data is inflated to about 2 to 3 times its previous size. Due to the fact that the
data origin and specification fields in the event data block descriptor structures were not present at the creation of this component the slice and patch
number that can be placed there for TPC data are not yet evaluated. Instead it is necessary to specify them using  command line parameters.
The slice and patch specifiers are placed at the beginning of the output data block together with the values for the minimum and maximum ADC padrows
contained in the data so that the next components also have access to these numbers. 

\subsection{The Cluster Finder}

Following the \texttt{ADC\-Un\-packer} component and processing its output data is the \texttt{Clus\-ter\-Finder} component. Using the unpacked ADC values, it calculates
three-dimensional space coordinates of charge distributions, called clusters, produced in the TPC by the passage of charged particles. % \cite{}. 
For each space point the produced output data contains the three cartesian coordinate values of the distribution's center-of-gravity, the cluster's width, 
and the amount of 
charge contained in it. The array of space points in the output data is preceeded by the originating data's patch, slice numbers, as well as minimum and maximum 
numbers of the padrows read out. Additionally, the number of clusters found in the ADC values is also contained in this preceeding data block.

\subsection{The Vertex Finder}

One of two components that accept cluster data as its input is the \texttt{Ver\-tex\-Finder} component that uses the clusters from a slice's outermost patch 
to provide a first calculation of an event's reaction vertex position along the beampipe. It produces the cartesian coordinates
of the determined vertex together with calculation error information for each coordinate. This is preceeded again by the first four numbers, patch, 
slice, minimum and maximum padrow number, extracted from the cluster data's information block. 

\subsection{The Tracker}

The \texttt{Tracker} component requires one or two input data blocks, cluster data from one patch, optionally together with the vertex data that has been calculated
for the patch's slice by the \texttt{Ver\-tex\-Finder}. When the vertex data is omitted, a central vertex position in the middle of the detector is assumed corresponding
to the coordinates $(0, 0, 0 )$. Tracking is possible without vertex data although the result is more exact when it is available. 
The \texttt{Tracker} uses its input data
to calculate segments of tracks, called tracklets, corresponding to paths of particles throught the TPC detector. Each tracklet is determined from the model of a
helix, the path that charged particles follow in the TPC due to the magnetic field inside. 
The relevant parameters for this track model are stored for each found track. Among them are
the center coordinates and radius of the helix when it is projected as a circle, the initial transverse momentum of the particle, as well as the start
and end-point coordinates of the tracklet. Output track data is again preceeded by the patch, slice, and both padrow numbers as well as the number of tracks
that have been found. 

\subsection{The Patch Internal Track Merger}

The next step after tracking one patch's data, the \texttt{Int\-Track\-Merger}, is an optional component working on tracklet data from the \texttt{Tracker}.
Multiple tracklets are merged into one tracklet, if their parameters are corresponding so that they belong to the same track. The output
data format is identical to its input format as tracklets are read and produced. Due to this it is possible to insert this component transparently after
a tracking component, although it is not mandatory. The following components cannot distinguish whether they work on data that was produced directly by the
\texttt{Tracker} or by the \texttt{Int\-Track\-Merger}. For high track densities the reduction in the number of tracks from one patch can speed up the following merging
steps of multiple patches and slices.

\subsection{The Patch Track Merger}

Merging of the tracks of the six patches belonging to the same slice is the task of the \texttt{Patch\-Merger} component. It requires six blocks of tracklet data 
as its input, each block belonging to one patch. Using the tracklets from these patches the patch merger attempts to merge
them across patch boundaries if they belong to a track with the same parameters. As its output the merger also produces tracklet data using the same track
data structures as they are used for the output data of the tracker and patch internal track merger components. Unlike the previous components the patch merger's
output data is preceeded by only one of the four numbers, the slice number. As the ouput data does not belong to a patch subset but a whole slice, the other three 
numbers are not
needed anymore. In addition to this location specifier the data is also preceeded by the number of tracklets contained in the following data section of the
data block. Optionally, data from less than the slice's full six patches can be merged. The number of patches on which to operate has to be specified as 
a command line argument. Missing patches in this case are assumed to contain no tracklets.

\subsection{The Slice Track Merger}

The last step in the TPC processing chain is the \texttt{Slice\-Merger} component that performs track merging using the tracklet data of multiple 
slices up to the TPC's full number of 36. Tracklets are merged across the boundaries between adjacent slices to finally form full tracks passing 
through the whole detector. The format of the output data is still the same tracklet structure which contains all parameters to describe a
full track as well. Preceeding the track array is just the number of tracks contained in the output data.

\subsection{Future Steps}

Following the final \texttt{Slice\-Merger} processing component the next required component is a trigger decision component. This component needs to
analyse an event's data to make a trigger decision to be passed back along the analysis chain through its event done data structure. The 
approach is similar to the \texttt{Ali\-Ran\-dom\-Trig\-ger\-De\-cision\-Unit} component, although of course with a real analysis part to generate the trigger decision.

\subsection{The Whole Chain}

\begin{figure}[hbt]
\begin{center}
\resizebox*{0.60\columnwidth}{!}{
\includegraphics{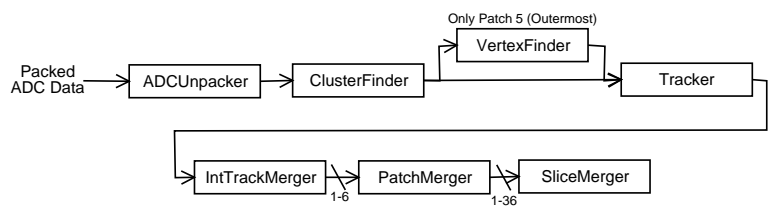}
}
\parbox{0.90\columnwidth}{
\caption[TPC analysis component sequence.]{\label{Fig:TPCAnalysisChain}A sketch of the sequence of analysis components for a TPC analysis chain.}
}
\end{center}
\end{figure}

Fig.~\ref{Fig:TPCAnalysisChain} shows a sketch of the sequence of data through the analysis components described in the preceeding sections.
The vertex finder component only runs on the data of the outermost patch 5, although its output data, the vertex location, is
used by all six trackers for the patches of the same slice.

\section{\label{Sec:FTComponents}Fault Tolerance Components}

\subsection{\label{Sec:FTConceptOverview}Framework Fault Tolerance Concept Overview}

As has been pointed out in the introductory sections, a major challenge of the framework is the behaviour in case of errors, especially with respect to
the intended operation of large clusters like the ALICE HLT. Errors can be single component failures on nodes
or in extreme cases even failures of complete nodes, and both can be software or hardware related. Although 
the fault tolerance (FT) part of the EU DataGrid fabric management software \cite{WP4Web} is intended to be responsible for
the handling of errors concerning the system software and
a node's hardware, the HLT system still has to
 be able to react to node failures. This reaction as well as its triggering should work closely together with the GRID software framework. 

In this section a set of components is presented that allows for such a reaction to the failure of any component in the system, which are handled on 
the granularity of complete nodes. 
On a failure the complete data stream to the node concerned is rerouted to other nodes and if possible a spare node is activated. The model is only
applicable for data distributed by a scatterer to multiple nodes, one or more of which may fail, and is then collected again by a gatherer. In the current
proof-of-concept implementation seven components are required: four data flow components, basically extensions of components described in 
section~\ref{Sec:DataFlowComponents}, one fault detection component, and two components that supervise and orchestrate the system's reaction to the fault condition. 

The four components extended with fault tolerance functionality are the \texttt{Pub\-lish\-er}- and \texttt{Sub\-scrib\-er\-Bridge\-Head} and the 
\texttt{Event\-Scat\-terer}- and -\texttt{Gath\-erer} components,
to form the \texttt{Tol\-e\-rant\-Pub\-lish\-er\-Bridge\-Head}, \texttt{Tol\-e\-rant\-Sub\-scrib\-er\-Bridge\-Head}, \texttt{Tol\-e\-rant\-Event\-Scat\-terer}, and \texttt{Tol\-e\-rant\-Event\-Gath\-erer} 
respectively. For the two bridge head components
the added functionality is primarily the capability to perform remotely triggered connect and disconnect operations from their respective remote partners. The scatterer's
fault tolerant capability is to activate and deactivate output publisher paths, also remotely triggered. Similarly the gatherer is able to activate and deactivate its
input subscribers for event done messages and to handle the case of multiple subscribers receiving the same event, which can happen  if
events are redistributed by a scatterer. In the following discussion of the components' principles, a worker or spare worker node can also
be a group of nodes connected together. One node in this group receives the data from the scatterer and passes it to the next
one for processing, which continues until the last one sends its data to the gatherer. The two nodes connected to the scatterer and gatherer act
as endpoints to the FT components.  

\texttt{Tol\-e\-rance\-De\-tec\-tion\-Sub\-scrib\-er}, the fault detection component, basically consists of a simple subscriber object that receives events and immediately releases
them again. For every event a retriggerable timer is started. The timeout used is configured by the command line. 
When the timer expires, indicating that no event has been received
in that time, the detection component sets its own status accordingly and informs the first of the two supervising components of the status change. 

In this supervision component, the \texttt{Tol\-e\-rance\-Super\-visor}, the status data of multiple fault detection subscriber components is checked 
regularly. When a change in the status of one of the subscribers is detected, the supervisor sends commands 
to the scatterer and gatherer components to deactivate the publishers and subscribers concerned. When an error is removed the
publishers or subscribers can also be activated instead. 
After this a command is sent to the second supervision component, the \texttt{Bridge\-Tol\-e\-rance\-Man\-ag\-er}. 

In response to this message the bridge tolerance manager
searches through its list of active and spare nodes and tries to activate a spare node if one is available. This activation is done by sending disconnect messages to the
two bridge head components in contact with partners on the failed node. Once the disconnect is complete, another command is sent to the two
bridge heads to reset their internal state by removing all event data left over from the severed connection. The reset step is necessary as the 
event data has already been resent 
to other nodes by the scatterer. Finally, a third command is sent to initiate a new connection to their new bridge head partners on the spare node.
As soon as it detects this connection as established in the participating bridge heads, the bridge tolerance manager sends its final commands to the scatterer
and gatherer components to reactivate their output publisher and input subscriber objects for the failed data path. 

In a summary, the sequence of events is as follows:
\begin{enumerate}
\item A node fails.
\item The \texttt{Tol\-e\-rance\-De\-tec\-tion\-Sub\-scrib\-er} on a receiving node detects that no data arrives from the publisher bridge head
connected to this node and sends a message to the \texttt{Tol\-e\-rance\-Super\-visor}.
\item The \texttt{Tol\-e\-rance\-Super\-visor} checks the status of all configured \texttt{Tol\-e\-rance\-De\-tec\-tion\-Sub\-scrib\-er}s and detects
that the path between scatterer and gatherer containing the faulty node is broken. 
\item The \texttt{Tol\-e\-rance\-Super\-visor} sends messages to the \texttt{Tol\-e\-rant\-Event\-Scat\-terer} and -\texttt{Gath\-erer} components
on the sending and receiving nodes to disable the path concerned. A message is also sent to the \texttt{Bridge\-Tol\-e\-rance\-Man\-ag\-er} to inform
it of the failure.
\item The scatterer and gatherer disable the path concerned. The Scatterer distributes all events that have been sent to that path and not received
back among the remaining nodes. Incoming events for the failed path are also distributed among the remaining paths. 
\item The \texttt{Bridge\-Tol\-e\-rance\-Man\-ag\-er} sends messages to the subscriber and publisher bridge heads on the sending and receiving nodes that communicate
with the failed node,  instructing them to disconnect from that node and to reset their internal state.
\item Once the bridge heads are disconnected and reset the bridge tolerance manager determines an available spare node and sends commands to the bridge heads to
connect to that node. (In a more complex real system it would also have to be ensured that the requires processes are available on the spare node. In
this setup the worker and spare nodes are configured identically. )
\item When this new connection is established on both sides the manager sends commands to the scatterer and gatherer components to reactivate the broken path.
\item The status change of the path is detected by the path's tolerance detector and the tolerance supervisor. 
\item The system functions normally as before, although with the number of available spare nodes reduced by one. 
\end{enumerate}

\begin{figure}[hbt]
\begin{center}
\resizebox*{0.50\columnwidth}{!}{
\includegraphics{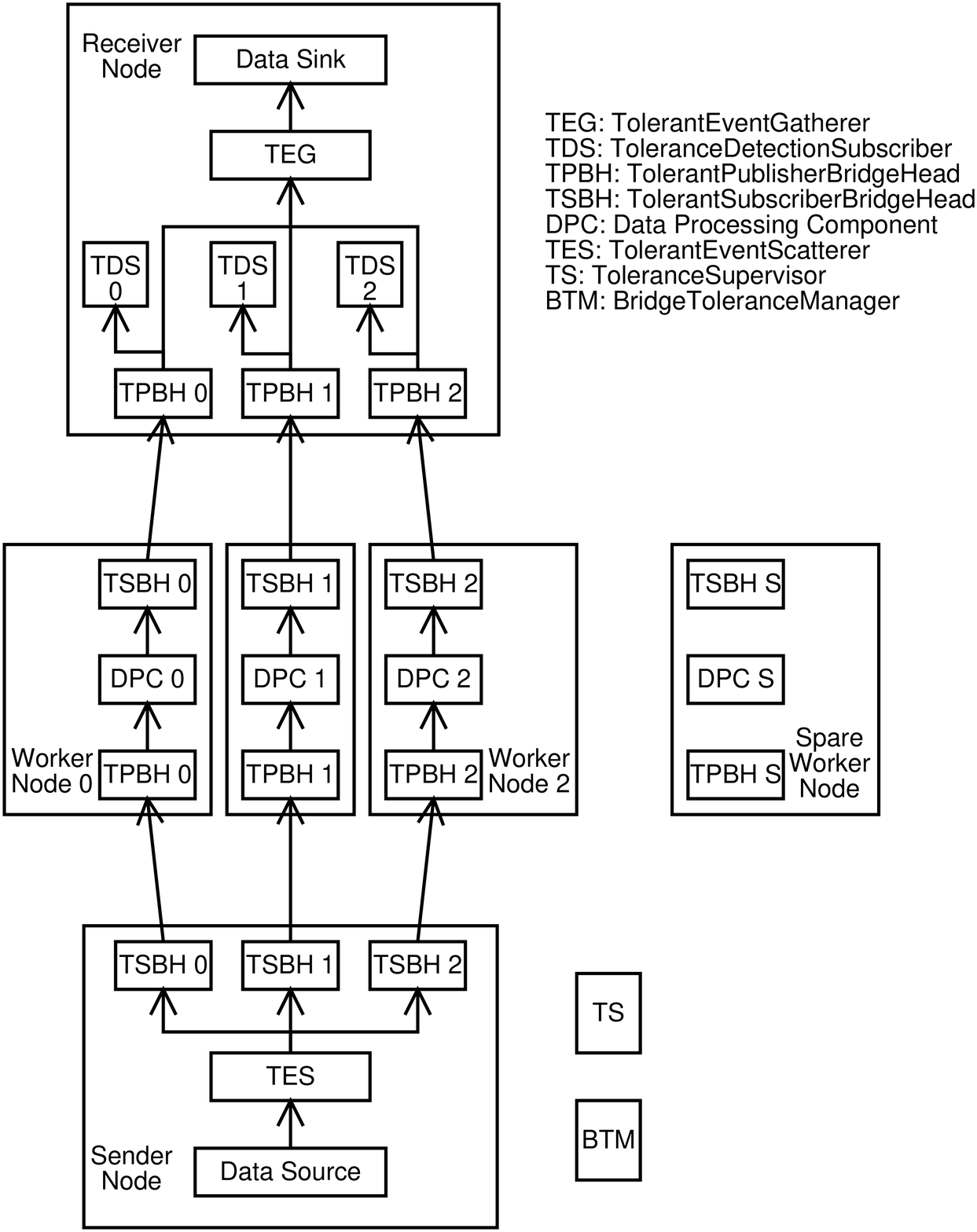}
}
\parbox{0.90\columnwidth}{
\caption[Sample fault tolerance component setup.]{\label{Fig:FaultTolerancePrinciple2}A sample fault tolerance component setup. The arrows show the normal flow of data through the system.}
}
\end{center}
\end{figure}

Fig.~\ref{Fig:FaultTolerancePrinciple2} shows the components in a sample setup using one data source and sink each, three processing or worker nodes, and one spare worker node.
The \texttt{Tol\-e\-rance\-Super\-visor} and \texttt{Bridge\-Tol\-e\-rance\-Man\-ag\-er} components can run on a separate node or on either the sink or source node. 
%{\large \bf Worker/spare Node can mean group of nodes}

%Ich hab dich ganz doll lieb!!!

%BridgeToleranceManager
%ToleranceSupervisor
%ToleranceDetectionSubscriber
%TolerantEventGatherer
%TolerantEventScatterer
%TolerantSubscriberBridgeHead
%TolerantPublisherBridgeHead

For implementations that exceed this prototype a number of extensions to the above concept will be desirable or even necessary, mainly on the 
supervisor level of the concept. At least the two existing supervisor components, \texttt{Tol\-e\-rance\-Super\-visor} and \texttt{Bridge\-Tol\-e\-rance\-Man\-ag\-er}, 
should be merged into one component. To avoid single points of failure in the system this supervisor component
should exist in multiple instances in a system setup, with these instances ideally monitoring each other for failure. 
In addition the granularity of the system should be made finer, so that not only whole nodes can be replaced but also faults in single components
can be recovered, e.g. by terminating and restarting the component and reattaching it to its communication partners. 

%A third possibility %that is also feasible
%is to place the functionality of the two supervisor components into the EU DataGrid fault tolerance system so that it controls the components of a chain.

% Decentralized, merger supervisors, interface with ft

\subsection{\label{Sec:SCClasses}Control and Monitoring Communication Classes}

%Among the central parts 
A central role in the fault tolerance functionality is taken by the classes that enable communication between a supervising and a supervised component.
This is provided by two primary and a number of auxiliary classes, that together allow to send commands to supervised components and to query their status.
An important characteristic supported by the classes is that supervised components are not purely passive but can also send interrupts, called
{\em LookAtMe} or LAM, to supervising components to indicate a special condition. Of the primary classes \texttt{Ali\-HLT\-SC\-Con\-troller} is used in the supervisor
and \texttt{Ali\-HLT\-SC\-Inter\-face} in the controlled component. Fig.~\ref{Fig:SCClasses} shows the six most important classes. 
The underlying mechanism used for communication between components by these classes are the BCL message communication classes. Communication is performed
primarily without explicit connects, although supervisor initiated connections between components are possible as well. 

\begin{figure}[hbt]
\begin{center}
\resizebox*{0.50\columnwidth}{!}{
\includegraphics{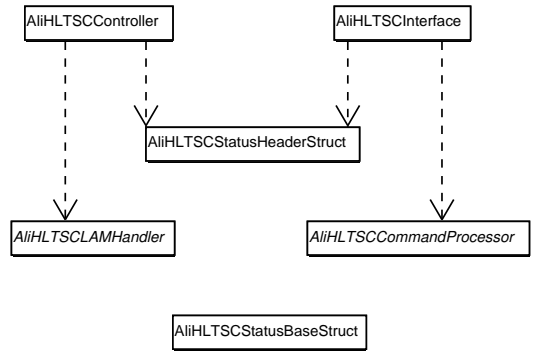}
}
\parbox{0.90\columnwidth}{
\caption{\label{Fig:SCClasses}The six main classes for monitoring and control of components.}
}
\end{center}
\end{figure}

\subsubsection{The Status Structures}

Two datatypes exist that help to define structures used to hold status data for components, \texttt{Ali\-HLT\-SC\-Status\-Header} and \texttt{Ali\-HLT\-SC\-Status\-Base}.
The first of these structures basically is a container
for structures derived from the second. As both are expected to be communicated over the network they make use of the data format
translation mechanism defined in section~\ref{Sec:DataFormatTranslation}. They are consequently derived from the \texttt{BCL\-Net\-work\-Data\-Struct} type. 

\texttt{Ali\-HLT\-SC\-Status\-Header} contains three fields, a 64~bit long ID, the number of actual status structures it contains, and the offset in bytes of the 
first status structure, counted from the beginning of the status header. The ID field holds an identifier specific to the combination of
status structures contained in the header structure. \texttt{Ali\-HLT\-SC\-Status\-Base} itself defines only two fields, another 64~bit long type ID and the offset of the
following status structure in bytes, also counted from the beginning of the status header structure. 

Actual structures containing status data are derived as \texttt{vstdl} types from \texttt{Ali\-HLT\-SC\-Status\-Base} and contain the status information as their fields.
One example is the \texttt{Ali\-HLT\-SC\-Pro\-cess\-Status} structure which defines common status data for all components, like the current state and logging level
of a process as well as the last update time of the status structure. This type is used as the first status structure contained in an \texttt{Ali\-HLT\-SC\-Status\-Header} 
container by most components.

\subsubsection{The Controlled Component Interface Class}

The primary class used in supervised and controlled components is the \texttt{Ali\-HLT\-SC\-Inter\-face} class. It contains functions to provide a status header
structure for readout by supervising components, to attach command processor objects handling received commands, and to send LAM messages to controllers. A number
of commands are defined to be processed by the interface class itself, other commands are forwarded to the registered command processor classes. Internally
the class uses two threads, one as the communication listening thread and the other for command processing. Its main data structures are pointers to the communication
object with its local address, a list of addresses of connected supervisors, a pointer to the registered status header, and the list of registered command handler
objects. 

To use the class in the monitored component it first has to be bound to a listening address, to enable communication, and the two background threads have to be
started using the \texttt{Start} function. 
Binding is done by calling the class's \texttt{Bind} function, which accepts the local address and an error callback object as its parameters. The object's local
address is passed in the form of a string holding an address URL as defined in section~\ref{Sec:BCLAddressURLClasses}. To release the bound address and stop the two threads
the \texttt{Un\-bind} and \texttt{Stop} functions are used. Status data for a component is defined by specifying the address of the status header structure holding the component's
data to the class's \texttt{Set\-Status\-Data} function. Status data can only be unset by passing a \texttt{NULL} pointer to the function. 

For the handling of command processor objects two functions exist in the \texttt{Ali\-HLT\-SC\-Inter\-face} class. \texttt{Add\-Com\-mand\-Pro\-cessor} adds an object to the list
and \texttt{Del\-Com\-mand\-Pro\-ces\-sor} removes it. Both functions accept the pointer to the handler object to be added or removed as the only parameter.  
Command handler objects are described in more detail in the following section about the \texttt{Ali\-HLT\-SC\-Com\-mand\-Pro\-ces\-sor} class. 
For {\em LookAtMe} notifications to supervising objects
two functions exist as well, differing in their parameters and the intended receivers of the notifications. The first \texttt{Look\-At\-Me} function sends the interrupt
message to all supervisors that have established connections to this interface object, while the second version sends it  to that supervisor object
only whose address is specified as the function's parameter.

\subsubsection{The Command Processor Classes}

Command processor objects are instances of classes derived from \texttt{Ali\-HLT\-SC\-Com\-mand\-Pro\-ces\-sor}. This class just defines one abstract function, 
\texttt{Pro\-cess\-Cmd}. This function is called for registered handlers by their interface class instance when commands are received. 
The function receives as its only parameter the command structure that was sent by the controlling instance. Similar to the status structures the command structure
\texttt{Ali\-HLT\-SC\-Com\-mand\-Struct}
also makes use of the data translation mechanism and is thus derived from \texttt{BCL\-Net\-work\-Data\-Struct}. It contains four fields: three 32~bit 
numbers holding the command itself as well as two parameters and a variable length array of 32~bit items. This array is available for 
holding additional required data which does not fit into the two numerical parameters.

\subsubsection{The LookAtMe Handler Class}

Similar to the command processor class, the \texttt{Ali\-HLT\-SC\-LAM\-Han\-dler} class is also an abstract class defining only one abstract function, \texttt{Look\-At\-Me}.
This function is called by the controller class for registered LAM handlers when a LAM request is received from a monitored component. 
%Thus LAM handler objects have to be instances of a class derived from \texttt{Ali\-HLT\-SC\-LAM\-Han\-dler}, similar to the command processor objects. 
The only parameter passed
to the \texttt{Look\-At\-Me} function is the address of the LAM's originating component in the form of a BCL address structure
% of the communication object used by the controlling interface class. 

\subsubsection{The Controller Class}

The \texttt{Ali\-HLT\-SC\-Con\-troller} class is the main class to be used in supervising components, providing functions to register LAM handlers, establish connections
to controlled components, and interact with supervised components. This interaction includes
sending commands to, querying the status of, and setting the logging verbosity of components. Like the interface class described
above the controller class also uses a BCL message communication object for communication with controlled components. One thread is used to  
receive and handle messages
in this communication object. The class's main data structures are a list of received messages that have to be handled, primarily replies from supervised components, 
the list of 
registered LAM handler objects, pointers to the communication object with its local address, and the address of a controlled object to which a connection
has been established. 

LAM handlers can be added or removed from a controller object using the two \texttt{Add\-LAM\-Han\-dler} and \texttt{Del\-LAM\-Han\-dler} functions respectively. Both functions 
require the pointer to the handler object to be added as their single parameter. As for the interface class, to be able to use an instance of this class 
its communication object
first has to be bound to a valid address, and the background listening thread has to be started by calling the class's \texttt{Start} function. Binding is performed
analogous to the interface class by calling the \texttt{Bind} function with a string URL specifying the address and an optional error callback object
as parameters. To release the bound address and stop the thread, the functions \texttt{Un\-bind} and \texttt{Stop} are available. 

Once the controller object is ready, a connection can be established to one supervised component by calling the \texttt{Con\-nect} function with the component's address.
Termination of a connection is achieved with the \texttt{Dis\-con\-nect} function, also requiring the remote address as its parameter. All functions in the class that 
interact with a remote controlled component exist in two versions, one which requires the remote address of the component and one without an address. The second
versions perform the corresponding task with the component to which the connection is established. If no connection is established they fail.

Three function pairs are available that operate on the remote controlled components. The first of these are the \texttt{Set\-Ver\-bosity} functions that allow to set
the logging verbosity as described in~\ref{Sec:LoggingSystem}. As their only parameter, besides the remote address in one of the versions, 
they use the 32~bit large value for the verbosity, corresponding
to the list of set flags for each verbosity level. This flag value is directly assigned to the global verbosity specifier in the remote component and takes effect
immediately after it is received. Sending commands to remote components is the purpose of the second set of functions, called \texttt{Send\-Com\-mand}, 
with the command to be sent as the only
(additional) parameter. It is specified in the form of a pointer to the network transparent \texttt{Ali\-HLT\-SC\-Com\-mand\-Struct} structure 
passed to the command handler objects in the receiving components. 

The final interaction function set consists of the two \texttt{Get\-Status} functions for querying a remote component's status data. They return a pointer reference
to a status header structure containing the data that has been read out from the monitored component. Memory for the structure is allocated 
with the required size in the function when the reply message containing the status data is received. To release the allocated memory the \texttt{Free\-Status} 
function in the class has to be called.

\subsection{\label{Sec:ToleranceDetector}Fault Tolerance Detection Subscriber}

In the described situation a fault will be noticed first by the \texttt{Tol\-e\-rance\-De\-tec\-tion\-Sub\-scrib\-er}. This is a simple data sink component
that notices when no events are received for a specified amount of time and signals an error condition to supervising components. Its internal main
parts are a subscriber object of the \texttt{Tol\-e\-rance\-De\-tec\-tion\-Sub\-scrib\-er} class, a status data structure, and an instance of the \texttt{Ali\-HLT\-SC\-Inter\-face} 
class for communication with supervisor components. No threads are started explicitly by the component besides those started by 
its constituent objects. 

In addition to its header two structure members are present in the component's status data. The first of these 
is the common component status data, with the component type and status update time  of main importance for the detection
subscriber. Following this is an \texttt{Ali\-HLT\-SC\-Tol\-e\-rance\-De\-tec\-tor\-Status} structure containing three fields specific to this component: the 
index of the path between scatterer and gatherer to which the subscriber is attached, the current state of the subscriber, and the ID of 
the last event that was received. A 32~bit unsigned integer is used as the state specifier field, holding either a 0 or a 1 for a faulty or functioning path respectively.
The value contained in the path index field has to be specified to the component on its command line. 

\subsubsection{The Tolerance Detection Subscriber Class}

As the primary class of the fault detection subscriber component the \texttt{Tol\-e\-rance\-De\-tec\-tion\-Sub\-scrib\-er} class is used. It is derived from the subscriber interface
class and implements all its functions. Except for the \texttt{New\-Event} and \texttt{Ping} method all functions are implemented as empty function
bodies only, without any functionality. In the \texttt{Ping} method the calling publisher's \texttt{Ping\-Ack} function is called to acknowledge the received ping.
The most important data structures in the class are the pointer to the component status structure and a list of supervisor addresses to which {\em LookAtME} 
messages are sent when an error is detected.

In the class's \texttt{New\-Event} function the component's status information is updated with the received information, including the timestamp, the
event's ID, and the state of the event path. If the component has not been paused as described below, a timer is set with a timeout value specified
on the component's command line. 
%When this timer expires the class's \texttt{Timer\-Ex\-pired} function is called by the timer object used. 
As the last action 
of the \texttt{New\-Event} function an event done message is sent back to the event's originating publisher. 
When the timer started in the \texttt{New\-Event} function expires, then the class's \texttt{Timer\-Ex\-pired} function is called. In this function the status data 
is updated by setting the last update time to the current time and the path's state to faulty. Following this, a {\em LookAtMe} message is sent
to each configured supervisor component address. Any error occuring during the send is ignored. 

The last function of the
class containing important functionality is the \texttt{Pro\-cess\-Cmd} function called when a command message is received for the component. 
Using these commands it is possible to initiate
a paused mode for the component when no events are expected to arrive, to suppress raising of alarms. This pause state is necessary if the chain is 
still functioning, but the component delivering events to the fault detection subscriber cannot send events.
Reasons for this might be errors in some readout hardware that 
have to be handled in a different manner or configuration changes in parts of the chain before this component.

\subsection{Fault Tolerance Supervisor}

As described above, the \texttt{Tol\-e\-rance\-Super\-visor} component is used as 
the location of the central supervising and decision making for the dataflow in a chain setup. 
When an interrupt is received from a fault tolerance detection component this component checks the state of all attached detection components to determine 
the status of the different data paths. A discovered faulty data path is removed from the active dataflow by sending the appropriate
commands to the components responsible for routing the data. Similarly it is possible to reactivate a path once it has recovered from a fault.

Two primary classes are used in this component, \texttt{Tol\-e\-rance\-Super\-visor} and \texttt{Ali\-HLT\-SC\-Con\-troller}. Apart from any background threads started by the 
controller class and its internal communication classes, no further threads are started by the component. The controller object is used in the supervisor object
to monitor and control the external detection and dataflow components. 

\subsubsection{The Tolerance Supervisor Class}

Inside the \texttt{Tol\-e\-rance\-Super\-visor} class the primary data structures are lists for the detection and dataflow component addresses, the 
current and previous states and the last received event IDs of each detection component. Of these, the lists for the last event IDs, previous and current states have 
to be of the same size. 

The main loop of the supervisor program is the class's \texttt{Super\-vise} function that runs in a loop until a signal is caught to terminate the program.
At the beginning of each loop iteration a wait is entered for the triggering of a signal object with a timeout of the interval between checks of the supervised detection
components. A timeout is used in waiting so that asynchronous loop iterations outside of the fixed intervals are possible by
triggering the signal object. Such a signal trigger is executed by the LAM handler function when an interrupt is received from a supervised component. 
When the signal's wait function returns, the \texttt{Check} function is called to determine the channel state data from the configured detection and dataflow components.
If a state change
in one of the components is detected, pause commands are sent to all detection components, and the path concerned by the change is set to enabled or disabled 
accordingly by calling the class's \texttt{Set} function. After these steps the next loop iteration is started. 

Inside the \texttt{Check} function the status of each supervised component is read out using the supervisor's controller class method \texttt{Get\-Status}. Depending
on whether or not a supervised component is a dataflow component or a tolerance detection subscriber, the channel state data in the status read is evaluated
differently. For detection subscribers the read channel state is accepted to be the current state, while for dataflow components a channel state is
only updated when the read state is faulty. 
%A dataflow component whose channel state does not yet reflect the channel's faulty state thus cannot incorrectly override 
%that faulty state in the supervisor. 
This is necessary as the dataflow components have very little ability to determine a faulty channel state, particularly
when the fault occurs on other nodes. 

The first task in the class's \texttt{Set} function is to send commands to the dataflow components, informing them that a specific monitored channel, or path,
has been reported as faulty and should not be used anymore. After building the appropriate command structure, it is sent to all configured dataflow components. 
In addition the function sends another command to the configured bridge tolerance manager component. This component  also has to be informed of the 
path's failure, to terminate bridge connections to any nodes concerned and if possible activate a spare node replacing the failed one.

\subsection{Bridge Fault Tolerance Manager}

Complementing the fault tolerance supervisor component from the previous section is the \texttt{Bridge\-Tol\-e\-rance\-Man\-ag\-er} component,
that controls the bridge connections for the specific paths. To do so it maintains a list of required connections between
data source,  sink, and worker nodes for each of the paths. Additionally, it maintains a list of spare nodes in the form of available connection endpoints
as well as lists of the supervised dataflow and detection subscriber components. The component contains two primary classes: 
\texttt{Bridge\-Config}, responsible for reading and storing the configuration and providing access to its parameters, and \texttt{Com\-mand\-Han\-dler}, 
mainly responsible for communication with outside components. Outside components include the supervisor as well as the bridge, dataflow, and detection subscriber 
components. No threads are created by the component apart from those created implicitly by its objects, such as the controller and interface
classes described in section~\ref{Sec:SCClasses}. 

In the component's main function the command line options are evaluated first and the necessary objects are created, configured, and activated as 
needed. After this a loop is entered in which the component remains until it receives a signal to terminate. During each loop iteration two different
types of status events with respect to the bridge components are checked. If a path has been deactivated and the bridge connections to its nodes have been terminated,
new connections have to be established to a spare node, if there is one available. When these conditions are met, the bridge head components on the data source
and sink node are checked whether the connections to the broken path have already been interrupted completely. This is necessary to ensure that 
the bridge heads have been able to reset their internal state and remove any old events from their internal lists. Once the connection
termination has completed successfully, commands are sent to the sink and source bridge head components,
containing the commands to connect to the corresponding partner components on the activated spare node(s). 

The second check performed by the main loop is executed prior to the first check described above. When a reconnection attempt has been started
it is necessary to periodically check the bridge components on the source and sink nodes whether the connection has been established successfully.  If 
this the case, then commands are sent to the dataflow components responsible for routing the data to reactivate the path concerned. Once this is done, 
the broken path has been handled, and the system functions as before.

\subsubsection{The Bridge Connection Configuration}

A configuration for the bridge fault tolerance manager is described in a file using six different types of entries, with each entry
consisting of one line:
\begin{enumerate}
\item Data source entries describe connection parameters for a connection from the data source node to a worker node. Each of these 
connections is identified by a unique number corresponding to its path. To account for the fact that multiple data sources may be 
present, each entry also contains a subnumber. Entries with identical major numbers must have different subnumbers. They belong to one path which has
multiple data sources that have to be merged in the path. 
%Data coming from multiple such sources in one path will {\bf (have to)} be merged before it is sent to the data sink node.
The number of data source entries with the same major identifier and different subnumbers must be identical for each of the different source numbers. 

Four parameters are specified for each data source entry: the control address URLs of the subscriber and publisher bridge head components as 
well as the message and blob-message address URLs of the publisher bridge head component. The subscriber bridge head component runs on 
the data source node, while the publisher bridge head runs on a worker node. 
It is not necessary here to specify the message and blob-message address URLs of the subscriber bridge head as they do not change, contrary
to the worker node address which can change due to a node's failure and replacement by a spare node. 

Entry format (On one line): \\
\verb|'source' <number> <subnumber> <subscriber-control-address-URL> \|\\
\verb|         <publisher-control-address-URL> <publisher-msg-address-URL> \|\\
\verb|         <publisher-blobmsg-address-URL>|

\item Data sink entries describe the connection parameters for the opposite chain ends from a worker node to the data sink  node. Each of these connections
is also identified by a unique number which corresponds to the path that feeds the connection. Unlike the data source connections described previously,
multiple data sink connections that belong to the same path are not supported. A one-to-one relation exists between a connection and a path, as it is
assumed that data has been merged before it is sent to the sink node. 

The parameters that have to be specified for a data sink entry are the control address URLs of the subscriber and publisher bridge heads as well
as the message and blob-message address URLs of the subscriber bridge head, in analogy to the data source entry parameters. Here the subscriber bridge
head component runs on the worker node while the publisher bridge head is located on the data sink node. Similar to the source entries' subscriber 
bridge head message and blob-message address, it is not necessary to specify these addresses for the publisher bridge head, since they do not change
either. 

Entry format (On one line): \\
\verb|'sink' <number> <publisher-control-address-URL> \|\\
\verb|       <subscriber-control-address-URL> <subscriber-msg-address-URL> \|\\
\verb|       <subscriber-blobmsg-address-URL>|

\item Spare data source entries contain the worker node parameters required for the connection of a data source node to a specific spare worker node. 
A data source worker node is identified like a normal data source entry by a unique major number in combination with a subnumber. This number is located in the 
same address space as the numbers for the normal source entries and is thus not allowed to conflict with them. 
For each spare source entry
the amount of subnumbers must also be identical to the one specified for the active source entries. Parameters that have to be specifed for a
spare data source entry are the three parameters for the publisher bridge head in the source node to worker node connection: its control,
message, and blob-message address URLs. It is not necessary to specify the control message of the subscriber bridge head on the source node as
this is obtained from an active source entry when a connection is established to the spare node. 
%Since it is not known beforehand which path fails it would even be impossible to know the source entry to use. 

Entry format (On one line): \\
\verb|'sparesource' <number> <subnumber> <publisher-control-address-URL> \|\\
\verb|              <publisher-msg-address-URL> <publisher-blobmsg-address-URL>|

\item Spare data sink entries are the analogue of the spare data source entries for the worker to data sink node connection.
They are identified similarly to the normal data sink entries by a unique number, located in the same address space as the normal 
sink entries. Three parameters for the subscriber bridge head on the worker node have to be specified: the control, message, and blob-message address URLs.
Analogous to the spare source entry the parameters for the publisher bridge head on the sink node do not have to be specified. 

Entry format (On one line): \\
\verb|'sparesink' <number> <subscriber-control-address-URL> \|\\
\verb|            <subscriber-msg-address-URL> <subscriber-blobmsg-address-URL>|

\item Target entries specify control address URLs of command targets to which command messages will be sent when a broken connection has been
reestablished using a spare node. Such a message instructs the targets to reactivate the path that has failed. Typically, these
are tolerant event scatterer and gatherer components controlling the data flow. The only parameter that needs to be specified here is the control
address URL used by the target component concerned. 

Entry format (On one line): \\
\verb|'target' <target-control-address-URL>|

\item Detector entries specify the fault detection subscriber components used for each path. A detection subscriber is identified by the number of the
path it belongs to. The parameter that has to be specified for such a component is the control address URL
used. Using these entries start commands are sent to detection subscribers when a faulty path is reactivated after a failed node has been
replaced by a spare node. Reactivation is required as the detection subscriber has been paused by the fault tolerance supervisor when the fault occured.

Entry format (On one line): \\
\verb|'detector' <nr> <tolerance-detector-control-address-URL>|

\end{enumerate}

%Configuration item list (config file lines)

\subsubsection{The Bridge Configuration Class}

Reading a configuration file, storing the read configuration, and providing easy access to its data is the task of the \texttt{Bridge\-Config} class in the bridge
tolerance manager component. To read a configuration the class's \texttt{Read\-Config} function has to be called with the name of the file in which the 
configuration is stored. Access to the configuration data is provided by a number of member functions that return different parts of the configuration in a 
structured manner. 

\texttt{Get\-Active\-Path} and \texttt{Get\-Path} return data about an active path or one path from the whole set of active and spare paths, respectively. In both cases the path is
selected by the number specified in the configuration file. For active paths only entries specified by the data source and sink entries are searched,
while for the whole set of paths the spare source and sink entries are searched as well. 
The information returned for a path includes the connection data for the data source connections and the data sink connection.
There may be multiple data source connections between the configured number of sources and worker nodes, but only a single 
data sink connection between one worker and data sink node.
%The information returned for a path includes the connection data for the data source
%connections, between the configured number of sources and worker nodes, the data for the single data sink connection between a worker and a data sink node.
Furthermore, the path's absolute and active path numbers and the type of the path are contained in the returned data field as well. A path's absolute
and active number can differ, e.g. for spare nodes that have been activated. The type of a path specifies whether it is active or down or whether it is a spare path.

The class's \texttt{Get\-Tar\-gets} function returns the list of address structures that have been specified in target component entries. 
\texttt{Get\-Tol\-e\-rance\-De\-tec\-tor} provides a structure for the fault detection subscriber component that has been configured for the 
path number specified to the function's call. Included in the returned data is the 
control address URL under which the detection subscriber can be addressed. 

Two functions are provided to set certain parameters of the configuration. The first of these, \texttt{Set\-Path\-Status}, is used to set the state of a specific
path in the stored configuration to either down or up. An active or spare path's state can be set to down, while only a down path's state can be set to up. 
When an active path is set to down, the list of spare paths is searched for an available path that can be used to replace the broken path. If a spare path is found,
the control addresses of each data sources's subscriber bridge head and of the data sink's publisher bridge head are copied into the spare path's data structure and reset 
in the original active path's data. The original path's active number is also copied into the spare path, and the states of the paths are set to down and active
respectively. A spare path which is set to down triggers no further action, while a down path which is available is placed into the list of spare paths. 

When a new spare node has just become available it can be used to reactivate a broken path by the \texttt{Set\-Spare\-Ac\-tive} function called
from the command handler class. 
This function searches for the path with the specified active number in the list of paths requiring replacement. It also
searches the list of spares for an available path to be used as a replacement. If both an active path to be replaced and an available spare are found, the
source subscriber bridge head's and sink publisher bridge head's control addresses are moved from the original path to the spare one. The active number
of the new path is set according to the old active path's one, and the new paths' state is set to active.

\subsubsection{The Bridge Command Handler Class}

The \texttt{Com\-mand\-Han\-dler} class in the component is derived from the \texttt{Ali\-HLT\-SC\-Com\-mand\-Pro\-ces\-sor} class described in section~\ref{Sec:SCClasses}. 
Its \texttt{Pro\-cess\-Cmd} function is called by the controller object in the component when a command is received, its four other functions are
called from the component's main loop. 

In the \texttt{Pro\-cess\-Cmd} function only the command to set a bridge node's state is handled, which specifies a change in the state of a path 
in the system. The command is handled by interfacing with the component's \texttt{Bridge\-Config} instance. When a path is set to up or available an internal list of
paths that need to be replaced is searched to determine whether a failed and unrecovered path exists. If such a path is found, 
it is placed into a list of paths that need to be connected to their source and sink node endpoints,
using the spare path's endpoints. 
The list of paths requiring reconnection can be queried by calling the class's \texttt{Get\-Con\-nects\-Needed} function. This is done in the 
component's main loop, as described above.

For a path whose state is set to down, different steps are taken in the command processing function. 
First the path's parameters are queried from the configuration object. Then three commands are sent to the bridge head components on the source
and sink nodes belonging to the path: A disconnect command to abort the connection to the broken path, preceeded and followed by 
a purge events command to 
clear all events that have remained in these components. Following this, the path's state is set to down, and a spare path to activate is searched. If no 
available spare path is found,
the path is placed into the list of path's that need to be replaced. The list is searched when a path becomes active again, as described above. Otherwise, if a spare path
is available, it is activated in the configuration and placed in the command handler object's list of connections that have to be 
established. This is the same list into which a newly activated path is placed when it has to replace a broken path (see above). 

When the program's main loop has retrieved a set of connections that have to be established, it calls the command handler's \texttt{Make\-Con\-nec\-tion} function for each
of them after the disconnection from their previous remote partners has completed. In this function connection command messages are assembled containing the
message and blob-message addresses of the new bridge head components in the path to be activated. After these messages have been sent successfully to the
bridge heads on the data source and sink nodes, the path is removed from the list of paths that have to be connected and placed into a list of paths to
which reconnect commands have been sent. This second list can be queried by calling the class's \texttt{Get\-Re\-con\-nect\-Paths} function. In the main loop 
this function is called, and the states of all connections in the subscriber and publisher bridge head components of the path are queried. 
If all connections are established successfully, the command handler's \texttt{Set\-Tar\-get} function is called to reactivate the replaced path. 
In this function a start command is sent to the fault detection subscriber monitoring the path concerned. Further commands are sent to all configured target 
components to set the path's state back to active, instructing the data flow components to send events to that path again.

\subsection{\label{Sec:TolerantEventScatterer}Fault Tolerant Event Scatterer}

One of the two components responsible for routing the dataflow in a fault tolerant chain setup, the
\texttt{Tol\-e\-rant\-Event\-Scat\-terer} component is an extension of the \texttt{Event\-Scat\-terer} from 
section~\ref{Sec:EventScatterer}. It replaces the \texttt{Ali\-HL\-Round\-Robin\-Event\-Scat\-terer} object in the original
scatterer with an \texttt{Ali\-HLT\-Tol\-e\-rant\-Round\-Robin\-Event\-Scat\-terer} object and adds a number of objects 
for the fault tolerance tasks in the component. Fig.~\ref{Fig:FTScattererClasses} gives an overview of the most important 
classes used. 
The component's main program does not differ significantly from the event scatterer's. Major differences
are the use of another central class and the configuration, creation, and setup of the required auxiliary
objects for the FT tasks. Primarily, these are an instance of the \texttt{Ali\-HLT\-SC\-Inter\-face} class from 
section~\ref{Sec:SCClasses} and a status structure derived from the ones described in the same section.  

\begin{figure}[hbt]
\begin{center}
\resizebox*{0.50\columnwidth}{!}{
\includegraphics{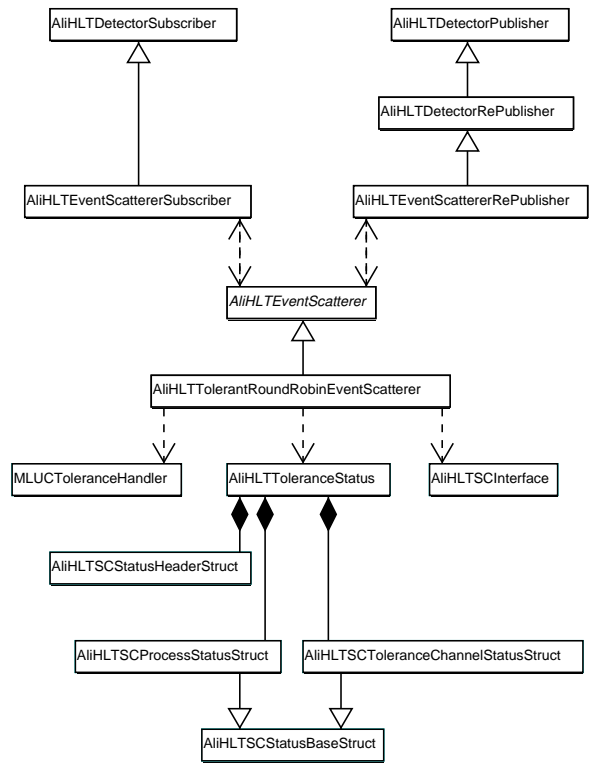}
}
\parbox{0.90\columnwidth}{
\caption{\label{Fig:FTScattererClasses}The main classes used in the \texttt{Event\-Scat\-terer} component.}
}
\end{center}
\end{figure}

\subsubsection{The Fault Tolerant Round Robin Event Scatterer Class}

Similar to the original \texttt{Ali\-HLT\-Round\-Robin\-Event\-Scat\-terer} class, the \texttt{Ali\-HLT\-Tol\-e\-rant\-Round\-Robin\-E\-vent\-Scat\-terer} class
is also derived from the \texttt{Ali\-HLT\-Event\-Scat\-terer} class, described in section~\ref{Sec:EventScatterer}, 
implementing the six abstract functions defined by that class. It also overwrites the two callback
functions provided by the base class. The main difference between the basic round-robin
scatterer class and this class is that \texttt{Ali\-HLT\-Tol\-e\-rant\-Round\-Robin\-Event\-Scat\-terer} uses an
\texttt{MLUC\-Fault\-Tol\-e\-rance\-Han\-dler}  object, described in section~\ref{Sec:FTHandler}. This object distributes the work load of 
events only between functional
output publisher paths, taking into account their respective status. In addition, it 
is possible to control this class to a certain degree from external components via a control and
monitoring interface class instance. 

The primary data structures on which the class operates are lists of events, one for each of the configured
output publishers and one for retry events. Events are entered into the retry list when all 
output publisher paths are marked as broken and no publisher is available to announce the event. An event is entered into
a specific publisher's list when it has been announced by that publisher. The data placed
into each of the lists contains everything required to announce, or reannounce, the event: its ID, sub-event descriptor, and
event trigger structure, plus the time at which it arrived. Storing this data in the component is necessary in case the event
has to be reannounced due to a failure of the output path to which it was assigned.

Beyond these lists 
two auxiliary objects are used in the class as well as objects of two proxy classes. The proxy class objects function as 
forwarders between a command processor and the scatterer object on one hand and the scatterer object and an
FT interface object, described in section~\ref{Sec:SCClasses}, on the other hand. In the first case a function in the scatterer
is called to handle received 
commands, while in the second LAM requests are forwarded from the scatterer to an interface object. 
The fault tolerance handler object is only used internally in the class, while the 
status structure is passed from the component's main program and is just updated in the class. 

Although all of the functions implemented in the class contain some code related to or affected by
the fault tolerance functionality of the component, the most important work of correctly distributing the 
events is done in two functions: \texttt{An\-nounce\-Event} and \texttt{Set\-Pub\-lish\-er}. 
\texttt{An\-nounce\-Event} is called whenever a new event has to be announced to a publisher.
%, that is determined for each event using the class's fault tolerance object. 
The publisher to be used is first determined with the help of
the fault tolerance handler object. When a publisher is found, the event is announced to it and is
placed with the required data in the event list for that specific publisher. If no publisher could
be found for the event, all publisher paths must be broken and the event is added to the list
of retry events. The retry list is accessed when at least one publisher becomes available again. Events are removed
from a publisher's list when the event is released by the scatterer's (re-)publisher or when it is 
cancelled by its original publisher from which the scatterer received it. In the first case an
 \texttt{Event\-Done} message is sent via the scatterer's subscriber to the event's originating publisher.

The second important new function, \texttt{Set\-Pub\-lish\-er}, is called by the command processor function when
the state of one of the output publisher paths in the system changes. If events have to be announced
or reannounced as a result of such a change, the function calls \texttt{An\-nounce\-Event} after updating
the concerned publisher's state. 
After a status change command has been received, a publisher with the given index is searched
and its status in the fault tolerance handler is compared to the specified state. If these states are
equal, the change has already been processed and the function performs no further action, otherwise
the state is updated in the handler object. 

When a publisher path has become functional again and the retry list contains events, these
events are processed. They are removed from the retry list and passed to the \texttt{An\-nounce\-Event}
function to be announced. If no events are contained in the retry list, the publisher is only used for the 
respective fraction of new incoming events. 
For a state change to non-functional the event list of the concerned publisher is accessed. Each
event is removed and handed to the announce event function. Due to the status change in the fault tolerance
handler object these events are assigned to a different, still functional, publisher able to
process them. After these events have been reannounced, they are aborted in the faulty publisher to
which they had been assigned originally, using the publisher's \texttt{Abort\-Event} function 
so that no event is contained in two publisher objects. 

One more function contains functionality related to the fault tolerance operation. This is
the \texttt{Pub\-lish\-er\-Down} method called by one of the scatterer's republisher objects when
an error occurs in one of its subscribers. In the function the state field in the status data structure
corresponding to this publisher's state is set to faulty, and a {\em LookAtMe} request is sent to a
supervising component. The decision to remove this publisher's path is not taken locally in this
component but instead in the supervisor component when it has checked the respective state.

\subsection{\label{Sec:TolerantEventGatherer}Fault Tolerant Event Gatherer}

The second component responsible for routing the dataflow in a fault tolerant chain setup
is the \texttt{Tol\-e\-rant\-Event\-Gath\-erer} component, an extension of the \texttt{Event\-Gath\-erer} component from
section~\ref{Sec:EventGatherer}. Similar to the fault tolerant event scatterer component, this component replaces
the \texttt{Ali\-HLT\-Event\-Gath\-erer} class from the original gatherer with the extended \texttt{Ali\-HLT\-Tol\-e\-rant\-Event\-Gath\-erer}
class and adds a number of additional helper objects. Fig.~\ref{Fig:FTGathererClasses} gives an overview of the most important 
classes used. 

\begin{figure}[hbt]
\begin{center}
\resizebox*{0.50\columnwidth}{!}{
\includegraphics{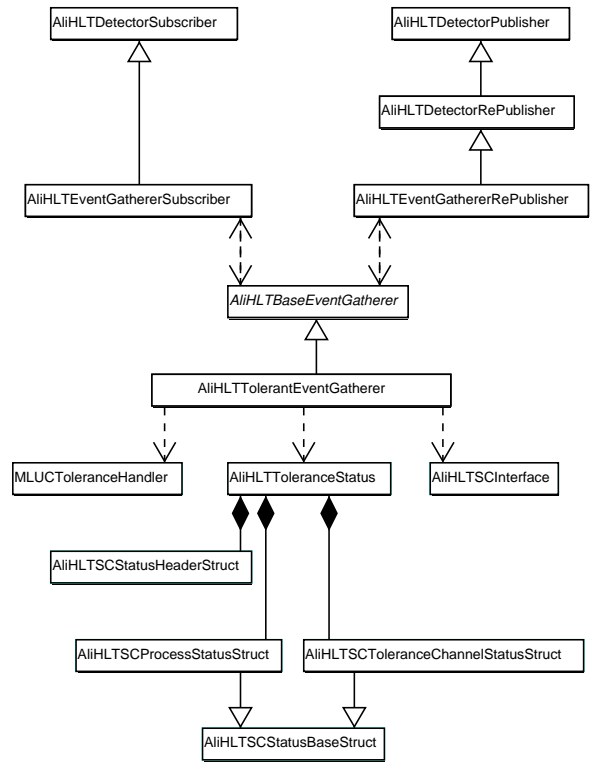}
}
\parbox{0.90\columnwidth}{
\caption{\label{Fig:FTGathererClasses}The main classes used in the \texttt{Event\-Gath\-erer} component.}
}
\end{center}
\end{figure}

As in the case of the fault tolerant scatterer component, the fault tolerant gatherer's main program does not differ significantly
from the original gatherer one's. The main differences are also the initialization of the added classes, 
primarily status structures and instances of the control and monitoring classes described in section~\ref{Sec:SCClasses}. 

\subsubsection{The Fault Tolerant Event Gatherer Class}

Identical to the \texttt{Ali\-HLT\-Event\-Gath\-erer} class, the \texttt{Ali\-HLT\-Tol\-e\-rant\-Event\-Gath\-erer} class is also derived
from the \texttt{Ali\-HLT\-Base\-Event\-Gath\-erer} class and implements its five abstract defined functions. 
Compared to the standard gatherer class it uses a number of additional objects or object pointers. One of these
is an \texttt{MLUC\-Fault\-Tol\-e\-rance\-Han\-dler} object, as described in section~\ref{Sec:FTHandler}. 
Unlike the scatterer, the fault tolerant gatherer does not use this object to determine where to send an event or event done data, but
only to keep track of the states of the paths associated with its input subscribers. To select an output path for an 
event done data structure, it uses the information from which subscriber the respective event has been received. 

The class's primary data structures are event lists, one for active events that have been received and announced but not yet released and a backlog list
holding a configurable amount of event done data structures for already released events.
%for the last events that have been released. {\bf The number} of released events that are stored in this backlog list can be configured
%via a command line parameter to the component and via a constructor parameter to a class object. 
In the first list \texttt{Event\-Gath\-erer\-Data}
structures are stored, similar to the original gatherer class's event data list. Included in this data structure is a pointer for each event to the
subscriber object from which it has been received and the index number associated with that subscriber. 
Events are added to the active event list in the \texttt{New\-Event}
function when a new event is received and in the \texttt{Event\-Done} function when an error occurs sending the event done data back to an event's
original publisher. 
%In the \texttt{Event\-Done} function events are also added to the backlog list after an event has been successfully released
%and its event done data message has been sent to its original publisher. 
Additionally, events are added to the backlog list in the \texttt{Event\-Done} function after their event done data structure
has been sent successfully to its originating publisher. 

Fault tolerance functionality provided by the class is primarily located in three of its functions: \texttt{New\-Event}, \texttt{Event\-Done}, and
\texttt{Set\-Sub\-scrib\-er}. \texttt{New\-Event} is called by one of the component's subscriber objects when a new event has been received and \texttt{Event\-Done}
is called by the republisher after an event has been released. \texttt{Set\-Sub\-scrib\-er} is called
by the \texttt{Pro\-cess\-Cmd} function upon receipt of a command indicating that the state of one of the input paths has changed. 

When the \texttt{New\-Event} function is called with a new event by one of the component's subscribers, the list of active events is first searched
whether the event has been received already.
If the event is found in that list, the subscriber pointer and index in
its data structure are changed to the corresponding ones of the subscriber from which it has just been received again. When a flag is set in the 
event's data structure to signal
 that the event has been released already, the respective event done data that has been received will be sent 
back to the subscriber from which the event has just been received. This is done immediately without announcing the event again through the component's 
output publisher.

If the event is not contained in the active event list, the backlog list is searched
as well and when the event is found in that list, its event done data contained in the list is sent back immediately as well. 
Events not contained in either list have been received for the first time or have already been removed from the backlog list. This last case
should not happen if the backlog list is large enough. An \texttt{Event\-Gath\-erer\-Data} structure for the event
is created, filled, and added to the active event list. The pointer to that data is added to the signal object used to notify the component's main loop
about new events. After adding the event's data to the signal object it is triggered to wake up the main loop and initiate the republishing of the event.

In the class's \texttt{Event\-Done} function, called when an event has been released, the data structure for it is searched in the list of 
active events first. 
%Events that cannot be found in that list are considered as fatal internal errors of the component. 
Once its data has been found, the 
pointer to its subscriber and the subscriber's index number are extracted. The subscriber index is used together with the fault tolerance handler object
to determine whether the subscriber and its associated path are in a working state. If this is the case, then the received event done data
is passed to the subscriber's \texttt{Event\-Done} function to signal the event's original publisher that it can be freed. Following the successful 
completion of this, the event's data is removed from the list and any resources occupied by its data are released. Finally, the event done data is added to the 
end of the backlog list of released events. That list's first element is then removed if it has become too long. 
When an error has occured in \texttt{Event\-Done}, the data structure remains in the list of active events. 
For an event whose subscriber is marked as faulty in the fault tolerance handler object, a flag is set in the event's data to indicate 
that it is already released and the received
done data is stored in its data structure as well. In this case neither of the two lists is modified. When the event is subsequently received
again through a different subscriber, it is not forwarded through the republisher again, but the event done data is immediately sent back, as written in
the description of the \texttt{New\-Event} function. 

The last of the important fault tolerance related functions is the \texttt{Set\-Sub\-scrib\-er} function, called when a command is received 
to change a subscriber's 
and its associated path's state. Unlike the \texttt{Tol\-e\-rant\-Event\-Scat\-terer}'s function of the analogue name, this function does not contain much 
functionality. It searches for the subscriber specified in its arguments to determine its index. Using the index it sets the state
of that subscriber in the fault tolerance handler object to the one specified. As the events concerned will be resent through other paths
by the event scatterer, they will be received again through different subscribers as well. When an event is received again, it can be
released using the new subscriber, thus no further action is necessary in this function. 

A fourth, additional, function that performs a task related to the fault tolerance is the \texttt{Sub\-scrib\-er\-Error} function. This function is called for a subscriber object 
when an error occurs while sending an event done message back to that specific subscriber's publisher in \texttt{Event\-Done}. The function determines the subscriber's index number 
and then sets the subscriber's state to faulty in the component's status data structure. To inform a supervising instance of this change it triggers a {\em LookAtMe}
interrupt for all specified supervisor components, which will later cause the corresponding path to become disabled.

\subsection{\label{Sec:TolerantBridgeComponents}Fault Tolerant Bridge Components}

The last fault tolerance components are the \texttt{Tol\-e\-rant\-Sub\-scrib\-er\-Bridge\-Head} and the \texttt{Tol\-e\-rant\-Pub\-lish\-er\-Bridge\-Head}, extensions of the
\texttt{Sub\-scrib\-er\-Bridge\-Head} and \texttt{Pub\-lish\-er\-Bridge\-Head} components from section~\ref{Sec:BridgeComponents} respectively. Unlike the fault tolerant scatterer and 
gatherer components they do not replace their central \texttt{Ali\-HLT\-Sub\-scrib\-er\-Bridge\-Head} and \texttt{Ali\-HLT\-Pub\-lish\-er\-Bridge\-Head} classes. Instead
the necessary functionality is contained in additional classes used in these components. Parts of the functionality of the two bridge head classes have not
been covered in the classes' sections in~\ref{Sec:BridgeComponents} as they are used in conjunction with the control and monitoring classes
from~\ref{Sec:SCClasses}. These parts of the classes are explained in the following paragraphs. Differences between each component and its respective basic counterpart
are principally identical in the two bridge head types with only minor deviations. As this section focuses on the differences
between the basic and fault tolerance components, both bridge head types are described together with comments on their respective deviations. 

Compared to the basic bridge components the major additional tasks in the two components' main programs  are the parsing of  
command line parameters for the fault tolerant relevant configuration as well as the creation, configuration, and activation of required additional 
objects. In addition to the objects of the classes described below, this includes primarily an instance of the \texttt{Ali\-HLT\-SC\-Inter\-face} class, described in 
section~\ref{Sec:SCClasses}, to allow the external control of the components. 
%The core functionality of both components and classes is unchanged compared to their basic counterparts.

\subsubsection{Additional Bridge Head Class Functionality}

The ability to change remote message and blob-message addresses during runtime of the component is a feature contained in both bridge head classes.
To ensure that this does not happen at times when the communication classes or addresses are in use, locks have been introduced
in the classes to protect the regions in the class methods where they are used. These locks have to be acquired by an external entity before it attempts to 
modify the remote addresses of a bridge head object. An additional feature that should also be used before changing addresses, is the ability to pause
processing in the classes. If a bridge head object is paused, no events are processed, new events are not announced to the publisher bridge
head component, and neither are event done messages sent to the subscriber bridge head. Pausing the objects before modifying the communication addresses ensures that 
no communication attempt is made during the process of modification. 

To support communication with the fault tolerance parts without introducing customized bridge head classes or adding optional 
functionality to the bridge head classes themselves, three additional classes and structures can be attached to bridge head objects. The first of these
external structures is a LAM proxy  object derived from a common abstract base class, \texttt{Ali\-HLT\-Bridge\-Head\-LAM\-Proxy}. Its
only function, \texttt{Look\-At\-Me}, is called by the bridge head objects when an error occurs in their \texttt{Con\-nect} function while trying to establish a connection to their 
remote partner. In the LAM proxy derived class used in the fault tolerant bridge components the \texttt{Look\-At\-Me} function
sends LAM requests to a configured list of supervisor component addresses. This address list is provided to the component via command line parameters and can
contain multiple target components. 

Next to the LAM proxy object an instance of the \texttt{Ali\-HLT\-Bridge\-Status\-Data} structure is the second external object used in the bridge classes. This 
structure contains status data for a bridge component and consists of a header field, a generic process status field, and a bridge status field. 
In this last field two elements are contained, signalling the connection status of the message communication object as well as the combined 
connection status of the blob and blob-message 
objects. Both communication status fields are updated in the bridge classes' \texttt{Con\-nect} and \texttt{Dis\-con\-nect} functions as required. 

The third external entity that can be used is an error callback, derived from the \texttt{Ali\-HLT\-Bridge\-Head\-Error\-Call\-back} template class. It contains
two abstract functions, \texttt{Con\-nec\-tion\-Error} and \texttt{Con\-nec\-tion\-Es\-tab\-lished}. The first is called when a communication error occurs, including
message sends, blob transfers, or connection or disconnection attempts. In the derived class used in the bridge head components this 
function pauses the bridge object in the component and calls its \texttt{Dis\-con\-nect} function to abort the connection. A later communication attempt
 automatically initiates a reconnection attempt of the communication objects by calling the classes' \texttt{Con\-nect} function. When this function completes
successfully with a new established connection, the callback object's second function, \texttt{Con\-nec\-tion\-Es\-tab\-lished}, is called to signal the new connection. The
fault tolerant bridge component callback object then acknowledges the connection by restarting the paused object. 

In addition to the address change ability and the use of the above three external classes one more function is contained in the class used for the fault 
tolerance functionality, \texttt{PURGE\-ALL\-EVENTS}. 
%To show the potential danger in using this function it spelled in all capitals unlike the other function names in the framework. 
When this function is called it will access all data fields in the corresponding bridge object and remove any contained event data structures. After
this function has been called, the object is in a state of not having received any event, making this function
inherently dangerous since calling it can cause events to be lost in a system. This functionality is required to bring the objects into a 
known clean state after a connection to one remote partner has been aborted and before a new connection to another remote bridge component is established.
In this situation no events will be lost as the scatterer takes care to resend them.

\subsubsection{Fault Tolerant Bridge Command Handler}

In the two fault tolerant bridge components the most important objects in addition to the three external ones attached
 to the bridge objects are the command processors that
handle commands received from external supervisor components. The \texttt{Pro\-cess\-Cmd} functions are able to handle five commands related to the bridge's
connections: \texttt{Dis\-con\-nect}, \texttt{Con\-nect}, \texttt{Re\-con\-nect}, \texttt{New\-Con\-nec\-tion}, and \texttt{New\-Re\-mote\-Ad\-dress}. For the first two just the
bridge objects' \texttt{Dis\-con\-nect} and \texttt{Con\-nect} functions are called respectively. A \texttt{Re\-con\-nect} command causes \texttt{Dis\-con\-nect} 
and \texttt{Con\-nect} calls 
in succession, with the bridge object being paused before and restarted after the calls. For the \texttt{Re\-con\-nect} command 
the communication lock is acquired before and released
after the two connection function calls, the two simple \texttt{Con\-nect} and \texttt{Dis\-con\-nect} commands do not use the locks. 

The last two commands both contain new message and blob-message addresses for the remote partner bridge head of the component. It is not necessary to transmit
the remote blob address as it is queried using the blob-message object. After extracting the two addresses from the command structure the bridge head object
is paused, its communication lock is acquired, and a currently established connection is terminated using the \texttt{Dis\-con\-nect} function. 
With the connection interrupted the new address is passed to the bridge head object, and in case of the \texttt{New\-Con\-nec\-tion} command, \texttt{Con\-nect} 
is called to reestablish the connection. Following this, the communication lock is released again and the object is restarted. 

In addition to the five connection-related commands, one more command is available for processing by the command handler objects. The \texttt{PURGE\-ALL\-EVENTS}
command clears all events from the bridge object by calling the object's function of the same name described above.

\clearpage

%AliHLTSCController
%AliHLTSCLAMHandler
%AliHLTSCStatusHeaderStruct
%AliHLTSCInterface
%AliHLTSCCommandProcessor

%%%%%%%%%%%%%%%%%%%%%%%%%%%%%%%%%%%%%%%%%%%%%%%%%%%%%%%%%%%%%%%%%%%%%%%%%%%%%%%%%%%%%%%%%%%%%%%%%%%%%%%%%%%%%%%%%%%%%%%%%%%%%
%%%%%%%%%%%%%%%%%%%%%%%%%%%%%%%%%%%%%%%%%%%%%%%%%%%%%%%%%%%%%%%%%%%%%%%%%%%%%%%%%%%%%%%%%%%%%%%%%%%%%%%%%%%%%%%%%%%%%%%%%%%%%

\chapter{\label{Chap:BenchmarksTests}Benchmarks and System Tests}

Results of tests executed with the developed software are presented in this section. First
short micro benchmarks are presented, followed by network reference tests. These reference tests are used for comparison 
and evaluation of the subsequent benchmarks of the two TCP communication classes. In the next section benchmarks and 
scalability evaluations of the basic publisher-subscriber interface are presented. The final section contains descriptions 
and results of two tests of the complete framework: A performance test using simulated ALICE TPC data with
ALICE analysis components and a test of the framework's fault tolerance functionality. 
The operating system used in all tests except for the logging overhead measurement (section~\ref{Sec:LoggingBenchmark}) was 
a SuSE Linux \cite{SuSEWebDe} \cite{SuSEWebInt} version 7.2 running a Linux kernel version 2.4.18 \cite{Linux.2.4.18} with the 
precise accounting \cite{PrecAccPatch} (cf. section~\ref{Sec:PreciseAccounting} below) and bigphysarea patches \cite{bigphysareapatchWeb} 
applied. For the logging overhead measurement a standard SuSE Linux 8.0 was used. 
The corresponding data can be found in appendix~\ref{Chap:BenchmarkTables}.
%Results only contained in graphs in this section are contained as tables in appendix~\ref{Chap:BenchmarkTables}.

\section{Micro-Benchmarks}
\subsection{\label{Sec:LoggingBenchmark}Logging Overhead}
The logging system available in the MLUC class library 
was designed so that logging calls for messages with deactivated severity levels impose as little
overhead for the calling program as possible (see also section~\ref{Sec:LoggingSystem}). For calling the logging system with multiple severity levels two major variants are possible.
The first is a
function call using all required parameters. Whether a specified message has to be logged is decided
inside the function based upon its severity and the activated severity levels. The other variant for calling the system 
is by using an \texttt{if}-statement to decide whether logging takes place followed by a function call to execute the 
actual logging process in the case of a positive decision. 

To evaluate the effects of these two variants, a small program has been written
to determine the amounts of time required to execute an \texttt{if}-statement and a function call. The program itself is listed in 
appendix~\ref{Sec:LoggingOverheadProgram}. Results obtained from the program are shown in Table~\ref{Tab:LoggingOverheadNoOpt} 
without compiler optimization and Table~\ref{Tab:LoggingOverheadOpt} with compiler
optimization level 2 (-O2) for a gcc 2.95.3 compiler for an i686 (-march=i686) processor. Execution of the program 
was performed on a 700~MHz Pentium III processor. Absolute execution times, though, are not as important
as the values relative to each other. The first column of each table, labeled
{\em Reference Loop}, includes only the time for a loop with just one pointer dereference increment (\texttt{*n++}). 
This pointer dereference increment is used as the test instruction for the \texttt{if}-statement tests and the function calls.
% and is therefore included in the reference loop. 
It is included in the reference loop to prevent the compiler from removing it during optimization and
is therefore also kept in each examined statement for comparison. Each of the different loop tests is executed $10^9$ times to obtain
good accuracy. In addition each test has been run ten times with the values shown averaged over these ten runs.
For the exact instructions executed in each case see appendix section~\ref{Sec:LoggingOverheadProgram}.

\begin{table}[hbt]
\begin{center}
\begin{tabular}{|l||c|c|c|c|c|}
\hline 
Measurement & Reference Loop & Loop w. \texttt{if} & Loop w.  & Loop w. \texttt{if} & Loop w. func. \\
            &                &                     & function &  \& func.           & cont. \texttt{if} \\
\hline \hline
Time per loop iteration / &   $12.14$ &  $14.686$ &  $32.627$ &  $28.321$ &   $34.0$ \\
$\mu \mathrm s$           & $\pm 0.72$ & $\pm 0.069$ & $\pm 0.094$ & $\pm 0.07$ & $\pm 0.15$ \\
\hline
Time per loop iteration w/o        & -        &  $2.546$ & $20.487$ & $16.181$ & $21.86$ \\
overhead / $\mu \mathrm s$ & & $\pm 0.789$ & $\pm 0.814$  & $\pm 0.79$  & $\pm 0.87$ \\
\hline
\end{tabular}
\parbox{0.90\columnwidth}{
\caption[Unoptimized logging test program results.]{\label{Tab:LoggingOverheadNoOpt}Logging overhead test program results without compiler optimization. 
Standard deviations are given as errors.}
}
\end{center}
\end{table}

%1               8030177.500000          187.390849   (1) Loop overhead
%2               10950021.400000         376.479858   (2) Loop with if
%3               21112340.200000         512.684870   (3) Loop with func
%4               25650656.800000         620.766084   (4) Loop with if and func
%5               23280450.900000         1001.890906  (5) Loop with func with if

\begin{table}[hbt]
\begin{center}
\begin{tabular}{|l||c|c|c|c|c|}
\hline 
Measurement & Reference Loop & Loop w. \texttt{if} & Loop w.  & Loop w. \texttt{if} & Loop w. func. \\
            &                &                     & function & \& func.             & cont. \texttt{if} \\
\hline \hline
Time per loop iteration / &    $8.76$ &   $10.04$  & $25.507$ &    $23.68$ &   $26.297$ \\
$\mu \mathrm s$           & $\pm 0.15$ & $\pm 0.55$ & $\pm 0.055$  & $\pm 0.51$  & $\pm 0.062$ \\
\hline
Time per loop iteration w/o         & - & $1.28$  & $16.747$ & $14.92$ & $17.537$ \\
overhead / $\mu \mathrm s$  &   & $\pm 0.7$ & $\pm 0.205$  & $\pm 0.66$  & $\pm 0.212$  \\
\hline
\end{tabular}
\parbox{0.90\columnwidth}{
\caption[Optimized logging test program results.]{\label{Tab:LoggingOverheadOpt}Logging overhead test program results with compiler optimization (-O2 for i686 processor w. gcc 2.95.3).
Standard deviations are given as errors.}
}
\end{center}
\end{table}

Taking into account the reference loop overhead one can compare the time required for a loop using an \texttt{if}-statement with the time
for a loop using a function call and an \texttt{if}-statement. As a result of these comparisons one can conclude that a function call 
containing an \texttt{if}-statement, corresponding to the first logging call option, is 8 to 14 times slower than 
just an \texttt{if}-statement, corresponding to the second option. 
The efforts in making the logging system handle calls with disabled levels efficiently are thus justified. 
Even with activated logging, corresponding to the case for an \texttt{if}-statement followed by a function call,
the approach chosen is more efficient than a function call containing an \texttt{if}-statement.

%\begin{verbatim}
%timm@bottom:~/src/c> gcc -Wall -O2 -o log_overhead_test log_overhead_test.c
%timm@bottom:~/src/c> ./log_overhead_test st.c
%Loop overhead:          9585621 us
%Loop with if:           8323386 us
%Loop with func:         21367664 us
%Loop with if and func:  21501342 us
%Loop with func with if: 23068236 us
%timm@bottom:~/src/c> ./log_overhead_test
%Loop overhead:          9596410 us
%Loop with if:           8194772 us
%Loop with func:         21745482 us
%Loop with if and func:  21493670 us
%Loop with func with if: 22975792 us
%timm@bottom:~/src/c> ./log_overhead_test
%Loop overhead:          9604453 us
%Loop with if:           8130253 us
%Loop with func:         21399943 us
%Loop with if and func:  21728740 us
%Loop with func with if: 23115828 us
%timm@bottom:~/src/c> gcc -Wall -o log_overhead_test log_overhead_test.c
%timm@bottom:~/src/c> ./log_overhead_test
%Loop overhead:          10068381 us
%Loop with if:           11665690 us
%Loop with func:         26933141 us
%Loop with if and func:  29596801 us
%Loop with func with if: 29800233 us
%timm@bottom:~/src/c> ./log_overhead_test
%Loop overhead:          9979354 us
%Loop with if:           11287266 us
%Loop with func:         27214764 us
%Loop with if and func:  29610459 us
%Loop with func with if: 29572226 us
%timm@bottom:~/src/c> ./log_overhead_test
%Loop overhead:          10083325 us
%Loop with if:           11284863 us
%Loop with func:         27124669 us
%Loop with if and func:  29819857 us
%Loop with func with if: 29462062 us
%timm@bottom:~/src/c>
%\end{verbatim}

\subsection{\label{Sec:CacheMeasurements}Cache and Memory Reference Tests}

In addition to the framework tests, three reference PCs, described below, have been examined with a cache testing program \cite{PattersonCache.c} to obtain
the different amounts of time required to access data stored in the level 1 cache, level 2 cache, and main memory respectively. 
These times are necessary for the evaluation of the scaling behaviour of the programs tested.
They are measured by accessing memory arrays of varying sizes with different distances between the array fields (strides) accessed. From the graphs obtained by plotting the
access times in dependance of array size and stride one can determine, amongst others, the different access times. 
% {\bf as well as for the determination of the effects they have on the programs}. 
Figures~\ref{Fig:Cache733MHz},~\ref{Fig:Cache800MHz}, and~\ref{Fig:Cache933MHz} contain the three different plots that have been obtained by running this program on the 
reference PCs. Table~\ref{Tab:ReferecePCCacheResults}
summarizes the values for the different access times.
During the tests the number of background processes, e.g. system daemons, has been restricted to a minimum to exclude outside interference
effects  as far as possible. 
The list of remaining processes is shown in appendix~\ref{Sec:BenchmarkProcessList}.
For similar reasons any networks in the machine have been disabled and unplugged for the duration of the tests.

All of the three reference PCs listed below have dual Pentium III processors with a 133~MHz front side bus using PC133 SDRAM memory in a two bus interleaved 
mode, doubling the theoretical memory bandwidth. 
\begin{itemize}
\item 733~MHz PC with the  Katmai  P3 version with 16~kB Level 1 data cache and 256~kB unified Level 2 cache running at half the CPU's clock frequency. 
This PC uses a Tyan Thunder 2500 motherboard with the Serverworks III HE chipset.
\item 800~MHz PC with the  Coppermine  P3 version with 16~kB Level 1 data cache and 256~kB unified Level 2 cache running at full CPU clock frequency. 
This PC uses a Tyan Thunder HEsl motherboard with the Serverworks III HEsl chipset, the successor to the III HE chipset.
\item 933~MHz PC with the  Coppermine  P3 version with 16~kB Level 1 data cache and 256~kB unified Level 2 cache running at full CPU clock frequency. 
This PC uses a Tyan Thunder HEsl motherboard with the Serverworks III HEsl chipset, 
%the successor to the III HE chipset, 
similar to the motherboard in the 800~MHz PC. 
\end{itemize}

%\begin{figure}[hbt]
\begin{figure}
\begin{center}
\resizebox*{0.90\columnwidth}{!}{
\includegraphics{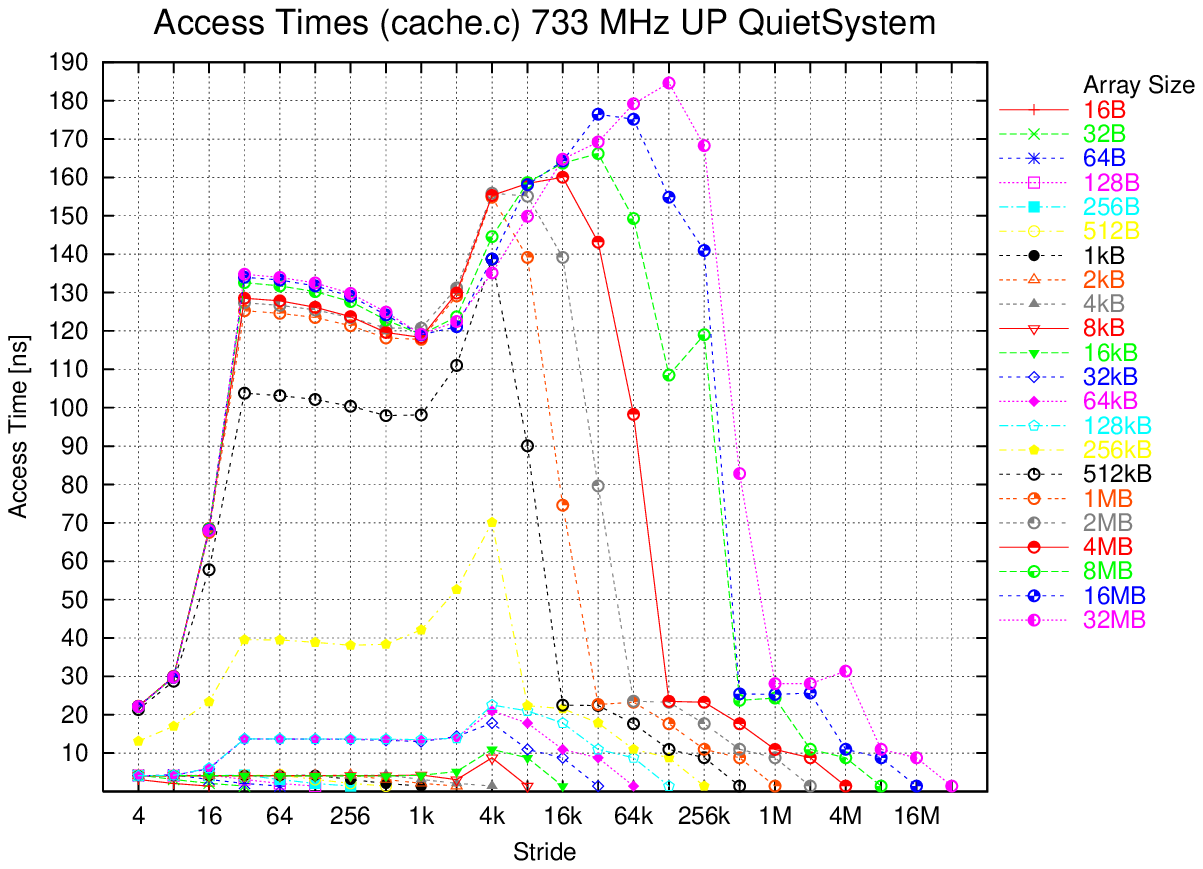}
}
\resizebox*{0.90\columnwidth}{!}{
\includegraphics{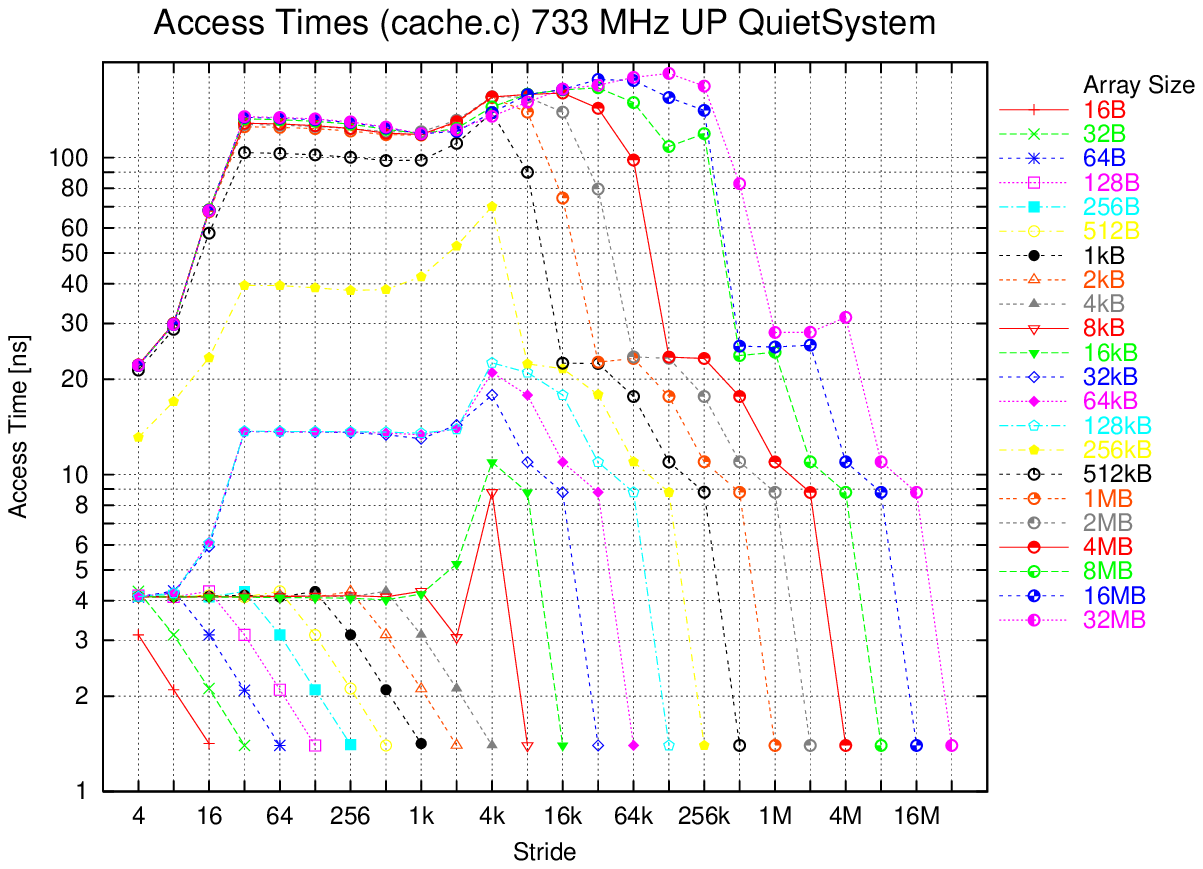}
}
\parbox{0.90\columnwidth}{
\caption[733~MHz cache and memory subsystem measurement plots.]{\label{Fig:Cache733MHz}Cache and memory subsystem measurement plots with linear (top) and logarithmic (bottom) scale
for the 733~MHz reference PCs.}
}
\end{center}
\end{figure}

%\begin{figure}[hbt]
\begin{figure}
\begin{center}
\resizebox*{0.90\columnwidth}{!}{
\includegraphics{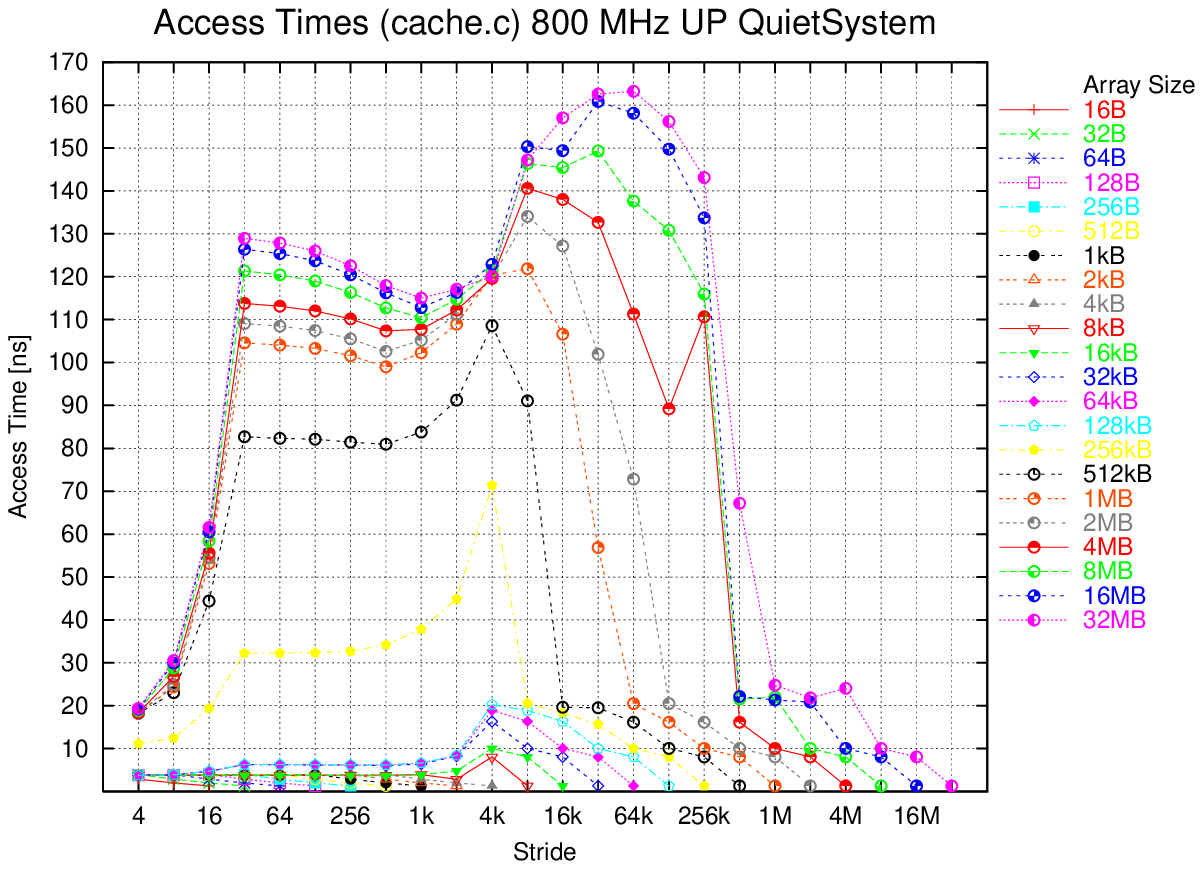}
}
\resizebox*{0.90\columnwidth}{!}{
\includegraphics{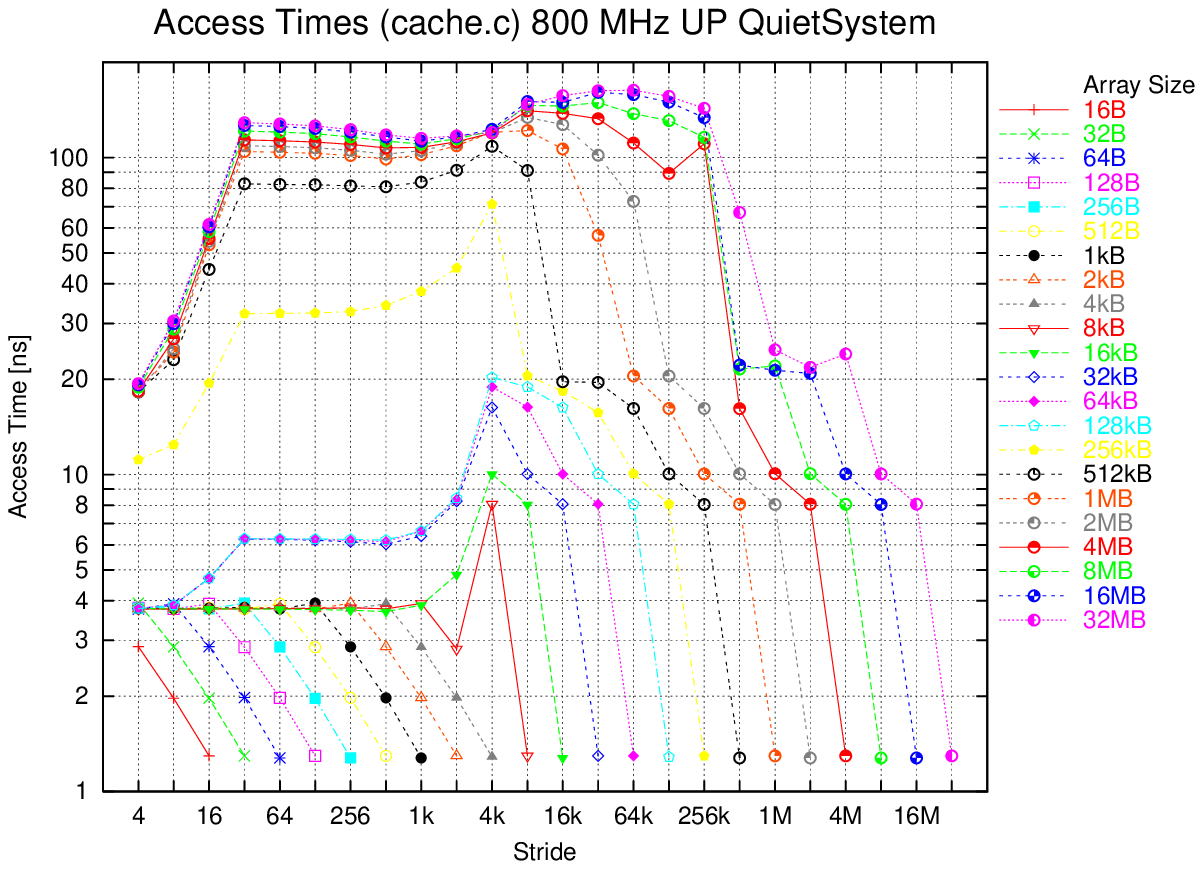}
}
\parbox{0.90\columnwidth}{
\caption[800~MHz cache and memory subsystem measurement plots.]{\label{Fig:Cache800MHz}Cache and memory subsystem measurement plots with linear (top) and logarithmic (bottom) scale
for the 800~MHz reference PCs.}
}
\end{center}
\end{figure}

%\begin{figure}[hbt]
\begin{figure}
\begin{center}
\resizebox*{0.90\columnwidth}{!}{
\includegraphics{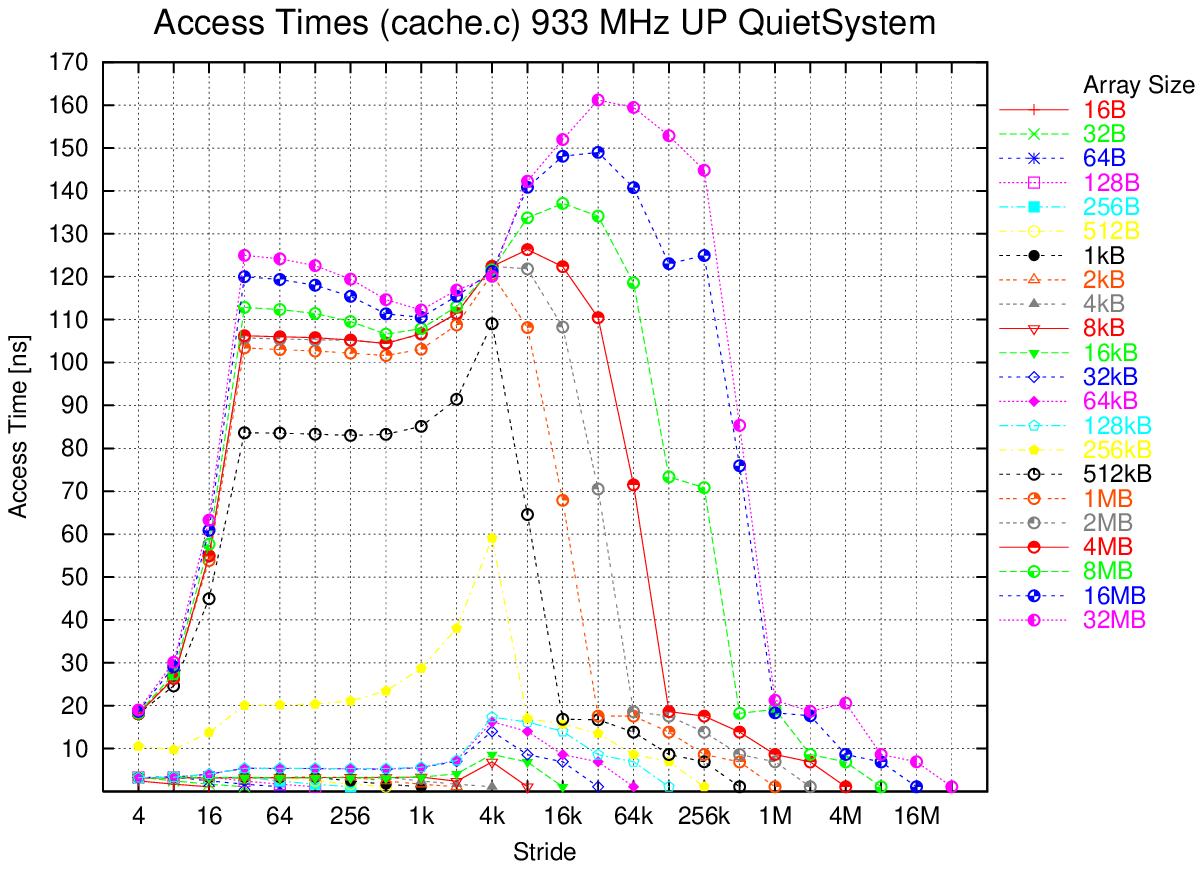}
}
\resizebox*{0.90\columnwidth}{!}{
\includegraphics{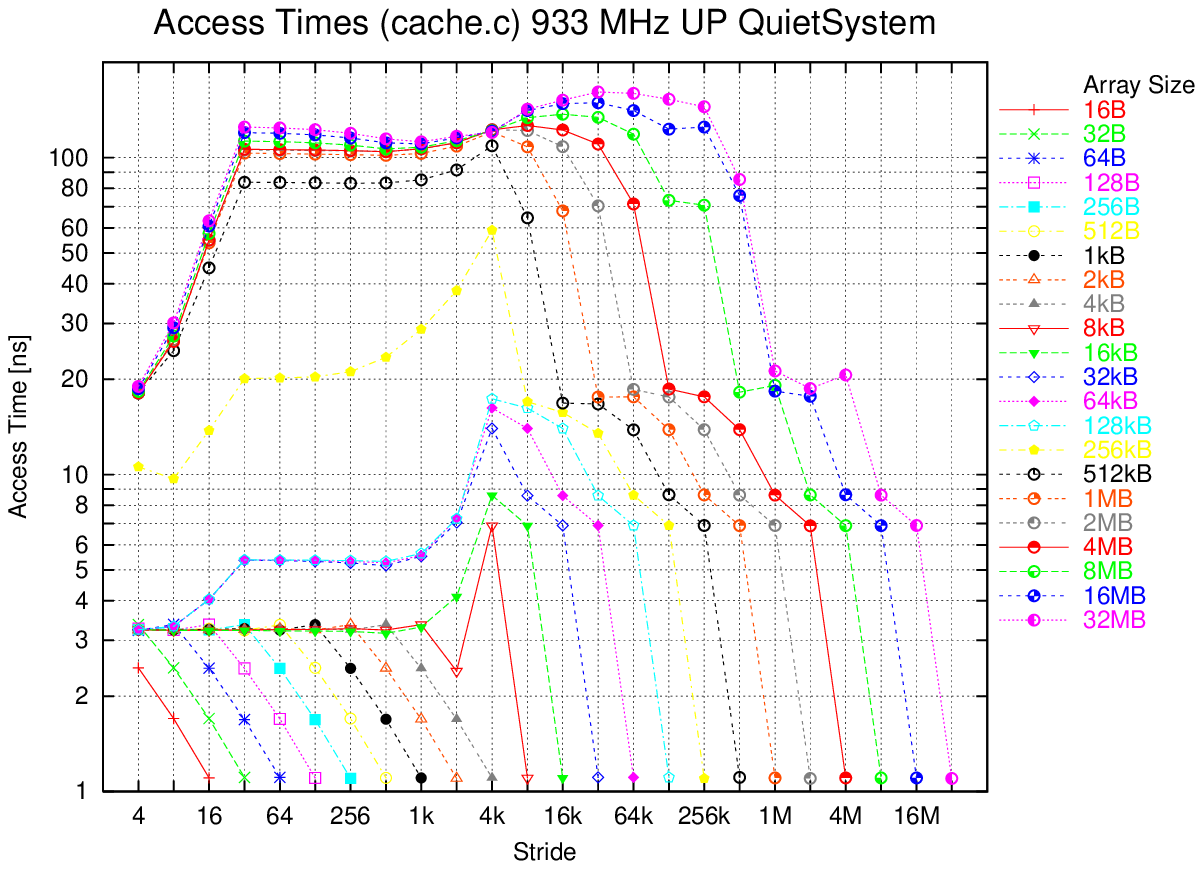}
}
\parbox{0.90\columnwidth}{
\caption[933~MHz cache and memory subsystem measurement plots.]{\label{Fig:Cache933MHz}Cache and memory subsystem measurement plots with linear (top) and logarithmic (bottom) scale
for the 933~MHz reference PCs.}
}
\end{center}
\end{figure}

\begin{table}[hbt]
\begin{center}
\begin{tabular}{|l||c|c|c|}
\hline  
 & 733~MHz PC & 800~MHz PC & 933~MHz PC \\
\hline \hline
Level 1 Cache access time / ns & 4.1 & 3.8 & 3.2 \\
\hline
Level 2 Cache access time / ns & 13.6 $\pm$ 0.06 & 6.2 $\pm$ 0.06 & 5.3 $\pm$ 0.06 \\
\hline
Memory access time / ns & 127 $\pm$ 8 & 114 $\pm$ 14 & 113 $\pm$ 12 \\
\hline
\end{tabular}
\parbox{0.90\columnwidth}{
\caption[Cache and memory access times for the reference PCs.]{\label{Tab:ReferecePCCacheResults}The different cache and memory access times for the three different measurement PCs, 
all values are in nanoseconds. Errors give the approximate value ranges measured in the tests.}
}
\end{center}
\end{table}

All cache measurements have been executed with the operating system running in single CPU mode, as exchanging the processes between the two CPUs would 
influence and distort the results. The effects of the three different types of PCs can clearly be seen in the measured access times shown in 
Table~\ref{Tab:ReferecePCCacheResults}. The level 1 
cache access times scale very well with the CPUs' clock frequencies. Level 2 cache times also show good scaling with the respective level 2 clock 
frequencies, taking into account the factor of 2 in the 733~MHz PC. 
Memory access times are primarily influenced by the chipsets as the access times on the two similar motherboards are 
basically identical, irrespective of the CPU clock frequencies. These measured times will be used later in the scaling evaluations of the 
publisher-subscriber communication properties. 

One results of the cache and memory benchmarks is not fully understood so far. This is the linear behaviour of the last four points of each
curve between 16~B and 4~kB array sizes, with decreasing times for larger strides. These behaviours can best be observed in the logarithmic plots 
for each reference PC. Due to the displayed behaviour cache properties (e.g. cache size, cache line size, or associativity) are unlikely to be the causes
for this phenomenon. Possible explanations include pipelining or queueing effects in the processor, e.g. branch prediction or data forwarding effects. 
However, an exact explanation would require a very intimate and detailed knowledge of the internal processor architecture, which could not be obtained
as part of this thesis. A qualitatively similar behaviour can be observed on a HP V-Class machine, while differing effects in this array size 
range can be observed on an AMD Athlon, a Pentium 4, and a SUN Enterprise 10000. 
For the measurements the plateau results before the decrease have been used. 
The actual value displayed on the plateau corresponds very well to the documented \cite{PProPerformance} 3 cycle load latency for a L1 cache hit of the Intel PentiumPro
system architecture on which the Pentium III is based \cite{P3Implementation}.
%The second unexplained results is the fact that on the plateau (before the decrease discussed above) all three reference PCs require three 
%clock cycles to access a piece of data in the Level 1 cache. According to \cite{} though, only one cycle should be necessary for a L1 cache hit. 
%This effect also could not be explained satisfyingly as part of this thesis. 

\section{\label{Sec:PreciseAccounting}CPU Usage Measurements}

For several measurements in this chapter CPU loads on PCs running Linux had to be measured. During some of these tests
strange behaviours could be observed which could be traced after some careful examination to the method of timeslice
(and thus CPU usage) accounting in the Linux kernel \cite{PrecAccLinMagDE}, \cite{PrecAccLinMagUK}. These values are exported
by the kernel via its \texttt{/proc} interface and are used as the basis for all CPU accounting and usage programs, e.g. 
\texttt{top} or \texttt{xosview} as well as the MLUC monitoring classes in section~\ref{Sec:MonitoringClasses}. 
As part of \cite{PrecAccLinMagDE} a Linux kernel precise accounting patch \cite{PrecAccPatch} has been written which allows
a global as well as process CPU usage accounting using the time stamp counter of the processor. Using this patch CPU usage accounting
can thus be done on the granularity of the CPU's clock frequency, much more accurate than the default kernel accounting based on
a 10~ms timeslice granularity. This patch has been used to determine the CPU load in all measurements in this chapter.

\section{\label{Sec:NetworkReferenceTests}Network Reference Tests}

To determine the influence of the network hardware and the operating system on the TCP communication class benchmarks, a number 
of measurements have been performed
using a C program that directly accesses the socket API to perform TCP communication. The tests have been executed four times, for Fast and Gigabit Ethernet 
with and without the \texttt{TCP\_\-NO\-DE\-LAY} socket option set respectively. For each of these four test types the message sending latency as well as the 
throughput in the mode of a continuous stream of packets to a receiver have been measured. 
%that can be achieved by streaming blocks to a receiver 
In a preparatory measurement the number of blocks has been determined which is to be sent in a stream for each block size in the throughput tests.
This block count has been determined by sending varying numbers of 32~byte large blocks in a continuous stream to a receiver and plotting 
the achieved sending rate for each block count.
%As a preparatory measurement 
%the sending rate has been determined as a function of the block count, using 32~B large blocks. This test is used to determine the number of blocks sent
%for each block size in the throughput tests.

The tests have been performed on four pairs of the 800~MHz reference PCs examined in section~\ref{Sec:CacheMeasurements},
 using the PC's onboard {\em Intel EEPro 100} Fast Ethernet interfaces as well as {\em 3Com 3C996T} Gigabit Ethernet adapter cards (based
on a Broadcom chip) in
64~bit/66~MHz PCI slots on the boards. As the maximum transmission unit (MTU) 1500~B has been specified for both interfaces. 
For the Fast Ethernet interfaces the standard kernel drivers were 
used while for Gigabit Ethernet a driver supplied by Broadcom \cite{BroadcomDrivers}, \cite{Broadcom5700Driver} was used in version 2.2.19. 
For each measuring point 10 measurements have been made with the average used
as the result for that point. 

\subsection{TCP Network Reference Throughput}

\subsubsection{\label{Sec:NetRefPlateauTest}Plateau Determination}

To determine the block count for the throughput measurement varying numbers of blocks of the same size have been sent
in direct succession to a remote receiver. Blocks of 32~bytes are transmitted in a varying number from 1 to $2^{23}$ (8388608 / 8~M). 
To calculate the sending rate the time required for all blocks to be sent has been measured as
the program's main output. 
The expected shape of the resulting curve is a rise that
flattens to slowly approach an asymptotic value. Actual obtained results are shown in Fig.~\ref{Fig:TCP-Ref-Count-Rates}.
None off the four tests displays the expected behaviour. The one that most closely follows the predicted form 
is the Gigabit Ethernet test without the \texttt{TCP\_\-NO\-DE\-LAY} option. Instead each test shows the same approximate form of its curve,
a steep rise to a maximum value followed by a decrease that levels off to approach an asymptotic value for large message sizes. For the
two tests with the \texttt{TCP\_\-NO\-DE\-LAY} option set the decreases even reach a minimum and rise again slighty towards their
asymptotic values. The respective peak and asymptotic values for the achieved sending rate are shown in Table~\ref{Tab:NetworkReferencePlateauResults} together with the
block count for each rate's peak value. An interesting fact observed is that both pairs of Gigabit and Fast Ethernet
tests reach approximately the same asymptotic value for large counts, showing no effect of the \texttt{TCP\_\-NO\-DE\-LAY} option for large block counts. This 
behaviour is explained by the fact that the Linux Kernel only evaluates this option if specific preconditions are met, e.g. all large blocks queued have been
actually sent. The results indicate that these preconditions are met only rarely in these tests, thereby reducing the option's effect on the measurements. 

One immediate fact that can be derived from this measurement is the overhead involved in doing a  write call for a 32~byte block (or message). As the overhead for 
writing the actual amount of data can be neglected, this approximates the minimal overhead of a write call. The overhead can be determined by
taking the inverse of the measured maximum sending rate. For Gigabit Ethernet this is approximately 1.8~$\mu \mathrm s$ and for Fast Ethernet it is about 
2.7~$\mu \mathrm s$. Note that this overhead only definitely includes the TCP protocol overhead until the network packet has been generated and enqueued
for transmission. The actual transmission until the packet reaches the physical network medium will mostly not be included here. 

\begin{figure}[ht!p]
\begin{center}
\resizebox*{0.50\columnwidth}{!}{
\includegraphics{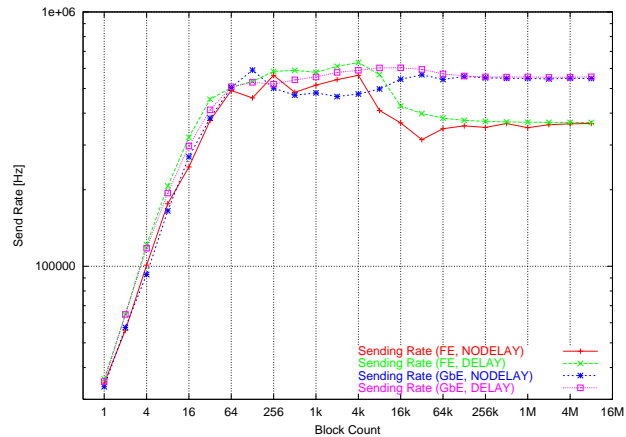}
}
\parbox{0.90\columnwidth}{
\caption{\label{Fig:TCP-Ref-Count-Rates}The measured network reference sending rates as a function of the block count.}
}
\end{center}
\end{figure}

\begin{table}[hbt]
\begin{center}
\begin{tabular}{|l||c|c|c|}
\hline  
Test Type & Peak Value / & Peak Value  & Asymptotic Value / \\
          & Hz           & (Block Count) & Hz                 \\
\hline \hline
Fast Ethernet              & 563600 & 4096  & 364000 \\
with \texttt{TCP\_\-NO\-DE\-LAY} &        & (4~k) &        \\
\hline
Fast Ethernet                 & 633000 & 4096  & 367800 \\
without \texttt{TCP\_\-NO\-DE\-LAY} &        & (4~k) &       \\
\hline
Gigabit Ethernet              & 590900 & 128 & 547000 \\
with \texttt{TCP\_\-NO\-DE\-LAY}    &        &     & \\
\hline
Gigabit Ethernet              & 603700 & 16384  & 555000\\
without \texttt{TCP\_\-NO\-DE\-LAY} &        & (16~k) & \\
\hline
\end{tabular}
\parbox{0.90\columnwidth}{
\caption[Network reference sending rate as a function of block count results.]{\label{Tab:NetworkReferencePlateauResults}Results obtained from measuring block sending rate as a function of block count in the
network reference tests.}
}
\end{center}
\end{table}

\subsubsection{Plateau Throughput Measurement}

A message count of 524288 (512~k) blocks has been chosen for the plateau throughput measurements as all four tests
have approached their asymptotic plateau value closely at this count in the prerequisite measurement. 
The results obtained from these tests are displayed in 
Fig.~\ref{Fig:TCP-Ref-Plateau-Rates} to~\ref{Fig:TCP-Ref-Plateau-CyclesPerNetBW}.

%\begin{figure}[hbt]
\begin{figure}[ht!p]
\begin{center}
\resizebox*{0.50\columnwidth}{!}{
\includegraphics{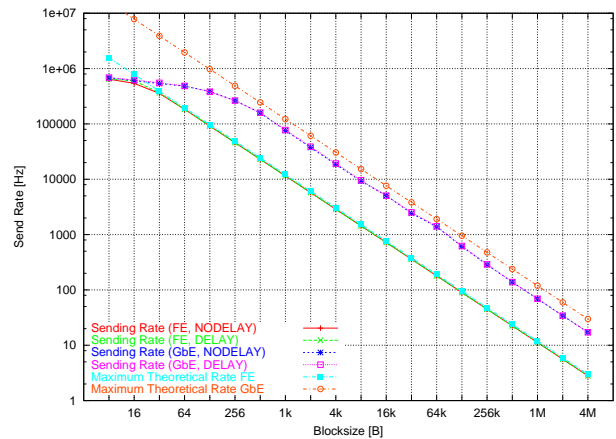}
}
\parbox{0.90\columnwidth}{
\caption[The measured network reference sending rates (plateau).]{\label{Fig:TCP-Ref-Plateau-Rates}The measured network reference sending rates (block count 512~k).}
}
\end{center}
\end{figure}

Fig.~\ref{Fig:TCP-Ref-Plateau-Rates} displays the block sending rate achieved in the tests. As can be seen the 
Fast and Gigabit Ethernet test pairs are almost identical with only slight deviations at small message sizes. Starting at about 64~B for Fast Ethernet (FE)
and 512~B for Gigabit Ethernet (GbE) the curves become basically linear with identical slopes. Both sets 
differ by about a factor of 6. The similarity of each pair can be explained by the same reasoning as for the limited effect of the \texttt{TCP\_\-NO\-DE\-LAY}
option in the first measurement in this section. In theory the difference between the two sets for Gigabit Ethernet (1~Gbps) and Fast Ethernet (100~Mbps) should be 
a factor of 10. The reason why this factor is only 6 is that the used GbE cards are not able to saturate the GbE link, whereas the 
FE cards are able to saturate their link. This is also shown in comparison with the curves showing the maximum theoretical sending rate for each of the two networks.
For FE the measured sending rate approaches the theoretical limit very quickly. In contrast for GbE the limit is approached for larger blocks and the measured rate is 
limited by another factor, as it runs parallel to but does not approach the theoretical curve. The factor between the theoretical and the measured curves is about 1.7, 
indicating that the used GbE adapter cards only utilize about 60~\% of the network's theoretical bandwidth.

\begin{figure}[ht!p]
\begin{center}
\resizebox*{0.50\columnwidth}{!}{
\includegraphics{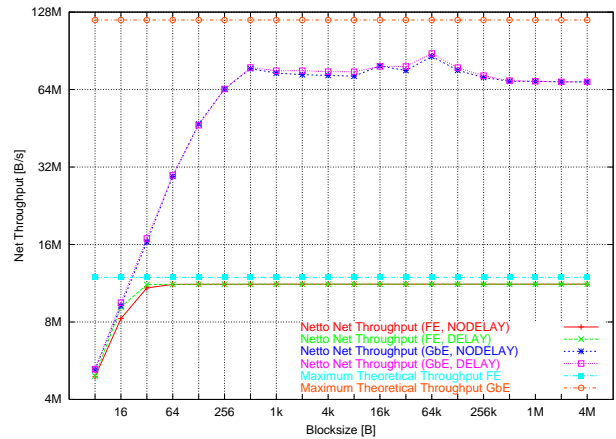}
}
\parbox{0.90\columnwidth}{
\caption[The application level network reference throughput (plateau).]{\label{Fig:TCP-Ref-Plateau-NetBW}The application level network reference throughput (block count 512~k).}
}
\end{center}
\end{figure}

Test results obtained for the network throughput can be seen in Fig.~\ref{Fig:TCP-Ref-Plateau-NetBW}. They
have been obtained by multiplying the measured sending rate with the respective block sizes so that they 
%do not differ in principle from the results obtained from those tests. As such they also 
reflect application level throughput that can be achieved by a program. In analogy with the rate test results, 
the initial rise levels off to a constant plateau at 64~B for FE and 512~B for GbE, and the same approximate factor of 6  between the two sets becomes apparent. 
%One of two effects of both Gigabit Ethernet tests that could not be observed so clearly in the rate curves are deviations between 16~kB and 64~kB blocks.
Two effects of both Gigabit Ethernet tests could not be observed so clearly in the rate curves. The first of these are deviations between 16~kB and 64~kB blocks.
In this interval they rise above the 
surrounding plateau level and display a peak at 64~kB blocks of around 86~MB/s. The second effect is that the network throughput actually drops slightly with increasing
block sizes, going from 77~MB/s for 512~B blocks to 68~MB/s for 4~MB blocks. This seems to indicate that for Fast Ethernet the limit is set by the saturated network link, while
for Gigabit some effect on the PC, e.g. from the network card, the operating system, or the PCs memory, limits the throughput. This can again be seen in comparison with
the theoretically achievable throughputs. For FE one can see in this figure that in the plateau the measured throughput is about 94~\% of the theoretical maximum. The
missing 6~\% are due to the TCP/IP protocol overhead, the protocol headers for each network packet which also require some of the available bandwidth. For GbE in contrast
only between about 72~\% and about 58~\% of the available bandwidth are used. Accounting for the 6~\% TCP protocol overhead, as determined from FE, one can see that 
with the GbE cards used in the test between 22~\% and 36~\% of the available bandwidth are not used.

\begin{figure}[ht!p]
\begin{center}
\resizebox*{1.0\columnwidth}{!}{
\includegraphics{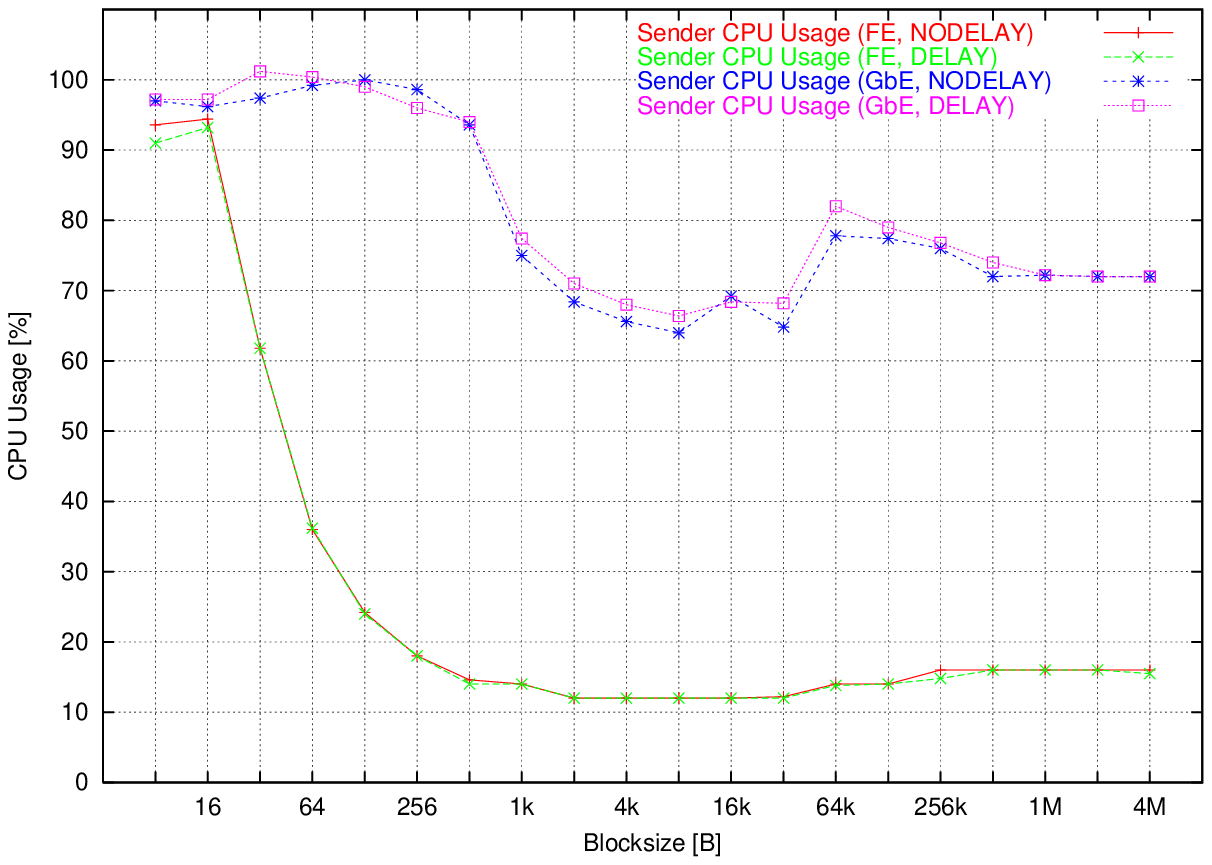}
\hfill
\includegraphics{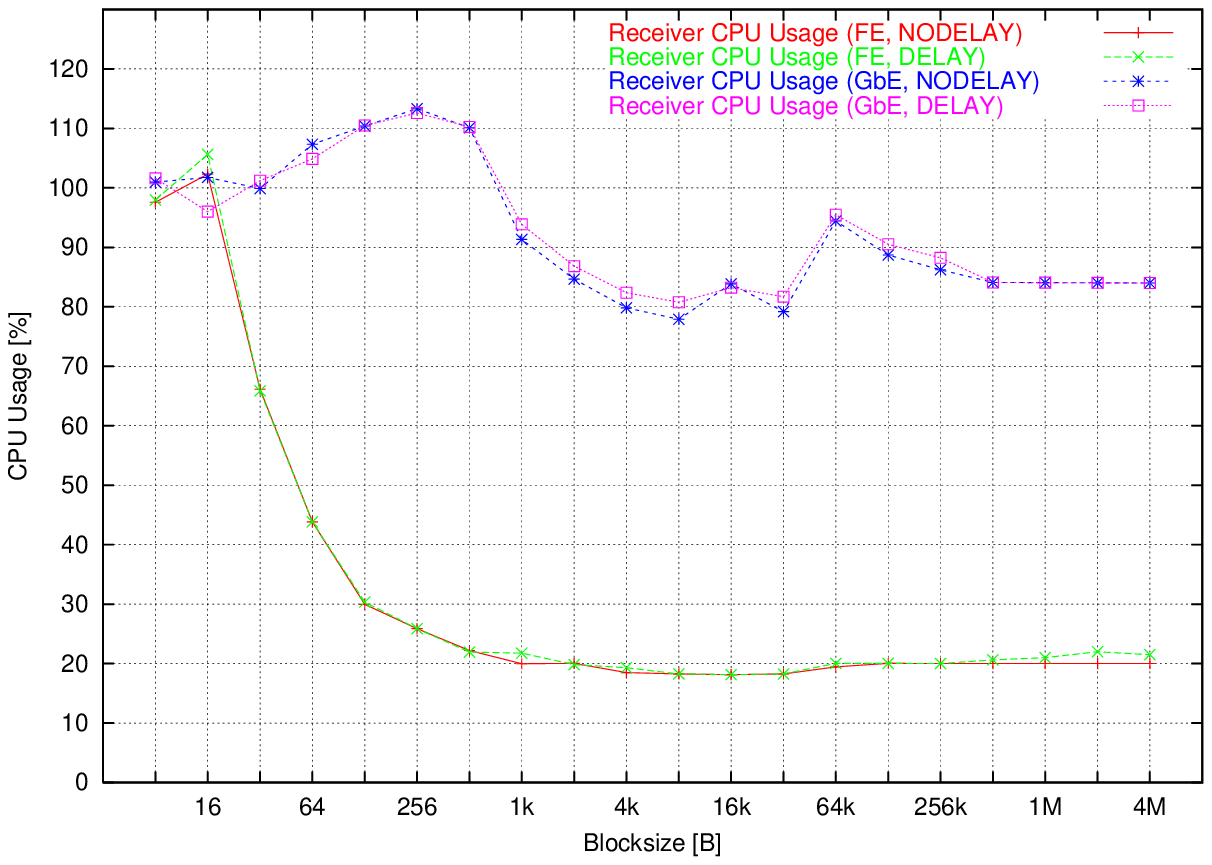}
}
\parbox{0.90\columnwidth}{
\caption[CPU usage during TCP reference sending (plateau).]{\label{Fig:TCP-Ref-Plateau-Cycles}The CPU usage on the sender (left) and receiver (right) 
during TCP reference message sending (block count 512~k).
The nodes are twin CPU nodes, 100~\% CPU usage corresponds to one CPU being fully used.}
}
\end{center}
\end{figure}

CPU usage measured during the tests is shown in Fig.~\ref{Fig:TCP-Ref-Plateau-Cycles}, on the left hand side for the sender and on the right for 
the receiver. The first fact to become apparent is the identity of the sets of Fast and Gigabit Ethernet respectively, with the FE curves lying 
practically on top of each other. For Fast Ethernet on the sender as well as on the receiver the shape of the curve is an initial steep decrease that levels off
to an almost flat plateau, with only a slight ``bathtub'' minimum at its center. An exception to the initial decrease
are the 8~B block measurements whose results are slightly lower than the ones of 
the following 16~B blocks. As can be expected, absolute usage values on the receiver are slightly higher than those on the sender. Maximum values are at 102~\% and 96~\% for
16~B blocks  on  receiver and sender respectively
and minimum values at 18~\% between 4~kB and 32~kB and 12~\% between 2~kB and 32~kB. 
For Gigabit Ethernet the absolute values are higher and the ``bathtub'' shape is more pronounced. Maximum, minimum,
and final values on receiver and sender respectively are at 114~\%, 78~\%, and 84~\% and 100~\%, 64~\%, and 72~\% respectively. Corresponding block
sizes are 256~B, 8~kB, 4~MB, 128~B, 8~kB, and 4~MB. 
In these GbE test results the curves on both nodes also display irregularities between 16~kB and 64~kB, 
with the 64~kB values being at a local maximum. 
One result from these tests that has to be considered is the fact
that the CPU usage during the GbE test reaches values larger than 100~\%. This means that a single CPU computer will be fully saturated in these parts of the 
test and will not be able to reach the data transfer rates displayed in Fig~\ref{Fig:TCP-Ref-Plateau-Rates} and~\ref{Fig:TCP-Ref-Plateau-NetBW}.

\begin{figure}[ht!p]
\begin{center}
\resizebox*{1.0\columnwidth}{!}{
\includegraphics{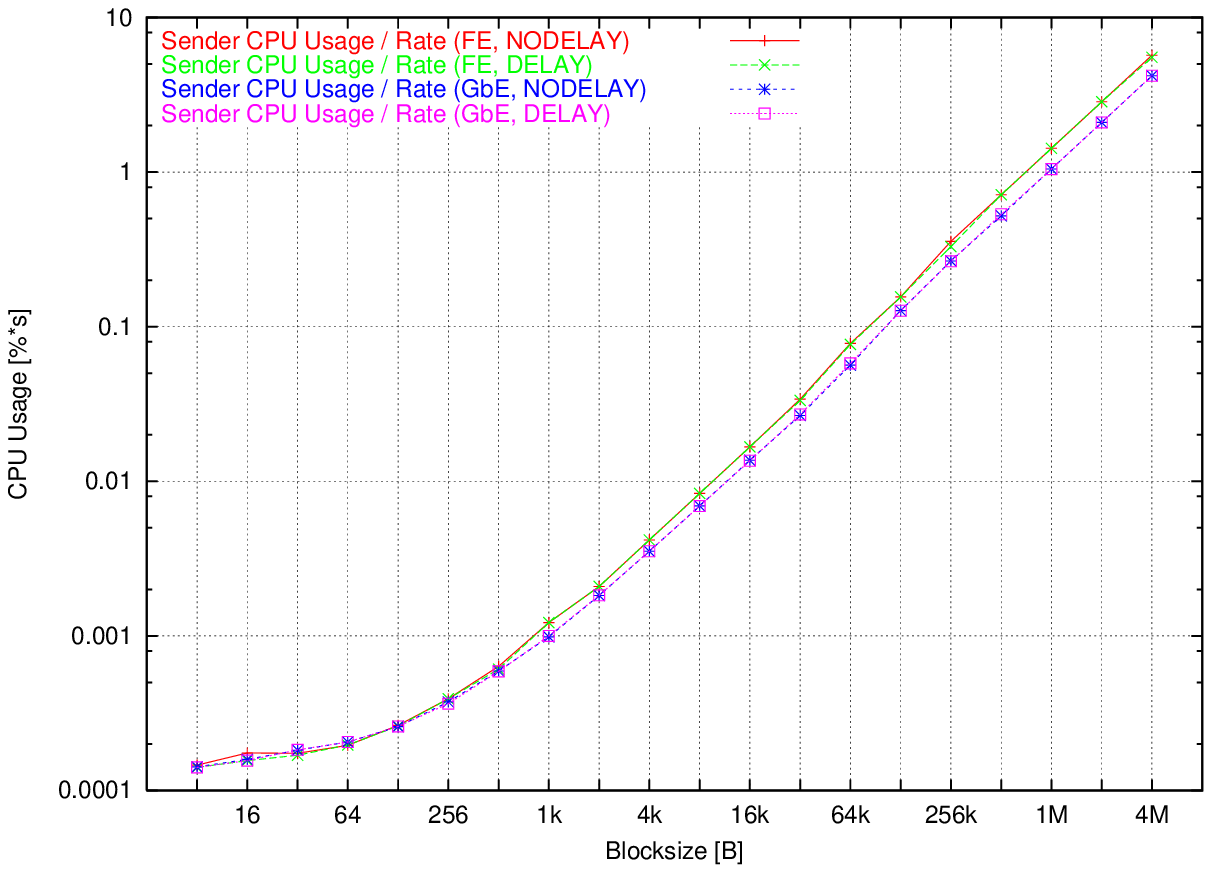}
\hfill
\includegraphics{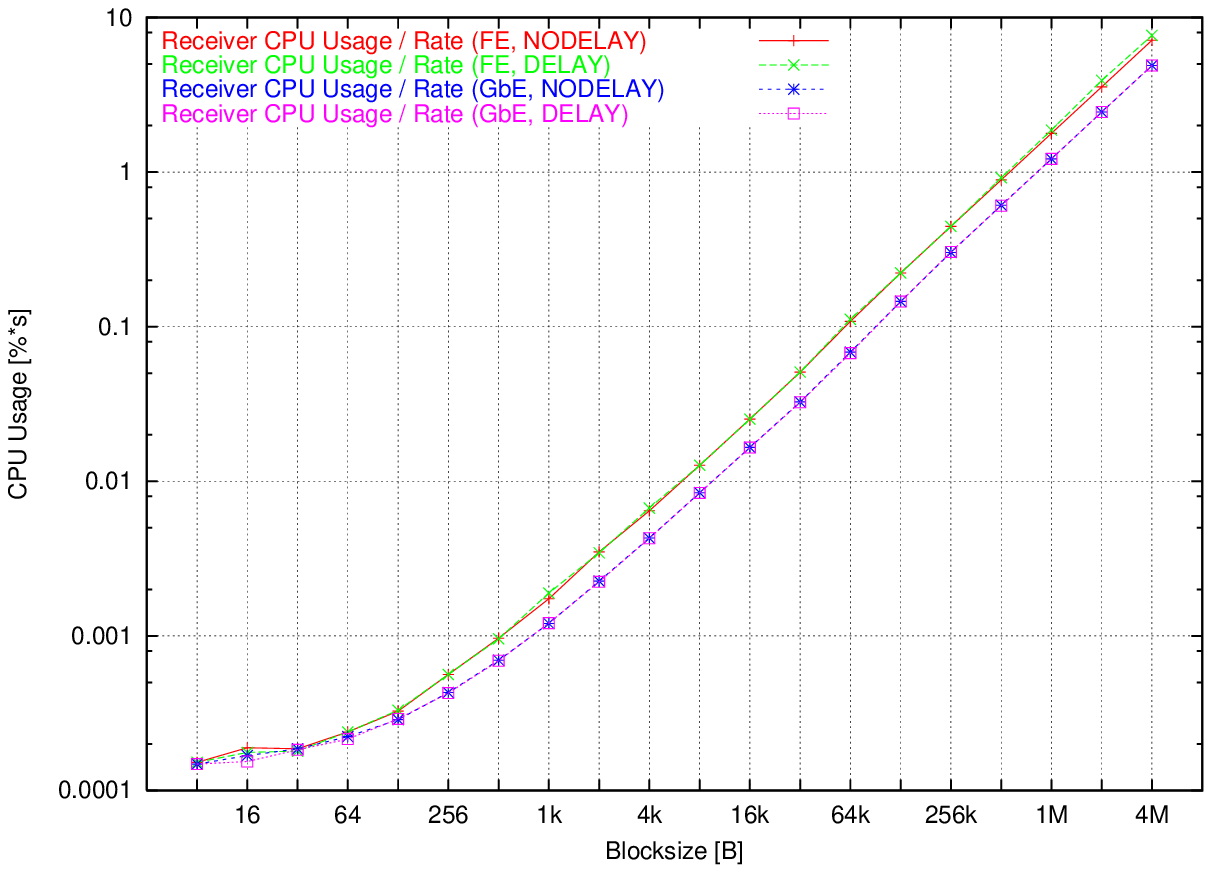}
}
\parbox{0.90\columnwidth}{
\caption[CPU usage divided by the sending rate during TCP reference sending (plateau).]{\label{Fig:TCP-Ref-Plateau-CyclesPerRate}The CPU usage on the 
sender (left) and receiver (right) divided by the sending rate 
during TCP reference message sending (block count 512~k).
The nodes are twin CPU nodes, 100~\% CPU usage corresponds to one CPU being fully used.}
}
\end{center}
\end{figure}

For a better comparison CPU usages of the different tests have been divided by the respective sending rates and network throughputs, with
the results displayed in Fig.~\ref{Fig:TCP-Ref-Plateau-CyclesPerRate} and Fig.~\ref{Fig:TCP-Ref-Plateau-CyclesPerNetBW}. 
The first of these two is a measure for the CPU overhead per send call or per message, while the second one measures the
overhead per transferred byte/s. 
In the plots in Fig.~\ref{Fig:TCP-Ref-Plateau-CyclesPerRate} showing the usage relative to the 
sending rate  no difference can be observed between the two pairs of FE and GbE curves each. 
In a comparison of the two different sets the Fast and Gigabit Ethernet curves are almost identical for small messages. At about 1~kB on the sender and 128~B on the receiver
the curves start to diverge and the values for GbE become smaller and thus better. The difference for the largest blocks is about a factor of 1.35 on the sender 
and 1.45 on the receiver. One can see that the minimum CPU overhead on the sender for each message (per second) is 
about $\mathrm{1.4 \times 10^{-4}~\%}$ for the smallest messages, 
increasing to about $\mathrm{5.7~\%}$ and $\mathrm{4.2~\%}$ for the largest messages on Fast and Gigabit Ethernet respectively. On the receiver the values are about
$\mathrm{1.5 \times 10^{-4}~\%}$ and  $\mathrm{7.1~\%}$ for FE and $\mathrm{1.5 \times 10^{-4}~\%}$ and  $\mathrm{4.9~\%}$ for GbE.
Using the results for a specific block size it is also possible to calculate the expected CPU usage resulting from transferring blocks of that size
with a given rate.

\begin{figure}[ht!p]
\begin{center}
\resizebox*{1.0\columnwidth}{!}{
\includegraphics{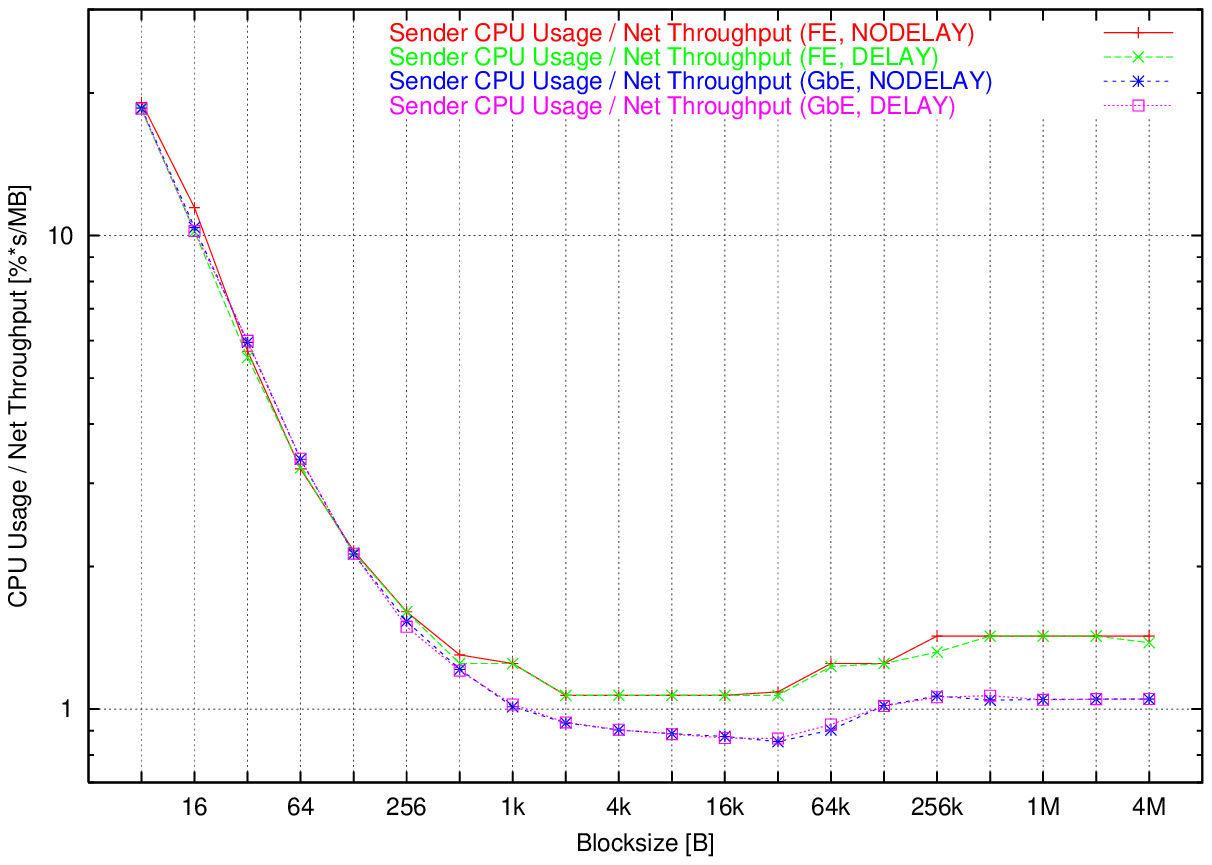}
\hfill
\includegraphics{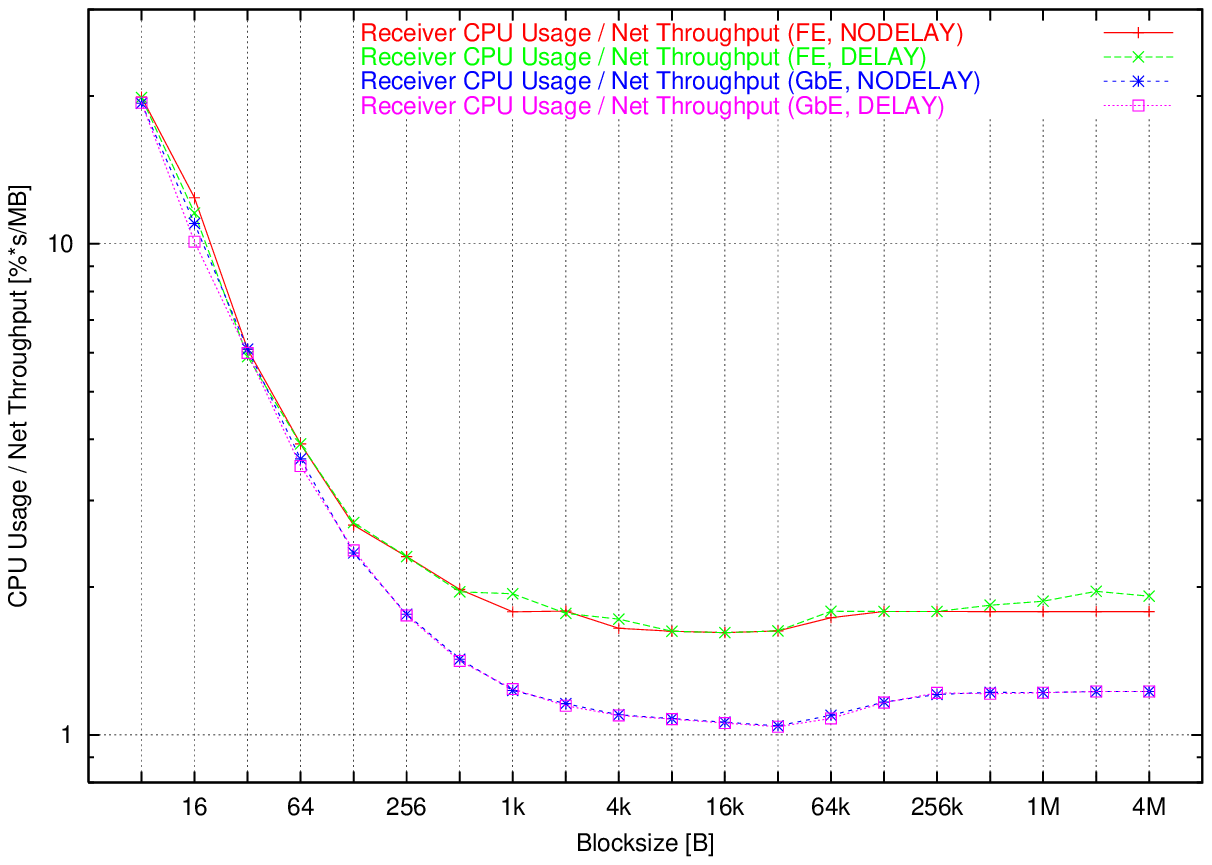}
}
\parbox{0.90\columnwidth}{
\caption[CPU usage per MB/s network throughput during TCP reference sending (plateau).]{\label{Fig:TCP-Ref-Plateau-CyclesPerNetBW}The CPU usage on the sender (left) and receiver (right) per MB/s network throughput 
during TCP reference message sending (block count 512~k).
The nodes are twin CPU nodes, 100~\% CPU usage corresponds to one CPU being fully used.}
}
\end{center}
\end{figure}

The graphs in Fig.~\ref{Fig:TCP-Ref-Plateau-CyclesPerNetBW} display CPU usage normalized with achieved throughput. They show
that up to 512~B blocks on the sender and about 64~B on the receiver the four respective curves are basically identical. They
start to diverge for increasing block sizes, 
with the Gigabit Ethernet curves at lower values than the Fast Ethernet ones. Seven of the eight curves exhibit an initial sharp decrease,
that flattens to a minimum in a slight ``bathtub'' and then develops into a plateau. The exception is Fast Ethernet
without the \texttt{TCP\_\-NO\-DE\-LAY} option on the receiver. In this test a pronounced plateau does not set in at all,
just an indication of it is showing at the last two block sizes of 2~MB and 4~MB. 
From these results one obvious conclusion can be drawn: Gigabit Ethernet is more efficient than Fast Ethernet concerning its use of CPU cycles
per megabyte of transferred data per second. In a second conclusion one can see that, since some of the measured values are larger than
$\mathrm{1~\%/(MB/s)}$, a single CPU machine will be fully busy already at values below 100~MB/s, before a GbE link will be
saturated. A third result to be deduced is that it is not necessarily the most efficient approach to send at the largest possible 
block sizes. The minimum CPU overhead per byte transferred can be found at the medium block sizes, between about 2~kB and 32~kB on the sender
and 8~kB and 32~kB on the receiver.

\subsubsection{Peak Throughput Measurement}

In addition to measuring the network transfer characteristics at the plateau values the same measurements should be performed at those block counts where the 
curves in Fig.~\ref{Fig:TCP-Ref-Count-Rates} show their peak values. 
Unlike the throughput plateau the peak values of the four curves do not show at identical message counts. Block counts of 4~k, 4~k,
128, and 16~k therefore have been used for Fast Ethernet with and without the \texttt{TCP\_\-NO\-DE\-LAY} option and Gigabit Ethernet with and without \texttt{TCP\_\-NO\-DE\-LAY}
respectively, as determined in section~\ref{Sec:NetRefPlateauTest}.
Results obtained from these tests are shown in Fig.~\ref{Fig:TCP-Ref-Peak-Rates} to~\ref{Fig:TCP-Ref-Peak-CyclesPerNetBW}. 

%\begin{figure}[hbt]
\begin{figure}[ht!p]
\begin{center}
\resizebox*{0.50\columnwidth}{!}{
\includegraphics{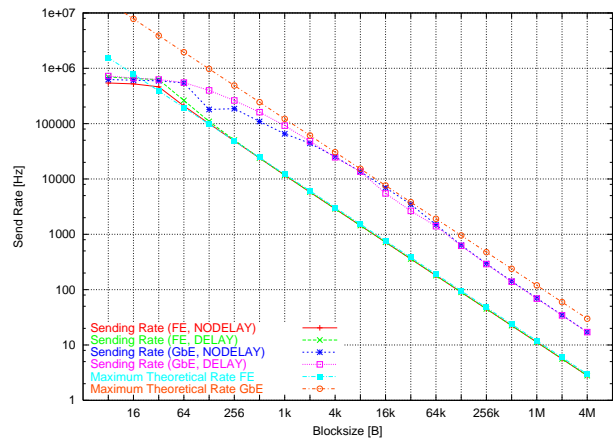}
}
\parbox{0.90\columnwidth}{
\caption[The measured network reference sending rates (peak).]{\label{Fig:TCP-Ref-Peak-Rates}The measured network reference sending rates (block counts 4~k, 4~k, 128, and 16~k).}
}
\end{center}
\end{figure}

Measured rates of these tests are shown in Fig.~\ref{Fig:TCP-Ref-Peak-Rates}. 
As can be seen, the Gigabit Ethernet rates are slightly higher  than for the respective plateau test, although more deviations
are present. Particularly noticeable is the drop at 128~B for the \texttt{TCP\_\-NO\-DE\-LAY} GbE test. For Fast Ethernet only
the values around 32~B for the \texttt{TCP\_\-NO\-DE\-LAY} test and the small block measurements of the test without  the \texttt{TCP\_\-NO\-DE\-LAY} option have
gained discernibly from the changed block count. Up to about 256~B the FE test using the \texttt{TCP\_\-NO\-DE\-LAY} option is at lower values compared
to the test without the option. Comparing measured and theoretical rates one can see that for GbE the theoretical limit is 
approached earlier and closer than in the plateau tests. For FE one can notice that for 32~B to 128~B block sizes the theoretical limit is actually
exceeded by the measured curve. This result indicates that the blocks written can all be stored in local buffers, by the operating system, the network card, or both,
and do not immediately reach the physical network medium. Only for larger blocks are the local buffers exceeded and the packets reach the network. This could also
explain the performance increase for Gigabit Ethernet, in particular for the curve with the \texttt{TCP\_\-NO\-DE\-LAY} option set, as this measurement
uses a very small block count of only 128. 
%The FE test without the \texttt{TCP\_\-NO\-DE\-LAY} option is approximately at the same values as in the plateau test. 

\begin{figure}[ht!p]
\begin{center}
\resizebox*{0.50\columnwidth}{!}{
\includegraphics{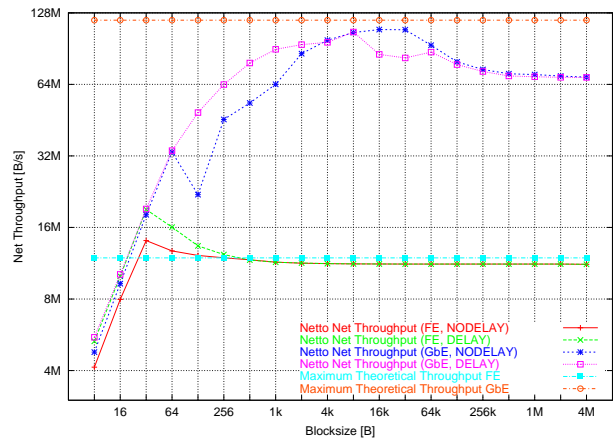}
}
\parbox{0.90\columnwidth}{
\caption[The application level network reference throughput for TCP sending (peak).]{\label{Fig:TCP-Ref-Peak-NetBW}The application level network reference throughput for TCP sending (block counts 4~k, 4~k, 128, and 16~k).}
}
\end{center}
\end{figure}

In the network throughput plots in Fig.~\ref{Fig:TCP-Ref-Peak-NetBW} the differences between the plateau and peek tests are more
clearly pronounced. For Fast Ethernet a peak at 32~B drops towards the same plateau as in the previous test, reached between 512~B and 1~kB
large blocks. Up to about 512~B the \texttt{TCP\_\-NO\-DE\-LAY} measurements provide values below the ones obtained without using the option. The Gigabit Ethernet 
curves for the \texttt{TCP\_\-NO\-DE\-LAY} test also display the drop at 128~B. Before that drop the values are above the ones from the plateau measurement and
afterwards below, up to between 1~kB and 2~kB. At larger values a local maximum is present with a peak at about 32~k, and from about 128~kB on the
two test's curves are again basically identical. For the GbE test without the \texttt{TCP\_\-NO\-DE\-LAY} option set the values are higher than for the plateau test up to about
32~kB. Between 1~kB and 8~kB the peak results are considerably higher in the local maximum present in the peak test, the difference for the values at 8~kB 
is 106~MB/s compared to 75~MB/s. The comparison of the measured and theoretical curves show more clearly that the FE measurements partly exceed the theoretical limits. 
One can also see that the GbE curves approach their network limit much more closely than in the plateau test. This behaviour though may be just a measuring artifact, 
similar to the detailed FE ``{\em superperformance}''. 

%In a comparison of Fig.~\ref{Fig:TCP-Ref-Peak-Rates} and~\ref{Fig:TCP-Ref-Peak-NetBW} with Fig.~\ref{Fig:TCP-Ref-Rates} and~\ref{Fig:TCP-Ref-NetBW} respectively,
%one can see that the peak tests display higher achieved rates and throughput values for smaller block sizes, before the respective plateau limit is reached.
%This increase is particularly noteable in the Fast Ethernet 32~B block measurement and the Gigabit Ethernet 8~kB to 32~kB measurements. 

\begin{figure}[ht!p]
\begin{center}
\resizebox*{1.0\columnwidth}{!}{
\includegraphics{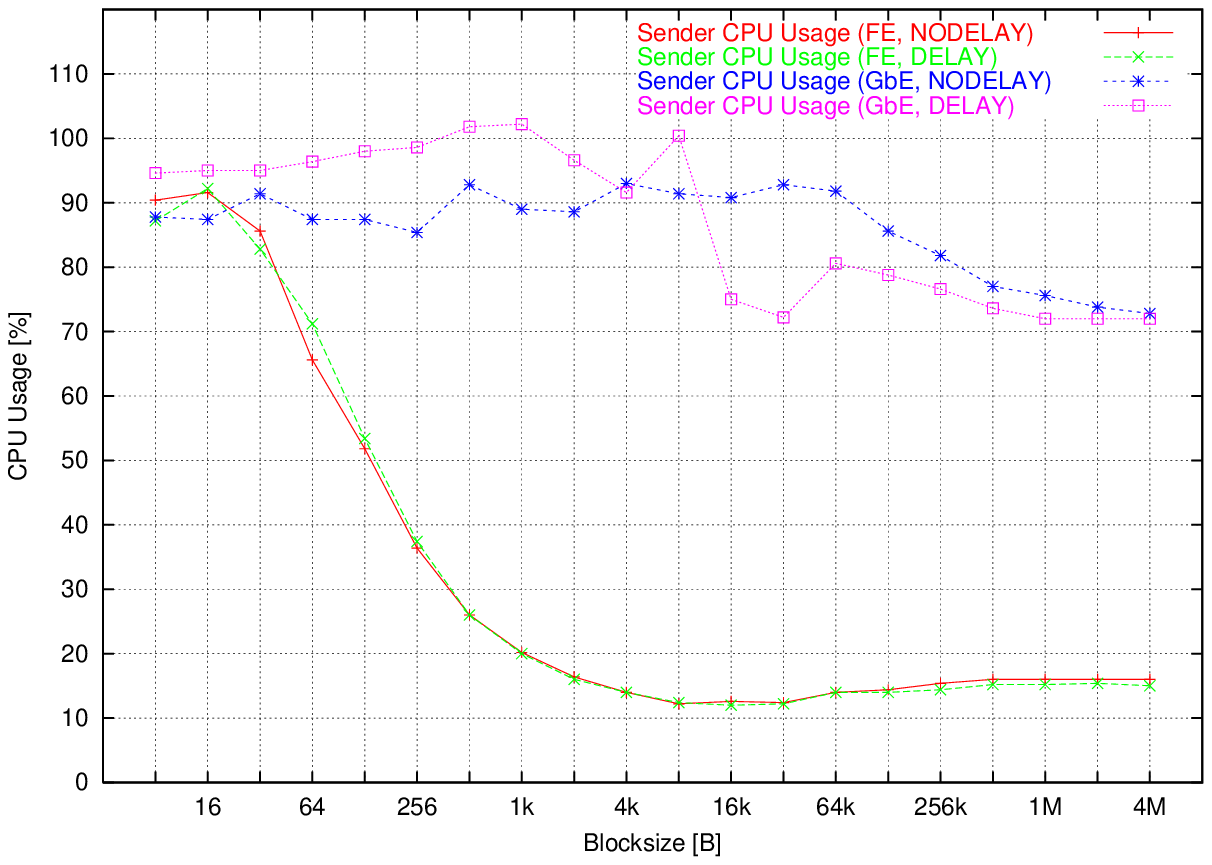}
\hfill
\includegraphics{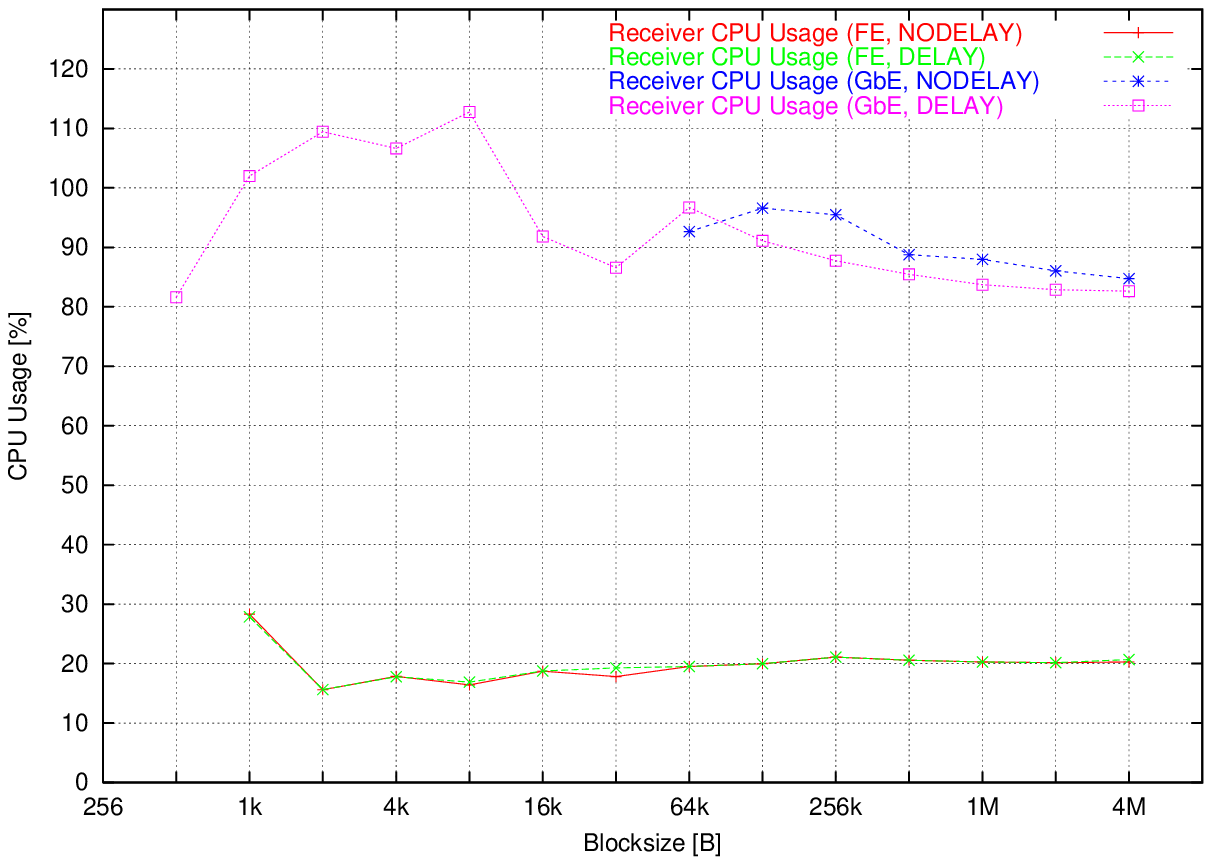}
}
\parbox{0.90\columnwidth}{
\caption[CPU usage during TCP reference sending (peak).]{\label{Fig:TCP-Ref-Peak-Cycles}The CPU usage on the sender (left) 
and receiver (right) during TCP reference sending (block counts 4~k, 4~k, 128, and 16~k).
The nodes are twin CPU nodes, 100~\% CPU usage corresponds to one CPU being fully used.}
}
\end{center}
\end{figure}

CPU usage for the sender and receiver is shown in Fig.~\ref{Fig:TCP-Ref-Peak-Cycles}. On the receiver the measured values for small blocks are not shown because 
of large inaccuracies in the respective measurements.
These inaccuracies are caused  
by the method of measuring the usage combined with the short running times of the tests due to the small block counts. For FE the limit seems to be between 1~kB and 2~kB,
while for GbE measurements up to 32~kB appear to be unreliable. At larger block sizes both FE and GbE values are basically
identical to the ones from the plateau tests (Fig.~\ref{Fig:TCP-Ref-Plateau-Cycles}). On the sending node the drop of the Fast Ethernet curves is less steep so that for block sizes
below 8~kB the usage is higher than for the corresponding plateau tests. GbE results on the sender are fairly irregular and depending 
on the block size can be higher or lower compared
to the corresponding results from the plateau measurements. Neither curve displays the ``bathtub'' shape as clearly as in the plateau test curves.

\begin{figure}[ht!p]
\begin{center}
\resizebox*{1.0\columnwidth}{!}{
\includegraphics{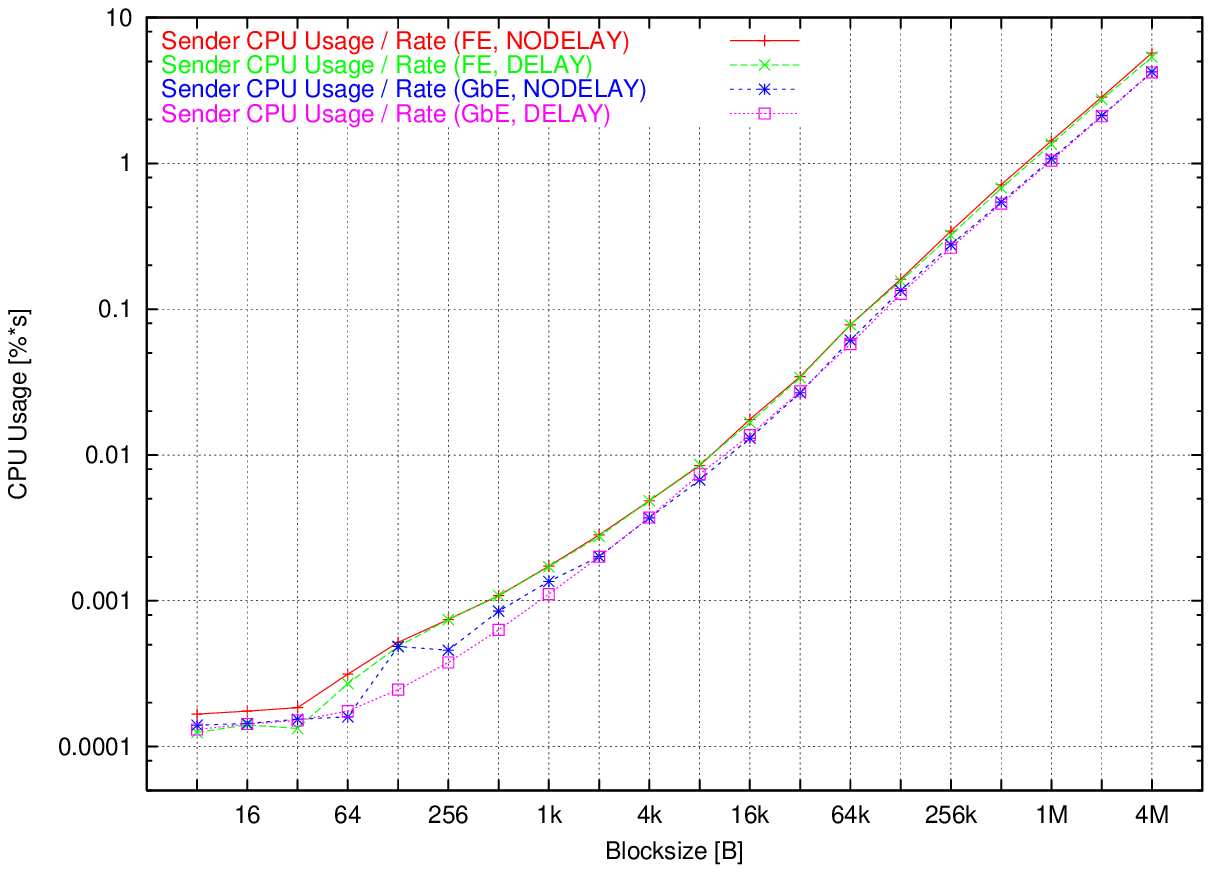}
\hfill
\includegraphics{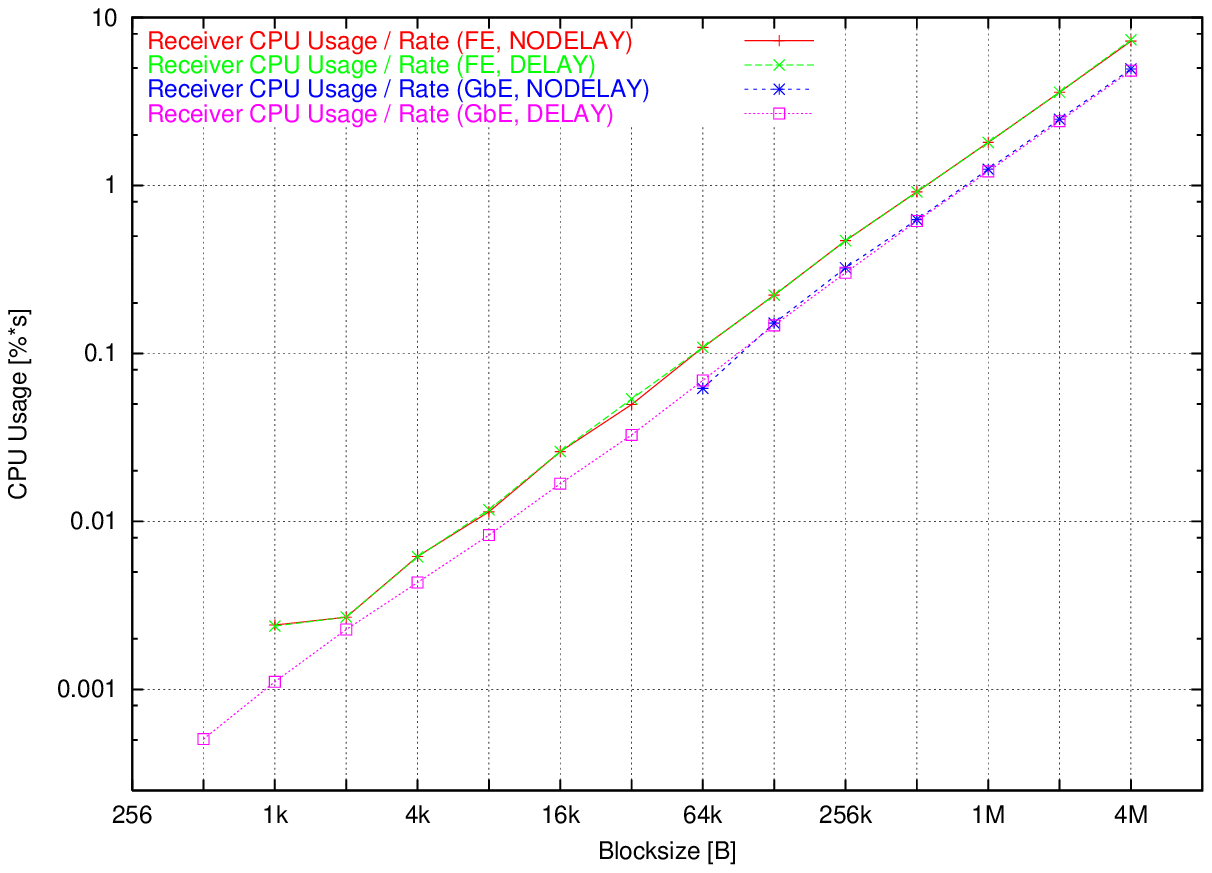}
}
\parbox{0.90\columnwidth}{
\caption[CPU usage divided by the sending rate during TCP reference sending (peak).]{\label{Fig:TCP-Ref-Peak-CyclesPerRate}The CPU usage on the sender (left) and receiver (right) divided by the sending rate 
during TCP reference sending (block counts 4~k, 4~k, 128, and 16~k).
The nodes are twin CPU nodes, 100~\% CPU usage corresponds to one CPU being fully used.}
}
\end{center}
\end{figure}

\begin{figure}[ht!p]
\begin{center}
\resizebox*{1.0\columnwidth}{!}{
\includegraphics{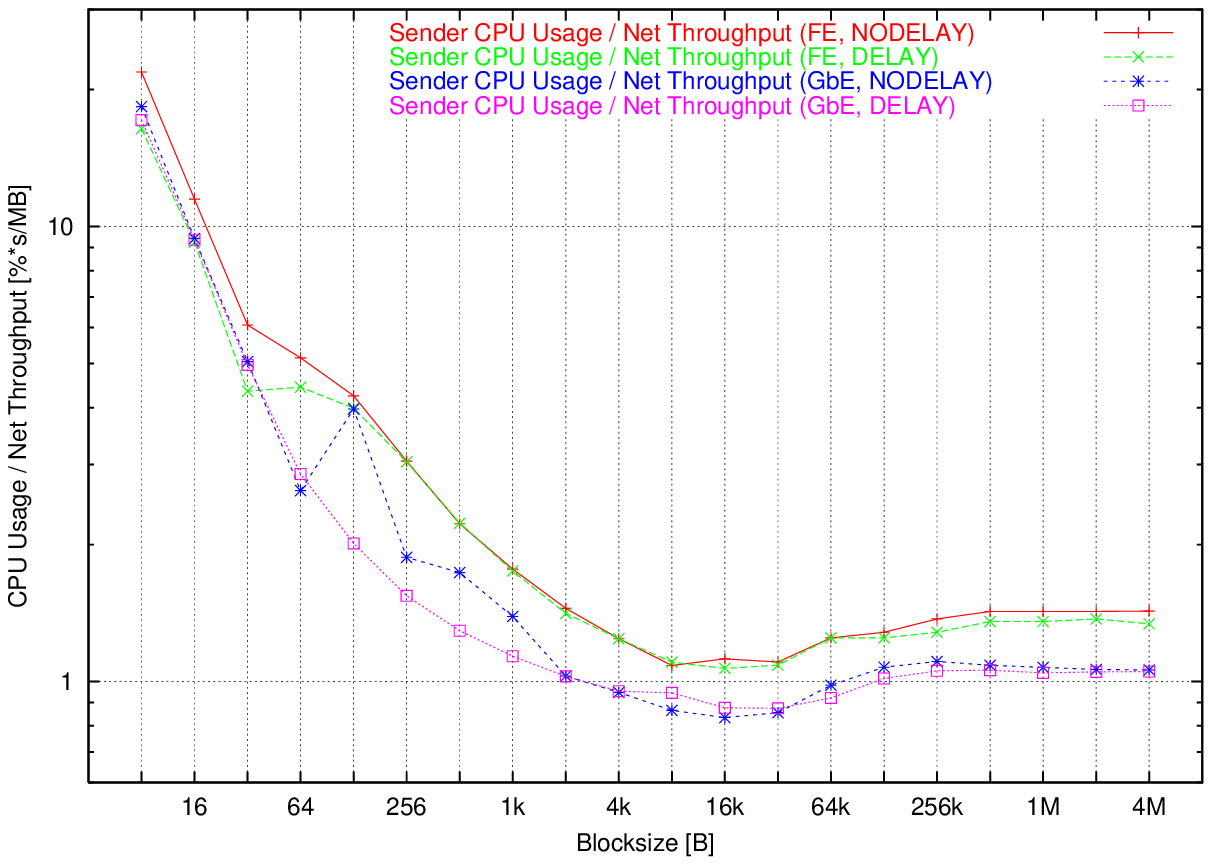}
\hfill
\includegraphics{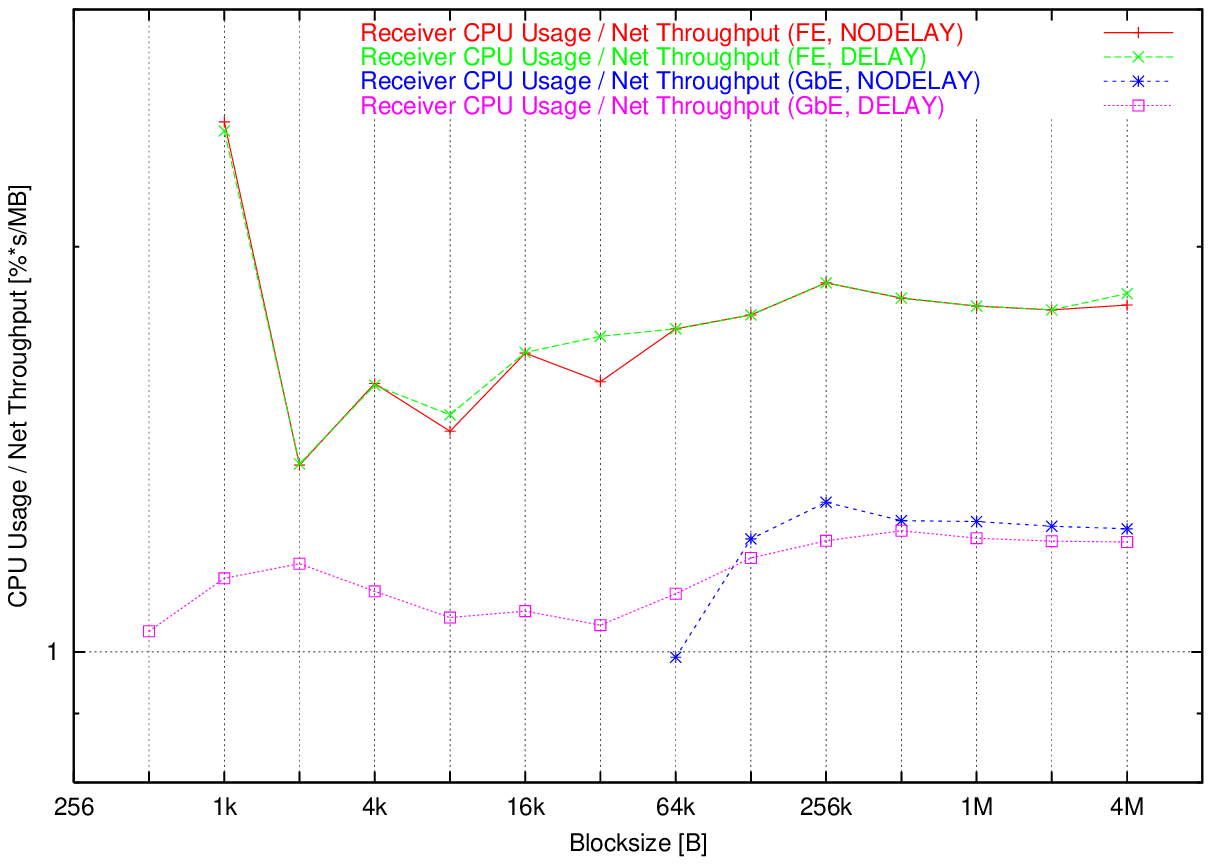}
}
\parbox{0.90\columnwidth}{
\caption[CPU usage per MB/s network throughput during TCP reference sending (peak).]{\label{Fig:TCP-Ref-Peak-CyclesPerNetBW}The CPU usage 
on the sender (left) and receiver (right) per MB/s network throughput 
during TCP reference sending (block counts 4~k, 4~k, 128, and 16~k).
The nodes are twin CPU nodes, 100~\% CPU usage corresponds to one CPU being fully used.}
}
\end{center}
\end{figure}

In the plots of CPU usage normalized to the sending rate and network throughput, respectively in Fig.~\ref{Fig:TCP-Ref-Peak-CyclesPerRate} 
and \ref{Fig:TCP-Ref-Peak-CyclesPerNetBW}, the results of the CPU usage test are reflected partially. 
On the receiver the measurements up to block sizes of about 2~kB for FE and 64~kB for GbE indicate unreliable values. With increasing block sizes the GbE results are 
approximately identical to the ones from the plateau tests, while the FE results display a slightly irregular behaviour, although at lower values than in
the plateau test. The previous tests' values are only approached for block sizes above 256~kB. On the sender the GbE curves, in particular the 
one from the \texttt{TCP\_\-NO\-DE\-LAY} measurement, display erratic behaviour for small message sizes, which  as
on the receiver might also be caused by measurement inaccuracies. For larger blocks the curves again become 
basically identical to the ones from the plateau test. For FE the \texttt{TCP\_\-NO\-DE\-LAY} curve is at higher values than its
counterpart without the option up to 128~B blocks. At higher block sizes they become identical and display higher values than the 
respective plateau test results. 
Starting between 8~kB and 16~kB the FE peak test curves also become basically identical to the ones from the plateau throughput tests. 
%The peak test values being lower than the plateau values is most likely due to the 
Therefore, as far as the CPU efficiency is concerned, there is no significant advantage over the plateau tests, the differences on the sending
and receiving nodes should approximately balance.

\subsection{TCP Network Reference Latency}

As the final network reference test the message latency has been determined by sending messages between an originating sender and a receiver.
The sender transmits a number of messages to the receiver and waits for an identical reply after each message before sending the next message. 
By measuring the time required to send all messages and receive all replies the average message latency is determined. The results are shown in 
Fig.~\ref{Fig:TCP-Ref-Latency}. 

\begin{figure}[ht!p]
\begin{center}
\resizebox*{0.50\columnwidth}{!}{
\includegraphics{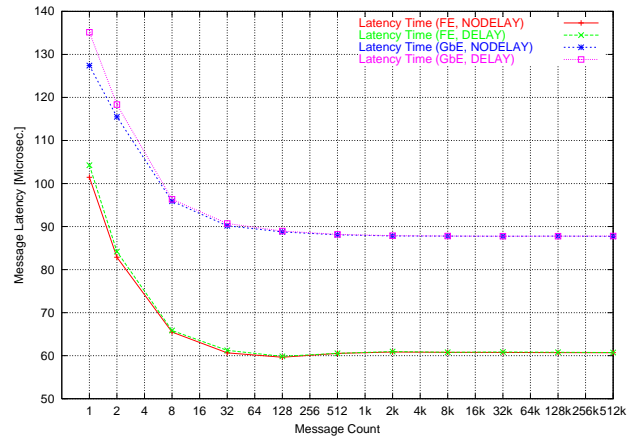}
}
\parbox{0.90\columnwidth}{
\caption[Network reference average latency measurements.]{\label{Fig:TCP-Ref-Latency}The measured average message latency (in $\mu \mathrm s$) as a function of the message count in the network reference test.}
}
\end{center}
\end{figure}

All four curves in the test display the same behaviour. With increasing message counts a drop to an asymptotically approached plateau is observed.
This reflects the decrease of the relative overhead per message due to infrastructure overhead, e.g. connection establishing. 
The values of the two Fast Ethernet and 
Gigabit Ethernet test pairs are identical in the plateaus. At smaller numbers the respective results obtained without the set
\texttt{TCP\_\-NO\-DE\-LAY} option are slightly higher than the ones obtained using the option. Latency times for Fast Ethernet are consistently smaller than
for Gigabit Ethernet, in the plateau the values are about $61~\mu \mathrm s$ compared to $88~\mu \mathrm s$. 

\subsection{Network Reference Summary}

As a summary Table~\ref{Tab:TCPReferenceMeasurementComparisonPlateau} and~\ref{Tab:TCPReferenceMeasurementComparisonPeak} list the parameters obtained from the network reference plateau and peak
throughput measurements respectively. Each entry holds the minimum and maximum values with their respective
block sizes as well as the average of all values. In this table the whole range covered by the tests from 8~B to 4~MB is 
included. One observation can be made regarding the rule of thumb that 1~\% of one CPU is used for every megabyte transferred per second.
With the tested
configuration this is an approximate lower bound, as almost every value of CPU usage divided by throughput is above that
limit. The only exceptions are the minimal Gigabit Ethernet values on the sender. 

\begin{table}[hbt!p]
\begin{center}
{\scriptsize
\begin{tabular}{|l||c|c|c|c|c|c|c|c|}
\hline  
                       & Rate / & Network                         & CPU Usage   & CPU Usage  & CPU Usage /          & CPU Usage /           & CPU Usage /                               & CPU Usage / \\
Measurement            & Hz     & Throughput /                    & Sender /    & Receiver / & Rate                 & Rate                  & Throughput                                & Throughput \\
Type                   &        & $\frac{\mathrm{MB}}{\mathrm s}$ & \%          & \%         & Sender /             & Receiver /            & Sender /                                  & Receiver /                                \\
                       &        &                                 &             &            & $\%\times \mathrm s$ & $\% \times \mathrm s$ & $\frac{\% \times \mathrm s}{\mathrm{MB}}$ & $\frac{\% \times \mathrm s}{\mathrm{MB}}$ \\
                       &        &                                 &             &            &                      &                       &                                                    & \\
\hline \hline
Reference              & 2.8 @ 4~M   & 4.9 @ 8                   & 12 @ 2~k     & 18.1 @ 16~k           & 0.000146 @ 8                     & 0.000152 @ 8                      & 1.07 @ 16~k                                                   & 1.616 @ 16~k \\
FE w.                  & 643000 @ 8 & 11.2 @ 4~M                 & 94.4 @ 16  & 102.4 @ 16           & 5.7 @ 4~M                     & 7.14 @ 4~M                      & 19.1 @ 8                                                   & 19.9 @ 8 \\
 \texttt{TCP\_\-NO\-DE\-LAY} & 95300       & 10.7                       & 26.2        & 32           & 0.568                     & 0.712                      & 3.04                                                   & 3.58 \\
\hline
Reference              & 2.8 @ 4~M   & 4.93 @ 8                  & 12 @ 2~k     & 18.1 @ 16~k           & 0.00014 @ 8                     & 0.000152 @ 8                      & 1.07 @ 32~k                                                   & 1.616 @ 16~k \\
FE w/o                 & 647000 @ 8 & 11.2 @ 2~M                 & 93.2 @ 16  & 105.6 @ 16           & 5.52 @ 4~M                     & 7.66 @ 4~M                      & 18.44 @ 8                                                   & 19.86 @ 8 \\
 \texttt{TCP\_\-NO\-DE\-LAY} & 99000       & 10.8                       & 26          & 32.6           & 0.558                     & 0.764                      & 1.46                                                   & 3.58 \\
\hline
Reference              & 17.1 @ 4~M  & 5.22 @ 8                  & 64 @ 8~k    & 78 @ 8~k           & 0.000142 @ 8                     & 0.000148 @ 8                      & 0.854 @ 32~k                                                   & 1.044 @ 32~k \\
GbE w.                 & 684000 @ 8 & 86.2 @ 64~k                & 100 @ 128   & 113.4 @ 256           & 4.2 @ 4~M                     & 4.9 @ 4~M                      & 18.58 @ 8                                                   & 19.34 @ 8 \\
 \texttt{TCP\_\-NO\-DE\-LAY} & 163000      & 60.2                       & 80.4        & 92.4           & 0.418                     & 0.488                      & 2.8                                                   & 3.04 \\
\hline
Reference              & 17.1 @ 4~M  & 5.24 @ 8                  & 66.4 @ 8~k  & 80.8 @ 8~k           & 0.000142 @ 8                     & 0.000148 @ 8                      & 0.866 @ 32~k                                                   & 1.038 @ 32~k \\
GbE w/o                & 687000 @ 8 & 88.4 @ 64~k                & 101.2 @ 32  & 112.6 @ 256           & 4.2 @ 4~M                     & 4.9 @ 4~M                      & 18.54 @ 8                                                   & 19.38 @ 8 \\
 \texttt{TCP\_\-NO\-DE\-LAY} & 166000      & 61.2                       & 81.6        & 92.8           & 0.42                     & 0.488                      & 2.78                                                   & 2.98 \\
\hline
\end{tabular}
}
\parbox{0.90\columnwidth}{
\caption[TCP reference plateau measurements summary.]{\label{Tab:TCPReferenceMeasurementComparisonPlateau}TCP reference plateau measurements summary. 
Shown are the minimum and maximum values with their respective block size in bytes as well as the average of all values. Note that
for the CPU related meausurements on the receiver not all measurement points are available. } %Sizes are in bytes.}
}
\end{center}
\end{table}

\begin{table}[hbt!p]
\begin{center}
{\scriptsize
\begin{tabular}{|l||c|c|c|c|c|c|c|c|}
\hline  
                       & Rate / & Network                         & CPU Usage   & CPU Usage  & CPU Usage /          & CPU Usage /           & CPU Usage /                               & CPU Usage / \\
Measurement            & Hz     & Throughput /                    & Sender /    & Receiver / & Rate                 & Rate                  & Throughput                                & Throughput \\
Type                   &        & $\frac{\mathrm{MB}}{\mathrm s}$ & \%          & \%         & Sender /             & Receiver /            & Sender /                                  & Receiver /                                \\
                       &        &                                 &             &            & $\%\times \mathrm s$ & $\% \times \mathrm s$ & $\frac{\% \times \mathrm s}{\mathrm{MB}}$ & $\frac{\% \times \mathrm s}{\mathrm{MB}}$ \\
                       &        &                                 &             &            &                      &                       &                                                    & \\
\hline \hline
Reference              & 2.8 @ 4~M   & 4.13 @ 8                  & 12.2 @ 8~k   & 15.6 @ 2~k        & 0.000166 @ 8                     & 0.00242 @ 1~k                      & 1.084 @ 8~k                                                   & 1.38 @ 2~k \\
FE w.                  & 542000 @ 8 & 14.1 @ 32                 & 91.6 @ 16  & 28.4 @ 1~k           & 5.72 @ 4~M                     & 7.24 @ 4~M                      & 21.8 @ 8                                                   & 2.48 @ 1~k \\
 \texttt{TCP\_\-NO\-DE\-LAY} & 96500       & 11.1                       & 32.2        & 19.7           & 0.568                     & 1.11                      & 3.58                                                   & 1.75 \\
\hline
Reference              & 2.8 @ 4~M   & 5.31 @ 8                  & 12 @ 16~k    & 15.6 @ 2~k           & 0.000126 @ 8                     & 0.00238 @ 1~k                      & 1.068 @ 16~k                                                   & 1.38 @ 2~k \\
FE w/o                 & 696000 @ 8 & 19 @ 32                   & 92.2 @ 16  & 27.8 @ 1~k           & 5.36 @ 4~M                     & 7.38 @ 4~M                      & 16.42 @ 8                                                   & 2.44 @ 1~k \\
 \texttt{TCP\_\-NO\-DE\-LAY} & 122000      & 11.7                       & 32          & 19.9           & 0.538                     & 1.12                      & 3.02                                                   & 1.77 \\
\hline
Reference              & 17.2 @ 4~M  & 4.78 @ 8                  & 72.8 @ 4~M  & 84.8 @ 4~M           & 0.00014 @ 8                     & 0.0619 @ 64~k                      & 0.834 @ 16~k                                                   & 0.991 @ 64~k \\
GbE w.                 & 627000 @ 8 & 109 @ 16~k                 & 93 @ 4~k  & 96.6 @ 128~k           & 4.24 @ 4~M                     & 4.94 @ 4~M                      & 18.36                                                   & 1.29 @ 256~k \\
 \texttt{TCP\_\-NO\-DE\-LAY} & 151000      & 70.8                       & 86.2        & 90.3           & 0.426                     & 1.4                      & 2.82                                                   & 1.21 \\
\hline
Reference              & 17.1 @ 4~M  & 5.52 @ 8                  & 72 @ 1~M    & 81.6 @ 512           & 0.00013 @ 8                     & 0.000506 @ 512                      & 0.872 @ 32~k                                                   & 1.04 @ 512 \\
GbE w/o                & 723000 @ 8 & 106 @ 8~k                  & 102.2 @ 1~k  & 113 @ 8~k           & 4.2 @ 4~M                     & 4.83 @ 4~M                      & 17.14 @ 8                                                   & 1.23 @ 512~k \\
 \texttt{TCP\_\-NO\-DE\-LAY} & 179000      & 66.4                       & 87.2        & 92.9           & 0.42                     & 0.69                      & 2.6                                                   & 1.14 \\
\hline
\end{tabular}
}
\parbox{0.90\columnwidth}{
\caption[TCP reference peak measurements summary.]{\label{Tab:TCPReferenceMeasurementComparisonPeak}TCP reference peak measurements summary. 
The table shows the minimum and maximum values with their respective block size in bytes as well as the average of all values. Note that
for the CPU related meausurements on the receiver not all measurement points are available. } %Sizes are in bytes.}
}
\end{center}
\end{table}

\section{Communication Class Benchmarks}

For the communication classes benchmarks have been carried out with the TCP message and blob class implementations. The SCI classes have not been 
tested due to the prototype status of the implementation.
Two different measurements have been executed for the message classes, measuring the message latency as well as the achievable continuous throughput
during message sending as a function of the message size. For the blob classes these two tests have been performed twice, using the
standard {\em on-demand} type allocation where a block is remotely allocated for each transfer as well as  the {\em preallocation} method where
the whole remote buffer is allocated before any transfer, and buffer management is executed locally in the sender. 
The hardware and system software used for the tests is identical to that used in the network reference tests, described in 
section~\ref{Sec:NetworkReferenceTests}.
For both network adapters sending has been performed twice, with and without explicit connect calls, to 
determine the influence of establishing implicit connections on the transfers. Again the results of the  measuring points
have been obtained as the average of ten measurements each.  
%{\bf \Large (See sec.~\ref{Sec:NetworkReferenceTests})}
%The hardware on which the tests have been performed are the 800~MHz reference PCs, the tests
%have been made both over the PCs onboard {\bf Intel EEPro 100???} Fast Ethernet adapters as well as over {\bf 3Com 3C996T Gigabit Ethernet adapters} in
%64~bit/66~MHz PCI slots of the boards. For both network adapters the sending has been performed twice, with explicit connect calls and without, to 
%determine the influence of implicit connection establishing on the sending. For each {\bf measuring point} 10 measurements have been made, the average
%of which has been sued as the result for that point. 
Each test's result is described first in detail, and then two summaries for the message and communication classes are given as well as an overall
summary for the TCP communication class implementations. 

\subsection{TCP Message Class Throughput}

In order to benchmark the TCP message class, measurements have been made to determine the maximum rate achievable by streaming a continuous 
sequence of messages to a target without
waiting for a reply. This test is relevant for the comunication classes' use in the framework which has been designed to not require 
a reply from a remote side anywhere. As for the network reference tests a prior measurement is used to determine the number of messages to be sent for each size
by measuring the sending rate for different numbers of messages streamed to the receiver.

\subsubsection{\label{Sec:NetworkMsgPlateauDetermination}Plateau Determination}

To determine the number of messages to be used for the following throughput measurements, a prerequisite measurement has been made
for each of the four test types (FE, GbE, explicit or implicit connects (cf. sections~\ref{Sec:BCLGeneralDesign} and~\ref{Sec:TCPComClasses})) 
in order to establish the influence of this number on the throughput. 
%Before the actual throughput measurement a prerequisite test has been made for each of the four test types (FE, GbE, explicit or implicit connects) 
%to observe the effect of the number of messages sent 
%on the throughput. The purpose of this test is to determine the number of messages with which the following
%throughput measurements are to be performed. 
For the test the smallest message of 32~bytes is transmitted in a varying number from 1 to $2^{22}$. %(1048576 / 1~M). 
A plateau with an asymptotic value was expected. The throughput tests were then to be performed using a  message count
on the plateau. To restrict the running times of the tests, the start of the plateau was intended to be used. %, {\bf and also at the number of messages showing the peak throughput value}. 

\begin{figure}[ht!p]
\begin{center}
\resizebox*{0.50\columnwidth}{!}{
\includegraphics{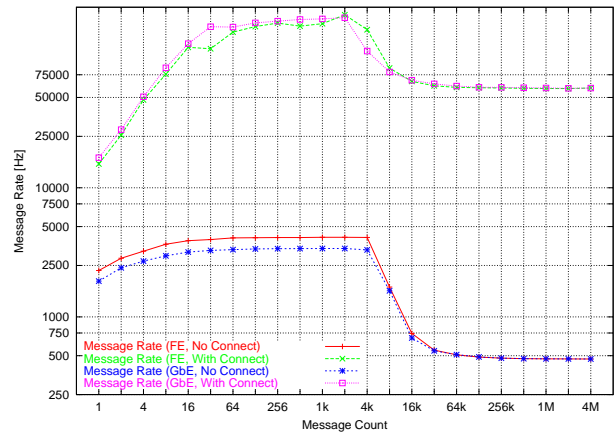}
}
\parbox{0.90\columnwidth}{
\caption{\label{Fig:TCP-Msg-Count-Rates}The measured message sending rates as a function of the message count.}
}
\end{center}
\end{figure}

The actual results of these tests are shown in Fig.~\ref{Fig:TCP-Msg-Count-Rates}.
Connected as well as unconnected tests display increasing curves to a first plateau followed by a steeper 
decrease to a second plateau. Peak values for all curves are reached at about 2048 (2~k) messages while asymptotic values are reached 
at 262144 (256~k) messages. These values are therefore used as the counts for two separate throughput measurements. 
The exact reason for the observed sudden decrease has not been determined yet. A possible explanation are overflows of system or
network interface buffers, e.g. socket send or receive buffers, causing packet loss and retransmits, but this hypothesis has not been verified yet. 
One test to determine or at least narrow the cause of this drop would be to modify the benchmark program to use different socket buffer sizes,
vary these over a certain range, and observe whether the drop occurs at different message counts. A variation of the message size to determine
its effect on the behaviour could also be performed in separate measurements as well as in combination with the buffer size variation. 
Due to the large parameter space and correspondingly large amount of measurements, and the time required for them, these investigations
have not been executed as part of this thesis. For the use of the framework the observed drops do not present a problem, as the communication classes
are only used with explicit connections. In this mode even the values after the drop are sufficiently high for the given requirements, as will be
detailed in the following sections. However, in the long run research into the phenomenon as well as modifications of the communication classes 
to work around it, if possible, are certainly desirable.

\subsubsection{Plateau Throughput Measurement}

At the message count of 262144 (256~k) messages the plateau has been reached for all four sending types and the first throughput measurement has been
performed using this message count. The message size varied from 32~B to 1~MB with the results obtained shown in 
Fig.~\ref{Fig:TCP-Msg-Rates} to~\ref{Fig:TCP-Msg-CyclesPerNetBW}.
%\ref{Fig:TCP-Msg-IntrPerSec}. 

%\begin{figure}[hbt]
\begin{figure}[ht!p]
\begin{center}
\resizebox*{0.50\columnwidth}{!}{
\includegraphics{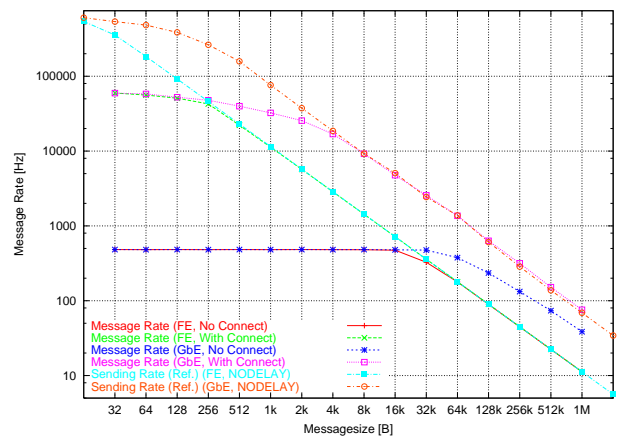}
}
\parbox{0.90\columnwidth}{
\caption[The measured message sending rates (plateau).]{\label{Fig:TCP-Msg-Rates}The measured message sending rates (message count 256~k).}
}
\end{center}
\end{figure}

Fig.~\ref{Fig:TCP-Msg-Rates} shows the message sending rates achieved in the four tests together with the results from the reference tests. 
As can be seen, the message sending rate is considerably
higher for the tests with explicit than with implicit connections, both for Fast and Gigabit Ethernet. This result could be expected, as
in the implicit connection measurements a new connection has to be established and terminated for each message, adding the connection
overhead every time. On the plateau of the two unconnected (implicitly connected) tests the rate is only limited by the
overhead of establishing the connection for all messages. Only for larger messages does the rate become limited by the network limit. For the GbE 
test the overhead is big enough that it does not even approach the limit fully but only starts to be limited by it. 
In the connected test the overhead introduced by the protocol between the sender and receiver communication objects also 
adds overhead, decreasing the achievable message rates in comparison with the reference tests. This decrease can be primarily seen for
small message sizes, for Fast Ethernet up to 256~B and for Gigabit Ethernet up to about 4~kB. For the smallest message sizes the decrease
is fairly significant, a factor of 6 for FE and almost one order of magnitude for GbE. This indicates that the message class code
still has some potential for optimizations. But even with these results the achieved rates in the connected mode are still easily high enough 
to allow the classes' use in the framework.

%Fig.~\ref{Fig:TCP-Msg-Rates} shows the message sending rates achieved in the four tests. As can be seen, the message sending rate is considerably
%higher for the tests with explicit than with implicit connections, both for Fast and Gigabit Ethernet. This result could be expected, as
%in the implicit connection measurements a new connection has to be established and terminated for each message, adding the connection
%overhead every time. For each of the
%two connection methods the rates for Fast and Gigabit Ethernet are basically identical for the smaller message rates, with the curves
%starting to diverge at 16~kB messages for implicit connections and already at 256~B for explicit connection calls. At these message sizes the sending rate
%is therefore not limited by the available network bandwidth but by another factor. 
%For 
%message sizes of 4~kB  and higher the connected Fast and the Gigabit Ethernet curves run parallel differing by a factor of about $7.5$. With larger message sizes the rates 
%approach the limit set by the underlying network technology so that the two Fast Ethernet curves are becoming identical from message sizes of about 32~kB. 
%The unconnected GbE curve
%also becomes linear between 64~kB and 128~kB although with a lower slope. In the measured message size range it is not yet dominated by the available bandwidth and
%has not approached the connected GbE curve. 

\begin{figure}[ht!p]
\begin{center}
\resizebox*{0.50\columnwidth}{!}{
\includegraphics{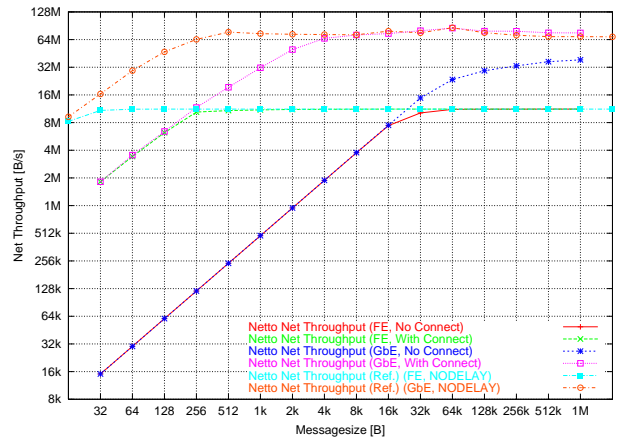}
}
\parbox{0.90\columnwidth}{
\caption[The application level network throughput for TCP message sending (plateau).]{\label{Fig:TCP-Msg-NetBW}The application level network throughput for TCP message sending (message count 256~k).}
}
\end{center}
\end{figure}

In order to show the network throughput that can be reached by an application, the achieved sending rates have been multiplied with the respective message sizes.
The resulting curves are shown in Fig.~\ref{Fig:TCP-Msg-NetBW}.
As was to be expected these curves correspond very closely to the rates from Fig.~\ref{Fig:TCP-Msg-Rates}. 
Both Fast Ethernet as well as the connection Gigabit Ethernet curve approach the curve from their respective reference measurement, while the unconnected Gigabit 
Ethernet curve still rises slowly towards it. Similar to the rate curves one can see the overhead from the communication classes by the fact that they
reach the hardware limit later than the corresponding reference measurement. An interesting point can be observed in the connected GbE curve. For the largest
measured block sizes, from about 128~kB on, this curve even exceeds the reference curve. This behaviour indicates that the communication approach used in 
the class is more effective at utilizing the systems' resources than the relatively simple reference program. Both graphs
in Fig.~\ref{Fig:TCP-Msg-Cycles} below support this thesis. In the receiver plot on the right hand side the receiver CPU usage of the message class is higher 
than the one from the reference benchmark, and in particular it is greater than 100~\%, indicating that due to its multi-threaded design it is able to utilize 
the system's two CPUs better. In the sender plot on the left hand side the CPU usage of the connected GbE curve is lower than the one from the GbE reference measurement
for most of the test. This in turn could indicate that on the sender the communication class uses the CPU or memory system more efficiently, being therefore less
constrained by it and allowing higher sending rates.

%%Network throughput for the measurements is shown in Fig.~\ref{Fig:TCP-Msg-NetBW}. The results have been derived from the message rate by multiplying it with 
%%the message size to show the network throughput that can be reached in an application. 
%As was to be expected these curves correspond very closely to the rates from Fig.~\ref{Fig:TCP-Msg-Rates}. 
%Both Fast Ethernet as well as the connection Gigabit Ethernet curve approach the limit set by the hardware while the unconnected Gigabit 
%Ethernet curve still rises slowly. At small message sizes the two connected and unconnected curve pairs each are identical, with the connected curves at higher values 
%than the unconnected ones. Consequently they reach the limit set by their respective hardware before the unconnected curves.

\begin{figure}[ht!p]
\begin{center}
\resizebox*{1.0\columnwidth}{!}{
\includegraphics{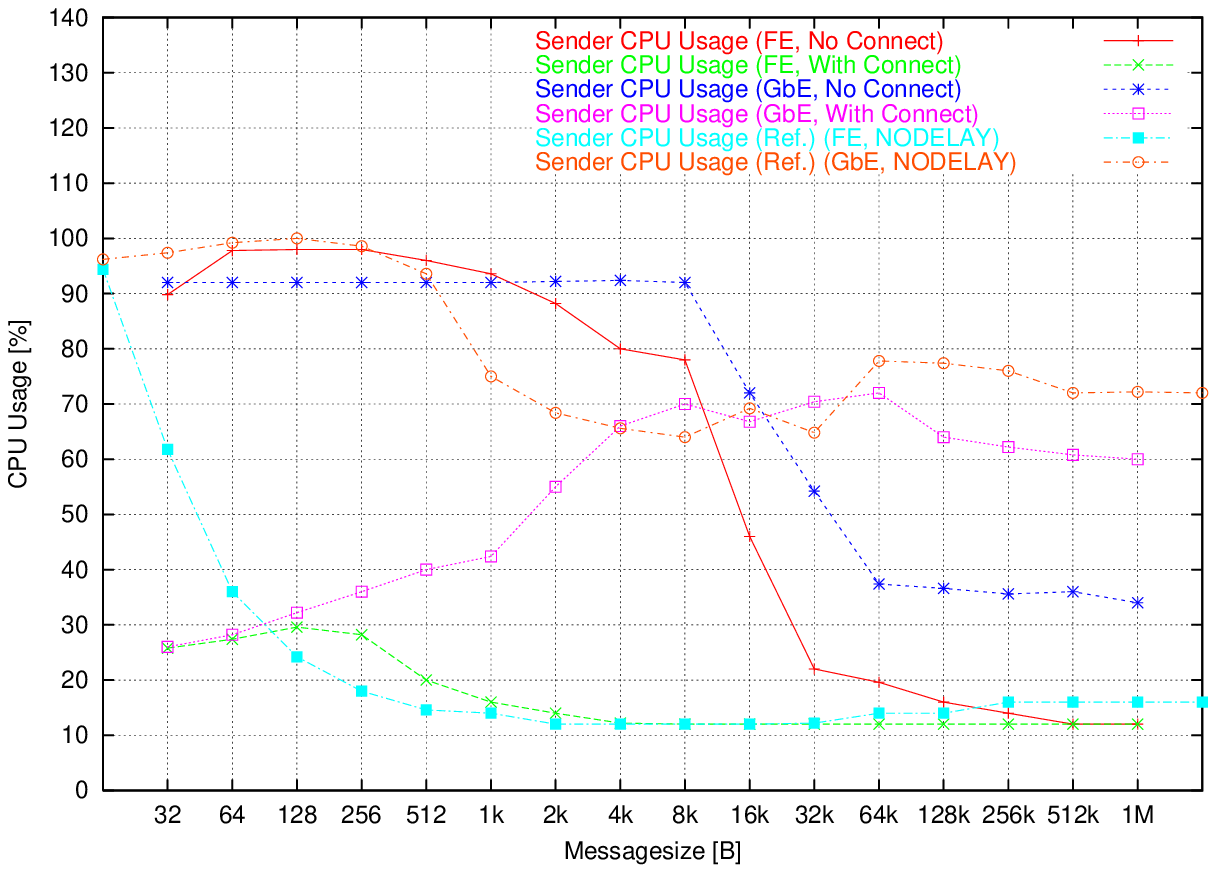}
\hfill
\includegraphics{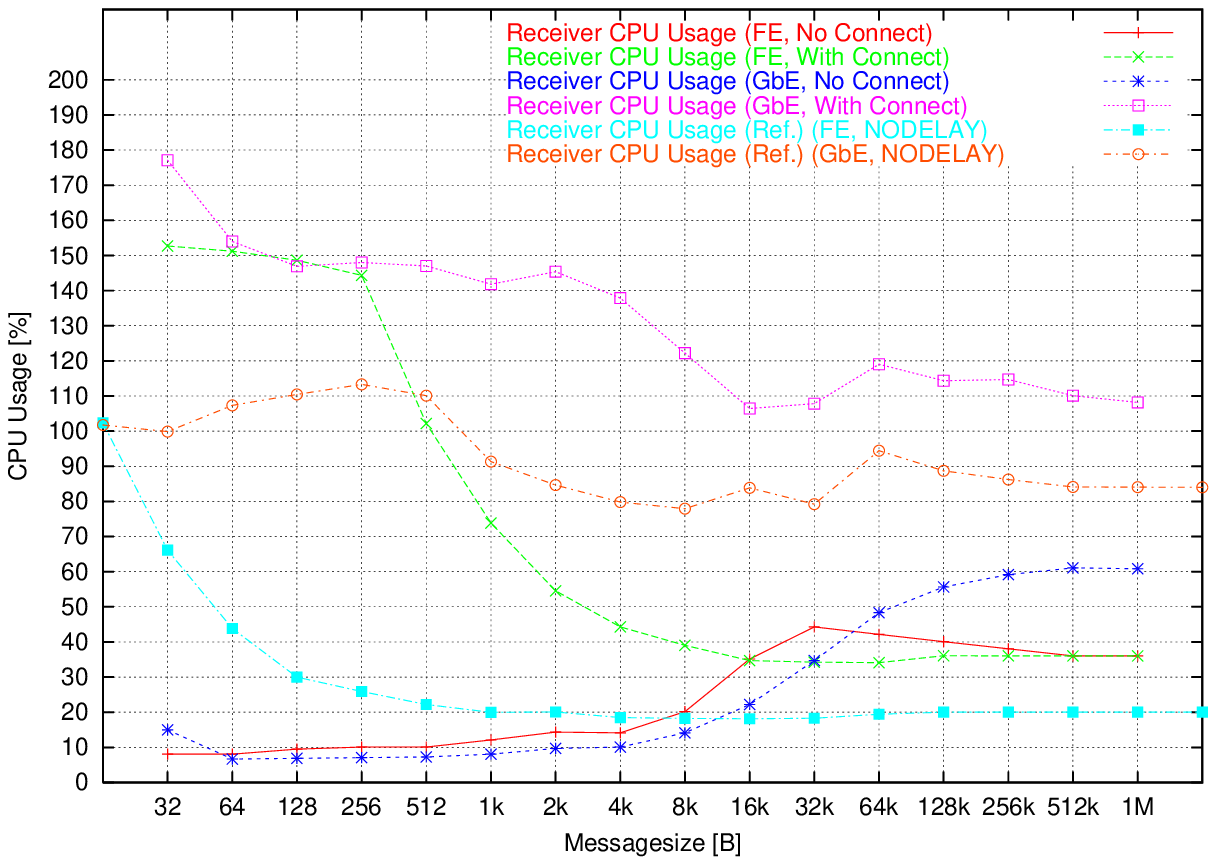}
}
\parbox{0.90\columnwidth}{
\caption[CPU usage during TCP message sending (plateau).]{\label{Fig:TCP-Msg-Cycles}The CPU usage on the sender (left) and receiver (right) during TCP message sending (message count 256~k).
The nodes are twin CPU nodes, 100~\% CPU usage corresponds to one CPU being fully used.}
}
\end{center}
\end{figure}

CPU usage for the sending and receiving nodes is displayed respectively on the left and right hand sides of Fig.~\ref{Fig:TCP-Msg-Cycles}. 
One obvious result that can be seen is the high CPU usage on the sending side and the very low usage on the receiver for the two unconnected measurements at small
block sizes, up to about 8~kB. The reason for the very low rates at small block sizes in the unconnected mode therefore seems to be the high CPU load
produced from initiating the connections on the sending node. Accepting connections on the receiving node does not seem to be so CPU intensive. 
A second interesting feature, as already remarked above, is the fact that on the sender the communication class CPU usage
in the connected GbE test is mostly lower than the reference GbE usage. 
For the lower block sizes this could, in addition to the potential reasons outlined above, also be caused by the 
lower sending rate of the communication class in that block range. At higher block sizes, however, the throughput achieved by the communication class was higher
and this reasoning cannot be applied. As outlined above at these rates it is therefore more likely that the sending approach used in the communication classes,
using \texttt{write} preceeded by \texttt{se\-lect} calls, is more efficient than the simple approach of using blocking \texttt{write} calls in the 
reference benchmark program. On the receiver the behaviour of the two respective curves is reversed, the communication class consistently uses between 10~\% 
and 20~\% more CPU cycles compared to the reference benchmark. Here the communication class introduces more overhead than the reference benchmark. One likely 
cause of this overhead are the allocation and deallocation calls of the memory for each message as well as its copying. Similar to  the reference test, 
the measured CPU load reaches more than 100~\% and therefore uses both of the nodes' CPUs, in particular on the  receiving side. As for the reference test this 
implies that single CPU nodes will only be able to handle lower throughputs than measured in this benchmark.

\begin{figure}[ht!p]
\begin{center}
\resizebox*{1.0\columnwidth}{!}{
\includegraphics{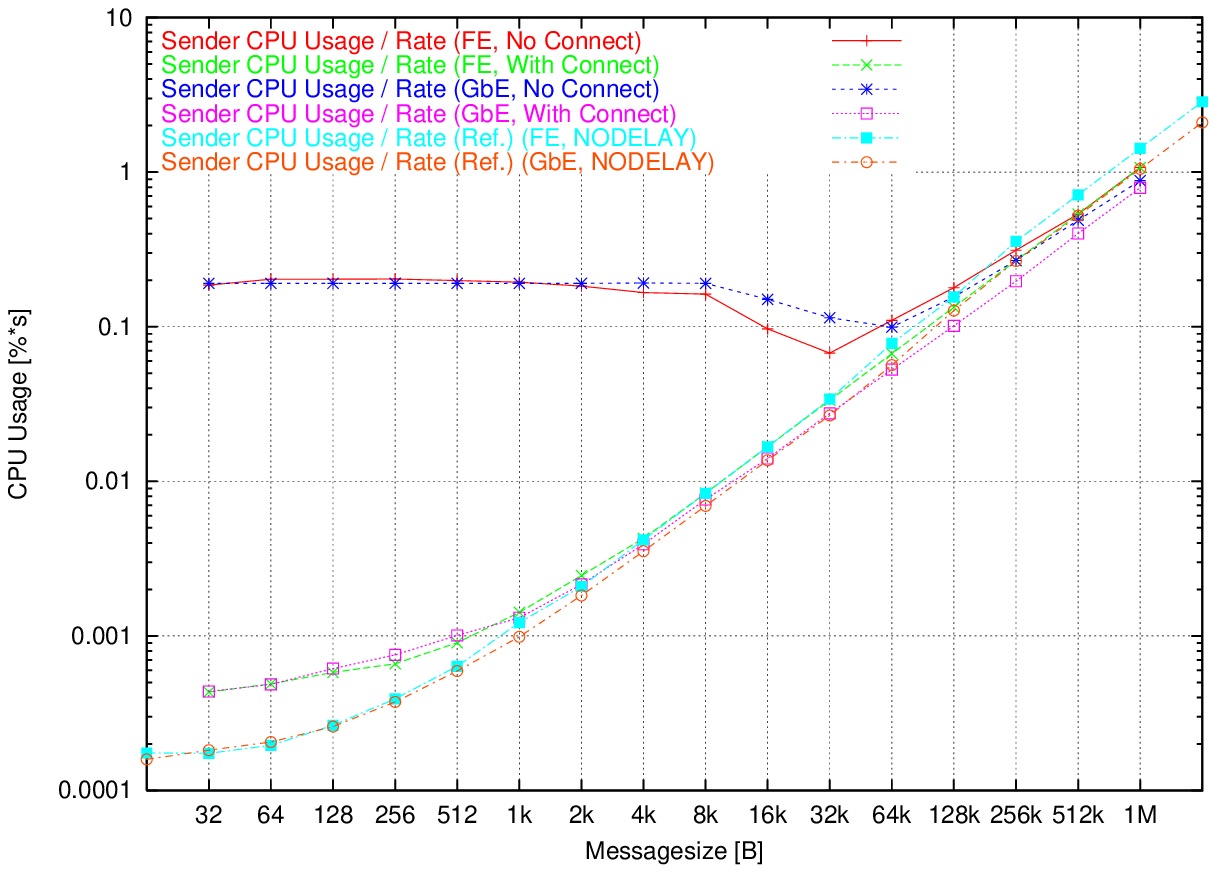}
\hfill
\includegraphics{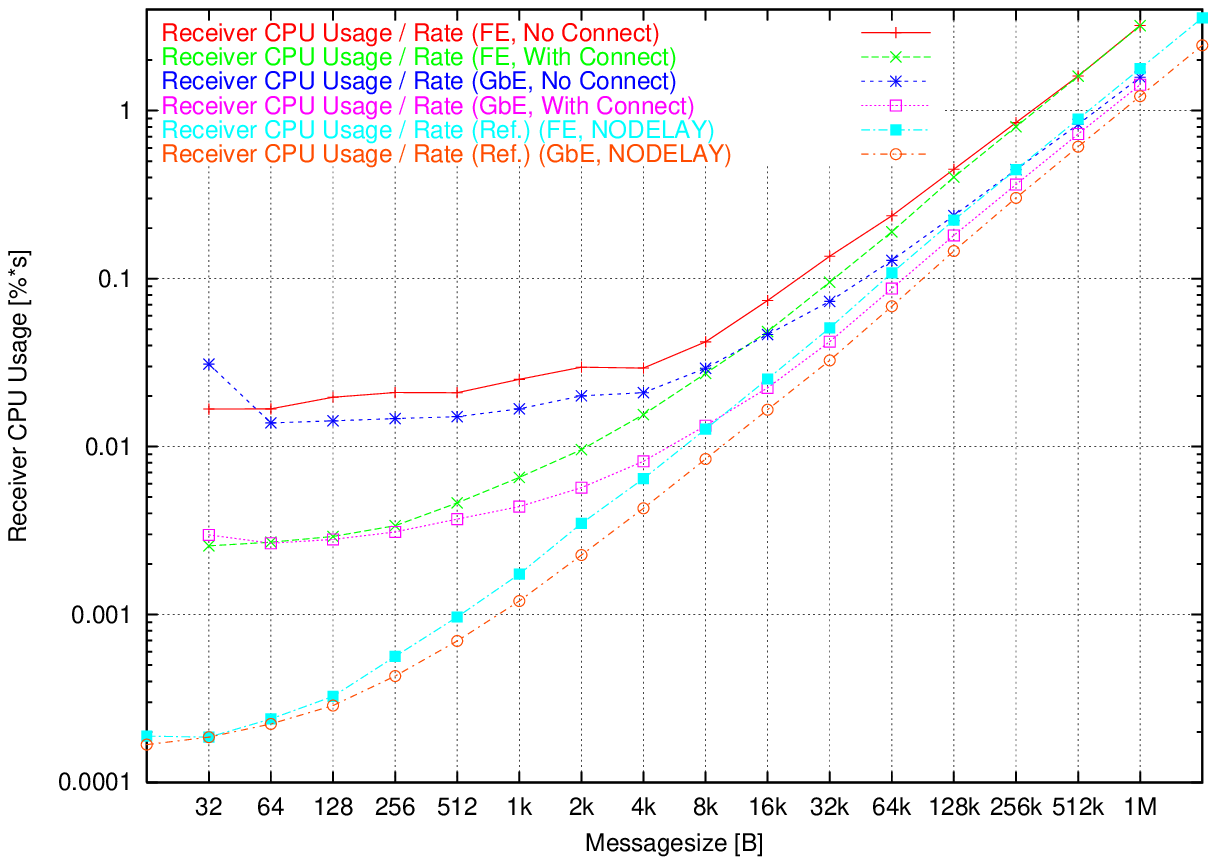}
}
\parbox{0.90\columnwidth}{
\caption[CPU usage divided by the sending rate during TCP message sending (plateau).]{\label{Fig:TCP-Msg-CyclesPerRate}The CPU usage on the sender (left) and receiver (right) divided by the sending rate 
during TCP message sending (message count 256~k).
The nodes are twin CPU nodes, 100~\% CPU usage corresponds to one CPU being fully used.}
}
\end{center}
\end{figure}

\begin{figure}[ht!p]
\begin{center}
\resizebox*{1.0\columnwidth}{!}{
\includegraphics{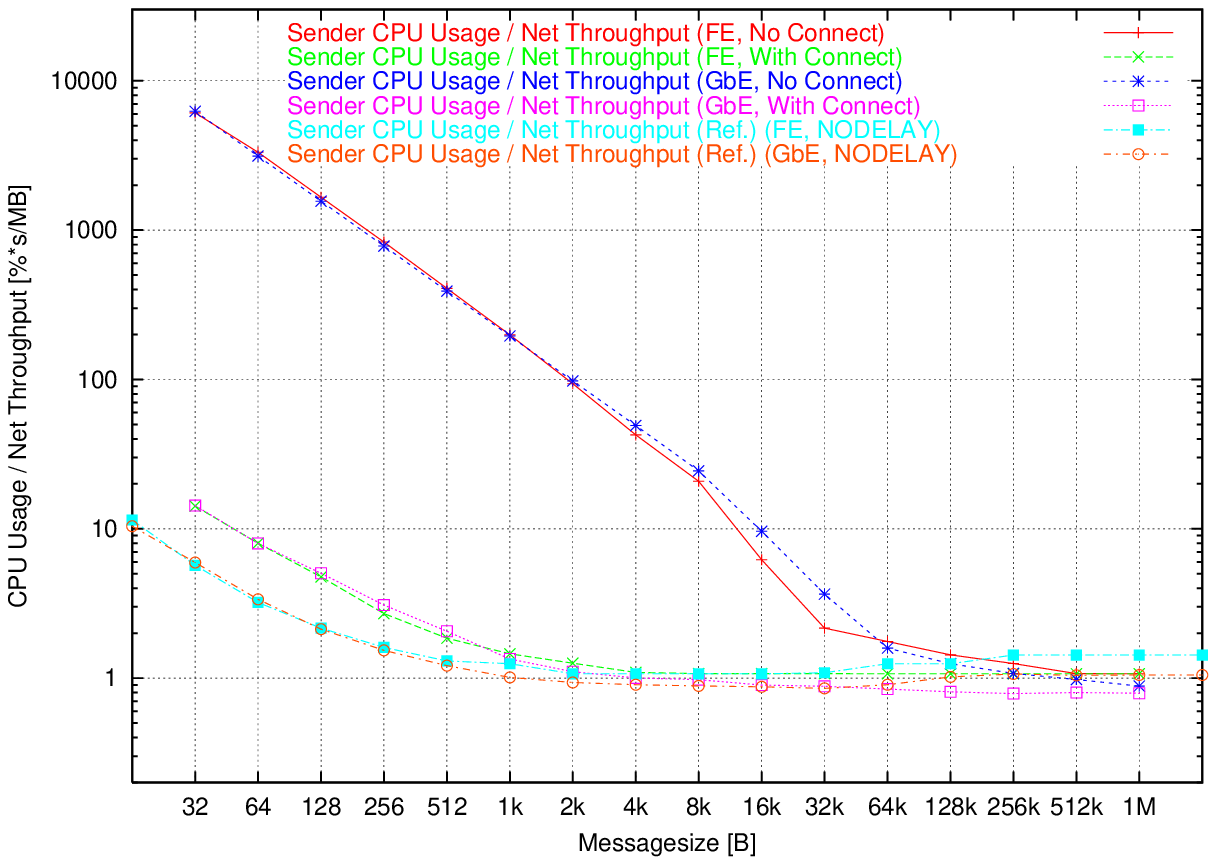}
\hfill
\includegraphics{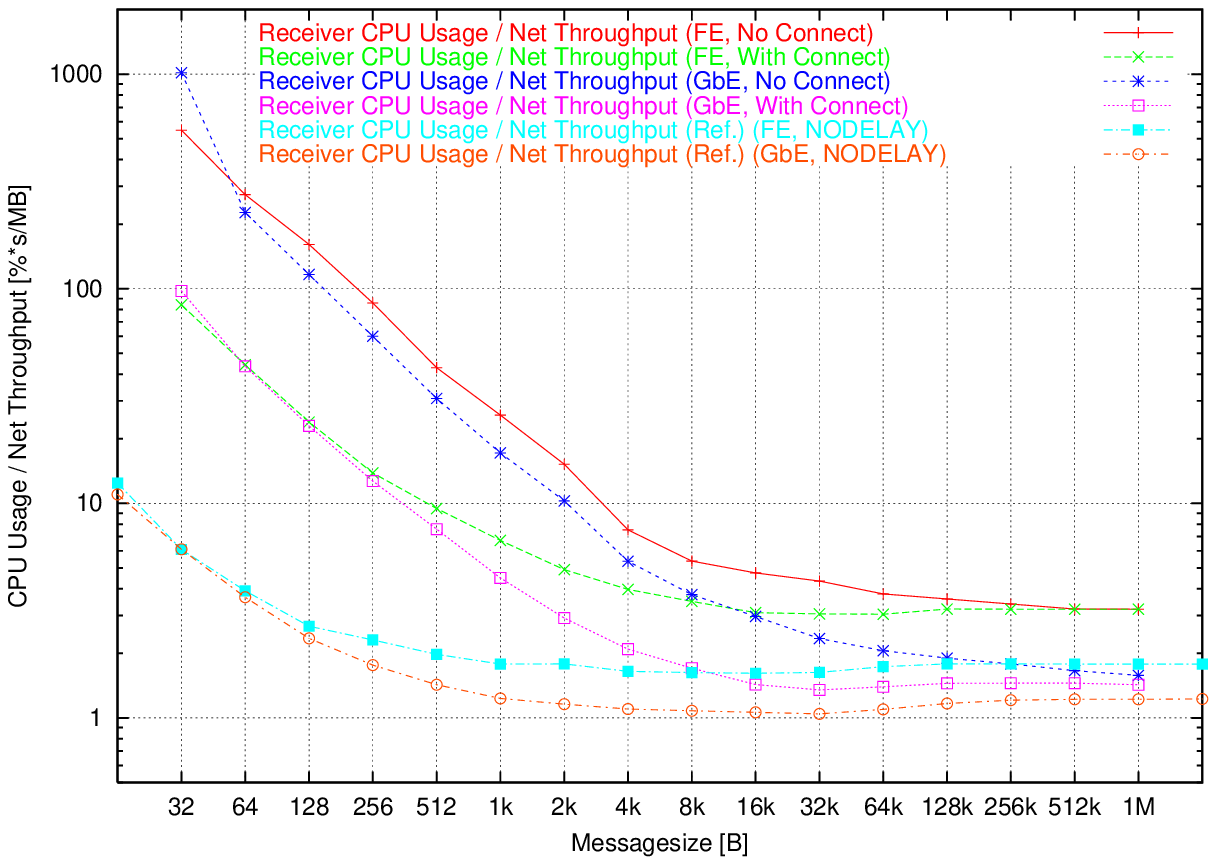}
}
\parbox{0.90\columnwidth}{
\caption[CPU usage per MB/s network throughput during TCP message sending (plateau).]{\label{Fig:TCP-Msg-CyclesPerNetBW}The CPU usage on the sender (left) and receiver (right) per MB/s network throughput 
during TCP message sending (message count 256~k).
The nodes are twin CPU nodes, 100~\% CPU usage corresponds to one CPU being fully used.}
}
\end{center}
\end{figure}

For a better comparison all CPU usage measurements should be evaluated with respect to the sending rates or network throughputs 
as shown in Fig.~\ref{Fig:TCP-Msg-CyclesPerRate} and~\ref{Fig:TCP-Msg-CyclesPerNetBW} respectively. The rate and
bandwidth plots are correlated by the message size due to the way the network throughput is determined, as detailed above. 
Fig.~\ref{Fig:TCP-Msg-CyclesPerRate} clearly shows the high relative overhead caused by establishing a connection for each message, 
particularly on the sender but on the receiver as well. This relative overhead per message can be several orders of magnitude 
above that for just sending a message over an established connection. In the sender graph one can also see that both connected message class
measurements initially are higher than their respective reference measurement. For larger message or block sizes, however, they fall below the
respective reference curve. This is a further indication for the behaviour already observed in the rate and sender side CPU usage plots (Fig.~\ref{Fig:TCP-Msg-Rates} 
and~\ref{Fig:TCP-Msg-Cycles}). In both of these single plots the message class showed better values (higher rate, smaller CPU usage) 
than the reference measurement for large messages. 
On the receiver side this behaviour is not present, here the CPU overhead outlined above is high enough that the per-message overhead 
of both connected message class curves is higher than the respective reference one. Similar to the reference measurement of CPU cycles per sending rate from
Fig.~\ref{Fig:TCP-Ref-Plateau-CyclesPerRate} one can again see that Gigabit Ethernet is more efficient in its use of CPU cycles per transfer than Fast Ethernet,
both on the sending and the receiving nodes. On both nodes the unconnected measurements approach the connected ones with increasing block sizes,
showing that the overhead per message for establishing the connections decreases with increasing message size, as could be expected.

The plots of CPU cycles per megabyte of data transferred per second in Fig.~\ref{Fig:TCP-Msg-CyclesPerNetBW} show mostly the same information as the
ones in Fig.~\ref{Fig:TCP-Msg-CyclesPerRate}. One additional item can be observed in Fig.~\ref{Fig:TCP-Msg-CyclesPerNetBW}. Unlike the reference
curves the connected message class measurement curves do not show a ``{\em bathtub}'' curve shape on the sender. At the points where the reference curves rise 
again the message class curves remain constant. The GbE curve even shows a slight drop. This behaviour again underlines the fact that the sending approach used
in the class makes a more efficient use of CPU cycles than the one used in the reference benchmark program. On the receiver the measured message class values show
the same behaviour as the reference curves, although at higher values. At small blocks the values are considerably higher, more than one magnitude for some 
message sizes. This shows again the high overhead added on the receiver side, presumably at least partly due to the message allocation and release calls.

\subsubsection{Peak Throughput Measurement}

At the message count of 2048 (2~k) all four sending types have reached their peak value for the measured rate. 
As for the plateau measurement the message size has been varied from 32~B to 1~MB with the results obtained shown in 
Fig.~\ref{Fig:TCP-Msg-Rates-Peak} to~\ref{Fig:TCP-Msg-CyclesPerNetBW-Peak}.
%\ref{Fig:TCP-Msg-IntrPerSec-Peak}. 
Due to the short measuring times involved and the details of the measurement the CPU 
related values for small message sizes, particularly on the receiving node, could not be measured accurately, similar to the problems in the network reference test.
These values have therefore been excluded from the measurement results. 

%\begin{figure}[hbt]
\begin{figure}[ht!p]
\begin{center}
\resizebox*{0.50\columnwidth}{!}{
\includegraphics{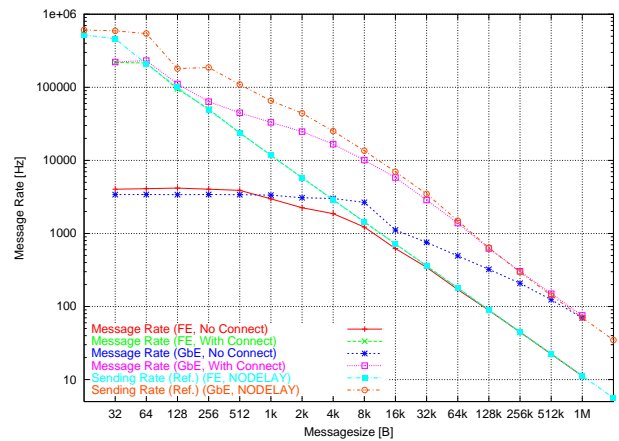}
}
\parbox{0.90\columnwidth}{
\caption[The measured message sending rates (peak).]{\label{Fig:TCP-Msg-Rates-Peak}The measured message sending rates (message count 2~k).}
}
\end{center}
\end{figure}

For the maximum achieved sending rate, shown in Fig.~\ref{Fig:TCP-Msg-Rates-Peak}, one can see that for smaller message sizes the achieved rates
are higher than for the plateau measurement in Fig.~\ref{Fig:TCP-Msg-Rates}, differing by factors of about 3.6 and 8 for the connected and 
unconnected tests respectively. 
The connected Gigabit Ethernet curve runs close to the FE one
up to the 256~B message size and starts to diverge for higher sizes. Both of these curves have reached 
their bandwidth limit at about 8~kB. 
For both Fast Ethernet tests as well 
as the connected Gigabit one the curves are identical with their corresponding plateau throughput curves 
for message sizes exceeding certain limits: 512~B for the connected GbE, 256~B for the connected
FE, and 16~kB for the unconnected FE curve. Below that limit each peak curve features higher values than its plateau counterpart. 
Compared to the peak reference tests one can see that the communication class's connected Fast Ethernet test reaches the reference sending rate earlier 
than in the plateau test, showing that it is less constrained by the limit at small messages encountered in that test.
In the connected Gigabit Ethernet measurement the communication class reaches the respective reference limit later than the FE and later
than its plateau counterpart. At about 256~B to 512~B it reaches the same values as the connected plateau GbE curve, and therefore already at
these sizes seems to be limited by the same factor as the plateau measurement. The two unconnected curves initially again run almost constant, 
but at higher values than in the plateau test and they start to decrease earlier. At small messages the limits between these two tests 
thus seem to be different while the limit that causes the later decrease, most likely the available bandwidth, is approached sooner. 

%For the maximum achieved sending rate shown in Fig.~\ref{Fig:TCP-Msg-Rates-Peak} one can see that for smaller message sizes the achieved rates
%are higher than for the plateau measurement in Fig.~\ref{Fig:TCP-Msg-Rates}, differing by factors of about 3.6 and 8 for the connected and 
%unconnected tests respectively. 
%For both Fast Ethernet peak tests the block size where the rates start to be
%limited by the available bandwidth 
%is smaller by a factor of 4 compared to the corresponding plateau tests. 
%The connected Gigabit Ethernet curve runs close to the FE one
%up to the 256~B message size and starts to diverge for higher sizes. Both of these curves have reached 
%their bandwidth limit at about 8~kB. After this point they run in parallel. 
%For both Fast Ethernet tests as well 
%as the connected Gigabit one the curves are identical with their corresponding plateau throughput curves 
%for message sizes exceeding certain limits: 512~B for the connected GbE, 256~B for the connected
%FE, and 16~kB for the unconnected FE curve. Below that limit each peak curve features higher values than its plateau counterpart. 
%Unlike these three curves the unconnected Gigabit one runs flat for a broad range and
%at a higher value than its counterpart, shows an unsteady decrease for a number of sizes, and then transitions into a linear decrease that differs in a lesser slope from 
%the plateau test. As a result of this it approaches the connected GbE curve at the 1~MB message size, reaching the same network limit as the connected
%test. 

\begin{figure}[ht!p]
\begin{center}
\resizebox*{0.50\columnwidth}{!}{
\includegraphics{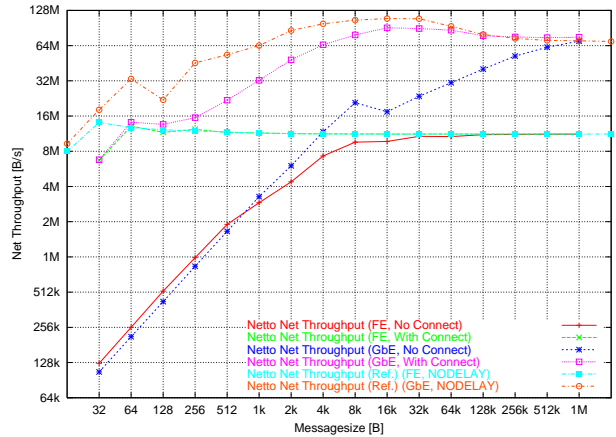}
}
\parbox{0.90\columnwidth}{
\caption[The application level network throughput for TCP message sending (peak).]{\label{Fig:TCP-Msg-NetBW-Peak}The application level network throughput for TCP message sending (message count 2~k).}
}
\end{center}
\end{figure}

Fig.~\ref{Fig:TCP-Msg-NetBW-Peak} shows the application level network throughput that has been achieved in these four tests. 
The respective plateau results are represented in Fig.~\ref{Fig:TCP-Msg-NetBW}. 
In consistency with the achieved rates one can see that the maximum network throughput is reached in the connected FE test for 
64~B messages already. It is closely approached by the unconnected one between 8~kB and 16~kB. In both cases this happens for smaller message sizes than for the related plateau test. All four 
tests also start at higher throughput values and correspondingly reach the bandwidth limit earlier. The connected GbE curve is identical to the plateau
curve for messages of at least 1~kB with the exception of the somewhat higher peak shifted towards the lower range between 16~kB to 32~kB. Apart from this peak the maximum 
values are not higher than the ones from the plateau tests, as expected. 
Similar to the FE reference peak measurement 
the message class curve reaches more than the theoretically possible network throughput. It must therefore be assumed that the data is buffered to a large degree as well. 
As detailed in the peak reference test this buffering is also a potential explanation for the performance increase in the peak tests. Further remarkable properties 
in this graph are the kinks in several of the curves. No clear explanation has been found for them yet, a possible explanation for at least some of them are buffer 
limits which are encountered. With full buffers the measurements then again display different behaviour as when all or a large amount of data can be buffered.
Similar to reference peak tests the peak message class tests are mostly at higher throughput values than the respective plateau measurements. An exception are the 
FE curves where they have already reached the hardware limit.

\begin{figure}[ht!p]
\begin{center}
\resizebox*{1.0\columnwidth}{!}{
\includegraphics{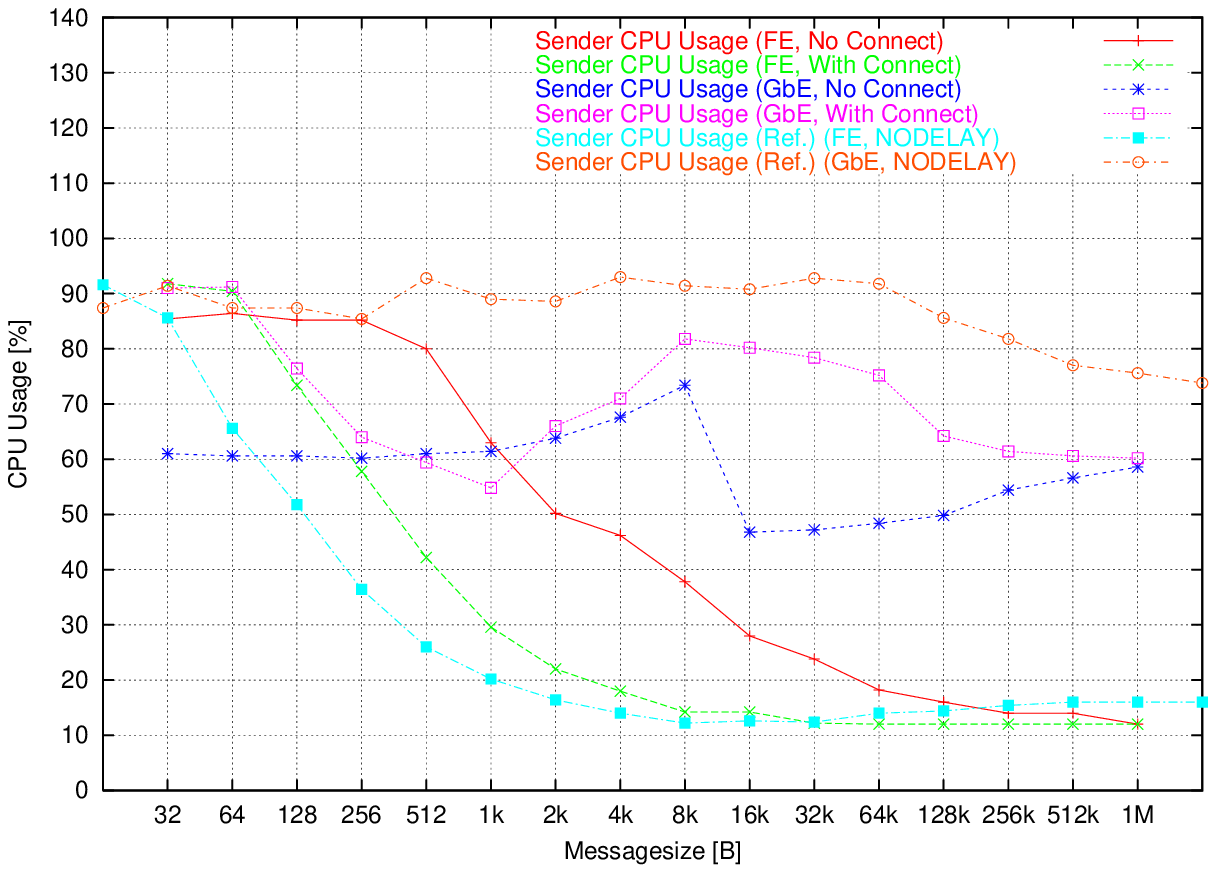}
\hfill
\includegraphics{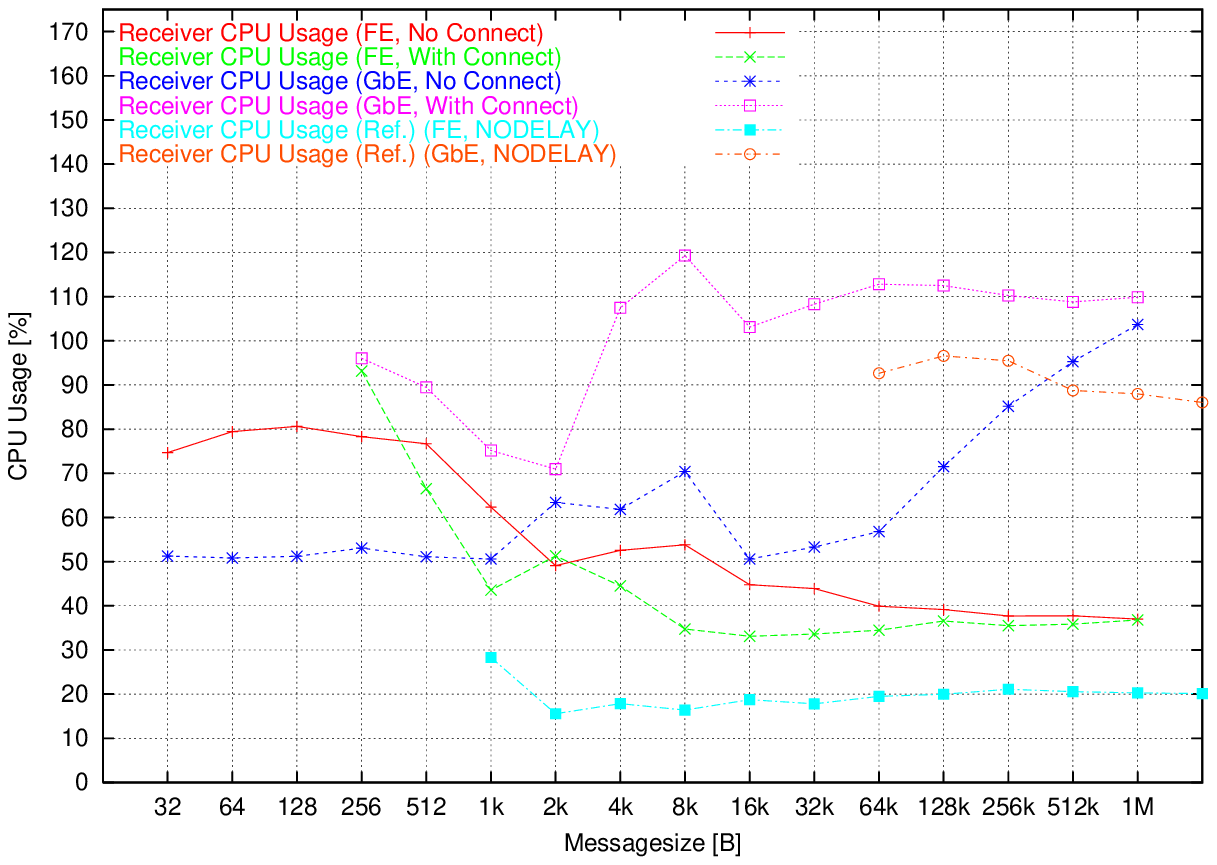}
}
\parbox{0.90\columnwidth}{
\caption[CPU usage during TCP message sending (peak).]{\label{Fig:TCP-Msg-Cycles-Peak}The CPU usage on the sender (left) and receiver (right) during TCP message sending (message count 2~k).
The nodes are twin CPU nodes, 100~\% CPU usage corresponds to one CPU being fully used.}
}
\end{center}
\end{figure}

CPU usages for the tests are shown in Fig.~\ref{Fig:TCP-Msg-Cycles-Peak}. As for the peak reference tests some values on the receiver could not be measured 
accurately. These values have therefore been excluded from the graphs. 
As can be seen in the figures,
CPU usage on the sender for the two explicit connection tests for small message sizes is much higher compared to the respective plateau test (Fig.~\ref{Fig:TCP-Msg-Cycles}). 
This continues up to the point where the usage levels become equal, for FE in the range between 32~kB to 64~kB and for GbE at about 64~kB messages sizes, 
similar to the observed behaviour for the sending rates and network throughputs. In the 
unconnected tests a different behaviour is displayed, with the FE test showing continuously lower CPU usage values coupled with an earlier decrease 
compared to the related plateau test. The unconnected GbE test initially displays lower values than its respective plateau throughput test, with both curves running 
basically flat at about 30~\% and 45~\% respectively. For messages sizes higher than 32~kB, though, the plateau test usage is lower. 
Both Fast as well as Gigabit Ethernet tests reach the same final value for the 1~MB message size.
On the receiver the Gigabit Ethernet tests are mostly separated and only approach similar values for the 1~MB message size as well. The two Fast Ethernet
measurements run much closer and are approximately similar. They also reach the identical values at 1~MB messages.
One can also see that, similar to the plateau tests, the connected message class measurements 
on the sender display lower CPU usage values than the reference test; the GbE one almost over the whole test range and the FE one for large messages only. 
%In general the usage on the sender is in parts higher as well as lower compared to the corresponding plateau measurement. 
In contrast on the receiver the 
usage of the reference measurements, where present, is considerably lower than the message class's, by at least 10~\%. This is again similar to the behaviour 
in the plateau tests. 
On the sender the unconnected measurements are mostly at lower values than in the plateau test, while they are higher on the receiver. 
The difference between the unconnected usage on the two nodes is therefore not as extreme as it was in the plateau measurement where most
of the load was produced on the sending node. 
Compared to the plateau test measurement all peak test curves can be lower or higher, depending on measurement type and message size.

\begin{figure}[ht!p]
\begin{center}
\resizebox*{1.0\columnwidth}{!}{
\includegraphics{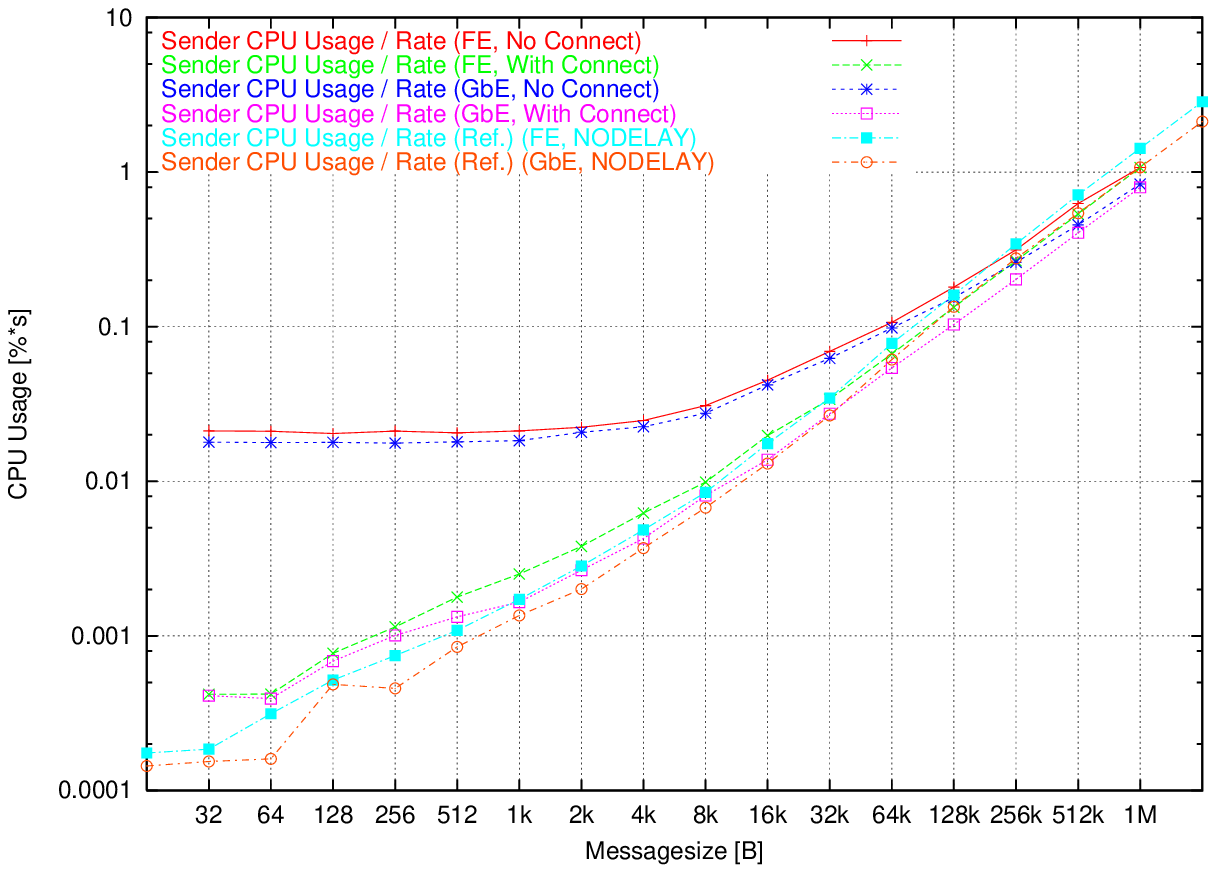}
\hfill
\includegraphics{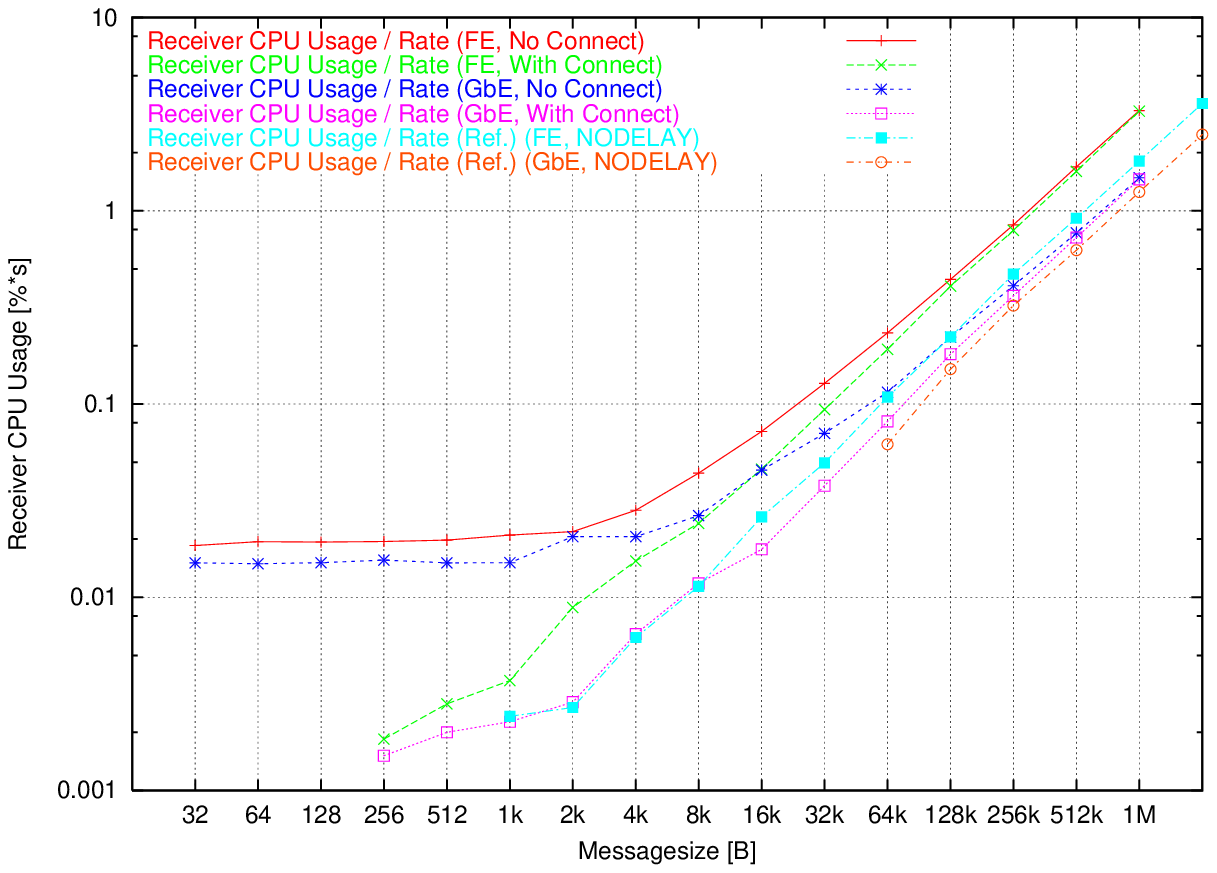}
}
\parbox{0.90\columnwidth}{
\caption[CPU usage divided by the sending rate during TCP message sending (peak).]{\label{Fig:TCP-Msg-CyclesPerRate-Peak}The CPU usage on the sender (left) and receiver (right) divided by the sending rate 
during TCP message sending (message count 2~k).
The nodes are twin CPU nodes, 100~\% CPU usage corresponds to one CPU being fully used.}
}
\end{center}
\end{figure}

\begin{figure}[ht!p]
\begin{center}
\resizebox*{1.0\columnwidth}{!}{
\includegraphics{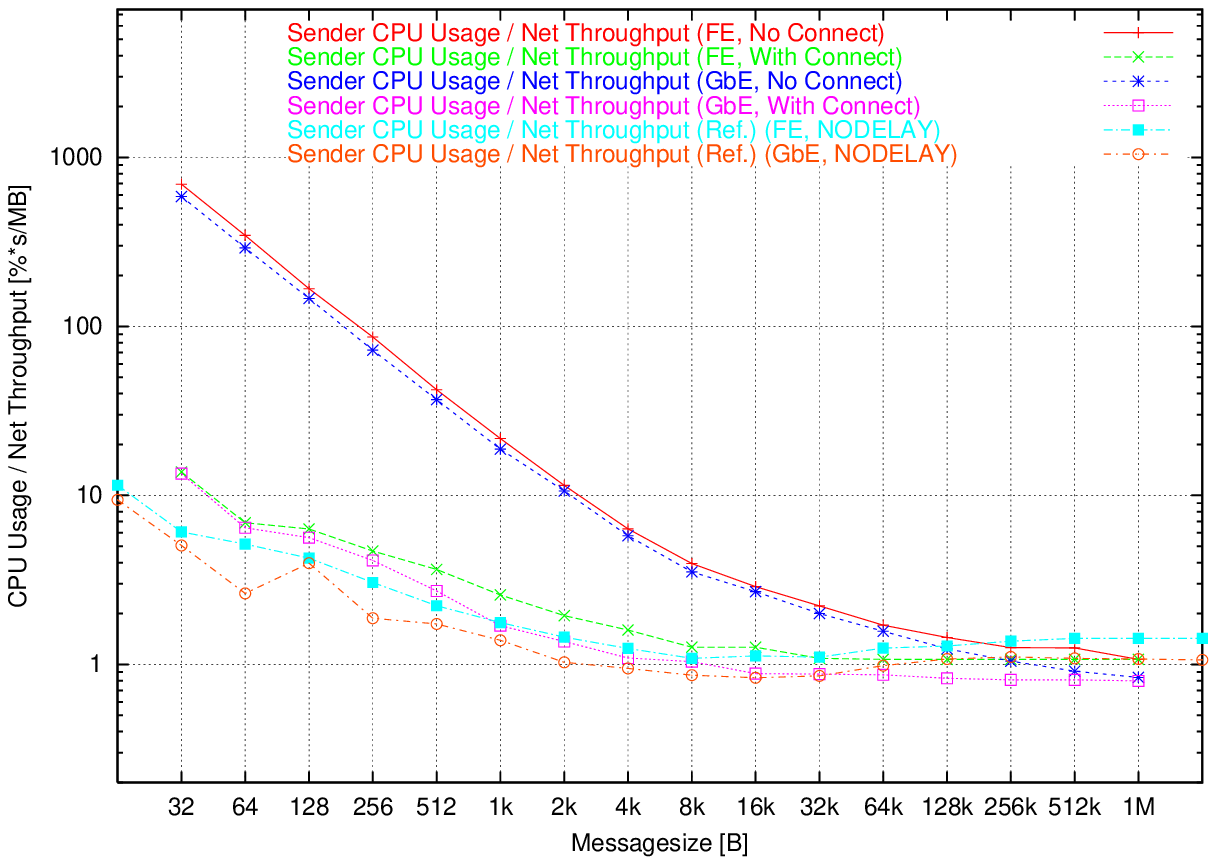}
\hfill
\includegraphics{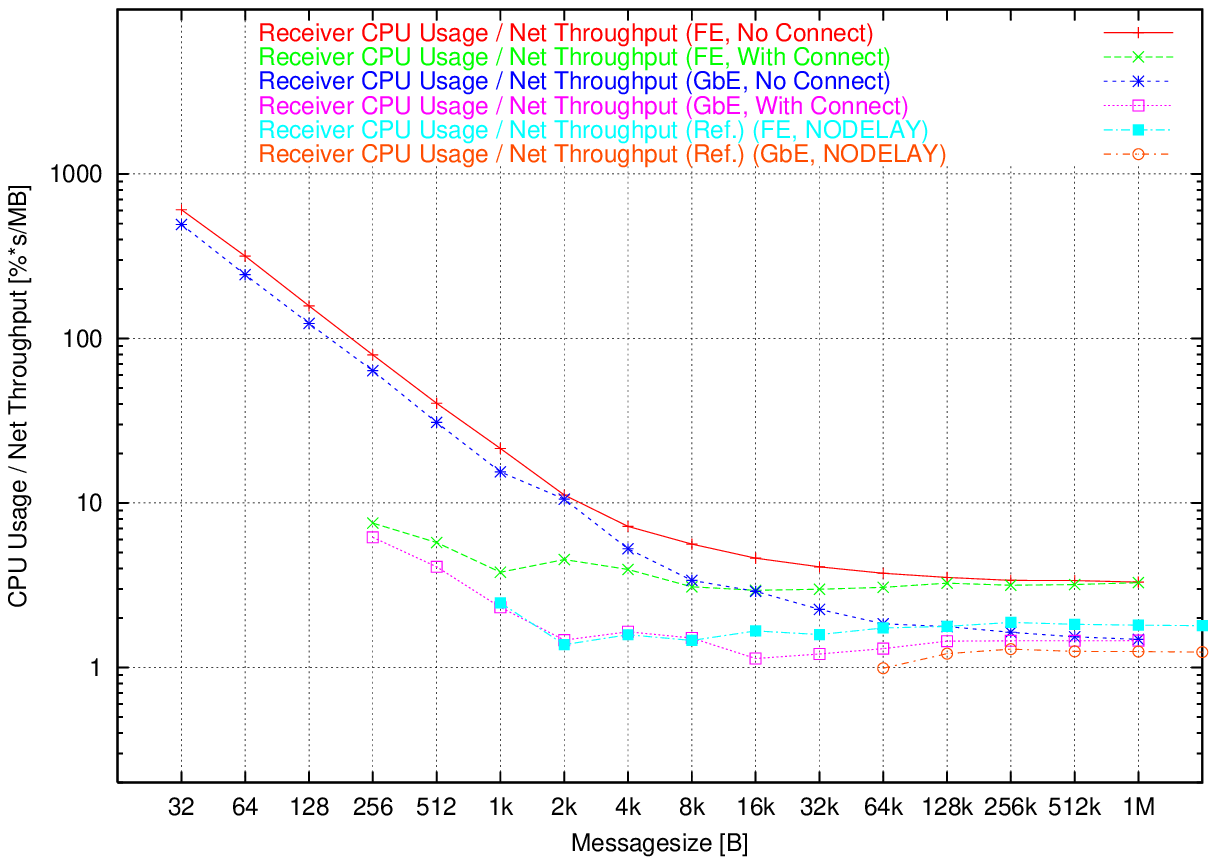}
}
\parbox{0.90\columnwidth}{
\caption[CPU usage per MB/s network throughput during TCP message sending (peak).]{\label{Fig:TCP-Msg-CyclesPerNetBW-Peak}The CPU usage on the sender (left) and receiver (right) per MB/s network throughput 
during TCP message sending (message count 2~k).
The nodes are twin CPU nodes, 100~\% CPU usage corresponds to one CPU being fully used.}
}
\end{center}
\end{figure}

For better comparison the CPU usage values of each test are normalized 
to the achieved sending rate as well as to the measured network throughput, as for the plateau tests. 
The resulting graphs are shown in 
Fig.~\ref{Fig:TCP-Msg-CyclesPerRate-Peak} and~\ref{Fig:TCP-Msg-CyclesPerNetBW-Peak}. 
For the CPU usage per rate measurements of the connected tests on the sender side one can see that the FE and GbE curves are almost identical.  
GbE values are slightly smaller than FE values, similar to the same measurement for the plateau tests from Fig.~\ref{Fig:TCP-Msg-CyclesPerRate}. Between 128~B and
2~kB inclusively the plateau curves are at lower and therefore more efficient values. From 4~kB on the curves for peak and plateau tests are almost identical,
although the FE peak curve runs at slightly higher values than the plateau one. In the unconnected tests on the sender the peak test curves are a factor of about 10
lower than the plateau ones for sizes from 32~B to about 4~kB. All four curves are basically constant for this whole interval. Between  8~kB and 32~kB the
peak and plateau curves show different behaviours, with the peak measurements still at lower values. From there on all four curves again show an identical linear 
rise.
Where reliable values are available on the receiver, the four curves rise continuously
with the peak curves at slightly lower values compared to the plateau ones. 
This trend continues up to 2~kB for FE and 32~kB for GbE, after which the respective curves in the peak and plateau tests are basically
identical. For the unconnected test the corresponding peak and plateau curves appear mostly similar, although  the peak curves are at slightly lower values for
sizes up to about 2~kB. All four curves
reach the same approximate final value for the largest messages as the respective plateau curve.
In Fig.~\ref{Fig:TCP-Msg-CyclesPerRate-Peak} one can see that over the whole test range each message class (and reference) GbE curve on the sender is at 
lower values than the corresponding FE curve.
This shows that in this mode, where at least part of the data is buffered, GbE makes more efficient use of the available CPU cycles compared with FE.
Also one can see again that at about 32~kB messages the connected message class measurements on the sender reach lower and better values than the corresponding reference
measurement. Therefore even in this partially buffered mode the sending method in the communication class is more efficient. On the receiver the behaviour is
also as before, the message class curves are higher than the reference measurements where present.

Comparing the CPU usage per network bandwidth in Fig.~\ref{Fig:TCP-Msg-CyclesPerNetBW-Peak} and~\ref{Fig:TCP-Msg-CyclesPerNetBW} it can be noticed  that
for the unconnected tests on the sender the plateau and peak test curves are almost identical in shape. However, the starting values of the two peak curves are
again a factor of 10 lower than the corresponding plateau values, and only for messages greater than 64~kB do the curves show identical values. 
Differing from this, the peak results in the connected tests on the sending
side show higher initial values. The respective curves are again almost identical after 32~kB. 
On the receiver node the results for the peak and plateau tests also have the same general shape, where values are available,
showing a steady decrease. The peak measurements
are at slightly lower values and display a slightly more unsteady behaviour. 
For messages exceeding 16~kB the corresponding peak and plateau test curves again run at basically identical
constant values. Results obtained for the unconnected tests are almost indistinguishable in the peak and plateau tests, both in behaviour and the
measured values. They show a constant decrease that flattens to constant values of about $3~\frac{\%}{\mathrm MB / \mathrm s}$ for Fast Ethernet and between $1.3$ and 
$1.6~\frac{\%}{\mathrm MB / \mathrm s}$ for Gigabit Ethernet. Apart from these values the same conclusions can be drawn from these graphs as 
already derived from Fig.~\ref{Fig:TCP-Msg-CyclesPerRate-Peak}.

\subsection{\label{Sec:TCPMsgLatency}TCP Message Class Latency}

To determine the latency of message send operations for the different configurations a number of messages are transmitted from a sender to a receiver. For each of these
messages the receiver sends a reply message to the originating sender. The sender in turn waits for this reply before sending its next message. 
A ping-pong message pattern is thus established between the two programs, similar to the network reference latency test principle.
Results obtained from this test are shown in Fig.~\ref{Fig:TCP-Msg-Latency}. 

\begin{figure}[ht!p]
\begin{center}
\resizebox*{0.50\columnwidth}{!}{
\includegraphics{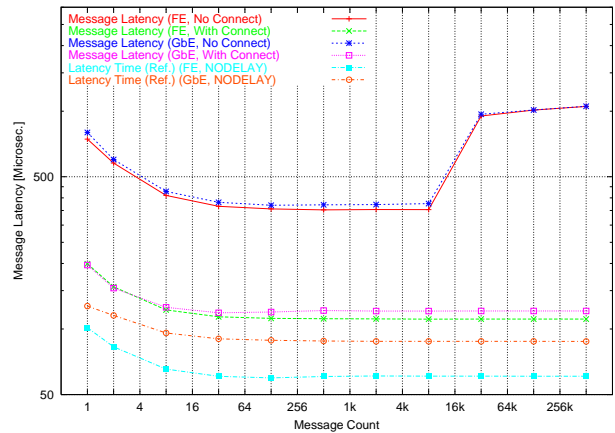}
}
\parbox{0.90\columnwidth}{
\caption[Message latency measurements.]{\label{Fig:TCP-Msg-Latency}The average latency (in $\mu \mathrm s$) of the message classes and the reference
measurements as a function of the message count. The measured times
include infrastructure overhead such as setting up the timing measurements.}
}
\end{center}
\end{figure}

As can be seen the measured latency decreases and approaches an asymptotic plateau value for all tests, although
the test remains on this plateau for the connected tests only. As expected the two unconnected tests have a latency much higher than the connected ones,
due to the overhead of establishing a new connection for each new message in the unconnected (or implicitly connected) tests.
The difference for the plateaus is almost 
a factor of 3. An unexpected and currently unexplained characteristic is displayed in both unconnected tests, starting between 8192 (8~k) 
and 32768 (32~k) messages, where the latency increases abruptly from about $370~\mu \mathrm s$ to $970~\mu \mathrm s$, a factor of about 2.5. It continues to increase
at a slower pace to about $1050~\mu \mathrm s$ for 524288 (512~k) messages. The initial higher start and decrease to the plateau value is most likely explained by
the measurement and infrastructure  overhead that dominates the timing results obtained from the measurements so that the plateau values reflect the actual latency 
present in applications. An obvious exception is the unexplained rise for high message counts in the unconnected tests, which
might be related to the drops shown in the graph of message rates as a function of message counts in section~\ref{Sec:NetworkMsgPlateauDetermination} 
and Fig.~\ref{Fig:TCP-Msg-Count-Rates}. But as for that graph the cause of the latency increase could not be determined in this thesis.

\begin{table}[hbt!p]
\begin{center}
\begin{tabular}{|l||c|c|c|c|c|c|}
\hline  
 & Fast & Fast & Gigabit & Gigabit     & Fast & Gigabit\\
 & Ethernet & Ethernet & Ethernet & Ethernet     & Ethernet & Ethernet\\
 & Implicit & Explicit & Implicit & Explicit & Reference & Reference\\
 & Connect & Connect & Connect & Connect &  & \\
\hline \hline
Average Message & 352.3 & 110.0 & 369.8 & 118.6 & 61 & 88 \\
Latency / $\mu \mathrm s$ &  &  &  &  &  &  \\
\hline
\end{tabular}
\parbox{0.90\columnwidth}{
\caption[Minimum message latency times.]{\label{Tab:TCP-Msg-Latency-Min}The minimum message latency times for the four configurations and the reference measurements.}
}
\end{center}
\end{table}

Table~\ref{Tab:TCP-Msg-Latency-Min} summarizes the minimum latency measured in
each of the four tests and the reference measurements. The unconnected tests, as already detailed, are higher than the connected ones by a factor of about 3. Gigabit Ethernet
tests are between 5.0~\% ($\approx 17.5~\mu \mathrm s$) and 7.8~\% ($\approx 8.6~\mu \mathrm s$) slower than Fast Ethernet ones for the unconnected and connected cases
respectively. One can see that the differences between the respective Fast and Gigabit Ethernet measurements are, to a first order, constant. The differences between a
message class measurement and the corresponding reference measurement can therefore be most probably attributed to overhead introduced in the class itself. 
This overhead therefore amounts to approximately $40~\mu \mathrm s$ and $285~\mu \mathrm s$ for explicit and implicit connection modes respectively.

\subsection{TCP Message Benchmark Summary}

One conclusion that can be drawn from these benchmarks is that the TCP message classes can be used in the data flow framework, especially in the ALICE HLT.
The framework is designed so that message sending rates do not have to be higher than the actual event rate, and all event related message transmissions are performed
over explicitly established connections. Therefore handling the 1~kHz maximum event rate required for the ALICE HLT presents no principal problem.
Fulfilling this requirement is in particular aided by the fact that no reply messages are needed and thus no restriction on the minimum latency exists. 
At the expected message sizes between 64~B and 512~B the CPU usage is expected to be between about 0.5~\% and 1.0~\% on the sender and 1.3~\% and 3.3~\% on the receiver
at the 1~kHz message rate required for proton-proton operation. These usage values are scaled to one 800~MHz CPU so that 100~\% corresponds to one fully used CPU. 
For the 200~Hz rate of
heavy-ion operation the CPU load is expected to be lower by the corresponding factor of 5. 
Message transfers are mainly dominated by the available memory bandwidth which
does not increase as fast as CPU power, as described in section~\ref{Sec:ComputingBackground}. CPU usage
 will most likely not decrease proportionally to the available CPU power in the future as a result.
As shown in the two different throughput tests, the classes seem to perform better when the messages can be sent in short bursts
% of up to 2048 messages 
than for a constant stream of messages to the receiver. 
%In addition to the rate, not only for the rate achieved but also for the number of CPU cycles consumed during the sending
%and receiving processes. 

A further result that can be inferred from the tests is related to the use of Fast or Gigabit Ethernet for message transfers. 
If latency is not the deciding factor, as for the classes' use in the framework, Gigabit Ethernet is the favored choice due to its lower CPU usage in sending the same
amount of messages or data (Fig.~\ref{Fig:TCP-Msg-CyclesPerRate} and~\ref{Fig:TCP-Msg-CyclesPerNetBW}), even if the network throughput of Fast Ethernet would already
be sufficient to fulfill the requirements. Fast Ethernet cards
should be chosen if latency is the primary concern,  due to their lower latencies compared with Gigabit Ethernet. This of course is only possible when they are able to fulfill the bandwidth requirements. 
These conclusions, however, should be treated with care as they 
depend on the respective network adapters used and the measurements should be repeated for adapters concerned. On the other hand, due
to their higher maximum throughput Gigabit Ethernet adapters are generally more efficient in their use of CPU cycles, and the technology
itself sets some restrictions on the minimum latency that can be reached. 
Cost differences between the adapters have to be considered as well, but due to its rapid evolution this is beyond the scope of this document. 
As a note on the cost calculation: Using a Gigabit Ethernet card to profit from the better efficiency does not necessitate a Gigabit Ethernet switch port as well.
If its bandwidth is sufficient, a Fast Ethernet switch can also be used, reducing the cost of this solution.
%{\Large Highly device specific, depending on optimization goal (throughput, CPU usage, usage / throughout, usage / rate) different block sizes
%can be optimal, not necessarily highest...}

In a direct comparison with the network reference tests several features can be noticed, also partially reflected
in the summaries of the peak and plateau throughput measurements in Table~\ref{Tab:TCPMsgMeasurementComparisonPlateau} and~\ref{Tab:TCPMsgMeasurementComparisonPeak}. 
A first difference can be observed in the behaviour 
of the sending rate in dependence of the block or message count. In the reference measurements the decrease after the initial plateau is not as obvious as in the message test.
Concerning the rate and throughput it can be seen that the results from the reference measurements are much better. The plateau values differ by a factor of 9
for Fast Ethernet and a factor of 6 for Gigabit Ethernet. Differences in the peak measurements are not as large, factors of 2 and 3 can only be observed here. 
These comparisons only apply to the connected message tests, as expected the results of the unconnected ones are far poorer. 
Examining the CPU usage as well as the efficiency of CPU usage divided by throughput it can be noticed
that the reference tests are better, except for the Gigabit Ethernet measurements on the sending nodes.
For this case the minimum values measured are actually below the ones from the reference
tests. The most likely explanation for this unexpected behaviour is that the use of \texttt{write} calls with timeouts followed by 
\texttt{se\-lect} calls in the message classes is more efficient for large blocks than the blocking \texttt{write} calls used in the reference program.
Looking at the message or block latencies it is again apparent that there is a certain time penalty associated with the functionality contained in the message classes,
as the results are almost a factor 2 worse for FE and 1.5 for GbE. 
Part of the overhead and performance decrease introduced by the message classes is certainly unavoidable due to the more elaborate checks and actions 
that have to be performed in them compared to the reference program. For example the reference program knows the block size in advance and 
can discard the data immediately after reading it. But even taking this into account, the results indicate that there still seem to be opportunities for improvement. 

%Other behaviour in dependence of message/block count, decrease not as obvious in reference
%Rate and throughput much better, plateau: factor of 6 FE, 9 GbE for small messages.
%                                 peak: factor of 2 FE, 3 GbE f. small messages
%Efficiency as best measure: plateau: sender better at large blocks, otherwise worse than reference, 
%                            factors small, large, avg: FE recv: 7, 1.7, 3.5; GbE recv: 7, 1.2, 3.7; FE sender: 2.6, 1/1.3, 1.6; GbE sender: 2.4, 1/1.3, 1.7
%                            peak: as for plateau, but GbE recv also lower
%                            factors small, large, avg: FE recv: 3.4, 1.8, 1.1; GbE recv: 1.0, 1/6.4, 1.7; FE sender: 3.2, 1/1.2, 1.6; GbE sender: 2.7, 1.3, 1.8; 
%                            (Recv FE from 2~kB, GbE from 32~kB)

\begin{table}[hbt!p]
\begin{center}
{\scriptsize
\begin{tabular}{|l||c|c|c|c|c|c|c|c|}
\hline  
                       & Rate / & Network                         & CPU Usage   & CPU Usage  & CPU Usage /          & CPU Usage /           & CPU Usage /                               & CPU Usage / \\
Measurement            & Hz     & Throughput /                    & Sender /    & Receiver / & Rate                 & Rate                  & Throughput                                & Throughput \\
Type                   &        & $\frac{\mathrm{MB}}{\mathrm s}$ & \%          & \%         & Sender /             & Receiver /            & Sender /                                  & Receiver /                                \\
                       &        &                                 &             &            & $\%\times \mathrm s$ & $\% \times \mathrm s$ & $\frac{\% \times \mathrm s}{\mathrm{MB}}$ & $\frac{\% \times \mathrm s}{\mathrm{MB}}$ \\
                       &        &                                 &             &            &                      &                       &                                                    & \\
\hline \hline
Reference              & 11.2 @ 1~M   & 10.8 @ 32                 & 12 @ 2~k    & 18.1 @ 16~k           & 0.000174 @ 32 & 0.000186 @ 32                      & 1.07 @ 16~k                                      & 1.62 @ 16~k \\
FE w.                  & 355~k @ 32 & 11.2 @ 1~M                 & 61.8 @ 32 & 66.2 @ 32           & 1.43 @ 1~M    & 1.98 @ 1~M                      & 5.70 @ 32                                     & 6.10 @ 32 \\
 \texttt{TCP\_\-NO\-DE\-LAY} & 45200        & 11.2                       & 19       & 25           & 0.175         & 0.222                      & 1.71                                          & 2.24 \\
\hline
Reference              & 11.2 @ 1~M   & 11.2 @ 32                 & 12 @ 2~k    & 18.1 @ 16~k           & 0.000168 @ 32 & 0.000180 @ 32                      & 1.07 @ 32~k                                    & 1.62 @ 16~k \\
FE w/o                 & 366~k @ 32 & 11.2 @ 1~M                 & 61.8 @ 32 & 65.8 @ 32           & 1.43 @ 1~M    & 1.87 @ 1~M                      & 5.52 @ 32                                     & 5.90 @ 32 \\
 \texttt{TCP\_\-NO\-DE\-LAY} & 45800        & 11.2                       & 18.9       & 25.3           & 0.173         & 0.230                      & 1.69                                           & 2.26 \\
\hline
Reference              & 68.9 @ 1~M   & 16.4 @ 32                 & 64 @ 8~k   & 78 @ 8~k           & 0.000182 @ 32 & 0.000186 @ 32                      & 0.854 @ 32~k                                    & 1.04 @ 32~k \\
GbE w.                 & 537~k @ 32 & 86.2 @ 64~k                & 100 @ 128  & 113.4 @ 256           & 1.04 @ 1~M    & 1.22 @ 1~M                      & 5.94 @ 32                                     & 6.10 @ 32 \\
 \texttt{TCP\_\-NO\-DE\-LAY} & 124~k       & 65.8                       & 79.5       & 92.2           & 0.130         & 0.151                      & 1.55                                           & 1.74 \\
\hline
Reference              & 69 @ 1~M     & 16.9 @ 32                 & 66.4 @ 8~k & 80.8 @ 8~k           & 0.000182 @ 32 & 0.000182 @ 32                      & 0.866 @ 32~k                                    & 1.04 @ 32~k \\
GbE w/o                & 553~k @ 32 & 88.4 @ 64~k                & 101.2 @ 32 & 112.6 @ 256           & 1.04 @ 1~M    & 1.22 @ 1~M                      & 6.00 @ 32                                     & 6.00 @ 32 \\
 \texttt{TCP\_\-NO\-DE\-LAY} & 125       & 67                         & 80.9       & 93.2           & 0.130         & 0.150                      & 1.55                                           & 1.73 \\
\hline
Msg Class              & 11.2 @ 1~M   & 1.82 @ 32                 & 12 @ 8~k     & 34.1 @ 64~k           & 0.000434 @ 32         & 0.00256 @ 32                      & 1.07 @ 1~M                                                   & 3.04 @ 64~k \\
FE                     & 59500 @ 32  & 11.2 @ 1~M                 & 29.6 @ 128 & 153 @ 32           & 1.07 @ 1~M                  & 3.20 @ 1~M                      & 14.2 @ 32                                                   & 84 @ 32 \\
w. Connect             & 15900        & 9.73                       & 16.8        & 72.4           & 0.134                   & 0.402                      & 2.74                                                   & 13.5 \\
\hline
Msg Class              & 11.2 @ 1~M   & 0.0147 @ 32               & 12 @ 512~k   & 8.07 @ 64           & 0.0674 @ 32~k           & 0.0167 @ 32                      & 1.07 @ 1~M                                                   & 3.22 @ 1~M \\
FE                     & 483 @ 128   & 11.2 @ 1~M                 & 98 @ 128   & 44.2 @ 32~k           & 1.07 @ 1~M               & 3.22 @ 1~M                      & 6100 @ 32                                                   & 548 @ 32 \\
w/o Connect            & 343          & 5.07                       & 60.1          & 23.6           & 0.254                    & 0.424                      & 794                                                   & 74.6 \\
\hline
Msg Class              & 75.8 @ 1~M   & 1.82 @ 32                 & 26 @ 32    & 106.4 @ 16~k           & 0.000438 @ 32          & 0.00266 @ 64                      & 0.788 @ 256~k                                                   & 1.35 @ 32~k \\
GbE                    & 59500 @ 32  & 85.3 @ 64~k                & 72 @ 64~k   & 177 @ 32           & 0.792 @ 1~M               & 1.43 @ 1~M                      & 14.3 @ 32                                                   & 97.6 @ 32 \\
w. Connect             & 21900        & 50.7                       & 53.3        & 131.4           & 0.100                     & 0.181                      & 2.66                                                   & 12.8 \\
\hline
Msg Class              & 38.5 @ 1~M   & 0.0147 @ 32               & 34 @ 1~M    & 6.66 @ 64           & 0.0994 @ 64~k            & 0.0138 @ 64                      & 0.882 @ 1~M                                                   & 1.58 @ 1~M \\
GbE                    & 483 @ 2~k    & 38.5 @ 1~M                 & 92.4 @ 4~k  & 61 @ 512~k           & 0.882 @ 1~M            & 1.58 @ 1~M                      & 6240 @ 32                                                   & 1020 @ 32 \\
w/o Connect            & 384          & 11.9                       & 70.9        & 26.6           & 0.242                     & 0.220                      & 780                                                   & 93.8 \\
\hline
\end{tabular}
}
\parbox{0.90\columnwidth}{
\caption[TCP reference and message class plateau measurements.]{\label{Tab:TCPMsgMeasurementComparisonPlateau}Comparison of the TCP reference and message class plateau measurements. 
Minimum and maximum values with their respective block size in bytes are shown as well as the average of all values.
For the reference tests only the block range from 32~B to 1~MB has been used, corresponding to the range
covered by the message class tests. } %Sizes are in bytes.}
}
\end{center}
\end{table}

\begin{table}[hbt!p]
\begin{center}
{\scriptsize
\begin{tabular}{|l||c|c|c|c|c|c|c|c|}
\hline  
                       & Rate / & Network                         & CPU Usage   & CPU Usage  & CPU Usage /          & CPU Usage /           & CPU Usage /                               & CPU Usage / \\
Measurement            & Hz     & Throughput /                    & Sender /    & Receiver / & Rate                 & Rate                  & Throughput                                & Throughput \\
Type                   &        & $\frac{\mathrm{MB}}{\mathrm s}$ & \%          & \%         & Sender /             & Receiver /            & Sender /                                  & Receiver /                                \\
                       &        &                                 &             &            & $\%\times \mathrm s$ & $\% \times \mathrm s$ & $\frac{\% \times \mathrm s}{\mathrm{MB}}$ & $\frac{\% \times \mathrm s}{\mathrm{MB}}$ \\
                       &        &                                 &             &            &                      &                       &                                                    & \\
\hline \hline
Reference              & 11.2 @ 1~M   & 11.2 @ 256~k                    & 12.2 @ 8~k   & 15.6 @ 2~k           & 0.000186 @ 32                     & 0.00242 @ 1~k                      & 1.08 @ 8~k                                                   & 1.38 @ 2~k \\
FE w.                  & 462~k @ 32 & 14.1 @ 32                      & 85.6 @ 32  & 28.3 @ 1~k           & 1.43 @ 1~M                     & 1.81 @ 1~M                      & 6.08 @ 32                                                   & 2.48 @ 1~k \\
 \texttt{TCP\_\-NO\-DE\-LAY} & 54200        & 11.7                            & 26.8        & 19.6           & 0.175                     & 0.329                      & 2.2                                                   & 1.74 \\
\hline
Reference              & 11.2 @ 1~M   & 11.2 @ 256~k                    & 12.0 @ 16~k    & 15.6 @ 2~k           & 0.000132 @ 32                     & 0.00238 @ 1~k                      & 1.07 @ 16~k                                                   & 1.38 @ 2~k \\
FE w/o                 & 624~k @ 32 & 19 @ 32                        & 82.8 @ 32  & 27.9 @ 1~k           & 1.35 @ 1~M                     & 1.81 @ 1~M                      & 4.44 @ 64                                                   & 2.44 @ 1~k \\
 \texttt{TCP\_\-NO\-DE\-LAY} & 68400        & 12.3                            & 26.4        & 19.8           & 0.166                     & 0.33                      & 2.02                                                   & 1.76 \\
\hline
Reference              & 70.4 @ 1~M   & 18.1 @ 32                      & 75.6 @ 1~M  & 88 @ 1~M           & 0.000154 @ 32                     & 0.0619 @ 64~k                      & 0.834 @ 16~k                                                   & 0.991 @ 64~k \\
GbE w.                 & 593~k @ 32 & 109 @ 16~k                      & 93.0 @ 4~k  & 96.6 @ 128~k           & 1.07 @ 1~M                     & 1.25 @ 1~M                      & 5.06 @ 32                                                   & 1.29 @ 256~k \\
 \texttt{TCP\_\-NO\-DE\-LAY} & 111~k       & 70.8                            & 87.6        & 92.3           & 0.134                     & 0.482                      & 1.66                                                   & 1.2 \\
\hline
Reference              & 68.9 @ 1~M   & 19.1 @ 32                      & 72.0 @ 1~M    & 81.6 @ 512           & 0.000152 @ 32                     & 0.000506 @ 512                      & 0.872 @ 32~k                                                   & 1.04 @ 512 \\
GbE w/o                & 627~k @ 32 & 106 @ 8~k                       & 102 @ 1~k  & 113 @ 8~k           & 1.04 @ 1~M                     & 1.21 @ 1~M                      & 4.96 @ 32                                                   & 1.23 @ 512~k \\
 \texttt{TCP\_\-NO\-DE\-LAY} & 137~k       & 73.4                            & 88.0          & 94.6           & 0.130                     & 0.201                      & 1.474                                                   & 1.13 \\
\hline
Msg Class              & 11.2 @ 1~M   & 6.69 @ 32                      & 12.0 @ 64~k    & 33.1 @ 16~k           & 0.000418 @ 32                     & 0.00184 @ 256                      & 1.07 @ 64~k                                                   & 2.95 @ 16~k \\
FE                     & 219~k @ 32 & 13.1 @ 64                      & 91.8 @ 32  & 93.1 @ 256           & 1.07 @ 1~M                     & 3.28 @ 1~M                      & 13.7 @ 32                                                   & 7.55 @ 256 \\
w. Connect             & 39200        & 11.2                            & 32.8        & 44.6           & 0.135                     & 0.497                      & 3.14                                                   & 3.89 \\
\hline
Msg Class              & 11.2 @ 1~M   & 0.123 @ 32                     & 12 @ 1~M     & 37 @ 1~M           & 0.0204 @ 128                     & 0.0185 @ 32                      & 1.07 @ 1~M                                                   & 3.3 @ 1~M \\
FE                     & 4180 @ 128  & 11.2 @ 1~M                      & 86.4 @ 64  & 80.6 @ 128           & 1.07 @ 1~M                     & 3.3 @ 1~M                      & 694 @ 32                                                   & 607 @ 32 \\
w/o Connect            & 1870         & 6.48                            & 46.6        & 55.5           & 0.164                     & 0.433                      & 86.8                                                   & 79.6 \\
\hline
Msg Class              & 75.5 @ 1~M   & 6.77 @ 32                      & 54.8 @ 1~k  & 70.9 @ 2~k           & 0.000392 @ 64                     & 0.00151 @ 256                      & 0.798 @ 1~M                                                   & 1.13 @ 16~k \\
GbE                    & 232~k @ 64 & 90.9 @ 16~k                     & 91.2 @ 64  & 119 @ 8~k           & 0.798 @ 1~M                     & 1.46 @ 1~M                      & 13.4 @ 32                                                   & 6.19 @ 256 \\
w. Connect             & 48100        & 54.2                            & 71.0        & 102           & 0.101                     & 0.222                      & 2.72                                                   & 2.05 \\
\hline
Msg Class              & 70 @ 1~M     & 0.104 @ 32                     & 46.8 @ 16~k & 50.6 @ 16~k           & 0.0177 @ 256                     & 0.0149 @ 64                      & 0.838 @ 1~M                                                   & 1.48 @ 1~M \\
GbE                    & 3410 @ 256  & 70 @ 1~M                        & 73.4 @ 8~k  & 104 @ 1~M           & 0.838 @ 1~M                     & 1.48 @ 1~M                      & 588 @ 32                                                   & 493 @ 32 \\
w/o Connect            & 2010         & 21.3                            & 58.2        & 63.8           & 0.131                     & 0.204                     & 74                                                   & 62.7 \\
\hline
\end{tabular}
}
\parbox{0.90\columnwidth}{
\caption[TCP reference and message class peak measurements.]{\label{Tab:TCPMsgMeasurementComparisonPeak}Comparison of the TCP reference and message class peak measurements. 
Minimum and maximum values with their respective block size in bytes are shown as well as the average of all values.
For the reference tests only the block range from 32~B to 1~MB has been used, corresponding to the range
covered by the message class tests. } %Sizes are in bytes.}
}
\end{center}
\end{table}

\subsection{TCP Blob Class Throughput with On-Demand Allocation}

Similar to the message classes the throughput benchmark using on-demand allocation for the blob class 
 also consists of two parts, 
the initial evaluation of the number of blocks or {\em blobs}
sent for each size and the actual throughput measurement as a function of the block size. 

\subsubsection{Plateau Determination}

\begin{figure}[ht!p]
\begin{center}
\resizebox*{0.50\columnwidth}{!}{
\includegraphics{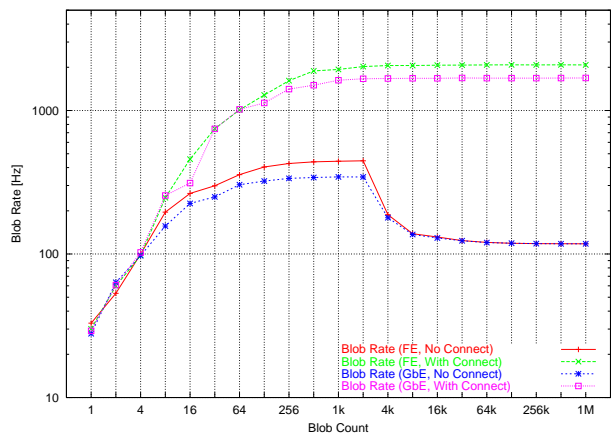}
}
\parbox{0.90\columnwidth}{
\caption[The blob sending rates in dependence of the number of blocks (on-demand alloc.).]{\label{Fig:TCP-Blob-Count-Rates}The blob sending rates in dependence of the number of blocks (on-demand allocation).}
}
\end{center}
\end{figure}

The results of the plateau measurements that have been made for the blob communication mechanism are shown in Fig.~\ref{Fig:TCP-Blob-Count-Rates}. 
It can be realized that for the two connected tests the rate rises steadily with the number of blocks to a plateau reached between
8~k and 32~k blocks. Unlike the message communication these tests show no peak over the measured spectrum. For the unconnected tests a similar 
rise is displayed which drops off again and reaches a lower plateau starting at about 8~kB. 
For the blob throughput tests' evaluation of the plateau value the number of 32~k blocks was chosen as a common 
point of reference.
A 2~k (2048) blob count  was chosen for the peak values of the 
unconnected tests, while for the connected tests the values at 32~k blobs were reused as no real peak exists for these tests.

\subsubsection{Plateau Throughput Measurement}

\begin{figure}[ht!p]
\begin{center}
\resizebox*{0.50\columnwidth}{!}{
\includegraphics{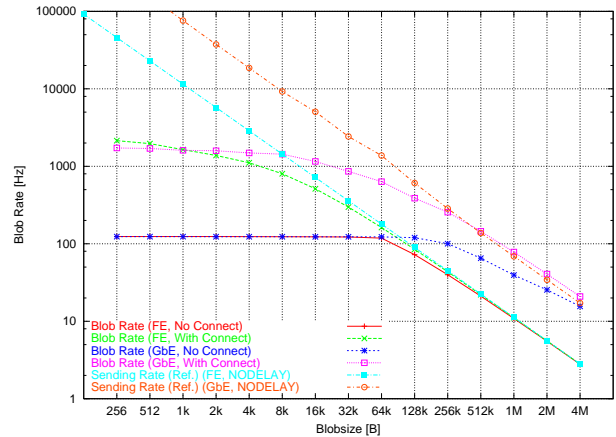}
}
\parbox{0.90\columnwidth}{
\caption[The measured blob sending rates (plateau, on-demand alloc.).]{\label{Fig:TCP-Blob-Rates}The measured blob sending rates (blob count 32~k, on-demand allocation).}
}
\end{center}
\end{figure}

Fig.~\ref{Fig:TCP-Blob-Rates} to \ref{Fig:TCP-Blob-CyclesPerNetBW} display
%\ref{Fig:TCP-Blob-IntrPerSec} 
the results that have been obtained during the on-demand allocation 
throughput tests using a blob count of 32768 (32~k) for each size, going from 256~B to 4~MB. The first result, the achievable sending rate, is shown
in Fig.~\ref{Fig:TCP-Blob-Rates}. As can be seen, the maximum values are about 1.5~kHz to 2~kHz for the connected tests and about 120~Hz for the unconnected tests.
In the Fast Ethernet test with explicit connections the maximum value is somewhat higher than 2~kHz, reached at the smallest block size of 256~B. With increasing block 
sizes the rate continously decreases. From about 16~kB to 32~kB the decrease in rate becomes linear with block size as the available network
bandwidth becomes the limiting factor, as can be seen in Fig.~\ref{Fig:TCP-Blob-NetBW}, too. 
For the connected Gigabit Ethernet test an initial plateau with only
a slight decrease can be observed from 256~B to 8~kB block sizes. At 8~kB blocks a steeper decrease starts which also develops into a linear decrease 
when the network starts to
become the bottleneck. 
%Absolute values of this test are higher than in the corresponding Fast Ethernet test, as could be expected due to GbE's higher bandwidth. 
Except for the initial two values at 256~B and 512~B, 
where it is slightly lower than Fast Ethernet, the connected GbE test constantly shows the highest achievable sending rate, as could be expected due to its higher
available bandwidth. The fact that the FE curve is higher at 256~B could be explained by its lower latency in exchanging the allocation messages which dominates
the rate at these small sizes. 
Between 64~kB and 128~kB the unconnected Gigabit Ethernet rate starts to exceed
the connected Fast Ethernet curve, which closely approaches the unconnected Fast Ethernet curve after 256~kB block sizes. 
For these sizes the overhead of establishing the connection
for each block thus is becoming negligible compared to the sending of the large blocks.
In comparison with the reference measurements one can see that the connected blob tests reach the reference values later than the corresponding message class curves.
This is most likely due to the overhead of allocating a block in the remote buffer and sending of an additional message to announce the block to the receiver. 
But, similar to the message class, for large blocks the achieved sending rate exceeds the one of the reference benchmark. The explanation for this is the same
as given in the message class section: The message class usage of preceeding the \texttt{read} and \texttt{write} calls with \texttt{se\-lect} class seems to be
more efficient than the simple use of blocking \texttt{read} and \texttt{write} calls.

\begin{figure}[ht!p]
\begin{center}
\resizebox*{0.50\columnwidth}{!}{
\includegraphics{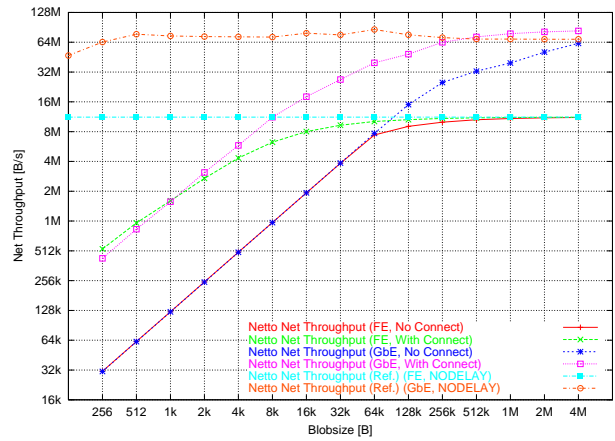}
}
\parbox{0.90\columnwidth}{
\caption[The application level network throughput for TCP blob sending (plateau, on-demand alloc.).]{\label{Fig:TCP-Blob-NetBW}The application level network throughput for TCP blob sending (blob count 32~k, on-demand allocation).}
}
\end{center}
\end{figure}

The measured network throughputs of the tests are shown in Fig.~\ref{Fig:TCP-Blob-NetBW}. One can
see that the blob class throughput curves rise linearly with three separate slopes to a specific point for each curve. At these points the increase
slows down as the hardware limit is approached, which can be seen by comparison with the reference measurements. In this comparison one can also see again that 
the connected GbE blob class throughput exceeds the respective reference throughput for large blocks, as already observed in the rate measurement above. 

%The measured network throughputs of the tests are shown in Fig.~\ref{Fig:TCP-Blob-NetBW}. One can
%see that the throughput curves rise linearly with three separate slopes to a specific point for each curve. At this point the increase
%slows down as the network limit is approached. Similar to the sending rate the two unconnected curves are identical up to 64~kB 
%blocks where the Fast Ethernet curve starts to saturate the network and approaches its connected counterpart. This curve starts at a higher
%throughput for small blocks compared to the connected Gigabit curve, but the latter reaches higher values from 1~kB on due to the former's lower slope. 
%Saturation for the connected Fast Ethernet test is already reached between 16~kB and 32~kB. Between 64~kB and 128~kB the unconnected 
%Gigabit Ethernet curve exceeds it as well. 
%At the largest block sizes from 512~kB the beginning of a plateau can be seen for the connected GbE measurement, while the unconnected 
%one still shows a linear increase, although with a less steep slope after 128~kB. 

\begin{figure}[ht!p]
\begin{center}
\resizebox*{1.0\columnwidth}{!}{
\includegraphics{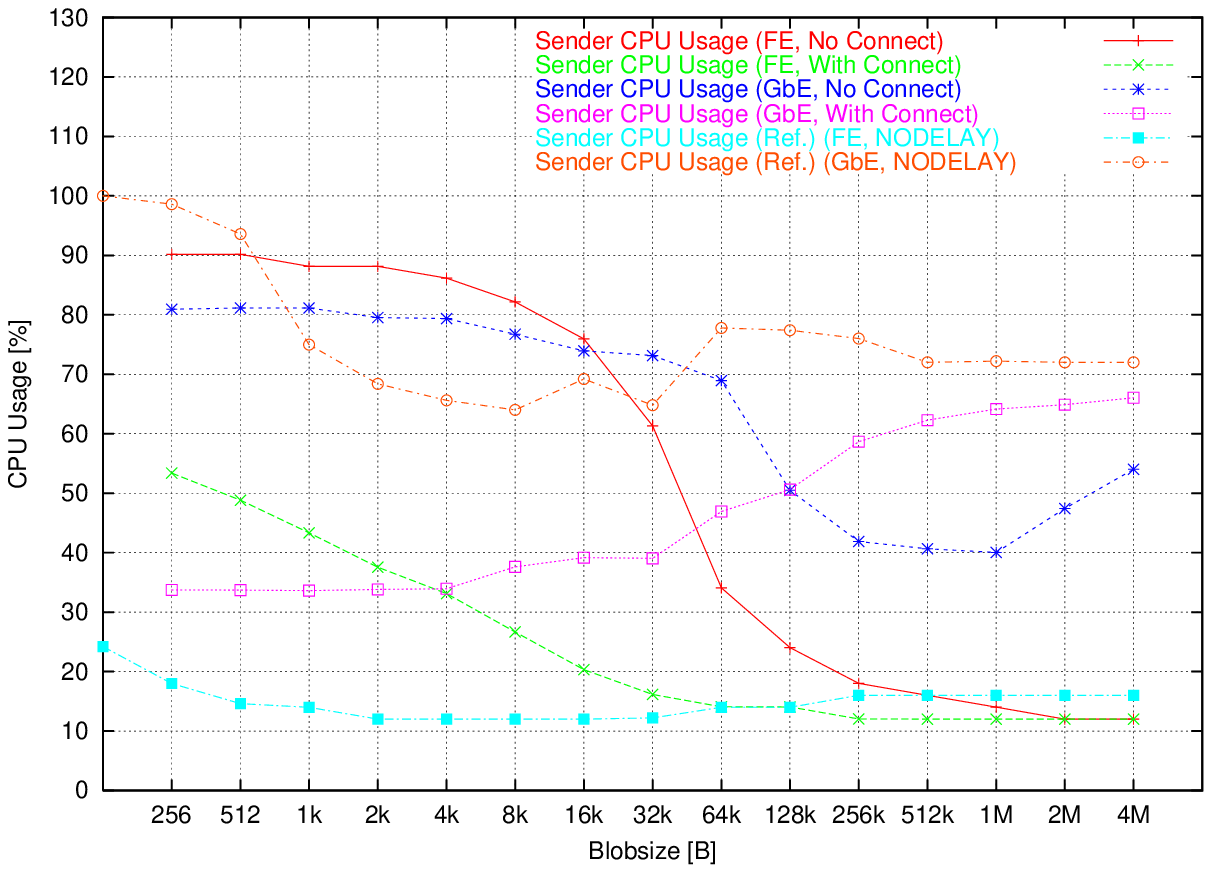}
\hfill
\includegraphics{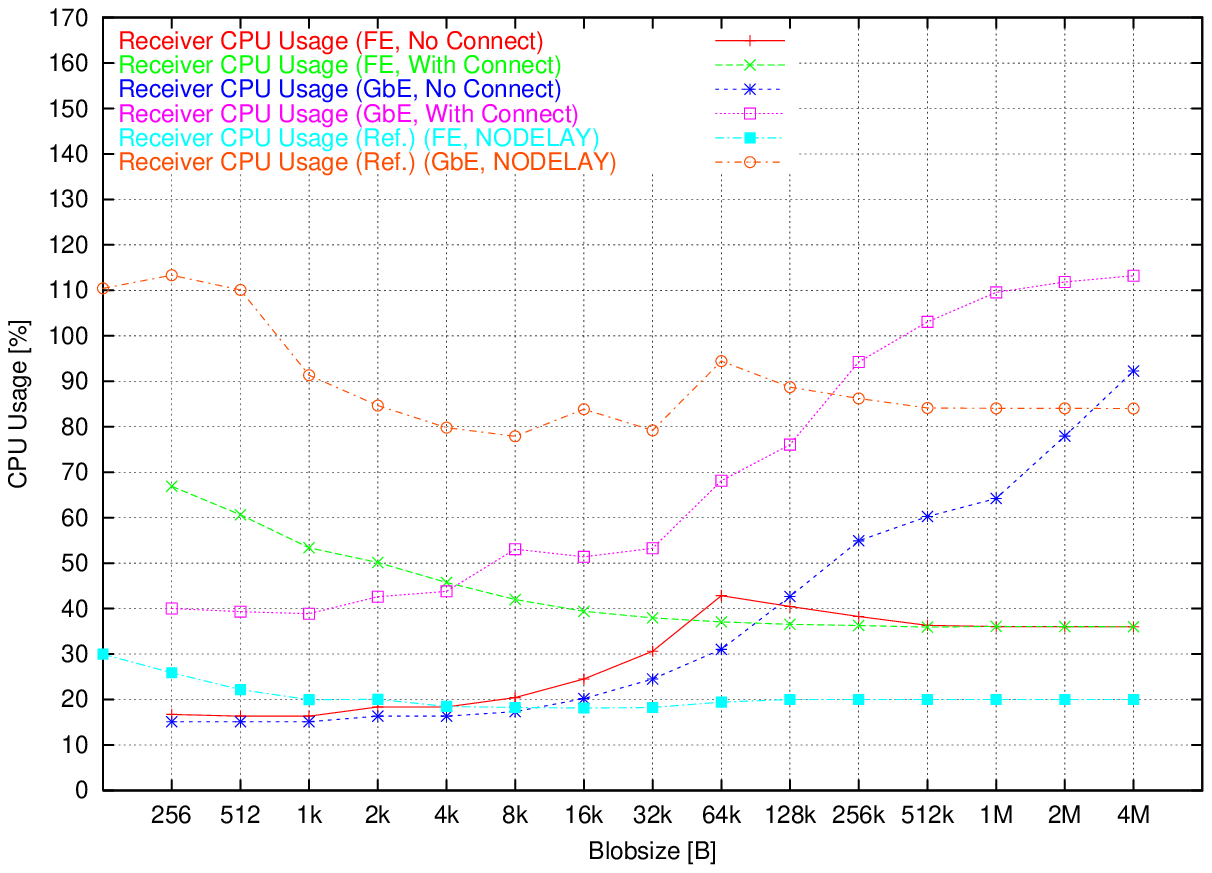}
}
\parbox{0.90\columnwidth}{
\caption[CPU usage during TCP blob sending (plateau, on-demand alloc.).]{\label{Fig:TCP-Blob-Cycles}The CPU usage on the sender (left) and receiver (right) during TCP blob sending (blob count 32~k, on-demand allocation).
The nodes are twin CPU nodes, 100~\% CPU usage corresponds to one CPU being fully used.}
}
\end{center}
\end{figure}

CPU usage during the blob transfers can be seen in Fig.~\ref{Fig:TCP-Blob-Cycles}, for the sender on the left and the receiver on the right hand side. 
On the sender one can see that the connected Fast and Gigabit Ethernet blob class measurements are lower than the corresponding reference measurements;
the FE one for large blocks and the GbE one for the whole test range. This corresponds to behaviour already observed in the message class tests. 
The reason why the GbE curve displays less CPU usage even at small block sizes is very likely related to the fact that the achieved rates and throughputs
at these block sizes are noticeably lower than in the reference measurements. The other notable feature in the sender graph is the very high load
of the two unconnected blob measurements at small blocks. This is also similar to the observed message class behaviour and is most likely caused by the 
connection establishing overhead. 
On the receiver node the blob class connected Fast Ethernet curve is constantly at higher values than its reference counterpart. The connected GbE blob class curve
is lower than the reference measurement at small block sizes and rises above it for large blocks, approximately where its rate and throughput also exceed the reference
ones. At small block sizes the low usage therefore seems to be, at least in part, caused by the low sending rate and throughput values.

\begin{figure}[ht!p]
\begin{center}
\resizebox*{1.0\columnwidth}{!}{
\includegraphics{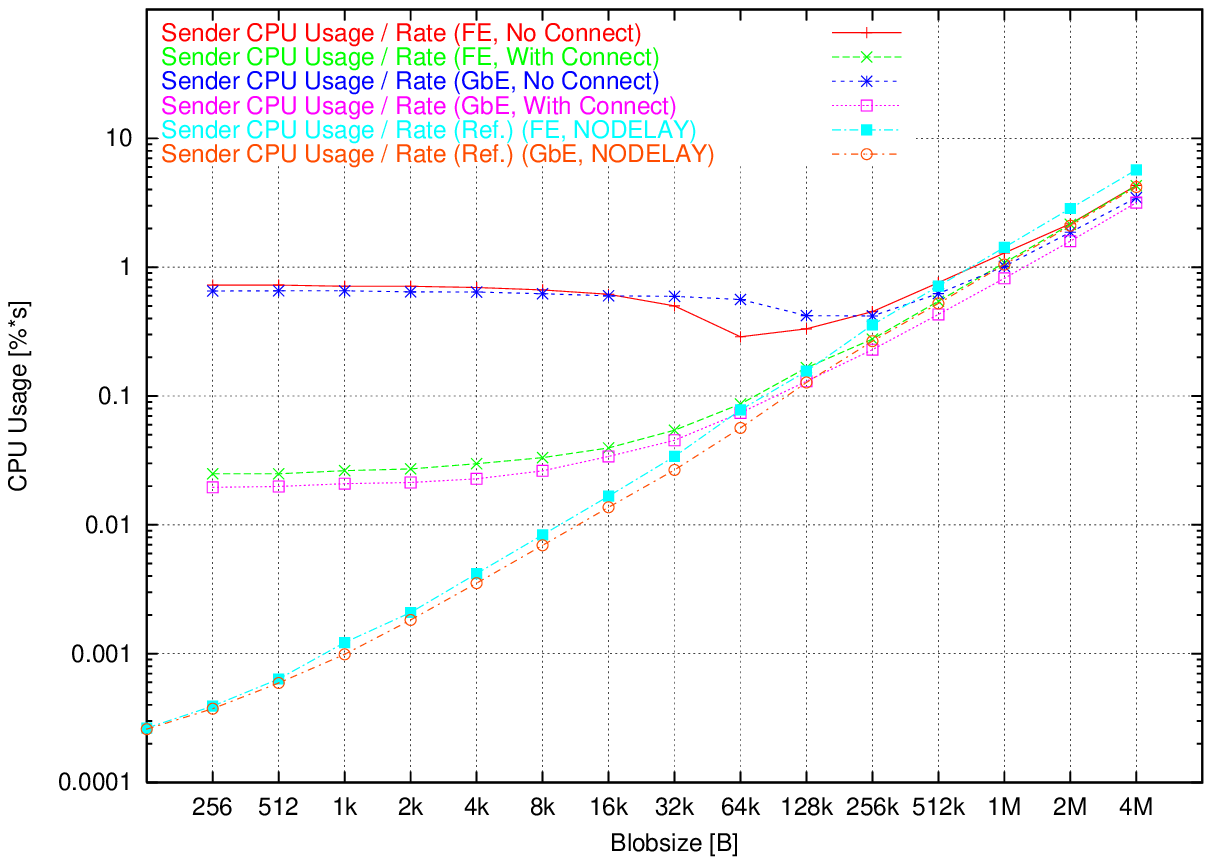}
\hfill
\includegraphics{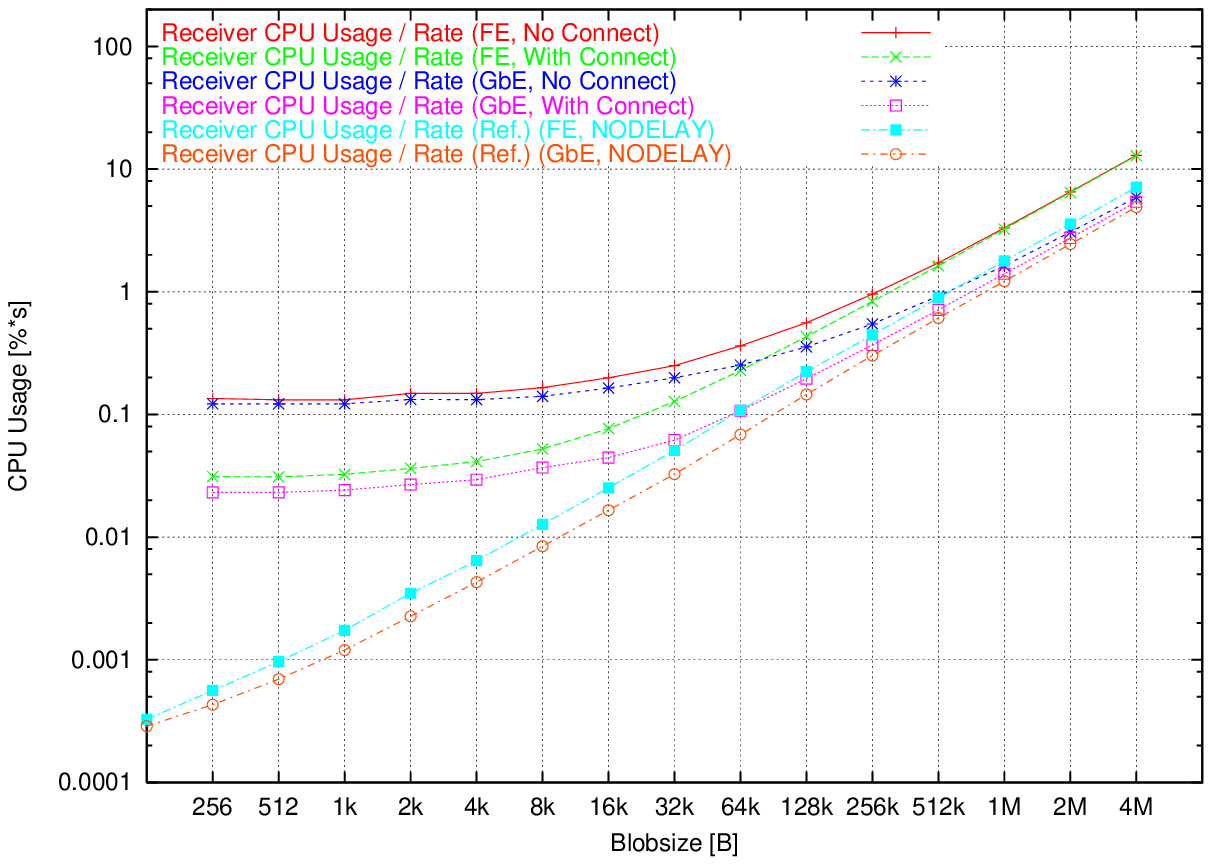}
}
\parbox{0.90\columnwidth}{
\caption[CPU usage divided by the sending rate during TCP blob sending (plateau, on-demand alloc.).]{\label{Fig:TCP-Blob-CyclesPerRate}The CPU usage on the sender (left) and receiver (right) divided by the sending rate 
during TCP blob sending (blob count 32~k, on-demand allocation).
The nodes are twin CPU nodes, 100~\% CPU usage corresponds to one CPU being fully used.}
}
\end{center}
\end{figure}

\begin{figure}[ht!p]
\begin{center}
\resizebox*{1.0\columnwidth}{!}{
\includegraphics{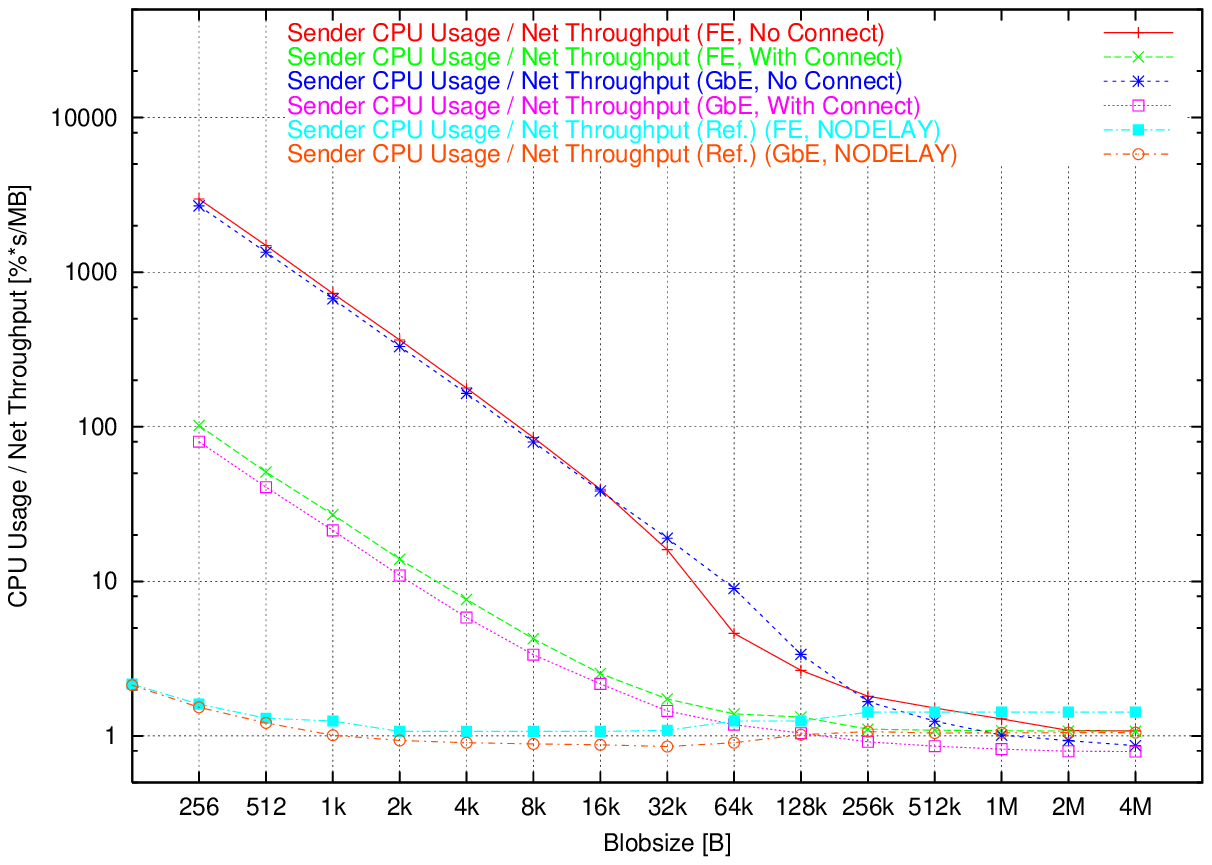}
\hfill
\includegraphics{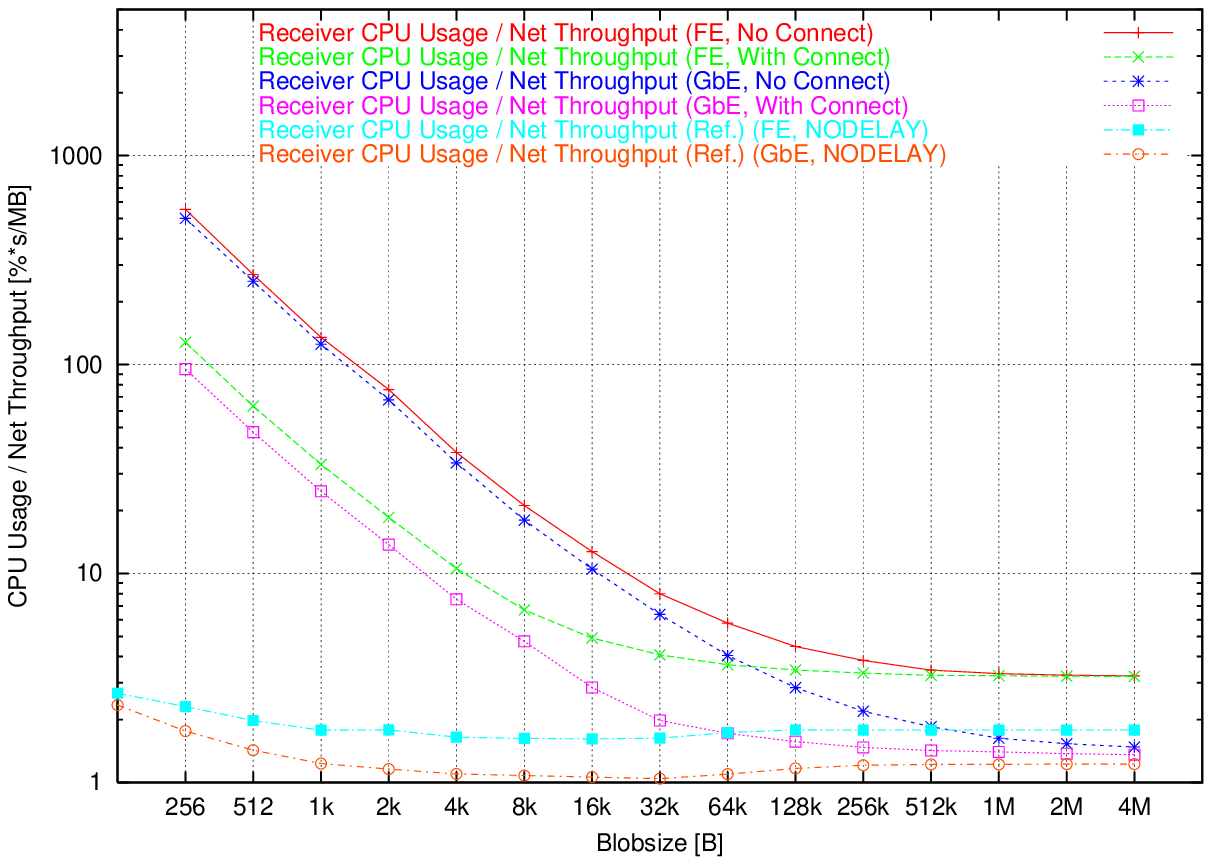}
}
\parbox{0.90\columnwidth}{
\caption[CPU usage per MB/s network throughput during TCP blob sending (plateau, on-demand alloc.).]{\label{Fig:TCP-Blob-CyclesPerNetBW}The CPU usage on the sender (left) and receiver (right) per MB/s network throughput 
during TCP blob sending (blob count 32~k, on-demand allocation).
The nodes are twin CPU nodes, 100~\% CPU usage corresponds to one CPU being fully used.}
}
\end{center}
\end{figure}

To allow a better interpretation of the CPU usage values they are normalized to the achieved blob sending rate and network throughput, 
similar to the message throughput tests. The results are displayed in Fig.~\ref{Fig:TCP-Blob-CyclesPerRate}
and Fig.~\ref{Fig:TCP-Blob-CyclesPerNetBW} respectively. These curves are very similar in appearance to
those for message sending in Fig.~\ref{Fig:TCP-Msg-CyclesPerRate} and Fig.~\ref{Fig:TCP-Msg-CyclesPerNetBW},
with the exception of the higher blob class usages compared to the corresponding message classes usages. 
This is the case for sender and receiver,
connected and unconnected, as well as Fast and Gigabit Ethernet tests. A possible explanation for this is that for each blob to be transferred three messages 
(block allocation request and reply as well as block announcement) have to be sent
as well, which speaks in favor of the preallocation method examined below in section~\ref{Sec:TCPBLobPreAllocThroughput}.

\subsubsection{Peak Throughput Measurement}

To measure the peak blob class throughput the implicit connection tests have been run with a blob count of 2048 (2~k) where the throughput for 256~B blocks
has reached its peak value. Unlike these unconnected tests, the connected ones have not been rerun as the respective curves from Fig.~\ref{Fig:TCP-Blob-Count-Rates} 
do not show a 
peak value. Instead the 32~k count measurements from the previous section have been used  for this measurement as well.
The discussion of the measurements and their differences is therefore primarily
focussed on the implicit connection tests. Results that have been obtained from these measurements are shown in 
Fig.~\ref{Fig:TCP-Blob-Rates-Peak} to~\ref{Fig:TCP-Blob-CyclesPerNetBW-Peak}. 

\begin{figure}[ht!p]
\begin{center}
\resizebox*{0.50\columnwidth}{!}{
\includegraphics{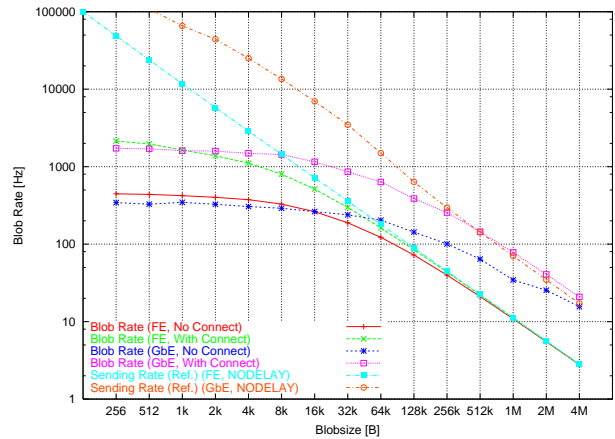}
}
\parbox{0.90\columnwidth}{
\caption[The measured blob sending rates (peak, on-demand alloc.).]{\label{Fig:TCP-Blob-Rates-Peak}The measured blob sending rates (blob counts 32~k and 2~k, on-demand allocation).}
}
\end{center}
\end{figure}

For the achieved blob sending rates shown in Fig.~\ref{Fig:TCP-Blob-Rates-Peak} it can be seen that the results for the peak tests
are higher than those from the previous plateau tests of Fig.~\ref{Fig:TCP-Blob-Rates}.
The results of the tests differ by  a factor of about 3 and 4 for the Fast and Gigabit Ethernet measurements respectively. 
A comparison of the connected tests with the respective reference measurements yields the same qualitative results as for the
plateau test. The reference rates are reached only for relatively large blocks, but the connected GbE measurement exceeds the 
GbE reference curve for the largest blocks, as before. Both blob class Fast Ethernet measurements initially have higher rates 
than their respective Gigabit Ethernet counterpart. As for the plateau test this is again presumed to be due to Fast Ethernet's lower
latency. Since at the operating system level messages have to be exchanged for connection establishing and termination the increased
latency should influence the unconnected tests more than the connected ones. This is reflected in the graph, as the connected GbE curve
exceeds the connected FE curve earlier than the unconnected GbE curve exceeds the unconnected FE curve. 
One interesting point to be found in the plot is the transition from 1~MB to 2~MB
blocks for the unconnected Gigabit Ethernet curve, where the new curve displays a bend and for large blocks becomes identical to the curve
from the previous plateau measurement. The reason for this bend are presumably buffers filled by the larger blocks which causes the same behaviour for the peak
test as for the plateau test. 

%For the achieved blob sending rates shown in Fig.~\ref{Fig:TCP-Blob-Rates-Peak} it can be seen that the results for the peak tests
%are higher than those from the previous plateau tests of Fig.~\ref{Fig:TCP-Blob-Rates}.
%The results of the tests differ by  a factor of about 3 and 4 for the Fast and Gigabit Ethernet measurements respectively. 
%Both tests are missing the initial flat plateau from the previous test. Instead they start directly
% with a slight decrease in rate which gradually
%steepens as the available network bandwidth increasingly becomes the bottleneck that defines the sustainable rate with growing block sizes.
%In the limit of large blocks both new tests also approach the results from the plateau throughput tests as well as those from their
%respective connected tests. For the Fast Ethernet tests the curves actually become identical after block sizes of around 256~kB. Also unlike the 
%plateau throughput tests the Fast and Gigabit Ethernet  curves initially are not identical, instead the FE one is higher by about 125~Hz, a third of the
%GbE value. This effect is probably caused by the lower message latency measured for Fast Ethernet in section~\ref{Sec:TCPMsgLatency}, which affects the 
%negotiation messages exchanged between the two blob objects. One interesting point to be found in the plot is the transition from 1~MB to 2~MB
%blocks for the unconnected Gigabit Ethernet curve, where the new curve displays a bend and from then on becomes identical to the curve
%from the previous plateau measurement. 

\begin{figure}[ht!p]
\begin{center}
\resizebox*{0.50\columnwidth}{!}{
\includegraphics{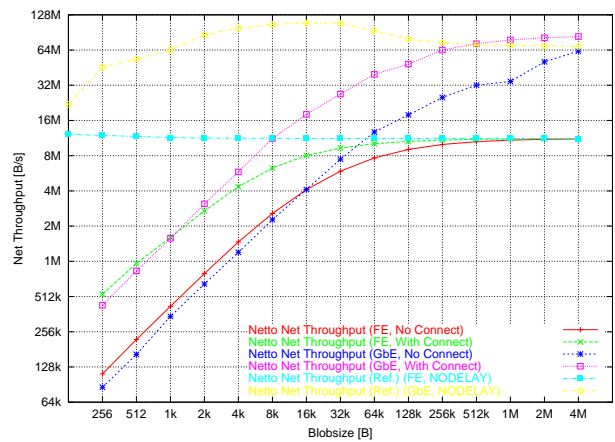}
}
\parbox{0.90\columnwidth}{
\caption[The application level network throughput for TCP blob sending (peak, on-demand alloc.).]{\label{Fig:TCP-Blob-NetBW-Peak}The application level network throughput for TCP blob sending (blob counts 32~k and 2~k, on-demand allocation).}
}
\end{center}
\end{figure}

From the measurements of the application level network throughput in Fig.~\ref{Fig:TCP-Blob-NetBW-Peak} the same tendencies as for the blob sending rate
can be derived. The throughput for the unconnected tests is higher by factors of about 3 to 4 for small blocks up to the point where
the available bandwidth becomes the limit. One can also see the bend in the unconnected GbE curve that marks the transition from the
peak to the plateau throughput measurement. This bend appears even more pronounced than in the rate plot of Fig.~\ref{Fig:TCP-Blob-Rates-Peak}.

\begin{figure}[ht!p]
\begin{center}
\resizebox*{1.0\columnwidth}{!}{
\includegraphics{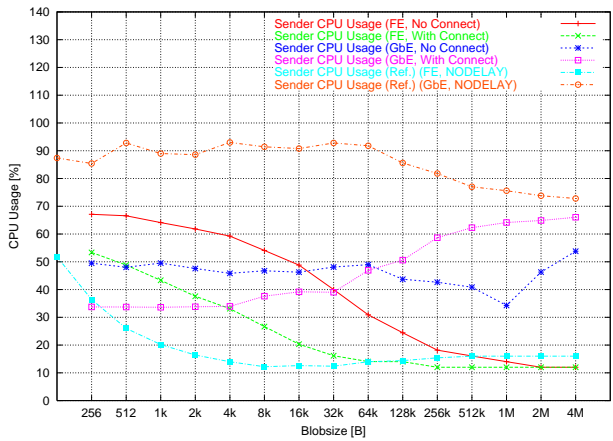}
\hfill
\includegraphics{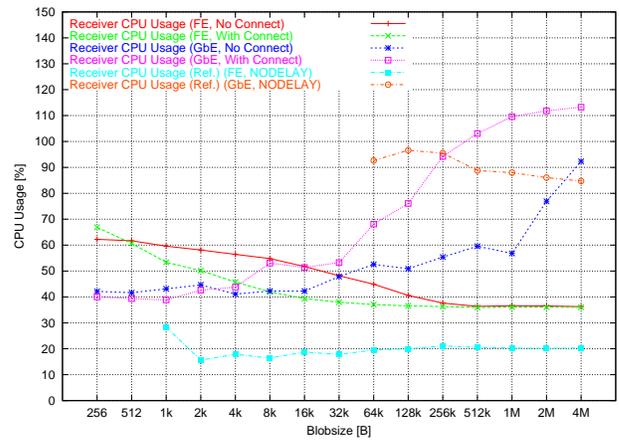}
}
\parbox{0.90\columnwidth}{
\caption[CPU usage during TCP blob sending (peak, on-demand alloc.).]{\label{Fig:TCP-Blob-Cycles-Peak}The CPU usage on the sender (left) and receiver (right) during TCP blob sending (blob counts 32~k and 2~k, on-demand allocation).
The nodes are twin CPU nodes, 100~\% CPU usage corresponds to one CPU being fully used.}
}
\end{center}
\end{figure}

In the CPU usage measurements of the peak throughput tests in  Fig.~\ref{Fig:TCP-Blob-Cycles-Peak} one can see that compared to the plateau
tests the sender CPU usage in general reaches lower values. For the receiver on the other hand the measured peak test usages are higher than the
ones from the corresponding plateau tests. As for the previous communication class measurements it can be seen that the connected CPU usages on the 
sender are still lower than the reference ones; for FE only for large blocks but for GbE over the whole test range. On the receiver the plateau behaviour is
also repeated. Connected FE blob class usage is always higher than the reference measurement and connected GbE usage exceeds the respective reference usage
only for large blocks, approximately in the range where the class reaches a higher throughput than the reference measurement. 
One can also see, on the sender as well as on the receiver, that the unconnected GbE blob class measurement transitions between 512~kB and 1~MB blocks from
the peak test behaviour to the same behaviour that has been exhibited in the plateau test. This is similar to what was seen in the rate and throughput measurements
(Fig.~\ref{Fig:TCP-Blob-Rates-Peak} and~\ref{Fig:TCP-Blob-NetBW-Peak}).

\begin{figure}[ht!p]
\begin{center}
\resizebox*{1.0\columnwidth}{!}{
\includegraphics{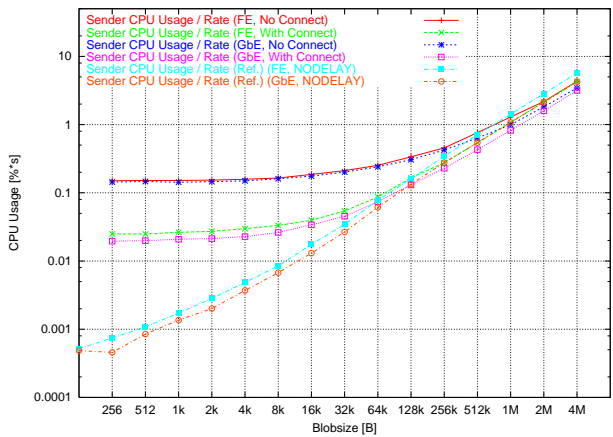}
\hfill
\includegraphics{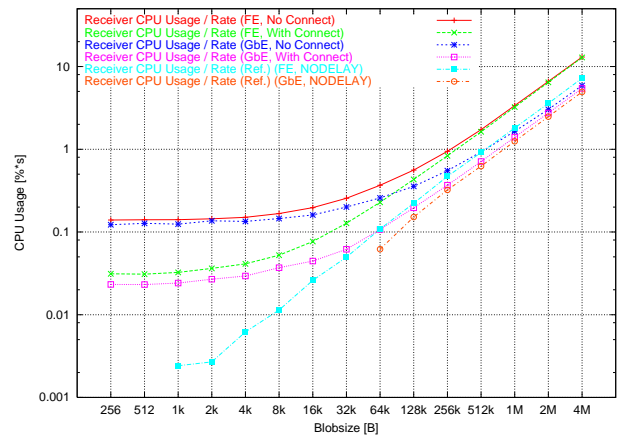}
}
\parbox{0.90\columnwidth}{
\caption[CPU usage divided by the sending rate during TCP blob sending (peak, on-demand alloc.).]{\label{Fig:TCP-Blob-CyclesPerRate-Peak}The CPU usage on the sender (left) and receiver (right) divided by the sending rate 
during TCP blob sending (blob counts 32~k and 2~k, on-demand allocation).
The nodes are twin CPU nodes, 100~\% CPU usage corresponds to one CPU being fully used.}
}
\end{center}
\end{figure}

For a better comparison normalized CPU usages, i.e. CPU usage divided by the achieved rate respectively 
network throughput, have been plotted in 
Fig.~\ref{Fig:TCP-Blob-CyclesPerRate-Peak} and~\ref{Fig:TCP-Blob-CyclesPerNetBW-Peak} instead of absolute 
CPU usage. As can be expected due to the higher sending rates, both
peak test curves show significantly lower CPU usage per rate values on the sending node compared to the corresponding plateau test 
results in Fig.~\ref{Fig:TCP-Blob-CyclesPerRate}. The improvement between the two tests is about a factor of 5 for each of the curves. 
An interesting point can be seen for the two curves at the end of the initial constant plateau
in Fig.~\ref{Fig:TCP-Blob-CyclesPerRate}. After a short transition at 32~kB for FE and 128~kB for GbE the curve from the previous plateau throughput
measurement becomes identical to the smoothly increasing curve from the peak throughput measurement. In this case the bend is clearly
seen in the curve of the plateau tests and not the in peak tests', as for the Gigabit Ethernet rate, network throughput, and CPU usage measurements. 
The reason for this behaviour has not been determined and no plausible explanation could be found in the course of this thesis. 
On the receiver node the curves from the two different throughput tests are nearly indistinguishable. 
Compared to the reference benchmarks the behaviour already observed in the preceeding communication class tests is seen: On the sender the connected FE and 
GbE blob class tests are better for larger blocks, while on the receiver they are less efficient over the whole test range.

\begin{figure}[ht!p]
\begin{center}
\resizebox*{1.0\columnwidth}{!}{
\includegraphics{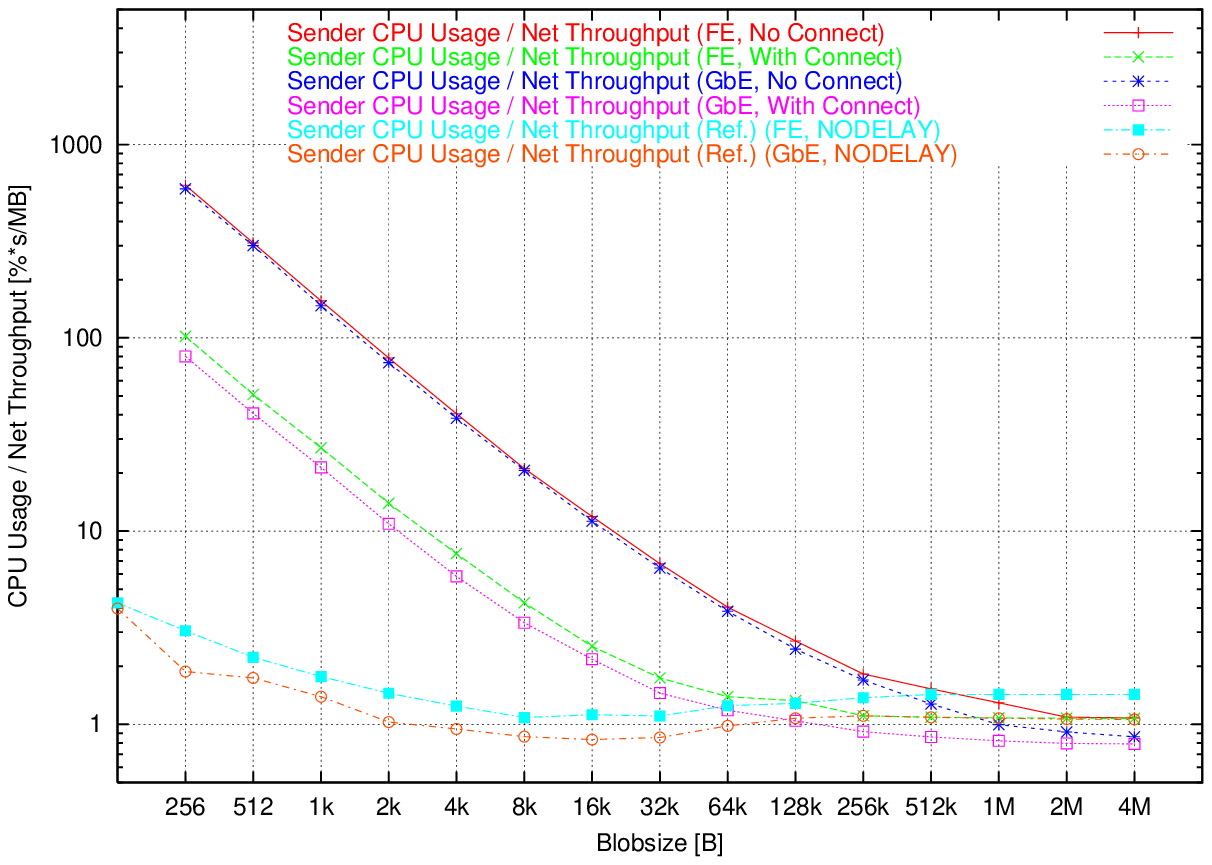}
\hfill
\includegraphics{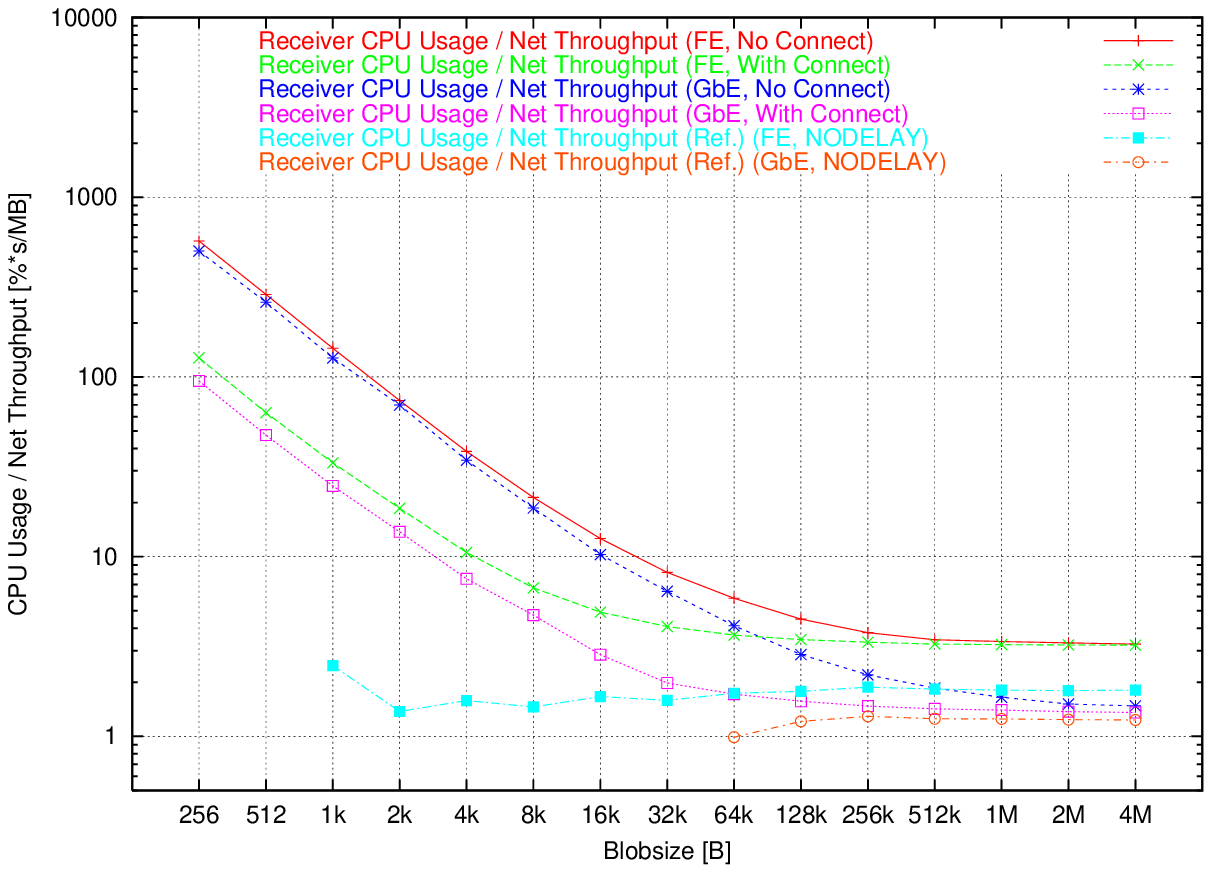}
}
\parbox{0.90\columnwidth}{
\caption[CPU usage per MB/s network throughput during TCP blob sending (peak, on-demand alloc.).]{\label{Fig:TCP-Blob-CyclesPerNetBW-Peak}The CPU usage on the sender (left) and receiver (right) per MB/s network throughput 
during TCP blob sending (blob counts 32~k and 2~k, on-demand allocation).
The nodes are twin CPU nodes, 100~\% CPU usage corresponds to one CPU being fully used.}
}
\end{center}
\end{figure}

As before, the measurement of the CPU usage per network throughput shown in Fig.~\ref{Fig:TCP-Blob-CyclesPerNetBW-Peak} results in a similar improvement  factor of roughly 5  
compared with the plateau throughput test in Fig.~\ref{Fig:TCP-Blob-CyclesPerNetBW}.
The 
transition in the plateau test curve towards the smooth curve from the peak test can also be seen, although less pronounced, between the 32~kB and 64~kB 
and 128~kB and 256~kB blocks for Fast and Gigabit Ethernet respectively. The comparison between the blob class and the reference measurements leads to the same
conclusions as discussed for the usage per rate graphs. 

%{\bf In the peak measurements the same relation between the connected and unconnected and Fast and Gigabit Ethernet tests respectively can be seen as for the 
%plateau measurements. For the small blocks the connected (explicit connection) measurements are far more efficient, a factor of 10 or more, in their
%usage of CPU cycles, while for larger blocks the difference starts to diminish as the network used for sending becomes the limiting factor. In the comparison of
%the two network types again the use of Gigabit Ethernet seems to suggest itself, even when its higher bandwidth is not needed, due to its lower
%CPU consumption per data transferred. 
%}

\subsection{\label{Sec:TCPBLobPreAllocThroughput}TCP Blob Class Throughput with Preallocation}

As for the previous throughput measurements the benchmark for the blob class in preallocation mode 
also splits in two parts, the initial determination of the number of blocks (blobs)
sent for each block size and the actual throughput measurement in dependence of the block size. 

\subsubsection{Plateau Determination}

\begin{figure}[ht!p]
\begin{center}
\resizebox*{0.50\columnwidth}{!}{
\includegraphics{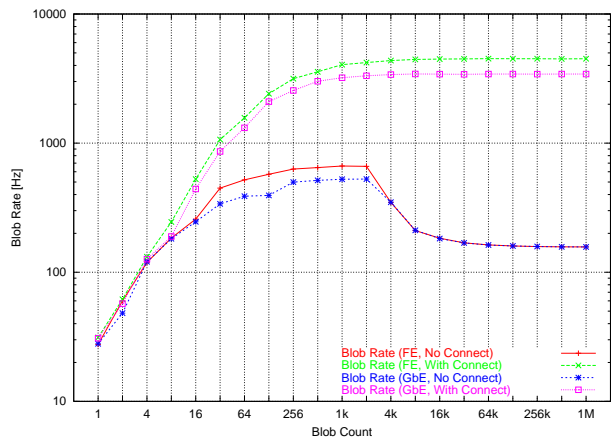}
}
\parbox{0.90\columnwidth}{
\caption[The blob sending rates in dependence of the number of blocks (prealloc.).]{\label{Fig:TCP-Blob-PreAlloc-Count-Rates}The blob sending rates in dependence of the number of blocks in preallocation mode.}
}
\end{center}
\end{figure}

The plot of the blob sending rate in dependence of the number of blocks sent in preallocation mode is shown in Fig.~\ref{Fig:TCP-Blob-PreAlloc-Count-Rates}.
The same shapes of the curves as in the corresponding on-demand allocation tests from the previous section
can be made out, although the absolute rates are higher
by a factor of 2 for the connected final values, 1.5 for the peaks of the unconnected curves, and about 1.2 for their final values. As the final plateaus for
the unconnected tests are reached at higher values than in the on-demand allocation rate measurements, the blob counts of 131072 (128~k) and 2048 (2~k)
have been chosen for the plateau and peak measurements respectively. Since the connected curves display no peak the plateau results at 128~k messages
are reused as the values for the peak tests.

\subsubsection{Plateau Throughput Measurement}

Results of the throughput plateau measurements with block counts of 128~k are displayed in Fig.~\ref{Fig:TCP-Blob-PreAlloc-Rates} to 
Fig.~\ref{Fig:TCP-Blob-PreAlloc-CyclesPerNetBW}. The rate measurements in Fig.~\ref{Fig:TCP-Blob-PreAlloc-Rates} show the same
basic shape as the equivalent curves in on-demand allocation mode from Fig.~\ref{Fig:TCP-Blob-Rates}. Achieved rates
are higher than in the on-demand tests though, initially about 160~Hz unconnected and 4.8~kHz connected. These rates are higher by a factor of 1.3 for the unconnected and
about 3 for the connected tests respectively. For the two Gigabit Ethernet curves 
even the final results where the available network bandwidth already influences the rate are higher than the ones from the on-demand tests. 
%A special feature observed in the connected GbE curve is the missing initial plateau between block sizes of 256~B to 4~kB. Instead, a slow decrease can be observed. 
In the Fast Ethernet measurements the two tests produce identical results from 64~kB blocks on. At these sizes the network sets the absolute
limit and is not only a limiting influence as for GbE. Comparing the two connected curves one sees that in this test Gigabit Ethernet
has a higher transfer rate than Fast Ethernet for all block sizes. The higher initial rate for FE in on-demand mode is thus due to the lower message latency 
that influences the round-trip time for the allocation messages as presumed. Since these messages are not required in preallocation mode the effect is not seen
in this test.
Comparing the blob class and reference measurements one can see that for larger blocks both connected curves reach the respective 
reference curve and in the case of Gigabit Ethernet
even exceed it, as also observed in the previous communication class measurements. Absolute rates are not considerably different than in the on-demand allocation 
blob class measurement, since the hardware is the primary limit for large blocks, but the block sizes where the reference curves are reached or exceeded are about
a factor of 2 smaller than in the on-demand allocation measurement. 
The performance increases, compared to the on-demand allocation measurements, should be due to the use of the preallocation mode, with its lack of the allocation
request-reply message sequence.

%Results of the throughput plateau measurements with block counts of 128~k are displayed in Fig.~\ref{Fig:TCP-Blob-PreAlloc-Rates} to 
%Fig.~\ref{Fig:TCP-Blob-PreAlloc-CyclesPerNetBW}. The rate measurements in Fig.~\ref{Fig:TCP-Blob-PreAlloc-Rates} show the same
%basic shape as the equivalent curves in on-demand allocation mode from Fig.~\ref{Fig:TCP-Blob-Rates}. Achieved rates
%are higher than in the on-demand tests though, initially about 160~Hz unconnected and 4.8~kHz connected. These rates are higher by a factor of 1.3 for the unconnected and
%about 3 for the connected tests respectively. For the two Gigabit Ethernet curves 
%even the final results where the available network bandwidth already influences the rate are higher than the ones from the on-demand tests. 
%A special feature observed in the connected GbE curve is the missing initial plateau between block sizes of 256~B to 4~kB. Instead, a slow decrease can be observed. 
%In the Fast Ethernet measurements the two tests produce identical results from 64~kB blocks on. At these sizes the network sets the absolute
%limit and is not only a limiting influence as for GbE. Comparing the two connected curves one sees that in this test Gigabit Ethernet
%has a higher transfer rate than Fast Ethernet for all block sizes. The higher initial rate for FE in on-demand mode is thus due to the lower message latency 
%that influences the round-trip time for the allocation messages as presumed. Since these messages are not required in preallocation mode the effect is not seen
%in this test.

\begin{figure}[ht!p]
\begin{center}
\resizebox*{0.50\columnwidth}{!}{
\includegraphics{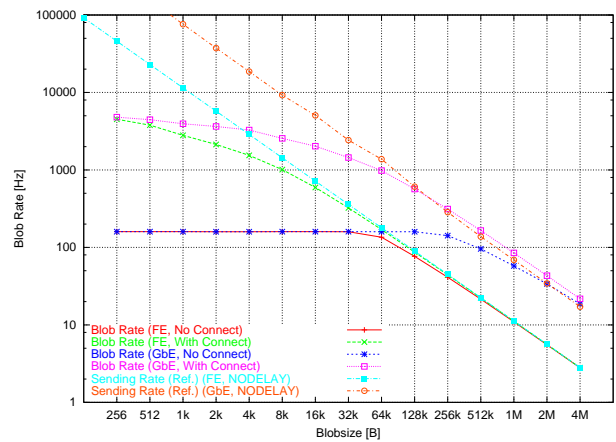}
}
\parbox{0.90\columnwidth}{
\caption[The measured blob sending rates (plateau, prealloc.).]{\label{Fig:TCP-Blob-PreAlloc-Rates}The measured blob sending rates (blob count 128~k, preallocation).}
}
\end{center}
\end{figure}

\begin{figure}[ht!p]
\begin{center}
\resizebox*{0.50\columnwidth}{!}{
\includegraphics{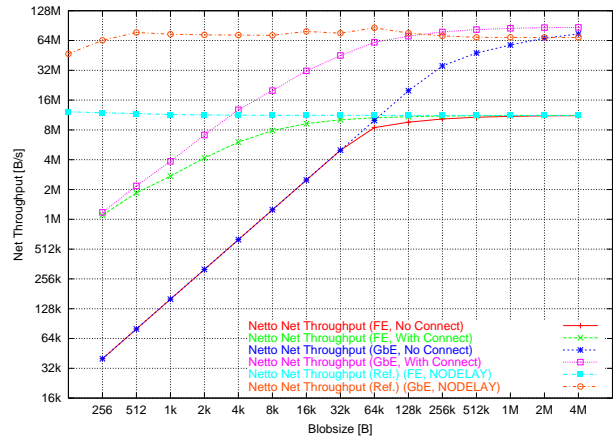}
}
\parbox{0.90\columnwidth}{
\caption[The application level network throughput for TCP blob sending (plateau, prealloc.).]{\label{Fig:TCP-Blob-PreAlloc-NetBW}The application level network throughput for TCP blob sending (blob count 128~k, preallocation).}
}
\end{center}
\end{figure}

Similar results can be deduced from the network throughput measurement in Fig.~\ref{Fig:TCP-Blob-PreAlloc-NetBW}. The achieved throughput is higher 
for all curves, with Fast Ethernet at the smaller blocksizes up to about 64~kB and with Gigabit Ethernet over the whole test range. 
The available bandwidth starts to limit the throughput already at about 256~kB blocks for the unconnected Fast, at 16~kB  for the connected Fast, 
and at 256~kB for the connected Gigabit Ethernet test. Unconnected Gigabit Ethernet is not limited by the available bandwidth up to the maximum
tested blocksizes of 4~MB.

\begin{figure}[ht!p]
\begin{center}
\resizebox*{1.0\columnwidth}{!}{
\includegraphics{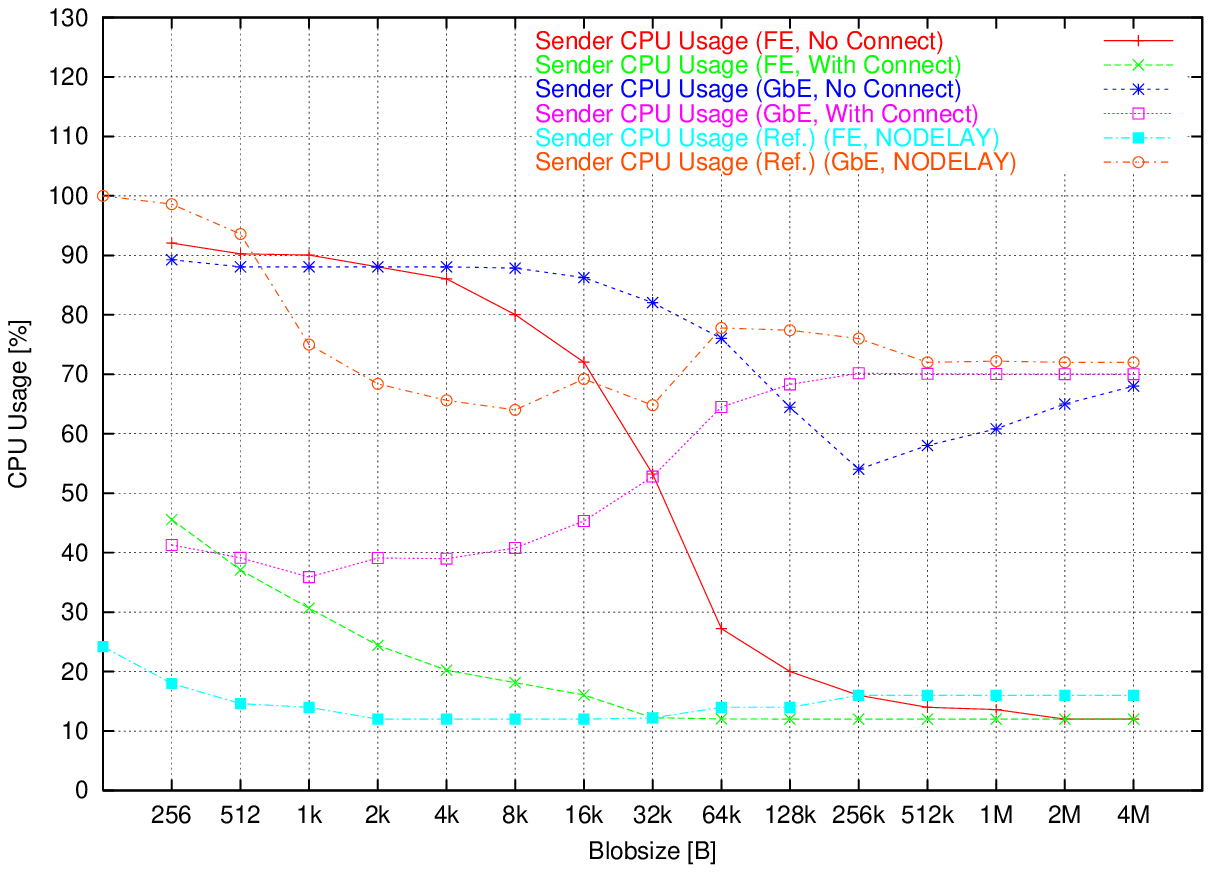}
\hfill
\includegraphics{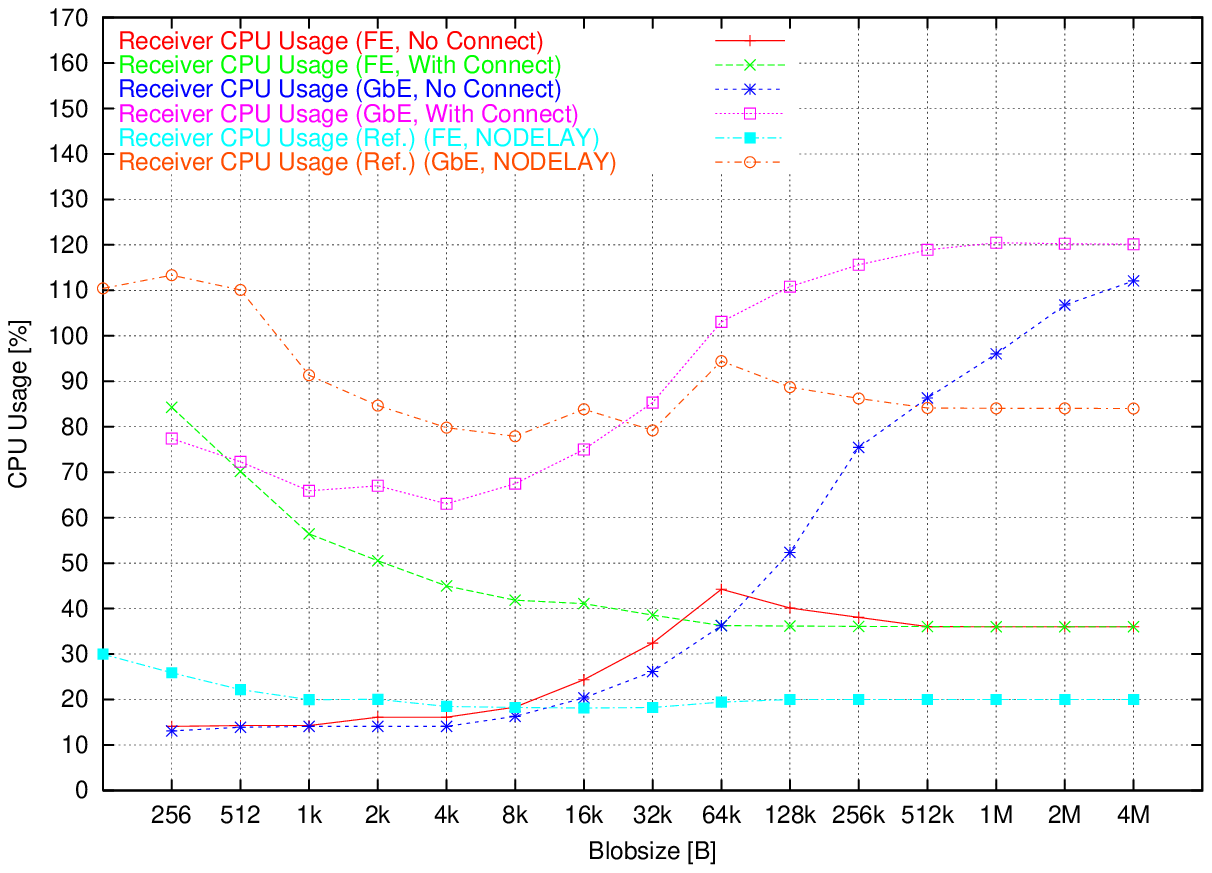}
}
\parbox{0.90\columnwidth}{
\caption[CPU usage during TCP blob sending (plateau, prealloc.).]{\label{Fig:TCP-Blob-PreAlloc-Cycles}The CPU usage on the sender (left) and receiver (right) during TCP blob sending (blob count 128~k, preallocation).
The nodes are twin CPU nodes, 100~\% CPU usage corresponds to one CPU being fully used.}
}
\end{center}
\end{figure}

As can be seen in Fig.~\ref{Fig:TCP-Blob-PreAlloc-Cycles} the CPU usage during blob transfers on the
sending node increases for both Gigabit Ethernet tests by about 5~\% to 10~\% over the whole
test range, compared to the equivalent on-demand allocation tests. This is presumably due to the absolute higher sending
rates observed in this mode. 
For the two Fast Ethernet measurements
on the sender the opposite effect is observed. CPU usage is lower than for the equivalent on-demand allocation tests by about 5~\% for smaller block sizes.
This decrease is present up to the largest block sizes where the network limits the throughput and thus the CPU usage too. The FE network limit also starts to 
affect the tests for smaller block sizes than for the on-demand tests. The reason for this decrease is presumably caused again by the lack of the
allocation request-reply messages. In addition to increasing the rate the CPU usage decreases as the additional messages do not have to be sent and 
received on each node. 
In comparison with the respective reference measurements one again can see that the communication class GbE curve is at lower values over the whole test range
while the FE curve is higher, but only for small block sizes. This is identical to previously observed behaviour, and the presumed causes are similar as well.
However, the communication class GbE curve approaches the reference curve closer than the GbE blob on-demand allocation measurement, due to its higher 
absolute usage values.

On the receiver the effect on the connected Gigabit Ethernet test is identical to that on the sender, except that on the receiver the increase is between 5~\% up to almost
20~\% for small blocks. For the Fast Ethernets test the opposite effect compared to the sender sets in for small block sizes, CPU usage is increased by almost
10~\%. 
The curves for FE are again identical to those from the on-demand test at large blocks. The unconnected GbE test 
only shows a small increase at small block sizes of less than 5~\%. With growing block sizes, though, the difference grows to about 10~\% as well.
These observed increases in CPU usage are most probably caused by the increases in blob rates/network throughput, as correspondingly more data has
to be handled by the receiver. On the sender the increased rate does not necessarily cause a CPU usage increase, as less data, practically none,
has to be received there without the allocation reply messages from the blob receiver node.

\begin{figure}[ht!p]
\begin{center}
\resizebox*{1.0\columnwidth}{!}{
\includegraphics{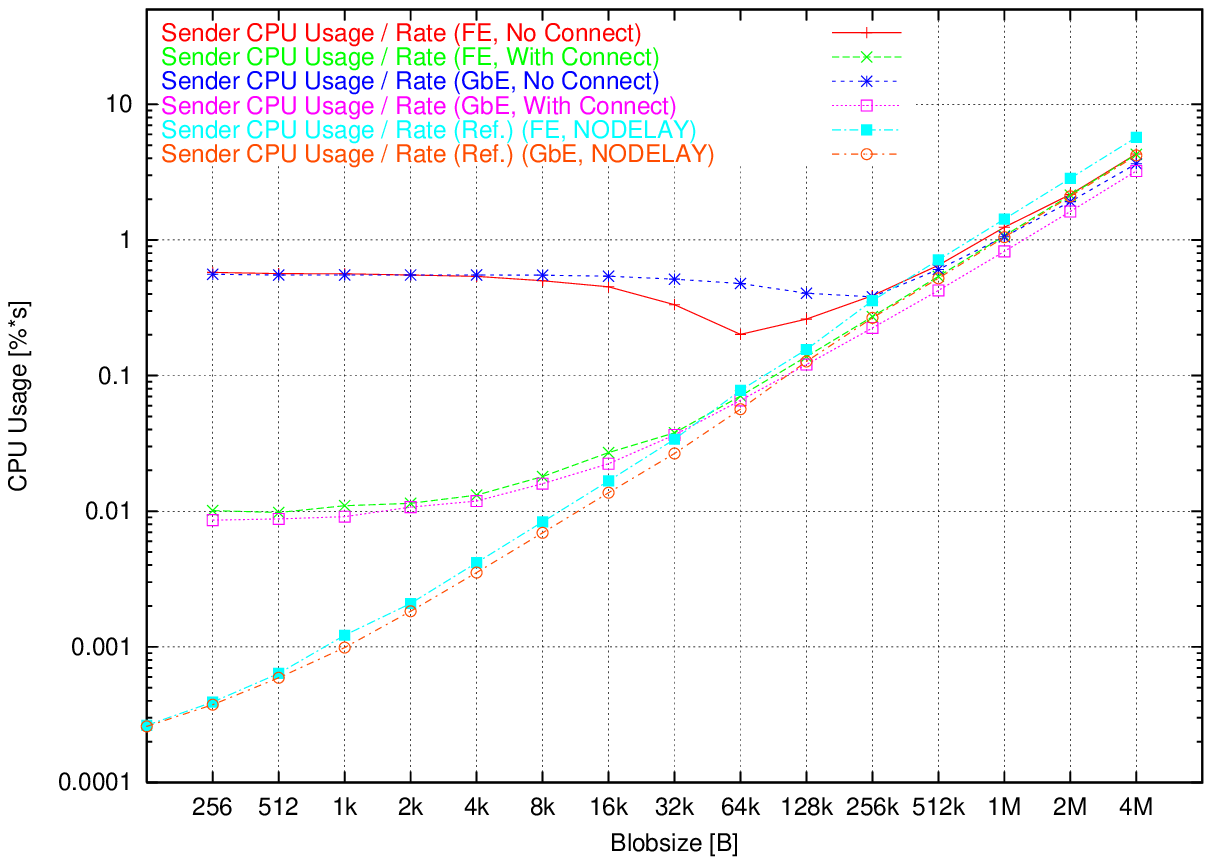}
\hfill
\includegraphics{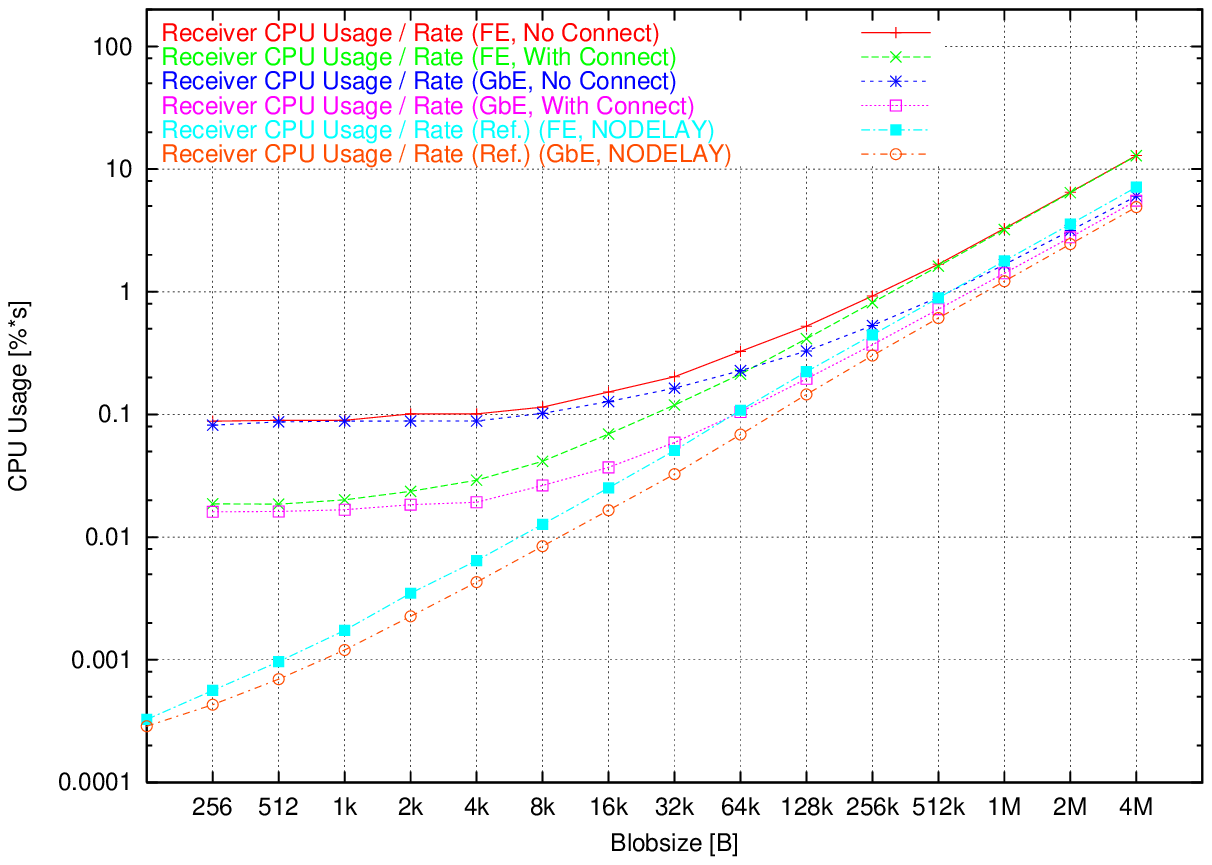}
}
\parbox{0.90\columnwidth}{
\caption[CPU usage divided by the sending rate during TCP blob sending (plateau, prealloc.).]{\label{Fig:TCP-Blob-PreAlloc-CyclesPerRate}The CPU usage on the sender (left) and receiver (right) divided by the sending rate 
during TCP blob sending (blob count 128~k, preallocation).
The nodes are twin CPU nodes, 100~\% CPU usage corresponds to one CPU being fully used.}
}
\end{center}
\end{figure}

\begin{figure}[ht!p]
\begin{center}
\resizebox*{1.0\columnwidth}{!}{
\includegraphics{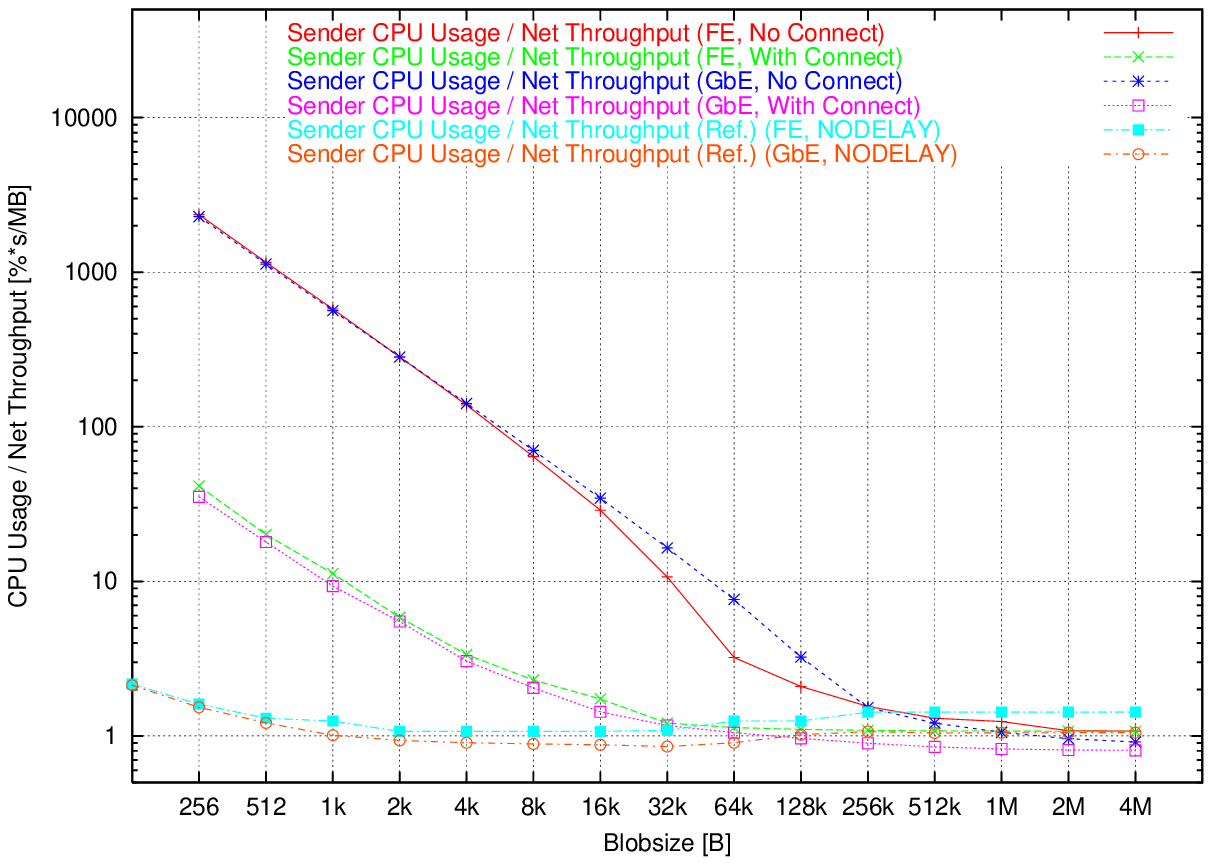}
\hfill
\includegraphics{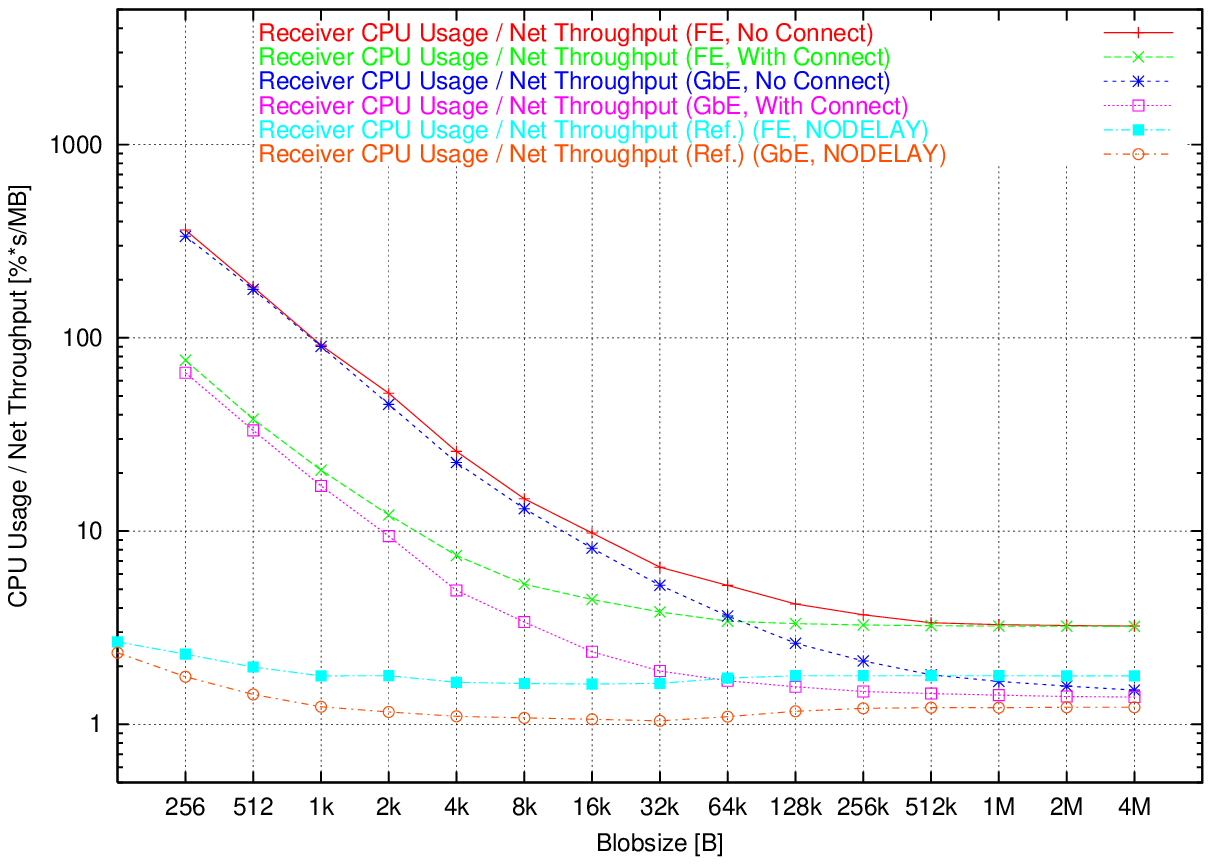}
}
\parbox{0.90\columnwidth}{
\caption[CPU usage per MB/s network throughput during TCP blob sending (plateau, prealloc.).]{\label{Fig:TCP-Blob-PreAlloc-CyclesPerNetBW}The CPU usage on the sender (left) and receiver (right) per MB/s network throughput 
during TCP blob sending (blob count 128~k, preallocation).
The nodes are twin CPU nodes, 100~\% CPU usage corresponds to one CPU being fully used.}
}
\end{center}
\end{figure}

Inspecting the CPU usage normalized with the event rate and network throughput in Fig.~\ref{Fig:TCP-Blob-PreAlloc-CyclesPerRate} and 
\ref{Fig:TCP-Blob-PreAlloc-CyclesPerNetBW} respectively, the comparison with the on-demand allocation mode is more favorable for the preallocation tests.
The basic forms of the curves are identical for both measurements on the sender as well as on the receiver. 
In a comparison of the two transfer modes, the CPU usage to rate (or throughput) ratios of the preallocation tests are better 
by factors of 2.5 to 2.2 for the connected Fast and Gigabit Ethernet tests respectively and 1.25 to 1.17 for the unconnected FE and GbE measurements.
These ratios are measured for 256~B blocks, towards the largest blocks the ratios become almost equal with a relative difference of only a few percent. 
This respective approach of the two tests' curves can be expected, as for these large blocks the main influence is by the actual transfer itself and the 
allocation message exchange becomes negligible. 
Qualitatively the behaviour relative to the reference tests is similar to the ones of the on-demand allocation measurements. Due to the lower
values in preallocation mode, however, the relative  differences are distinct; when the communication class measurements are higher than the reference,
the difference has become smaller and when the class measurements are lower, the difference has become larger.

%Inspecting the CPU usage normalized with the event rate and network throughput in Fig.~\ref{Fig:TCP-Blob-PreAlloc-CyclesPerRate} and 
%\ref{Fig:TCP-Blob-PreAlloc-CyclesPerNetBW} respectively, the comparison with the on-demand allocation mode is more favorable for the preallocation tests.
%The basic forms of the curves are identical for both measurements on the sender as well as on the receiver, including the bends in the unconnected 
%curves on the sending node. In a comparison of the two transfer modes, the CPU usage to rate (or throughput) ratios of the preallocation tests are better 
%by factors of 2.5 to 2.2 for the connected Fast and Gigabit Ethernet tests respectively and 1.25 to 1.17 for the unconnected FE and GbE measurements.
%These ratios are measured for 256~B blocks, towards the largest blocks the ratios become almost equal with a relative difference of only a few percent. This respective approach of the two tests' curves can be 
%expected, as for these large blocks the main influence is by the actual transfer itself and the 
%allocation message exchange becomes negligible. 

\subsubsection{Peak Throughput Measurement}

The final blob class throughput measurement is performed in preallocation mode with a block count of 2048 (2~k) for the unconnected tests, 
with results shown in Fig.~\ref{Fig:TCP-Blob-PreAlloc-Rates-Peak} to~\ref{Fig:TCP-Blob-PreAlloc-CyclesPerNetBW-Peak}. As for the on-demand allocation 
peak throughput tests the connected plateau throughput tests with a 128~k block count from the previous section are reused here. The following discussion
will therefore focus on the two unconnected tests. 

\begin{figure}[ht!p]
\begin{center}
\resizebox*{0.50\columnwidth}{!}{
\includegraphics{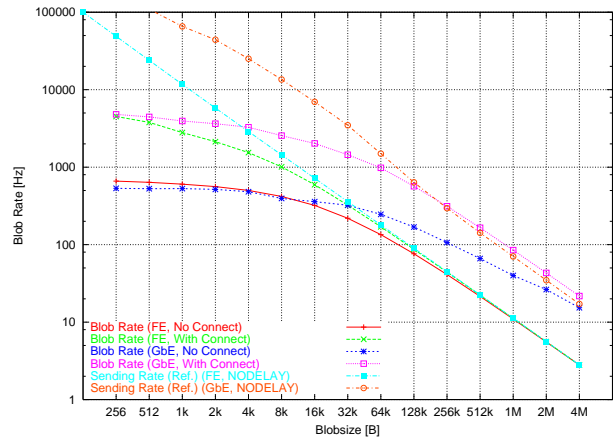}
}
\parbox{0.90\columnwidth}{
\caption[The measured blob sending rates (peak, prealloc.).]{\label{Fig:TCP-Blob-PreAlloc-Rates-Peak}The measured blob sending rates (blob counts 128~k and 2~k, preallocation).}
}
\end{center}
\end{figure}

\begin{figure}[ht!p]
\begin{center}
\resizebox*{0.50\columnwidth}{!}{
\includegraphics{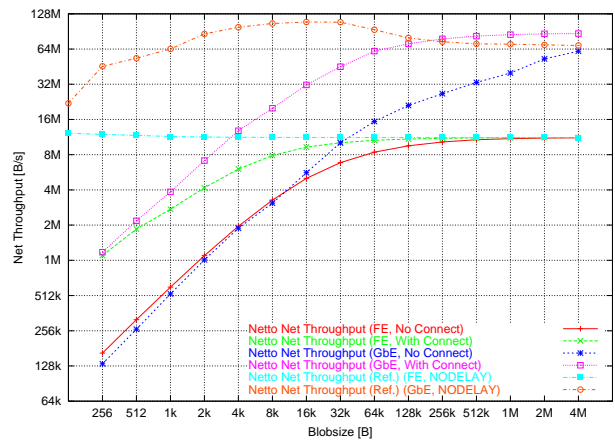}
}
\parbox{0.90\columnwidth}{
\caption[The application level network throughput for TCP blob sending (peak, prealloc.).]{\label{Fig:TCP-Blob-PreAlloc-NetBW-Peak}The application level network throughput for TCP blob sending (blob counts 128~k and 2~k, preallocation).}
}
\end{center}
\end{figure}

For the measured rates and network throughputs, shown in Fig.~\ref{Fig:TCP-Blob-PreAlloc-Rates-Peak} and~\ref{Fig:TCP-Blob-PreAlloc-NetBW-Peak}
respectively, the relation to the preallocation plateau throughput tests is in principle identical to the relation of the plateau and peak
on-demand allocation tests. The rates (and therefore also throughput values) achieved are the highest of all four unconnected 
block transfer tests: a maximum rate of more than 650~Hz and 533~Hz at 256~B blocks for Fast and Gigabit Ethernet respectively. 
Network throughput for Fast Ethernet is higher
up to about 64~kB blocks where the network bandwidth becomes the limit for all FE tests. 
The achieved results differ by factors of 3 
to more than 4 relative to the preallocation plateau tests, and for the on-demand peak tests the factors are 1.3 to 1.4. At larger block sizes this effect is 
decreasing, and at the largest blocks this test has only a small advantage for Gigabit Ethernet and none for Fast Ethernet. 
The reasons for this increase relative to the on-demand peak test are again the lack of the request-reply allocation message sequence.
With regard to the preallocation plateau test the increase is presumably caused by buffers which are able to accept a large part of the small blobs,
analogous to the other peak test increases. 
As for the other comunication class tests the connected FE curve approaches its appropriate reference curve and the connected GbE curve exceeds it 
for large blocks.

%For the measured rates and network throughputs, shown in Fig.~\ref{Fig:TCP-Blob-PreAlloc-Rates-Peak} and~\ref{Fig:TCP-Blob-PreAlloc-NetBW-Peak}
%respectively, the relation to the preallocation plateau throughput tests is in principle identical to the relation of the plateau and peak
%on-demand allocation tests. In both cases the initial flat curve of the plateau test is replaced by a decrease that becomes steeper and
%develops into a linear decrease with increasing block sizes. The rates (and therefore also throughput values) achieved are the highest of all four unconnected 
%block transfer tests: a maximum rate of more than 650~Hz and 533~Hz at 256~B blocks for Fast and Gigabit Ethernet respectively. 
%Network throughput for Fast Ethernet is higher
%up to about 64~kB blocks where the network bandwidth becomes the limit for all FE tests. 
%They differ by factors of 3 
%to more than 4 relative to the preallocation plateau tests, and for the on-demand peak tests the factors are 1.3 to 1.4. At larger block sizes this effect is 
%decreasing, and at the largest blocks this test has only a small advantage for Gigabit Ethernet and none for Fast Ethernet. 

\begin{figure}[ht!p]
\begin{center}
\resizebox*{1.0\columnwidth}{!}{
\includegraphics{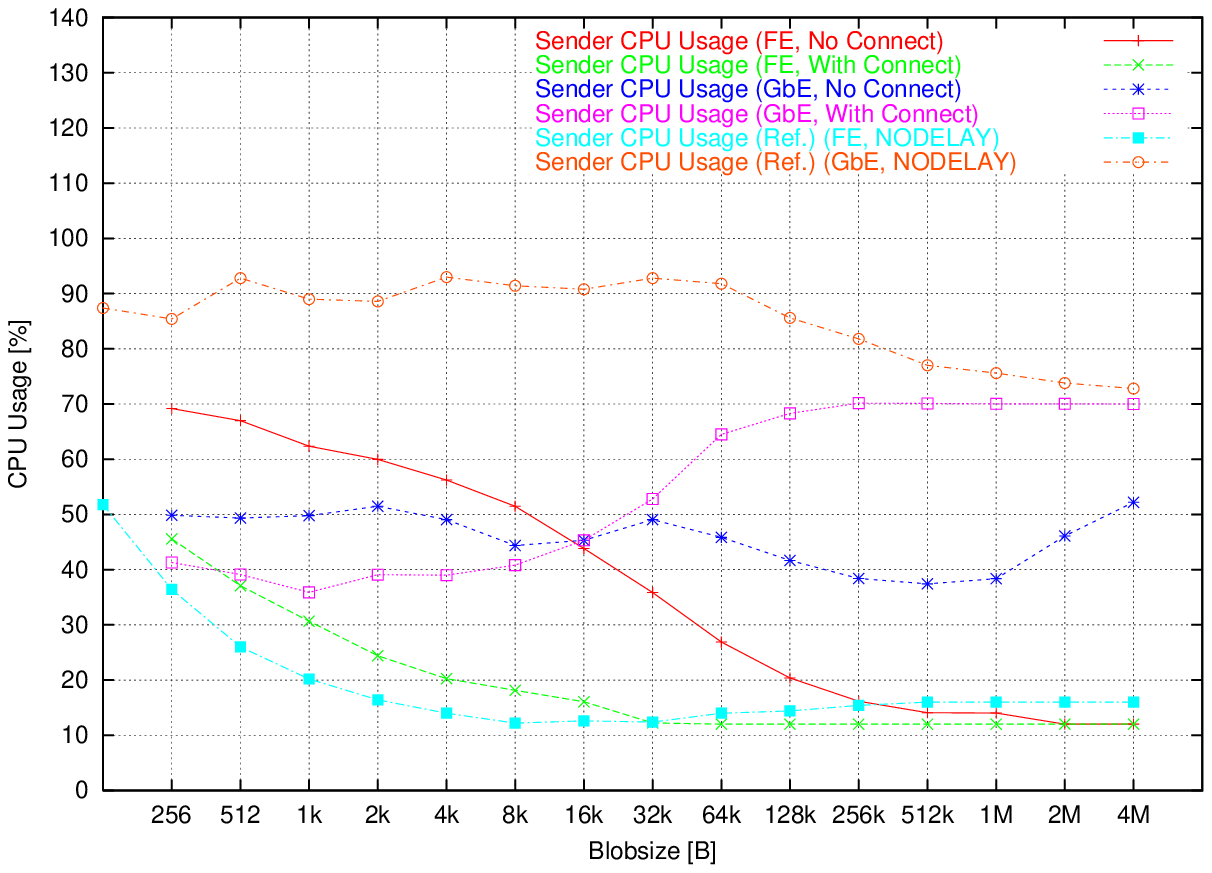}
\hfill
\includegraphics{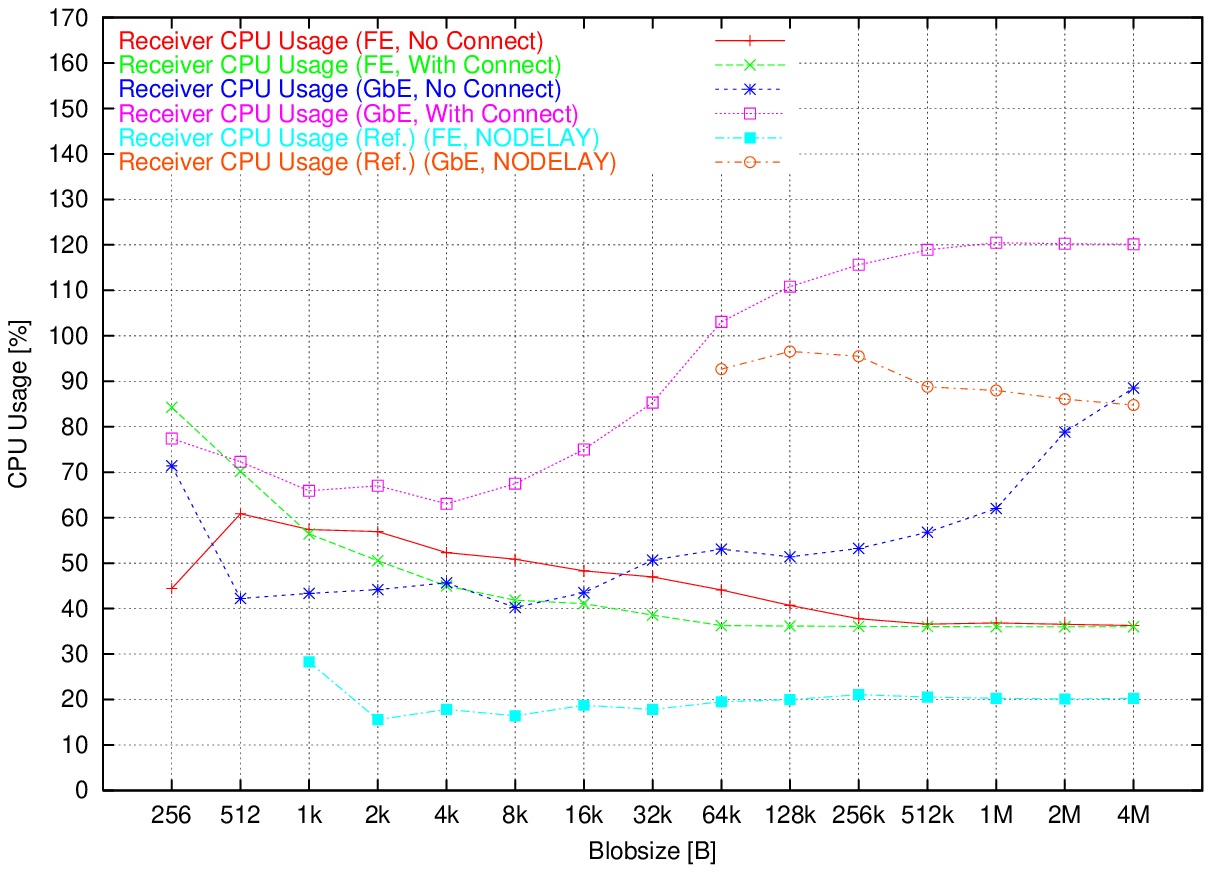}
}
\parbox{0.90\columnwidth}{
\caption[CPU usage during TCP blob sending (peak, prealloc.).]{\label{Fig:TCP-Blob-PreAlloc-Cycles-Peak}The CPU usage on the sender (left) and receiver (right) during TCP blob sending (blob counts 128~k and 2~k, preallocation).
The nodes are twin CPU nodes, 100~\% CPU usage corresponds to one CPU being fully used.}
}
\end{center}
\end{figure}

CPU usage on the sending node is displayed in Fig.~\ref{Fig:TCP-Blob-PreAlloc-Cycles-Peak}. For the connected measurements it is approximately equal to the
one from the preallocation plateau test described above, despite the higher rates. Concerning the unconnected tests, their usage is considerably higher
at small blocks than in the plateau test, reflecting the already observed overhead of establishing connections on the initiating (or sending) node, coupled with the higher
rates in this test. Compared to the on-demand peak test the usage is slightly higher, most likely because of the higher sending rates. 

%As for the on-demand peak tests the CPU usage measurements on the receiver at small block sizes 
%suffer from a high inaccuracy due to the short runnning times. Measurements are therefore not present for all values. 
On the receiver measured connected usage is again roughly the same as in the preallocation plateau test and the unconnected usage is again considerably
higher at small blocks. The factor for the unconnected measurements is between 3 and 4 for both tests relative to the plateau test. Compared to the on-demand allocation
peak tests the results are identical or higher, up to a factor of 2 for the connected GbE curve at small block sizes. This increase is again caused most probably by the 
increase in sending rate relative to the on-demand test. Where measurements are available both connected measurements considerably exceed their respective
reference measurement, at least in part due to the additional block announcement message which has to be received by the communication class, as already discussed.

\begin{figure}[ht!p]
\begin{center}
\resizebox*{1.0\columnwidth}{!}{
\includegraphics{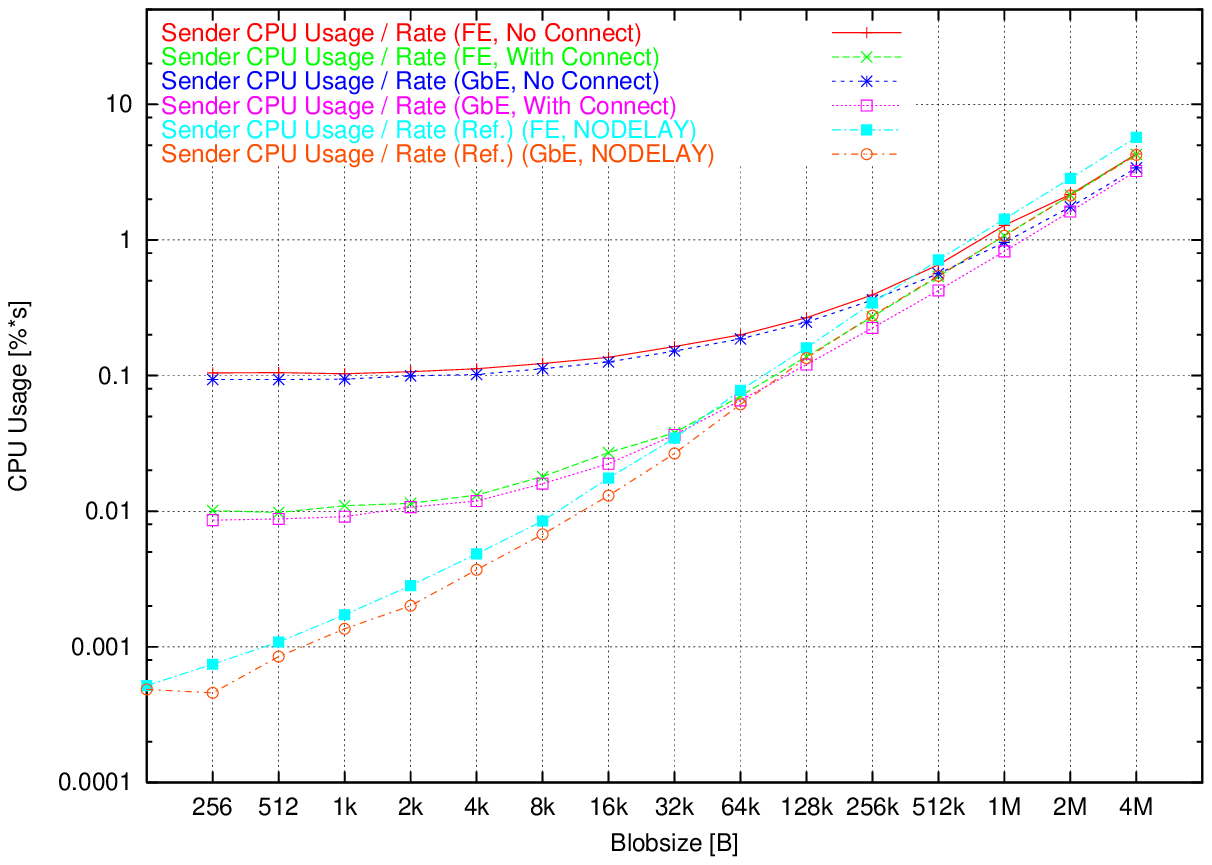}
\hfill
\includegraphics{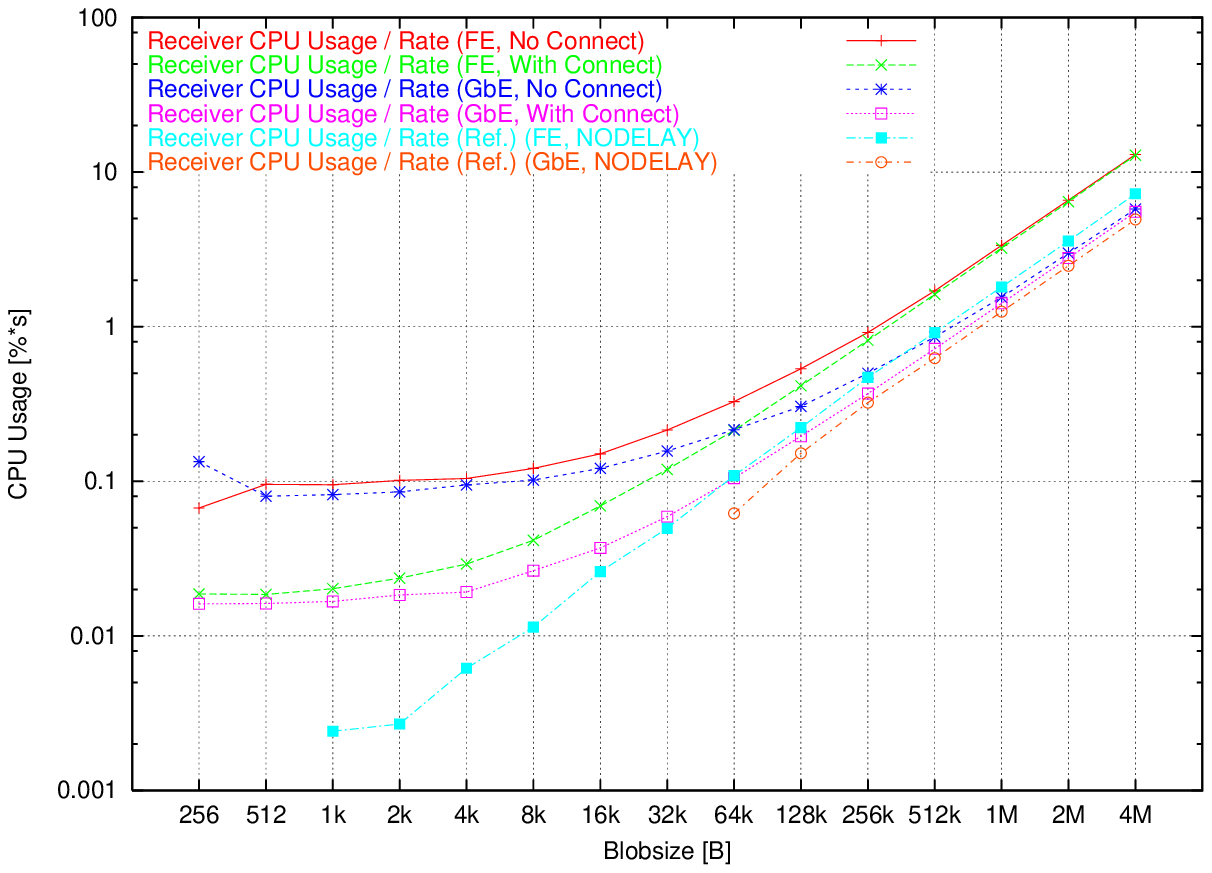}
}
\parbox{0.90\columnwidth}{
\caption[CPU usage divided by the sending rate during TCP blob sending (peak, prealloc.).]{\label{Fig:TCP-Blob-PreAlloc-CyclesPerRate-Peak}The CPU usage on the sender (left) and receiver (right) divided by the sending rate 
during TCP blob sending (blob counts 128~k and 2~k, preallocation).
The nodes are twin CPU nodes, 100~\% CPU usage corresponds to one CPU being fully used.}
}
\end{center}
\end{figure}

\begin{figure}[ht!p]
\begin{center}
\resizebox*{1.0\columnwidth}{!}{
\includegraphics{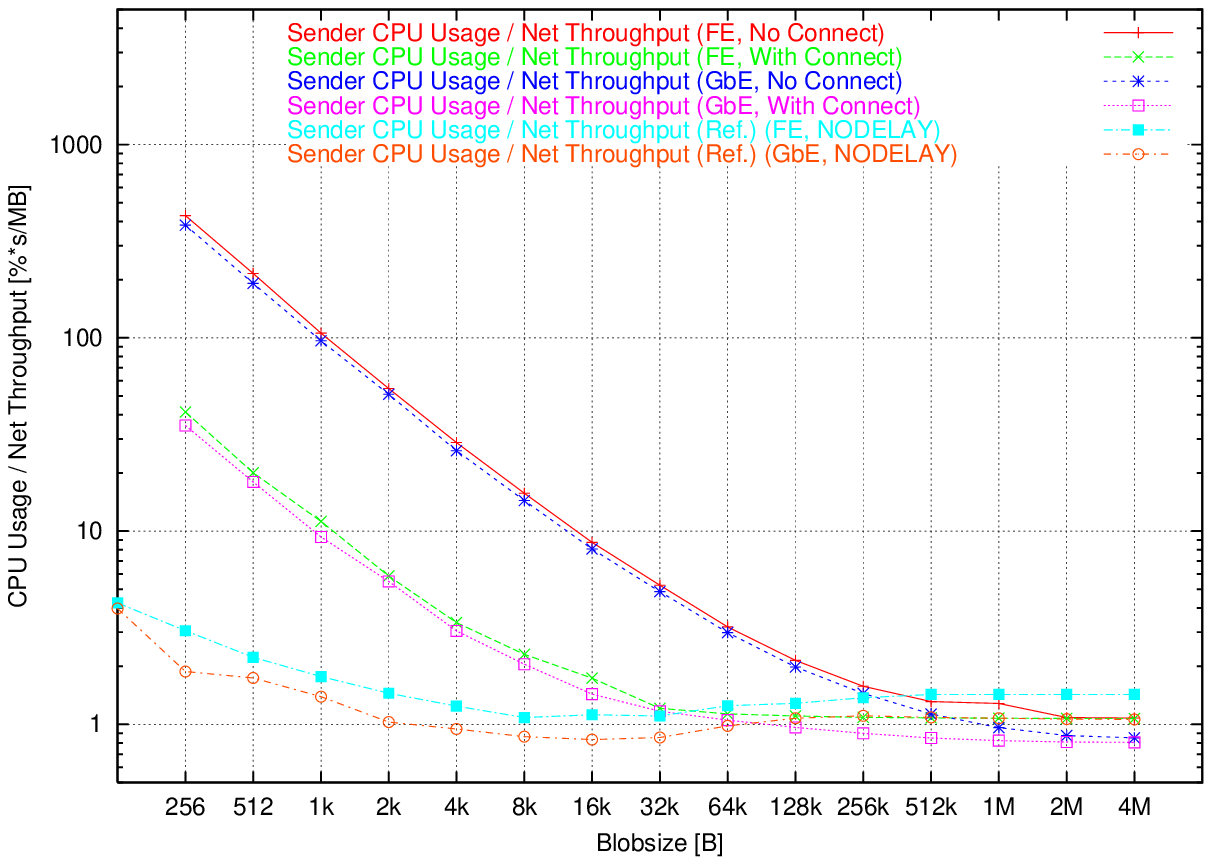}
\hfill
\includegraphics{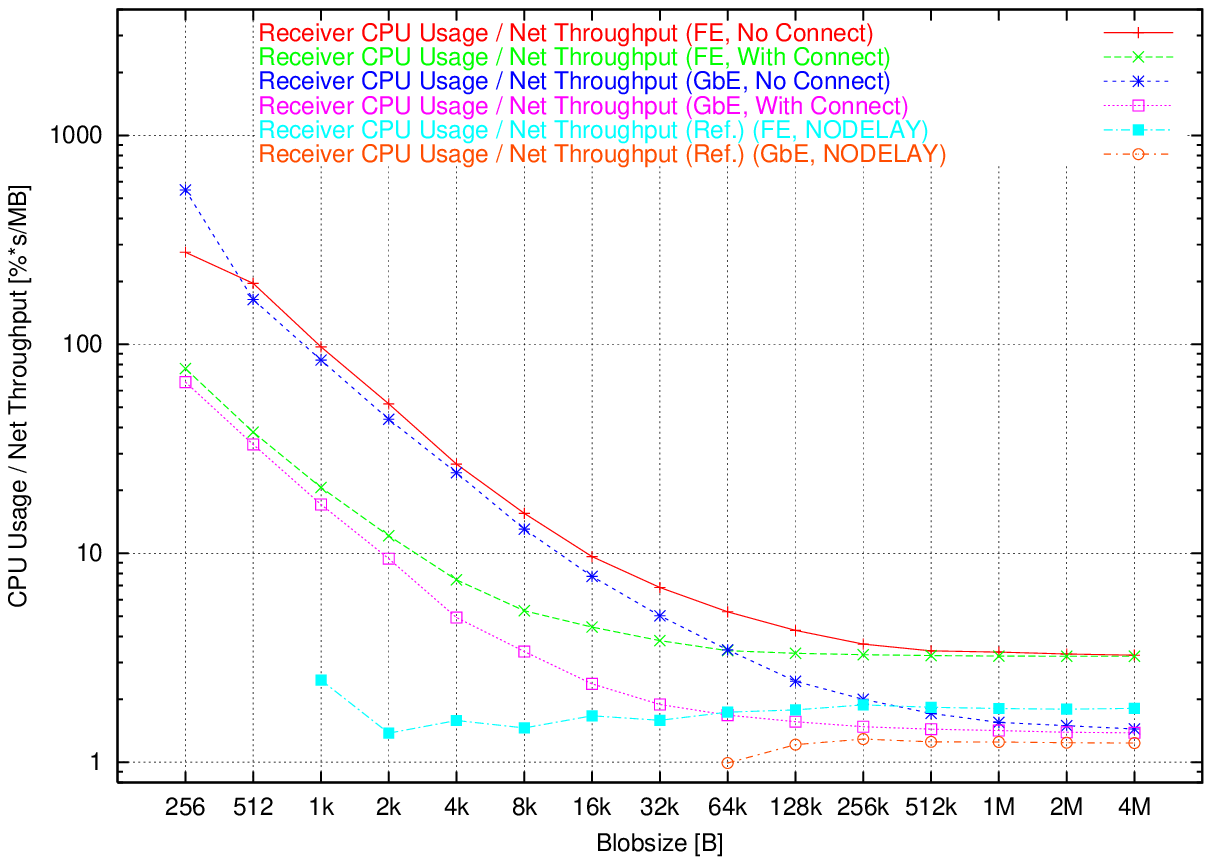}
}
\parbox{0.90\columnwidth}{
\caption[CPU usage per MB/s network throughput during TCP blob sending (peak, prealloc.).]{\label{Fig:TCP-Blob-PreAlloc-CyclesPerNetBW-Peak}The CPU usage on the sender (left) and receiver (right) per MB/s network throughput 
during TCP blob sending (blob counts 128~k and 2~k, preallocation).
The nodes are twin CPU nodes, 100~\% CPU usage corresponds to one CPU being fully used.}
}
\end{center}
\end{figure}

In the efficiency comparison of CPU usage per rate respectively network throughput, the curves on both sender and receiver have
almost identical forms in the two peak throughput tests, with the results of the
preallocation mode tests at lower (and better) values. For small block sizes the preallocation test results are better by a factor of about 1.5 for 
the two unconnected tests and by a factor of about 2.3 for the connected ones. With growing block sizes the difference between the tests becomes 
smaller. On a small scale the results are even reversed so that the on-demand test partially has better values. At the largest block sizes the difference between the tests is at most
1~\% for the unconnected tests in favour of the preallocation tests. 
For the connected tests at these block sizes the on-demand allocation
mode tests are better by about 0.5~\% to 2~\%. 
As the block sizes increase the block's actual transfer increasingly dominates the overhead and the difference caused by the different allocation messages
becomes smaller and smaller, causing the different tests' results to become similar. 
A possible explanation for the on-demand allocation test's efficiency  being higher than the preallocation test's could be that in preallocation mode buffer can be
filled more quickly, due to the missing allocation sequence latency. Therefore buffers can also overflow more quickly, leading to packet losses and retransmits.
These retransmits do not increase the throughput but still have to be processed, decreasing the transfer's efficiency. 
The effect is probably not seen in the plateau tests' due to the larger amount of data transferred, causing overflows in both allocation modes.
On the other hand, these differences are not large and could therefore just be noise respectively measurement uncertainties.

%In the efficiency comparison of CPU usage per rate respectively network throughput, the curves on both sender and receiver have
%almost identical forms in the two peak throughput tests, with the results of the
%preallocation mode tests at lower (and better) values. For small block sizes the preallocation test results are better by a factor of about 1.5 for 
%the two unconnected tests and by a factor of about 2.3 for the connected ones. With growing block sizes the difference between the tests becomes 
%smaller. On a small scale the results are even reversed so that the on-demand test partially has better values. At the largest block sizes the difference between the tests is at most
%1~\% for the unconnected tests in favour of the preallocation tests. 
%For the connected tests at these block sizes the on-demand allocation
%mode tests are better by about 0.5~\% to 2~\%. 

%FE NOC 1.44 - 1.002
%FE C 2.42 - 1/1.006
%GbE NOC 1.54 - 1.011
%GbE C 2.22 - 1/1.018

\subsection{\label{Sec:TCPBlobLatency}TCP Blob Class Latency with On-Demand Allocation}

To determine the latency of the TCP blob class, measurements similar to those for the message class from section~\ref{Sec:TCPMsgLatency} 
have been executed in on-demand mode. Corresponding measurements in preallocation mode are described in section~\ref{Sec:TCPBlobPreAllocLatency}.
In this test varying numbers of data blocks are transferred from sender to receiver. After each block the sender waits for the block to be sent back by the receiver
before continuing with the next block. Fig.~\ref{Fig:TCP-Blob-Latency} shows the results that have been obtained from this ping-pong pattern. 

\begin{figure}[ht!p]
\begin{center}
\resizebox*{0.50\columnwidth}{!}{
\includegraphics{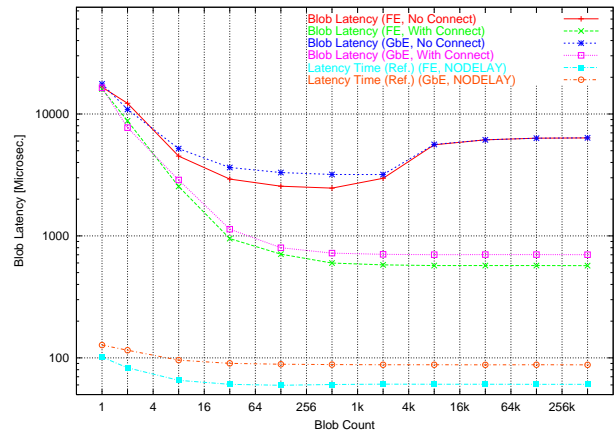}
}
\parbox{0.90\columnwidth}{
\caption[Minimum blob latency times (on-demand alloc.).]{\label{Fig:TCP-Blob-Latency}The blob latency (in $\mu \mathrm s$) as a function of the blob count with on-demand allocation.}
}
\end{center}
\end{figure}

As can be seen in the figure, the latency curves display the same general pattern as those for messages in Fig.~\ref{Fig:TCP-Msg-Latency}.
A sharp decrease turns into a plateau and rises sharply to a second plateau in the unconnected tests. 
The jump to the second plateau in the unconnected tests sets in at lower counts than for the message tests,
between 2~k and 8~k for the blobs compared to between 8~k and 32~k for messages. 
In comparison with the message tests the plateaus are at higher values. 
These higher values are to be expected as sending a blob with on-demand allocation requires the sending of three messages on both nodes, two for the buffer space
allocation and one for the notification that a blob is available. Similar to the message tests and also as expected the unconnected tests display again much 
higher latencies compared to the connected tests. The jump between the two unconnected test plateaus is just by a factor of 2 instead of 2.5 as found in the message test.

\begin{table}[hbt!p]
\begin{center}
\begin{tabular}{|l||c|c|c|c|c|c|}
\hline  
 & Fast & Fast & Gigabit & Gigabit & Fast & Gigabit \\
 & Ethernet & Ethernet & Ethernet & Ethernet & Ethernet & Ethernet \\
 & Implicit & Explicit & Implicit & Explicit & Reference & Reference\\
 & Connect & Connect & Connect & Connect &  & \\
\hline \hline
Minimum Average Blob & 2470 & 570 & 3190 & 700 & 61 & 88 \\
Latency / $\mu \mathrm s$ &  &  &  &  &  & \\
\hline
Minimum Average Blob & 1413 & 240 & 2080 & 344 & - & - \\
Latency / $\mu \mathrm s$ &  &  &  &  &  & \\
w/o Message Latencies &  &  &  &  &  & \\
\hline
\end{tabular}
\parbox{0.90\columnwidth}{
\caption[Minimum blob latency times in on-demand allocation mode.]{\label{Tab:TCP-Blob-Latency-Min}The minimum blob latency times of the four configurations in on-demand allocation mode.}
}
\end{center}
\end{table}

Table~\ref{Tab:TCP-Blob-Latency-Min} summarizes the minimum latency times measured for each of the four different configurations. 
Each connected test is about 4.5 times faster than the respective unconnected test, and the Fast Ethernet tests are between 23~\% (connected) and
29~\% (unconnected) faster than their corresponding Gigabit test counterparts, in each case a higher relative difference than in the message 
latency tests. Compared to the reference measurements the latencies are considerably increased, almost by an order of magnitude for the connected and about a factor of 40 for 
the unconnected tests. One part of the explanation for this is most definitely the fact that three times the respective message latency (allocation request, allocation reply, 
blob announcement) is included in these times. When these message latencies are subtracted the remaining ``pure'' blob latencies are considerably lower,
but still higher than the message and in particular the reference latencies. This can in part be caused by the larger block used in the blob
test, 256~B instead of 8~B for the reference and 32~B for the message class. A second potential cause can be the 32~bit value that is written back to the sender by the receiver blob class after 
each transfer, to indicate completion, which contributes approximately one respective reference latency to the blob latency.

\subsection{\label{Sec:TCPBlobPreAllocLatency}TCP Blob Class Latency with Preallocation}

The same test as in section~\ref{Sec:TCPBlobLatency} has been performed for the blob classes in preallocation
mode as well, with the results shown in Fig.~\ref{Fig:TCP-Blob-PreAlloc-Latency}. As can be seen the shape of the four curves is as good as identical to the ones in
the on-demand allocation latency measurements in Fig.~\ref{Fig:TCP-Blob-Latency}. A major difference between the plots is that the values in preallocation mode
are lower than those in on-demand allocation mode, which also can be seen when comparing Table~\ref{Tab:TCP-Blob-PreAlloc-Latency-Min} and 
Table~\ref{Tab:TCP-Blob-Latency-Min}. The unconnected tests are faster roughly by a factor of 1.5,  the connected ones even by a factor of about 1.8, compared to 
the values from the on-demand allocation tests. For the preallocation values themselves a comparison of the unconnected and connected tests yields factors of 5.1 and 5.4 for 
Fast and Gigabit Ethernet respectively. Compared to Gigabit Ethernet, Fast Ethernet is about 22~\% and 29~\% faster for connected and unconnected tests respectively,
basically identical to the on-demand allocation latency differences. 
Relative to the reference latency results the measured latencies are still fairly high. Taking into account the latency corresponding to the one remaining message 
(blob announcement), the resulting ``pure'' times are identical to a first approximation with the respective times in on-demand allocation mode. 
In preallocation mode, however, all values are slightly lower. These time differences could be due
to the added allocation and release functionality that has to be executed locally on the receiver in on-demand allocation mode. 

\begin{figure}[ht!p]
\begin{center}
\resizebox*{0.50\columnwidth}{!}{
\includegraphics{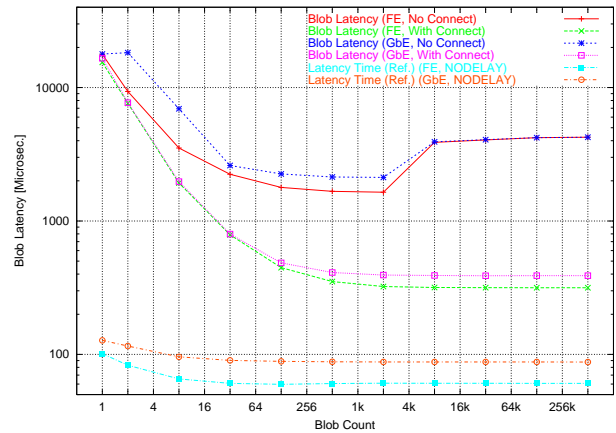}
}
\parbox{0.90\columnwidth}{
\caption[Minimum blob latency times (prealloc.).]{\label{Fig:TCP-Blob-PreAlloc-Latency}The blob latency (in $\mu \mathrm s$) as a function of the blob count with preallocation.}
}
\end{center}
\end{figure}

\begin{table}[hbt!p]
\begin{center}
\begin{tabular}{|l||c|c|c|c|c|c|}
\hline  
 & Fast & Fast & Gigabit & Gigabit & Fast & Gigabit \\
 & Ethernet & Ethernet & Ethernet & Ethernet & Ethernet & Ethernet \\
 & Implicit & Explicit & Implicit & Explicit & Reference & Reference\\
 & Connect & Connect & Connect & Connect & & \\
\hline \hline
Minimum Average Blob & 1640 & 320 & 2120 & 390 & 61 & 88 \\
Latency / $\mu \mathrm s$ &  &  &  &   & & \\
\hline
Minimum Average Blob & 1290 & 210 & 1750 & 271 & - & - \\
Latency / $\mu \mathrm s$ &  &  &  &  & & \\
w/o Message Latencies &  &  &  &  &  & \\
\hline
\end{tabular}
\parbox{0.90\columnwidth}{
\caption[Minimum blob latency times in preallocation mode.]{\label{Tab:TCP-Blob-PreAlloc-Latency-Min}The minimum blob latency times of the four configurations in preallocation mode.}
}
\end{center}
\end{table}

\subsection{TCP Blob Benchmark Summary}

The primary conclusion to be drawn for the TCP blob classes is that, just as the message classes, they are able to handle
the requirements presented by the ALICE HLT within the scope of the current hardware used in the tests. 
The requirements are particularly fulfilled in preallocation mode, as it is used in the framework's bridge components.
In heavy-ion mode average block sizes of around 300~kB are expected for the largest parts of data, the ADC values read-out from the detector.
At these block sizes the classes are still able to handle
a rate of more than the required 200~Hz over Gigabit Ethernet. The 1~kHz rate required for operation in pp mode is
possible up to 64~kB large blocks. These evaluations all refer to the connected mode, as  the blob classes will 
not be used with implicit connections in the HLT. 
A reduction of CPU usage during transfers is still desirable, and it should also be achievable
to a certain degree by more powerful CPUs and more efficient network adapters. But 
optimization potential in this respect should also still exist in the communication library itself so that
further tuning measures of its classes can be undertaken as well. 

Just as for the message classes, if latency is of lesser importance, the use of Gigabit Ethernet recommends itself due to  the lower relative CPU usage
values per network throughput, even when the absolute throughput required does not necessitate its use. If latency is a concern, the use
of Fast Ethernet is suggested whenever possible because its latencies are lower than those of Gigabit. For the framework components this
is of no concern as the conditions allow to treat latency with secondary priority. For the HLT  the amount of data to be transferred over the network, however, implies the
use of at least Gigabit Ethernet even without its efficiency advantages.  Since the sizes of the 
different types of data passed between the HLT's stages are not yet known, predictions of the CPU usage incurred by the transfers cannot be made at the current
stage.

%Comparison w. network reference tests

A direct comparison with the network reference measurements reveals similar results as in the message class -- reference comparison.
Again the first noticeable details are the different characteristics in the graphs showing rate as a function of count. For the connected tests the 
decrease is not present at all, while in the unconnected ones it is again  more obvious than in the reference tests. 
As far as latency is concerned, this is considerably increased in the blob classes 
%presumably 
because of the 
need to send the blob data itself as well as the message informing the receiver about the transmission.
Results from the plateau and peak throughput tests are shown in a summary in Tables~\ref{Tab:TCPBlobMeasurementComparisonPlateau} 
and~\ref{Tab:TCPBlobMeasurementComparisonPeak} respectively. In the following
discussion only the connected tests are regarded, as the results of the unconnected ones are considerably poorer in turn.
Concerning the achieved block sending rate the reference tests are mostly much faster, particularly for small block sizes. The respective 
differences are actually between one and two orders of magnitude. An exception are the Gigabit Ethernet measurements
at the largest block sizes, where the minimum achieved rates, with the blob measurements are slightly higher than those for the reference's. 
For the network throughput the reference tests consistently show better values than each of the
blob tests, at least for the minimum and maximum values listed in the tables. 
As a further exception the GbE measurements of CPU usage and efficiency for sender and receiver differ from the expected characteristics that the
reference tests always results in lower values. These differences also vary between the plateau and peak throughput tests. 
In the plateau measurements
the absolute CPU usages for GbE are always lower on the sender, while for the efficiency the minimum values at large block sizes are lower.
On the receiver the absolute CPU usage minima are lower, and maxima are roughly equal or slightly higher. As expected, 
efficiency values on the receiver are always higher than
in the reference tests.  In the peak tests only values on the sender are better in the blob tests. Absolute CPU usage is always better 
for the blob classes, while for the efficiency only the minimum values are lower. 
The conclusion that can be drawn is also quite similar to the one for the message classes. Some of the overhead and performance loss in the blob classes
certainly has to be accepted as part of the added functionality and in particular flexibility compared to the simple reference test program.
However, the potential for optimization is definitely greater than in the message classes, as can be seen in the considerably lower sending rates of the blob classes
compared with the message class rates, so that the need for tuning measures is a definite must. 
The efficiency measurements that, at least on the sender, indicate better results compared to the reference program, again stand out 
positively, with the same possible explanation as for the message class.

\begin{table}[hbt!p]
\begin{center}
{\scriptsize
\begin{tabular}{|l||c|c|c|c|c|c|c|c|}
\hline  
                            & Rate / & Network                         & CPU Usage   & CPU Usage  & CPU Usage /          & CPU Usage /           & CPU Usage /                               & CPU Usage / \\
Measurement                 & Hz     & Throughput /                    & Sender /    & Receiver / & Rate                 & Rate                  & Throughput                                & Throughput \\
Type                        &        & $\frac{\mathrm{MB}}{\mathrm s}$ & \%          & \%         & Sender /             & Receiver /            & Sender /                                  & Receiver /                                \\
                            &        &                                 &             &            & $\%\times \mathrm s$ & $\% \times \mathrm s$ & $\frac{\% \times \mathrm s}{\mathrm{MB}}$ & $\frac{\% \times \mathrm s}{\mathrm{MB}}$ \\
                            &        &                                 &             &            &                      &                       &                                                    & \\
\hline \hline
Reference                   & 2.8 @ 4~M     & 11.2 @ 256              & 12 @ 2~k     & 18.1 @ 16~k           & 0.000392                     & 0.000563                      & 1.07 @ 16~k                                                   & 1.62 @ 16~k \\
FE w.                       & 45900 @ 256  & 11.2 @ 4~M               & 18 @ 256    & 25.9 @ 256           &  @ 256                     &  @ 256                      & 1.61 @ 256                                                   & 2.31 @ 256 \\
 \texttt{TCP\_\-NO\-DE\-LAY}      & 6130          & 11.2                     & 14.3        & 20           & 5.7 @ 4~M                     & 7.13 @ 4~M                      & 1.28                                                   & 1.79 \\
                                  &               &                          &             &              & 0.756                     & 0.95                      &                                                         &       \\
\hline                                                                                                                                                                                                                                                              %
Reference                   & 2.8 @ 4~M     & 11.2 @ 256              & 12 @ 2~k     & 18.1 @ 16~k           & 0.000392                     & 0.000563                      & 1.07 @ 32~k                                                   & 1.62 @ 16~k \\
FE w/o                      & 45900 @ 256  & 11.2 @ 2~M               & 18 @ 256    & 25.9 @ 256           &  @ 256                     &  @ 256                      & 1.61 @ 256                                                   & 2.31 @ 256 \\
 \texttt{TCP\_\-NO\-DE\-LAY}      & 6130          & 11.2                     & 14.1        & 20.6           & 5.52 @ 4~M                     & 7.67 @ 4~M                      & 1.26                                                   & 1.83 \\
                                  &               &                          &             &                & 0.744                     & 1.02                      &                                                        &       \\
\hline                                                                                                                                                                                                                                                              %
Reference                   & 17.1 @ 4~M    & 64.3 @ 256              & 64 @ 8~k    & 77.9 @ 8~k           & 0.000374                     & 0.00043                      & 0.854 @ 32~k                                                   & 1.04 @ 32~k \\
GbE w.                      & 264~k @ 256 & 86.2 @ 64~k              & 98.6 @ 256 & 113 @ 256           &  @ 256                     &  @ 256                      & 1.53 @ 256                                                   & 1.76 @ 256 \\
 \texttt{TCP\_\-NO\-DE\-LAY}      & 38200         & 73.1                     & 74.6        & 88.4           & 4.2 @ 4~M                     & 4.9 @ 4~M                      & 1.03                                                   & 1.21 \\
                                  &               &                          &             &                & 0.558                     & 0.651                      &                                                         &       \\
\hline                                                                                                                                                                                                                                                              %
Reference                   & 17.1 @ 4~M    & 64.3 @ 256              & 66.4 @ 8~k  & 80.8 @ 8~k           & 0.000364                     & 0.000427                      & 0.866 @ 32~k                                                   & 1.04 @ 32~k \\
GbE w/o                     & 263~k @ 256 & 88.4 @ 64~k              & 96 @ 256   & 113 @ 256           &  @ 256                     &  @ 256                      & 1.49 @ 256                                                   & 1.75 @ 256 \\
 \texttt{TCP\_\-NO\-DE\-LAY}      & 38600         & 74.4                     & 75.8        & 89.5             & 4.2 @ 4~M                     & 4.9 @ 4~M                      & 1.03                                                   & 1.21 \\
                                  &               &                          &             &                  & 0.56                     & 0.651                      &                                                         &       \\
\hline                                                                                                                                                                                                                                                              %
Blob Class                  & 2.8 @ 4~M     & 0.522 @ 256             & 12 @ 4~M     & 35.9 @ 512~k           & 0.0248                     & 0.0309                      & 1.07 @ 4~M                                                   & 3.22 @ 4~M \\
On-Demand                   & 2140 @ 256   & 11.2 @ 4~M               & 53.2 @ 256 & 66.9 @ 256          &  @ 256                     &  @ 512                      & 102 @ 256                                                   & 128 @ 256 \\
Alloc.                      & 678           & 7.32                     & 24.4        & 43.3                 & 4.28 @ 4~M                     & 12.9 @ 4~M                      & 14.5                                                   & 19.5 \\
FE w. Connect               &                &                         &             &                      & 0.59                     & 1.74                      &                                                    & \\
\hline                                                                                                                                                                                                                                                              %
Blob Class                  & 2.78 @ 4~M    & 0.0303 @ 256            & 12 @ 4~M     & 16.3 @ 1~k           & 0.288                     & 0.132                      & 1.08 @ 4~M                                                   & 3.24 @ 4~M \\
On-Demand                   & 124 @ 512    & 11.1 @ 4~M               & 90.4 @ 512 & 42.8 @ 64~k          &  @ 64~k                     &  @ 512                      & 2980 @ 256                                                   & 553 @ 256 \\
Alloc.                      & 84.1          & 5.18                     & 53        & 28.5                 & 4.32 @ 4~M                     & 12.9 @ 4~M                      & 392                                                   & 76.1 \\
FE w/o Connect              &        &                                 &             &                      & 0.998                     & 1.85                      &                                                    & \\
\hline                                                                                                                                                                                                                                                              %
Blob Class                  & 20.8 @ 4~M    & 0.42 @ 256              & 33.6 @ 1~k  & 38.9 @ 1~k           & 0.0198                     & 0.0232                      & 0.792 @ 4~M                                                   & 1.36 @ 4~M \\
On-Demand                   & 1720 @ 256   & 83.4 @ 4~M               & 66 @ 4~M    & 113 @ 4~M           &  @ 256                     &  @ 512                      & 80.2 @ 256                                                   & 95.1 @ 256 \\
Alloc.                      & 874           & 35.7                     & 26.6        & 69.2                 & 3.16 @ 4~M                     & 5.43 @ 4~M                      & 11.5                                                   & 13.9 \\
GbE w. Connect              &        &                                 &             &                      & 0.444                     & 0.749                      &                                                    & \\
\hline                                                                                                                                                                                                                                                              %
Blob Class                  & 15.6 @ 4~M    & 0.0302 @ 256            & 40 @ 1~M    & 15.1 @ 1~k           & 0.418                     & 0.122                      & 0.866 @ 4~M                                                   & 1.48 @ 4~M \\
On-Demand                   & 124 @ 256    & 62.4 @ 4~M               & 81.4 @ 512 & 92.2 @ 4~M           &  @ 256~k                     &  @ 256                      & 2680 @ 256                                                   & 500 @ 256 \\
Alloc.                      & 98.6          & 16.1                     & 64.8        & 37.6                 & 3.46 @ 4~M                     & 5.91 @ 4~M                      & 356                                                   & 68.5 \\
GbE w/o Connect             &        &                                 &             &                      & 0.896                     & 0.921                      &                                                    & \\
\hline                                                                                                                                                                                                                                                              %
Blob Class                  & 2.8 @ 4~M   & 1.1 @ 256                 & 12 @ 4~M     & 36 @ 4~M             & 0.0098                     & 0.0185                      & 1.07 @ 4~M                                                   & 3.21 @ 4~M \\
Prealloc.                  & 4510 @ 256 & 11.2 @ 4~M                 & 45.6 @ 256 & 84.3 @ 256          &  @ 512                     &  @ 512                      & 41.4 @ 256                                                   & 76.5 @ 256 \\
FE w. Connect               & 1140        & 8.03                       & 19.2        & 45.4                 & 4.28 @ 4~M                     & 12.8 @ 4~M                      & 6.32                                                   & 12.7 \\
                            &             &                            &             &                      & 0.578                     & 1.73                      &                                                        &      \\
\hline                                                                                                                                                                                                                                                              %
Blob Class                  & 2.79 @ 4~M  & 0.039 @ 256               & 12 @ 4~M     & 14.1 @ 256          & 0.202                     &  0.0882                      & 1.08 @ 4~M                                                   & 3.23 @ 4~M \\
Prealloc.                  & 160 @ 1~k   & 11.2 @ 4~M                 & 92 @ 256   & 44.3 @ 64~k          &  @ 64~k                     &  @ 256                      & 2360 @ 256                                                   & 361 @ 256 \\
FE w/o Connect              & 105         & 5.48                       & 51.6        & 27.8                 & 2.3 @ 4~M                     & 12.9 @ 4~M                      & 308                                                   & 51.4 \\
                            &             &                            &             &                      & 0.886                     & 1.8                      &                                                       &      \\
\hline                                                                                                                                                                                                                                                              %
Blob Class                  & 21.7 @ 4~M  & 1.17 @ 256                & 35.8 @ 1~k  & 63.1 @ 4~k           & 0.0086                     & 0.0161                      & 0.806 @ 4~M                                                   & 1.38 @ 4~M \\
Prealloc.                  & 4810 @ 256 & 86.8 @ 4~M                 & 70.2 @ 256~k & 120 @ 1~M          &  @ 256                     &  @ 256                      & 35.2 @ 256                                                   & 65.9 @ 256 \\
GbE w. Connect              & 1890        & 45                         & 54.4        & 92.2                 & 3.22 @ 4~M                     & 5.53 @ 4~M                      & 5.46                                                   & 9.91 \\
                            &             &                            &             &                      & 0.442                     & 0.756                      &                                                        &      \\
\hline                                                                                                                                                                                                                                                              %
Blob Class                  & 18.6 @ 4~M  & 0.039 @ 256               & 54 @ 256~k  & 13.1 @ 256          & 0.38                     & 0.0819                      & 0.912 @ 4~M                                                   & 1.5 @ 4~M \\
Prealloc.                  & 160 @ 1~k   & 74.5 @ 4~M                 & 89.2 @ 256 & 112 @ 4~M             &  @ 256~k                     &  @ 256                      & 2280 @ 256                                                   & 336 @ 256 \\
GbE w/o Connect             & 130         & 21.5                       & 76.2        & 46.5                 & 3.64 @ 4~M                     & 6.02 @ 4~M                      & 302                                                   & 47.6 \\
                            &             &                            &             &                      & 0.858                     & 0.91                      &                                                       &      \\
\hline                                                                                                                                                                                                                                                              %
\end{tabular}
}
\parbox{0.90\columnwidth}{
\caption[TCP reference and blob class plateau measurements.]{\label{Tab:TCPBlobMeasurementComparisonPlateau}Comparison of the TCP reference and blob class plateau measurements. 
Shown are the minimum and maximum values with their respective block size in bytes as well as the average of all values.
For the reference tests only the block range from 256~B to 4~MB has been used, corresponding to the range
covered by the blob class tests. } %Sizes are in bytes.}
}
\end{center}
\end{table}

\begin{table}[hbt!p]
\begin{center}
{\scriptsize
\begin{tabular}{|l||c|c|c|c|c|c|c|c|}
\hline  
                            & Rate / & Network                         & CPU Usage   & CPU Usage  & CPU Usage /          & CPU Usage /           & CPU Usage /                               & CPU Usage / \\
Measurement                 & Hz     & Throughput /                    & Sender /    & Receiver / & Rate                 & Rate                  & Throughput                                & Throughput \\
Type                        &        & $\frac{\mathrm{MB}}{\mathrm s}$ & \%          & \%         & Sender /             & Receiver /            & Sender /                                  & Receiver /                                \\
                            &        &                                 &             &            & $\%\times \mathrm s$ & $\% \times \mathrm s$ & $\frac{\% \times \mathrm s}{\mathrm{MB}}$ & $\frac{\% \times \mathrm s}{\mathrm{MB}}$ \\
                            &        &                                 &             &            &                      &                       &                                                    & \\
\hline \hline
Reference                   & 2.8 @ 4~M     & 11.2 @ 4~M               & 12.2 @ 8~k   & 15.6 @ 2~k           & 0.000746                    & 0.00242 @ 1~k                      & 1.08 @ 8~k                                                   & 1.38 @ 2~k \\
FE w.                       & 48800 @ 256  & 11.9 @ 256              & 36.4 @ 256 & 28.3 @ 1~k           &  @ 256                   & 7.24 @ 4~M                      & 3.06 @ 256                                                   & 2.48 @ 1~k \\
 \texttt{TCP\_\-NO\-DE\-LAY}      & 6400          & 11.3                     & 17.2         &19.7           & 5.72 @ 4~M                & 1.11                      & 1.51                                                   & 1.75 \\
                                  &               &                          &             &                & 0.758                     &                            &                                                         &      \\
\hline                                                                                                                                                                                                                                                              %
Reference                   & 2.8 @ 4~M     & 11.2 @ 4~M               & 12 @ 16~k    & 15.6 @ 2~k           & 0.000742                    & 0.00238 @ 1~k                      & 1.07 @ 16~k                                                   & 1.38 @ 2~k \\
FE w/o                      & 50400 @ 256  & 12.3 @ 256              & 37.4 @ 256 & 27.9 @ 1~k           &  @ 256                   & 7.38 @ 4~M                      & 3.04 @ 256                                                   & 2.44 @ 1~k \\
 \texttt{TCP\_\-NO\-DE\-LAY}      & 6500          & 11.3                     & 16.9        & 19.9           & 5.36 @ 4~M                & 1.12                      & 1.48                                                   & 1.77 \\
                                  &               &                          &             &                & 0.7.18                     &                            &                                                         &       \\
\hline                                                                                                                                                                                                                                                              %
Reference                   & 17.2 @ 4~M    & 45.6 @ 256              & 72.8 @ 4~M  & 84.8 @ 4~M           & 0.000458                     & 0.0619 @ 64~k                      & 0.834 @ 16~k                                                   & 0.991 @ 64~k \\
GbE w.                      & 187~k @ 256 & 109 @ 16~k               & 93 @ 4~k  & 96.6 @ 128~k           &  @ 256                     & 4.94 @ 4~M                      & 1.88 @ 256                                                   & 1.29 @ 256~k \\
 \texttt{TCP\_\-NO\-DE\-LAY}      & 30500         & 79.8                     & 85.4        & 90.3           & 4.24 @ 4~M                     & 1.4                      & 1.13                                                   & 1.21 \\
                                  &               &                          &             &                & 0.568                     &                            &                                                         &     \\
\hline                                                                                                                                                                                                                                                              %
Reference                   & 17.1 @ 4~M    & 63.9 @ 256              & 72 @ 1~M    & 81.6 @ 512           & 0.000376                     & 0.000506                      & 0.872 @ 32~k                                                   & 1.04 @ 512 \\
GbE w/o                     & 262~k @ 256 & 106 @ 8~k                & 102 @ 1~k  & 113 @ 8~k           &  @ 256                    &  @ 512                      & 1.54 @ 256                                                   & 1.23 @ 512~k \\
 \texttt{TCP\_\-NO\-DE\-LAY}      & 40800         & 80.7                     & 84.2        & 92.9           & 4.2 @ 4~M                     & 4.83 @ 4~M                      & 1.06                                                   & 1.14 \\
                                  &               &                          &             &                & 0.56                     & 0.69                      &                                                         &       \\
\hline                                                                                                                                                                                                                                                              %
Blob Class                        & 2.8 @ 4~M     & 0.522 @ 256             & 12 @ 4~M     & 35.9 @ 512~k           & 0.0248                       & 0.0309                      & 1.07 @ 4~M                                                   & 3.22 @ 4~M \\
On-Demand                         & 2140 @ 256   & 11.2 @ 4~M               & 53.2 @ 256 & 66.9 @ 256           &  @ 256                     &  @ 512                      & 102 @ 256                                                   & 128 @ 256 \\
Alloc.                            & 678           & 7.32                     & 24.4        & 43.3               & 4.28 @ 4~M                     & 12.9 @ 4~M                      & 14.5                                                   & 19.5 \\
FE w. Connect                     &                &                        &             &                       & 0.59                       & 1.74                      &                                                    & \\
\hline                                                                                                                                                                                                                                                              %
Blob Class                        & 2.78 @ 4~M    & 0.122 @ 256             & 12 @ 4~M  & 36.3 @ 4~M           & 0.15                     & 0.139                      & 1.08 @ 4~M                                                   & 3.26 @ 4~M  \\
On-Demand                         & 500 @ 256    & 11.1 @ 4~M               & 75.2 @ 256 & 62.2 @ 256           &  @ 256                     &  @ 256                     & 616 @ 256                                                   & 571 @ 256 \\
Alloc.                            & 229           & 5.83                     & 42.4        & 48.1                & 4.32 @ 4~M                     & 13 @ 4~M                      & 83.6                                                   & 79 \\
FE w/o Connect                    &               &                          &             &                      & 0.364                     & 1.86                      &                                                    & \\
\hline                                                                                                                                                                                                                                                              %
Blob Class                        & 20.8 @ 4~M    & 0.42 @ 256              & 33.6 @ 1~k  & 38.9 @ 1~k           & 0.0196                     & 0.0232                      & 0.792 @ 4~M                                                   & 1.36 @ 4~M \\
On-Demand                         & 1720 @ 256   & 83.4 @ 4~M               & 66 @ 4~M    & 113 @ 4~M           &  @ 256                     &  @ 512                      & 80.2 @ 256                                                   & 95.1 @ 256 \\
Alloc.                            & 874           & 35.7                     & 46.6        & 69.2                 & 3.16 @ 4~M                     & 5.43 @ 4~M                      & 11.5                                                   & 13.9 \\
GbE w. Connect                    &               &                          &             &                      & 0.444                     & 0.749                      &                                                    & \\
\hline                                                                                                                                                                                                                                                              %
Blob Class                        & 15.7 @ 4~M    & 0.0915 @ 256            & 34.6 @ 1~M  & 41.1 @ 4~k           & 0.153                     & 0.123                      & 0.86 @ 4~M                                                   & 1.48 @ 4~M \\
On-Demand                         & 378 @ 1~k     & 62.8 @ 4~M               & 54 @ 4~M    & 92.3 @ 4~M           &  @ 1~k                     & @ 256                       & 590 @ 256                                                   &  502 @ 256 \\
Alloc.                            & 216           & 17.1                     & 48.6        & 52.6                 & 3.44 @ 4~M                     & 5.91 @ 4~M                      & 80                                                   & 69.7 \\
GbE w/o Connect                   &               &                          &             &                      & 0.61                           & 0.922                      &                                                    & \\
\hline                                                                                                                                                                                                                                                              %
Blob Class                        & 2.8 @ 4~M   & 1.1 @ 256                 & 12 @ 4~M     & 36 @ 4~M             & 0.0098                     & 0.0185                      & 1.07 @ 4~M                                                   & 3.21 @ 4~M \\
Prealloc.                        & 4510 @ 256 & 11.2 @ 4~M                 & 45.6 @ 256 & 84.3 @ 256          &  @ 512                     &  @ 512                      & 41.4 @ 256                                                   & 76.5 @ 256 \\
FE w. Connect                     & 1140        & 8.03                       & 19.2        & 45.4                 & 4.28 @ 4~M                     & 12.8 @ 4~M                      & 6.32                                                   & 12.7 \\
                                  &             &                            &             &                      & 0.578                     & 1.73                      &                                                        &      \\
\hline                                                                                                                                                                                                                                                              %
Blob Class                        & 2.79 @ 4~M  & 0.161 @ 256               & 12 @ 4~M     & 36.3 @ 4~M          & 0.1.03                     & 0.0673                      & 1.08 @ 4~M                                                   & 3.25 @ 4~M \\
Prealloc.                        & 660 @ 256  & 11.2 @ 4~M                 & 69.2 @ 256 & 60.9 @ 512          &  @ 1~k                     &  @ 256                      & 428 @ 256                                                   & 276 @ 256 \\
FE w/o Connect                    & 281         & 6.09                       & 37.4        & 45.8                 & 4.3 @ 4~M                     & 13 @ 4~M                      & 58.4                                                   & 47 \\
                                  &             &                            &             &                      & 0.682                     & 1.83                      &                                                        &      \\
\hline                                                                                                                                                                                                                                                              %
Blob Class                        & 21.7 @ 4~M  & 1.17 @ 256                & 35.8 @ 1~k  & 63.1 @ 4~k           & 0.0086                     & 0.0161                      & 0.806 @ 4~M                                                   & 1.38 @ 4~M \\
Prealloc.                        & 4810 @ 256 & 86.8 @ 4~M                 & 70.2 @ 256~k & 120 @ 1~M          &  @ 256                     &  @ 256                      & 35.2 @ 256                                                   & 65.9 @ 256 \\
GbE w. Connect                    & 1890        & 45                         & 54.4        & 92.2                 & 3.22 @ 4~M                       & 5.53 @ 4~M                      & 5.46                                                   & 9.91 \\
                                  &             &                            &             &                      & 0.442                       & 0.756                      &                                                        &      \\
\hline                                                                                                                                                                                                                                                              %
Blob Class                        & 15.3 @ 4~M  & 0.13 @ 256                & 37.4 @ 512~k & 40.2 @ 8~k          & 0.0934                     & 0.08                      & 0.85 @ 4~M                                                   & 1.44 @ 4~M \\
Prealloc.                        & 533 @ 256  & 61.4 @ 4~M                 & 52.2 @ 4~M  & 88.5 @ 4~M           &  @ 256                     &  @ 512                      & 382 @ 256                                                   & 548 @ 256 \\
GbE w/o Connect                   & 289         & 18.2                       & 45.8        & 55                 & 3.4 @ 4~M                     & 5.77 @ 4~M                      & 52.4                                                   & 60.3 \\
                                  &             &                            &             &                      & 0.556                     & 0.87                      &                                                             &  \\
\hline                                                                                                                                                                                                                                                              %
\end{tabular}
}
\parbox{0.90\columnwidth}{
\caption[TCP reference and blob class peak measurements.]{\label{Tab:TCPBlobMeasurementComparisonPeak}Comparison of the TCP reference and blob class peak measurements. 
Shown are the minimum and maximum values with their respective block size in bytes as well as the average of all values.
For the reference tests only the block range from 256~B to 4~MB has been used, corresponding to the range
covered by the blob class tests. } %Sizes are in bytes.}
}
\end{center}
\end{table}

\subsection{\label{Sec:TCPComClassesBenchmarkSummary}TCP Communication Class Benchmark Summary}

As an overall summary for the two types of TCP communication classes one can repeat the separate communication classes' conclusions that they are suited 
for use in the ALICE High Level Trigger in the present version, even though some room for optimization is still present. The separation
into different classes, optimized for small and large transfers, does not produce significant advantages in the tested version if both
types of communication are handled by the same physical communication medium. An advantage might be seen when the message communication is run
over Fast Ethernet, utilizing its lower latency, and the blob communication over Gigabit Ethernet using the higher bandwidth and better efficiency. One further reason why the 
separation does not produce any clearly visible effects could also be found in the fact that the current
implementations of the communication code are
not yet optimized enough for each of their specific tasks. Additionally, the combination of Gigabit Ethernet and the TCP network protocol does not provide 
enough features that allow to optimize for transfer efficiency. Also, TCP is not able to take enough advantage of many features  
provided by the networking hardware. The use 
of other network protocols and technologies could therefore provide clearer effects of the separation. 
Possible optimization measures for the communication classes
are the use of the \texttt{writev} Linux system call that allows to pass several blocks to write into a connection socket in one system call as well
as the reduction of memory allocation and release calls in the message classes. Preliminary tests of these modifications in the publisher-subscriber interface 
from section~\ref{Chap:PubSubInterface} indicate good benefits from these measures as described below in section~\ref{Sec:PubSubFutureOptimization}.

%{\Large Highly device specific, depending on optimization goal (throughput, CPU usage, usage / throughout, usage / rate) different block sizes
%can be optimal, not necessarily highest...}

Although all the above measurements are of course highly specific for each network device on which the corresponding test was executed, the results indicate for
message as well as for blob classes that depending on the optimization goal, e.g. throughput, absolute CPU usage, or CPU usage relative to throughput,
different block or message sizes are the optimum choice. The largest block size is not necessarily always the best choice.

\section{Publisher-Subscriber Interface Benchmarks}

\subsection{\label{Sec:PubSubTimingMeasurements}Timing Measurements}

To evaluate the performance of the publisher-subscriber framework and provide data and estimates of the current and expected future overhead 
incurred by the framework, a number of measurements have been performed. A set of benchmark publisher and subscriber programs has been written to execute 
the basic functions associated with announcing and freeing events only. No additional functionality, e.g. shared memory mapping or 
accessing, is contained in these programs. The tests have been performed on the three reference PCs evaluated
and described in section~\ref{Sec:CacheMeasurements} to obtain measurements about the scaling properties of the software. All benchmarks 
have been performed with almost no user processes or daemons running on the system to exclude interference effects from other processes, e.g. (de-) 
scheduling, as far as possible. The list of remaining processes is shown in appendix~\ref{Sec:BenchmarkProcessList}.

In the two benchmark processes four different parts of the framework have been instrumented for timing measurements using 
the \texttt{get\-time\-of\-day} system 
call that delivers a microsecond resolution. The timing overhead of a \texttt{get\-time\-of\-day} call itself is small. In a test program on an 800~MHz PC the time 
needed to execute 100000 calls was 61275 $\mu \mathrm s$, so one call requires about 600 ns.

The four benchmarked parts of the framework are executed for each event, 
as it is announced to a subscriber and released again, as detailed in chapter~\ref{Chap:PubSubInterface}:
\begin{itemize}
\item The main publisher object's \texttt{An\-nounce\-Event} function that stores an event's management data into the 
publisher's internal tables and dispatches the data's descriptor to the write threads for each subscriber.
% (only one thread present)
\item The write thread's function that writes the data into the named pipe.
\item The \texttt{New\-Event} function in the subscriber that writes the data release message (\texttt{Event\-Done}) into the pipe 
to the publisher.
\item The publisher's \texttt{Event\-Done} function that releases the event management data from its internal tables. 
\end{itemize}

One of the principal problems of measuring a program's (processing) overhead is that a program's running time for a particular code section does not provide an adequate measure
of its overhead. This inadequacy results from the fact that a program may be suspended while executing the section, increasing the section's runtime but not the 
overhead. Reasons for a suspension might be that the operating system deschedules it, allowing other programs to execute, or because it has to wait
for an operation to complete, e.g. disk or network I/O, or for a lock to become available. While the case of explicit sleeps can in principle be accounted for 
during measurement by deducting the corresponding sleep time from the runtime, the other cases cannot be predicted. Even in the explicit sleep case there are problems, as the
operating system may let a program sleep longer than the specified time. Therefore a way has to be found to exclude these sleep times from the overhead measurement
for a given program, as these represent time where the active thread sleeps and thus does not use the CPU producing overhead. 

In the publisher-subscriber measurements lock calls have been excluded from the timing, by starting and stopping the timing measurement before and after it explicitly. 
For the other cases the presumption can be made that code section runtimes during which a program has been suspended, which therefore are unsuitable for overhead
measurements, will be significantly longer than those where the section could be executed uninterrupted. These longer values can then be excluded from the 
overhead measurement. A precondition for this is of course that the examined code sections are sufficiently short so that a minimal descheduling will actually
be longer than a normal section execution. 

$50 \cdot 10^6$ events have been processed on each of the three 
PCs and the four timing values for each event have been entered into a runtime histogram for later analysis. For the analysis a cut-off has been made so that 
the contents of the bins used for the analysis amount to at least 90~\% of the histogram entries. This 
cut-off is made under the assumption, detailed above, that longer times only occur when the examined process is inactive, which has no 
influence on the framework's overhead. The mean values with and without the cut-off are shown in Table~\ref{Tab:PubSubBenchmarkResults}, 
all values are in microseconds. For 
both mean values the scaling constants from 733~MHz to 800~MHz and 800~MHz to 933~MHz are shown as well. 
The complete timing analysis plots are shown in Fig.~\ref{Fig:AnnounceEventTime},~\ref{Fig:ThreadAnnounceTime},~\ref{Fig:EventDoneTime}, and
~\ref{Fig:SubscriberEventDoneTime}, containing the superimposed time distribution for the three reference PCs, 
733~MHz values in green, 800~MHz values in blue, and 933~MHz values in red. 800~MHz and 933~MHz values are scaled with factors of $10^3$ and $10^6$ 
respectively for clarity. Each plot also shows the respective bins where the 90~\% cut-off was made. As can be seen from the times which make up the majority 
of the measurements, the values presumed to be descheduled are indeed significantly longer than the majority.

%\begin{figure}[hbt]
%\begin{center}
%\resizebox*{0.90\columnwidth}{!}{
%\includegraphics{BenchPublisher-1Subscriber-8192EventsInTransit-All-QuietSystem-AnnouceEventTimeNoLocks.eps}
%}
%\end{center}
%\caption{\label{Fig:AnnounceEventTime}Publisher Announce Event compute time distributions, 733~MHz values are shown in green unscaled, 800~MHz values in blue scaled
%with a factor of $10^3$, and 933~MHz are shown in red scaled with a factor of $10^6$.}
%}
%\end{center}
%\end{figure}

%\begin{figure}[hbt]
\begin{figure}
\begin{center}
\resizebox*{0.90\columnwidth}{!}{
\includegraphics{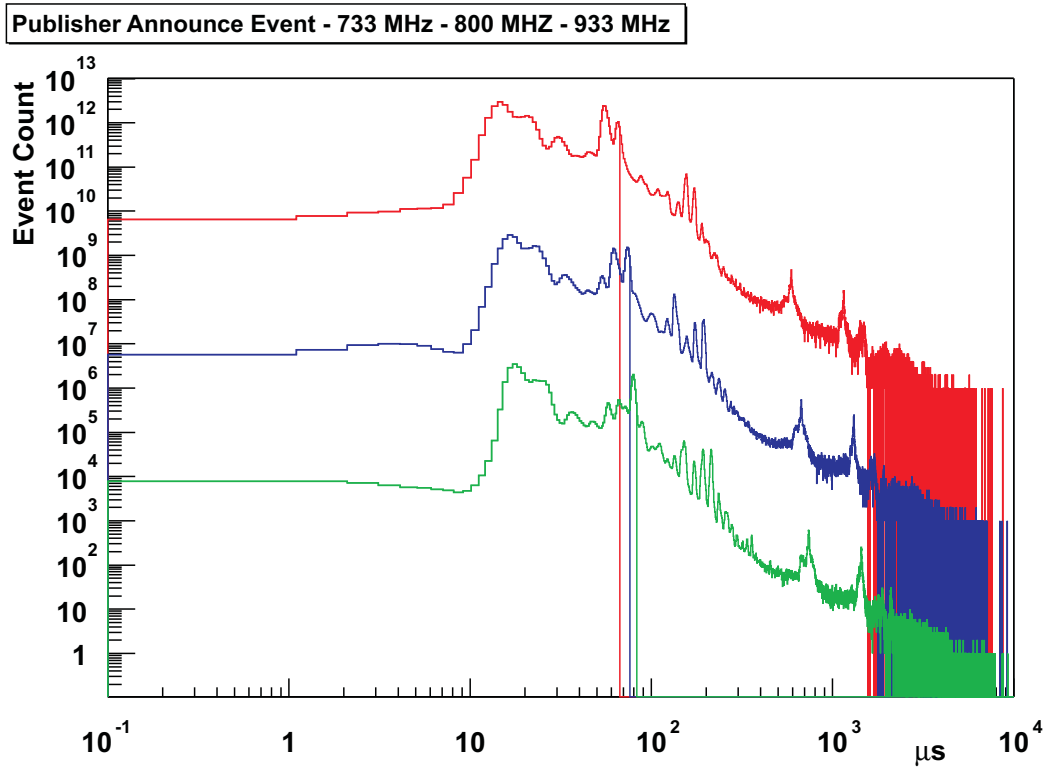}
}
\parbox{0.90\columnwidth}{
\caption[Publisher Announce Event compute time distributions.]{\label{Fig:AnnounceEventTime}Publisher Announce Event compute time distributions. 733~MHz values are shown in green unscaled, 800~MHz values in blue scaled
with a factor of $10^3$, and 933~MHz in red scaled with a factor of $10^6$.}
}
\end{center}
\end{figure}

%\begin{figure}[hbt]
\begin{figure}
\begin{center}
\resizebox*{0.90\columnwidth}{!}{
\includegraphics{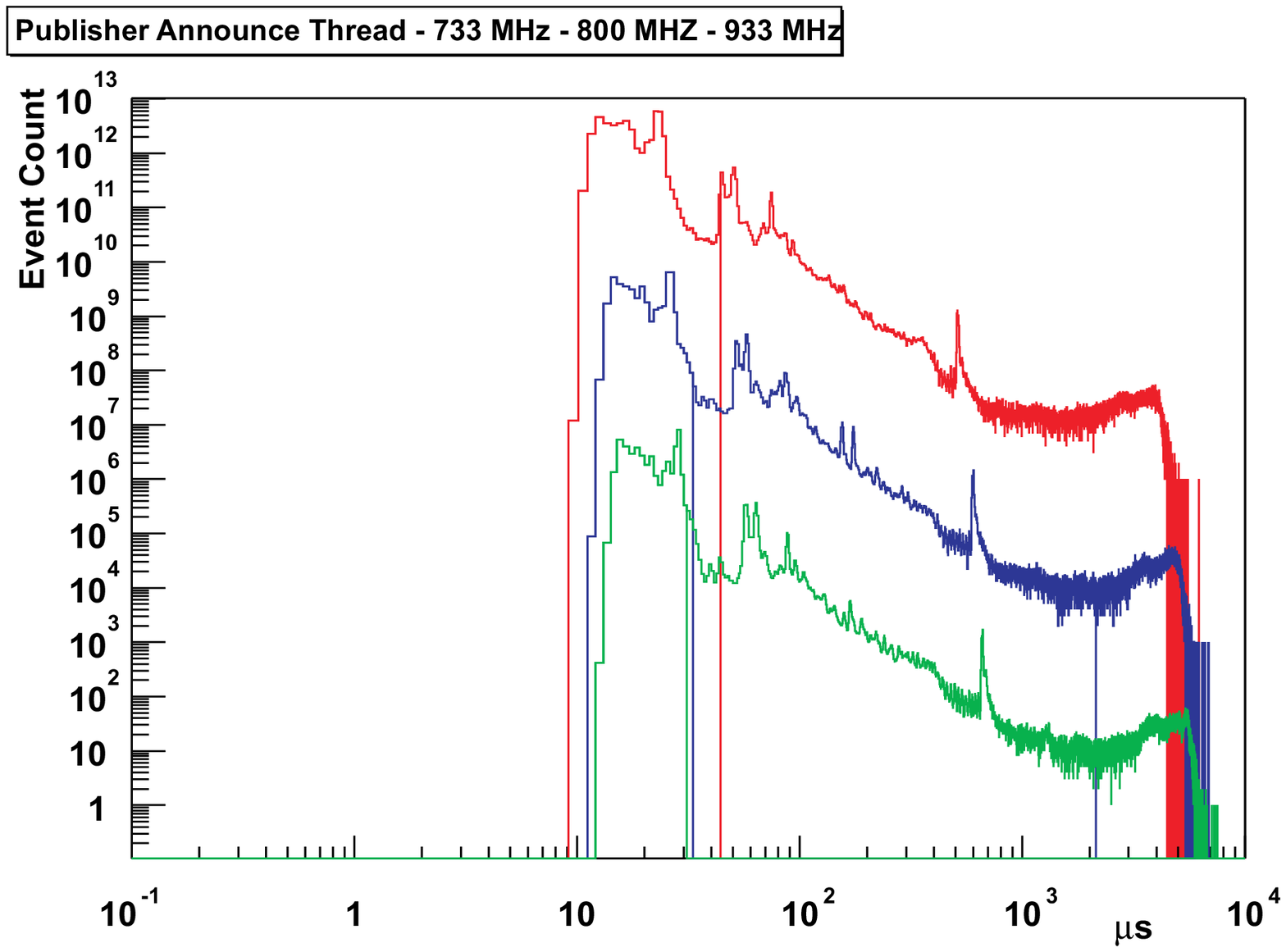}
}
\parbox{0.90\columnwidth}{
\caption[Publisher Announce Thread compute time distributions.]{\label{Fig:ThreadAnnounceTime}Publisher Announce Thread compute time distributions. 733~MHz values are shown in green unscaled, 800~MHz values in blue scaled
with a factor of $10^3$, and 933~MHz in red scaled with a factor of $10^6$.}
}
\end{center}
\end{figure}

%\begin{figure}[hbt]
\begin{figure}
\begin{center}
\resizebox*{0.90\columnwidth}{!}{
\includegraphics{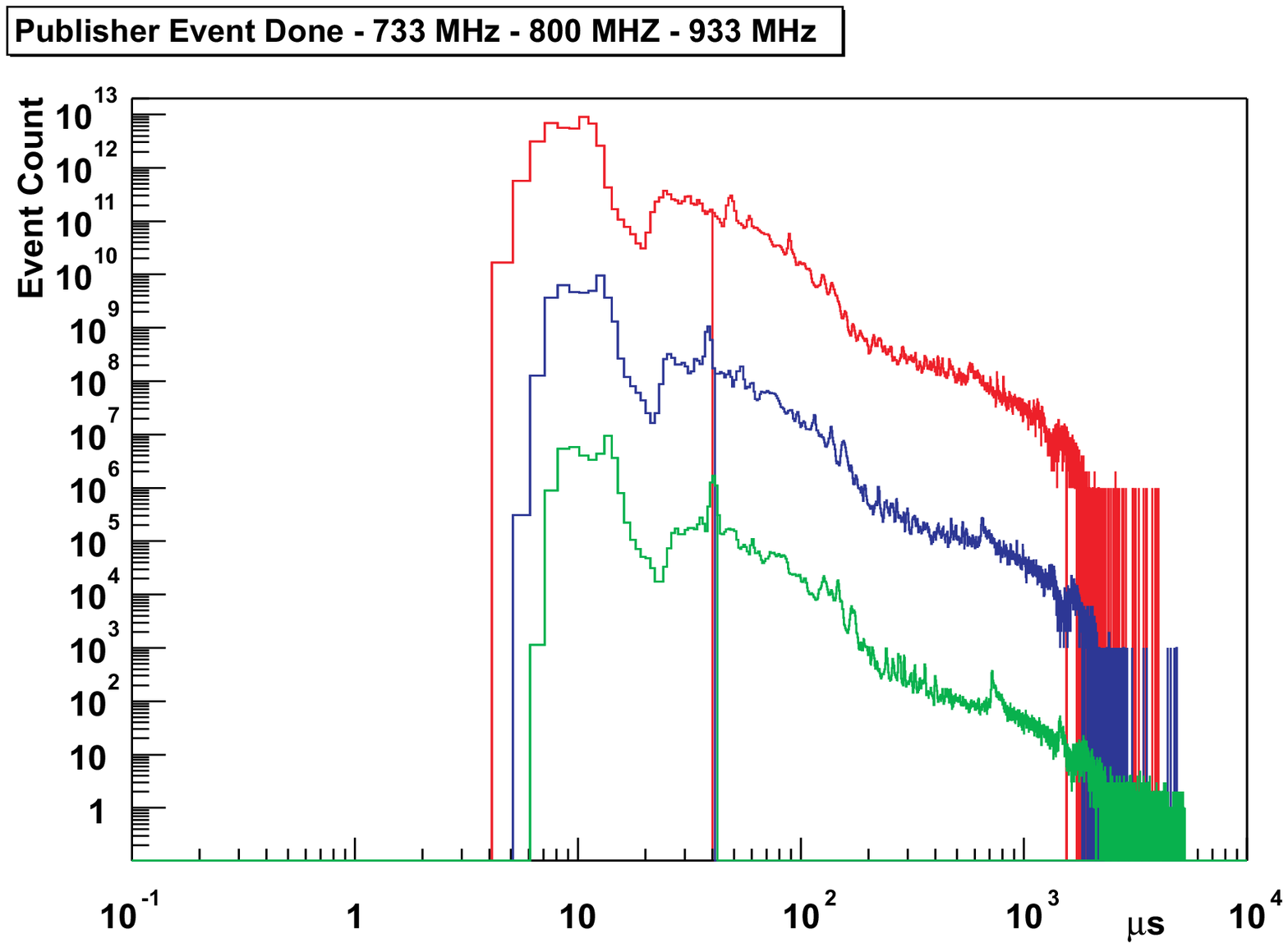}
}
\parbox{0.90\columnwidth}{
\caption[Publisher Event Done compute time distributions.]{\label{Fig:EventDoneTime}Publisher Event Done compute time distributions. 733~MHz values are shown in green unscaled, 800~MHz values in blue scaled
with a factor of $10^3$, and 933~MHz in red scaled with a factor of $10^6$.}
}
\end{center}
\end{figure}

%\begin{figure}[hbt]
\begin{figure}
\begin{center}
\resizebox*{0.90\columnwidth}{!}{
\includegraphics{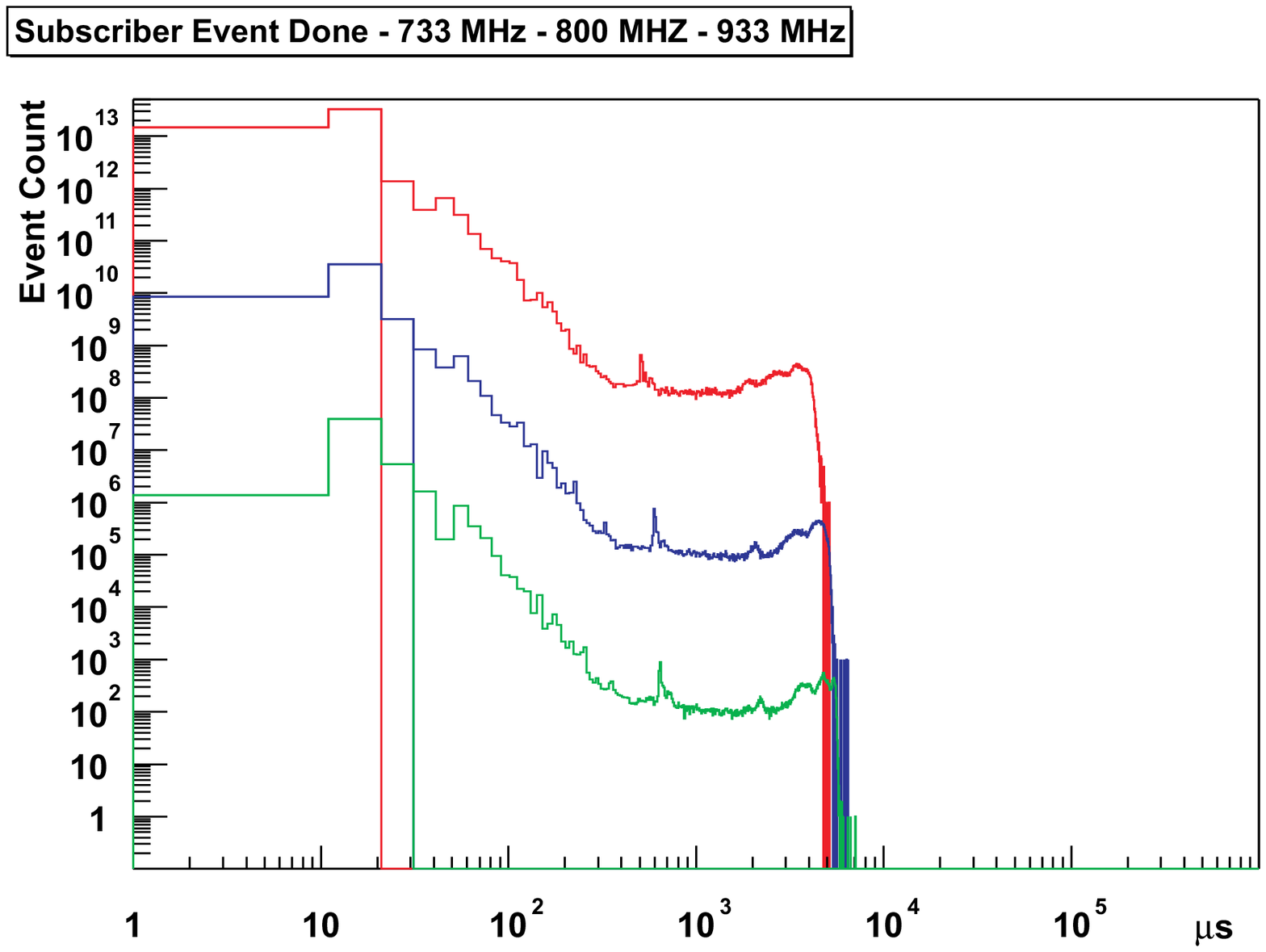}
}
\parbox{0.90\columnwidth}{
\caption[Subscriber Event Done compute time distributions.]{\label{Fig:SubscriberEventDoneTime}Subscriber Event Done compute time distributions. 733~MHz values are shown in green unscaled, 800~MHz values in blue scaled
with a factor of $10^3$, and 933~MHz in red scaled with a factor of $10^6$.}
}
\end{center}
\end{figure}

\begin{table}[hbt!p]
\begin{center}
\begin{tabular}{|l||c|c|c|c|c|}
\hline  
&733~MHz & 800~MHz & 933~MHz & scaling & scaling \\
&PC & PC & PC & 733-800 & 800-933 \\
\hline \hline
event announce / & 50.20 & 47.42 & 42.55 & 1.06 & 1.11 \\
$\mu \mathrm s$ &  &  &  &  &  \\
\hline
event announce & 40.89 & 39.21 & 35.73 & 1.04 & 1.10 \\
(cut-off) / $\mu \mathrm s$ &  &  &  &  &  \\
\hline
announce thread / & 35.73 & 32.51 & 27.95 & 1.00 & 1.16 \\
$\mu \mathrm s$ &  &  &  &  &  \\
\hline
announce thread  & 21.98 & 20.68 & 18.41 & 1.06 & 1.12 \\
(cut-off) / $\mu \mathrm s$  &  &  &  &  &  \\
\hline
publisher event done / & 22.82 & 20.46 & 17.85 & 1.12 & 1.15 \\
 $\mu \mathrm s$ &  &  &  &  &  \\
\hline
publisher event done & 15.52 & 13.82 & 11.56 & 1.12 & 1.20 \\
(cut-off) / $\mu \mathrm s$ &  &  &  &  &  \\
\hline
subscriber event done / & 26.67 & 22.06 & 18.52 & 1.21 & 1.19 \\
 $\mu \mathrm s$ &  &  &  &  &  \\
\hline
subscriber event done & 15.86 & 13.88 & 11.88 & 1.14 & 1.17 \\
 (cut-off) / $\mu \mathrm s$ &  &  &  &  &  \\
\hline
total / $\mu \mathrm s$ & 135.45 & 122.45 & 106.87 & 1.11 & 1.15 \\
\hline
total (cut-off) / $\mu \mathrm s$ & 94.25 & 87.59 & 77.58 & 1.08 & 1.13 \\
\hline
\end{tabular}
\parbox{0.90\columnwidth}{
\caption[Times and scaling properties of different parts of the framework.]{\label{Tab:PubSubBenchmarkResults}The times and scaling properties of the different parts of the framework, all values are in microseconds.}
}
\end{center}
\end{table}

A final measurement that has been performed is the global average announce rate that can be sustained
over the $50 \cdot 10^6$ events. These values are shown for the three different 
PCs with the derived time overheads ($2 \cdot \mathrm{rate}^{-1}$) in Table~\ref{Tab:PubSubAverageRates}. 
The average processing overhead is scaled by a factor 2 with respect to the transaction rate period as there are two processors in the tested computers,
which both have been fully busy during the tests. 
It should be noted, however, that these averages include overhead introduced by waits from the operating system, which would be present even 
in an idle system, but become less likely in case of a system operating at a much lower transaction rate and performing trigger algorithms. 
The numbers stated here should be taken as an all inclusive upper limit. As can be seen the achieved rates on the reference PCs are already
high enough to easily allow the use of the framework in the ALICE HLT. No performance problem should therefore be encountered in running
the interface on PCs available when the HLT becomes operational. 

\begin{table}[hbt!p]
\begin{center}
\begin{tabular}{|l||c|c|c|}
\hline
& 733~MHz PC & 800~MHz PC & 933~MHz PC  \\
\hline \hline
Average event rate / kHz & 11.86 & 12.73 & 14.41 \\
\hline
Average time overhead / $\mu \mathrm s$ & 168.7 & 157.1 & 138.8 \\
\hline
\end{tabular}
\parbox{0.90\columnwidth}{
\caption[Global average event rates and resulting time overheads.]{\label{Tab:PubSubAverageRates}Global average rates and resulting time overheads.}
}
\end{center}
\end{table}

These performance tests were specifically made using a multi-processor architecture in order to include scheduling effects of the Linux system. For instance, 
even the best data locality in the communication algorithm
% (\cf to Table~\ref{Tab:PubSubBenchmarkResults}) 
can be destroyed, if the rescheduling results 
in a job often being assigned to different CPUs and thus requiring the cache coherency protocols to copy the cached data between the CPUs across their 
front side bus. The results seem to indicate that this problem only occurs with a negligible frequency in the framework interface.

\subsection{Scaling Behaviour}

To gain an impression of how the publisher-subscriber framework will perform on future CPUs with their large expected increases in clock frequency, an 
analysis of the software's scaling behavior on the three reference PCs has been made. Since the software handles to a large fraction inter-process communication 
one would expect that a high fraction of the data accesses address the system's main memory accessible by both the reference systems' CPUs. Such a behavior 
would result in a very bad scaling behavior with regard to the clock frequency, as the memory bandwidth and access latency increase much slower than the CPU 
frequency. This effect can also be seen in the access time measurements of the three test PCs in Table~\ref{Tab:ReferecePCCacheResults}.

%\begin{figure}[hbt]
\begin{figure}
\begin{center}
\resizebox*{0.90\columnwidth}{!}{
\includegraphics{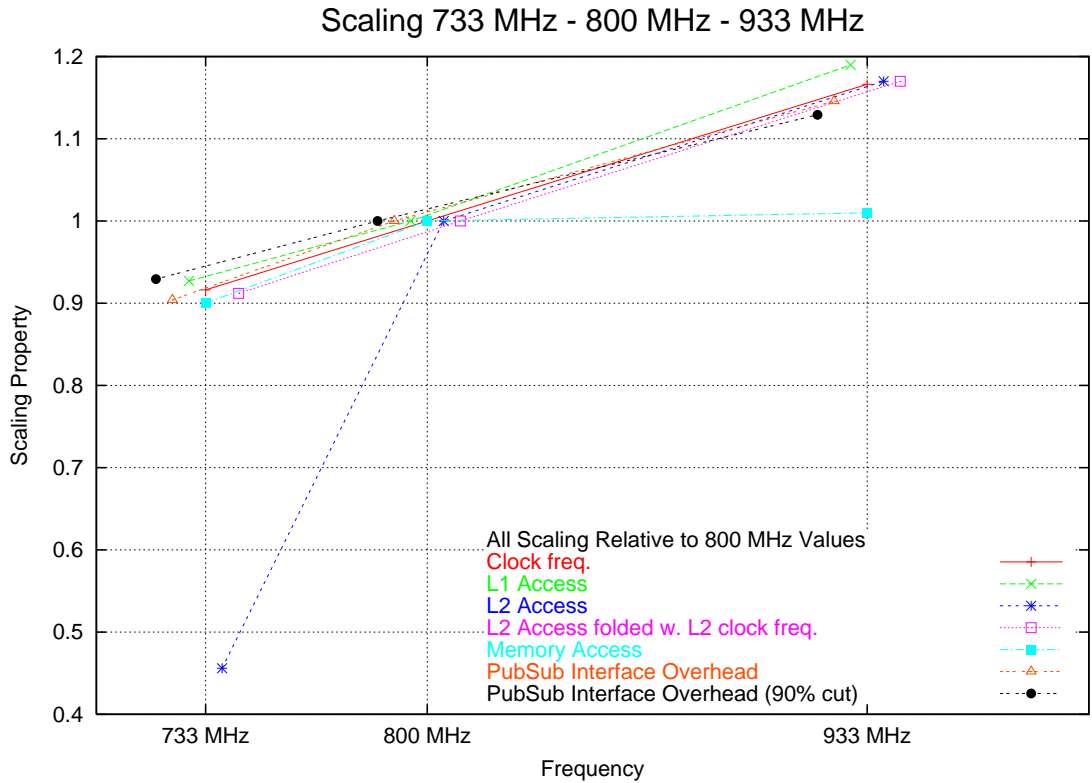}
}
\resizebox*{0.90\columnwidth}{!}{
\includegraphics{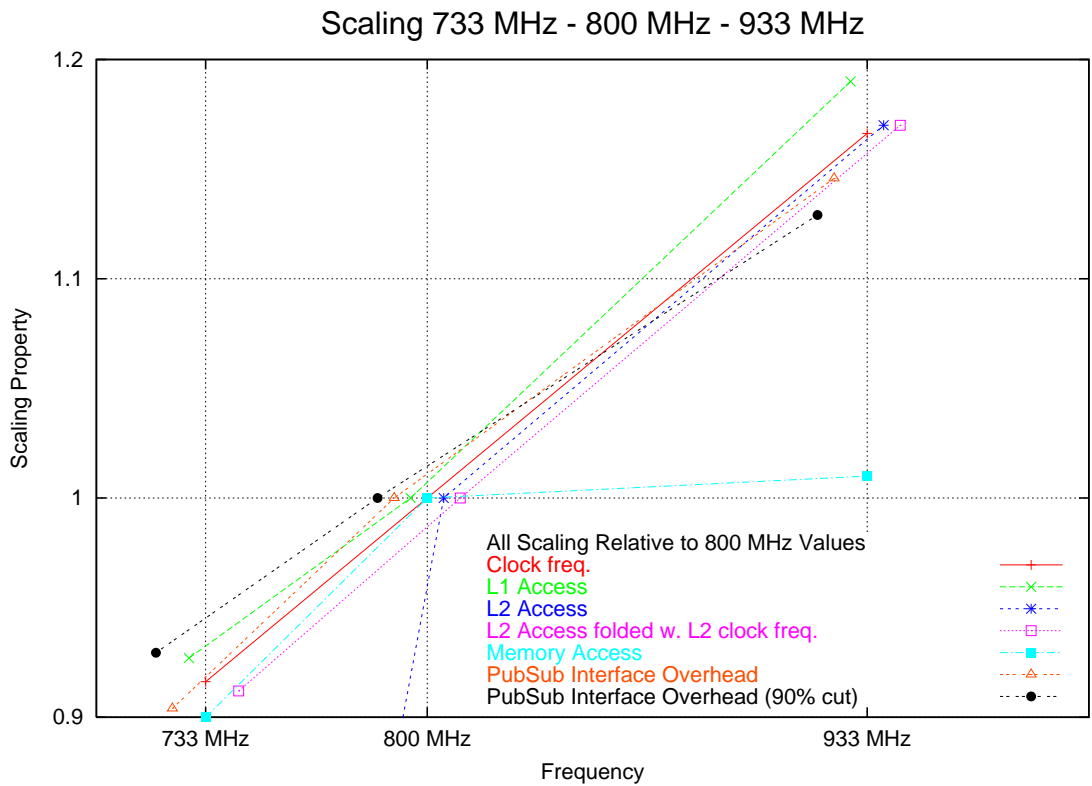}
}
\parbox{0.90\columnwidth}{
\caption[Scaling behaviour of the publisher-subscriber interface as well as the cache and memory systems.]{\label{Fig:PubSubScaling}
The scaling behaviour of the publisher-subscriber interface as well as the cache and memory systems. 
All values are scaled relative to the values of the 800~MHz PC, the red reference curve shows the clock 
frequencies. For clarity the curves have been offset somewhat to prevent overlapping.}
}
\end{center}
\end{figure}

For comparison a scaling plot has been produced, shown in Fig.~\ref{Fig:PubSubScaling}, in which the relative values of various measurements are plotted as determined for the 
three different PCs over their clock frequencies. All values are scaled relative to the values of the 800~MHz PC. The red reference curve shows the clock 
frequencies. For clarity the curves have been offset slightly to prevent overlapping.

The green curve for the level 1 cache access times shows a perfect scaling, which can be expected as this cache works with the CPU's core frequency. In the blue level 2 
access curve one can see the influence  of the level 2 frequency for the 733~MHz CPUs, which is only half the CPU's frequency unlike for the other CPUs. 
Folding in this factor of 2 for the 733~MHz PC the pink curve again 
shows the same perfect scaling property. The cyan memory access curve shows the influence of the chipset (733~MHz to 800~MHz transition) and that for identical 
motherboards the CPU frequency basically has no influence on the memory access time (800~MHz to 933~MHz transition).

The orange curve shows the scaling behavior of the sum of the times measured in the previous section~\ref{Sec:PubSubTimingMeasurements} without the 90~\% cut-off, 
while the black curve shows the same sum using 90~\% cut-off. One can see, as also shown 
in Table~\ref{Tab:PubSubBenchmarkResults}, that both values somewhat under-scale compared to the theoretical values of 1.091 and 1.166 for 733~MHz to 800~MHz and 
800~MHz to 933~MHz respectively. But even taking into account this scaling behavior, the results indicate that the framework can utilize and profit from more than 90~\% of CPU performance 
increases.

Based on those measurements it is assumed that the processing overhead for a complete event announce and release loop is going to drop to $15~\mu \mathrm s/\mathrm{event}$ or less
during the next four to five years, before ALICE (and its HLT) starts to operate. 
The value of $15~\mu \mathrm s$ per event announcement is a useful metric that can be used to calculate the overhead in a more complex chain of multiple processes.
Even given all scaling uncertainties, however, the existing framework is fast enough to fulfill all ALICE HLT requirements to operate at full 
speed, already to date. 
%The only uncertainty left is the CPU time required to perform the communication, which is a cost factor and outside of the scope of the framework itself. 
On the other hand, scaling uncertainties are minimized for CPU bound processes, and the interface's architecture is optimized for smalls amounts of data exchange, 
making it CPU bound as much as possible.

\subsection{\label{Sec:PubSubFutureOptimization}Future Optimization Options}

Preliminary tests with two optimizations of the low-level pipe communication and the pipe proxy classes used in the publisher-subscriber interface show very 
promising results. The optimizations in question are the replacing of multiple \texttt{write} calls with one \texttt{writev} call that allows to specify
multiple blocks to be written with one system call as well as the reduction of \texttt{new} and \texttt{de\-lete} allocation and release calls in the 
communication functions. Measurements of the interface with these optimizations in place are currently very preliminary and by far not as exhaustive
as the ones presented above, but the performances measured so far indicate that a factor of 4 improvement of the maximum performance, and 
by deduction also overhead, in favour of the new optimized version could be possible. 

\section{\label{Sec:FrameworkSystemTests}Framework System Tests}

\subsection{ALICE HLT Proton-Proton Performance Test}

\begin{figure}
\begin{center}
\resizebox*{0.90\columnwidth}{!}{
\includegraphics{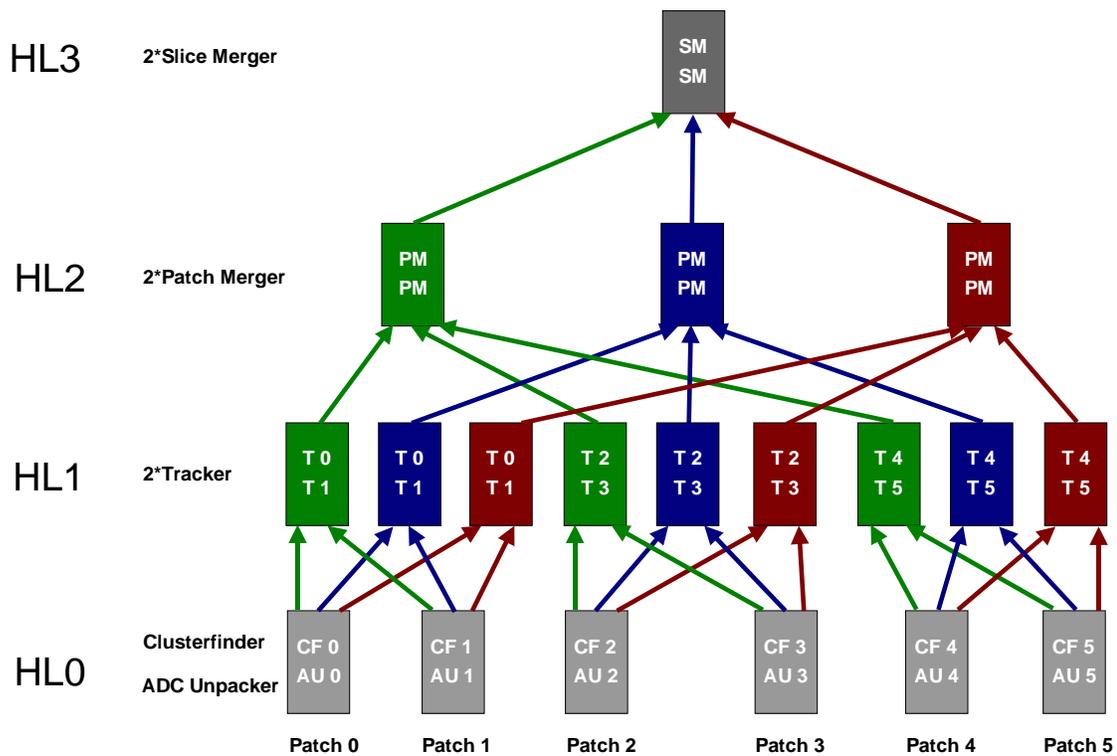}
}
\parbox{0.90\columnwidth}{
\caption{\label{Fig:19NodeTestConfig}The 19 node setup used in the proton-proton performance test.}
}
\end{center}
\end{figure}

In a large-scale test of the framework in an ALICE High Level Trigger configuration, a setup with 19 nodes has been used to simulate the readout and 
online data processing of one slice of the ALICE Time Projection Chamber. A slice is one of 18 sectors in one readout plane and therefore represents 
$\frac{1}{36}$ of the total data volume of the TPC, as described in section~\ref{Sec:ALICEHLT}. For the test simulated piled-up proton-proton events have been used as the 
processing power available for the test  would not have been sufficient to enable the intended operation at the maximum readout frequency of the TPC of 200~Hz 
using simulated Pb-Pb events. The limit in this case is the time required to reconstruct the tracks in the event. A schematic view of the cluster 
configuration used for this test is shown in Figure~\ref{Fig:19NodeTestConfig}.

The data sources for the HLT are the FEPs (see section~\ref{Sec:ALICEHLT}) connected via optical fibers with the readout electronics mounted on the 
detectors. Data from each TPC slice is shipped  to the HLT over six fibers, the sub-sectors associated
with each fiber are called patches. In the present test setup these FEPs 
are replaced by software in the form of \texttt{Mult\-File\-Pub\-lish\-er} components. For each patch
the zero-suppressed and run-length encoded simulated ADC data is read from files by 
the \texttt{Mult\-File\-Pub\-lish\-er}s and published into the start of the chain. This type of data is similar to the data shipped from the detector. 
On average the size of the encoded ADC data files is about 14 kB per patch. Encoded ADC 
data is expanded to sequences of ADC values by the \texttt{ADC\-Un\-packer} components, increasing the size of the data by a factor of about 2 to 3. These values 
in turn form the input for the \texttt{Clus\-ter\-Finder}, which reconstructs the three-dimensional coordinates of deposited charges in the detector, called space points. 
Together with each space point the amount of charge associated with that respective cluster is stored. The \texttt{Mult\-File\-Pub\-lish\-er}, \texttt{ADC\-Un\-packer}, and \texttt{Clus\-ter\-Finder} components run on one node for each patch, called 
Hierarchy-Level (HL) 0. From this node, the space point data are shipped via bridge components to the next Hierarchy-Level, responsible for combining the space 
points into track segments, performed by the \texttt{Tracker} component. Since tracking is the most time consuming process in the chain, data is distributed to three trackers on separate nodes using an \texttt{Event\-Scat\-terer}. 
Due to the usage of two-processor machines it is possible to run two trackers in parallel 
on each node, each processing data belonging to the same event but from different patches. At the output of each Hierarchy-Level 1 node the data stream is merged by an 
\texttt{Event\-Merger} component and forwarded to Hierarchy-Level 2. On this level the data streams of the six patches are merged into a data stream consisting of events with six blocks of 
track segments per event. For load balancing reasons this data stream is processed by six \texttt{Patch\-Merger} components running on three nodes. 
Each \texttt{Patch\-Merger} combines the track segments of tracks crossing boundaries between the patches. Since Hierarchy-Level 2 contains three nodes running \texttt{Patch\-Merger} components 
the data streams of these nodes are merged in Hierarchy-Level 3 using a \texttt{Slice\-Merger} component. Again, for load balancing reasons two \texttt{Slice\-Merger} processes 
are running on each node. The output obtained from running the processing chain described are reconstructed tracks of one TPC slice.

Operation at a sustained rate of more than 430 events/s has been achieved by
using the described setup consisting of 19 nodes with twin CPUs operating at 733~MHz and 800~MHz and connected via Fast Ethernet. The bottlenecks in this setup were the nodes in HL0, 
especially the \texttt{ADC\-Un\-packer} components. In the final setup of the HLT, these steps will 
be performed by FPGAs implemented on the RORC cards and thus will not consume time on the FEP CPUs. The 
maximum TPC readout rate intended for p-p mode in ALICE is 1~kHz and CPUs with more than 3 times the clock frequency relative to those 
used in the test are already available today. Therefore
the use of the framework with these software-only analysis components for online tracking in p-p mode seems to be a practicable option for the ALICE High Level Trigger. 
Given the necessary increases in CPU processing power and an adequate number of CPUs and thus financial resources, the use in Pb-Pb mode is possible as well.

\subsection{Framework Fault Tolerance Test}

\subsubsection{Fault Tolerance Test Setup}

%\begin{figure}
%\begin{center}
%\resizebox*{0.90\columnwidth}{!}{
%%\includegraphics{scaling-overhead-0.4-1.2.eps}
%}
%\end{center}
%\caption{\label{Fig:FTTestConfig}}
%}
%\end{center}
%\end{figure}

\begin{figure}[hbt]
\begin{center}
\resizebox*{0.50\columnwidth}{!}{
\includegraphics{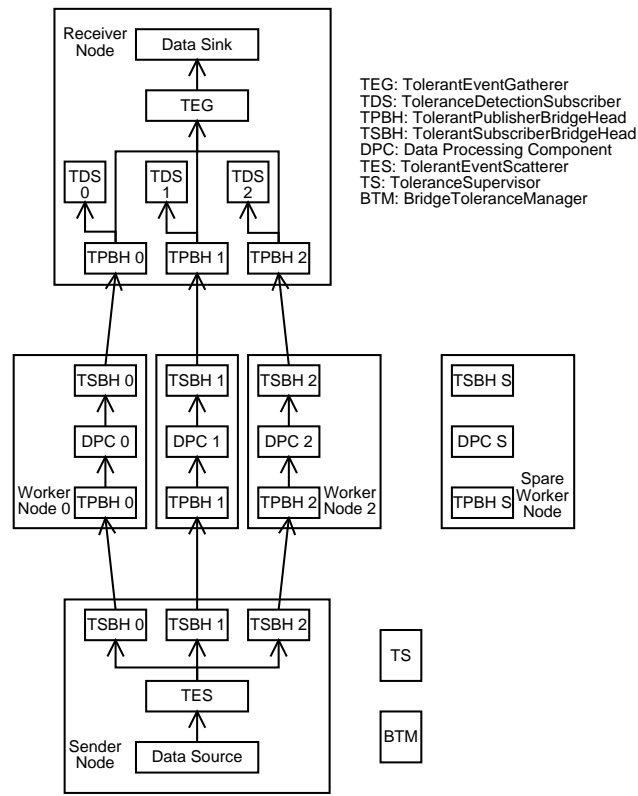}
}
\parbox{0.90\columnwidth}{
\caption[Fault tolerance test setup.]{\label{Fig:FTTestConfig}The fault tolerance component setup used during the test. The arrows show the normal flow of data through the system.}
}
\end{center}
\end{figure}

In order to demonstrate the fault tolerance capabilities of the framework described in section~\ref{Sec:FTComponents} a test setup has been created using seven computers.
The setup used is the sample setup shown in Fig.~\ref{Fig:FaultTolerancePrinciple2} and detailed in section~\ref{Sec:FTConceptOverview}. It is briefly
described and shown again here for convenience. Two of the seven computers function as a data source and sink, one supervisor hosts two control programs, three 
perform identical worker tasks, and one acts as a spare worker node, as displayed in Fig.~\ref{Fig:FTTestConfig}.

On the data source computer one process publishes data from a file to a \texttt{Tol\-e\-rant\-Event\-Scat\-terer} component (see section~\ref{Sec:TolerantEventScatterer}) 
with three output publishers. Each 
of these publishers in turn supplies its data to one \texttt{Tol\-e\-rant\-Sub\-scrib\-er\-Bridge\-Head} (section~\ref{Sec:TolerantBridgeComponents}) that sends the received data to a 
\texttt{Tol\-e\-rant\-Pub\-lish\-er\-Bridge\-Head} 
(section~\ref{Sec:TolerantBridgeComponents}) on one of the worker nodes. Attached to this \texttt{Tol\-e\-rant\-Pub\-lish\-er\-Bridge\-Head} is a dummy processing component (section~\ref{Sec:DummyLoad})
that publishes any received data unchanged to a further \texttt{Tol\-e\-rant\-Sub\-scrib\-er\-Bridge\-Head}. This \texttt{Tol\-e\-rant\-Sub\-scrib\-er\-Bridge\-Head} on the worker node in turn sends its input data to a 
\texttt{Tol\-e\-rant\-Pub\-lish\-er\-Bridge\-Head} on the data sink computer. 

Each of the three \texttt{Tol\-e\-rant\-Pub\-lish\-er\-Bridge\-Head} processes on the sink node has two subscribers attached. The first one is an instance 
of the \texttt{Tol\-e\-rance\-De\-tec\-tion\-Sub\-scrib\-er}
%Fault Tolerance Detection Subscriber
(section~\ref{Sec:ToleranceDetector}), controlling each of the three data streams for continuous operation and the second is one 
of three subscribers belonging to a \texttt{Tol\-e\-rant\-Event\-Gath\-erer} component (section~\ref{Sec:TolerantEventGatherer}). 
These gatherer-subscribers merge the three parts of the data stream 
into one stream again. Attached to the gatherers output publisher is one subscriber process checking for lost events in the data stream, an instance of the 
\texttt{Event\-Super\-visor} component (section~\ref{Sec:EventSupervisor}).

\subsubsection{Normal Setup Operation}

During normal operation the flow of data is basically the one outlined above. Data is published from a file, split up by the scatterer components, and distributed 
evenly to the three worker nodes, which send their data to the data sink node for data collection and merging into one data stream. This data stream is then checked 
for lost data.

\subsubsection{Node Failure Scenario}

If during normal operation of the setup described above one of the three worker nodes fails, the following sequence of events should take place. Due to the node's 
failure no more events arrive at the corresponding \texttt{Tol\-e\-rant\-Pub\-lish\-er\-Bridge\-Head} on the data sink node. This causes the timeout of the fault detection component attached to that 
\texttt{Tol\-e\-rant\-Pub\-lish\-er\-Bridge\-Head} to expire after the specified interval, resulting in a message being sent to one of the control programs. In this supervisor the status of all 
configured fault detection programs is now checked to determine which of the three data streams is broken. It subsequently sends messages to the scatterer and 
gatherer components, informing them about the broken link.
Now the scatterer marks the output publisher concerned as bad and checks for events that have been sent to that publisher's path and have not been received back. 
These events are presumed 
to be lost and are distributed evenly to the remaining output publishers. All new events arriving after this are also distributed to the remaining publishers.
The publisher associated with the broken path does not receive any new events until further notice.

No special action is taken by the Gatherer component upon receiving the notification other than marking the path concerned as broken. 
\texttt{Event\-Done} messages for events 
received from that path are now processed by just marking the event as already done, as the gatherer expects the scatterer to send these events again. Any new event 
received is first checked against the backlog of \texttt{Event\-Done} messages that have already been received as well as the list of events marked as done. 
An event is entered
 internally into this last list when event done data is received and it belongs 
to the broken path. If such an event is found, the \texttt{Event\-Done} message is sent back immediately and the event is removed from the internal tables. 
An event that cannot 
be found in these two tables is presumed to be a new event and is handled as usual.

After notifying the scatterer and gatherer components about the failure of the broken data stream, the first control program also informs the second supervisor program of 
the failure. This program now sends disconnect messages to the corresponding bridge head components on the data source and sink nodes and checks whether a spare node is available. 
If there is an available spare node it waits for the bridge heads on the sink and source nodes to be properly disconnected and then sends connect commands to them with the addresses 
of the corresponding bridge head components on the spare node. After it detects that this new connection has been properly established it sends a message to the 
scatterer and gatherer components 
to reintegrate the broken path. From this point on the system functions as before with the role of the broken node taken over by the spare. As soon as the functioning 
node is available again it can be reintegrated into the system as a new spare node.

\subsubsection{Fault Tolerance Test Results}

To test the fault-tolerance functionality of the system described above,  the test setup has been activated with communication between the computers being done via Fast 
Ethernet. When the data flow chain had been running for a time the network cable was unplugged from one of the worker nodes. This caused the corresponding
\texttt{Tol\-e\-rant\-Sub\-scrib\-er\-Bridge\-Head}, that was trying to send from the data source to that node, to block in the TCP code until the specified sending timeout expired. 
The \texttt{Tol\-e\-rant\-Pub\-lish\-er\-Bridge\-Head} on 
the data sink did not block while trying to send its accumulated event done messages back to the node. This was presumably because the messages were small enough that they could 
be placed in buffers of the kernel's network code or the network interface hardware.

After the timeout in the fault detector component for the broken node's data path expired, but before the TCP network timeout expired, the first control 
program was informed of the failure. It subsequently notified the gatherer and scatterer components on the data source and sink, causing lost events to be resent along the 
remaining two nodes. At the same time, the second control program was notified as well, which then sent disconnect commands to the appropriate bridge head components on the source 
and sink, and waited for them to become disconnected. Because of mutex semaphores regulating access to the communication classes, the disconnection only happened after 
the network send timeout expired. When the bridge heads had disconnected from their partners on the "broken" node, commands were sent to them to clear all 
events from their internal 
data structures and to reconnect them to the BridgeHeads on the spare node. As soon as the second control program had determined that the new connection was properly 
established, it sent commands to the scatterer and gatherer to reactivate the broken path and send new events again to all three data streams.

\begin{figure}
\begin{center}
\resizebox*{0.95\columnwidth}{!}{
\includegraphics{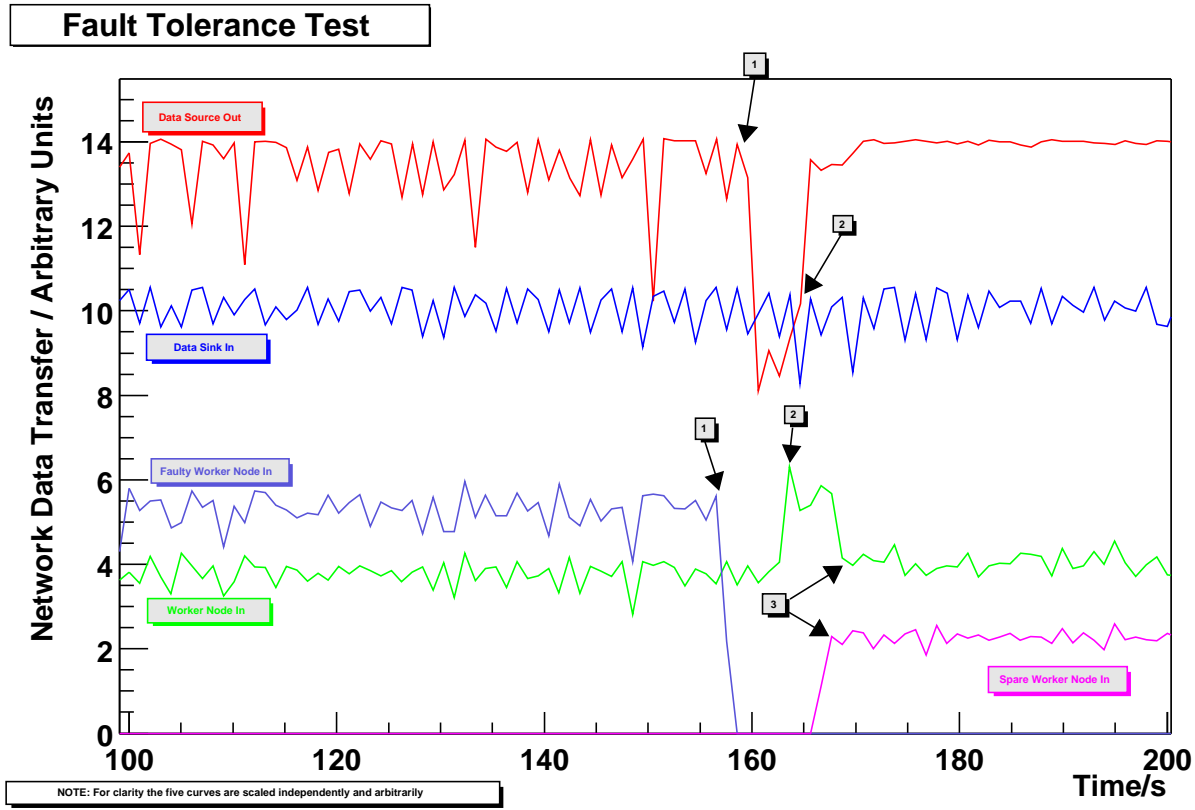}
}
\parbox{0.90\columnwidth}{
\caption[Results of the fault tolerance test.]{\label{Fig:FTTestResults}The results of the fault tolerance test. The curves are scaled arbitrarily and independently.}
}
\end{center}
\end{figure}

Fig.~\ref{Fig:FTTestResults} shows results of measurements that were made on the different nodes during the test. In the curves the amount of network traffic 
going in or coming out of the corresponding nodes is displayed. The measurements were made locally on each of the nodes. Note that the five curves shown are scaled independently 
to arbitrary values for a better visualization. Real values of the plateaus for the data source and sink node are between 11~MB/s and 12~MB/s and the four
other nodes' (two normal worker/one faulty worker/one spare worker) plateaus are at about 4~MB/s with the peak in the worker node's curve going to about 6~MB/s. This shows that the network 
load going out from the data source is evenly distributed to the three active nodes at first and after recovery as well as to the two remaining nodes 
during the recovery process.

At the points marked  {\em 1}  the cable is unplugged from the  {\em faulty}  node, causing its incoming network traffic to fall to zero immediately. At the same 
time or shortly afterwards the network traffic going out of the data source decreases to about two thirds as the \texttt{Tol\-e\-rant\-Sub\-scrib\-er\-Bridge\-Head} sending 
to the unplugged node blocks. 
The reason for the decrease to somewhat less than two thirds of the previous might be due to buffers filling up on the source as they do not get freed by the ``faulty''  
node. At point {\em 2} the faulty node has been taken out of the path, and events are passed only to the two remaining worker nodes. The amount of data leaving the data 
source increases to its previous value, and the amount of data going into the worker node increases by about a factor 1.5, as expected. Finally, at point {\em 3} the spare 
node has been connected into the chain and the third path has been activated again. Data starts to go into the spare node at the same rate as for the faulty node before 
the cable was unplugged, and the amount of data going into the regular worker node decreases back to the value before the simulated failure. At this point the data flow 
chain has fully recovered. During the running time of the test, including some time after the recovery, the event supervisor component has issued no warning about missing events,
so not a single event was lost due to this simulated node failure. 

\subsubsection{Fault Tolerance Test Summary}

The presented framework includes a number of components that make a data flow chain tolerant against faults in components of the chain, even against hardware 
faults of complete nodes. Failures that occur while spare nodes are available cause no further impact except for a short performance decrease until 
the spare node is activated. 
With no spares to activate, the system will continue to run with at most a performance decrease corresponding to the processing power loss due to the 
failure(s). Even in the case of multiple failures no events will be lost. 
If no output path is available the scatterer component stores events in a list. As soon as a path becomes available again, all events will
be sent via this path. Neither of these cases results in the loss of just one single event in the chain. Every event inserted at the 
beginning arrives at the end of the chain.

\subsection{System Tests Summary}

In the two system tests described above the framework has been demonstrated to be operational and usable in its current form. It has been shown
to handle data rates, including processing, within a factor of 2.4 of  the highest requirements for operation
in the ALICE High Level Trigger. Furthermore, the fault tolerance test has proven that the current concept to ensure fault tolerance
works and can in principle also be used in production systems already, despite its proof-of-concept status. 

\clearpage

%%%%%%%%%%%%%%%%%%%%%%%%%%%%%%%%%%%%%%%%%%%%%%%%%%%%%%%%%%%%%%%%%%%%%%%%%%%%%%%%%%%%%%%%%%%%%%%%%%%%%%%%%%%%%%%%%%%%%%%%%%%%%
%%%%%%%%%%%%%%%%%%%%%%%%%%%%%%%%%%%%%%%%%%%%%%%%%%%%%%%%%%%%%%%%%%%%%%%%%%%%%%%%%%%%%%%%%%%%%%%%%%%%%%%%%%%%%%%%%%%%%%%%%%%%%

\chapter{\label{Chap:Conclusion}Conclusion and Outlook}

%Framework constructed, data flow oriented applications, particularly High-Energy \& Nuclear Physics Triggers (high level), specifically
%ALICE HLT.
%
%Component principle allowed to quickly and easily adapt framework to new uses/tasks not foreseen by writing new components.
%
%Separation of communication classes from rest into library with abstract interface allows to not yet specify network technology
%and protocol, can be added by adding classes to library, without modifying framework components

%Minimal requirements fulfilled, operation thus ensured, optimization in places (pub sub interface, communication classes) still possible, 

%Next steps: Configuration creation, process starting and supervision, DCS, state machine, integration with global ALICE DCS

In this thesis a framework has been presented, that has been developed for data flow oriented applications with a particular 
emphasis on its use in trigger systems of high-energy and nuclear physics. Design and implementation of the framework
have been carried out for the data transport software to be used in the High Level Trigger of the future ALICE heavy-ion experiment. 
To allow flexible configurations the framework is composed of distinct components, that communicate via a defined interface
and can be combined in various configurations. Configuration changes are even possible during the runtime of a system. 

A first conclusion to be drawn from the framework's development and its use in a number of setups is, that the composition
into multiple independent modules has been proven to be highly functional and efficient. 
As requirements for specific tests have evolved the modularity has enabled to add functionality in new as well as in existing 
components and to provide proof-of-concept implementations and prototypes of new characteristics quite fast and easy. 
Furthermore it has allowed to vary the configurations of tests in a simple and rapid way and therefore to change test
setups and to introduce new ones easily. In the two system tests described in section~\ref{Sec:FrameworkSystemTests} it has been
demonstrated that the framework is able to operate in conditions closely approaching the ones expected for the operation
of the ALICE High Level Trigger. In addition, the current fault tolerance capability also has been shown to be
able to handle failures of complete nodes in a running system. The performance impact caused by such a failure
is only temporary provided that enough spare nodes are available, otherwise it is at most proportional to the amount
of processing power lost.

The separation of the network code into an individual communication class library has turned out
to be advantageous, too, since it has allowed to implement and test the communication related functionality of the framework 
without the need  to decide upon a network technology at the current stage. For the tests and developments
the currently widespread available and comparatively cheap Gigabit Ethernet TCP/IP solution could be used, 
however, at the obvious processing overhead. 
%which therefore also has shown its availability as a viable solution, even if only as a fallback.  
Once the decisions for a network
technology and protocol have been made, the appropriate classes have to be implemented only in the communication
class library used. 
%The split up of the
%communication classes into two types, optimized for small message respectively  large data block transfers, so far has not
%proved to be particularly advantageous when both transfers are performed using the same network technology. But as pointed
%out in section~\ref{Sec:TCPComClassesBenchmarkSummary} this might change when different communication mechanism are
%used or when the existing TCP based classes become more optimized for their specific tasks. 

Concerning the performance the framework already meets the requirements set by the conditions of the ALICE High Level Trigger
in the existing implementation and with the tested hardware. 
As the available CPU power does not yet reach the projected level a correspondingly, and quite considerably, larger
number of nodes and CPUs would be required to perform the necessary analysis steps of the HLT would it be built today. However, in principle it could be realized at the moment and
will be able to operate at the start of the LHC and ALICE. 
With the potential optimizations discussed in section~\ref{Sec:TCPComClassesBenchmarkSummary} and~\ref{Sec:PubSubFutureOptimization}
it should be possible to further enhance the performance and particularly the efficiency of the framework,
reducing the CPU power required for the operation at a given rate. 

In addition to these performance improvements a number of further tasks will be useful for a full working trigger system.
The first of these is a configuration program that provides a plain manner to graphically connect the functional components for a system.
%to generate a configuration that operates a specified processing chain on a cluster. 
%Ideally, just the actual worker components, data source, processing, and sink components should have to be specified together
%with the nodes on which to run them. 
The required framework components should be automatically inserted by this configuration program. 
A more pressing need exists for a process startup, control, and supervision system that can monitor and control
the components in a framework configuration. It also has to react to changes in their state by sending appropriate commands, 
effectively functioning as a 
Detector Control System (DCS). For this task the framework components have to be modified using the monitoring and control classes described
in section~\ref{Sec:SCClasses} so that they can react as finite state machines (FSM), shifting between states as a result of received
commands or other external stimuli. Supervising processes can monitor the states of a number of components, e.g. all components
on one node, and react to changes by sending commands. A summary status can be derived from the supervised components' states  and
is reported to a further supervisor component that controls multiple nodes. This component in turn can send commands to its subordinate
supervisors, which translate them into appropriate commands for the actual framework components. Furthermore, for the use in the ALICE HLT such a system requires 
an interface to the global ALICE DCS, translating and forwarding its commands and providing it with status information. A final item
needed for the framework is a good packaging and distribution mechanism that allows an easy installation of the framework by users 
not involved in its development. With these enhancements in place, the framework will provide a toolbox from which cluster applications, in particular
trigger related ones, can be constructed easily.

\begin{appendix}

%%%%%%%%%%%%%%%%%%%%%%%%%%%%%%%%%%%%%%%%%%%%%%%%%%%%%%%%%%%%%%%%%%%%%%%%%%%%%%%%%%%%%%%%%%%%%%%%%%%%%%%%%%%%%%%%%%%%%%%%%%%%%
%%%%%%%%%%%%%%%%%%%%%%%%%%%%%%%%%%%%%%%%%%%%%%%%%%%%%%%%%%%%%%%%%%%%%%%%%%%%%%%%%%%%%%%%%%%%%%%%%%%%%%%%%%%%%%%%%%%%%%%%%%%%%

%\chapter{\label{Chap:Installation}Installation}
%\section{Installing the Utility Classes}
%\section{Installing the Communication Classes}
%\section{Installing the Publisher Subscriber Framework}

%%%%%%%%%%%%%%%%%%%%%%%%%%%%%%%%%%%%%%%%%%%%%%%%%%%%%%%%%%%%%%%%%%%%%%%%%%%%%%%%%%%%%%%%%%%%%%%%%%%%%%%%%%%%%%%%%%%%%%%%%%%%%%
%%%%%%%%%%%%%%%%%%%%%%%%%%%%%%%%%%%%%%%%%%%%%%%%%%%%%%%%%%%%%%%%%%%%%%%%%%%%%%%%%%%%%%%%%%%%%%%%%%%%%%%%%%%%%%%%%%%%%%%%%%%%%%

%\chapter{\label{Chap:ComponentUsage}Component Usage}

%%%%%%%%%%%%%%%%%%%%%%%%%%%%%%%%%%%%%%%%%%%%%%%%%%%%%%%%%%%%%%%%%%%%%%%%%%%%%%%%%%%%%%%%%%%%%%%%%%%%%%%%%%%%%%%%%%%%%%%%%%%%%
%%%%%%%%%%%%%%%%%%%%%%%%%%%%%%%%%%%%%%%%%%%%%%%%%%%%%%%%%%%%%%%%%%%%%%%%%%%%%%%%%%%%%%%%%%%%%%%%%%%%%%%%%%%%%%%%%%%%%%%%%%%%%

\chapter{\label{Chap:BenchmarkStuff}Benchmark Supplement}

\section{Microbenchmark Programs}

\subsection{\label{Sec:LoggingOverheadProgram}Logging Overhead}
\begin{verbatim}

#include <sys/time.h>
#include <stdio.h>

#define COUNT 1000000000

void test_function1( int* a )
    {
    (*a)++;
    }

void test_function2( unsigned long flags, int* a )
    {
    if ( flags & 1 )
        (*a)++;
    }

unsigned long long calc_tdiff( struct timeval* start, struct timeval* stop )
    {
    unsigned long long tmp;
    tmp = (stop->tv_sec - start->tv_sec);
    tmp *= 1000000;
    tmp += (stop->tv_usec - start->tv_usec);
    return tmp;
    }

int main( int argc, char** argv )
    {
    struct timeval start, stop;
    unsigned long i; 
    int n;
    int *p = &n;
    unsigned long flags = 1;
    unsigned long long loopoverhead;
    unsigned long long loop_if;
    unsigned long long loop_iffunc;
    unsigned long long loop_func;
    unsigned long long loop_funcif;
    
    gettimeofday( &start, NULL );
    for ( i = 0; i < COUNT; i++ )
        {
        (*p)++;
        }
    gettimeofday( &stop, NULL );
    loopoverhead = calc_tdiff( &start, &stop );
    printf( "Loop overhead:          %Lu us\n", loopoverhead );

    gettimeofday( &start, NULL );
    for ( i = 0; i < COUNT; i++ )
        {
        if ( flags & 1 )
            (*p)++;
        }
    gettimeofday( &stop, NULL );
    loop_if = calc_tdiff( &start, &stop );
    printf( "Loop with if:           %Lu us\n", loop_if );

    gettimeofday( &start, NULL );
    for ( i = 0; i < COUNT; i++ )
        {
        test_function1( &n );
        }
    gettimeofday( &stop, NULL );
    loop_func = calc_tdiff( &start, &stop );
    printf( "Loop with func:         %Lu us\n", loop_func );
    
    gettimeofday( &start, NULL );
    for ( i = 0; i < COUNT; i++ )
        {
        if ( flags & 1 )
            test_function1( &n );
        }
    gettimeofday( &stop, NULL );
    loop_iffunc = calc_tdiff( &start, &stop );
    printf( "Loop with if and func:  %Lu us\n", loop_iffunc );

    gettimeofday( &start, NULL );
    for ( i = 0; i < COUNT; i++ )
        {
        test_function2( flags, &n );
        }
    gettimeofday( &stop, NULL );
    loop_funcif = calc_tdiff( &start, &stop );
    printf( "Loop with func with if: %Lu us\n", loop_funcif );

    return 0;
    }

\end{verbatim}

\section{\label{Sec:BenchmarkProcessList}Minimal Benchmark Process List}

\begin{verbatim}

  PID TTY      STAT   TIME COMMAND
    1 ?        S      0:05 init [3] 
    2 ?        SW     0:00 [keventd]
    3 ?        SWN    0:00 [ksoftirqd_CPU0]
    4 ?        SW     0:00 [kswapd]
    5 ?        SW     0:00 [bdflush]
    6 ?        SW     0:00 [kupdated]
   34 ?        SW     0:00 [kreiserfsd]
  229 ?        S      0:00 /sbin/syslogd
  233 ?        S      0:00 /sbin/klogd -c 1
  264 ?        SW     0:00 [khubd]
  508 tty1     S      0:00 login -- root     
  509 tty2     S      0:00 /sbin/mingetty tty2
  510 tty3     S      0:00 /sbin/mingetty tty3
  511 tty4     S      0:00 /sbin/mingetty tty4
  512 tty5     S      0:00 /sbin/mingetty tty5
  513 tty6     S      0:00 /sbin/mingetty tty6
  965 tty1     S      0:00 -bash
 1259 tty1     R      0:00 ps x

\end{verbatim}

%%%%%%%%%%%%%%%%%%%%%%%%%%%%%%%%%%%%%%%%%%%%%%%%%%%%%%%%%%%%%%%%%%%%%%%%%%%%%%%%%%%%%%%%%%%%%%%%%%%%%%%%%%%%%%%%%%%%%%%%%%%%%
%%%%%%%%%%%%%%%%%%%%%%%%%%%%%%%%%%%%%%%%%%%%%%%%%%%%%%%%%%%%%%%%%%%%%%%%%%%%%%%%%%%%%%%%%%%%%%%%%%%%%%%%%%%%%%%%%%%%%%%%%%%%%

\chapter{\label{Chap:BenchmarkTables}Benchmark Result Tables}

The following tables were generated automatically. Layout and number formats may therefore not be optimal.
Errors are given as standard deviations where present.
%\clearpage

\section{Micro-Benchmarks}
\subsection{\label{Sec:CacheMeasurementsTables}Cache and Memory Reference Tests}

\begin{table}[h!t!p!b]
\begin{center}
{\scriptsize
\begin{tabular}{|l| c| c| c| c| c| c| c| c| }
\hline
 Stride  /  &  Access Time   &  Access Time   &  Access Time   &  Access Time   &  Access Time   &  Access Time   &  Access Time   &  Access Time  \\
 B   &  16B /  &  32B /  &  64B /  &  128B /  &  256B /  &  512B /  &  1kB /  &  2kB / \\
   &  ns   &  ns   &  ns   &  ns   &  ns   &  ns   &  ns   &  ns  \\
\hline \hline
         4  & 3.119937  & 4.284117  & 4.107164  & 4.149105  & 4.125885  & 4.121354  & 4.116948  & 4.112793  \\
\hline
         8  & 2.095478  & 3.119937  & 4.293430  & 4.111821  & 4.153762  & 4.135198  & 4.116697  & 4.116948  \\
\hline
        16  & 1.415612  & 2.114104  & 3.119937  & 4.284117  & 4.107164  & 4.153762  & 4.130542  & 4.116697  \\
\hline
        32  & -  & 1.396985  & 2.086165  & 3.119937  & 4.274803  & 4.107164  & 4.158419  & 4.125885  \\
\hline
        64  & -  & -  & 1.396985  & 2.095478  & 3.119937  & 4.284117  & 4.107164  & 4.153762  \\
\hline
       128  & -  & -  & -  & 1.396985  & 2.095478  & 3.115280  & 4.274803  & 4.107164  \\
\hline
       256  & -  & -  & -  & -  & 1.406298  & 2.114104  & 3.119937  & 4.274803  \\
\hline
       512  & -  & -  & -  & -  & -  & 1.396985  & 2.095478  & 3.119937  \\
\hline
      1024  & -  & -  & -  & -  & -  & -  & 1.415612  & 2.104791  \\
\hline
      2048  & -  & -  & -  & -  & -  & -  & -  & 1.396985  \\
\hline
\end{tabular}
}
\end{center}
\caption [733~MHz cache and memory test access times from 16~B to 2~kB.] {\label{Tab:Cache733MHz1 }Cache and memory test access times for the 733~MHz reference PC from 16~B to 2~kB block sizes. The columns show the different block sizes.  }
\end{table}

\begin{table}[h!t!p!b]
\begin{center}
{\scriptsize
\begin{tabular}{|l| c| c| c| c| c| c| c| c| }
\hline
 Stride  /  &  Access Time   &  Access Time   &  Access Time   &  Access Time   &  Access Time   &  Access Time   &  Access Time   &  Access Time  \\
 B   &  4kB /  &  8kB /  &  16kB /  &  32kB /  &  64kB /  &  128kB /  &  256kB /  &  512kB / \\
   &  ns   &  ns   &  ns   &  ns   &  ns   &  ns   &  ns   &  ns  \\
\hline \hline
         4  & 4.123116  & 4.120470  & 4.119836  & 4.127914  & 4.125392  & 4.120289  & 13.103088  & 21.351235  \\
\hline
         8  & 4.117451  & 4.109139  & 4.106486  & 4.227147  & 4.226086  & 4.223951  & 16.996937  & 28.737954  \\
\hline
        16  & 4.112291  & 4.117451  & 4.109139  & 5.919679  & 6.121432  & 6.067987  & 23.380403  & 57.773931  \\
\hline
        32  & 4.121354  & 4.112291  & 4.103478  & 13.675857  & 13.689239  & 13.697381  & 39.531339  & 103.797231  \\
\hline
        64  & 4.135198  & 4.121354  & 4.093662  & 13.650541  & 13.675857  & 13.689239  & 39.471520  & 103.155772  \\
\hline
\end{tabular}
}
\end{center}
\caption [733~MHz cache and memory test access times from 4~kB to 512~kB.] {\label{Tab:Cache733MHz2 }Cache and memory test access times for the 733~MHz reference PC from 4~kB to 512~kB block sizes. The columns show the different block sizes.  }
\end{table}

\begin{table}[h!t!p!b]
\begin{center}
{\scriptsize
\begin{tabular}{|l| c| c| c| c| c| c| }
\hline
 Stride  /  &  Access Time   &  Access Time   &  Access Time   &  Access Time   &  Access Time   &  Access Time  \\
 B   &  1MB /  &  2MB /  &  4MB /  &  8MB /  &  16MB /  &  32MB / \\
   &  ns   &  ns   &  ns   &  ns   &  ns   &  ns  \\
\hline \hline
         4  & 22.190313  & 22.165477  & 22.156164  & 22.095628  & 22.114255  & 22.129389  \\
\hline
         8  & 29.951334  & 29.976169  & 29.914081  & 29.765069  & 29.755756  & 29.758085  \\
\hline
        16  & 67.502260  & 68.073471  & 68.470836  & 68.414956  & 68.144873  & 68.014488  \\
\hline
        32  & 125.269095  & 127.355258  & 128.522515  & 132.583082  & 134.017318  & 134.753063  \\
\hline
        64  & 124.573708  & 126.659870  & 127.851963  & 131.726265  & 133.253634  & 133.961439  \\
\hline
\end{tabular}
}
\end{center}
\caption [733~MHz cache and memory test access times from 1~MB to 32~MB.] {\label{Tab:Cache733MHz3 }Cache and memory test access times for the 733~MHz reference PC from 1~MB to 32~MB block sizes. The columns show the different block sizes.  }
\end{table}

\begin{table}[h!t!p!b]
\begin{center}
{\scriptsize
\begin{tabular}{|l| c| c| c| c| c| c| c| c| }
\hline
 Stride  /  &  Access Time   &  Access Time   &  Access Time   &  Access Time   &  Access Time   &  Access Time   &  Access Time   &  Access Time  \\
 B   &  16B /  &  32B /  &  64B /  &  128B /  &  256B /  &  512B /  &  1kB /  &  2kB / \\
   &  ns   &  ns   &  ns   &  ns   &  ns   &  ns   &  ns   &  ns  \\
\hline \hline
         4  & 2.863822  & 3.930211  & 3.767229  & 3.799854  & 3.790599  & 3.776744  & 3.772317  & 3.772778  \\
\hline
         8  & 1.969749  & 2.859166  & 3.911585  & 3.767229  & 3.809167  & 3.781285  & 3.776744  & 3.772317  \\
\hline
        16  & 1.294540  & 1.969749  & 2.863822  & 3.911585  & 3.767229  & 3.799854  & 3.790599  & 3.776744  \\
\hline
        32  & -  & 1.294540  & 1.979062  & 2.854509  & 3.930211  & 3.767229  & 3.809167  & 3.790599  \\
\hline
        64  & -  & -  & 1.275913  & 1.974406  & 2.859166  & 3.911585  & 3.767229  & 3.809167  \\
\hline
       128  & -  & -  & -  & 1.294540  & 1.965093  & 2.854509  & 3.930211  & 3.767229  \\
\hline
       256  & -  & -  & -  & -  & 1.275913  & 1.974406  & 2.859166  & 3.930211  \\
\hline
       512  & -  & -  & -  & -  & -  & 1.294540  & 1.974406  & 2.859166  \\
\hline
      1024  & -  & -  & -  & -  & -  & -  & 1.275913  & 1.974406  \\
\hline
      2048  & -  & -  & -  & -  & -  & -  & -  & 1.294540  \\
\hline
\end{tabular}
}
\end{center}
\caption [800~MHz cache and memory test access times from 16~B to 2~kB.] {\label{Tab:Cache800MHz1 }Cache and memory test access times for the 800~MHz reference PC from 16~B to 2~kB block sizes. The columns show the different block sizes.  }
\end{table}

\begin{table}[h!t!p!b]
\begin{center}
{\scriptsize
\begin{tabular}{|l| c| c| c| c| c| c| c| c| }
\hline
 Stride  /  &  Access Time   &  Access Time   &  Access Time   &  Access Time   &  Access Time   &  Access Time   &  Access Time   &  Access Time  \\
 B   &  4kB /  &  8kB /  &  16kB /  &  32kB /  &  64kB /  &  128kB /  &  256kB /  &  512kB / \\
   &  ns   &  ns   &  ns   &  ns   &  ns   &  ns   &  ns   &  ns  \\
\hline \hline
         4  & 3.783017  & 3.770882  & 3.783903  & 3.781973  & 3.782783  & 3.774961  & 11.146069  & 18.232635  \\
\hline
         8  & 3.763462  & 3.759722  & 3.766221  & 3.867887  & 3.875471  & 3.871955  & 12.411225  & 23.032938  \\
\hline
        16  & 3.772317  & 3.763462  & 3.769040  & 4.707776  & 4.703051  & 4.707598  & 19.400351  & 44.405460  \\
\hline
        32  & 3.776744  & 3.772317  & 3.763462  & 6.261545  & 6.269267  & 6.275401  & 32.205430  & 82.680157  \\
\hline
        64  & 3.781285  & 3.776744  & 3.758345  & 6.246043  & 6.261545  & 6.269267  & 32.252214  & 82.333883  \\
\hline
\end{tabular}
}
\end{center}
\caption [800~MHz cache and memory test access times from 4~kB to 512~kB.] {\label{Tab:Cache800MHz2 }Cache and memory test access times for the 800~MHz reference PC from 4~kB to 512~kB block sizes. The columns show the different block sizes.  }
\end{table}

\begin{table}[h!t!p!b]
\begin{center}
{\scriptsize
\begin{tabular}{|l| c| c| c| c| c| c| }
\hline
 Stride  /  &  Access Time   &  Access Time   &  Access Time   &  Access Time   &  Access Time   &  Access Time  \\
 B   &  1MB /  &  2MB /  &  4MB /  &  8MB /  &  16MB /  &  32MB / \\
   &  ns   &  ns   &  ns   &  ns   &  ns   &  ns  \\
\hline \hline
         4  & 18.291175  & 18.272549  & 18.328428  & 18.845312  & 19.182917  & 19.352883  \\
\hline
         8  & 24.363399  & 24.947027  & 26.803464  & 28.768554  & 29.928051  & 30.521769  \\
\hline
        16  & 53.147475  & 54.488579  & 55.581331  & 58.412552  & 60.498714  & 61.551109  \\
\hline
        32  & 104.606152  & 109.076500  & 113.770366  & 121.332705  & 126.417726  & 128.941610  \\
\hline
        64  & 104.052680  & 108.480453  & 113.149484  & 120.475888  & 125.393271  & 127.870589  \\
\hline
\end{tabular}
}
\end{center}
\caption [800~MHz cache and memory test access times from 1~MB to 32~MB.] {\label{Tab:Cache800MHz3 }Cache and memory test access times for the 800~MHz reference PC from 1~MB to 32~MB block sizes. The columns show the different block sizes.  }
\end{table}

\begin{table}[h!t!p!b]
\begin{center}
{\scriptsize
\begin{tabular}{|l| c| c| c| c| c| c| c| c| }
\hline
 Stride  /  &  Access Time   &  Access Time   &  Access Time   &  Access Time   &  Access Time   &  Access Time   &  Access Time   &  Access Time  \\
 B   &  16B /  &  32B /  &  64B /  &  128B /  &  256B /  &  512B /  &  1kB /  &  2kB / \\
   &  ns   &  ns   &  ns   &  ns   &  ns   &  ns   &  ns   &  ns  \\
\hline \hline
         4  & 2.454040  & 3.362087  & 3.222401  & 3.264335  & 3.241101  & 3.236543  & 3.232084  & 3.232479  \\
\hline
         8  & 1.699665  & 2.454040  & 3.366744  & 3.231714  & 3.264335  & 3.245758  & 3.231887  & 3.232084  \\
\hline
        16  & 1.103618  & 1.699665  & 2.449383  & 3.362087  & 3.231714  & 3.259679  & 3.245758  & 3.236543  \\
\hline
        32  & -  & 1.108274  & 1.685695  & 2.444726  & 3.362087  & 3.222401  & 3.264335  & 3.241101  \\
\hline
        64  & -  & -  & 1.108274  & 1.695009  & 2.444726  & 3.357431  & 3.231714  & 3.264335  \\
\hline
       128  & -  & -  & -  & 1.103618  & 1.685695  & 2.454040  & 3.362087  & 3.231714  \\
\hline
       256  & -  & -  & -  & -  & 1.098961  & 1.699665  & 2.444726  & 3.362087  \\
\hline
       512  & -  & -  & -  & -  & -  & 1.103618  & 1.690352  & 2.444726  \\
\hline
      1024  & -  & -  & -  & -  & -  & -  & 1.103618  & 1.690352  \\
\hline
      2048  & -  & -  & -  & -  & -  & -  & -  & 1.098961  \\
\hline
\end{tabular}
}
\end{center}
\caption [933~MHz cache and memory test access times from 16~B to 2~kB.] {\label{Tab:Cache933MHz1 }Cache and memory test access times for the 933~MHz reference PC from 16~B to 2~kB block sizes. The columns show the different block sizes.  }
\end{table}

\begin{table}[h!t!p!b]
\begin{center}
{\scriptsize
\begin{tabular}{|l| c| c| c| c| c| c| c| c| }
\hline
 Stride  /  &  Access Time   &  Access Time   &  Access Time   &  Access Time   &  Access Time   &  Access Time   &  Access Time   &  Access Time  \\
 B   &  4kB /  &  8kB /  &  16kB /  &  32kB /  &  64kB /  &  128kB /  &  256kB /  &  512kB / \\
   &  ns   &  ns   &  ns   &  ns   &  ns   &  ns   &  ns   &  ns  \\
\hline \hline
         4  & 3.233268  & 3.239510  & 3.242679  & 3.239688  & 3.238363  & 3.240411  & 10.579824  & 18.083623  \\
\hline
         8  & 3.227821  & 3.223950  & 3.225526  & 3.317331  & 3.314486  & 3.318148  & 9.700175  & 24.608203  \\
\hline
        16  & 3.232084  & 3.227821  & 3.223950  & 4.036569  & 4.035853  & 4.039091  & 13.765835  & 44.916357  \\
\hline
        32  & 3.231887  & 3.232084  & 3.223163  & 5.367039  & 5.374323  & 5.379582  & 20.095280  & 83.616802  \\
\hline
        64  & 3.245758  & 3.241200  & 3.213455  & 5.351755  & 5.367039  & 5.374323  & 20.181100  & 83.525976  \\
\hline
\end{tabular}
}
\end{center}
\caption [933~MHz cache and memory test access times from 4~kB to 512~kB.] {\label{Tab:Cache933MHz2 }Cache and memory test access times for the 933~MHz reference PC from 4~kB to 512~kB block sizes. The columns show the different block sizes.  }
\end{table}

\begin{table}[h!t!p!b]
\begin{center}
{\scriptsize
\begin{tabular}{|l| c| c| c| c| c| c| }
\hline
 Stride  /  &  Access Time   &  Access Time   &  Access Time   &  Access Time   &  Access Time   &  Access Time  \\
 B   &  1MB /  &  2MB /  &  4MB /  &  8MB /  &  16MB /  &  32MB / \\
   &  ns   &  ns   &  ns   &  ns   &  ns   &  ns  \\
\hline \hline
         4  & 18.166999  & 18.086284  & 18.095598  & 18.291175  & 18.675346  & 18.971041  \\
\hline
         8  & 26.648243  & 26.524067  & 26.300550  & 27.352944  & 29.071234  & 30.100346  \\
\hline
        16  & 53.842862  & 54.885944  & 54.948032  & 57.648867  & 60.880557  & 63.255429  \\
\hline
        32  & 103.414059  & 105.748574  & 106.245279  & 112.839043  & 120.010227  & 125.002116  \\
\hline
        64  & 103.030886  & 105.500221  & 105.996927  & 112.354755  & 119.358301  & 124.201179  \\
\hline
\end{tabular}
}
\end{center}
\caption [933~MHz cache and memory test access times from 1~MB to 32~MB.] {\label{Tab:Cache933MHz3 }Cache and memory test access times for the 933~MHz reference PC from 1~MB to 32~MB block sizes. The columns show the different block sizes.  }
\end{table}

\clearpage

\section{\label{Sec:NetworkReferenceTestTables}Network Reference Tests}

\subsection{TCP Network Reference Throughput}
\subsubsection{\label{Sec:NetRefPlateauTestTables}Plateau Determination}

\begin{table}[h!t!p!b]
\begin{center}
{\scriptsize
\begin{tabular}{|l| c| c| c| c| }
\hline
 Block Count   &  Send Rate   &  Send Rate   &  Send Rate   &  Send Rate  \\
   &  FE, NODELAY /  &  FE, DELAY /  &  GbE, NODELAY /  &  GbE, DELAY / \\
   &  Hz   &  Hz   &  Hz   &  Hz  \\
\hline \hline
         1  & 34762.024000  & 36024.591000  & 33636.267000  & 35230.800000  \\
  & $\pm$ 1171.626428  & $\pm$ 1379.099518  & $\pm$ 1608.692948  & $\pm$ 836.383433  \\
\hline
         2  & 56049.666000  & 64545.192000  & 57497.668000  & 64732.369000  \\
  & $\pm$ 1246.266239  & $\pm$ 3641.027099  & $\pm$ 1227.956107  & $\pm$ 3753.324588  \\
\hline
         4  & 101077.648000  & 121682.192000  & 92851.441000  & 117885.076000  \\
\hline
\end{tabular}
}
\end{center}
\caption {\label{Tab:TCPRefPlateauRates }TCP reference measurement plateau determination. }
\end{table}

\clearpage
\subsubsection{Plateau Throughput Measurement}

\begin{table}[h!t!p!b]
\begin{center}
{\scriptsize
\begin{tabular}{|l| c| c| c| c| }
\hline
 Block Size  /  &  Send Rate   &  Send Rate   &  Send Rate   &  Send Rate  \\
 B   &  FE, NODELAY /  &  FE, DELAY /  &  GbE, NODELAY /  &  GbE, DELAY / \\
   &  Hz   &  Hz   &  Hz   &  Hz  \\
\hline \hline
         8  & 642587.764937  & 646578.653660  & 684377.610693  & 687405.107604  \\
  & $\pm$ 7860.841550  & $\pm$ 5965.302372  & $\pm$ 7122.679994  & $\pm$ 5121.024119  \\
\hline
        16  & 540255.800438  & 599165.742889  & 606062.147356  & 623164.961234  \\
  & $\pm$ 107688.853272  & $\pm$ 4129.651092  & $\pm$ 11968.191890  & $\pm$ 6217.346900  \\
\hline
        32  & 355438.647634  & 366263.323020  & 536517.926126  & 553240.164277  \\
\hline
\end{tabular}
}
\end{center}
\caption [Maximum reference sending rate (plateau).] {\label{Tab:TCPRefPlateauThroughputRate }Maximum reference sending rate (block count 512~k). }
\end{table}

\begin{table}[h!t!p!b]
\begin{center}
{\scriptsize
\begin{tabular}{|l| c| c| c| c| }
\hline
 Block Size  /  &  Network Throughput   &  Network Throughput   &  Network Throughput   &  Network Throughput  \\
 B   &  FE, NODELAY /  &  FE, DELAY /  &  GbE, NODELAY /  &  GbE, DELAY / \\
   &  $\mathrm{B}/\mathrm s$   &  $\mathrm{B}/\mathrm s$   &  $\mathrm{B}/\mathrm s$   &  $\mathrm{B}/\mathrm s$  \\
\hline \hline
         8  & 5140702.119496  & 5172629.229280  & 5475020.885544  & 5499240.860832  \\
  & $\pm$ 62886.732400  & $\pm$ 47722.418976  & $\pm$ 56981.439952  & $\pm$ 40968.192952  \\
\hline
        16  & 8644092.807008  & 9586651.886224  & 9696994.357696  & 9970639.379744  \\
  & $\pm$ 1723021.652352  & $\pm$ 66074.417472  & $\pm$ 191491.070240  & $\pm$ 99477.550400  \\
\hline
        32  & 11374036.724288  & 11720426.336640  & 17168573.636032  & 17703685.256864  \\
\hline
\end{tabular}
}
\end{center}
\caption [Reference network throughput (plateau).] {\label{Tab:TCPRefPlateauThroughputThroughput }Reference network throughput (block count 512~k). }
\end{table}

\begin{table}[h!t!p!b]
\begin{center}
{\scriptsize
\begin{tabular}{|l| c| c| c| c| }
\hline
 Block Size  /  &  CPU Usage   &  CPU Usage   &  CPU Usage   &  CPU Usage  \\
 B   &  FE, NODELAY /  &  FE, DELAY /  &  GbE, NODELAY /  &  GbE, DELAY / \\
   &  \%   &  \%   &  \%   &  \%  \\
\hline \hline
         8  & 93.600000  & 91.000000  & 97.000000  & 97.200000  \\
  & $\pm$ 2.154066  & $\pm$ 1.612452  & $\pm$ 1.341640  & $\pm$ 1.326650  \\
\hline
        16  & 94.400000  & 93.200000  & 96.200000  & 97.200000  \\
  & $\pm$ 4.543126  & $\pm$ 0.979796  & $\pm$ 2.088062  & $\pm$ 1.326650  \\
\hline
        32  & 61.800000  & 61.800000  & 97.400000  & 101.200000  \\
\hline
\end{tabular}
}
\end{center}
\caption [Reference sender CPU usage (plateau).] {\label{Tab:TCPRefPlateauThroughputUsage }CPU usage on the sender during reference transmission (block count 512~k). The nodes are twin CPU nodes, 100~\% CPU usage corresponds to one CPU being fully used. }
\end{table}

\begin{table}[h!t!p!b]
\begin{center}
{\scriptsize
\begin{tabular}{|l| c| c| c| c| }
\hline
 Block Size  /  &  CPU Usage   &  CPU Usage   &  CPU Usage   &  CPU Usage  \\
 B   &  FE, NODELAY /  &  FE, DELAY /  &  GbE, NODELAY /  &  GbE, DELAY / \\
   &  \%   &  \%   &  \%   &  \%  \\
\hline \hline
         8  & 97.584911  & 97.972927  & 100.962679  & 101.611486  \\
  & $\pm$ 5.452952  & $\pm$ 1.244379  & $\pm$ 0.776517  & $\pm$ 1.579818  \\
\hline
        16  & 102.316781  & 105.639370  & 101.767603  & 96.024573  \\
  & $\pm$ 37.730304  & $\pm$ 0.745266  & $\pm$ 8.206925  & $\pm$ 5.253103  \\
\hline
        32  & 66.124514  & 65.880738  & 99.890350  & 101.211413  \\
\hline
\end{tabular}
}
\end{center}
\caption [Reference receiver CPU usage (plateau).] {\label{Tab:TCPRefPlateauThroughputUsageServer }CPU usage on the receiver during reference transmission (block count 512~k). The nodes are twin CPU nodes, 100~\% CPU usage corresponds to one CPU being fully used. }
\end{table}

\begin{table}[h!t!p!b]
\begin{center}
{\scriptsize
\begin{tabular}{|l| c| c| c| c| }
\hline
 Block Size  /  &  CPU Usage / Rate   &  CPU Usage / Rate   &  CPU Usage / Rate   &  CPU Usage / Rate  \\
 B   &  FE, NODELAY /  &  FE, DELAY /  &  GbE, NODELAY /  &  GbE, DELAY / \\
   &  $\% \times \mathrm s$   &  $\% \times \mathrm s$   &  $\% \times \mathrm s$   &  $\% \times \mathrm s$  \\
\hline \hline
         8  & 0.000146  & 0.000141  & 0.000142  & 0.000141  \\
  & $\pm$ 0.000005  & $\pm$ 0.000004  & $\pm$ 0.000003  & $\pm$ 0.000003  \\
\hline
        16  & 0.000175  & 0.000156  & 0.000159  & 0.000156  \\
  & $\pm$ 0.000043  & $\pm$ 0.000003  & $\pm$ 0.000007  & $\pm$ 0.000004  \\
\hline
        32  & 0.000174  & 0.000169  & 0.000182  & 0.000183  \\
\hline
\end{tabular}
}
\end{center}
\caption [Reference sender CPU usage divided by rate (plateau).] {\label{Tab:TCPRefPlateauThroughputUsagePerRate }CPU usage on the sender divided by the sending rate during reference transmission (block count 512~k). The nodes are twin CPU nodes, 100~\% CPU usage corresponds to one CPU being fully used. }
\end{table}

\begin{table}[h!t!p!b]
\begin{center}
{\scriptsize
\begin{tabular}{|l| c| c| c| c| }
\hline
 Block Size  /  &  CPU Usage / Rate   &  CPU Usage / Rate   &  CPU Usage / Rate   &  CPU Usage / Rate  \\
 B   &  FE, NODELAY /  &  FE, DELAY /  &  GbE, NODELAY /  &  GbE, DELAY / \\
   &  $\% \times \mathrm s$   &  $\% \times \mathrm s$   &  $\% \times \mathrm s$   &  $\% \times \mathrm s$  \\
\hline \hline
         8  & 0.000152  & 0.000152  & 0.000148  & 0.000148  \\
  & $\pm$ 0.000010  & $\pm$ 0.000003  & $\pm$ 0.000003  & $\pm$ 0.000003  \\
\hline
        16  & 0.000189  & 0.000176  & 0.000168  & 0.000154  \\
  & $\pm$ 0.000108  & $\pm$ 0.000002  & $\pm$ 0.000017  & $\pm$ 0.000010  \\
\hline
        32  & 0.000186  & 0.000180  & 0.000186  & 0.000183  \\
\hline
\end{tabular}
}
\end{center}
\caption [Reference receiver CPU usage divided by rate (plateau).] {\label{Tab:TCPRefPlateauThroughputUsagePerRateServer }CPU usage on the receiver divided by the sending rate during reference transmission (block count 512~k). The nodes are twin CPU nodes, 100~\% CPU usage corresponds to one CPU being fully used. }
\end{table}

\begin{table}[h!t!p!b]
\begin{center}
{\scriptsize
\begin{tabular}{|l| c| c| c| c| }
\hline
 Block Size  /  &  CPU Usage / Throughput   &  CPU Usage / Throughput   &  CPU Usage / Throughput   &  CPU Usage / Throughput  \\
 B   &  FE, NODELAY /  &  FE, DELAY /  &  GbE, NODELAY /  &  GbE, DELAY / \\
   &  $\% / (\mathrm{MB} / \mathrm s)$   &  $\% / (\mathrm{MB} / \mathrm s)$   &  $\% / (\mathrm{MB} / \mathrm s)$   &  $\% / (\mathrm{MB} / \mathrm s)$  \\
\hline \hline
         8  & 19.092082  & 18.447177  & 18.577439  & 18.533755  \\
  & $\pm$ 0.672930  & $\pm$ 0.497064  & $\pm$ 0.450297  & $\pm$ 0.391032  \\
\hline
        16  & 11.451240  & 10.194099  & 10.402503  & 10.222171  \\
  & $\pm$ 2.833674  & $\pm$ 0.177429  & $\pm$ 0.431214  & $\pm$ 0.241506  \\
\hline
        32  & 5.697361  & 5.528979  & 5.948735  & 5.994000  \\
\hline
\end{tabular}
}
\end{center}
\caption [Reference sender CPU usage divided by throughput (plateau).] {\label{Tab:TCPRefPlateauThroughputUsagePerThroughput }CPU usage on the sender divided by the network throughput during reference transmission (block count 512~k). The nodes are twin CPU nodes, 100~\% CPU usage corresponds to one CPU being fully used. }
\end{table}

\begin{table}[h!t!p!b]
\begin{center}
{\scriptsize
\begin{tabular}{|l| c| c| c| c| }
\hline
 Block Size  /  &  CPU Usage / Throughput   &  CPU Usage / Throughput   &  CPU Usage / Throughput   &  CPU Usage / Throughput  \\
 B   &  FE, NODELAY /  &  FE, DELAY /  &  GbE, NODELAY /  &  GbE, DELAY / \\
   &  $\% / (\mathrm{MB} / \mathrm s)$   &  $\% / (\mathrm{MB} / \mathrm s)$   &  $\% / (\mathrm{MB} / \mathrm s)$   &  $\% / (\mathrm{MB} / \mathrm s)$  \\
\hline \hline
         8  & 19.904905  & 19.860703  & 19.336372  & 19.374922  \\
  & $\pm$ 1.355764  & $\pm$ 0.435491  & $\pm$ 0.349963  & $\pm$ 0.445572  \\
\hline
        16  & 12.411589  & 11.554702  & 11.004551  & 10.098556  \\
  & $\pm$ 7.050887  & $\pm$ 0.161155  & $\pm$ 1.104761  & $\pm$ 0.653203  \\
\hline
        32  & 6.096039  & 5.894065  & 6.100834  & 5.994676  \\
\hline
\end{tabular}
}
\end{center}
\caption [Reference receiver CPU usage divided by throughput (plateau).] {\label{Tab:TCPRefPlateauThroughputUsagePerThroughputServer }CPU usage on the receiver divided by the network throughput during reference transmission (block count 512~k). The nodes are twin CPU nodes, 100~\% CPU usage corresponds to one CPU being fully used. }
\end{table}

\clearpage
\subsubsection{Peak Throughput Measurement}

\begin{table}[h!t!p!b]
\begin{center}
{\scriptsize
\begin{tabular}{|l| c| c| c| c| }
\hline
 Block Size  /  &  Send Rate   &  Send Rate   &  Send Rate   &  Send Rate  \\
 B   &  FE, NODELAY /  &  FE, DELAY /  &  GbE, NODELAY /  &  GbE, DELAY / \\
   &  Hz   &  Hz   &  Hz   &  Hz  \\
\hline \hline
         8  & 541915.324000  & 696167.471000  & 626658.268000  & 723372.680000  \\
  & $\pm$ 35008.367322  & $\pm$ 14870.187772  & $\pm$ 9065.183291  & $\pm$ 15520.764632  \\
\hline
        16  & 522420.247000  & 653932.195000  & 608307.026000  & 665579.900000  \\
  & $\pm$ 33832.129742  & $\pm$ 4171.639445  & $\pm$ 11849.314835  & $\pm$ 15945.381829  \\
\hline
        32  & 461676.648000  & 624047.986000  & 592984.818000  & 627173.696000  \\
\hline
\end{tabular}
}
\end{center}
\caption [Maximum reference sending rate (peak).] {\label{Tab:TCPRefPeakThroughputRate }Maximum reference sending rate (block counts 4~k, 4~k, 128, and 16~k). }
\end{table}

\begin{table}[h!t!p!b]
\begin{center}
{\scriptsize
\begin{tabular}{|l| c| c| c| c| }
\hline
 Block Size  /  &  Network Throughput   &  Network Throughput   &  Network Throughput   &  Network Throughput  \\
 B   &  FE, NODELAY /  &  FE, DELAY /  &  GbE, NODELAY /  &  GbE, DELAY / \\
   &  $\mathrm{B}/\mathrm s$   &  $\mathrm{B}/\mathrm s$   &  $\mathrm{B}/\mathrm s$   &  $\mathrm{B}/\mathrm s$  \\
\hline \hline
         8  & 4335322.592000  & 5569339.768000  & 5013266.144000  & 5786981.440000  \\
  & $\pm$ 280066.938576  & $\pm$ 118961.502176  & $\pm$ 72521.466328  & $\pm$ 124166.117056  \\
\hline
        16  & 8358723.952000  & 10462915.120000  & 9732912.416000  & 10649278.400000  \\
  & $\pm$ 541314.075872  & $\pm$ 66746.231120  & $\pm$ 189589.037360  & $\pm$ 255126.109264  \\
\hline
        32  & 14773652.736000  & 19969535.552000  & 18975514.176000  & 20069558.272000  \\
\hline
\end{tabular}
}
\end{center}
\caption [Reference network throughput (peak).] {\label{Tab:TCPRefPeakThroughputThroughput }Reference network throughput (block counts 4~k, 4~k, 128, and 16~k). }
\end{table}

\begin{table}[h!t!p!b]
\begin{center}
{\scriptsize
\begin{tabular}{|l| c| c| c| c| }
\hline
 Block Size  /  &  CPU Usage   &  CPU Usage   &  CPU Usage   &  CPU Usage  \\
 B   &  FE, NODELAY /  &  FE, DELAY /  &  GbE, NODELAY /  &  GbE, DELAY / \\
   &  \%   &  \%   &  \%   &  \%  \\
\hline \hline
         8  & 90.400000  & 87.200000  & 87.800000  & 94.600000  \\
  & $\pm$ 5.200000  & $\pm$ 7.704544  & $\pm$ 6.095900  & $\pm$ 6.514598  \\
\hline
        16  & 91.600000  & 92.200000  & 87.400000  & 95.000000  \\
  & $\pm$ 0.800000  & $\pm$ 6.539114  & $\pm$ 0.916516  & $\pm$ 4.753946  \\
\hline
        32  & 85.600000  & 82.800000  & 91.400000  & 95.000000  \\
\hline
\end{tabular}
}
\end{center}
\caption [Reference sender CPU usage (peak).] {\label{Tab:TCPRefPeakThroughputUsage }CPU usage on the sender during reference transmission (block counts 4~k, 4~k, 128, and 16~k). The nodes are twin CPU nodes, 100~\% CPU usage corresponds to one CPU being fully used. }
\end{table}

\begin{table}[h!t!p!b]
\begin{center}
{\scriptsize
\begin{tabular}{|l| c| c| c| c| }
\hline
 Block Size  /  &  CPU Usage   &  CPU Usage   &  CPU Usage   &  CPU Usage  \\
 B   &  FE, NODELAY /  &  FE, DELAY /  &  GbE, NODELAY /  &  GbE, DELAY / \\
   &  \%   &  \%   &  \%   &  \%  \\
\hline \hline
       512  & -  & -  & -  & 81.601356  \\
  &  &  &  & $\pm$ 1.689649  \\
\hline
      1024  & 28.312258  & 27.861712  & -  & 101.990701  \\
  & $\pm$ 0.179137  & $\pm$ 1.578916  &  & $\pm$ 4.337756  \\
\hline
      2048  & 15.584871  & 15.636630  & -  & 109.437158  \\
\hline
\end{tabular}
}
\end{center}
\caption [Reference receiver CPU usage (peak).] {\label{Tab:TCPRefPeakThroughputUsageServer }CPU usage on the receiver during reference transmission (block counts 4~k, 4~k, 128, and 16~k). The nodes are twin CPU nodes, 100~\% CPU usage corresponds to one CPU being fully used. }
\end{table}

\begin{table}[h!t!p!b]
\begin{center}
{\scriptsize
\begin{tabular}{|l| c| c| c| c| }
\hline
 Block Size  /  &  CPU Usage / Rate   &  CPU Usage / Rate   &  CPU Usage / Rate   &  CPU Usage / Rate  \\
 B   &  FE, NODELAY /  &  FE, DELAY /  &  GbE, NODELAY /  &  GbE, DELAY / \\
   &  $\% \times \mathrm s$   &  $\% \times \mathrm s$   &  $\% \times \mathrm s$   &  $\% \times \mathrm s$  \\
\hline \hline
         8  & 0.000167  & 0.000125  & 0.000140  & 0.000131  \\
  & $\pm$ 0.000020  & $\pm$ 0.000014  & $\pm$ 0.000012  & $\pm$ 0.000012  \\
\hline
        16  & 0.000175  & 0.000141  & 0.000144  & 0.000143  \\
  & $\pm$ 0.000013  & $\pm$ 0.000011  & $\pm$ 0.000004  & $\pm$ 0.000011  \\
\hline
        32  & 0.000185  & 0.000133  & 0.000154  & 0.000151  \\
\hline
\end{tabular}
}
\end{center}
\caption [Reference sender CPU usage divided by rate (peak).] {\label{Tab:TCPRefPeakThroughputUsagePerRate }CPU usage on the sender divided by the sending rate during reference transmission (block counts 4~k, 4~k, 128, and 16~k). The nodes are twin CPU nodes, 100~\% CPU usage corresponds to one CPU being fully used. }
\end{table}

\begin{table}[h!t!p!b]
\begin{center}
{\scriptsize
\begin{tabular}{|l| c| c| c| c| }
\hline
 Block Size  /  &  CPU Usage / Rate   &  CPU Usage / Rate   &  CPU Usage / Rate   &  CPU Usage / Rate  \\
 B   &  FE, NODELAY /  &  FE, DELAY /  &  GbE, NODELAY /  &  GbE, DELAY / \\
   &  $\% \times \mathrm s$   &  $\% \times \mathrm s$   &  $\% \times \mathrm s$   &  $\% \times \mathrm s$  \\
\hline \hline
       512  & -  & -  & -  & 0.000506  \\
  &  &  &  & $\pm$ 0.000013  \\
\hline
      1024  & 0.002418  & 0.002381  & -  & 0.001108  \\
  & $\pm$ 0.000017  & $\pm$ 0.000140  &  & $\pm$ 0.000049  \\
\hline
      2048  & 0.002689  & 0.002694  & -  & 0.002272  \\
\hline
\end{tabular}
}
\end{center}
\caption [Reference receiver CPU usage divided by rate (peak).] {\label{Tab:TCPRefPeakThroughputUsagePerRateServer }CPU usage on the receiver divided by the sending rate during reference transmission (block counts 4~k, 4~k, 128, and 16~k). The nodes are twin CPU nodes, 100~\% CPU usage corresponds to one CPU being fully used. }
\end{table}

\begin{table}[h!t!p!b]
\begin{center}
{\scriptsize
\begin{tabular}{|l| c| c| c| c| }
\hline
 Block Size  /  &  CPU Usage / Throughput   &  CPU Usage / Throughput   &  CPU Usage / Throughput   &  CPU Usage / Throughput  \\
 B   &  FE, NODELAY /  &  FE, DELAY /  &  GbE, NODELAY /  &  GbE, DELAY / \\
   &  $\% / (\mathrm{MB} / \mathrm s)$   &  $\% / (\mathrm{MB} / \mathrm s)$   &  $\% / (\mathrm{MB} / \mathrm s)$   &  $\% / (\mathrm{MB} / \mathrm s)$  \\
\hline \hline
         8  & 21.864870  & 16.417715  & 18.364271  & 17.141109  \\
  & $\pm$ 2.670212  & $\pm$ 1.801270  & $\pm$ 1.540676  & $\pm$ 1.548198  \\
\hline
        16  & 11.490936  & 9.240131  & 9.416045  & 9.354129  \\
  & $\pm$ 0.844514  & $\pm$ 0.714285  & $\pm$ 0.282157  & $\pm$ 0.692193  \\
\hline
        32  & 6.075553  & 4.347727  & 5.050711  & 4.963474  \\
\hline
\end{tabular}
}
\end{center}
\caption [Reference sender CPU usage divided by throughput (peak).] {\label{Tab:TCPRefPeakThroughputUsagePerThroughput }CPU usage on the sender divided by the network throughput during reference transmission (block counts 4~k, 4~k, 128, and 16~k). The nodes are twin CPU nodes, 100~\% CPU usage corresponds to one CPU being fully used. }
\end{table}

\begin{table}[h!t!p!b]
\begin{center}
{\scriptsize
\begin{tabular}{|l| c| c| c| c| }
\hline
 Block Size  /  &  CPU Usage / Throughput   &  CPU Usage / Throughput   &  CPU Usage / Throughput   &  CPU Usage / Throughput  \\
 B   &  FE, NODELAY /  &  FE, DELAY /  &  GbE, NODELAY /  &  GbE, DELAY / \\
   &  $\% / (\mathrm{MB} / \mathrm s)$   &  $\% / (\mathrm{MB} / \mathrm s)$   &  $\% / (\mathrm{MB} / \mathrm s)$   &  $\% / (\mathrm{MB} / \mathrm s)$  \\
\hline \hline
       512  & -  & -  & -  & 1.036229  \\
  &  &  &  & $\pm$ 0.026840  \\
\hline
      1024  & 2.475530  & 2.437862  & -  & 1.134133  \\
  & $\pm$ 0.017473  & $\pm$ 0.143825  &  & $\pm$ 0.050518  \\
\hline
      2048  & 1.376598  & 1.379464  & -  & 1.163072  \\
\hline
\end{tabular}
}
\end{center}
\caption [Reference receiver CPU usage divided by throughput (peak).] {\label{Tab:TCPRefPeakThroughputUsagePerThroughputServer }CPU usage on the receiver divided by the network throughput during reference transmission (block counts 4~k, 4~k, 128, and 16~k). The nodes are twin CPU nodes, 100~\% CPU usage corresponds to one CPU being fully used. }
\end{table}

\clearpage

\subsection{TCP Network Reference Latency}

\begin{table}[h!t!p!b]
\begin{center}
{\scriptsize
\begin{tabular}{|l| c| c| c| c| }
\hline
 Block Count   &  Latency   &  Latency   &  Latency   &  Latency  \\
   &  FE, NODELAY /  &  FE, DELAY /  &  GbE, NODELAY /  &  GbE, DELAY / \\
   &  $\mu \mathrm s$   &  $\mu \mathrm s$   &  $\mu \mathrm s$   &  $\mu \mathrm s$  \\
\hline \hline
         1  & 101.450000  & 104.250000  & 127.400000  & 135.150000  \\
  & $\pm$ 1.588238  & $\pm$ 2.795085  & $\pm$ 1.445683  & $\pm$ 5.258564  \\
\hline
         2  & 82.875000  & 84.175000  & 115.500000  & 118.350000  \\
  & $\pm$ 0.550568  & $\pm$ 1.491853  & $\pm$ 2.258318  & $\pm$ 5.703289  \\
\hline
         8  & 65.481300  & 65.843600  & 95.887400  & 96.331400  \\
  & $\pm$ 0.794003  & $\pm$ 0.723955  & $\pm$ 1.250443  & $\pm$ 1.589698  \\
\hline
        32  & 60.659400  & 61.229900  & 90.182900  & 90.710900  \\
  & $\pm$ 0.693309  & $\pm$ 0.767954  & $\pm$ 0.397236  & $\pm$ 0.475232  \\
\hline
       128  & 59.650400  & 59.847200  & 88.748600  & 88.963600  \\
  & $\pm$ 0.671058  & $\pm$ 0.544253  & $\pm$ 0.141511  & $\pm$ 0.951703  \\
\hline
       512  & 60.527300  & 60.549000  & 88.080100  & 88.195400  \\
  & $\pm$ 0.860857  & $\pm$ 0.812871  & $\pm$ 0.105079  & $\pm$ 0.146712  \\
\hline
      2048  & 60.849700  & 60.958000  & 87.815600  & 87.884100  \\
  & $\pm$ 0.293932  & $\pm$ 0.252560  & $\pm$ 0.106643  & $\pm$ 0.092248  \\
\hline
      8192  & 60.775000  & 60.796500  & 87.779000  & 87.798300  \\
  & $\pm$ 0.267876  & $\pm$ 0.181381  & $\pm$ 0.061992  & $\pm$ 0.056462  \\
\hline
     32768  & 60.764300  & 60.865900  & 87.743200  & 87.768800  \\
  & $\pm$ 0.161018  & $\pm$ 0.060476  & $\pm$ 0.054468  & $\pm$ 0.035165  \\
\hline
    131072  & 60.715100  & 60.803500  & 87.791700  & 87.765400  \\
  & $\pm$ 0.058996  & $\pm$ 0.071744  & $\pm$ 0.101963  & $\pm$ 0.041962  \\
\hline
    524288  & 60.703600  & 60.729100  & 87.719200  & 87.800300  \\
  & $\pm$ 0.062303  & $\pm$ 0.049068  & $\pm$ 0.066490  & $\pm$ 0.079381  \\
\hline
\end{tabular}
}
\end{center}
\caption [Average network reference sending latency.] {\label{Tab:TCPRefLatency }Average network reference sending latency. }
\end{table}

\clearpage

\section{Communication Class Benchmarks}

\subsection{TCP Message Class Throughput}
\subsubsection{\label{Sec:NetworkMsgPlateauDeterminationTables}Plateau Determination}

\begin{table}[h!t!p!b]
\begin{center}
{\scriptsize
\begin{tabular}{|l| c| c| c| c| }
\hline
 Message Count   &  Send Rate   &  Send Rate   &  Send Rate   &  Send Rate  \\
   &  FE, No Connect /  &  FE, With Connect /  &  GbE, No Connect /  &  GbE, With Connect / \\
   &  Hz   &  Hz   &  Hz   &  Hz  \\
\hline \hline
         1  & 2283.367000  & 15298.840000  & 1889.928000  & 17081.710000  \\
  & $\pm$ 139.709282  & $\pm$ 1016.872883  & $\pm$ 212.282220  & $\pm$ 548.797062  \\
\hline
         2  & 2841.808000  & 25556.770000  & 2399.285000  & 28170.090000  \\
  & $\pm$ 91.582126  & $\pm$ 593.480682  & $\pm$ 64.967467  & $\pm$ 1044.034633  \\
\hline
         4  & 3225.942000  & 47914.240000  & 2701.397000  & 50610.360000  \\
\hline
\end{tabular}
}
\end{center}
\caption {\label{Tab:TCPMsgPlateauRates }TCP message measurement plateau determination. }
\end{table}

\clearpage
\subsubsection{Plateau Throughput Measurement}

\begin{table}[h!t!p!b]
\begin{center}
{\scriptsize
\begin{tabular}{|l| c| c| c| c| }
\hline
 Message Size  /  &  Send Rate   &  Send Rate   &  Send Rate   &  Send Rate  \\
 B   &  FE, No Connect /  &  FE, With Connect /  &  GbE, No Connect /  &  GbE, With Connect / \\
   &  Hz   &  Hz   &  Hz   &  Hz  \\
\hline \hline
        32  & 482.731600  & 59540.300000  & 482.603500  & 59483.970000  \\
  & $\pm$ 4.758178  & $\pm$ 467.242511  & $\pm$ 4.548058  & $\pm$ 444.569252  \\
\hline
        64  & 481.701100  & 56041.100000  & 481.711300  & 57976.970000  \\
  & $\pm$ 5.518463  & $\pm$ 491.973491  & $\pm$ 5.296564  & $\pm$ 299.133318  \\
\hline
       128  & 482.843100  & 50947.740000  & 482.794900  & 52432.320000  \\
\hline
\end{tabular}
}
\end{center}
\caption [Maximum message sending rate (plateau).] {\label{Tab:TCPMsgPlateauThroughputRates }Maximum message sending rate (message count 256~k). }
\end{table}

\begin{table}[h!t!p!b]
\begin{center}
{\scriptsize
\begin{tabular}{|l| c| c| c| c| }
\hline
 Message Size  /  &  Network Throughput   &  Network Throughput   &  Network Throughput   &  Network Throughput  \\
 B   &  FE, No Connect /  &  FE, With Connect /  &  GbE, No Connect /  &  GbE, With Connect / \\
   &  $\mathrm{B}/\mathrm s$   &  $\mathrm{B}/\mathrm s$   &  $\mathrm{B}/\mathrm s$   &  $\mathrm{B}/\mathrm s$  \\
\hline \hline
        32  & 15447.411200  & 1905289.600000  & 15443.312000  & 1903487.040000  \\
  & $\pm$ 152.261696  & $\pm$ 14951.760352  & $\pm$ 145.537856  & $\pm$ 14226.216064  \\
\hline
        64  & 30828.870400  & 3586630.400000  & 30829.523200  & 3710526.080000  \\
  & $\pm$ 353.181632  & $\pm$ 31486.303424  & $\pm$ 338.980096  & $\pm$ 19144.532352  \\
\hline
       128  & 61803.916800  & 6521310.720000  & 61797.747200  & 6711336.960000  \\
\hline
\end{tabular}
}
\end{center}
\caption [Message sending network throughput (plateau).] {\label{Tab:TCPMsgPlateauThroughputThroughput }Network throughput during message sending (message count 256~k). }
\end{table}

\begin{table}[h!t!p!b]
\begin{center}
{\scriptsize
\begin{tabular}{|l| c| c| c| c| }
\hline
 Message Size  /  &  CPU Usage   &  CPU Usage   &  CPU Usage   &  CPU Usage  \\
 B   &  FE, No Connect /  &  FE, With Connect /  &  GbE, No Connect /  &  GbE, With Connect / \\
   &  \%   &  \%   &  \%   &  \%  \\
\hline \hline
        32  & 89.818182  & 25.800000  & 92.000000  & 26.000000  \\
  & $\pm$ 25.873180  & $\pm$ 0.600000  & $\pm$ 0.000000  & $\pm$ 0.000000  \\
\hline
        64  & 97.800000  & 27.400000  & 92.000000  & 28.200000  \\
  & $\pm$ 0.600000  & $\pm$ 0.916516  & $\pm$ 0.000000  & $\pm$ 0.600000  \\
\hline
       128  & 98.000000  & 29.600000  & 92.000000  & 32.200000  \\
\hline
\end{tabular}
}
\end{center}
\caption [Message sending sender CPU usage (plateau).] {\label{Tab:TCPMsgPlateauThroughputUsage }CPU usage on the sender during message sending (message count 256~k). The nodes are twin CPU nodes, 100~\% CPU usage corresponds to one CPU being fully used. }
\end{table}

\begin{table}[h!t!p!b]
\begin{center}
{\scriptsize
\begin{tabular}{|l| c| c| c| c| }
\hline
 Message Size  /  &  CPU Usage   &  CPU Usage   &  CPU Usage   &  CPU Usage  \\
 B   &  FE, No Connect /  &  FE, With Connect /  &  GbE, No Connect /  &  GbE, With Connect / \\
   &  \%   &  \%   &  \%   &  \%  \\
\hline \hline
        32  & 8.073643  & 152.713391  & 14.951497  & 177.083368  \\
  & $\pm$ 0.160694  & $\pm$ 3.380133  & $\pm$ 33.345408  & $\pm$ 212.136454  \\
\hline
        64  & 8.073484  & 151.196134  & 6.660626  & 153.972530  \\
  & $\pm$ 0.186001  & $\pm$ 3.134093  & $\pm$ 1.072097  & $\pm$ 2.408296  \\
\hline
       128  & 9.486549  & 148.620184  & 6.862606  & 146.936067  \\
\hline
\end{tabular}
}
\end{center}
\caption [Message sending receiver CPU usage (plateau).] {\label{Tab:TCPMsgPlateauThroughputUsageServer }CPU usage on the receiver during message sending (message count 256~k). The nodes are twin CPU nodes, 100~\% CPU usage corresponds to one CPU being fully used. }
\end{table}

\begin{table}[h!t!p!b]
\begin{center}
{\scriptsize
\begin{tabular}{|l| c| c| c| c| }
\hline
 Message Size  /  &  CPU Usage / Rate   &  CPU Usage / Rate   &  CPU Usage / Rate   &  CPU Usage / Rate  \\
 B   &  FE, No Connect /  &  FE, With Connect /  &  GbE, No Connect /  &  GbE, With Connect / \\
   &  $\% \times \mathrm s$   &  $\% \times \mathrm s$   &  $\% \times \mathrm s$   &  $\% \times \mathrm s$  \\
\hline \hline
        32  & 0.186062  & 0.000433  & 0.190633  & 0.000437  \\
  & $\pm$ 0.055431  & $\pm$ 0.000013  & $\pm$ 0.001797  & $\pm$ 0.000003  \\
\hline
        64  & 0.203030  & 0.000489  & 0.190986  & 0.000486  \\
  & $\pm$ 0.003572  & $\pm$ 0.000021  & $\pm$ 0.002100  & $\pm$ 0.000013  \\
\hline
       128  & 0.202964  & 0.000581  & 0.190557  & 0.000614  \\
\hline
\end{tabular}
}
\end{center}
\caption [Message sending sender CPU usage divided by rate (plateau).] {\label{Tab:TCPMsgPlateauThroughputUsagePerRate }CPU usage on the sender divided by the sending rate during message sending (message count 256~k). The nodes are twin CPU nodes, 100~\% CPU usage corresponds to one CPU being fully used. }
\end{table}

\begin{table}[h!t!p!b]
\begin{center}
{\scriptsize
\begin{tabular}{|l| c| c| c| c| }
\hline
 Message Size  /  &  CPU Usage / Rate   &  CPU Usage / Rate   &  CPU Usage / Rate   &  CPU Usage / Rate  \\
 B   &  FE, No Connect /  &  FE, With Connect /  &  GbE, No Connect /  &  GbE, With Connect / \\
   &  $\% \times \mathrm s$   &  $\% \times \mathrm s$   &  $\% \times \mathrm s$   &  $\% \times \mathrm s$  \\
\hline \hline
        32  & 0.016725  & 0.002565  & 0.030981  & 0.002977  \\
  & $\pm$ 0.000498  & $\pm$ 0.000077  & $\pm$ 0.069387  & $\pm$ 0.003589  \\
\hline
        64  & 0.016760  & 0.002698  & 0.013827  & 0.002656  \\
  & $\pm$ 0.000578  & $\pm$ 0.000080  & $\pm$ 0.002378  & $\pm$ 0.000055  \\
\hline
       128  & 0.019647  & 0.002917  & 0.014214  & 0.002802  \\
\hline
\end{tabular}
}
\end{center}
\caption [Message sending receiver CPU usage divided by rate (plateau).] {\label{Tab:TCPMsgPlateauThroughputUsagePerRateServer }CPU usage on the receiver divided by the sending rate during message sending (message count 256~k). The nodes are twin CPU nodes, 100~\% CPU usage corresponds to one CPU being fully used. }
\end{table}

\begin{table}[h!t!p!b]
\begin{center}
{\scriptsize
\begin{tabular}{|l| c| c| c| c| }
\hline
 Message Size  /  &  CPU Usage / Throughput   &  CPU Usage / Throughput   &  CPU Usage / Throughput   &  CPU Usage / Throughput  \\
 B   &  FE, No Connect /  &  FE, With Connect /  &  GbE, No Connect /  &  GbE, With Connect / \\
   &  $\% / (\mathrm{MB} / \mathrm s)$   &  $\% / (\mathrm{MB} / \mathrm s)$   &  $\% / (\mathrm{MB} / \mathrm s)$   &  $\% / (\mathrm{MB} / \mathrm s)$  \\
\hline \hline
        32  & 6096.808444  & 14.199026  & 6246.605106  & 14.322646  \\
  & $\pm$ 1816.265084  & $\pm$ 0.441636  & $\pm$ 58.954244  & $\pm$ 0.107043  \\
\hline
        64  & 3326.417469  & 8.010579  & 3129.145267  & 7.969179  \\
  & $\pm$ 58.535515  & $\pm$ 0.338274  & $\pm$ 34.376855  & $\pm$ 0.210675  \\
\hline
       128  & 1662.679629  & 4.759449  & 1561.041826  & 5.030912  \\
\hline
\end{tabular}
}
\end{center}
\caption [Message sending sender CPU usage divided by throughput (plateau).] {\label{Tab:TCPMsgPlateauThroughputUsagePerThroughput }CPU usage on the sender divided by the network throughput during message sending (message count 256~k). The nodes are twin CPU nodes, 100~\% CPU usage corresponds to one CPU being fully used. }
\end{table}

\begin{table}[h!t!p!b]
\begin{center}
{\scriptsize
\begin{tabular}{|l| c| c| c| c| }
\hline
 Message Size  /  &  CPU Usage / Throughput   &  CPU Usage / Throughput   &  CPU Usage / Throughput   &  CPU Usage / Throughput  \\
 B   &  FE, No Connect /  &  FE, With Connect /  &  GbE, No Connect /  &  GbE, With Connect / \\
   &  $\% / (\mathrm{MB} / \mathrm s)$   &  $\% / (\mathrm{MB} / \mathrm s)$   &  $\% / (\mathrm{MB} / \mathrm s)$   &  $\% / (\mathrm{MB} / \mathrm s)$  \\
\hline \hline
        32  & 548.034415  & 84.045793  & 1015.174973  & 97.550094  \\
  & $\pm$ 16.301859  & $\pm$ 2.519800  & $\pm$ 2273.663588  & $\pm$ 117.588880  \\
\hline
        64  & 274.598959  & 44.203231  & 226.544199  & 43.511867  \\
  & $\pm$ 9.473856  & $\pm$ 1.304329  & $\pm$ 38.953463  & $\pm$ 0.905077  \\
\hline
       128  & 160.949916  & 23.896967  & 116.443641  & 22.957218  \\
\hline
\end{tabular}
}
\end{center}
\caption [Message sending receiver CPU usage divided by throughput (plateau).] {\label{Tab:TCPMsgPlateauThroughputUsagePerThroughputServer }CPU usage on the receiver divided by the network throughput during message sending (message count 256~k). The nodes are twin CPU nodes, 100~\% CPU usage corresponds to one CPU being fully used. }
\end{table}

\clearpage
\subsubsection{Peak Throughput Measurement}

\begin{table}[h!t!p!b]
\begin{center}
{\scriptsize
\begin{tabular}{|l| c| c| c| c| }
\hline
 Message Size  /  &  Send Rate   &  Send Rate   &  Send Rate   &  Send Rate  \\
 B   &  FE, No Connect /  &  FE, With Connect /  &  GbE, No Connect /  &  GbE, With Connect / \\
   &  Hz   &  Hz   &  Hz   &  Hz  \\
\hline \hline
        32  & 4032.240000  & 219126.300000  & 3405.114000  & 221919.700000  \\
  & $\pm$ 47.627298  & $\pm$ 6737.940131  & $\pm$ 5.959390  & $\pm$ 7884.281033  \\
\hline
        64  & 4101.872000  & 215002.000000  & 3406.505000  & 232095.100000  \\
  & $\pm$ 119.112664  & $\pm$ 19130.135650  & $\pm$ 16.928688  & $\pm$ 4684.250238  \\
\hline
       128  & 4179.501000  & 95122.690000  & 3395.496000  & 111265.600000  \\
\hline
\end{tabular}
}
\end{center}
\caption [Maximum message sending rate (peak).] {\label{Tab:TCPMsgPeakThroughput }Maximum message sending rate (message count 2~k). }
\end{table}

\begin{table}[h!t!p!b]
\begin{center}
{\scriptsize
\begin{tabular}{|l| c| c| c| c| }
\hline
 Message Size  /  &  Network Throughput   &  Network Throughput   &  Network Throughput   &  Network Throughput  \\
 B   &  FE, No Connect /  &  FE, With Connect /  &  GbE, No Connect /  &  GbE, With Connect / \\
   &  $\mathrm{B}/\mathrm s$   &  $\mathrm{B}/\mathrm s$   &  $\mathrm{B}/\mathrm s$   &  $\mathrm{B}/\mathrm s$  \\
\hline \hline
        32  & 129031.680000  & 7012041.600000  & 108963.648000  & 7101430.400000  \\
  & $\pm$ 1524.073536  & $\pm$ 215614.084192  & $\pm$ 190.700480  & $\pm$ 252296.993056  \\
\hline
        64  & 262519.808000  & 13760128.000000  & 218016.320000  & 14854086.400000  \\
  & $\pm$ 7623.210496  & $\pm$ 1224328.681600  & $\pm$ 1083.436032  & $\pm$ 299792.015232  \\
\hline
       128  & 534976.128000  & 12175704.320000  & 434623.488000  & 14241996.800000  \\
\hline
\end{tabular}
}
\end{center}
\caption [Message sending network throughput (peak).] {\label{Tab:TCPMsgPeakThroughputThroughput }Network throughput during message sending (message count 2~k). }
\end{table}

\begin{table}[h!t!p!b]
\begin{center}
{\scriptsize
\begin{tabular}{|l| c| c| c| c| }
\hline
 Message Size  /  &  CPU Usage   &  CPU Usage   &  CPU Usage   &  CPU Usage  \\
 B   &  FE, No Connect /  &  FE, With Connect /  &  GbE, No Connect /  &  GbE, With Connect / \\
   &  \%   &  \%   &  \%   &  \%  \\
\hline \hline
        32  & 85.400000  & 91.800000  & 61.000000  & 91.000000  \\
  & $\pm$ 0.916516  & $\pm$ 1.400000  & $\pm$ 1.000000  & $\pm$ 4.404544  \\
\hline
        64  & 86.400000  & 90.400000  & 60.600000  & 91.200000  \\
  & $\pm$ 2.653300  & $\pm$ 10.947146  & $\pm$ 1.280624  & $\pm$ 6.523802  \\
\hline
       128  & 85.200000  & 73.400000  & 60.600000  & 76.400000  \\
\hline
\end{tabular}
}
\end{center}
\caption [Message sending sender CPU usage (peak).] {\label{Tab:TCPMsgPeakThroughputUsage }CPU usage on the sender during message sending (message count 2~k). The nodes are twin CPU nodes, 100~\% CPU usage corresponds to one CPU being fully used. }
\end{table}

\begin{table}[h!t!p!b]
\begin{center}
{\scriptsize
\begin{tabular}{|l| c| c| c| c| }
\hline
 Message Size  /  &  CPU Usage   &  CPU Usage   &  CPU Usage   &  CPU Usage  \\
 B   &  FE, No Connect /  &  FE, With Connect /  &  GbE, No Connect /  &  GbE, With Connect / \\
   &  \%   &  \%   &  \%   &  \%  \\
\hline \hline
        32  & 74.697741  & -  & 51.252759  & -  \\
  & $\pm$ 10.768214  &  & $\pm$ 0.220740  &  \\
\hline
        64  & 79.450134  & -  & 50.817927  & -  \\
  & $\pm$ 2.694902  &  & $\pm$ 1.323226  &  \\
\hline
       128  & 80.622564  & -  & 51.216815  & -  \\
\hline
\end{tabular}
}
\end{center}
\caption [Message sending receiver CPU usage (peak).] {\label{Tab:TCPMsgPeakThroughputUsageServer }CPU usage on the receiver during message sending (message count 2~k). The nodes are twin CPU nodes, 100~\% CPU usage corresponds to one CPU being fully used. }
\end{table}

\begin{table}[h!t!p!b]
\begin{center}
{\scriptsize
\begin{tabular}{|l| c| c| c| c| }
\hline
 Message Size  /  &  CPU Usage / Rate   &  CPU Usage / Rate   &  CPU Usage / Rate   &  CPU Usage / Rate  \\
 B   &  FE, No Connect /  &  FE, With Connect /  &  GbE, No Connect /  &  GbE, With Connect / \\
   &  $\% \times \mathrm s$   &  $\% \times \mathrm s$   &  $\% \times \mathrm s$   &  $\% \times \mathrm s$  \\
\hline \hline
        32  & 0.021179  & 0.000419  & 0.017914  & 0.000410  \\
  & $\pm$ 0.000477  & $\pm$ 0.000019  & $\pm$ 0.000325  & $\pm$ 0.000034  \\
\hline
        64  & 0.021064  & 0.000420  & 0.017789  & 0.000393  \\
  & $\pm$ 0.001259  & $\pm$ 0.000088  & $\pm$ 0.000464  & $\pm$ 0.000036  \\
\hline
       128  & 0.020385  & 0.000772  & 0.017847  & 0.000687  \\
\hline
\end{tabular}
}
\end{center}
\caption [Message sending sender CPU usage divided by rate (peak).] {\label{Tab:TCPMsgPeakThroughputUsagePerRate }CPU usage on the sender divided by the sending rate during message sending (message count 2~k). The nodes are twin CPU nodes, 100~\% CPU usage corresponds to one CPU being fully used. }
\end{table}

\begin{table}[h!t!p!b]
\begin{center}
{\scriptsize
\begin{tabular}{|l| c| c| c| c| }
\hline
 Message Size  /  &  CPU Usage / Rate   &  CPU Usage / Rate   &  CPU Usage / Rate   &  CPU Usage / Rate  \\
 B   &  FE, No Connect /  &  FE, With Connect /  &  GbE, No Connect /  &  GbE, With Connect / \\
   &  $\% \times \mathrm s$   &  $\% \times \mathrm s$   &  $\% \times \mathrm s$   &  $\% \times \mathrm s$  \\
\hline \hline
        32  & 0.018525  & -  & 0.015052  & -  \\
  & $\pm$ 0.002889  &  & $\pm$ 0.000091  &  \\
\hline
        64  & 0.019369  & -  & 0.014918  & -  \\
  & $\pm$ 0.001219  &  & $\pm$ 0.000463  &  \\
\hline
       128  & 0.019290  & -  & 0.015084  & -  \\
\hline
\end{tabular}
}
\end{center}
\caption [Message sending receiver CPU usage divided by rate (peak).] {\label{Tab:TCPMsgPeakThroughputUsagePerRateServer }CPU usage on the receiver divided by the sending rate during message sending (message count 2~k). The nodes are twin CPU nodes, 100~\% CPU usage corresponds to one CPU being fully used. }
\end{table}

\begin{table}[h!t!p!b]
\begin{center}
{\scriptsize
\begin{tabular}{|l| c| c| c| c| }
\hline
 Message Size  /  &  CPU Usage / Throughput   &  CPU Usage / Throughput   &  CPU Usage / Throughput   &  CPU Usage / Throughput  \\
 B   &  FE, No Connect /  &  FE, With Connect /  &  GbE, No Connect /  &  GbE, With Connect / \\
   &  $\% / (\mathrm{MB} / \mathrm s)$   &  $\% / (\mathrm{MB} / \mathrm s)$   &  $\% / (\mathrm{MB} / \mathrm s)$   &  $\% / (\mathrm{MB} / \mathrm s)$  \\
\hline \hline
        32  & 694.004258  & 13.727710  & 587.012587  & 13.436788  \\
  & $\pm$ 15.642760  & $\pm$ 0.631471  & $\pm$ 10.651260  & $\pm$ 1.127739  \\
\hline
        64  & 345.105808  & 6.888836  & 291.462459  & 6.437968  \\
  & $\pm$ 20.619350  & $\pm$ 1.447161  & $\pm$ 7.607385  & $\pm$ 0.590461  \\
\hline
       128  & 166.995627  & 6.321234  & 146.204121  & 5.624998  \\
\hline
\end{tabular}
}
\end{center}
\caption [Message sending sender CPU usage divided by throughput (peak).] {\label{Tab:TCPMsgPeakThroughputUsagePerThroughput }CPU usage on the sender divided by the network throughput during message sending (message count 2~k). The nodes are twin CPU nodes, 100~\% CPU usage corresponds to one CPU being fully used. }
\end{table}

\begin{table}[h!t!p!b]
\begin{center}
{\scriptsize
\begin{tabular}{|l| c| c| c| c| }
\hline
 Message Size  /  &  CPU Usage / Throughput   &  CPU Usage / Throughput   &  CPU Usage / Throughput   &  CPU Usage / Throughput  \\
 B   &  FE, No Connect /  &  FE, With Connect /  &  GbE, No Connect /  &  GbE, With Connect / \\
   &  $\% / (\mathrm{MB} / \mathrm s)$   &  $\% / (\mathrm{MB} / \mathrm s)$   &  $\% / (\mathrm{MB} / \mathrm s)$   &  $\% / (\mathrm{MB} / \mathrm s)$  \\
\hline \hline
        32  & 607.032205  & -  & 493.213355  & -  \\
  & $\pm$ 94.675767  &  & $\pm$ 2.988037  &  \\
\hline
        64  & 317.346096  & -  & 244.414488  & -  \\
  & $\pm$ 19.979422  &  & $\pm$ 7.578535  &  \\
\hline
       128  & 158.023658  & -  & 123.566162  & -  \\
\hline
\end{tabular}
}
\end{center}
\caption [Message sending receiver CPU usage divided by throughput (peak).] {\label{Tab:TCPMsgPeakThroughputUsagePerThroughputServer }CPU usage on the receiver divided by the network throughput during message sending (message count 2~k). The nodes are twin CPU nodes, 100~\% CPU usage corresponds to one CPU being fully used. }
\end{table}

\clearpage

\subsection{\label{Sec:TCPMsgLatencyTables}TCP Message Class Latency}

\begin{table}[h!t!p!b]
\begin{center}
{\scriptsize
\begin{tabular}{|l| c| c| c| c| }
\hline
 Message Count   &  Latency   &  Latency   &  Latency   &  Latency  \\
   &  FE, No Connect /  &  FE, With Connect /  &  GbE, No Connect /  &  GbE, With Connect / \\
   &  $\mu \mathrm s$   &  $\mu \mathrm s$   &  $\mu \mathrm s$   &  $\mu \mathrm s$  \\
\hline \hline
         1  & 744.150000  & 199.250000  & 798.750000  & 196.600000  \\
  & $\pm$ 15.979753  & $\pm$ 9.263504  & $\pm$ 29.400893  & $\pm$ 10.617909  \\
\hline
         2  & 577.575000  & 156.200000  & 599.900000  & 154.150000  \\
  & $\pm$ 11.560736  & $\pm$ 2.428477  & $\pm$ 16.118002  & $\pm$ 4.752894  \\
\hline
         8  & 410.287500  & 122.393600  & 428.237700  & 125.774800  \\
\hline
\end{tabular}
}
\end{center}
\caption [Average message sending latency.] {\label{Tab:TCPMsgLatency }Average message sending latency. }
\end{table}

\clearpage

\subsection{TCP Blob Class Throughput with On-Demand Allocation}
\subsubsection{Plateau Determination}

\begin{table}[h!t!p!b]
\begin{center}
{\scriptsize
\begin{tabular}{|l| c| c| c| c| }
\hline
 Blob Count   &  Send Rate   &  Send Rate   &  Send Rate   &  Send Rate  \\
   &  FE, No Connect /  &  FE, With Connect /  &  GbE, No Connect /  &  GbE, With Connect / \\
   &  Hz   &  Hz   &  Hz   &  Hz  \\
\hline \hline
         1  & 32.968482  & 30.100536  & 27.855153  & 29.192819  \\
  & $\pm$ 3.967261  & $\pm$ 3.582491  & $\pm$ 10.212522  & $\pm$ 3.365419  \\
\hline
         2  & 53.236797  & 59.894586  & 63.488033  & 60.731204  \\
  & $\pm$ 19.133391  & $\pm$ 9.065262  & $\pm$ 6.751473  & $\pm$ 6.708980  \\
\hline
         4  & 99.930049  & 97.718278  & 97.487266  & 102.543068  \\
\hline
\end{tabular}
}
\end{center}
\caption [TCP blob measurement plateau determination (on-demand alloc.).] {\label{Tab:TCPBlobOnDemandPlateauRates }TCP blob measurement plateau determination (on-demand allocation). }
\end{table}

\clearpage
\subsubsection{Plateau Throughput Measurement}

\begin{table}[h!t!p!b]
\begin{center}
{\scriptsize
\begin{tabular}{|l| c| c| c| c| }
\hline
 Blob Size  /  &  Send Rate   &  Send Rate   &  Send Rate   &  Send Rate  \\
 B   &  FE, No Connect /  &  FE, With Connect /  &  GbE, No Connect /  &  GbE, With Connect / \\
   &  Hz   &  Hz   &  Hz   &  Hz  \\
\hline \hline
       256  & 123.999560  & 2143.739402  & 123.666624  & 1724.287357  \\
  & $\pm$ 4.162359  & $\pm$ 6.477876  & $\pm$ 3.228727  & $\pm$ 17.078924  \\
\hline
       512  & 124.014143  & 1961.446397  & 123.646522  & 1697.099006  \\
  & $\pm$ 0.855639  & $\pm$ 5.690365  & $\pm$ 1.937501  & $\pm$ 8.262662  \\
\hline
      1024  & 123.940865  & 1643.022150  & 123.631986  & 1609.702545  \\
\hline
\end{tabular}
}
\end{center}
\caption [Maximum blob sending rate (plateau, on-demand alloc.).] {\label{Tab:TCPBlobOnDemandPlateauThroughputRate }Maximum blob sending rate (blob count 32~k, on-demand allocation).  }
\end{table}

\begin{table}[h!t!p!b]
\begin{center}
{\scriptsize
\begin{tabular}{|l| c| c| c| c| }
\hline
 Blob Size  /  &  Network Throughput   &  Network Throughput   &  Network Throughput   &  Network Throughput  \\
 B   &  FE, No Connect /  &  FE, With Connect /  &  GbE, No Connect /  &  GbE, With Connect / \\
   &  $\mathrm{B}/\mathrm s$   &  $\mathrm{B}/\mathrm s$   &  $\mathrm{B}/\mathrm s$   &  $\mathrm{B}/\mathrm s$  \\
\hline \hline
       256  & 31802.857562  & 547722.298543  & 31717.309365  & 440721.825650  \\
  & $\pm$ 37.579287  & $\pm$ 58.042948  & $\pm$ 29.149823  & $\pm$ 153.145206  \\
\hline
       512  & 63613.209246  & 1002460.379066  & 63424.288560  & 867566.717411  \\
  & $\pm$ 15.449974  & $\pm$ 102.006814  & $\pm$ 34.984556  & $\pm$ 148.181656  \\
\hline
      1024  & 127151.103025  & 1679927.681031  & 126833.636535  & 1645909.784860  \\
\hline
\end{tabular}
}
\end{center}
\caption [Blob network throughput (plateau, on-demand alloc.).] {\label{Tab:TCPBlobOnDemandPlateauThroughputThroughput }Network throughput during blob transmission (blob count 32~k, on-demand allocation). }
\end{table}

\begin{table}[h!t!p!b]
\begin{center}
{\scriptsize
\begin{tabular}{|l| c| c| c| c| }
\hline
 Blob Size  /  &  CPU Usage   &  CPU Usage   &  CPU Usage   &  CPU Usage  \\
 B   &  FE, No Connect /  &  FE, With Connect /  &  GbE, No Connect /  &  GbE, With Connect / \\
   &  \%   &  \%   &  \%   &  \%  \\
\hline \hline
       256  & 90.173693  & 53.369219  & 80.955519  & 33.734391  \\
  & $\pm$ 6.047987  & $\pm$ 0.312357  & $\pm$ 5.204845  & $\pm$ 0.651094  \\
\hline
       512  & 90.173714  & 48.836100  & 81.155879  & 33.707043  \\
  & $\pm$ 1.243113  & $\pm$ 0.275130  & $\pm$ 3.542855  & $\pm$ 0.319908  \\
\hline
      1024  & 88.169753  & 43.327786  & 81.155861  & 33.619134  \\
\hline
\end{tabular}
}
\end{center}
\caption [Blob throughput sender CPU usage (plateau, on-demand alloc.).] {\label{Tab:TCPBlobOnDemandPlateauThroughputUsage }CPU usage on the sender during blob transmission (blob count 32~k, on-demand allocation). The nodes are twin CPU nodes, 100~\% CPU usage corresponds to one CPU being fully used. }
\end{table}

\begin{table}[h!t!p!b]
\begin{center}
{\scriptsize
\begin{tabular}{|l| c| c| c| c| }
\hline
 Blob Size  /  &  CPU Usage   &  CPU Usage   &  CPU Usage   &  CPU Usage  \\
 B   &  FE, No Connect /  &  FE, With Connect /  &  GbE, No Connect /  &  GbE, With Connect / \\
   &  \%   &  \%   &  \%   &  \%  \\
\hline \hline
       256  & 16.741902  & 66.883981  & 15.107717  & 40.023931  \\
  & $\pm$ 1.928896  & $\pm$ 0.325113  & $\pm$ 1.780904  & $\pm$ 1.753174  \\
\hline
       512  & 16.333601  & 60.656706  & 15.107665  & 39.349298  \\
  & $\pm$ 0.223170  & $\pm$ 1.562763  & $\pm$ 1.468792  & $\pm$ 0.335732  \\
\hline
      1024  & 16.333401  & 53.371427  & 15.107634  & 38.868490  \\
\hline
\end{tabular}
}
\end{center}
\caption [Blob throughput receiver CPU usage (plateau, on-demand alloc.).] {\label{Tab:TCPBlobOnDemandPlateauThroughputUsageServer }CPU usage on the receiver during blob transmission (blob count 32~k, on-demand allocation). The nodes are twin CPU nodes, 100~\% CPU usage corresponds to one CPU being fully used. }
\end{table}

\begin{table}[h!t!p!b]
\begin{center}
{\scriptsize
\begin{tabular}{|l| c| c| c| c| }
\hline
 Blob Size  /  &  CPU Usage / Rate   &  CPU Usage / Rate   &  CPU Usage / Rate   &  CPU Usage / Rate  \\
 B   &  FE, No Connect /  &  FE, With Connect /  &  GbE, No Connect /  &  GbE, With Connect / \\
   &  $\% \times \mathrm s$   &  $\% \times \mathrm s$   &  $\% \times \mathrm s$   &  $\% \times \mathrm s$  \\
\hline \hline
       256  & 0.727210  & 0.024895  & 0.654627  & 0.019564  \\
  & $\pm$ 0.073185  & $\pm$ 0.000221  & $\pm$ 0.059179  & $\pm$ 0.000571  \\
\hline
       512  & 0.727124  & 0.024898  & 0.656354  & 0.019862  \\
  & $\pm$ 0.015041  & $\pm$ 0.000213  & $\pm$ 0.038938  & $\pm$ 0.000285  \\
\hline
      1024  & 0.711386  & 0.026371  & 0.656431  & 0.020885  \\
\hline
\end{tabular}
}
\end{center}
\caption [Blob throughput sender CPU usage divided by rate (plateau, on-demand alloc.).] {\label{Tab:TCPBlobOnDemandPlateauThroughputUsagePerRate }CPU usage on the sender divided by the sending rate during blob transmission (blob count 32~k, on-demand allocation). The nodes are twin CPU nodes, 100~\% CPU usage corresponds to one CPU being fully used. }
\end{table}

\begin{table}[h!t!p!b]
\begin{center}
{\scriptsize
\begin{tabular}{|l| c| c| c| c| }
\hline
 Blob Size  /  &  CPU Usage / Rate   &  CPU Usage / Rate   &  CPU Usage / Rate   &  CPU Usage / Rate  \\
 B   &  FE, No Connect /  &  FE, With Connect /  &  GbE, No Connect /  &  GbE, With Connect / \\
   &  $\% \times \mathrm s$   &  $\% \times \mathrm s$   &  $\% \times \mathrm s$   &  $\% \times \mathrm s$  \\
\hline \hline
       256  & 0.135016  & 0.031200  & 0.122165  & 0.023212  \\
  & $\pm$ 0.020088  & $\pm$ 0.000246  & $\pm$ 0.017590  & $\pm$ 0.001247  \\
\hline
       512  & 0.131708  & 0.030924  & 0.122184  & 0.023186  \\
  & $\pm$ 0.002708  & $\pm$ 0.000886  & $\pm$ 0.013794  & $\pm$ 0.000311  \\
\hline
      1024  & 0.131784  & 0.032484  & 0.122198  & 0.024146  \\
\hline
\end{tabular}
}
\end{center}
\caption [Blob throughput receiver CPU usage divided by rate (plateau, on-demand alloc.).] {\label{Tab:TCPBlobOnDemandPlateauThroughputUsagePerRateServer }CPU usage on the receiver divided by the sending rate during blob transmission (blob count 32~k, on-demand allocation). The nodes are twin CPU nodes, 100~\% CPU usage corresponds to one CPU being fully used. }
\end{table}

\begin{table}[h!t!p!b]
\begin{center}
{\scriptsize
\begin{tabular}{|l| c| c| c| c| }
\hline
 Blob Size  /  &  CPU Usage / Throughput   &  CPU Usage / Throughput   &  CPU Usage / Throughput   &  CPU Usage / Throughput  \\
 B   &  FE, No Connect /  &  FE, With Connect /  &  GbE, No Connect /  &  GbE, With Connect / \\
   &  $\% / (\mathrm{MB} / \mathrm s)$   &  $\% / (\mathrm{MB} / \mathrm s)$   &  $\% / (\mathrm{MB} / \mathrm s)$   &  $\% / (\mathrm{MB} / \mathrm s)$  \\
\hline \hline
       256  & 2978.683745  & 101.971475  & 2681.356618  & 80.135095  \\
  & $\pm$ 299.749932  & $\pm$ 0.905043  & $\pm$ 242.373941  & $\pm$ 2.340451  \\
\hline
       512  & 1489.145457  & 50.991138  & 1344.219018  & 40.676468  \\
  & $\pm$ 30.808465  & $\pm$ 0.435175  & $\pm$ 79.744363  & $\pm$ 0.584119  \\
\hline
      1024  & 728.458913  & 27.003682  & 672.187296  & 21.386558  \\
\hline
\end{tabular}
}
\end{center}
\caption [Blob throughput sender CPU usage divided by throughput (plateau, on-demand alloc.).] {\label{Tab:TCPBlobOnDemandPlateauThroughputUsagePerThroughput }CPU usage on the sender divided by the network throughput during blob transmission (blob count 32~k, on-demand allocation). The nodes are twin CPU nodes, 100~\% CPU usage corresponds to one CPU being fully used. }
\end{table}

\begin{table}[h!t!p!b]
\begin{center}
{\scriptsize
\begin{tabular}{|l| c| c| c| c| }
\hline
 Blob Size  /  &  CPU Usage / Throughput   &  CPU Usage / Throughput   &  CPU Usage / Throughput   &  CPU Usage / Throughput  \\
 B   &  FE, No Connect /  &  FE, With Connect /  &  GbE, No Connect /  &  GbE, With Connect / \\
   &  $\% / (\mathrm{MB} / \mathrm s)$   &  $\% / (\mathrm{MB} / \mathrm s)$   &  $\% / (\mathrm{MB} / \mathrm s)$   &  $\% / (\mathrm{MB} / \mathrm s)$  \\
\hline \hline
       256  & 553.030820  & 127.793855  & 500.388083  & 95.075721  \\
  & $\pm$ 82.277122  & $\pm$ 1.007469  & $\pm$ 72.045900  & $\pm$ 5.106409  \\
\hline
       512  & 269.736120  & 63.333364  & 250.234621  & 47.485341  \\
  & $\pm$ 5.547440  & $\pm$ 1.815429  & $\pm$ 28.249146  & $\pm$ 0.636370  \\
\hline
      1024  & 134.946636  & 33.263298  & 125.131562  & 24.725896  \\
\hline
\end{tabular}
}
\end{center}
\caption [Blob throughput receiver CPU usage divided by throughput (plateau, on-demand alloc.).] {\label{Tab:TCPBlobOnDemandPlateauThroughputUsagePerThroughputServer }CPU usage on the receiver divided by the network throughput during blob transmission (blob count 32~k, on-demand allocation). The nodes are twin CPU nodes, 100~\% CPU usage corresponds to one CPU being fully used. }
\end{table}

\clearpage
\subsubsection{Peak Throughput Measurement}

\begin{table}[h!t!p!b]
\begin{center}
{\scriptsize
\begin{tabular}{|l| c| c| c| c| }
\hline
 Blob Size  /  &  Send Rate   &  Send Rate   &  Send Rate   &  Send Rate  \\
 B   &  FE, No Connect /  &  FE, With Connect /  &  GbE, No Connect /  &  GbE, With Connect / \\
   &  Hz   &  Hz   &  Hz   &  Hz  \\
\hline \hline
       256  & 446.349567  & 2143.739402  & 343.941384  & 1724.287357  \\
  & $\pm$ 3.776770  & $\pm$ 6.477876  & $\pm$ 5.432648  & $\pm$ 17.078924  \\
\hline
       512  & 439.308878  & 1961.446397  & 327.669986  & 1697.099006  \\
  & $\pm$ 2.927866  & $\pm$ 5.690365  & $\pm$ 48.361721  & $\pm$ 8.262662  \\
\hline
      1024  & 422.772317  & 1643.022150  & 346.293234  & 1609.702545  \\
\hline
\end{tabular}
}
\end{center}
\caption [Maximum blob sending rate (peak, on-demand alloc.).] {\label{Tab:TCPBlobOnDemandPeakThroughputRate }Maximum blob sending rate (blob counts 32~k and 2~k, on-demand allocation).  }
\end{table}

\begin{table}[h!t!p!b]
\begin{center}
{\scriptsize
\begin{tabular}{|l| c| c| c| c| }
\hline
 Blob Size  /  &  Network Throughput   &  Network Throughput   &  Network Throughput   &  Network Throughput  \\
 B   &  FE, No Connect /  &  FE, With Connect /  &  GbE, No Connect /  &  GbE, With Connect / \\
   &  $\mathrm{B}/\mathrm s$   &  $\mathrm{B}/\mathrm s$   &  $\mathrm{B}/\mathrm s$   &  $\mathrm{B}/\mathrm s$  \\
\hline \hline
       256  & 114265.489152  & 548797.286912  & 88048.994304  & 441417.563392  \\
  & $\pm$ 966.853120  & $\pm$ 1658.336256  & $\pm$ 1390.757888  & $\pm$ 4372.204544  \\
\hline
       512  & 224926.145536  & 1004260.555264  & 167767.032832  & 868914.691072  \\
  & $\pm$ 1499.067392  & $\pm$ 2913.466880  & $\pm$ 24761.201152  & $\pm$ 4230.482944  \\
\hline
      1024  & 432918.852608  & 1682454.681600  & 354604.271616  & 1648335.406080  \\
\hline
\end{tabular}
}
\end{center}
\caption [Blob network throughput (peak, on-demand alloc.).] {\label{Tab:TCPBlobOnDemandPeakThroughputThroughput }Network throughput during blob transmission (blob counts 32~k and 2~k, on-demand allocation). }
\end{table}

\begin{table}[h!t!p!b]
\begin{center}
{\scriptsize
\begin{tabular}{|l| c| c| c| c| }
\hline
 Blob Size  /  &  CPU Usage   &  CPU Usage   &  CPU Usage   &  CPU Usage  \\
 B   &  FE, No Connect /  &  FE, With Connect /  &  GbE, No Connect /  &  GbE, With Connect / \\
   &  \%   &  \%   &  \%   &  \%  \\
\hline \hline
       256  & 67.113551  & 53.369219  & 49.505616  & 33.734391  \\
  & $\pm$ 2.412333  & $\pm$ 0.312357  & $\pm$ 2.370739  & $\pm$ 0.651094  \\
\hline
       512  & 66.563892  & 48.836100  & 48.022929  & 33.707043  \\
  & $\pm$ 0.843512  & $\pm$ 0.275130  & $\pm$ 18.832624  & $\pm$ 0.319908  \\
\hline
      1024  & 64.106253  & 43.327786  & 49.532322  & 33.619134  \\
\hline
\end{tabular}
}
\end{center}
\caption [Blob throughput sender CPU usage (peak, on-demand alloc.).] {\label{Tab:TCPBlobOnDemandPeakThroughputUsage }CPU usage on the sender during blob transmission (blob counts 32~k and 2~k, on-demand allocation). The nodes are twin CPU nodes, 100~\% CPU usage corresponds to one CPU being fully used. }
\end{table}

\begin{table}[h!t!p!b]
\begin{center}
{\scriptsize
\begin{tabular}{|l| c| c| c| c| }
\hline
 Blob Size  /  &  CPU Usage   &  CPU Usage   &  CPU Usage   &  CPU Usage  \\
 B   &  FE, No Connect /  &  FE, With Connect /  &  GbE, No Connect /  &  GbE, With Connect / \\
   &  \%   &  \%   &  \%   &  \%  \\
\hline \hline
       256  & 62.236010  & 66.883981  & 42.134169  & 40.023931  \\
  & $\pm$ 2.408188  & $\pm$ 0.325113  & $\pm$ 1.334580  & $\pm$ 1.753174  \\
\hline
       512  & 61.686996  & 60.656706  & 41.643720  & 39.349298  \\
  & $\pm$ 2.307401  & $\pm$ 1.562763  & $\pm$ 12.469649  & $\pm$ 0.335732  \\
\hline
      1024  & 59.607163  & 53.371427  & 43.124014  & 38.868490  \\
\hline
\end{tabular}
}
\end{center}
\caption [Blob throughput receiver CPU usage (peak, on-demand alloc.).] {\label{Tab:TCPBlobOnDemandPeakThroughputUsageServer }CPU usage on the receiver during blob transmission (blob counts 32~k and 2~k, on-demand allocation). The nodes are twin CPU nodes, 100~\% CPU usage corresponds to one CPU being fully used. }
\end{table}

\begin{table}[h!t!p!b]
\begin{center}
{\scriptsize
\begin{tabular}{|l| c| c| c| c| }
\hline
 Blob Size  /  &  CPU Usage / Rate   &  CPU Usage / Rate   &  CPU Usage / Rate   &  CPU Usage / Rate  \\
 B   &  FE, No Connect /  &  FE, With Connect /  &  GbE, No Connect /  &  GbE, With Connect / \\
   &  $\% \times \mathrm s$   &  $\% \times \mathrm s$   &  $\% \times \mathrm s$   &  $\% \times \mathrm s$  \\
\hline \hline
       256  & 0.150361  & 0.024895  & 0.143936  & 0.019564  \\
  & $\pm$ 0.006677  & $\pm$ 0.000221  & $\pm$ 0.009166  & $\pm$ 0.000571  \\
\hline
       512  & 0.151520  & 0.024898  & 0.146559  & 0.019862  \\
  & $\pm$ 0.002930  & $\pm$ 0.000213  & $\pm$ 0.079105  & $\pm$ 0.000285  \\
\hline
      1024  & 0.151633  & 0.026371  & 0.143036  & 0.020885  \\
\hline
\end{tabular}
}
\end{center}
\caption [Blob throughput sender CPU usage divided by rate (peak, on-demand alloc.).] {\label{Tab:TCPBlobOnDemandPeakThroughputUsagePerRate }CPU usage on the sender divided by the sending rate during blob transmission (blob counts 32~k and 2~k, on-demand allocation). The nodes are twin CPU nodes, 100~\% CPU usage corresponds to one CPU being fully used. }
\end{table}

\begin{table}[h!t!p!b]
\begin{center}
{\scriptsize
\begin{tabular}{|l| c| c| c| c| }
\hline
 Blob Size  /  &  CPU Usage / Rate   &  CPU Usage / Rate   &  CPU Usage / Rate   &  CPU Usage / Rate  \\
 B   &  FE, No Connect /  &  FE, With Connect /  &  GbE, No Connect /  &  GbE, With Connect / \\
   &  $\% \times \mathrm s$   &  $\% \times \mathrm s$   &  $\% \times \mathrm s$   &  $\% \times \mathrm s$  \\
\hline \hline
       256  & 0.139433  & 0.031200  & 0.122504  & 0.023212  \\
  & $\pm$ 0.006575  & $\pm$ 0.000246  & $\pm$ 0.005815  & $\pm$ 0.001247  \\
\hline
       512  & 0.140418  & 0.030924  & 0.127090  & 0.023186  \\
  & $\pm$ 0.006188  & $\pm$ 0.000886  & $\pm$ 0.056813  & $\pm$ 0.000311  \\
\hline
      1024  & 0.140991  & 0.032484  & 0.124530  & 0.024146  \\
\hline
\end{tabular}
}
\end{center}
\caption [Blob throughput receiver CPU usage divided by rate (peak, on-demand alloc.).] {\label{Tab:TCPBlobOnDemandPeakThroughputUsagePerRateServer }CPU usage on the receiver divided by the sending rate during blob transmission (blob counts 32~k and 2~k, on-demand allocation). The nodes are twin CPU nodes, 100~\% CPU usage corresponds to one CPU being fully used. }
\end{table}

\begin{table}[h!t!p!b]
\begin{center}
{\scriptsize
\begin{tabular}{|l| c| c| c| c| }
\hline
 Blob Size  /  &  CPU Usage / Throughput   &  CPU Usage / Throughput   &  CPU Usage / Throughput   &  CPU Usage / Throughput  \\
 B   &  FE, No Connect /  &  FE, With Connect /  &  GbE, No Connect /  &  GbE, With Connect / \\
   &  $\% / (\mathrm{MB} / \mathrm s)$   &  $\% / (\mathrm{MB} / \mathrm s)$   &  $\% / (\mathrm{MB} / \mathrm s)$   &  $\% / (\mathrm{MB} / \mathrm s)$  \\
\hline \hline
       256  & 615.878859  & 101.971475  & 589.563130  & 80.135095  \\
  & $\pm$ 27.348065  & $\pm$ 0.905043  & $\pm$ 37.543167  & $\pm$ 2.340451  \\
\hline
       512  & 310.312495  & 50.991138  & 300.152686  & 40.676468  \\
  & $\pm$ 6.001039  & $\pm$ 0.435175  & $\pm$ 162.007747  & $\pm$ 0.584119  \\
\hline
      1024  & 155.272082  & 27.003682  & 146.468630  & 21.386558  \\
\hline
\end{tabular}
}
\end{center}
\caption [Blob throughput sender CPU usage divided by throughput (peak, on-demand alloc.).] {\label{Tab:TCPBlobOnDemandPeakThroughputUsagePerThroughput }CPU usage on the sender divided by the network throughput during blob transmission (blob counts 32~k and 2~k, on-demand allocation). The nodes are twin CPU nodes, 100~\% CPU usage corresponds to one CPU being fully used. }
\end{table}

\begin{table}[h!t!p!b]
\begin{center}
{\scriptsize
\begin{tabular}{|l| c| c| c| c| }
\hline
 Blob Size  /  &  CPU Usage / Throughput   &  CPU Usage / Throughput   &  CPU Usage / Throughput   &  CPU Usage / Throughput  \\
 B   &  FE, No Connect /  &  FE, With Connect /  &  GbE, No Connect /  &  GbE, With Connect / \\
   &  $\% / (\mathrm{MB} / \mathrm s)$   &  $\% / (\mathrm{MB} / \mathrm s)$   &  $\% / (\mathrm{MB} / \mathrm s)$   &  $\% / (\mathrm{MB} / \mathrm s)$  \\
\hline \hline
       256  & 571.119278  & 127.793855  & 501.776456  & 95.075721  \\
  & $\pm$ 26.931322  & $\pm$ 1.007469  & $\pm$ 23.817263  & $\pm$ 5.106409  \\
\hline
       512  & 287.577019  & 63.333364  & 260.281384  & 47.485341  \\
  & $\pm$ 12.673940  & $\pm$ 1.815429  & $\pm$ 116.353221  & $\pm$ 0.636370  \\
\hline
      1024  & 144.374813  & 33.263298  & 127.519063  & 24.725896  \\
\hline
\end{tabular}
}
\end{center}
\caption [Blob throughput receiver CPU usage divided by throughput (peak, on-demand alloc.).] {\label{Tab:TCPBlobOnDemandPeakThroughputUsagePerThroughputServer }CPU usage on the receiver divided by the network throughput during blob transmission (blob counts 32~k and 2~k, on-demand allocation). The nodes are twin CPU nodes, 100~\% CPU usage corresponds to one CPU being fully used. }
\end{table}

\clearpage

\subsection{\label{Sec:TCPBLobPreAllocThroughputTables}TCP Blob Class Throughput with Preallocation}
\subsubsection{Plateau Determination}

\begin{table}[h!t!p!b]
\begin{center}
{\scriptsize

}
\end{center}
\caption [TCP blob measurement plateau determination (prealloc.).] {\label{Tab:TCPBlobPreAllocPlateauRates }TCP blob measurement plateau determination (preallocation). }
\end{table}

\clearpage
\subsubsection{Plateau Throughput Measurement}

\begin{table}[h!t!p!b]
\begin{center}
{\scriptsize
%
}
\end{center}
\caption [Maximum blob sending rate (plateau, prealloc.).] {\label{Tab:TCPBlobPreAllocPlateauThroughputRate }Maximum blob sending rate (blob count 128~k, preallocation).  }
\end{table}

\begin{table}[h!t!p!b]
\begin{center}
{\scriptsize
%
}
\end{center}
\caption [Blob network throughput (plateau, prealloc.).] {\label{Tab:TCPBlobPreAllocPlateauThroughputThroughput }Network throughput during blob transmission (blob count 128~k, preallocation). }
\end{table}

\begin{table}[h!t!p!b]
\begin{center}
{\scriptsize
%
}
\end{center}
\caption [Blob throughput sender CPU usage (plateau, prealloc.).] {\label{Tab:TCPBlobPreAllocPlateauThroughputUsage }CPU usage on the sender during blob transmission (blob count 128~k, preallocation). The nodes are twin CPU nodes, 100~\% CPU usage corresponds to one CPU being fully used. }
\end{table}

\begin{table}[h!t!p!b]
\begin{center}
{\scriptsize
%
}
\end{center}
\caption [Blob throughput receiver CPU usage (plateau, prealloc.).] {\label{Tab:TCPBlobPreAllocPlateauThroughputUsageServer }CPU usage on the receiver during blob transmission (blob count 128~k, preallocation). The nodes are twin CPU nodes, 100~\% CPU usage corresponds to one CPU being fully used. }
\end{table}

\begin{table}[h!t!p!b]
\begin{center}
{\scriptsize
%
}
\end{center}
\caption [Blob throughput sender CPU usage divided by rate (plateau, prealloc.).] {\label{Tab:TCPBlobPreAllocPlateauThroughputUsagePerRate }CPU usage on the sender divided by the sending rate during blob transmission (blob count 128~k, preallocation). The nodes are twin CPU nodes, 100~\% CPU usage corresponds to one CPU being fully used. }
\end{table}

\begin{table}[h!t!p!b]
\begin{center}
{\scriptsize
%
}
\end{center}
\caption [Blob throughput receiver CPU usage divided by rate (plateau, prealloc.).] {\label{Tab:TCPBlobPreAllocPlateauThroughputUsagePerRateServer }CPU usage on the receiver divided by the sending rate during blob transmission (blob count 128~k, preallocation). The nodes are twin CPU nodes, 100~\% CPU usage corresponds to one CPU being fully used. }
\end{table}

\begin{table}[h!t!p!b]
\begin{center}
{\scriptsize
%
}
\end{center}
\caption [Blob throughput sender CPU usage divided by throughput (plateau, prealloc.).] {\label{Tab:TCPBlobPreAllocPlateauThroughputUsagePerThroughput }CPU usage on the sender divided by the network throughput during blob transmission (blob count 128~k, preallocation). The nodes are twin CPU nodes, 100~\% CPU usage corresponds to one CPU being fully used. }
\end{table}

\begin{table}[h!t!p!b]
\begin{center}
{\scriptsize
%
}
\end{center}
\caption [Blob throughput receiver CPU usage divided by throughput (plateau, prealloc.).] {\label{Tab:TCPBlobPreAllocPlateauThroughputUsagePerThroughputServer }CPU usage on the receiver divided by the network throughput during blob transmission (blob count 128~k, preallocation). The nodes are twin CPU nodes, 100~\% CPU usage corresponds to one CPU being fully used. }
\end{table}

\clearpage
\subsubsection{Peak Throughput Measurement}

\begin{table}[h!t!p!b]
\begin{center}
{\scriptsize
%
}
\end{center}
\caption [Maximum blob sending rate (peak, prealloc.).] {\label{Tab:TCPBlobPreAllocPeakThroughputRate }Maximum blob sending rate (blob counts 128~k and 2~k, preallocation). }
\end{table}

\begin{table}[h!t!p!b]
\begin{center}
{\scriptsize
%
}
\end{center}
\caption [Blob network throughput (peak, prealloc.).] {\label{Tab:TCPBlobPreAllocPeakThroughputThroughput }Network throughput during blob transmission (blob counts 128~k and 2~k, preallocation). }
\end{table}

\begin{table}[h!t!p!b]
\begin{center}
{\scriptsize
%
}
\end{center}
\caption [Blob throughput sender CPU usage (peak, prealloc.).] {\label{Tab:TCPBlobPreAllocPeakThroughputUsage }CPU usage on the sender during blob transmission (blob counts 128~k and 2~k, preallocation). The nodes are twin CPU nodes, 100~\% CPU usage corresponds to one CPU being fully used. }
\end{table}

\begin{table}[h!t!p!b]
\begin{center}
{\scriptsize
%
}
\end{center}
\caption [Blob throughput receiver CPU usage (peak, prealloc.).] {\label{Tab:TCPBlobPreAllocPeakThroughputUsageServer }CPU usage on the receiver during blob transmission (blob counts 128~k and 2~k, preallocation). The nodes are twin CPU nodes, 100~\% CPU usage corresponds to one CPU being fully used. }
\end{table}

\begin{table}[h!t!p!b]
\begin{center}
{\scriptsize
%
}
\end{center}
\caption [Blob throughput sender CPU usage divided by rate (peak, prealloc.).] {\label{Tab:TCPBlobPreAllocPeakThroughputUsagePerRate }CPU usage on the sender divided by the sending rate during blob transmission (blob counts 128~k and 2~k, preallocation). The nodes are twin CPU nodes, 100~\% CPU usage corresponds to one CPU being fully used. }
\end{table}

\begin{table}[h!t!p!b]
\begin{center}
{\scriptsize
%
}
\end{center}
\caption [Blob throughput receiver CPU usage divided by rate (peak, prealloc.).] {\label{Tab:TCPBlobPreAllocPeakThroughputUsagePerRateServer }CPU usage on the receiver divided by the sending rate during blob transmission (blob counts 128~k and 2~k, preallocation). The nodes are twin CPU nodes, 100~\% CPU usage corresponds to one CPU being fully used. }
\end{table}

\begin{table}[h!t!p!b]
\begin{center}
{\scriptsize
%
}
\end{center}
\caption [Blob throughput sender CPU usage divided by throughput (peak, prealloc.).] {\label{Tab:TCPBlobPreAllocPeakThroughputUsagePerThroughput }CPU usage on the sender divided by the network throughput during blob transmission (blob counts 128~k and 2~k, preallocation). The nodes are twin CPU nodes, 100~\% CPU usage corresponds to one CPU being fully used. }
\end{table}

\begin{table}[h!t!p!b]
\begin{center}
{\scriptsize
%
}
\end{center}
\caption [Blob throughput receiver CPU usage divided by throughput (peak, prealloc.).] {\label{Tab:TCPBlobPreAllocPeakThroughputUsagePerThroughputServer }CPU usage on the receiver divided by the network throughput during blob transmission (blob counts 128~k and 2~k, preallocation). The nodes are twin CPU nodes, 100~\% CPU usage corresponds to one CPU being fully used. }
\end{table}

\clearpage

\subsection{\label{Sec:TCPBlobLatencyTables}TCP Blob Class Latency with On-Demand Allocation}

\begin{table}[h!t!p!b]
\begin{center}
{\scriptsize
%
}
\end{center}
\caption [Average blob sending latency (on-demand alloc.). ] {\label{Tab:TCPBlobOnDemandLatency }Average blob sending latency (on-demand allocation) }
\end{table}

\subsection{\label{Sec:TCPBlobPreAllocLatencyTables}TCP Blob Class Latency with Preallocation}

\begin{table}[h!t!p!b]
\begin{center}
{\scriptsize
%
}
\end{center}
\caption [Average blob sending latency (prealloc.).] {\label{Tab:TCPBlobPreAllocLatency }Average blob sending latency (preallocation) }
\end{table}

\clearpage

\section{Publisher-Subscriber Interface Benchmarks}

\subsection{Scaling Behaviour}

\begin{table}[hbt!p]
\begin{center}
%
\parbox{0.90\columnwidth}{
\caption[Scaling Properties.]{\label{Tab:ScalingResults}Scaling properties of the memory subsystems
and the publisher-subscriber interface overheads. The values are scaled relative to the 800~MHz
results.}
}
\end{center}
\end{table}

%%%%%%%%%%%%%%%%%%%%%%%%%%%%%%%%%%%%%%%%%%%%%%%%%%%%%%%%%%%%%%%%%%%%%%%%%%%%%%%%%%%%%%%%%%%%%%%%%%%%%%%%%%%%%%%%%%%%%%%%%%%%%
%%%%%%%%%%%%%%%%%%%%%%%%%%%%%%%%%%%%%%%%%%%%%%%%%%%%%%%%%%%%%%%%%%%%%%%%%%%%%%%%%%%%%%%%%%%%%%%%%%%%%%%%%%%%%%%%%%%%%%%%%%%%%

\chapter{\label{Chap:ObsoleteComponents}Obsolete Framework Components}

\section{\label{Sec:DATEComponents}ALICE DAQ Interface Components}

For a previous software only version of an interface between the ALICE High Level Trigger and the ALICE Data Acquisition system DATE  
%at the end of a chain, 
for the recording
of events two data sink components have been created. Both interfaces contain a subscriber object of a class derived from the common 
\texttt{Ali\-HLT\-DATE\-Base\-Sub\-scrib\-er} parent class. This class contains functionality to initialize a DATE interface library, pass events for 
recording to DATE, and query already recorded events that can be released. 

\subsection{The DATE Subscriber Base Class}

Basic functionality for interfacing from the publisher-subscriber interface framework to the DATE system is defined in the \texttt{Ali\-HLT\-DATE\-Base\-Sub\-scrib\-er} 
class. It is derived from \texttt{Ali\-HLT\-Sub\-scrib\-er\-Inter\-face} but defines none of the functions needed to implement the
interface, which has to be done by its own derived classes, \texttt{Direct\-DATE\-Sub\-scrib\-er} and \texttt{Trig\-gered\-DATE\-Sub\-scrib\-er}. 
In the class four primary and one auxiliary functions are provided for the interface with the DAQ system. Fig.~\ref{Fig:DATEClasses}
shows the relation of the classes, more detailed explanations are contained in the following paragraphs. 

\begin{figure}[hbt]
\begin{center}
\resizebox*{0.40\columnwidth}{!}{
\includegraphics{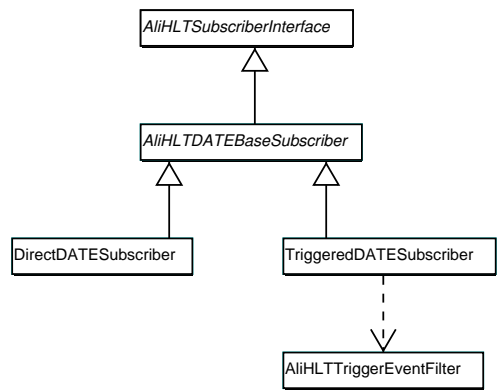}
}
\parbox{0.90\columnwidth}{
\caption[Classes used in the DATE interface components.]{\label{Fig:DATEClasses}The classes used in the DATE interface components. 
%See the text for more details.
}
}
\end{center}
\end{figure}

The first of the primary functions is \texttt{Run}, which has the purpose of initializing the library provided by DATE and starting
the background thread that runs the \texttt{DATE\-Events\-Done} function described later. It also interacts with the DATE run-control
to inform it about the activation of a recording program. \texttt{Run}'s counterpart is the \texttt{Stop} method called at the end
of a program. It terminates the started background thread, informs the run-control about the program's end, and deinitializes the DATE
library. To pass an event to the DATE system for recording the third function, \texttt{Send\-Event\-To\-DATE}, has to be called. Its main parameters 
are the ID of the event concerned, its corresponding sub-event descriptor, and a 32~bit unsigned integer containing additional event flags
to pass to DATE. 

In the current version this function imposes one restriction on the event data. The program is currently
only able to create one type of DATE events, called streamlined events, in which the event data has to be prefixed directly by the DATE event header. 
As a consequence space has to be available in front of the data in shared memory
for the appropriate amount of bytes so that an event header can be created there. In addition to the event ID
the header contains the number of the GDC node where the event will be assembled as well as the additional flags that have been passed in the function's
parameter. The GDC ID is obtained by calling the class's \texttt{Get\-GDC\-ID} helper function. After calling the DATE function to record the event 
described by the constructed header, the function updates the run-control's event and byte counters appropriately. 

Running as a continuous loop in a background thread the \texttt{DATE\-Events\-Done} function's task is to query DATE via its library periodically
for events that have already been recorded. For these finished events the publisher proxy's \texttt{Event\-Done} function is called
to allow the event's original publisher to release it. This polling is done in configurable intervals, by default in half a seconds intervals. 
Next to the periodic check for recorded events the loop also queries run-control status flags to determine whether
the program should continue running or whether it should terminate. 

\texttt{Get\-GDC\-ID} is a helper function to encapsulate the determining of the GDC ID to be used for an event. 
For tests the function currently implements a round-robin scheme 
based on an event's ID and the number of active GDCs. 

%Since no clear interface
%to determine an event's destination could be found in the DATE source code used as a reference, 

\subsection{The Direct DATE Subscriber}

In the \texttt{Direct\-DATE\-Sub\-scrib\-er} component each event received by the components's subscriber object is directly passed to DATE for 
recording. The \texttt{New\-Event} method implemented by the \texttt{Direct\-DATE\-Sub\-scrib\-er} class directly calls the \texttt{Send\-Event\-To\-DATE} method 
provided by its \texttt{Ali\-HLT\-DATE\-Base\-Sub\-scrib\-er} parent class. Apart from setting up all necessary objects there is not much more functionality contained
in this component.

\subsection{The Triggered DATE Subscriber}

Unlike the previous \texttt{Direct\-DATE\-Sub\-scrib\-er}, the \texttt{Trig\-gered\-DATE\-Sub\-scrib\-er} component does not forward each event to DATE immediately. Instead it uses
an approach similar to the \texttt{Trig\-gered\-Fil\-ter} component from section~\ref{Sec:TriggerFilterComponent}. A new received event is entered into a list, and
only upon receipt of event done data for it a decision is made which parts of the
event are to be passed to DATE. It is possible that an event is not announced at all or only as an empty event with no data. 
Event done data for an event is received from another subscriber via the publisher component. 
The event trigger decision is made using an object of the \texttt{Ali\-HLT\-Trig\-ger\-Event\-Fil\-ter} class, also described in
\ref{Sec:TriggerFilterComponent}. In the  \texttt{Event\-Done\-Data} function implemented by \texttt{Trig\-gered\-DATE\-Sub\-scrib\-er} the filter object's
\texttt{Fil\-ter\-Event\-De\-scrip\-tor} function is called to determine the data blocks to record. Similar to the trigger filter component the triggered DATE subscriber
can also be configured to either forward untriggered events as empty events with no data blocks or to simply release them without invoking DATE.

%%%%%%%%%%%%%%%%%%%%%%%%%%%%%%%%%%%%%%%%%%%%%%%%%%%%%%%%%%%%%%%%%%%%%%%%%%%%%%%%%%%%%%%%%%%%%%%%%%%%%%%%%%%%%%%%%%%%%%%%%%%%%
%%%%%%%%%%%%%%%%%%%%%%%%%%%%%%%%%%%%%%%%%%%%%%%%%%%%%%%%%%%%%%%%%%%%%%%%%%%%%%%%%%%%%%%%%%%%%%%%%%%%%%%%%%%%%%%%%%%%%%%%%%%%%

\chapter{\label{Chap:Glossar}Glossar}

%\begin{description}
%\item[API] Application Programmer's Interface
%\item[BCL] Basic Communication Library - C++ communication class library 
%\item[$dN/d\eta$] ???? Particle space distribution
%\item[$dE/dx$] Specific energy loss
%\item[DMA] Direct Memory Access
%\item[gcc] GNU C Compiler or GNU Compiler Collection
%\item[I/O] Input \& Output
%\item[MLUC] More or Less Useful Class Library - C++ utility class library
%\item[Mutex] Mutual Exclusion Semaphore
%\item[OO] Object Oriented
%\item[OOP] Object Oriented Programming
%\item[PIO] Programmed I/O
%\item[$\mathrm{PbWO}_4$???]
%\item[Pseudo-rapidity]
%\item[PSI] PCI and Shared Memory Interface
%\item[$p_t$] Transversal Momentum
%\item[SAN] System Area Network
%\item[SMP] Symmetric Multi-Processor system - A system with multiple processors (CPUs) accessing the same memory
%\item[kB,~MB] As $2^{10}$ and $2^{20}$??
%\end{description}

%\input{Glossar}
%\clearpage

\begin{description}
\item[ACM] Association for Computing Machinery 
\item[ADC] Analog to Digital Converter
\item[ALICE] A Large Ion Collider Experiment --- Future heavy-ion experiment at CERN's LHC Collider
\item[API] Application Programmer's Interface
\item[ATLAS] A Toroidal LHC ApparatuS --- Future general purpose experiment at CERN's LHC Collider
\item[ATOLL] ATOmically Low Latencies --- A SAN for the PCI bus being developed at the University of Mannheim
\item[BAR] Base Address Register
\item[BCL] Basic Communication Library --- C++ communication class library covered in this thesis 
\item[BLOB] Binary Large OBject
\item[BNL] Brookhaven National Laboratory
\item[C] Procedural programming language well suited to system programming
\item[C++] Programming language based on C with object-oriented extensions
\item[CAMAC] Computer Automated Measurement and Control --- Industry standard instrumentation bus
\item[CBM] Compressed-Baryonic-Matter --- Planned experiment at the future HESR accelerator at GSI
%\item[CDR] Comprehensive Design Report
\item[CERN] European Organisation for Nuclear Research in Geneva, Switzerland
%\item[CHF] Swiss Franc
\item[CMS] Compact Muon Solenoid --- Future general purpose experiment at CERN's LHC Collider
\item[COTS] Commodity-Off-The-Shelf
\item[COW] Cluster Of Workstations
%\item[CPU] Central Processing Unit
\item[CRC] Cyclic Redundancy Check % --- Error che
\item[CSMA/CD] Carrier Sense Multiple Access with Collision Detection --- Ethernet technology for regulating access to physical transmission medium
\item[CSR] Configuration Space Register
\item[DAQ] Data AcQuisition
\item[DARPA] Defense Advanced Research Projects Agency
\item[DATE] ALICE Data Acquisition and Test Environment
\item[DCS] Detector Control System
\item[DDL] Detector Data Link
\item[$dE/dx$] Specific energy loss of charged particles per distance travelled
\item[DMA] Direct Memory Access
\item[$dN/d\eta$] Distribution of particles per unit of pseudo-rapidity
\item[EDM] Event Destination Manager
\item[EG] Event Gatherer
\item[EM] Event Merger
\item[ES] Event Scatterer
%\item[fast-RICH]
\item[FE] Fast Ethernet or Front End
\item[FEE] Front End Electronics
\item[FEP] Front End Processor
\item[FIFO] First-In-First-Out
\item[FMD] Forward Multiplicity Detector --- One of ALICE's detectors
\item[FPGA] Field Programmable Gate Array
\item[FSM] Finite State Machine
\item[FT] Fault Tolerance
\item[gcc] GNU C Compiler or GNU Compiler Collection
\item[GDC] Global Data Concentrator
\item[GNU] GNU's Not Unix --- Project to provide a freely available version of a Unix like operating system
\item[GSI] Gesellschaft f\"ur Schwerionenforschung in Darmstadt, Germany
\item[GbE] Gigabit Ethernet
\item[HADES] High Acceptance Di-Electron Spectrometer --- Detector at GSI
\item[HESR] High Energy Storage Ring --- Future Accelerator at GSI
\item[HI] Heavy-Ion
\item[HL] Hierarchy Level
\item[HLT] High Level Trigger
\item[HMPID] High Momentum Particle IDentification --- One of ALICE's detectors
\item[HPC] High Performance Computing
\item[I/O] Input \& Output
\item[IEEE] Institute of Electrical and Electronics Engineers, Inc.
\item[IETF] Internet Engineering Task Force
\item[IIS-A] Fraunhofer-Institut f\"ur Integrierte Schaltungen
\item[IP] Internet Protocol
\item[ISA] Industry Standard Architecture --- PC extension bus 
\item[ITS] Inner Tracking System  --- One of ALICE's detectors
\item[k] As a prefix usually $10^3$, however in this thesis when used as prefix for bits or bytes or when used for counts of multiples of 2 (e.g. 32 or 128), means $2^{10}$
\item[kB] $2^{10}$ Bytes
\item[L0] Level 0 Trigger
\item[L1] Level 1 Trigger
\item[L2] Level 2 Trigger
\item[L3] Level 3 Trigger
\item[LAM] Look At Me --- An interrupt signal
\item[LAM/MPI] MPI implementation
\item[LDC] Local Data Concentrator
\item[LHC] Large Hadron Collider --- Future accelerator at CERN
\item[LHCb] Large Hadron Collider beauty experiment --- Future experiment dedicated to b-physics at CERN's LHC Collider
\item[LSF] Load Sharing Facility
\item[M] As a prefix usually $10^6$, however in this thesis when used as prefix for bits or bytes or when used for counts of multiples of 2 (e.g. 32 or 128), means $2^{20}$
\item[MB] $2^{20}$ Bytes
\item[MCP] Micro Channel Plate --- A technology for particle detectors
\item[MLUC] More or Less Useful Class Library --- C++ utility class library covered in this thesis
\item[MP3] MPEG Audio Layer 3 --- Compressed audio file format
\item[Moli\`{e}re radius] Material characteristic used to describe the transversal dimension of electromagnetic particle showers
\item[MPEG] Motion Picture Experts Group
\item[MPI] Message Passing Interface -- Standard for parallel program communication
\item[MPICH] MPI implementation
%\item[MSGC] Micro Strip Gas Chamber --- A technology for particle detectors
\item[MTU] Maximum Transmission Unit
\item[Mutex] Mutual Exclusion Semaphore
\item[MWPC] Multi Wire Proportional Chamber --- A technology for particle detectors
%\item[NASA] 
\item[NOW] Network Of Workstations
\item[Ogg Vorbis] Compressed audio file format
\item[OO] Object Oriented
\item[OOP] Object Oriented Programming
\item[PANDA] Proton-ANtiproton-at-DArmstadt --- Planned experiment at the future HESR accelerator at GSI
\item[PBH] Publisher Bridge Head
%\item[$\mathrm{PbWO}_4$???]
\item[PC133] Specification for SDRAM modules with 133~MHz clock frequency
\item[PCI] Peripheral Component Interconnect --- PC extension bus
%\item[PCI-X] 
\item[PCISIG] Peripheral Component Interconnect Special Interest Group --- PCI standardization body
\item[PDS] Permanent Data Storage
\item[PHOS] PHOton Spectrometer --- One of ALICE's detectors
\item[PID] Particle IDentification
\item[PIO] Programmed I/O
\item[PM] Patch Merger
\item[PMD] Photon Multiplicity Detector  --- One of ALICE's detectors
%\item[POSIX]
\item[PPC] Parallel Plate Counters --- A technology for particle detectors
\item[Pseudo-rapidity] Variable for particles in a collision. Defined as $\eta=-\ln \tan(\theta/2)$, where $\theta$ is the angle between the particle and
the direction of the undeflected beam. Approximates the relativistic rapidity of a particle.
\item[PSI] PCI and Shared memory Interface --- Driver and library for PCI hardware and shared memory access covered in this thesis
\item[$p_t$] Transversal Momentum
\item[PVM] Parallel Virtual Machine --- Library for parallel program communication
\item[QGP] Quark-Gluon Plasma
%\item[RAM] 
\item[RFC] Request For Comment --- Informal Internet standard
\item[RHIC] Relativistic Heavy Ion Collider --- Accelerator at BNL
\item[RICH] Ring Image \v{C}erenkov (or Cherenkov) Counter --- A technology for particle detectors
\item[RORC] Read Out and Receiver Card
\item[RPC] Resistive Plate Chamber --- A technology for particle detectors (In computing also {\em Remote Procedure Call}, but not used as such in this thesis)
\item[SAN] System Area Network
\item[SBH] Subscriber Bridge Head
\item[SCI] Scalable Coherent Interface --- SAN technology
\item[SDD] Silicon Drift Detector --- A technology for particle detectors
%\item[SDRAM] 
\item[SI95] SpecInt95 --- Unit to measure computing speed
\item[SISCI] Software Infrastructure for SCI --- SCI programming API
\item[SM] Slice Merger
\item[SMP] Symmetric Multi-Processor system - A system with multiple processors (CPUs) accessing the same memory
\item[SPD] Silicon Pixel Detector --- A technology for particle detectors
\item[SPS] Super Proton Synchrotron --- Accelerator at CERN
\item[SSD] Silicon Strip Detector --- A technology for particle detectors
\item[SSI] Single System Image
\item[SSIC] Single System Image Cluster
\item[STAR] Solenoidal Tracker at RHIC --- Detector at the RHIC accelerator at BNL
\item[STL] Standard Template Library
\item[ShM] Shared Memory
\item[Si] Silicon
%\item[$T_0$]
\item[TCP] Transmission Control Protocol
%\item[TCP/IP]
\item[TDR] Technical Design Report
\item[TOF] Time Of Flight --- One of ALICE's detectors
\item[TPC] Time Projection Chamber --- A technology for particle detectors as well as one of ALICE's detectors
\item[TRD] Transition Radiation Detector --- A technology for particle detectors as well as one of ALICE's detectors
\item[UML] Unified Modelling Language
%\item[URL] Uniform
\item[VITA] VMEbus International Trade Association
\item[VMEbus] VERSAmodule Eurocard extension bus --- Industry standard instrumentation bus
\item[WAN] Wide Area Network
\item[ZDC] Zero Degree Calorimeter --- One of ALICE's detectors
\item[ZN] Zero degree Neutron calorimeters
\item[ZP] Zero degree Proton calorimeters
\end{description}

%%%%%%%%%%%%%%%%%%%%%%%%%%%%%%%%%%%%%%%%%%%%%%%%%%%%%%%%%%%%%%%%%%%%%%%%%%%%%%%%%%%%%%%%%%%%%%%%%%%%%%%%%%%%%%%%%%%%%%%%%%%%%
%%%%%%%%%%%%%%%%%%%%%%%%%%%%%%%%%%%%%%%%%%%%%%%%%%%%%%%%%%%%%%%%%%%%%%%%%%%%%%%%%%%%%%%%%%%%%%%%%%%%%%%%%%%%%%%%%%%%%%%%%%%%%

\end{appendix}
\clearpage

\end{document}